\documentclass[fleqn,usenatbib]{mnras}

\usepackage{newtxtext,newtxmath}
\usepackage[T1]{fontenc}

\usepackage{graphicx}
\usepackage{float}
\usepackage{longtable}
\usepackage{wrapfig}
\usepackage{amsmath}

\usepackage{bm}
\usepackage[font=small,labelfont=bf]{caption}
\usepackage{citesort}
\usepackage{threeparttable}
\usepackage{comment}
\usepackage[dvipsnames]{xcolor}
\usepackage{pdflscape}
\usepackage{adjustbox}
\usepackage{multicol}

\def\apj{Astroph. J.}
\def\apjl{Astroph. J. Lett.}
\def\apjs{Astroph. J. Suppl.}
\def\mnras{MNRAS}
\def\nat{Nature}
\def\aap{Astron. Astroph.}

\def\aj{Astron. J.}
\def\pasp{Pub. Astron. Soc. Pacific}

\newcommand{\HI}{H\,{\sc i}}

\newcommand{\CI}{C\,{\sc i}}

\newcommand{\ZnII}{Zn\,{\sc ii}}
\newcommand{\SII}{S\,{\sc ii}}
\newcommand{\PII}{P\,{\sc ii}}

\newcommand{\DK}[1]{{\color[rgb]{0.19,0.55,0.91}DK: #1}}

\title[HD in Magellanic Clouds]{Cold diffuse interstellar medium of Magellanic Clouds: I. HD {\bf molecule} and {\bf cosmic-ray} ionization rate}
\author[D. N.~Kosenko and S. A.~Balashev]{D. N.~Kosenko,\thanks{E-mail: kosenkodn@yandex.ru}
S. A.~Balashev,\thanks{E-mail: s.balashev@gmail.com}
\\
Ioffe Institute, 26 Politeknicheskaya st., St.\ Petersburg, 194021, Russia
}

\date{Accepted XXX. Received YYY; in original form ZZZ}

\pubyear{2023}

\begin{document}

\label{firstpage}
\pagerange{\pageref{firstpage}--\pageref{lastpage}}
\maketitle

\begin{abstract}
HD molecule is one the most abundant molecules in the Universe and due to its sensibility to the conditions in the medium, it can be used to constrain physical parameters in the medium where HD resides. Lately we have shown that HD abundance can be enhanced in the low metallicity medium. Large and Small Magellanic Clouds give us an opportunity to study low metallicity galaxies in details towards different sightlines due to their proximity to our Galaxy. We revisited FUSE space telescope archival spectra towards bright stars in Magellanic Clouds to search for HD molecules, associated with the medium of these galaxies. We reanalysed H$_2$ absorption lines and constrained HD column density at the positions of H$_2$ components. We detected HD towards 24 sightlines (including 19 new detections). We try to measure cosmic ray ionization rate for several systems using measured $N({\rm HD})/N({\rm H_2})$, and  in most cases get loose constraints due to insufficient quality of the FUSE spectra. 
\end{abstract}

\begin{keywords}
galaxies: ISM; ISM: molecules; cosmic rays
\end{keywords}

\section{Introduction}

The physical state of interstellar medium (ISM) is in the close relationship with the galaxy evolution, since it determines the star-formation, a key process of galaxy formation. On the other hand, stars influence environment around, by enrichment of the ISM with metals and dust, supply of the ultraviolet (UV) radiation and cosmic rays, energy and momentum injection etc. These change rates of reactions in the ISM, and hence impact on the abundances and level populations of the species. Therefore spectroscopic analysis of latter gives us opportunity to constrain the physical conditions in the observed systems, using comparison with the models.
%Interstellar medium is a crucially important constituent of any galaxy which influences many processes and eventually defines galaxy evolution.  %Molecular hydrogen, H$_2$ is the most abundant molecule in the Universe and it is used as a tracer of a cold diffuse molecular gas which is an intermediate step of the gas on its way to dense molecular cloud where start-formation occurs. Therefore properties of interstellar clouds define properties of the stars. 

%Deuterated hydrogen, HD, is an isotopologue of H$_2$. 

Star-formation occurs in the molecular gas, which predominantly consists of H$_2$, which is hard to observe in the emission in cold ISM. Hence usually other tracers of H$_2$ are used, out of which CO is most abundant ($n({\rm CO})/n({\rm H_2})\lesssim 10^{-4}$, where $n({\rm X})$ -- is a number density of species) and exhaustively studied. However, H$_2$ can be easily observed in absorption using UV resonant transition \citep[see recent note by ][]{Shull2022}. Unlike CO, H$_2$ in UV band predominantly probes diffuse molecular and translucent medium, due to saturation effects and associated extinction. However, such gas is certainly in intimate connection with the dense molecular gas, either providing its supply, or being the remnants of the intermittent molecular clouds.

Similar to H$_2$, HD -- an isotopologue of H$_2$ with abundance $n({\rm HD})/n({\rm H_2})\sim 10^{-5}$, can be detected in the absorption using resonant transitions in UV band in our Galaxy \citep[e.g.][]{Snow2008} and in optical band at high redshifts \citep[e.g.][and references therein]{Varshalovich2001,Tumlinson2010,Ivanchik2015,Noterdaeme2017,Kosenko2021}. The importance of HD for ISM studies has been discussed starting from its detection \citep[e.g.][]{Black1973,Hartquist1978}, since HD provides an opportunity to constrain physical conditions in the medium, associated with absorption systems. Indeed, at the moderate presence of H$_2$, HD molecules form through a fast ion-molecular reaction\footnote{unlike H$_2$, which forms in the cold ISM mainly on the surface of dust grains}:
\begin{equation}
    {\rm H_2 + D^{+}\longrightarrow HD + H^{+}},
    \label{eq:HD_form}
\end{equation}
and destructed by photodissociation. This makes relative abundance of HD/H$_2$ sensitive to 
the main physical conditions in the medium, namely, number density, $n$, UV field intensity, $\chi$, abundance of metals (and their depletion), and cosmic ray ionisation rate (CRIR), $\zeta$. Regarding the latter, cosmic rays determine H$^+$ formation, which defines the ionized deuterium, D$^+$ abundance through charge exchange reaction, and therefore CRIR directly affects HD production rate.

Using a balance equation between HD formation (adding formation on the dust grains, which may be sufficient on the edge of the cloud, where there H$_2$ abundance is small $n({\rm H_2})/n_{\rm tot}\lesssim 10^{-4}$, and rate of reaction~\ref{eq:HD_form} is negligible) and destruction by UV radiation (taking into account self-shielding) we have recently obtained a simple semi-analytical formalism for dependence of HD/H$_2$ abundance on the aforementioned parameters \citep{Balashev2020} which allows us to constrain the physical conditions in the medium. 
One of the most important results of this recent study is that measured HD/H$_2$ ratios allow us to get constraints on CRIR in diffuse molecular medium of high-redshift galaxies.\footnote{Before that CRIR have been exhaustively explored in different environments, from diffuse neutral medium to dense prestellar clouds, in our Galaxy \citep[e.g.][]{Padovani2009, Indriolo2012b}, but was constrained only for several sightlines in other galaxies, mostly representing the dense molecular gas \citep[e.g.][]{Muller2016, Shaw2016, Indriolo2018}.}, therefore we have applied our method to a sample of all known HD/H$_2$ absorption systems at high redshifts \citep{Kosenko2021}.
More intriguing, we also showed that HD abundance is really enhanced in the low-metallicity medium  probed by high-z Damped Lyman alpha (DLA) systems in comparison with high metallicities \citep{Balashev2020,Kosenko2021}, as a consequence of enhanced ionization fractions. This behaviour of HD at low metallicity can be tested on the local dwarf galaxies, out of which the Large and Small Magellanic Clouds (LMC and SMC, respectively), seem to be the most appropriate, since they provide individually resolved stars due to their proximity to the Milky Way. Additionally, studies of diffuse gas with HD/H$_2$ provide us with valuable constraints on the physical conditions, which are known to differ much from our Galaxy.

LMC and SMC are the closest dwarf galaxies to the Milky Way. Distance to LMC is about 50 kpc \citep{Pietrzynski2019} and it is a spiral galaxy with bar and bright dominant north spiral arm and a weak south arm. Average metallicity in the LMC is about 0.5 relative to solar \citep{Russell1992}. 
Most of the star-formation activity occurs in the bar and in the dominant arm, e.g. one of the most studied star-forming complex, 30 Dor, is located in the northeast of the bar. Distance to the SMC is about 62 kpc \citep{Graczyk2020} and it is an irregular galaxy and can be divided to a bar (most dense, turbulent and UV exposed region of the SMC) and more quiescent and less dense wing on the east side. Metallicity of the SMC is about 0.2 relative to solar \citep{Russell1992} and it is comparable with the average metallicity in the high redshift DLAs. From the east of the wing of SMC to the west side of the LMC there has been discovered Magellanic Bridge, a stream of neutral hydrogen with several stars \citep{Hindman1963}, which, as believed, was created due to the interaction between LMC and SMC about 200 Myr ago. Abundances in the Magellanic Bridge differ significantly from the abundances in both LMC and SMC and may be lower even then in the SMC \citep[e.g.][]{Lehner2008, Dufton2008, Lee2005, Ramachandran2021}. 

Proximity of Magellanic Clouds to our Galaxy gives us a perfect opportunity to study low metallicity galaxies in detail and to observe systems along different lines of sight within one galaxy.
As we mentioned above, H$_2$ and HD are available in local Universe only using UV space telescopes.
Fortunately, Far Ultraviolet Spectroscopic Explorer (FUSE) have been successfully used to study cold diffuse medium with HD and H$_2$ in the Milky Way \citep{Snow2008, Shull2021} and with H$_2$ in Magellanic Clouds \citep{Welty2012}.

The main purpose of the work is to search for HD molecules in Magellanic Clouds using archival data obtained by FUSE telescope, to constrain HD/H$_2$ relative abundances in this satellite galaxies. %We refit H$_2$ lines to get self-consistent results and obtain rotation level population. For some systems we also used Hubble space telescope archival data to estimate \CI\ fine-structure level population and metallicity. 
Using measured HD and H$_2$ column density %and \CI\ column densities and level populations for a sample of the sightlines we constrained the number density, UV field intensity, and CRIR.
we tried to constrain cosmic ray ionization rates in the analysed systems. 
To our knowledge there was no attempt made to estimate CRIR in the Magellanic Clouds using the abundance matching. However, some constraints of cosmic-ray flux in LMC and SMC were obtained using gamma-ray radiation, that found average cosmic-ray density in both LMC and SMC lower then in Solar neighborhood \citep[see e.g.][]{Ackermann2016, Lopez2018}. Nevertheless, one should bear in mind, that only cosmic rays with energies $>280$\,MeV contribute into gamma-ray flux due to pion decay, while ISM is ionized mostly by low-energy cosmic rays $\lesssim 100$ MeV.  Also low-energy cosmic rays may be trapped in the region where they were accelerated and their flux (and therefore CRIR) depends on the distance to the star-forming regions and supernova remnants.
This paper is accompanied by another one (Kosenko in prep.), which is focused on the constraints on the number density and UV flux in LMC and SMC using the population of the rotational levels of H$_2$ (measured and presented in this paper) and \CI\ fine-structure levels. 

The paper is organized as follows: in Section~\ref{sect:Data} we describe a sample of data which was used in the work, in Section~\ref{sect:analysis} we provide a description of the method to analyse observations, including line profile fitting. The results of data analysis, including new HD detections, are presented in Section~\ref{sect:results}. In the Section~\ref{sect:phys_properties} we the describe constraint on the cosmic ray ionization rate obtained from measured HD/H$_2$ ratio. Section~\ref{sect:discussion} we discuss how high resolution observations may improve obtained constraints, before summarizing the obtained results in Section~\ref{sect:summary}.

\section{Data}
\label{sect:Data}

HD and H$_2$ absorption lines fall into UV part of electromagnetic spectrum $(\lambda_{\rm H_2} \rm \lesssim 110\,nm)$, therefore observations of these molecules are limited due to the atmospheric absorption. The progress of H$_2$ observations in the Milky Way and the local Universe was certainly attributed to availability of the space UV telescopes. The best resolution and quality observations were performed by Far Ultraviolet Spectroscopic Explorer (FUSE, \citealt{Moos2000, Sahnow2000}), which covers 907-1187\,\AA\, and have nominal spectral resolution $R \sim 20000$. 

%\subsection{FUSE data}
\citealt{Blair2009} collected all of the FUSE observations in Magellanic Clouds and reprocessed them with the final calibration pipeline (CalFUSE 3.2) to get uniform data (FUSE Magellanic Clouds Legacy Project\footnote{\url{https://archive.stsci.edu/prepds/fuse_mc/}}). Using these data \citealt{Welty2012} have found H$_2$ in 80 sightlines in LMC and in 65 sightlines in SMC. We selected those systems where $\log N_{\rm H_2}\gtrsim 18$, since the quality of FUSE spectra does not allow detection of $\log N_{\rm HD} \lesssim 13$ (typically measured relative abundance $N({\rm HD})/N({\rm H_2})\lesssim 10^{-5}$). This cut the sample to 48 sightlines in the LMC and 46 sightlines in the SMC. 

The fully reduced spectra of the systems towards bright stars in LMC are available in the FUSE Magellanic Clouds Legacy Project \citep{Blair2009}. Typical observation of a star consists of the several exposures, obtained in different FUSE channels. We used spectra from 1A LiF channel, as it covers most of HD and H$_2$ absorption lines and has relatively high sensitivity \citep{Dixon2007}.
Unfortunately, most of FUSE spectra suffer from inappropriate calibration issues, so we made an attempt to improve the quality of calibration. %before coadding the exposures, which may improve the quality of the spectrum. 
First, we find a zero-order spectrum, which is obtained by coadding of exposures, using the constant velocity shifts, that was found from the procedure of cross-correlation \citep{Simkin1974, Tonry1979}. 
From the zero-order spectrum we obtained rough estimates on H$_2$ column densities, Doppler parameters and redshifts. Then, using H$_2$ synthetic spectrum, constructed from these estimates, as a template, we obtained a wavelength dependent shifts for each exposure. We took only narrow, unblended lines and cross-correlated them with the template. The wavelength dependency of the estimated shifts was interpolated with piece-wise linear function, that was finally applied to correct the exposures, during coadding.

\section{Analysis}
\label{sect:analysis}

We model line profiles with standard multicomponent Voigt profile fitting using Monte Carlo Markov Chain (MCMC) sampler to obtain posterior distribution function of the absorption system parameters: column densities ($N$, measured throughout the paper in cm$^{-2}$), Doppler parameters, ($b$) and redshifts ($z$). 
The continuum was independently constructed for each considered fit (H$_2$, HD, etc) by B-splain interpolation of the neighboring regions of absorption lines, clear from the evident absorptions. We note that in the case of the saturated H$_2$ lines, which show Lorentzian wings this is a bit ambiguous task, therefore we iteratively adjusted continuum in the case of outlying lines. To construct the synthetic spectrum we used HD and H$_2$ transition lines and molecular data, collected by \citealt{Ubachs2019}. %\DK{the wavelengths are the the same that we used, but oscillator strengths and gamma a bit differ} \SB{I do not know why there is inconsistency for some oscillator strength. My database is a compilation have not references to data inside. May be I need to update the data, but for lines that we used it is consistent.} \DK{OK, then it is not a problem}

To estimate parameters and its uncertainties we used maximum aposteriori probability and $68.3\%$ credible intervals, respectively. We judge between detection and non detection of HD based on the constrained column density posterior probability function. For the cases where it shows a solitary peak we reported a measurement, while if there is a significant amount of the posterior near the lower end of column density range (taken to be $\log N(\rm HD)=13$) we suggest an upper limit. Upper limits on column densities (where necessary) were found using one-sided 3$\sigma$ (99.7$\%$) credible interval. We note that reported in our study uncertainties are statistical ones, derived in particular model assumptions. Systematic uncertainties, arisen from continuum placement, particular choice of fitting spectral pixels, and component decomposition in some cases may dominate the statistical ones, therefore the latter should be taken with caution, especially in some systems, where it is found relatively small (in comparison to other systems). From the other side, since we used MCMC sampler to constrain the posterior distribution, we were able to explore a multimodal structure of the likelihood function and hence in some cases our constraints on column density are quite wide, since both non-saturated and saturated solutions provide the similar quality of the fit. 

%\subsection{FUSE data}
Firstly we refitted H$_2$ absorption lines, that typically corresponds to several lower rotational levels ($J\lesssim5$) of the lowest vibrational state. %\SB{Check next sentence!}
In comparison with previous studies, where either only curve of growth analyses was used, or profiles were fitted only for the lower, $J=0,1$ levels, we performed the full joint profile fitting of all available levels. We used the same redshifts for all H$_2$ rotational levels, but allowed Doppler parameters to vary independently (except upper levels in several systems where the values were found to be unreasonably large). Additionally, we used the penalty functions \citep[for details see, e.g.][]{Noterdaeme2019, Kosenko2021}  to favour a regular smooth shape of H$_2$ excitation diagram and an effect of increase of Doppler parameters with an increase of J, as was noticed before \citealt[e.g.][]{Balashev2009, Noterdaeme2007}. 

The nominal resolution of spectra, obtained by FUSE is $R = \lambda/\Delta\lambda =20000$, but we noticed that it can be reduced by a procedure of exposure coadding, calibration and other systematic effects of observations. So we made $R$ to be an additional independent parameter during the fitting procedure. In the most cases, we found $R$ below 20000, in the range $\sim 11000 - 18000$. 

Using obtained velocity decomposition from H$_2$ profiles, we constrained the column density of HD molecules at the position of H$_2$ components. To fit HD column density we used priors on Doppler parameters from H$_2$ $\rm J=0$ (except systems where we detected HD) and fixed redshifts. One should note that many HD lines are blended, mostly by H$_2$, but also by metals; moreover some blends are attributed to the stellar absorptions.

\begin{comment}
\subsection{HST data}

 Using HST spectra we fit \CI\ absorption lines from three fine-structure levels (denoted below as \CI, \CI$^{*}$ and \CI$^{**}$) with an assumption of tied Doppler parameters between them. This is reasonable approximation, since \CI\ levels are populated by collisions and hence typically cospatial within the cloud. 
 
 An important ingredient of the physical conditions determination is metallicity. To obtain metallicity in the systems we used \ZnII\ or (if possible) \PII\ absorption lines, which are believed to be weakly depleted \citep{DeCia2018}, otherwise, if there were no \ZnII\ nor \PII, we used \SII\ absorption lines. Using constrained column densities of metals and $N_{\rm HI}$ obtained by \citealt{Welty2012, Roman_Duval2019}, and the Solar values from \citealt{Asplund2009} we estimated metallicity as, $Z \equiv [{\rm X/H}] \equiv \log\left({\rm X/H}\right) - \log\left({\rm X/H}\right)_{\odot}$. %\SB{Check this equation, that it is was you used! Since it was not properly written.} \DK{corrected}
\end{comment}

 \section{Results}
 \label{sect:results}

\begin{figure*}
    \centering
    \includegraphics[width=1.0\textwidth]{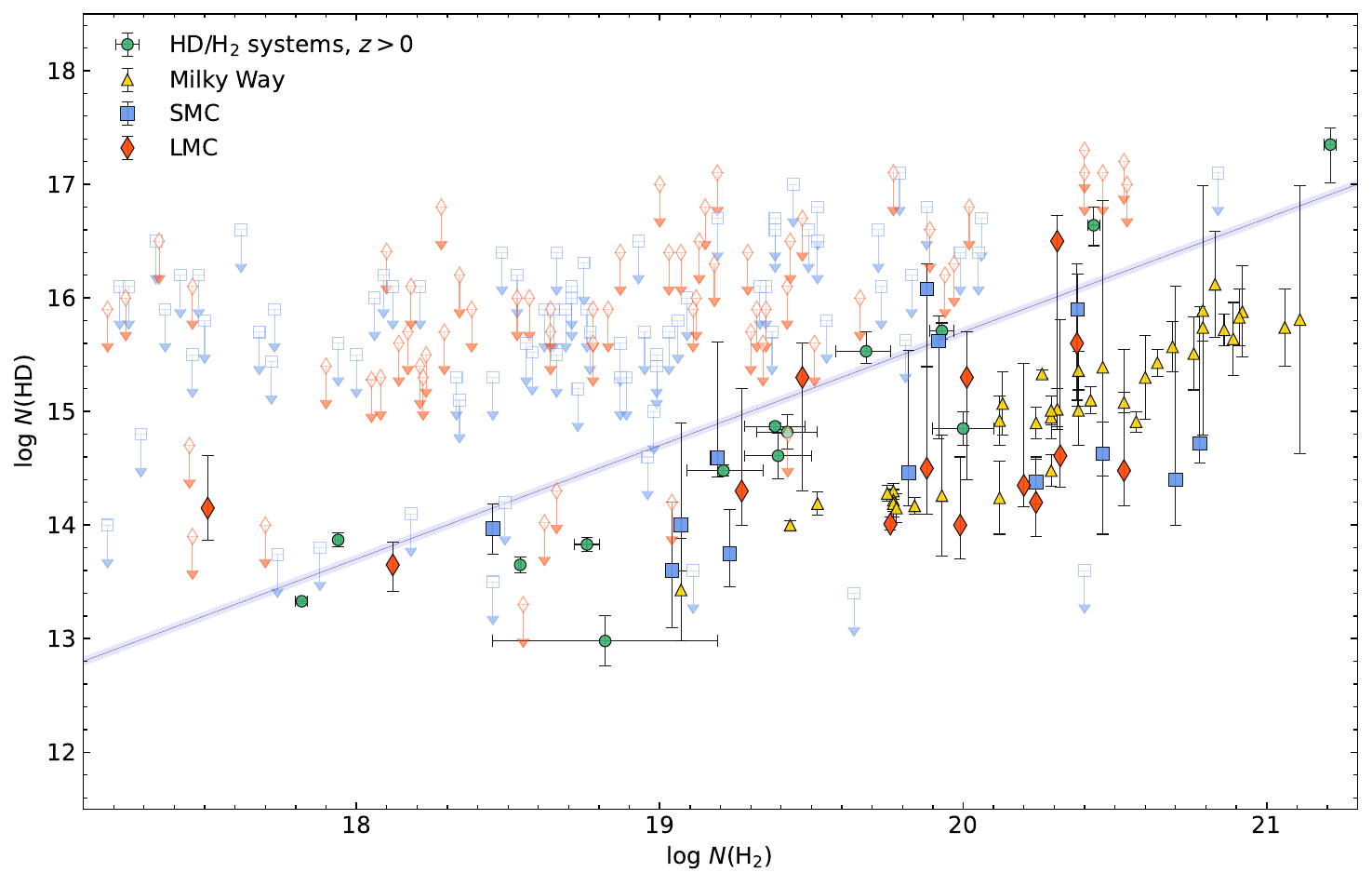}
    \caption{ The HD versus H$_2$ column densities. The measurements in Small and Large Magellanic Clouds (this work) are shown by filled blue squares and red diamonds, respectively. Unfilled symbols represent the 3$\sigma$ upper limits on HD column densities. Green circles show the measurements towards quasars at high redshifts \citep[see][and references therein]{Kosenko2021}, yellow triangles indicate the measurements in our Galaxy \citep{Snow2008} and solid the blue line and stripe show the primordial D/H ratio \citep{Planck2018}.
    }
    \label{fig:H2_HD}
\end{figure*}

In the spectra of stars towards Magellanic Clouds there are typically two groups of the components: one from Milky Way disk and/or halo and another from LMC or SMC. In most of the cases each group can be resolved into few sub-components, that was fitted with profile fitting. In total, HD was detected towards 24 sightlines (among which in 19 there were new detections), 
%Unfortunately, in most of the systems  \SB{It is always good to be positive!} 
and we placed upper limits on HD column densities in all others. The total H$_2$ and HD column densities derived from line profiles are shown in Tables~\ref{tab:HD_H2_LMC} and \ref{tab:HD_H2_SMC} for LMC and SMC, respectively.  Details on analysis of each systems are shown in Appendix~\ref{appendix:fitting} (see Supplementary materials).
We show an example of line profile fitting of HD lines for the system towards Sk 191 in the SMC in the Figure~\ref{fig:lines_HD_Sk191}; in the Appendix~\ref{sect:New_HD} we show all of the new HD detections. We also show an example of non-detection of HD lines in the Figure~\ref{fig:lines_HD_AV14}\, in the system towards AV 14 in the SMC.
%Also we show examples of line profile fitting of HD lines for several systems in the Figures~\ref{fig:lines_HD_J0534}-\ref{fig:lines_HD_Sk191}.
In Figure~\ref{fig:H2_HD} we show measured HD and H$_2$ column densities in the components and compared it with known measurement obtained in Milky Way and at high-redshifts. We note that since the resolution and spectral quality is not enough we predominantly obtained conservative limits on HD at the level $\log N({\rm HD}) \lesssim 15-16$, which we also show in Figure~\ref{fig:H2_HD}.

We found a wide ($\approx 2$ dex) range of measured HD/H$_2$ ratios in Magellanic Clouds from the values below the Milky-Way measurements, to the values higher that the isotopic D/H ratio. However, at the intermediate H$_2$ column densities, HD abundance is very sensitive to the physical conditions \citep[see][]{Balashev2020}, due to the sensitivity of the position of D/HD transition, and therefore is not a problem to explain such wide range. We additionally, note that there is an obvious selection effect in Fig.~\ref{fig:H2_HD}, that is we were limited with $\log N({\rm HD}) \gtrsim 13.5$. 

%\DK{added comparison of our results with Welty2012 below}

\begin{figure}
    \centering
    \includegraphics[width=\linewidth]{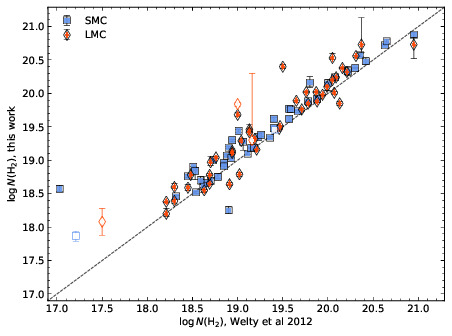}    \caption{Comparison between the total H$_2$ column density estimates obtained by \citealt{Welty2012} 
    (x-axis) and our results (y-axis). 
    The blue squares and red diamonds show measurements in SMC and LMC, respectively. 
    Open symbols show systems with low Doppler parameters (for details see text). 
    Dashed black line shows the one-to-one ratio. 
    }
    \label{fig:H2_comparison}
\end{figure}

In Figure~\ref{fig:H2_comparison} we present comparison of the total H$_2$ column density obtained by \citealt{Welty2012} with the our results, obtained using the sum over all components associated with LMC and SMC.
%(blue squares show SMC and red diamonds show LMC). 
One can see that $N({\rm H_2})$ values that we obtained are systematically higher ($\sim 0.3$ dex) then that obtained by \citealt{Welty2012}. It may arise, firstly, from differences in methods: for most of the systems \citealt{Welty2012} fitted only J = 0 and 1 lines and for only about 40 systems in their sample they provided detailed fit assuming multiple components. This may lead to underestimation of their results. Secondly, the discrepancy may arise from difference and difficulties in the calibration of data and uncertainty in resolution of combined spectra after reduction. Thirdly, as was noted above, in our analysis we used MCMC method, which in some cases provide the results with relatively low Doppler parameters and therefore may overestimate column densities (in the Tables~\ref{tab:HD_H2_LMC} and \ref{tab:HD_H2_SMC} such systems are denoted by item $e$) in the case of the low column densities $\log N \lesssim 19$, at which Lorentz wings are not yet evident. Such low ($b\gtrsim 1\rm\,km\,s^{-1}$) Doppler parameters are not surprising for cold clouds, and sometimes confirmed in high-resolution observations \citealt[e.g.][]{Carswell2011}. The Doppler parameters, under FUSE resolution (corresponding $b\approx20\rm\,km\,s^{-1}$) mainly estimated from joint fit of lines with different oscillator strengths. However, since the FUSE spectrum has limited spectral range and number of available H$_2$ bands is typically limited to few. Therefore, high-resolution observations with larger wavelength range are needed to confirm a realm of such low Doppler parameter values. And the last, \citealt{Welty2012} have not provided uncertainties on their column densities and we do not know real inconsistency between our results.

\begin{figure*}
    \centering
    \includegraphics[width=\linewidth]{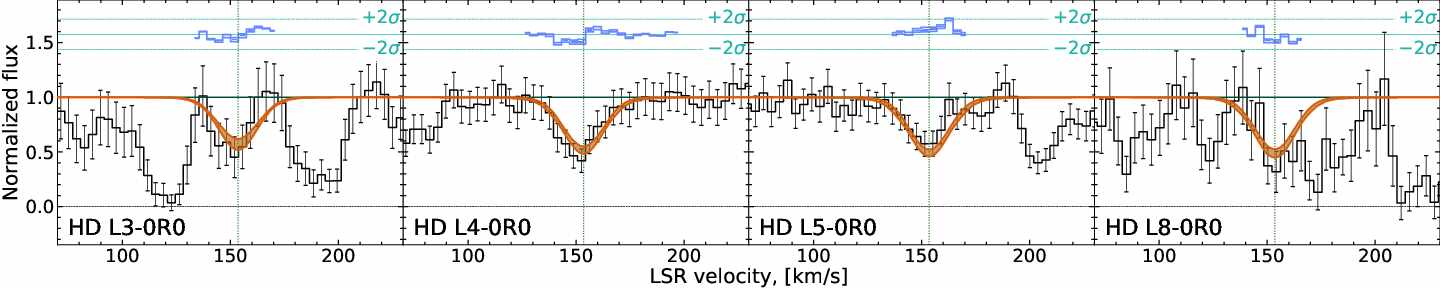}
    \caption{Fit to HD absorption lines towards Sk 191 in SMC.
    Here black line shows spectrum, while the coloured stripes show 0.683 credible interval of the line profiles sampled from posterior probability distributions of fitting parameters. The red represents the total line profile, while the green ones indicate individual components. In this particular spectrum, the green and red coincides, since there is only one component.
    %    red line represents obtained synthetic spectrum for HD. {\bf Green lines represent individual components. In this particular spectrum, the green } {\bf Regions} between red {\bf and green} lines {\bf were} sampled from posterior probability distributions of fitting parameters. 
    Blue points at the top of each panel show residuals. Here we show only components found in Magellanic Clouds.    
    }
    \label{fig:lines_HD_Sk191}
\end{figure*}

\begin{figure*}
    \centering
    \includegraphics[width=\linewidth]{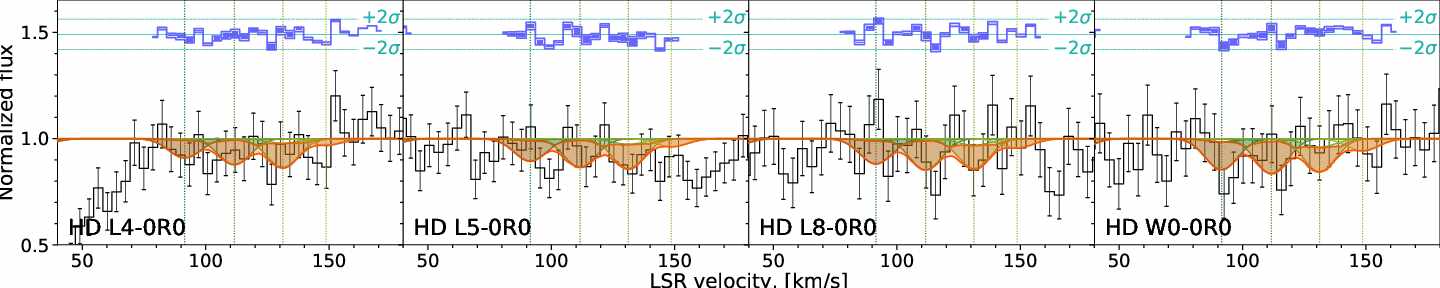}
    \caption{Fit to HD absorption lines towards AV 14 in SMC, which shows an example of non-detection of HD molecule. Lines are the same as for \ref{fig:lines_HD_Sk191}.}
    \label{fig:lines_HD_AV14}
\end{figure*}

%\CI\ was detected in the most of the sightlines, where spectra were available, and in the most cases we were able to measure the excitation of the fine-structure levels, which were used below to constrain the physical conditions. The results of \CI\ lines analysis are given in Tables~\ref{tab:CI_LMC} and \ref{tab:CI_SMC} for LMC and SMC, respectively. 

%We found an average metallicity in our sample to be $Z\sim 0.1Z_{\odot}$ and $Z\sim 0.3Z_{\odot}$ for SMC and LMC, respectively. These values are slightly lower then reported in literature ($0.2Z_{\odot}$ for SMC and $0.5Z_{\odot}$ for LMC, \citealt{Russell1992}), which can be connected with the depletion of metals and usage of the different metals, or some systematic effects in our sample. Meanwhile, recent studies, indicate, that even in Solar vicinity in Milky-Way the value of metallicity is a debatable question \citep{DeCia2021,Ritchey2023}.

Below we provide some details regarding several interesting sightlines.

\subsection{Sk-67 5}

The star is located in diffuse H\,{\sc ii} region on the Western edge of LMC
and the system have previously been studied by \citealt{Andre2004}, who used data, obtained by FUSE, HST/STIS and VLT/UVES.%, towards Sk-67 5 was also found high velocity component (high velocity cloud, HVC) from Na\,{\sc ii} and Ca\,{\sc ii} \citep{Andre2004} and Fe\,{\sc ii} \citep{Lehner2009}.\citealt{Andre2004} found high value of Cl\,{\sc i}/H$_2$ and CO and concluded the presence of dense molecular clumps towards Sk-67 5. 
They found $\log N({\rm H_2}) = 19.44\pm 0.05$ which is in agreement with our value $\log N({\rm H_2}) = 19.47\pm 0.01$. Their value of HD column density is $13.62^{+0.09}_{-0.12}$, while we obtained a wider range $15.5^{+0.3}_{-2.1}$, which includes saturated solution for the fit (see Sect.~\ref{sect:analysis}). Indeed, to obtain HD column density \citealt{Andre2004} used an average value between Apparent Optical Depth (AOD, \citealt{Savage1991}) method and line profile fitting, however it is known that AOD is not appropriate for lines with large optical depth and/or low resolution (above it was noted than resolution of FUSE spectra may be degraded after data reduction), therefore HD column density given by \citealt{Andre2004} may be underestimated.

\subsection{Sk-67 105}

Sk-67 105 is one of the most massive eclipsing binary system \citep{Niemela1986} in the LMC, which probably reveals a contact configuration with 48.3$M_{\odot}$ and 31.4$M_{\odot}$ \citep{Ostrov2003}.

From H$_2$ lines (mostly from saturated $\rm J = 0$ and 1) we found an evidence of partial coverage of absorption system towards Sk-67 105 which may arise from the configuration of the background radiation source. To get more reliable results we added a covering factor \citep[see e.g.][]{Balashev2011} as an independent parameter during the fit to H$_2$ lines, which value was found to be $ 0.888^{+0.002}_{-0.002}$.
We could not detect HD in this system and placed upper limit on the column density in both H$_2$ components to be $\log N({\rm HD}) \lesssim 14.8$ and $\lesssim 13.9$.

\subsection{Sk-68 135}

The star is located in the north of 30 Dor complex, one of the most studied star-forming regions and was previously studied by \citealt{Andre2004}. %\citealt{Andre2004} found HVC towards Sk-68 135 with large amount of C\,{\sc ii} relative to Na\,{\sc i}. \citealt{Lehner2009} also found high amount of Fe\,{\sc ii} in this HVC, which probably indicates a presence of a shock wave. \citealt{Andre2004} report a large radiation field in this system (several thousand times over a mean Galactic value). Therefore they suggested that the envelope of molecular cloud should be destroyed by radiation from the star or swept away by the shock wave. 

Our HD column density estimate, $\log N({\rm HD}) = 14.0^{+0.6}_{-0.3}$ is consistent with the result obtained by \citealt{Andre2004}, $\log N({\rm HD}) = 14.15^{+0.11}_{-0.15}$, while we found H$_2$ column density to be a bit higher (our $\log N(\rm H_2) = 19.99\pm 0.01$ versus their $19.87\pm 0.05$; note that \citealt{Andre2004} fitted H$_2$ assuming a single-component model, while we fitted H$_2$ with 2-components model). \citealt{Roman_Duval2019} found \HI\ column density in this system to be significantly lower then \citealt{Welty2012}.   \citealt{Roman_Duval2019} used Ly$\alpha$ from HST, while \citealt{Welty2012} used Ly$\beta$ line from FUSE spectra, where the difficulties in \HI\ column density estimation partly arises from large amount of H$_2$ absorption lines in the wings of Ly$\beta$. Additionally \citealt{Welty2012} did not provide uncertainties of the $N(\rm HI)$ measurements. Analysis of Ly$\alpha$ provides more reliable result, therefore we took value from \citealt{Roman_Duval2019}. 

\subsection{Sk-69 246}

This star belongs to 30 Dor complex and the system on the line-of-sight have previously been studied by \citealt{Bluhm2001} and \citealt{Andre2004}. %Interestingly, that CO molecules were found in this system. 

\citealt{Bluhm2001} used curve-of-growth method to estimate H$_2$ and HD column densities.
Nevertheless, we got $\log N({\rm H_2}) = 19.76\pm 0.01$ and $\log N({\rm HD}) = 13.89\pm 0.06$ which is in agreement with results obtained by \citealt{Bluhm2001}. Comparing with \citealt{Andre2004} we obtained that $\log N({\rm H_2})$ is close to their result ($19.66\pm 0.04$). They found HD to be lower then both values found by \citealt{Bluhm2001} and us. \citealt{Andre2004} used only L5, L8 and L14 HD lines, while we used L3, L5, L7 and W0 (we did not use neither L14, since spectra from 1A LiF channel do not cover this line, moreover this line is weak and noizy nor L8 since it is blended by H$_2$ and \SII\ absorptions). Also \citealt{Andre2004} reported blend of L3 line, but we found that this line does not disagree with other HD lines so we included it in the analysis.

\subsection{AV 95}

The star AV 95 belongs to the bar of SMC and H$_2$ and HD in the system towards AV 95 have previously been studied by \citealt{Andre2004}.

Both our HD ($\log N({\rm HD}) = 13.75^{+0.39}_{-0.29}$) and H$_2$ ($\log N({\rm H_2}) = 19.48^{+0.03}_{-0.02}$) column densities are consistent with values $\log N({\rm H_2}) = 19.43\pm 0.04$ and $\log N({\rm HD}) = 13.82^{+0.96}_{-0.18}$) found by \citealt{Andre2004}, respectively. 

\subsection{AV 242}

The star AV 242 is located to the south-west of the star-formation region NGC 346 in the SMC and it is a bow shock-producing star, moving towards NGC 346 \citep{Gvaramadze2011}. Among the absorption systems towards AV 242 we found an evidence of the high (or intermediate) velocity cloud (HVC) at $v = 98.8$ km/s with $\log N({\rm H_2}) = 17.18^{+0.14}_{-0.16}$. To our knowledge, it is the first found HVC towards SMC, which contains H$_2$ molecules. Unfortunately, we could place only upper limits on the HD column density towards AV 242.

\subsection{AV 488}
The star AV 488 is located within Knot 1 region, in the wing of the SMC (a more quiescent region of SMC comparing with the bar). 

Metals and H$_2$, HD and CO molecules have previously been studied by \citealt{Andre2004}. They have found H$_2$ and HD column densities to be $19.21\pm 0.06$ and $13.85^{+0.11}_{-0.14}$, respectively, which consist with our results within errors (total H$_2$ column density is $19.30^{+0.01}_{-0.02}$ and HD $13.6^{+0.6}_{-0.5}$).

\renewcommand{\arraystretch}{1.35}
\setlength{\tabcolsep}{3pt}
\begin{table}
    \centering
    \caption{Fit results for HD and H$_2$ towards LMC. %\SB{Fill $\Delta V$} \DK{Done}
    }
\begin{tabular}{lcccc}
    \hline
      Star & $\log N (\mbox{HI})^{a}$ & $\Delta V^{b}, \rm [km/s]$ & $\log N (\mbox{H}_2)$ & $\log N (\mbox{HD})^{c}$   \\
      \hline 
      Sk-67 2 & $21.00$ & 260.5 & $20.46^{+0.15}_{-0.46}$ & $\lesssim 17.1$  \\
              & '-' & 280.3 & $20.40^{+0.14}_{-0.20}$ & $\lesssim 17.3$ \\
      Sk-67 5 & $21.00$ & 294.7 & $19.47^{+0.01}_{-0.01}$ & $15.5^{+0.3}_{-2.1}$ \\
      Sk-66 1 & & 286.3 & $19.32^{+0.01}_{-0.01}$ & $\lesssim 15.9$ \\
      Sk-67 20 & & 283.2 & $19.04^{+0.01}_{-0.01}$ & $\lesssim 14.2$ \\
      Sk-66 18 & & 274.5 & $17.70^{+0.11}_{-0.12}$ & $\lesssim 14.0$ \\
      			& & 291.2 & $18.17^{+0.05}_{-0.05}$ & $\lesssim 15.7$\\
      PGMW 3070 & & 264.5 & $18.14^{+0.04}_{-0.04}$ & $\lesssim 15.6$ \\
                & & 286.5 & $19.12^{+0.02}_{-0.01}$ & $\lesssim 16.0$ \\
      LH 10-3073$^{e, f}$ & & 265.5 & $19.27^{+0.01}_{-0.03}$ & $14.3^{+0.9}_{-0.3}$ \\
                 & & 285.8 & $18.05^{+0.17}_{-0.14}$ & $\lesssim 15.3$ \\
      LH 10-3102$^{f}$ &  & 264.0 & $17.51^{+0.11}_{-0.06}$ & $14.2^{+0.4}_{-0.3}$ \\
                 &  & 285.0 & $18.62^{+0.03}_{-0.01}$ & $\lesssim 15.6$ \\
                 &  & 299.7 & $14.65^{+0.07}_{-0.05}$ & $\lesssim 15.7$ \\
      LH 10-3120 & $21.48^d$ & 272.2 & $18.22^{+0.03}_{-0.04}$ & $\lesssim 15.3$ \\
                 & '-' & 289.1 & $17.90^{+0.07}_{-0.05}$ & $\lesssim 15.4$ \\
      PGMW 3157 & & 266.9 & $17.35^{+0.25}_{-0.38}$ & $\lesssim 16.5$ \\
                & & 287.2 & $19.13^{+0.03}_{-0.03}$ & $\lesssim 16.5$ \\
      PGMW 3223 & $21.4^d$ & 267.8 & $18.78^{+0.02}_{-0.01}$ & $\lesssim 15.9$ \\
                & '-' & 287.8 & $15.27^{+0.22}_{-0.17}$ & $\lesssim 15.7$ \\
      Sk-66 35 & $20.83^d$ & 264.4 & $18.34^{+0.24}_{-0.18}$ & $\lesssim 16.2$ \\
               & '-' & 275.4 & $19.35^{+0.07}_{-0.02}$ & $\lesssim 15.9$ \\
               & '-' & 286.0 & $18.64^{+0.18}_{-0.13}$ & $\lesssim 15.7$ \\
      Sk-69 52 & & 248.8 & $18.64^{+0.03}_{-0.02}$ & $\lesssim 15.9$ \\
      Sk-65 21 & $< 20.50$ & 247.8 & $18.38^{+0.01}_{-0.01}$ & $\lesssim 15.9$ \\
      Sk-66 51$^{e}$ & & 311.5 & $18.08^{+0.20}_{-0.20}$ & $\lesssim 15.3$ \\
      Sk-70 79$^f$ & $21.26^d$ & 234.6 & $20.38^{+0.01}_{-0.01}$ & $15.6^{+0.7}_{-0.5}$ \\
      Sk-68 52 & $21.30^d$ & 248.8 & $19.51^{+0.01}_{-0.01}$ & $\lesssim 15.6$ \\
      Sk-71 8 & $21.04$ & 186.6 & $18.21^{+0.04}_{-0.07}$ & $\lesssim 16.4$ \\
              & '-' & 220.4 & $18.66^{+0.04}_{-0.03}$ & $\lesssim 14.3$ \\
      Sk-70 85 &   & 256.7 & $18.57^{+0.05}_{-0.04}$ & $\lesssim 16.0$ \\
              &    & 283.3 & $17.18^{+0.18}_{-0.30}$ & $\lesssim 15.9$ \\
      Sk-69 106 &  & 253.4 & $18.78^{+0.02}_{-0.01}$ & $\lesssim 15.6$ \\
      Sk-68 73$^f$ & $21.66^d$ & 294.2 & $20.24^{+0.01}_{-0.01}$ & $14.2^{+0.1}_{-0.1}$ \\
      Sk-67 105 & $21.25^d$ & 301.9 & $19.42^{+0.05}_{-0.01}$ & $\lesssim 14.8$ \\
                & '-' & 310.3 & $17.46^{+0.29}_{-0.85}$ & $\lesssim 13.9$ \\
      BI 184 & $21.12^d$ & 240.4 & $19.89^{+0.01}_{-0.02}$ & $\lesssim 16.7$ \\
             & '-' & 281.2 & $16.78^{+0.11}_{-0.20}$ & $\lesssim 15.8$ \\
      Sk-71 38 & & 201.3 & $17.46^{+0.07}_{-0.07}$ & $\lesssim 16.1$ \\
               & & 243.0 & $18.53^{+0.06}_{-0.04}$ & $\lesssim 16.0$ \\
               & & 280.1 & $17.45^{+0.06}_{-0.05}$ & $\lesssim 14.7$ \\
      Sk-71 45 & $21.09$ & 254.1 & $18.55^{+0.01}_{-0.01}$ & $\lesssim 13.3$ \\
      Sk-71 46$^f$ &  & 243.4 & $20.32^{+0.03}_{-0.04}$ & $14.60^{+1.08}_{-0.30}$ \\
      Sk-69 191 & $20.78^d$ & 229.2 & $19.11^{+0.01}_{-0.02}$ & $\lesssim 15.9$ \\
                & '-' & 257.7 & $15.34^{+0.92}_{-0.30}$ & $\lesssim 15.1$ \\
    \hline 
    \end{tabular}
    
    \label{tab:HD_H2_LMC}
\end{table}

\setlength{\tabcolsep}{3pt}
\begin{table}
    \centering
    \contcaption{}
   
    \begin{tabular}{lcccc}
    \hline
      Star & $\log N (\mbox{HI})^{a}$ & $\Delta V^{b}, \rm [km/s]$ & $\log N (\mbox{H}_2)$ & $\log N (\mbox{HD})^{c}$    \\
      \hline 
      J\,0534-6932$^f$ & $21.60$ & 270.6 & $19.88^{+0.01}_{-0.01}$ & $14.5^{+0.9}_{-0.4}$ \\
      			   & '-' & 294.3 & $17.24^{+0.26}_{-0.41}$ & $\lesssim 16.0$ \\
      BI 237$^f$ & $21.62$ & 296.2 & $20.20^{+0.01}_{-0.01}$ & $14.4^{+1.0}_{-0.2}$  \\
      Sk-68 129 & $21.72$ & 276.0 & $20.53^{+0.07}_{-0.03}$ & $\lesssim 17.2$ \\
      Sk-69 220 & $21.28$ & 283.3 & $19.29^{+0.01}_{-0.01}$ & $\lesssim 16.4$ \\
      Sk-69 223 & $21.9$ & 273.0 & $19.18^{+0.08}_{-0.10}$ & $\lesssim 16.3$ \\
      			& '-' & 308.3 & $19.94^{+0.01}_{-0.02}$ & $\lesssim 16.2$ \\
      Sk-66 172 & $21.25$ & 289.3 & $18.18^{+0.08}_{-0.04}$ & $\lesssim 16.1$ \\
                & '-' & 301.5 & $16.62^{+0.49}_{-0.19}$ & $\lesssim 16.1$ \\
                & '-' & 331.0 & $16.35^{+0.30}_{-0.79}$ & $\lesssim 14.9$ \\
                & '-' & 375.3 & $15.97^{+0.38}_{-0.67}$ & $\lesssim 15.5$ \\ 
      Sk-69 228 & $21.63$ & 262.2 & $18.28^{+0.03}_{-0.13}$ & $\lesssim 16.8$ \\
                & '-' & 296.3 & $18.87^{+0.06}_{-0.05}$ & $\lesssim 16.4$ \\
      BI 253$^f$ & $21.67^d$ & 267.3 & $20.01^{+0.01}_{-0.01}$ & $15.3^{+0.4}_{-0.9}$ \\
             & '-' & 279.9 & $18.12^{+0.19}_{-0.35}$ & $13.65^{+0.20}_{-0.23}$ \\
      Sk-68 135 & $21.46^d$ & 250.7 & $18.83^{+0.11}_{-0.06}$ & $\lesssim 15.9$ \\
      			& '-' & 272.4 & $19.99^{+0.01}_{-0.01}$ & $14.0^{+0.6}_{-0.3}$ \\
      Sk-68 137$^f$ & $21.50$ & 278.5 & $20.31^{+0.05}_{-0.32}$ & $16.5^{+0.2}_{-1.6}$ \\
                & '-' & 302.9 & $20.53^{+0.05}_{-0.11}$ & $14.5^{+1.1}_{-0.3}$ \\
      Brey 77 & $21.79$ & 277.1 & $19.30^{+0.04}_{-0.06}$ & $\lesssim 15.7$ \\
              & '-' & 296.1 & $19.42^{+0.05}_{-0.04}$ & $\lesssim 16.1$ \\
              & '-' & 316.4 & $18.10^{+0.11}_{-0.11}$ & $\lesssim 16.4$ \\
      Sk-69 243$^{e}$ & $21.80$ & 271.0 & $18.29^{+0.27}_{-0.19}$ & $\lesssim 15.7$ \\
                            & '-' & 280.3 & $19.34^{+0.07}_{-0.05}$ & $\lesssim 15.6$ \\
                            & '-' & 290.0 & $19.66^{+0.02}_{-0.04}$ & $\lesssim 16.0$ \\
                            & '-' & 312.4 & $16.80^{+0.45}_{-0.16}$ & $\lesssim 16.1$ \\
      Sk-69 246 & $21.47^d$ & 273.0 & $19.76^{+0.01}_{-0.01}$ & $13.89^{+0.06}_{-0.06}$ \\
      Sk-68 140 & $21.47^d$ & 275.6 & $20.40^{+0.03}_{-0.03}$ & $\lesssim 17.1$ \\
      Sk-71 50  & $21.18$ & 238.2 & $19.47^{+0.03}_{-0.04}$ & $\lesssim 16.7$ \\
                & '-' & 267.5 & $19.43^{+0.05}_{-0.05}$ & $\lesssim 16.5$ \\
                & '-' & 292.1 & $19.15^{+0.05}_{-0.07}$ & $\lesssim 16.8$ \\
      Sk-69 279 & $21.59^{d}$ & 249.4 & $19.19^{+0.23}_{-0.44}$ & $\lesssim 17.1$ \\
      		& '-' & 265.9 & $20.54^{+0.01}_{-0.03}$ & $\lesssim 17.0$ \\
      Sk-68 155 & $21.44^d$ & 289.5 & $20.02^{+0.03}_{-0.05}$ & $\lesssim 16.8$ \\
                & '-' & 307.7 & $19.03^{+0.40}_{-0.42}$ & $\lesssim 16.4$ \\
                & '-' & 333.7 & $19.00^{+0.22}_{-0.63}$ & $\lesssim 17.0$ \\
      Sk-69 297 & & 249.6 & $19.77^{+0.03}_{-0.03}$ & $\lesssim 16.3$ \\
      		& & 295.1 & $19.07^{+0.05}_{-0.05}$ & $\lesssim 16.4$ \\
      Sk-70 115 & $21.13^d$ & 215.6 & $19.97^{+0.01}_{-0.01}$ & $\lesssim 16.3$ \\
                & '-' & 292.1 & $18.23^{+0.03}_{-0.02}$ & $\lesssim 15.5$ \\
    \hline 
    \end{tabular}
    
    \begin{tablenotes}
    \item $a$ Total \HI\ column density, taken from \citealt{Welty2012}, the typical uncertainty is about $10\%$ (but the values were not shown by \citealt{Welty2012})   except systems denoted by $d$
    \item $b$ Kinematic shift of the component relative to the local standard of rest.
    \item $c$ Upper limits were constrained from $3\sigma$ confidence interval.
    \item $d$ Taken from \citealt{Roman_Duval2019}.
    \item $e$ Low Doppler parameter, hence $\log N (\rm H_2)$ may be overestimated.
    \item $f$ Systems with new HD detections.
    \end{tablenotes}

\end{table}

\renewcommand{\arraystretch}{1.35}
\setlength{\tabcolsep}{3pt}
\begin{table}
\centering
\caption{Fit results for HD and H$_2$ towards SMC.}
\begin{tabular}{lcccc}
     \hline
      Star & $\log N (\mbox{HI})^{a}$ & $\Delta V^{b}, \rm [km/s]$ & $\log N (\mbox{H}_2)$ & $\log N (\mbox{HD})^{c}$    \\
      \hline 
      AV 6  & $21.54$ & 107.4 & $16.82^{+0.26}_{-0.46}$ & $\lesssim 15.9$ \\
            & '-' & 132.0 & $19.09^{+0.03}_{-0.02}$ & $\lesssim 16.0$ \\
            & '-' & 150.5 & $17.42^{+0.15}_{-0.67}$ & $\lesssim 16.2$ \\
      AV 14 & $21.76$ & 91.5 & $15.69^{+0.44}_{-0.29}$ & $\lesssim 14.4$ \\
            & '-' & 111.6 & $18.18^{+0.07}_{-0.05}$ & $\lesssim 14.1$ \\
            & '-' & 131.2 & $18.33^{+0.07}_{-0.07}$ & $\lesssim 15.3$ \\
            & '-' & 148.7 & $16.60^{+0.25}_{-0.87}$ & $\lesssim 15.4$ \\
      AV 15 & $21.58$ & 119.9 & $16.69^{+0.13}_{-0.12}$ & $\lesssim 15.0$ \\
            &         & 137.7 & $18.45^{+0.01}_{-0.03}$ & $\lesssim 15.3$ \\
      AV 16 &         & 128.0 & $20.06^{+0.10}_{-0.07}$ & $\lesssim 16.7$ \\
            &         & 152.1 & $19.99^{+0.10}_{-0.11}$ & $\lesssim 16.4$ \\
      AV 18$^f$ & $22.04$ & 103.4 & $16.04^{+0.22}_{-0.15}$ & $\lesssim 15.6$ \\
            & '-' & 130.6 & $20.46^{+0.03}_{-0.02}$ & $14.63^{+0.28}_{-0.20}$\\
            & '-' & 154.2 & $19.19^{+0.17}_{-0.26}$ & $14.6^{+1.0}_{-0.2}$\\
            & '-' & 176.0 & $19.83^{+0.05}_{-0.06}$ & $\lesssim 16.2$\\
      AV 22 & $< 20.80$ & 112.4 & $18.06^{+0.06}_{-0.08}$ & $\lesssim 16.0$ \\
            & '-' & 133.4 & $17.50^{+0.18}_{-0.03}$ & $\lesssim 15.8$ \\
            & '-' & 153.8 & $17.48^{+0.14}_{-0.08}$ & $\lesssim 16.2$ \\
      AV 26$^f$ & $21.70$ & 119.3 & $19.19^{+0.24}_{-0.22}$ & $\lesssim 16.7$ \\
            & '-' & 132.5 & $20.70^{+0.02}_{-0.01}$ & $14.4^{+1.7}_{-0.4}$ \\
            & '-' & 152.0 & $18.66^{+0.40}_{-0.77}$ & $\lesssim 16.4$ \\
      AV 39a & $21.74$ & 103.7 & $17.22^{+0.21}_{-0.26}$ & $\lesssim 16.1$ \\
             & '-' & 117.1 & $18.66^{+0.04}_{-0.04}$ & $\lesssim 15.5$ \\
             & '-' & 129.8 & $17.08^{+0.44}_{-0.22}$ & $\lesssim 16.3$ \\
      AV 47 & $21.32$ & 116.5 & $17.72^{+0.07}_{-0.02}$ & $\lesssim 15.4$ \\
            & '-' & 129.1 & $18.45^{+0.02}_{-0.02}$ & $\lesssim 13.5$ \\
      AV 60a & $21.81$ & 125.0 & $18.76^{+0.13}_{-0.13}$ & $\lesssim 15.9$ \\
             & '-' & 143.6 & $19.72^{+0.02}_{-0.03}$ & $\lesssim 16.6$ \\
             & '-' & 217.7 & $18.12^{+0.15}_{-0.05}$ & $\lesssim 16.1$ \\
      AV 69 & $21.59$ & 112.8 & $18.34^{+0.05}_{-0.05}$ & $\lesssim 15.1$ \\
                        & '-' & 128.4 & $18.89^{+0.02}_{-0.02}$ & $\lesssim 15.3$\\
      AV 75 & $21.79$ & 113.4 & $18.87^{+0.01}_{-0.02}$ & $\lesssim 15.3$ \\
            & '-' & 126.7 & $17.74^{+0.11}_{-0.08}$ & $\lesssim 13.7$ \\
      AV 80$^f$ & $21.81$ & 118.8 & $20.24^{+0.01}_{-0.01}$ &  $14.50^{+0.03}_{-0.04}$ \\
            & '-' & 135.0 & $18.75^{+0.24}_{-0.86}$ & $\lesssim 15.4$ \\
      AV 81 &         & 111.9 & $15.99^{+0.18}_{-0.10}$ & $\lesssim 15.1$ \\
            &         & 139.8 & $18.58^{+0.03}_{-0.02}$ & $\lesssim 15.5$ \\
            &         & 177.3 & $17.88^{+0.04}_{-0.08}$ & $\lesssim 13.8$ \\
      AV 95 & $21.49$ & 93.7 & $15.46^{+0.05}_{-0.06}$ & $\lesssim 13.4$ \\
            & '-' & 124.5 & $19.23^{+0.03}_{-0.04}$ & $13.75^{+0.39}_{-0.29}$ \\
            & '-' & 139.0 & $19.11^{+0.05}_{-0.05}$ & $\lesssim 13.6$ \\
      AV 104 & $21.45$ & 119.4 & $19.35^{+0.01}_{-0.01}$ & $\lesssim 16.1$ \\
             & '-' & 147.5 & $16.77^{+0.28}_{-0.44}$ & $\lesssim 15.9$ \\
      
      \hline 
\end{tabular}
\label{tab:HD_H2_SMC}
\end{table}

\setlength{\tabcolsep}{3pt}
\begin{table}
\centering
\contcaption{}
\begin{tabular}{lcccc}
     \hline
      Star & $\log N (\mbox{HI})^{a}$ & $\Delta V^{b}, \rm [km/s]$ & $\log N (\mbox{H}_2)$ & $\log N (\mbox{HD})^{c}$     \\
      \hline 
      AV 170 & $21.14$ & 138.1 & $19.73^{+0.01}_{-0.01}$ & $\lesssim 16.1$ \\
      AV 175 &         & 121.7 & $19.49^{+0.17}_{-0.12}$ & $\lesssim 16.6$ \\
             &         & 138.5 & $20.05^{+0.10}_{-0.09}$ & $\lesssim 16.4$ \\
      NGC 346-12 & $21.81$ & 138.9 & $19.64^{+0.17}_{-0.07}$ & $\lesssim 13.4$ \\
                 & '-' & 166.5 & $18.87^{+0.46}_{-0.51}$ & $\lesssim 15.6$ \\
                 & '-' & 183.3 & $20.40^{+0.04}_{-0.06}$ & $\lesssim 13.6 $ \\
      AV 207 & $21.43$ & 160.1 & $19.52^{+0.02}_{-0.04}$ & $\lesssim 16.8$ \\
             & '-' & 181.2 & $18.93^{+0.08}_{-0.12}$ & $\lesssim 16.5$ \\
      AV 208$^f$ & $21.85$ & 160.1 & $19.82^{+0.09}_{-0.12}$ & $14.5^{+1.0}_{-0.3}$ \\
             & '-' & 181.2 & $19.88^{+0.07}_{-0.16}$ & $\lesssim 16.8$ \\
      AV 210 & $21.85$ & 129.3 & $19.33^{+0.01}_{-0.02}$ & $\lesssim 16.1$ \\
             & '-' & 164.8 & $17.62^{+0.26}_{-0.34}$ & $\lesssim 16.6$ \\
             & '-' & 182.1 & $17.34^{+0.16}_{-0.22}$ & $\lesssim 16.5$ \\
      AV 215 & $21.86$ & 126.2 & $19.38^{+0.05}_{-0.10}$ & $\lesssim 16.6$ \\
             & '-' & 145.7 & $19.52^{+0.06}_{-0.05}$ & $\lesssim 16.5$ \\ 
      AV 216 & $21.64$ & 124.7 & $17.68^{+0.16}_{-0.22}$ & $\lesssim 15.7$ \\
             & '-' & 139.1 & $18.71^{+0.05}_{-0.02}$ & $\lesssim 16.0$ \\
             & '-' & 162.5 & $15.99^{+0.31}_{-0.22}$ & $\lesssim 16.2$ \\
      NGC 346-637 & $21.65$ & 162.4 & $19.44^{+0.02}_{-0.02}$ & $\lesssim 17.0$ \\
      AV 243 & $21.52$ & 128.4 & $18.95^{+0.02}_{-0.02}$ & $\lesssim 15.7$ \\
             & '-' & 147.9 & $17.25^{+0.18}_{-0.30}$ & $\lesssim 16.1$ \\
      AV 242$^e$ & $21.32$ & 98.8 & $17.18^{+0.14}_{-0.16}$ & $\lesssim 14.0$ \\
             & '-' & 132.2 & $15.00^{+0.16}_{-0.04}$ & $\lesssim 15.7$ \\
             & '-' & 163.9 & $17.73^{+0.08}_{-0.11}$ & $\lesssim 15.9$ \\
             & '-' & 182.2 & $16.75^{+0.14}_{-0.17}$ & $\lesssim 14.9$ \\
      AV 261 & $21.43$ & 93.2 & $16.54^{+1.12}_{-0.37}$ & $\lesssim 16.5$ \\
             & '-' & 125.3 & $19.38^{+0.03}_{-0.03}$ & $\lesssim 16.7$ \\
      AV 266 &         & 106.4 & $17.01^{+0.27}_{-0.68}$ & $\lesssim 15.2$ \\
             &         & 127.3 & $19.81^{+0.01}_{-0.02}$ & $\lesssim 15.6$ \\
             &         & 132.5 & $17.46^{+0.44}_{-0.23}$ & $\lesssim 15.5$ \\
      AV 304 & $21.48$ & 121.0 & $19.55^{+0.03}_{-0.03}$ & $\lesssim 15.8$ \\
             & '-' & 137.6 & $18.77^{+0.19}_{-0.27}$ & $\lesssim 15.7$ \\
      AV 372$^f$ & $21.57$ & 122.6 & $15.52^{+0.15}_{-0.05}$ & $\lesssim 15.6$ \\
             & '-' & 140.9 & $18.62^{+0.04}_{-0.21}$ & $\lesssim 16.2$ \\
             & '-' & 147.1 & $18.45^{+0.10}_{-0.18}$ & $13.97^{+0.22}_{-0.23}$ \\
      AV 374 & $21.14$ & 86.5 & $18.09^{+0.03}_{-0.04}$ & $\lesssim 16.2$ \\
             & '-' & 133.1 & $18.65^{+0.02}_{-0.02}$ & $\lesssim 15.9$ \\
      AV 423 & $21.49$ & 141.8 & $18.00^{+0.36}_{-0.34}$ & $\lesssim 15.5$ \\
             & '-' & 145.6 & $18.49^{+0.10}_{-0.11}$ & $\lesssim 14.2$ \\
             & '-' & 184.1 & $17.03^{+0.28}_{-0.20}$ & $\lesssim 16.1$ \\
      AV 435$^f$ & $21.54$ & 177.4 & $19.92^{+0.01}_{-0.01}$ & $15.6^{+0.2}_{-0.8}$ \\
      AV 440 & $21.38$ & 130.5 & $17.29^{+0.04}_{-0.03}$ & $\lesssim 14.8$ \\
             & '-' & 173.5 & $19.06^{+0.01}_{-0.01}$ & $\lesssim 15.8$ \\
      AV 472$^f$ &         & 137.8 & $20.38^{+0.01}_{-0.01}$ &  $15.9^{+0.3}_{-0.8}$ \\
      AV 476 & $21.85$ & 157.4 & $20.84^{+0.03}_{-0.05}$ & $\lesssim 17.1$ \\
             & '-' & 167.1 & $19.79^{+0.21}_{-0.35}$ & $\lesssim 17.1$ \\
      \hline 
\end{tabular}
\end{table}

\setlength{\tabcolsep}{3pt}
\begin{table}
\centering
\contcaption{}
\begin{tabular}{lcccc}
     \hline
      Star & $\log N (\mbox{HI})^{a}$ & $\Delta V^{b}, \rm [km/s]$ & $\log N (\mbox{H}_2)$ & $\log N (\mbox{HD})^{c}$  \\
      \hline 
      AV 479 & $21.42$ & 152.1 & $18.73^{+0.01}_{-0.02}$ & $\lesssim 15.2$ \\
             & '-' & 167.8 & $18.99^{+0.01}_{-0.01}$ & $\lesssim 15.4$ \\
      AV 480 & $21.42$ & 183.8 & $18.21^{+0.10}_{-0.16}$ & $\lesssim 16.1$ \\
             & '-' & 193.4 & $18.53^{+0.07}_{-0.07}$ & $\lesssim 16.2$ \\
      AV 483 & $21.13$ & 145.9 & $18.71^{+0.03}_{-0.04}$ & $\lesssim 16.1$ \\
             & '-' & 157.2 & $18.48^{+0.06}_{-0.04}$ & $\lesssim 16.4$ \\
      AV 486$^f$ & $21.18$ & 145.5 & $18.56^{+0.07}_{-0.07}$ & $\lesssim 15.6$ \\
             & '-' & 157.6 & $19.07^{+0.02}_{-0.03}$ & $14.0^{+0.9}_{-0.4}$ \\
      AV 488 & $21.15$ & 146.6 & $19.04^{+0.01}_{+0.03}$ & $13.6^{+0.6}_{-0.5}$ \\
             & '-' & 158.5 & $18.96^{+0.03}_{-0.02}$ & $\lesssim 14.6$\\
      AV 490$^f$ & $21.46$ & 132.0 & $19.88^{+0.01}_{-0.01}$ & $16.1^{+0.2}_{-0.7}$ \\
             & '-' & 162.7 & $17.94^{+0.29}_{-0.39}$ & $\lesssim 15.6$\\
      AV 491 & $21.40$ & 97.7 & $15.66^{+0.47}_{-0.71}$ & $\lesssim 14.7$ \\
             & '-' & 143.6 & $18.99^{+0.05}_{-0.10}$ & $\lesssim 15.5 $ \\
             & '-' & 155.6 & $18.98^{+0.05}_{-0.26}$ & $\lesssim 15.0$ \\
             & '-' & 184.4 & $15.09^{+0.81}_{-0.06}$ & $\lesssim 14.6$ \\
      AV 506 & $21.35$ & 132.3 & $16.36^{+0.29}_{-0.12}$ & $\lesssim 14.0$ \\
             & '-' & 147.5 & $19.03^{+0.01}_{-0.01}$ & $\lesssim 15.7$ \\
      Sk 191$^f$ & $21.51$ & 153.5 & $20.78^{+0.02}_{-0.03}$ & $14.7^{+1.2}_{-0.2}$ \\
      WR 9$^{e}$ & $21.48$ & 87.7 & $16.06^{+0.34}_{-0.25}$ & $\lesssim 15.7$ \\
                       & '-' & 117.8 & $18.69^{+0.08}_{-0.05}$ & $\lesssim 15.9$ \\
                       & '-' & 135.5 & $19.37^{+0.01}_{-0.02}$ & $\lesssim 15.7$ \\
                       & '-' & 149.9 & $16.43^{+0.42}_{-0.19}$ & $\lesssim 15.7$ \\
                       & '-' & 167.9 & $17.37^{+0.11}_{-0.44}$ & $\lesssim 15.9$ \\
      \hline 
\end{tabular}

\begin{tablenotes}
 \item $a$ Total \HI\ column density, taken from \citealt{Welty2012}
    \item $b$ Kinematic shift of the component relative to the local standard of rest.
    \item $c$ Upper limits were constrained from $3\sigma$ confidence interval
    \item $e$ low Doppler parameter, $\log N(\rm H_2)$ may be overestimated 
    \item $f$ Systems with new HD detections.
\end{tablenotes}
\end{table}

\clearpage

 %\section{Physical conditions}
 %\label{sect:phys_properties}

%We used column densities and metallicities, reported in previous sections, to constrain physical conditions in the medium of LMC and SMC, associated with absorption systems. We followed the procedure developed in \citealtalp{Kosenko2021}, that includes two steps. Firstly, we estimate number density ($n$) and UV field intensity ($\chi$, below we expressed it in the values relative to Mathis field, \citealt{Mathis1983}) from joint comparison of the observed population of \CI\ fine structure and H$_2$ low rotational levels, with the results of \Meudon\ code \citep{LePetit2006} (details will be presented in the accompanied paper (Kosenko et al in prep.)). Secondly, we used constraints on $n$ and $\chi$ to estimate CRIR, using measured HD and total H$_2$ column densities and the formalism presented in \citealt{Balashev2020}.

\section{Cosmic-ray ionization rate}%, $\zeta$}
\label{sect:phys_properties}

We used column densities, reported in the previous sections, to constrain physical conditions in the medium of LMC and SMC, associated with absorption systems. 
Among the parameters, listed above, in this work we focus mainly on CRIR (estimation of $n$, $\chi$ and metallicity in Magellanic Clouds will be presented in the accompanied paper, Kosenko et al in prep.)
We followed the procedure developed and used for high-redshift systems in \citealt{Kosenko2021}. 
%that includes two steps. Firstly, we estimate number density ($n$) and UV field intensity ($\chi$, below we expressed it in the values relative to Mathis field, \citealt{Mathis1983}) from joint comparison of the observed population of \CI\ fine structure and H$_2$ low rotational levels, with the results of \Meudon\ code \citep{LePetit2006} \citep[using method, described by][]{Klimenko2020}
%(details on $n$ and $\chi$ estimation in Magellanic Clouds will be presented in the accompanied paper (Kosenko et al in prep.)). Secondly, we used constraints on $n$ and $\chi$ to estimate CRIR, using measured HD and total H$_2$ column densities and the formalism presented in \citealt{Balashev2020}.

This method is based on the balance equation between formation and destruction of HD molecules in the plane-parallel steady-state cloud, therefore leading to solve only one differential equation \citep{Balashev2020}:
$$\dfrac{{\rm d}N({\rm HD})}{{\rm d}N({\rm H_2})} = f(\zeta, n, \chi, Z, N({\rm H_2})),$$
except modelling of the cloud using full chemical network.
Certainly to get reasonable constrain on CRIR this method requires an estimates of number density, UV field intensity and metallicity to remove degeneracy of parameters, which will be presented in Kosenko et al in prep. We fixed metallicity and used constraints on $n$ and $\chi$ as priors during the Bayesian inference. For $\zeta$ we used flat priors in log space, emulating a wide distribution. We used Markov Chain Monte-Carlo method to estimate posterior distribution function on CRIR. Results are shown in the last column of Table~\ref{tab:phys_conditions}.

Unfortunately, in the most sightlines except one, we get only upper limits on CRIR, $\zeta$, due to large uncertainties in HD column density estimates (which is connected with insufficient quality of the data). 
In the systems, denoted by $\dagger$ in the Table~\ref{tab:phys_conditions}, we could not constrain CRIR at all within the physically reasonable range $\lesssim10^{-13}$, where the model from \citep{Balashev2020} is appropriate\footnote{For example, a very high CRIR may change chemistry of the cloud, due to destruction of molecules even in the self-shielded core, which is not considered in \citealt{Balashev2020}}. %i.e. obtained CRIR should be high enough to get sufficient amount of ionized hydrogen to provide reaction~\ref{eq:HD_form}, otherwise, 
For these systems we got too wide posterior probabilities in this region (also due to the large uncertainties in $N({\rm HD})$) encompassing all the range for $\zeta$.

\begin{figure*}
    \includegraphics[width=1\linewidth]{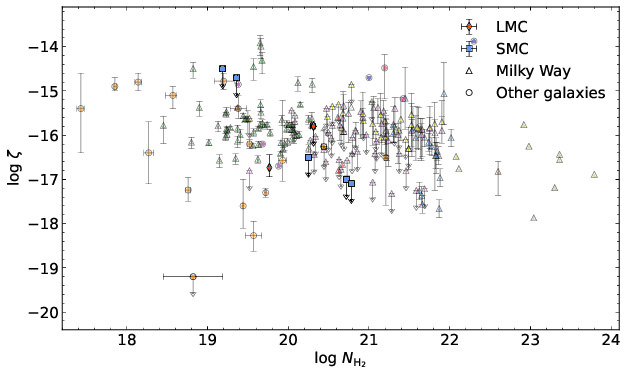}
    \caption{The constraints on CRIR as a function of H$_2$ column density. Here blue squares and red diamonds are values obtained in this work (for SMC and LMC, respectively), circles are values found for other galaxies (orange - \citealt{Kosenko2021}, blue - \citealt{Muller2016}, violet - \citealt{Shaw2016}, pink - \citealt{Indriolo2018}), triangles are values for our Galaxy (yellow - \citealt{Indriolo2007}, light green - table 6 from \citealt{Padovani2009}, blue - \citealt{Caselli1998}, cyan - \citealt{Shaw2008}, brown - \citealt{Maret2007}, violet - \citealt{Indriolo2012b}, dark green - \citealt{Indriolo2015})
    }
    \label{fig:zeta}
\end{figure*}

Constraints on CRIR as a function of H$_2$ column density are shown in the Fig.~\ref{fig:zeta} (red diamonds for the LMC and blue squares for the SMC) and compared to other known CRIR estimations in the Milky Way and other galaxies. Note that %the value $\zeta\lesssim 18.1$ s$^{-1}$ 
constraint on $\zeta$
in the system towards Sk 191 in the LMC should be taken with caution since we got metallicity about an order lower than average value in SMC and it may be attributed to the large depletion in this system (one of the highest value of H$_2$ column density also may point at this; details will be provided in Kosenko et al. in prep.).

\renewcommand{\arraystretch}{1.35}
\setlength{\tabcolsep}{3pt}
\begin{table}
    \centering
    \caption{Physical conditions derived in the absorption systems associated with LMC and SMC.}
    \label{tab:phys_conditions}
    
    \begin{tabular}{cccccc}
    \hline 
      Star & $\rm [X/H]$ & $\rm X$ & $\log n [\rm cm^{-3}]$ & $\log \chi$ & $\log \zeta [\rm s^{-1}]$ \\
      \hline 
      \multicolumn{6}{c}{LMC} \\
      \hline
    
       Sk-67 5$^{\dagger}$  & $-0.58^{+0.05}_{-0.05}$ & Zn & $1.99^{+0.17}_{-0.18}$ & $0.40^{+0.19}_{-0.18}$ & --  \\
       Sk-70 79$^{\dagger}$ & $-0.63^{+0.06}_{-0.05}$ & Zn & $2.75^{+0.36}_{-0.31}$ & $1.81^{+0.21}_{-0.18}$ & --  \\
       Sk-71 46 & -- & -- & $2.50^{+0.32}_{-0.28}$ & $1.24^{+0.27}_{-0.20}$ & $\lesssim -15.8$ \\
       Sk-68 135$^{\dagger}$ & $-0.67^{+0.06}_{-0.05}$ & Zn & $2.00^{+0.20}_{-0.19}$ & $1.62^{+0.31}_{-0.27}$ & -- \\
       Sk-69 246 & $-0.65^{+0.05}_{-0.05}$ & Zn & $2.37^{+0.19}_{-0.17}$ & $1.66^{+0.22}_{-0.22}$ & $-16.73^{+0.29}_{-0.21}$ \\
     \hline 
      \multicolumn{6}{c}{SMC} \\
      \hline 
      AV 26 & $-0.96^{+0.06}_{-0.06}$ & Zn &  $2.02^{+0.22}_{-0.16}$ & $0.35^{+0.61}_{-0.69}$ & $\lesssim -17.0$ \\
      AV 80 & $-1.16^{+0.05}_{-0.05}$ & Zn & $3.89^{+0.15}_{-0.17}$ & $ 2.01^{+0.31}_{-0.29}$ & $\lesssim -16.5$ \\
      AV 372$^{\dagger}$ & $-1.06^{+0.05}_{-0.05}$ & Zn & $1.41^{+0.08}_{-0.10}$ & $1.75^{+0.14}_{-0.10}$ & -- \\
      AV 488 & $-0.81^{+0.05}_{-0.06}$ & Zn & $1.95^{+0.40}_{-0.35}$ & $1.03^{+0.66}_{-0.55}$ & $\lesssim -14.7$ \\
      AV 490$^{\dagger}$ & $-1.06^{+0.05}_{-0.06}$ & P & $2.01^{+0.11}_{-0.13}$ & $1.38^{+0.18}_{-0.17}$ & -- \\
      Sk 191 & $-1.51^{+0.06}_{-0.06}$ & Zn & $3.70^{+0.48}_{-0.68}$ & $1.43^{+0.20}_{-0.26}$ & $\lesssim -17.1$ \\
      \hline  
    \end{tabular}
    \begin{tablenotes}
     \item The columns are: (i) name of the star; (ii) estimated metallicity; (iii) species that is used to derive metallicity; (iv) the number density; (v) the UV field strength in the units of Mathis field; (vi) the cosmic ray ionization rate. %\SB{Be careful of very large $\zeta$ values, where contours formally suggests that CRIR unconstrained.} \DK{added to the text}
     \item Upper limits were constrained from $3\sigma$ credible interval 
     \item In the case of Sk-71 46 we have not found \HI\, column density in the literature therefore we could not obtain metallicity for this system and to estimate CRIR we used an average metallicity of LMC ($Z=0.5 Z_{\odot}$)
     \item $\dagger$ In these systems we could not constrain $\zeta$  (see the text)
    \end{tablenotes}
\end{table}

\section{Discussion}

\label{sect:discussion}
\begin{figure*}
    \begin{tabular}{ccc}
    \includegraphics[width=0.33\linewidth]{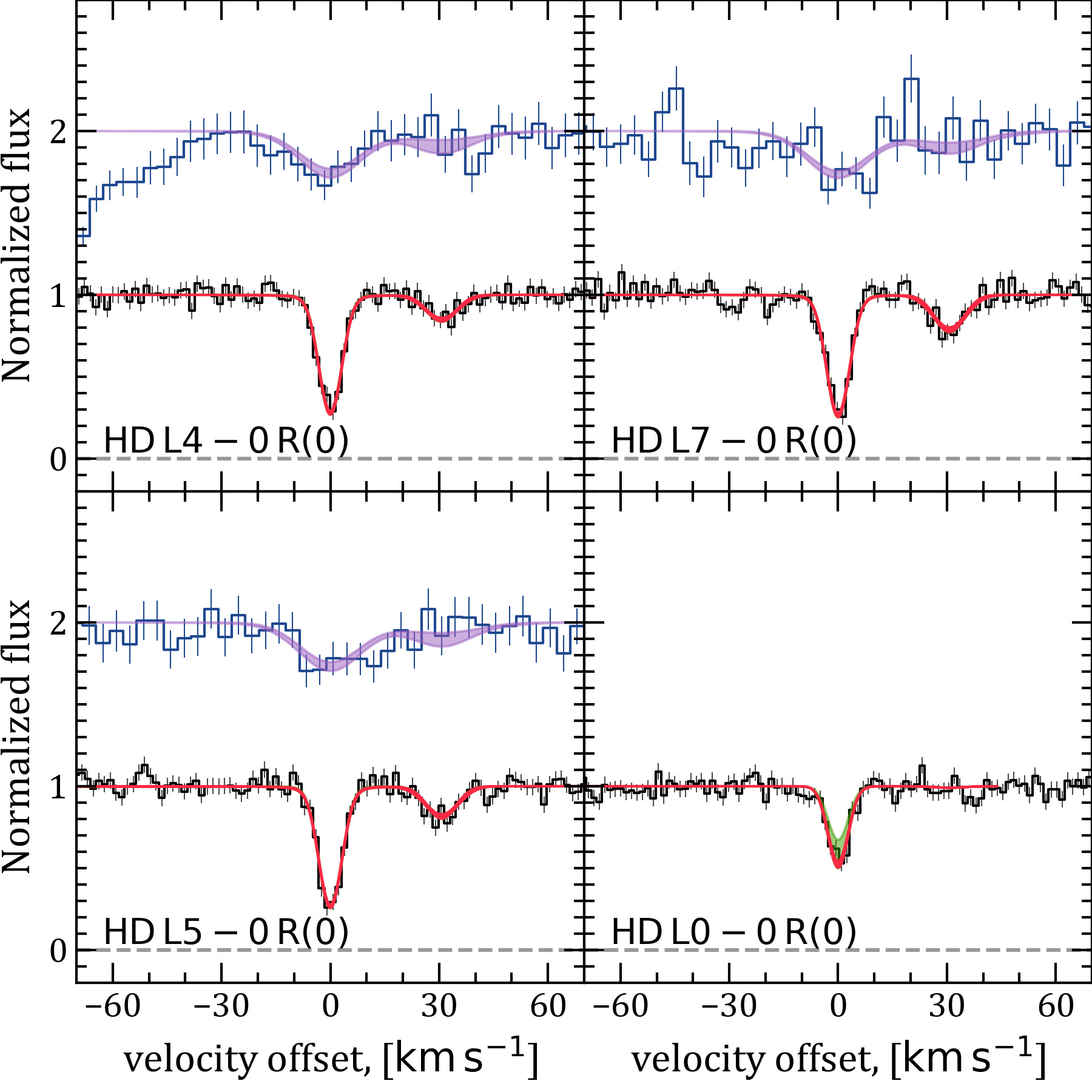} & 
    \includegraphics[width=0.33\linewidth]{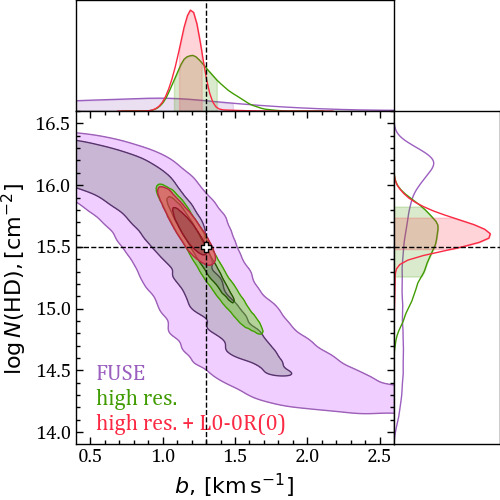} &
    \includegraphics[width=0.33\linewidth]{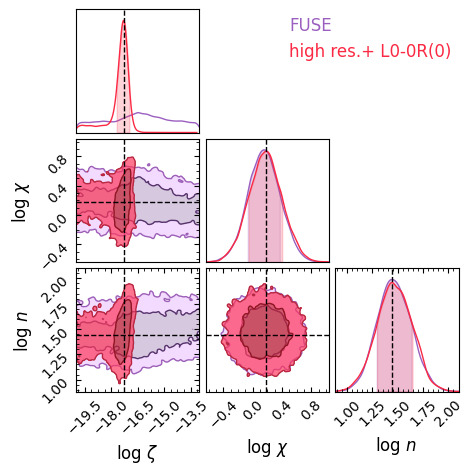} \\
    \end{tabular}
    \caption{{\sl Left:} The profiles of selected HD lines. The blue and black lines show the real FUSE spectrum of AV\,490 (shifted by 1 vertically for the presentation purposes) and the mock high resolution spectrum. To generate the mock spectrum we assume, that the $b=1.3\rm\,km\,s^{-1}$ and $\log N(\rm HD)=15.5$ (shown by cross at the middle panel). The fits to HD line profiles using two component Voigt function are shown by lavender, green and red lines and regions and correspond to FUSE (using L4-0R(0), L5-0R(0), L7-0R(0)), mock (the same sample of HD lines) and mock with additionally taking into account L0-0R(0) line, respectively. {\sl Middle:} The 2d posterior in the Doppler parameter - column density space, obtained during the fit to HD lines. The color correspond to the fit shown in left panel. {\sl Right:} The constraints on the CRIR using estimates of HD column density from FUSE spectrum (lavender lines and regions) and from high-resolution mock spectrum (red).}
        \label{fig:highres}
\end{figure*}

Nevertheless that we detected HD towards two dozens of LMC and SMC sightlines, we could not obtain reasonable constraint on CRIR using HD/H$_2$. The reason is that we get insufficient constraints on the HD column density due to low quality of the FUSE spectra. To illustrate this in Fig.~\ref{fig:highres} we compare line profiles and estimates of the fitting parameters of HD obtained from the FUSE spectra towards AV\,490 and mock high resolution spectrum (with $R=50000$ and SNR=20). The original FUSE data indicates inability to remove degeneracy between high and low column density solutions, resulting in the wide ($\sim 2$ dex) constraint on the column density. One can see that the usage of the higher resolution not only allow to decompose the two velocity components in HD lines, but significantly reduces the constrained interval for HD column density. Moreover, the inclusion of the weakest HD line L0-0R(0)\footnote{note that L1-0(0) and L2-0R(0) are usually blended with H$_2$ absorption lines} allows one to further shrink the interval and to get the uncertainties comparable to what is obtained for HD in high redshift DLAs \citealt[see e.g.][]{Kosenko2021}. This situation is similar to the case of the galactic sightlines, where HD measurement are also shallow, while typically galactic sightlines have higher SNR, which can reduce the uncertainty, but cannot remove the degeneracy. The lower uncertainty of HD column density will allow to get more strict constraints on CRIR, instead of upper limits obtained using FUSE real data, as is shown in the right panel of Fig.~\ref{fig:highres}. Therefore, the next generation of the space UV telescopes, equipped with high resolution spectrograph will allow to constrain the CRIR using HD/H$_2$ ratio, that will provide important insights on the cold diffuse medium of the Magellanic clouds, as well as in our Galaxy. 

\section{Summary}
\label{sect:summary}

We have provided a systematic search for HD molecules in the LMC and SMC, the closest low-metallicity dwarf galaxies, using FUSE archival data. To make analysis homogeneous we reanalyzed H$_2$ using multicomponent Voight profile fitting of H$_2$ $\rm J\lesssim 5$ rotational levels. On the positions of H$_2$ components we looked for HD molecules. Although the quality of the FUSE data allows to obtain in the most cases only upper limits on HD column densities, we detected HD towards 24 sightlines (including 19 new detection). 

In six systems (out of 24 where we have detected HD) we obtained constraints on CRIR, and in five among them we could place only upper limits, mostly due to large uncertainties in HD column density estimation. In the five systems we could not obtain reasonable constraints on the CRIR due to the too wide form of posterior probability functions, which cover all of the range within reasonable range of CRIR.

We also discuss influence of the spectral resolution on the results of our analysis and show that increase of the resolution of data will result in decrease of the column density uncertainties, which allows to obtain strict constraints on CRIR. 

\section*{Acknowledgements} 
This work was supported by RSF grant 18-12-00301.

\section*{Data availability} 
The results of this paper are based on open data retrieved from the FUSE telescope archive. These data can be shared on reasonable requests to the authors.

\bibliographystyle{mnras}
\bibliography{references.bib}

\appendix
\section{New HD detections}
\label{sect:New_HD}
%\DK{I suggest to do as referee said: to put figures with all of the new HD detections in the appendix in the main part of the paper, leaving maybe one system, as an example, in the main body of the paper.}

\begin{figure*}
    \centering
    \includegraphics[width=\linewidth]{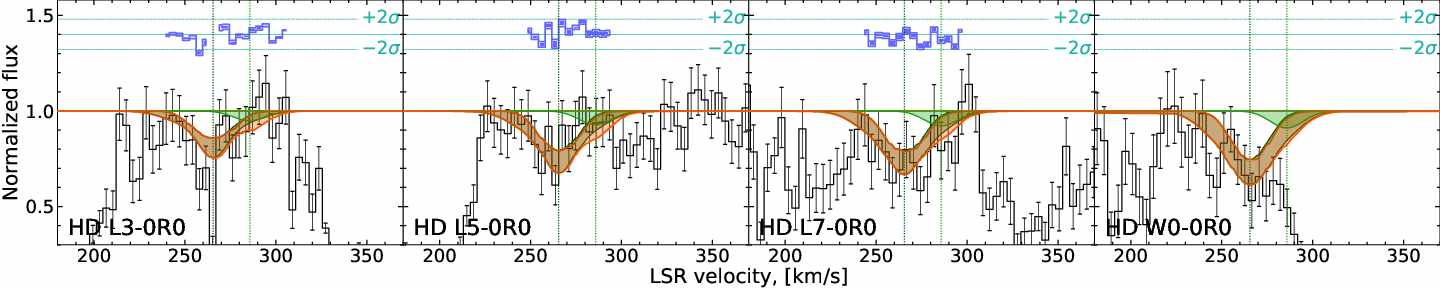}
    \caption{Fit to HD absorption lines towards LH10 3073 in LMC. Lines are the same as for \ref{fig:lines_HD_Sk191}.}
\end{figure*}
\begin{figure*}
    \centering
    \includegraphics[width=\textwidth]{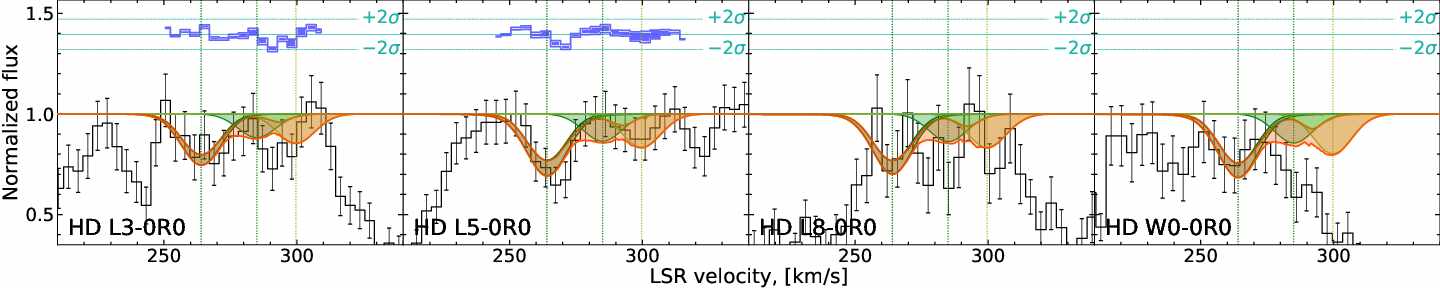}
    \caption{Fit to HD absorption lines towards LH10 3102 in LMC. Lines are the same as for \ref{fig:lines_HD_Sk191}.}
\end{figure*}
\begin{figure*}
    \centering
    \includegraphics[width=\textwidth]{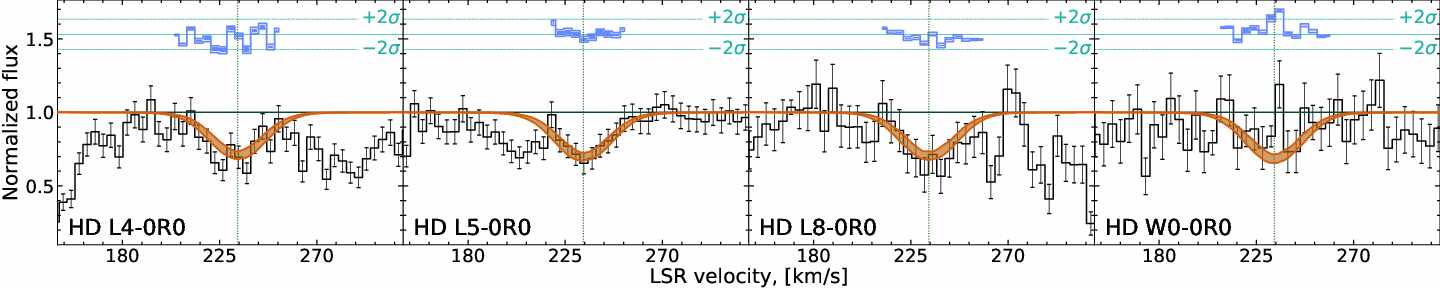}
    \caption{Fit to HD absorption lines towards Sk-70 79 in LMC. Lines are the same as for \ref{fig:lines_HD_Sk191}.}
\end{figure*}
\begin{figure*}
    \centering
    \includegraphics[width=\textwidth]{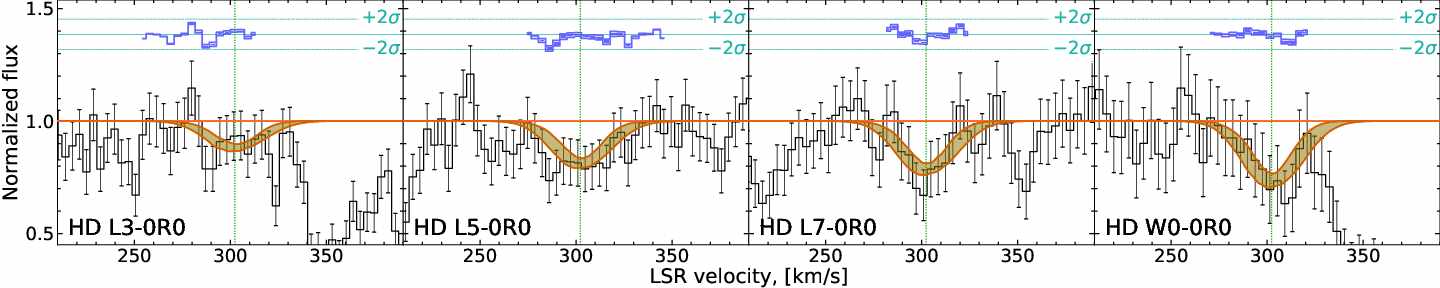}
    \caption{Fit to HD absorption lines towards Sk-68 73 in LMC. Lines are the same as for \ref{fig:lines_HD_Sk191}.}
\end{figure*}
\begin{figure*}
    \centering
    \includegraphics[width=\textwidth]{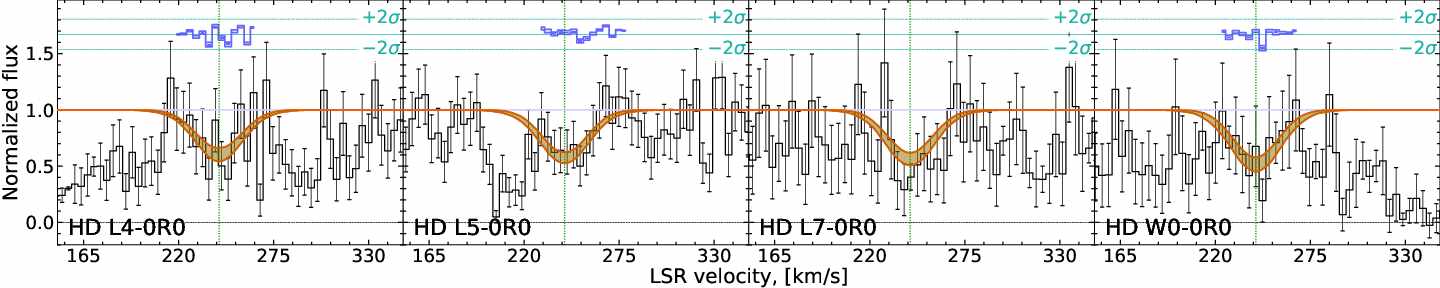}
    \caption{Fit to HD absorption lines towards Sk-71 46 in LMC. Lines are the same as for \ref{fig:lines_HD_Sk191}.}
\end{figure*}
\begin{figure*}
    \centering
    \includegraphics[width=\textwidth]{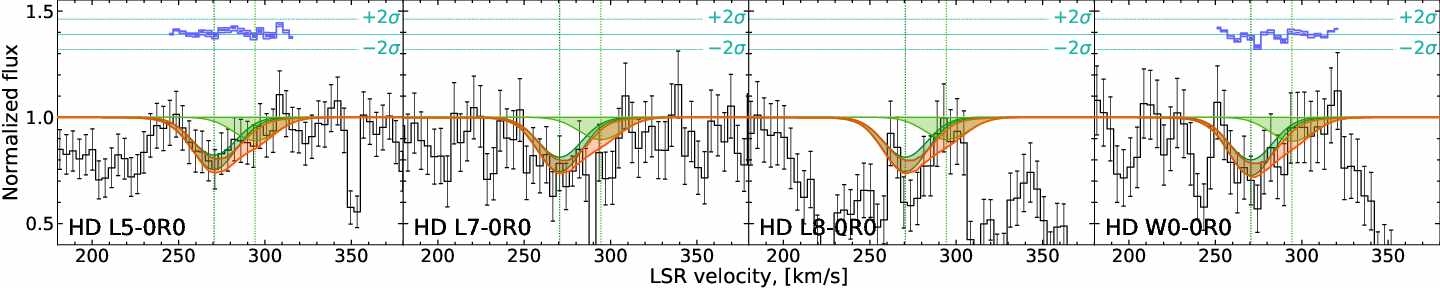}
    \caption{Fit to HD absorption lines towards J\,0534-6932 in LMC. Lines are the same as for \ref{fig:lines_HD_Sk191}.}
\end{figure*}
\begin{figure*}
    \centering
    \includegraphics[width=\textwidth]{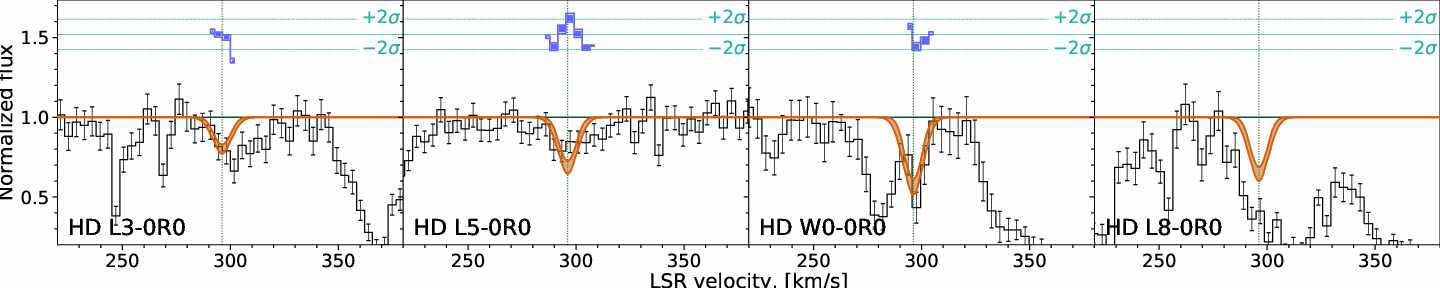}
    \caption{Fit to HD absorption lines towards BI 237 in LMC. Lines are the same as for \ref{fig:lines_HD_Sk191}.}
\end{figure*}
\begin{figure*}
    \centering
    \includegraphics[width=\textwidth]{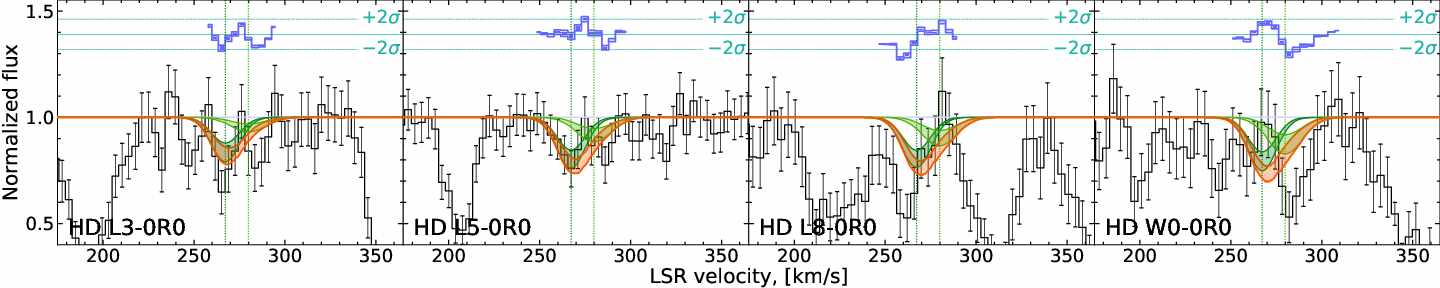}
    \caption{Fit to HD absorption lines towards BI 253 in LMC. Lines are the same as for \ref{fig:lines_HD_Sk191}.}
\end{figure*}
\begin{figure*}
    \centering
    \includegraphics[width=\textwidth]{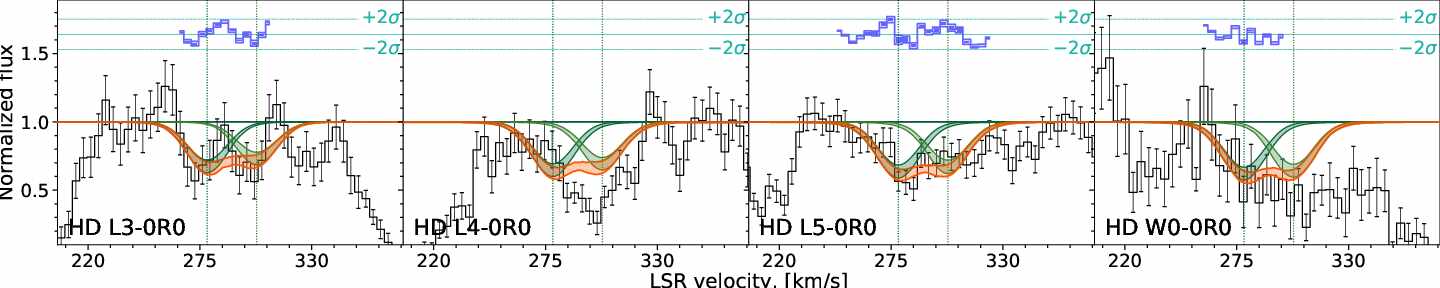}
    \caption{Fit to HD absorption lines towards Sk-68 137 in LMC. Lines are the same as for \ref{fig:lines_HD_Sk191}.}
\end{figure*}
\begin{figure*}
    \centering
    \includegraphics[width=\textwidth]{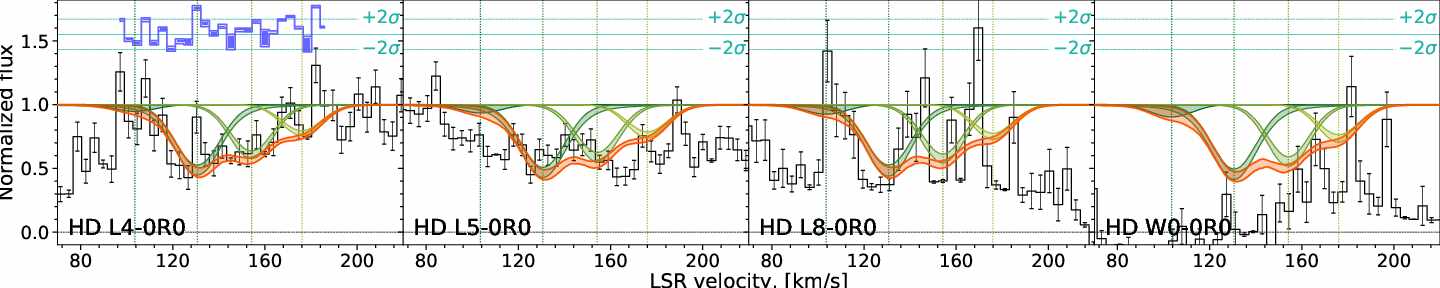}
    \caption{Fit to HD absorption lines towards AV 18 in SMC. Lines are the same as for \ref{fig:lines_HD_Sk191}.}
\end{figure*}
\begin{figure*}
    \centering
    \includegraphics[width=\textwidth]{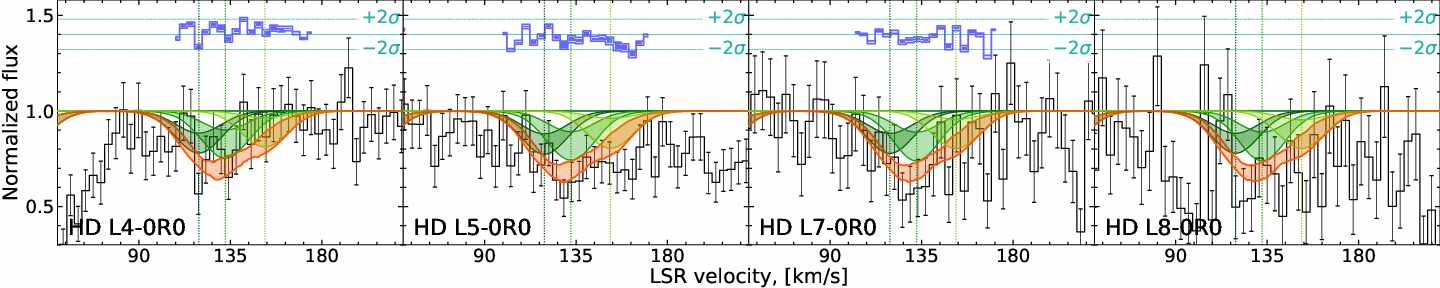}
    \caption{Fit to HD absorption lines towards AV 26 in SMC. Lines are the same as for \ref{fig:lines_HD_Sk191}.}
\end{figure*}
\begin{figure*}
    \centering
    \includegraphics[width=\textwidth]{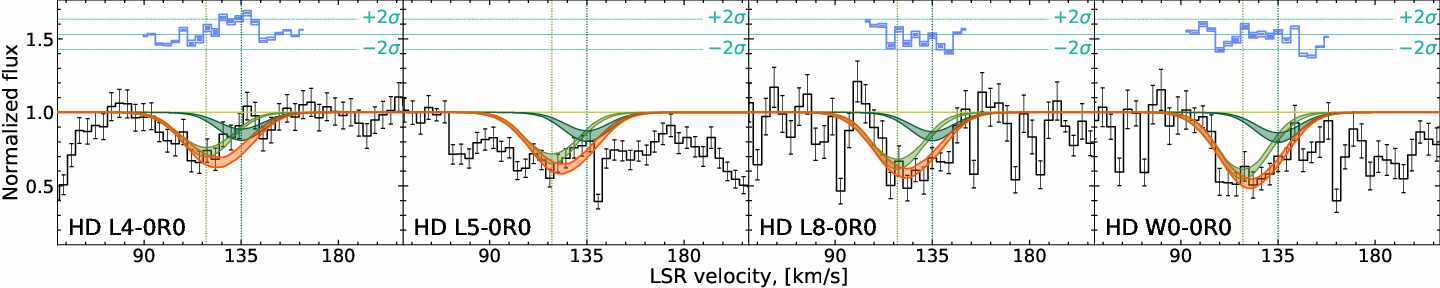}
    \caption{Fit to HD absorption lines towards AV 80 in SMC. Lines are the same as for \ref{fig:lines_HD_Sk191}.}
\end{figure*}
\begin{figure*}
    \centering
    \includegraphics[width=\textwidth]{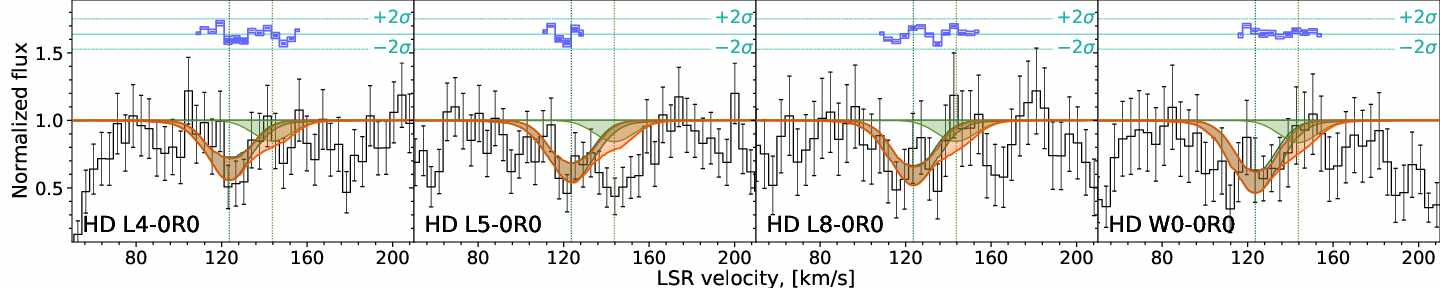}
    \caption{Fit to HD absorption lines towards AV 208 in SMC. Lines are the same as for \ref{fig:lines_HD_Sk191}.}
\end{figure*}
\begin{figure*}
    \centering
    \includegraphics[width=\textwidth]{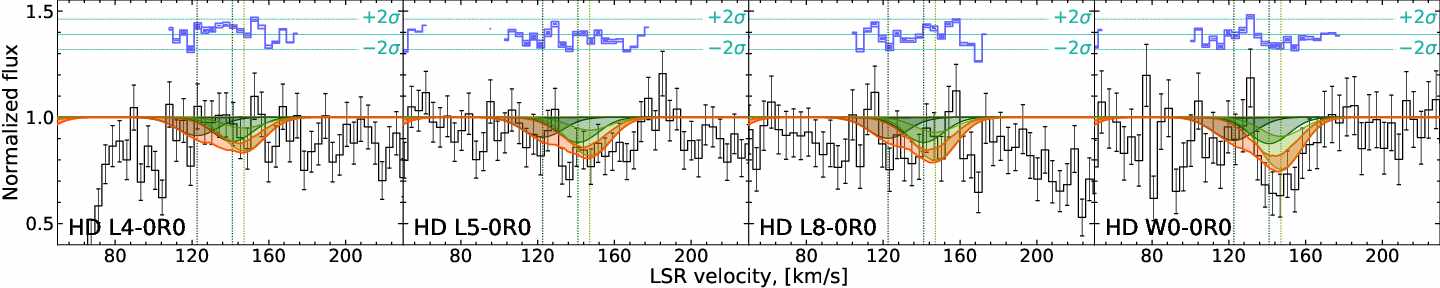}
    \caption{Fit to HD absorption lines towards AV 372 in SMC. Lines are the same as for \ref{fig:lines_HD_Sk191}.}
\end{figure*}
\begin{figure*}
    \centering
    \includegraphics[width=\textwidth]{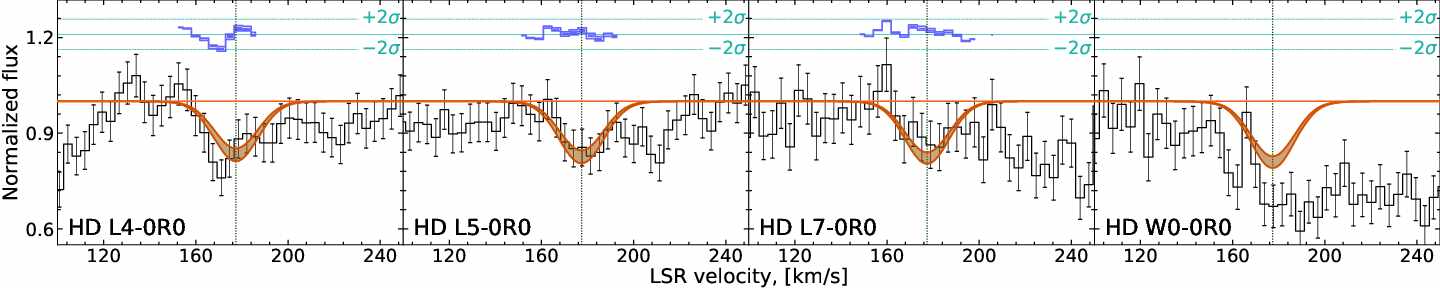}
    \caption{Fit to HD absorption lines towards AV 435 in SMC. Lines are the same as for \ref{fig:lines_HD_Sk191}.}
\end{figure*}
\begin{figure*}
    \centering
    \includegraphics[width=\textwidth]{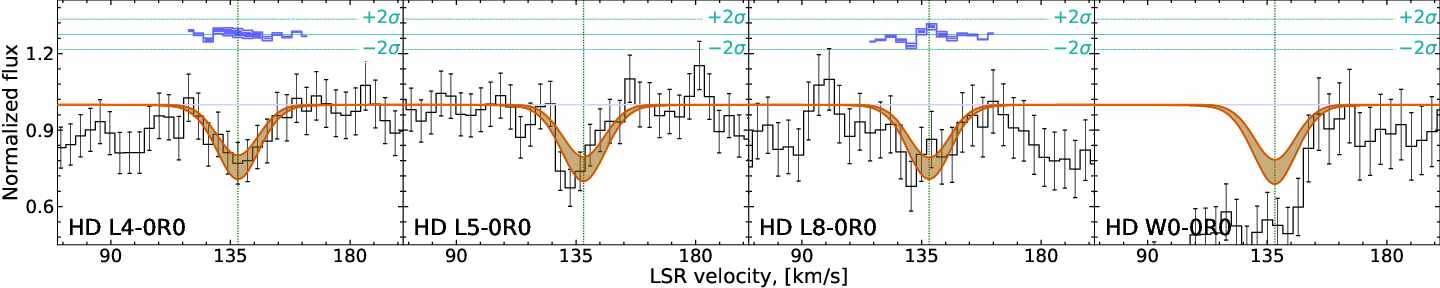}
    \caption{Fit to HD absorption lines towards AV 472 in SMC. Lines are the same as for \ref{fig:lines_HD_Sk191}.}
\end{figure*}
\begin{figure*}
    \centering
    \includegraphics[width=\textwidth]{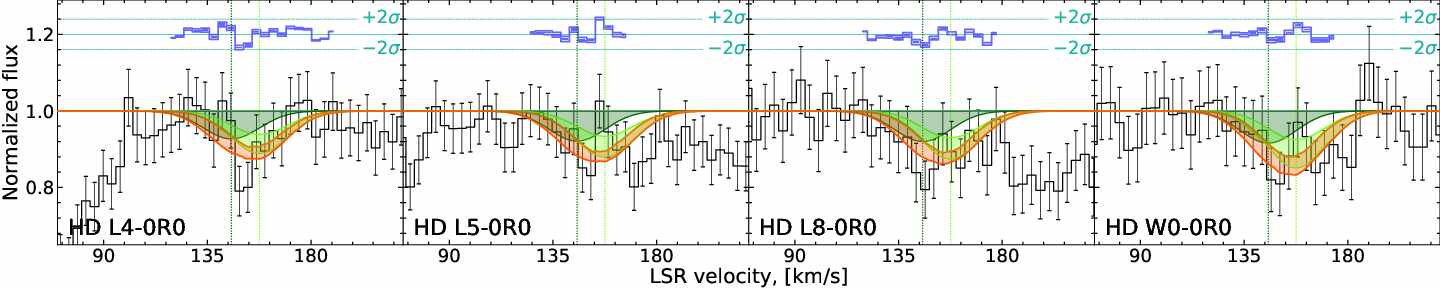}
    \caption{Fit to HD absorption lines towards AV 486 in SMC. Lines are the same as for \ref{fig:lines_HD_Sk191}.}
\end{figure*}
\begin{figure*}
    \centering
    \includegraphics[width=\textwidth]{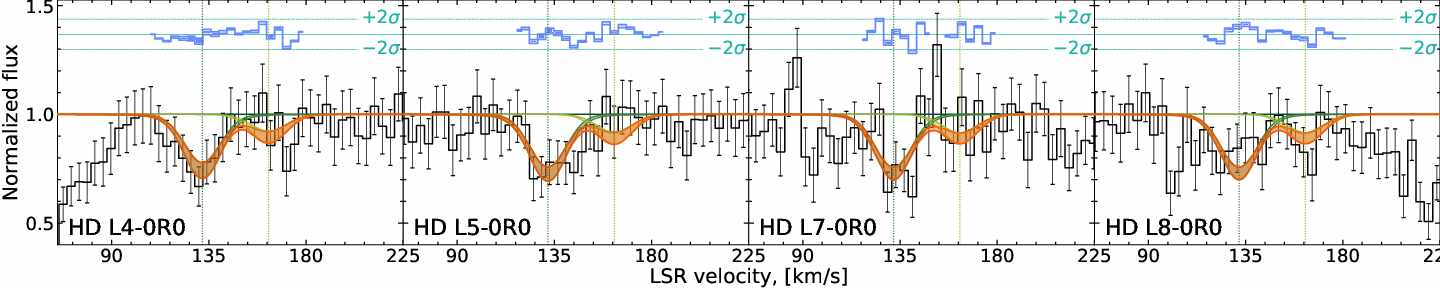}
    \caption{Fit to HD absorption lines towards AV 490 in SMC. Lines are the same as for \ref{fig:lines_HD_Sk191}.}
\end{figure*}

\clearpage
\section{Detailed fitting results}
\label{appendix:fitting}
\subsection{Large Magellanic Cloud}

\begin{table*}
    \caption{Fit results of H$_2$ lines towards Sk-67 2}
    \label{tab:Sk67_2}
    \begin{tabular}{ccccc}
    \hline
    \hline
    species & comp & 1 & 2 & 3   \\
            & z & $0.0000665(^{+25}_{-19})$ & $0.000869(^{+3}_{-6})$ & $0.000935(^{+6}_{-6})$ \\
    \hline 
    ${\rm H_2\, J=0}$ & b\,km/s & $0.91^{+0.72}_{-0.12}$ & $0.81^{+0.14}_{-0.31}$ & $0.98^{+0.23}_{-0.42}$ \\
                      & $\log N$ &  $19.66^{+0.13}_{-0.09}$ & $20.38^{+0.17}_{-0.65}$ & $20.16^{+0.21}_{-0.43}$ \\
    ${\rm H_2\, J=1}$ & b\,km/s & $1.7^{+0.4}_{-0.8}$ &$1.0^{+0.5}_{-0.4}$ & $1.6^{+0.3}_{-0.4}$ \\
                      & $\log N$ &  $19.4^{+0.4}_{-5.0}$ & $19.64^{+0.14}_{-0.06}$ & $20.01^{+0.10}_{-0.08}$\\
    ${\rm H_2\, J=2}$ & b\,km/s & $4.4^{+0.3}_{-1.5}$ & $2.0^{+0.5}_{-0.5}$ & $1.82^{+0.45}_{-0.30}$ \\
                      & $\log N$ & $17.73^{+0.09}_{-0.47}$ & $17.67^{+0.19}_{-0.23}$ & $18.61^{+0.04}_{-0.17}$\\
    ${\rm H_2\, J=3}$ & b\,km/s & $5.01^{+0.40}_{-0.31}$ & $2.43^{+0.21}_{-0.35}$ & $2.52^{+0.13}_{-0.41}$\\
                      & $\log N$ & $15.87^{+0.42}_{-0.21}$ & $17.65^{+0.09}_{-0.25}$ &$17.29^{+0.14}_{-0.27}$ \\
    ${\rm H_2\, J=4}$ & $\log N$ & $14.85^{+0.12}_{-0.14}$ & $16.96^{+0.18}_{-0.72}$ & $16.1^{+0.9}_{-0.3}$ \\
    \hline 
        & $\log N_{\rm tot}$ & $19.84^{+0.19}_{-0.20}$ & $20.46^{+0.15}_{-0.46}$ & $20.40^{+0.14}_{-0.20}$ \\
                       
    \hline
    ${\rm HD\, J=0}$ & b\,km/s &  $1.11^{+0.58}_{-0.29}$ & $0.78^{+0.19}_{-0.15}$ & $1.00^{+0.19}_{-0.34}$ \\
                     & $\log N$ & $\lesssim 16.9$ & $\lesssim 17.1$ & $\lesssim 17.3$ \\
    \hline                  
    \end{tabular}
    \begin{tablenotes}
    \item Doppler parameters of H$_2$ ${\rm J = 4}$ rotational levels were tied to H$_2$ ${\rm J = 3}$. 
    
    We have not fitted H$_2$ $\rm J > 4$ lines due to low signal-to-noise ratio.
    \end{tablenotes}
\end{table*}

\begin{figure*}
    \centering
    \includegraphics[width=\linewidth]{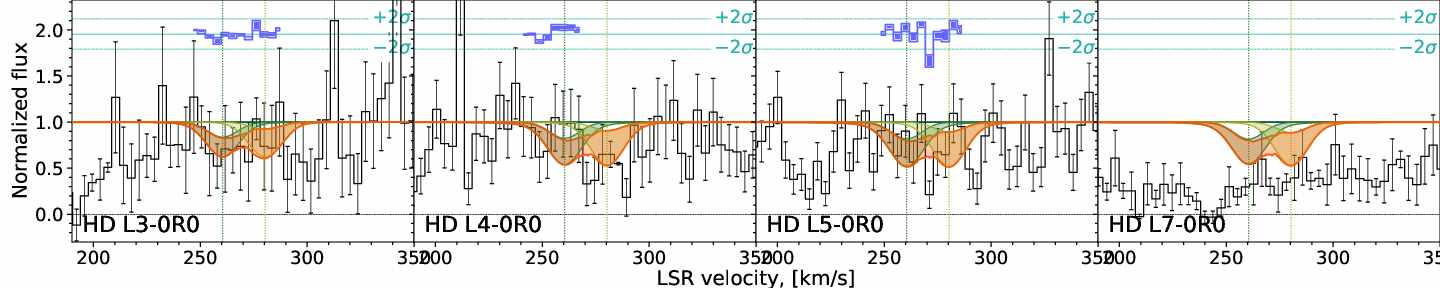}
    \caption{Fit to HD absorption lines towards Sk-67 2 in LMC.
     Here black line shows spectrum, red line represents obtained synthetic spectrum for HD. Green lines represent individual components. Regions between red  and green lines were sampled from posterior probability distributions of fitting parameters. Blue points at the top of each panel show residuals. Here we show only components found in Magellanic Clouds.
     %Lines are the same as for \ref{fig:lines_HD_J0534}.
    }
    \label{fig:lines_HD_Sk67_2}
\end{figure*}

\begin{figure*}
    \centering
    \includegraphics[width=\linewidth]{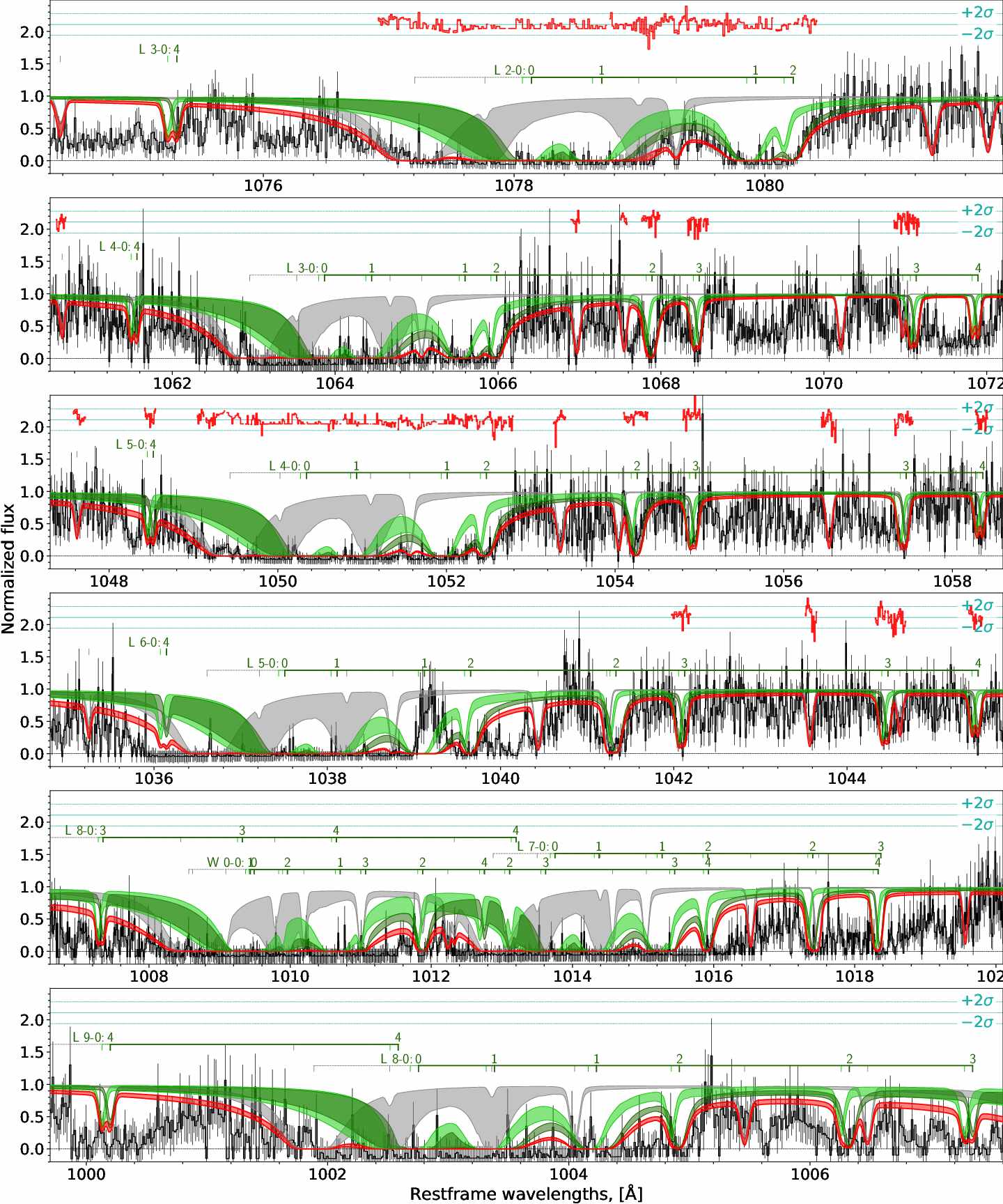}
    \caption{Fit to H$_2$ absorption lines towards Sk-67 2 in LMC. Here black points with uncertainties show spectrum, red line represents total systhetic spectrum for H$_2$, grey lines show Milky Way subcomponents, green lines show Magellanic Cloud subcomponents. Regions between red, grey and green lines were sampled from posterior probability distributions of fitting parameters. Red points at the top of each panel show residuals.
    }
    \label{fig:lines_H2_Sk67_2}
\end{figure*}

\begin{table*}
    \caption{Fit results of H$_2$ lines towards Sk-67 5}
    \label{tab:Sk67_5}
    \begin{tabular}{cccc}
    \hline
    \hline
    species & comp & 1 & 2   \\
            & z & $0.0000933(^{+12}_{-5})$ &  $0.00098317(^{+23}_{-34})$ \\
    \hline 
    ${\rm H_2\, J=0}$ & b\,km/s & $1.66^{+0.43}_{-0.19}$ & $1.0^{+1.0}_{-0.5}$\\
                      & $\log N$ &  $15.6^{+0.3}_{-0.5}$ & $19.315^{+0.004}_{-0.005}$ \\
    ${\rm H_2\, J=1}$ & b\,km/s & $5.01^{+0.34}_{-0.20}$ & $4.65^{+0.24}_{-0.36}$ \\
                      & $\log N$ &  $15.18^{+0.04}_{-0.06}$ & $18.952^{+0.007}_{-0.005}$ \\
    ${\rm H_2\, J=2}$ & b\,km/s & $5.68^{+0.30}_{-0.72}$ & $4.65^{+0.16}_{-0.27}$ \\
                      & $\log N$ & $14.395^{+0.037}_{-0.015}$ &   $16.88^{+0.14}_{-0.14}$ \\
    ${\rm H_2\, J=3}$ & b\,km/s & $6.08^{+0.26}_{-0.38}$ & $4.51^{+0.18}_{-0.20}$ \\
                      & $\log N$ & $14.352^{+0.019}_{-0.034}$ &  $15.97^{+0.14}_{-0.09}$ \\
    ${\rm H_2\, J=4}$ & $\log N$ & $13.20^{+0.11}_{-0.15}$ & $14.644^{+0.036}_{-0.013}$ \\
    ${\rm H_2\, J=5}$ & $\log N$ & $12.86^{+0.21}_{-0.32}$ & $13.90^{+0.06}_{-0.07}$ \\
    \hline 
        & $\log N_{\rm tot}$ & $15.81^{+0.23}_{-0.28}$ & $19.473^{+0.003}_{-0.005}$ \\
    \hline
    {\rm HD\, J=0} & b\,km/s & $9.81^{+0.19}_{-9.31}$ & $0.31^{+1.77}_{-0.30}$ \\
                   & $\log N_{\rm tot}$ & $\lesssim 13.9$ & $15.5^{+0.3}_{-2.1}$ \\

    \hline
    \end{tabular}
    \begin{tablenotes}
    \item Doppler parameters of H$_2$ ${\rm J = 4, 5}$ rotational levels were tied to H$_2$ ${\rm J = 3}$, while H$_2$ ${\rm J = 1, 2, 3}$ were varied using penalty function.
    \end{tablenotes}
\end{table*}

\begin{figure*}
    \centering
    \includegraphics[width=\linewidth]{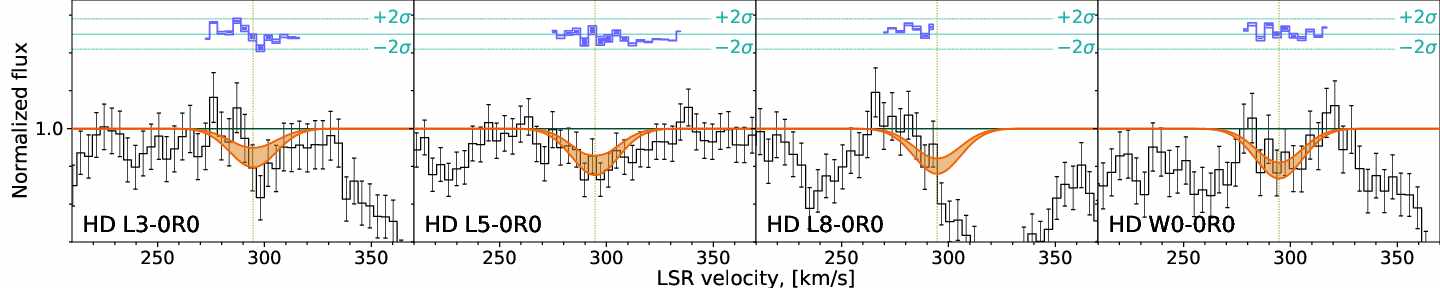}
    \caption{Fit to HD absorption lines towards Sk-67 5 in LMC. Lines are the same as for \ref{fig:lines_HD_Sk67_2}.
    }
    \label{fig:lines_HD_Sk67_5}
\end{figure*}

\begin{figure*}
    \centering
    \includegraphics[width=\linewidth]{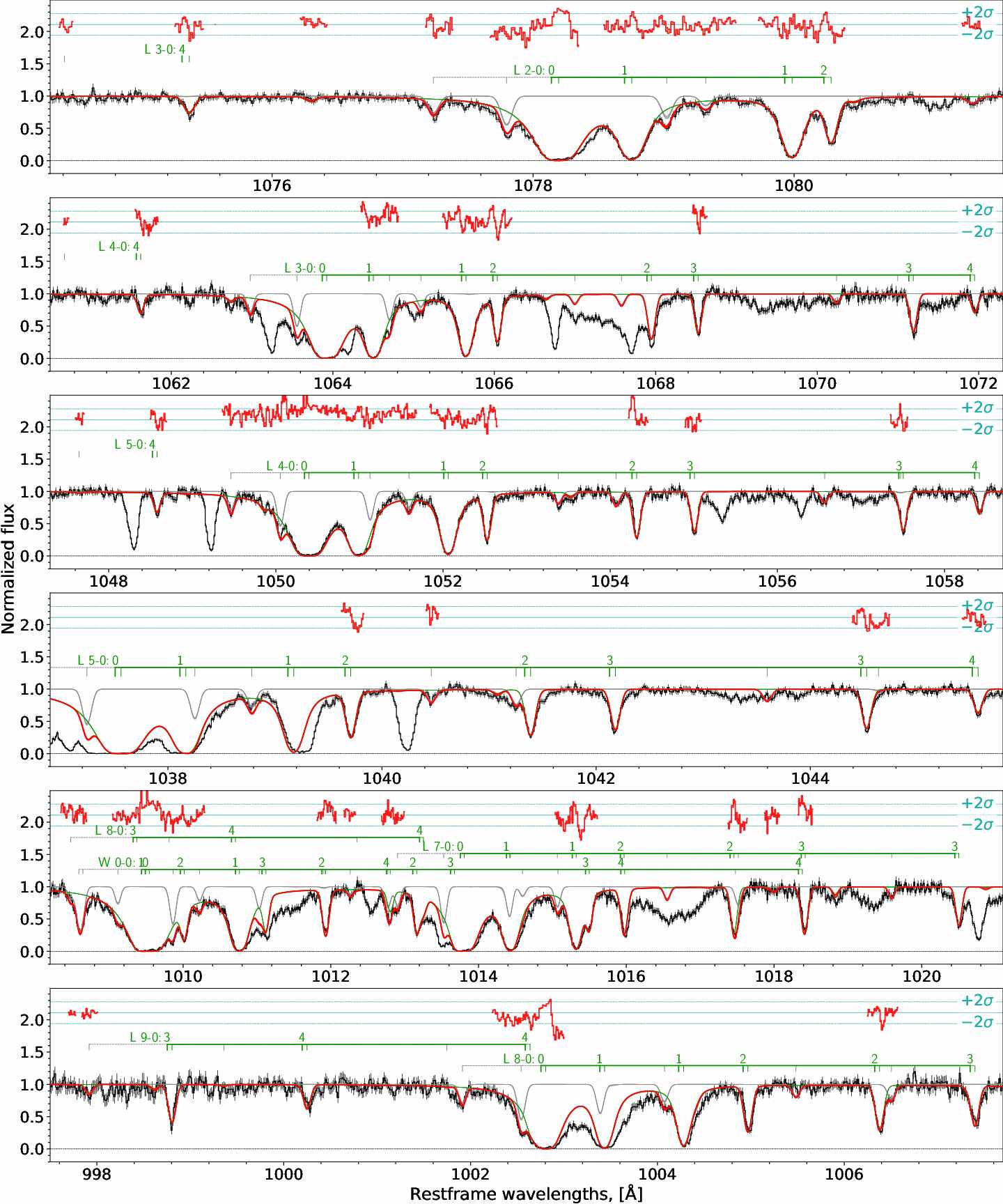}
    \caption{Fit to H2 absorption lines towards Sk-67 5 in LMC. Lines are the same as for \ref{fig:lines_H2_Sk67_2}.
    }
    \label{fig:lines_H2_Sk67_5}
\end{figure*}

\begin{table*}
    \caption{Fit results of H$_2$ lines towards Sk-66 1}
    \label{tab:Sk66_1}
    \begin{tabular}{ccccc}
    \hline
    \hline
    species & comp & 1 & 2 & 3 \\
            & z & $-0.0000044(^{+20}_{-9})$ & $0.0000860(^{+13}_{-8})$ & $0.0009551(^{+11}_{-3})$ \\
    \hline 
     ${\rm H_2\, J=0}$ & b\,km/s & $0.94^{+1.56}_{-0.22}$ & $7.3^{+0.4}_{-0.6}$ & $1.6^{+1.1}_{-1.1}$ \\
                       & $\log N$ & $18.602^{+0.032}_{-0.030}$ & $18.32^{+0.07}_{-0.06}$ & $19.128^{+0.017}_{-0.006}$  \\
    ${\rm H_2\, J=1}$ & b\,km/s & $2.9^{+0.6}_{-1.3}$ & $7.1^{+0.6}_{-0.6}$ & $3.8^{+0.4}_{-2.5}$ \\
                      & $\log N$ & $18.922^{+0.029}_{-0.037}$ & $17.21^{+0.47}_{-0.16}$ & $18.875^{+0.026}_{-0.013}$ \\
    ${\rm H_2\, J=2}$ & b\,km/s & $3.1^{+0.4}_{-0.3}$ & $6.8^{+0.6}_{-0.5}$ & $4.6^{+0.5}_{-1.0}$ \\
                      & $\log N$ & $16.87^{+0.29}_{-0.31}$ & $15.89^{+0.17}_{-0.17}$ & $15.60^{+0.38}_{-0.26}$ \\
    ${\rm H_2\, J=3}$ & b\,km/s & $3.3^{+0.6}_{-0.5}$ & $7.2^{+0.5}_{-0.6}$ & $8.2^{+1.5}_{-0.9}$ \\
                      & $\log N$ & $15.5^{+0.4}_{-0.4}$ & $15.50^{+0.15}_{-0.13}$ & $14.93^{+0.05}_{-0.05}$ \\
    ${\rm H_2\, J=4}$ & b\,km/s & $5.6^{+2.5}_{-2.8}$ & $7.3^{+1.2}_{-0.7}$ & $9.89^{+4.24}_{-4.34}$ \\
                      & $\log N$ & $14.39^{+0.09}_{-0.07}$ & $14.49^{+0.05}_{-0.05}$ & $14.37^{+0.06}_{-0.06}$ \\
    \hline 
         & $\log N_{\rm tot}$ & $19.09^{+0.02}_{-0.03}$ & $18.35^{+0.08}_{-0.06}$ & $19.32^{+0.01}_{-0.01}$ \\
    \hline
    {\rm HD\, J=0} & b\,km/s & $0.73^{+1.11}_{-0.23}$ & $7.3^{+0.5}_{-0.7}$ & $0.57^{+1.65}_{-0.07}$ \\
                    & $\log N$ & $\lesssim 16.1$ & $\lesssim 13.9$ & $\lesssim 15.9$ \\
                  
    \hline   
    \end{tabular}
\end{table*}

\begin{figure*}
    \centering
    \includegraphics[width=\linewidth]{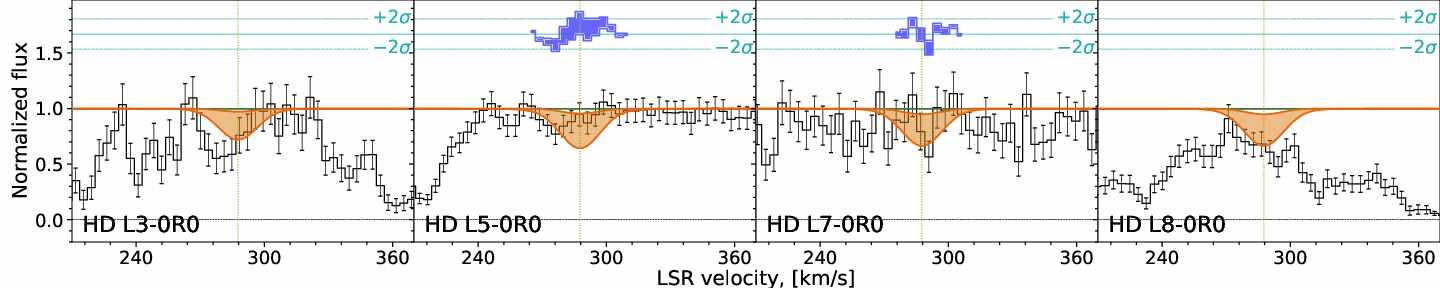}
    \caption{Fit to HD absorption lines towards Sk-66 1 in LMC. Lines are the same as for \ref{fig:lines_HD_Sk67_2}.
    }
    \label{fig:lines_HD_Sk66_1}
\end{figure*}

\begin{figure*}
    \centering
    \includegraphics[width=\linewidth]{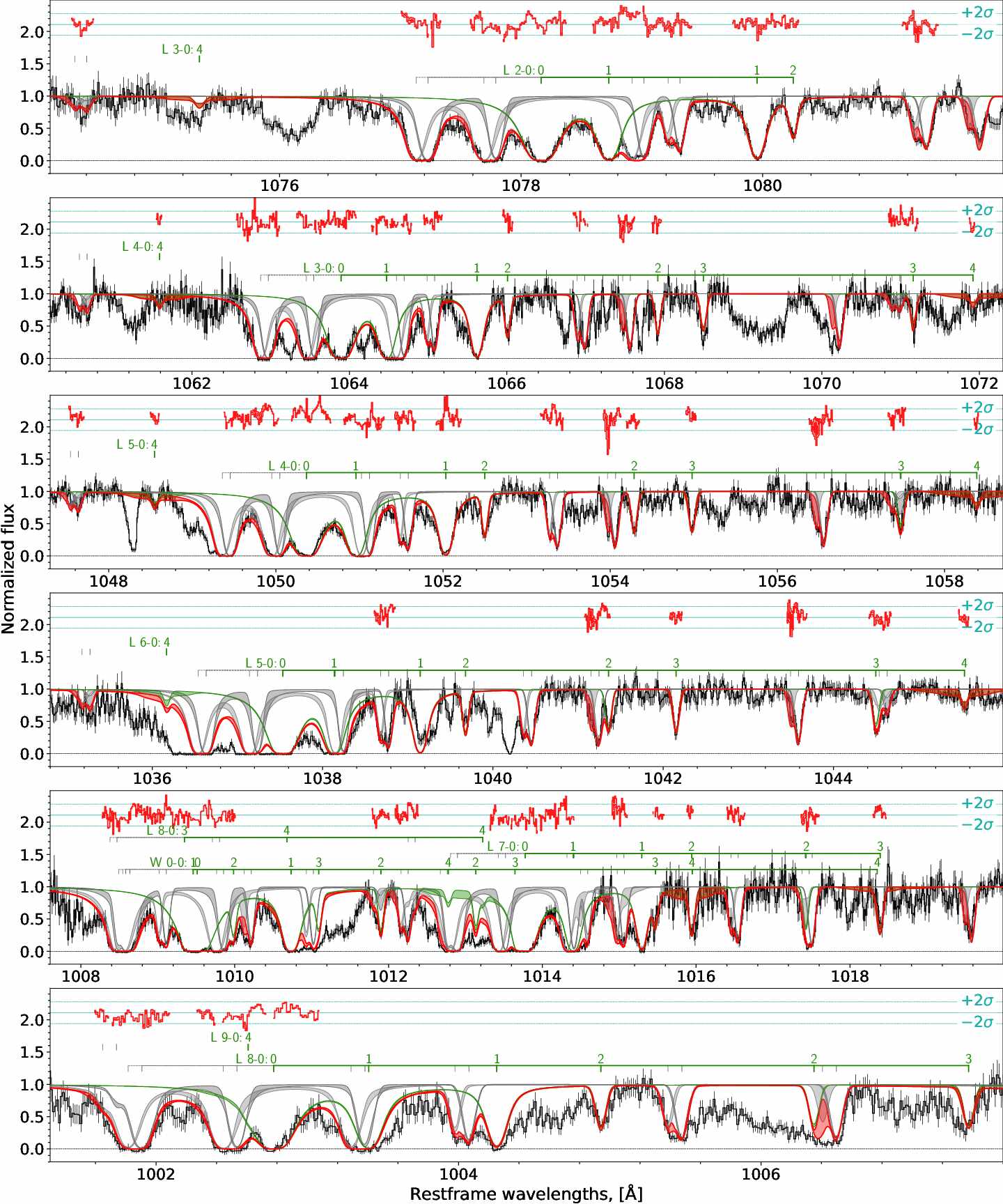}
    \caption{Fit to H2 absorption lines towards Sk-66 1 in LMC. Lines are the same as for \ref{fig:lines_H2_Sk67_2}.
    }
    \label{fig:lines_H2_Sk66_1}
\end{figure*}

\begin{table*}
    \caption{Fit results of H$_2$ lines towards Sk-67 20}
    \label{tab:Sk67_20}
    \begin{tabular}{ccccc}
    \hline
    \hline
    species & comp & 1 & 2 & 3 \\
            & z & $-0.0000199(^{+7}_{-8})$ & $0.0000597(^{+8}_{-12})$ & $0.00094472(^{+26}_{-47})$ \\
    \hline 
     ${\rm H_2\, J=0}$ & b\,km/s &  $1.8^{+0.4}_{-0.6}$ & $2.08^{+0.38}_{-0.26}$ & $1.6^{+0.6}_{-0.8}$ \\
                       & $\log N$ & $17.751^{+0.027}_{-0.048}$ & $17.05^{+0.09}_{-0.18}$ & $18.775^{+0.012}_{-0.006}$  \\
    ${\rm H_2\, J=1}$ & b\,km/s & $2.03^{+0.36}_{-0.22}$ & $2.8^{+0.4}_{-0.3}$ & $2.65^{+0.31}_{-0.21}$ \\
                      & $\log N$ & $18.120^{+0.024}_{-0.027}$ &  $16.89^{+0.24}_{-0.46}$ & $18.642^{+0.007}_{-0.009}$ \\
    ${\rm H_2\, J=2}$ & b\,km/s & $2.10^{+0.17}_{-0.26}$ & $3.5^{+0.4}_{-0.3}$ & $2.76^{+0.15}_{-0.15}$ \\
                      & $\log N$ & $16.95^{+0.10}_{-0.18}$ & $15.21^{+0.17}_{-0.16}$ & $17.41^{+0.05}_{-0.04}$ \\
    ${\rm H_2\, J=3}$ & b\,km/s & $2.49^{+0.08}_{-0.08}$ & $3.77^{+0.55}_{-0.25}$ & $2.73^{+0.11}_{-0.24}$ \\
                      & $\log N$ & $16.26^{+0.11}_{-0.11}$ & $14.92^{+0.09}_{-0.08}$ & $17.40^{+0.07}_{-0.06}$ \\
    ${\rm H_2\, J=4}$ & b\,km/s & -- & $4.3^{+0.5}_{-0.5}$ & $2.76^{+0.19}_{-0.16}$ \\
    				  & $\log N$ & $13.96^{+0.06}_{-0.09}$ & $13.71^{+0.11}_{-0.11}$ & $15.98^{+0.16}_{-0.12}$ \\
    ${\rm H_2\, J=5}$ & b\,km/s & -- & -- & $3.59^{+0.11}_{-0.30}$ \\
    				  & $\log N$ & $13.95^{+0.07}_{-0.15}$ & $13.88^{+0.12}_{-0.10}$ &$15.26^{+0.11}_{-0.11}$ \\
    ${\rm H_2\, J=6}$ & $\log N$ & -- & -- & $13.74^{+0.06}_{-0.09}$ \\	    
    \hline 
         & $\log N_{\rm tot}$ & $18.30^{+0.02}_{-0.02}$ & $17.28^{+0.12}_{-0.17}$ & $19.036^{+0.007}_{-0.006}$ \\
    \hline
    ${\rm HD\, J=0}$ & b\,km/s & $0.527^{+1.279}_{-0.027}$ & $2.05^{+0.45}_{-0.28}$ & $0.53^{+1.08}_{-0.03}$ \\
                    & $\log N$ & $\lesssim 15.3$ & $\lesssim 13.9$ & $\lesssim 14.2$ \\
                  
    \hline   
    \end{tabular}
    \begin{tablenotes}
    \item Doppler parameters of H$_2$ ${\rm J = 4, 5}$ rotational levels in the 1 component, H$_2$ $J=5$ in the 2 component and H$_2$ $J=6$ in the 3 component were tied to H$_2$ ${\rm J = 3}$. 
    \end{tablenotes}
\end{table*}

\begin{figure*}
    \centering
    \includegraphics[width=\linewidth]{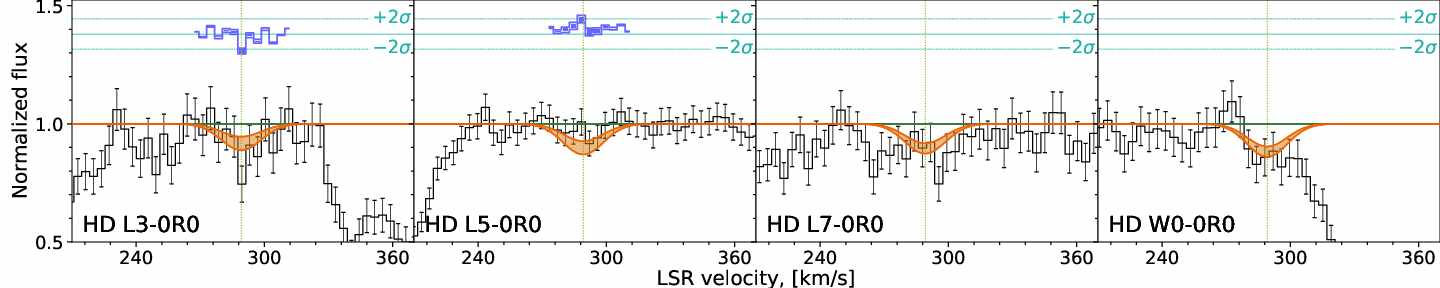}
    \caption{Fit to HD absorption lines towards Sk-67 20 in LMC. Lines are the same as for \ref{fig:lines_HD_Sk67_2}.
    }
    \label{fig:lines_HD_Sk67_20}
\end{figure*}

\begin{figure*}
    \centering
    \includegraphics[width=\linewidth]{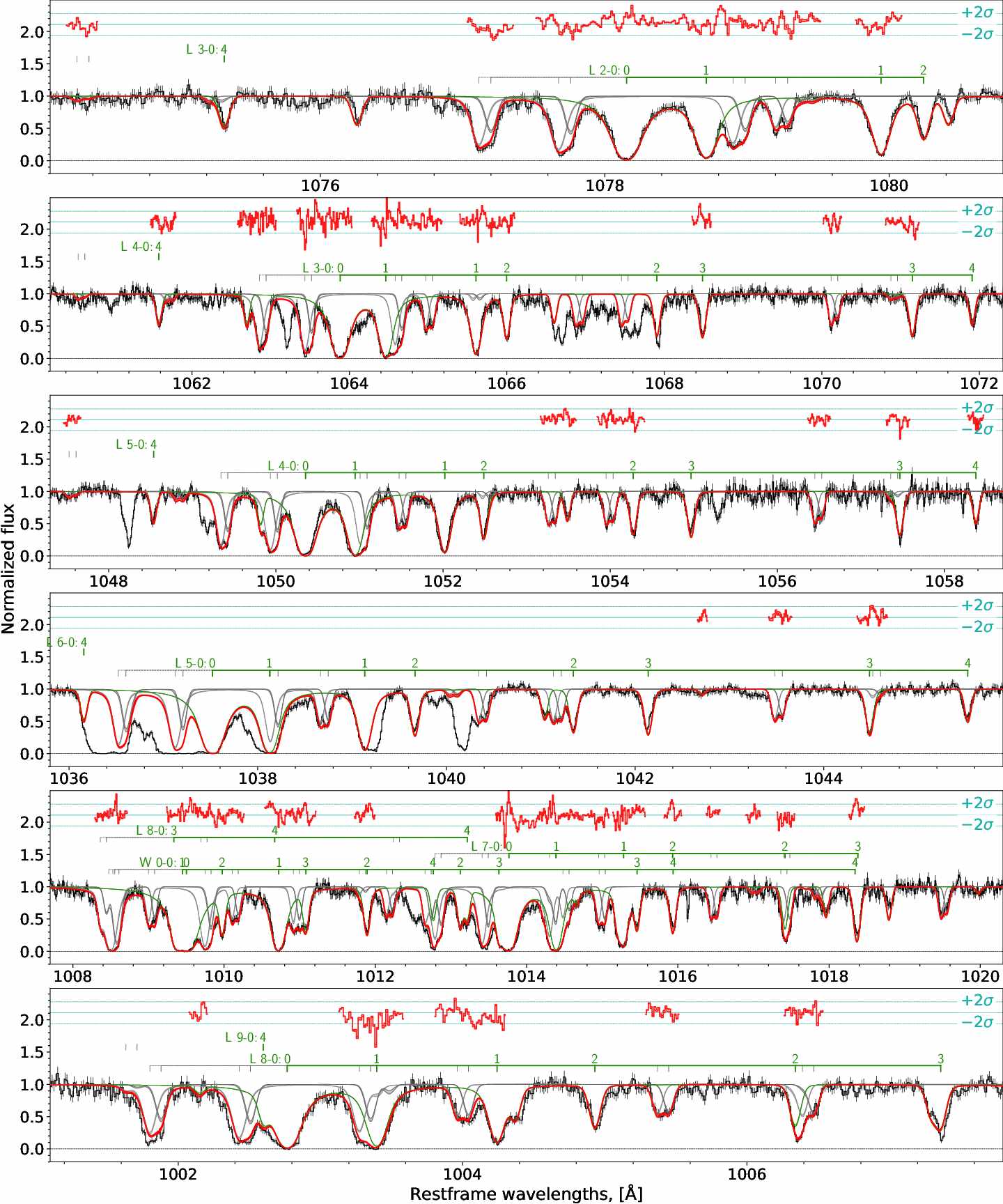}
    \caption{Fit to H2 absorption lines towards Sk-67 20 in LMC. Lines are the same as for \ref{fig:lines_H2_Sk67_2}.
    }
    \label{fig:lines_H2_Sk67_20}
\end{figure*}

\begin{table*}
    \caption{Fit results of H$_2$ lines towards Sk-66 18}
    \label{tab:Sk66_18}
    \begin{tabular}{cccccc}
    \hline
    \hline
    species & comp & 1 & 2 & 3 & 4 \\
            & z & $0.0000084(^{+14}_{-10})$ & $0.0000945(^{+8}_{-7})$ &  $0.0009155(^{+20}_{-31})$ & $0.0009712(^{+31}_{-23})$ \\
    \hline 
     ${\rm H_2\, J=0}$ & b\,km/s & $2.29^{+0.19}_{-0.16}$ & $2.8^{+0.9}_{-0.5}$ &  $2.9^{+0.3}_{-0.6}$ &  $0.73^{+0.47}_{-0.23}$ \\
                       & $\log N$ & $14.91^{+0.11}_{-0.14}$ & $17.70^{+0.08}_{-0.16}$ & $17.18^{+0.18}_{-0.17}$ & $17.91^{+0.07}_{-0.05}$\\
    ${\rm H_2\, J=1}$ & b\,km/s & $2.69^{+0.16}_{-0.13}$ & $3.89^{+0.21}_{-0.28}$ & $3.10^{+0.36}_{-0.27}$ & $1.1^{+0.5}_{-0.3}$\\
                      & $\log N$ & $16.10^{+0.13}_{-0.11}$ & $17.57^{+0.16}_{-0.04}$ & $17.53^{+0.13}_{-0.17}$ & $17.82^{+0.07}_{-0.09}$ \\
    ${\rm H_2\, J=2}$ & b\,km/s & $2.15^{+0.16}_{-0.07}$ & $4.03^{+0.31}_{-0.26}$ & $3.2^{+0.4}_{-0.4}$ &$3.5^{+0.7}_{-1.0}$ \\
                      & $\log N$ & $16.60^{+0.12}_{-0.11}$ & $16.32^{+0.17}_{-0.28}$ & $15.82^{+0.22}_{-0.30}$ & $15.05^{+0.25}_{-0.14}$\\
    ${\rm H_2\, J=3}$ & b\,km/s & $3.6^{+5.0}_{-0.9}$ & $4.20^{+0.56}_{-0.30}$ & $3.35^{+0.29}_{-0.50}$ & $8.5^{+2.5}_{-5.0}$\\
                      & $\log N$ &$14.59^{+0.06}_{-0.06}$ & $15.48^{+0.13}_{-0.23}$ & $15.24^{+0.11}_{-0.19}$ & $14.73^{+0.06}_{-0.08}$\\
    ${\rm H_2\, J=4}$ & b\,km/s & -- & -- & $10.7^{+5.3}_{-3.0}$ & $11.1^{+4.7}_{-2.4}$\\
    				  & $\log N$ & $13.78^{+0.10}_{-0.20}$ &$14.351^{+0.067}_{-0.022}$ & $14.03^{+0.16}_{-0.06}$ & $13.88^{+0.10}_{-0.13}$\\
    ${\rm H_2\, J=5}$ & $\log N$ & -- & -- & $13.62^{+0.17}_{-0.22}$ & $13.77^{+0.13}_{-0.16}$\\
    \hline 
         & $\log N_{\rm tot}$ & $16.73^{+0.10}_{-0.08}$ & $17.95^{+0.08}_{-0.08}$ & $17.70^{+0.11}_{-0.12}$ & $18.17^{+0.05}_{-0.05}$ \\
    \hline
    HD\, J=0 & b\,km/s & $2.28^{+0.21}_{-0.17}$ & $2.7^{+1.1}_{-0.5}$ & $2.98^{+0.28}_{-0.72}$ & $0.522^{+0.388}_{-0.022}$ \\
             & $\log N$ & $\lesssim 13.9$ & $\lesssim 13.6$ & $\lesssim 14.0$ & $\lesssim 15.7$ \\
    \hline   
    \end{tabular}
    \begin{tablenotes}
    \item Doppler parameters of H$_2$ ${\rm J = 4}$ rotational levels in the 1 and 2 components, H$_2$ $J=5$ in the 3 and 4 components were tied to H$_2$ ${\rm J = 3}$ and ${\rm J = 4}$, respectively. 
    \end{tablenotes}
\end{table*}

\begin{figure*}
    \centering
    \includegraphics[width=\linewidth]{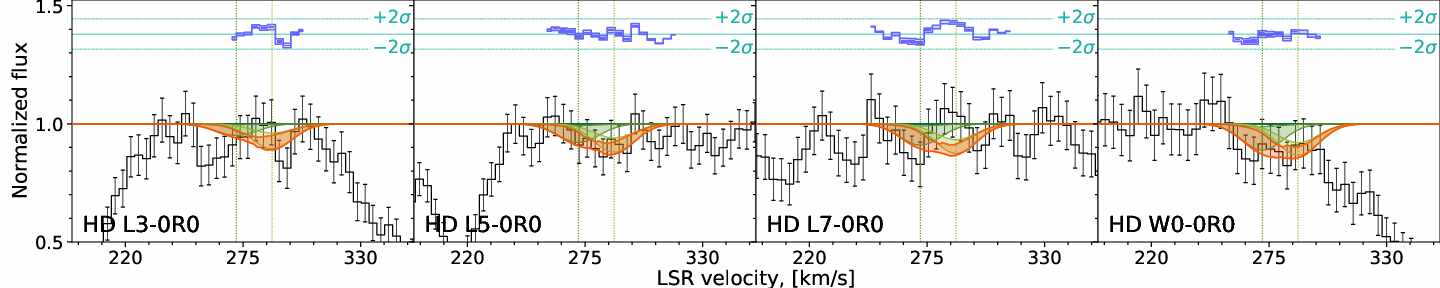}
    \caption{Fit to HD absorption lines towards Sk-66 18 in LMC. Lines are the same as for \ref{fig:lines_HD_Sk67_2}.
    }
    \label{fig:lines_HD_Sk66_18}
\end{figure*}

\begin{figure*}
    \centering
    \includegraphics[width=\linewidth]{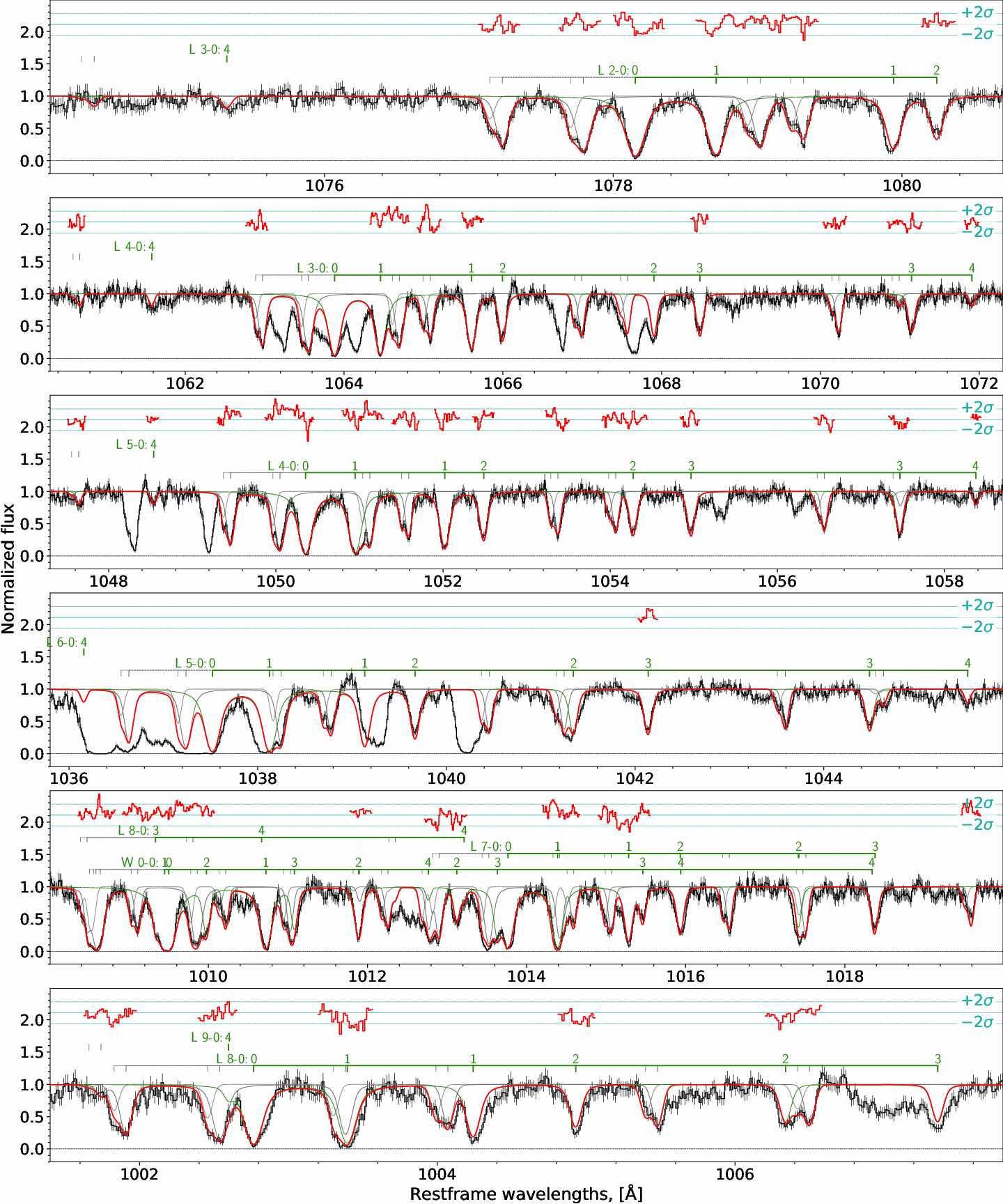}
    \caption{Fit to H2 absorption lines towards Sk-66 18 in LMC. Lines are the same as for \ref{fig:lines_H2_Sk67_2}.
    }
    \label{fig:lines_H2_Sk66_18}
\end{figure*}

\begin{table*}
    \caption{Fit results of H$_2$ lines towards PGMW 3070}
    \label{tab:Sk66_1}
    \begin{tabular}{cccccc}
    \hline
    \hline
    species & comp & 1 & 2 & 3 & 4\\
            & z & $-0.0000121(^{+5}_{-12})$ & $0.0000680(^{+9}_{-6})$ & $0.0008822(^{+13}_{-3})$ & $0.0009555(^{+7}_{-5})$ \\
    \hline 
     ${\rm H_2\, J=0}$ & b\,km/s & $1.0^{+0.5}_{-0.4}$ & $1.35^{+0.31}_{-0.70}$ & $0.62^{+0.30}_{-0.08}$ & $0.69^{+0.47}_{-0.15}$ \\
                       & $\log N$ & $18.686^{+0.032}_{-0.015}$ & $18.053^{+0.020}_{-0.051}$ & $17.67^{+0.09}_{-0.11}$ & $18.950^{+0.028}_{-0.007}$ \\
    ${\rm H_2\, J=1}$ & b\,km/s & $1.7^{+0.6}_{-0.8}$ & $2.87^{+0.21}_{-1.06}$ & $1.13^{+0.35}_{-0.27}$ & $1.72^{+0.08}_{-0.91}$ \\
                      & $\log N$ & $18.604^{+0.030}_{-0.022}$ & $18.40^{+0.03}_{-0.04}$ & $17.922^{+0.041}_{-0.030}$ &  $18.592^{+0.013}_{-0.017}$ \\
    ${\rm H_2\, J=2}$ & b\,km/s & $2.75^{+0.18}_{-0.20}$ & $2.85^{+0.13}_{-0.19}$ & $2.32^{+0.17}_{-0.20}$ & $1.91^{+0.22}_{-0.25}$ \\
                      & $\log N$ & $17.57^{+0.09}_{-0.06}$ &  $17.41^{+0.05}_{-0.14}$ & $16.73^{+0.18}_{-0.18}$ & $17.512^{+0.030}_{-0.070}$ \\
    ${\rm H_2\, J=3}$ & b\,km/s & $2.69^{+0.24}_{-0.21}$ & $3.25^{+0.10}_{-0.26}$ & $2.57^{+0.58}_{-0.21}$ &  $2.18^{+0.30}_{-0.16}$ \\
                      & $\log N$ & $16.99^{+0.18}_{-0.21}$ & $16.74^{+0.24}_{-0.10}$ & $15.86^{+0.24}_{-0.49}$ & $17.23^{+0.09}_{-0.12}$ \\
    ${\rm H_2\, J=4}$ & b\,km/s & $3.3^{+0.6}_{-0.6}$ & $5.2^{+0.8}_{-1.3}$ & $3.1^{+0.9}_{-0.7}$ & $2.42^{+0.39}_{-0.31}$ \\
    				  & $\log N$ & $14.56^{+0.07}_{-0.08}$  & $14.48^{+0.05}_{-0.04}$ & $13.92^{+0.05}_{-0.16}$ & $15.50^{+0.15}_{-0.53}$ \\
    ${\rm H_2\, J=5}$ & b\,km/s & $4.8^{+1.3}_{-0.6}$  & $18.0^{+1.9}_{-5.7}$ & $3.6^{+1.1}_{-0.9}$ &  $2.6^{+1.1}_{-0.4}$ \\
    				  & $\log N$ &  $14.21^{+0.04}_{-0.10}$ & $12.4^{+0.6}_{-0.9}$ & $14.044^{+0.097}_{-0.021}$ &  $14.41^{+0.10}_{-0.07}$ \\
    \hline 
         & $\log N_{\rm tot}$ & $19.97^{+0.02}_{-0.01}$ & $18.60^{+0.02}_{-0.03}$ & $18.14^{+0.04}_{-0.04}$ & $19.124^{+0.019}_{-0.007}$ \\
    \hline
    HD\, J=0 & b\,km/s & $0.525^{+0.587}_{-0.025}$ & $0.522^{+0.876}_{-0.022}$ & $0.62^{+0.25}_{-0.12}$ & $0.71^{+0.30}_{-0.20}$ \\
             & $\log N$ & $\lesssim 15.9$ & $\lesssim 15.5$  & $\lesssim 15.6$ & $\lesssim 16.0$ \\
                  
    \hline   
    \end{tabular}
   
\end{table*}

\begin{figure*}
    \centering
    \includegraphics[width=\linewidth]{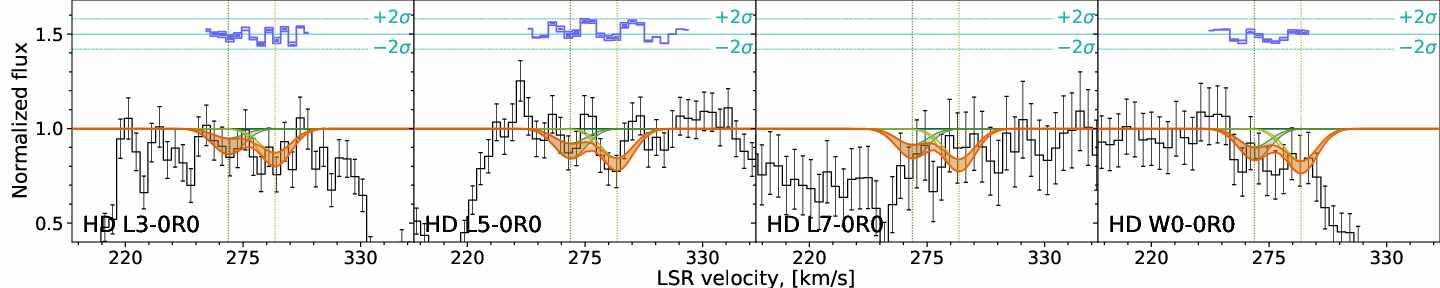}
    \caption{Fit to HD absorption lines towards PGMW 3070 in LMC. Lines are the same as for \ref{fig:lines_HD_Sk67_2}.
    }
    \label{fig:lines_HD_PGMW3070}
\end{figure*}

\begin{figure*}
    \centering
    \includegraphics[width=\linewidth]{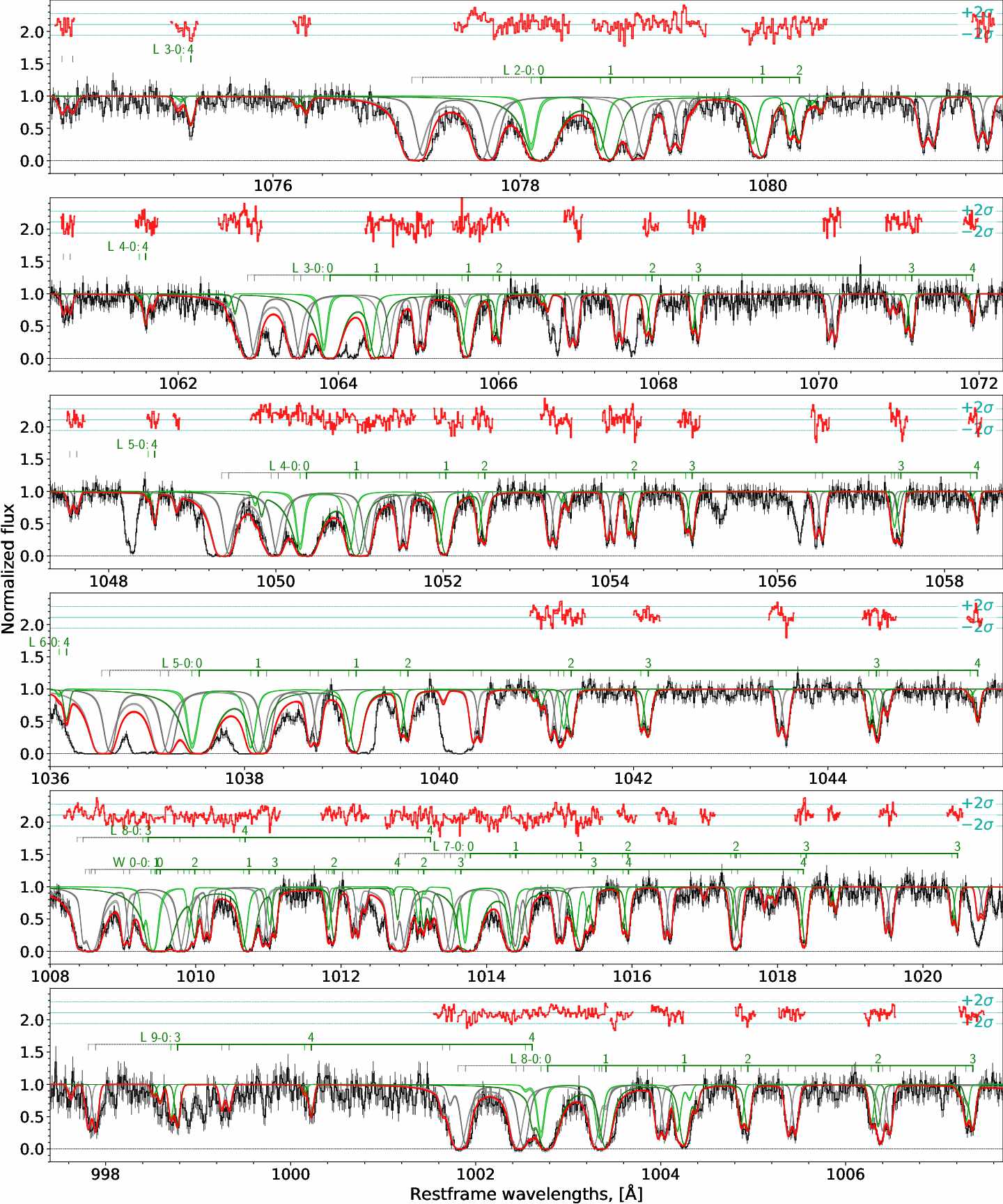}
    \caption{Fit to H2 absorption lines towards PGMW 3070 in LMC. Lines are the same as for \ref{fig:lines_H2_Sk67_2}.
    }
    \label{fig:lines_H2_PGMW3070}
\end{figure*}

\begin{table*}
    \caption{Fit results of H$_2$ lines towards LH10 3073}
    \label{tab:LH10_3073}
    \begin{tabular}{cccccc}
    \hline
    \hline
    species & comp & 1 & 2 & 3 & 4\\
            & z & $-0.0000064(^{+15}_{-16})$ & $0.0000763(^{+10}_{-21})$ & $0.0008855(^{+18}_{-9})$ & $0.0009532(^{+18}_{-18})$ \\
    \hline 
     ${\rm H_2\, J=0}$ & b\,km/s & $0.57^{+0.06}_{-0.06}$ & $0.518^{+0.052}_{-0.018}$ &$0.85^{+0.11}_{-0.31}$ & $0.74^{+0.35}_{-0.17}$\\
                       & $\log N$ & $19.108^{+0.024}_{-0.025}$ & $18.28^{+0.09}_{-0.05}$ & $19.006^{+0.013}_{-0.046}$ & $17.4^{+0.5}_{-3.7}$\\
    ${\rm H_2\, J=1}$ & b\,km/s & $0.60^{+0.07}_{-0.09}$ & $0.578^{+0.026}_{-0.063}$ &$1.01^{+0.18}_{-0.28}$ & $1.05^{+0.27}_{-0.31}$\\
                      & $\log N$ & $18.953^{+0.023}_{-0.029}$ & $18.28^{+0.06}_{-0.09}$ & $18.859^{+0.022}_{-0.022}$ & $17.87^{+0.05}_{-0.13}$\\
    ${\rm H_2\, J=2}$ & b\,km/s & $0.69^{+0.09}_{-0.11}$ & $0.587^{+0.076}_{-0.028}$ &  $1.24^{+0.15}_{-0.30}$ & $1.55^{+0.07}_{-0.33}$\\
                      & $\log N$ & $18.139^{+0.015}_{-0.043}$ &  $17.62^{+0.04}_{-0.05}$ & $17.750^{+0.045}_{-0.029}$ & $16.96^{+0.13}_{-0.11}$\\
    ${\rm H_2\, J=3}$ & b\,km/s & $0.825^{+0.030}_{-0.158}$ & $0.65^{+0.04}_{-0.05}$ & $1.35^{+0.11}_{-0.24}$ & $1.59^{+0.08}_{-0.07}$\\
                      & $\log N$ & $17.94^{+0.03}_{-0.04}$ &$17.62^{+0.04}_{-0.04}$ & $17.72^{+0.03}_{-0.05}$ & $16.64^{+0.16}_{-0.13}$\\
    ${\rm H_2\, J=4}$ & b\,km/s & $0.85^{+0.05}_{-0.08}$ &$0.687^{+0.025}_{-0.070}$ & $1.37^{+0.14}_{-0.12}$ & -- \\
    				  & $\log N$ & $16.75^{+0.06}_{-0.12}$ & $16.20^{+0.10}_{-0.15}$ &$16.24^{+0.16}_{-0.15}$ &  $14.72^{+0.16}_{-0.26}$\\
    ${\rm H_2\, J=5}$ & $\log N$ & $16.20^{+0.22}_{-0.26}$ &$15.37^{+0.39}_{-0.31}$ & $15.24^{+0.43}_{-0.25}$ & $12.8^{+0.3}_{-0.3}$\\
    \hline 
         & $\log N_{\rm tot}$ & $19.38^{+0.02}_{-0.02}$ & $18.67^{+0.05}_{-0.04}$ & $19.27^{+0.01}_{-0.03}$ & $18.05^{+0.17}_{-0.14}$ \\
    \hline
    HD\, J=0 & b\,km/s & $0.59^{+0.05}_{-0.06}$ & $0.505^{+0.073}_{-0.005}$ &  $2.5^{+11.5}_{-2.0}$ & $0.81^{+0.15}_{-0.10}$ \\
             & $\log N_{\rm tot}$ & $\lesssim 16.4$ & $\lesssim 16.0$ & $14.3^{+0.9}_{-0.3}$ & $\lesssim 15.3$ \\
                  
    \hline   
    \end{tabular}
    \begin{tablenotes}
     \item Doppler parameters of H$_2$ ${\rm J = 5}$ rotational levels  were tied to H$_2$ ${\rm J = 4}$, Doppler parameter of H$_2$ ${\rm J=4}$ was tied to H$_2$ ${\rm J = 3}$. 
    \end{tablenotes}
\end{table*}

\begin{figure*}
    \centering
    \includegraphics[width=\linewidth]{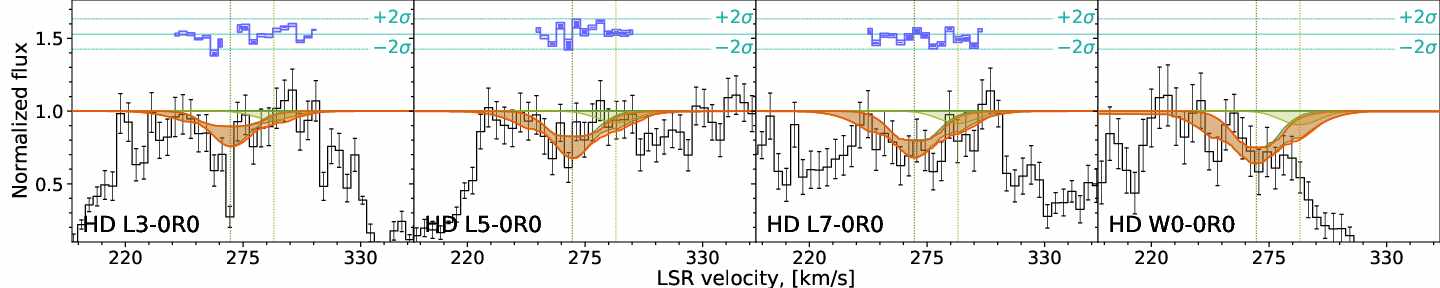}
    \caption{Fit to HD absorption lines towards LH10 3073 in LMC. Lines are the same as for \ref{fig:lines_HD_Sk67_2}.
    }
    \label{fig:lines_HD_LH10_3073}
\end{figure*}

\begin{figure*}
    \centering
    \includegraphics[width=\linewidth]{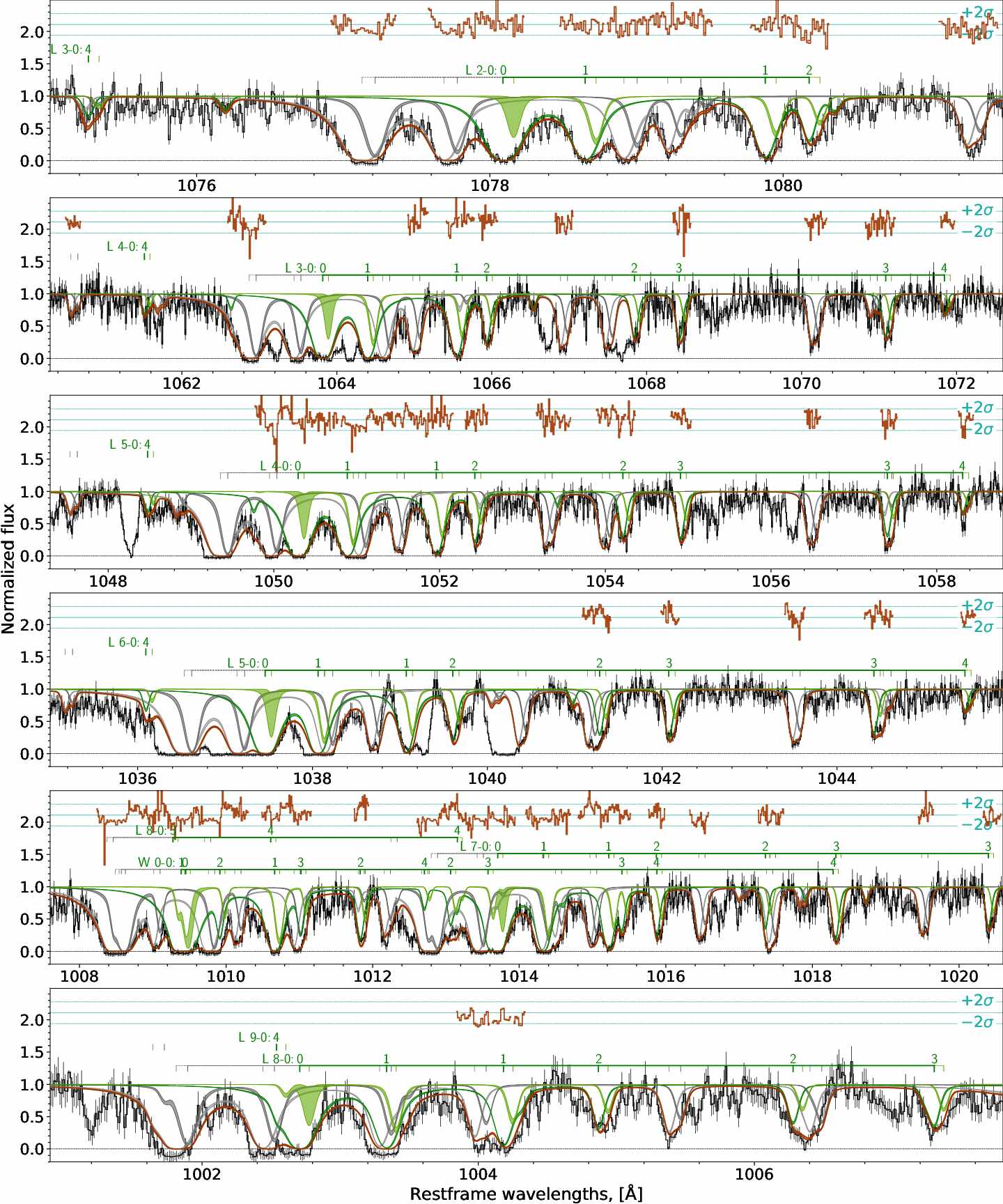}
    \caption{Fit to H2 absorption lines towardsLH10 3073 in LMC. Lines are the same as for \ref{fig:lines_H2_Sk67_2}.
    }
    \label{fig:lines_H2_LH10_3073}
\end{figure*}

\renewcommand{\arraystretch}{1.35}
\setlength{\tabcolsep}{1pt}
\begin{table*}
    \caption{Fit results of H$_2$ lines towards LH10 3102}
    \label{tab:LH10_3102}
    \begin{tabular}{cccccccccc}
    \hline
    \hline
    species & comp & 1 & 2 & 3 & 4 & 5 & 6 & 7 & 8 \\
            & z & $-0.0001336(^{+37}_{-22})$ & $-0.0000657(^{+34}_{-7})$ & $0.0000039(^{+11}_{-7})$ & $0.0000772(^{+9}_{-13})$ & $0.0001332(^{+7}_{-32})$ & $0.0008807(^{+20}_{-9})$ & $0.0009506(^{+10}_{-12})$ & $0.0009996(^{+3}_{-15})$ \\
    \hline 
     ${\rm H_2\, J=0}$ & b\,km/s &$0.64^{+0.49}_{-0.12}$ & $0.92^{+0.44}_{-0.20}$ & $1.49^{+0.61}_{-0.29}$ & $0.518^{+0.155}_{-0.018}$ & $1.8^{+1.0}_{-0.5}$  & $0.73^{+0.18}_{-0.17}$ & $0.61^{+0.25}_{-0.09}$ & $1.8^{+0.7}_{-0.7}$ \\
                       & $\log N$ &$13.0^{+0.4}_{-0.4}$ & $15.27^{+0.68}_{-0.25}$ & $18.79^{+0.04}_{-0.04}$ & $17.95^{+0.09}_{-0.09}$ & $13.06^{+0.21}_{-0.24}$ & $16.99^{+0.09}_{-0.19}$ & $18.428^{+0.050}_{-0.025}$ &  $12.90^{+0.13}_{-0.18}$\\
    ${\rm H_2\, J=1}$ & b\,km/s &$1.23^{+0.69}_{-0.30}$ & $1.32^{+0.25}_{-0.40}$ &  $2.06^{+0.19}_{-0.31}$ & $0.68^{+0.29}_{-0.11}$ & $2.7^{+0.4}_{-0.6}$ & $1.68^{+0.25}_{-0.11}$ & $0.92^{+0.32}_{-0.25}$ & $2.6^{+0.8}_{-0.9}$ \\
                      & $\log N$ &$14.44^{+0.35}_{-0.22}$ &$15.61^{+0.43}_{-0.26}$ &$18.594^{+0.077}_{-0.017}$ & $17.96^{+0.07}_{-0.07}$ & $14.27^{+0.16}_{-0.29}$ & $17.35^{+0.15}_{-0.06}$ & $18.166^{+0.018}_{-0.078}$ & $13.85^{+0.13}_{-0.19}$ \\
    ${\rm H_2\, J=2}$ & b\,km/s &$1.8^{+0.4}_{-0.7}$ & $2.62^{+0.21}_{-0.55}$ & $2.36^{+0.12}_{-0.31}$ & $2.48^{+0.29}_{-0.16}$ & $3.15^{+0.28}_{-0.50}$ & $2.05^{+0.21}_{-0.21}$ & $2.52^{+0.20}_{-0.19}$ &  $4.7^{+0.5}_{-1.0}$ \\
                      & $\log N$ &$13.45^{+0.11}_{-1.12}$ &  $14.56^{+0.13}_{-0.13}$ & $17.77^{+0.08}_{-0.05}$ & $16.59^{+0.14}_{-0.31}$ & $14.67^{+0.14}_{-0.19}$ & $15.65^{+0.29}_{-0.09}$ & $16.39^{+0.16}_{-0.15}$ & $14.27^{+0.11}_{-0.08}$ \\
    ${\rm H_2\, J=3}$ & b\,km/s &$2.8^{+0.3}_{-0.3}$ & $4.1^{+0.6}_{-0.4}$ & $2.45^{+0.21}_{-0.29}$ & $4.27^{+0.26}_{-0.61}$ & $2.91^{+0.44}_{-0.29}$ & $2.55^{+0.31}_{-0.48}$ & $2.92^{+0.19}_{-0.37}$ & $4.9^{+0.5}_{-1.0}$ \\
                      & $\log N$ &$14.06^{+0.12}_{-1.12}$ & $14.35^{+0.08}_{-0.11}$ & $17.44^{+0.14}_{-0.11}$ & $15.25^{+0.32}_{-0.13}$ & $14.27^{+0.08}_{-0.12}$ &$14.85^{+0.27}_{-0.14}$ & $15.38^{+0.24}_{-0.12}$ & $14.13^{+0.12}_{-0.06}$ \\
    ${\rm H_2\, J=4}$ & b\,km/s & $3.9^{+0.5}_{-0.4}$ & $5.4^{+0.4}_{-0.6}$ & $2.20^{+0.21}_{-0.26}$ & -- & -- & -- & -- & -- \\
    				  & $\log N$ & $13.91^{+0.09}_{-0.21}$ & $12.39^{+0.72}_{-0.22}$ & $15.57^{+0.18}_{-0.51}$ & $14.24^{+0.07}_{-0.10}$ & $13.77^{+0.13}_{-0.40}$ & $13.63^{+0.21}_{-0.23}$ & $14.14^{+0.08}_{-0.15}$ & $13.66^{+0.16}_{-0.27}$ \\
    ${\rm H_2\, J=5}$ & b\,km/s & -- & -- &  $4.82^{+0.14}_{-0.07}$ & -- & -- & -- & -- & -- \\
                      & $\log N$ & -- & -- & $13.933^{+0.021}_{-0.044}$ & -- & -- & -- & -- & -- \\
    \hline 
         & $\log N_{\rm tot}$ & $14.71^{+0.22}_{-0.16}$ & $15.82^{+0.40}_{-0.16}$ & $19.04^{+0.04}_{-0.02}$ & $18.26^{+0.06}_{-0.05}$ & $14.96^{+0.09}_{-0.11}$ & $17.51^{+0.11}_{-0.06}$ & $18.62^{+0.03}_{-0.03}$ & $14.65^{+0.07}_{-0.05}$ \\
    HD\, J=0 & b\,km/s &  $0.81^{+0.43}_{-0.30}$ &  $0.84^{+0.48}_{-0.25}$ & $1.3^{+0.8}_{-0.4}$ & $0.63^{+0.12}_{-0.10}$ & $1.6^{+1.4}_{-0.5}$ & $0.50^{+4.23}_{-0.00}$ & $0.67^{+0.17}_{-0.16}$ & $1.6^{+1.1}_{-0.5}$ \\
             & $\log N$ & $\lesssim 15.3$ & $\lesssim 15.6$ & $\lesssim 15.8$ & $\lesssim 15.6$ & $\lesssim 15.8$ & $14.15^{+0.46}_{-0.28}$ & $\lesssim 15.6$ & $\lesssim 15.7$ \\             
                  
    \hline   
    \end{tabular}
    \begin{tablenotes}
     \item Doppler parameters of H$_2$ ${\rm J = 4}$ rotational levels in 4, 5, 6, 7 and 8 components  were tied to H$_2$ ${\rm J = 3}$. 
    \end{tablenotes}
\end{table*}

\begin{figure*}
    \centering
    \includegraphics[width=\linewidth]{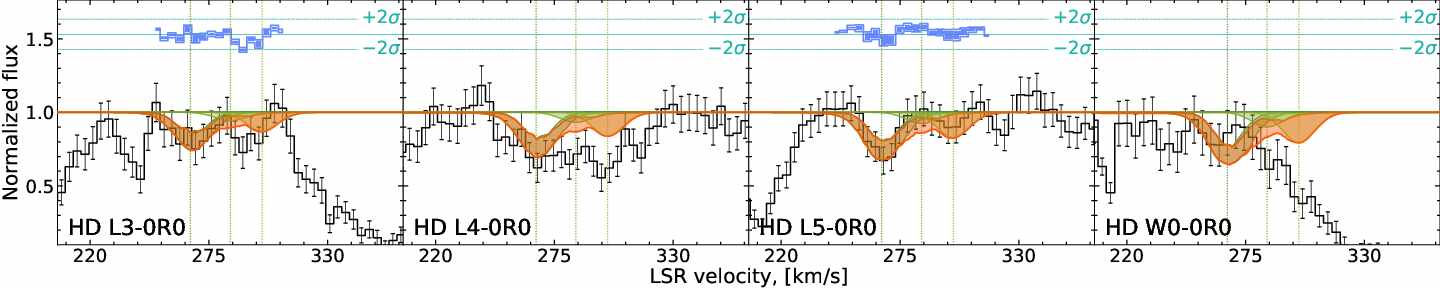}
    \caption{Fit to HD absorption lines towards LH10 3102 in LMC. Lines are the same as for \ref{fig:lines_HD_Sk67_2}.
    }
    \label{fig:lines_HD_LH10_3102_apendix}
\end{figure*}

\begin{figure*}
    \centering
    \includegraphics[width=\linewidth]{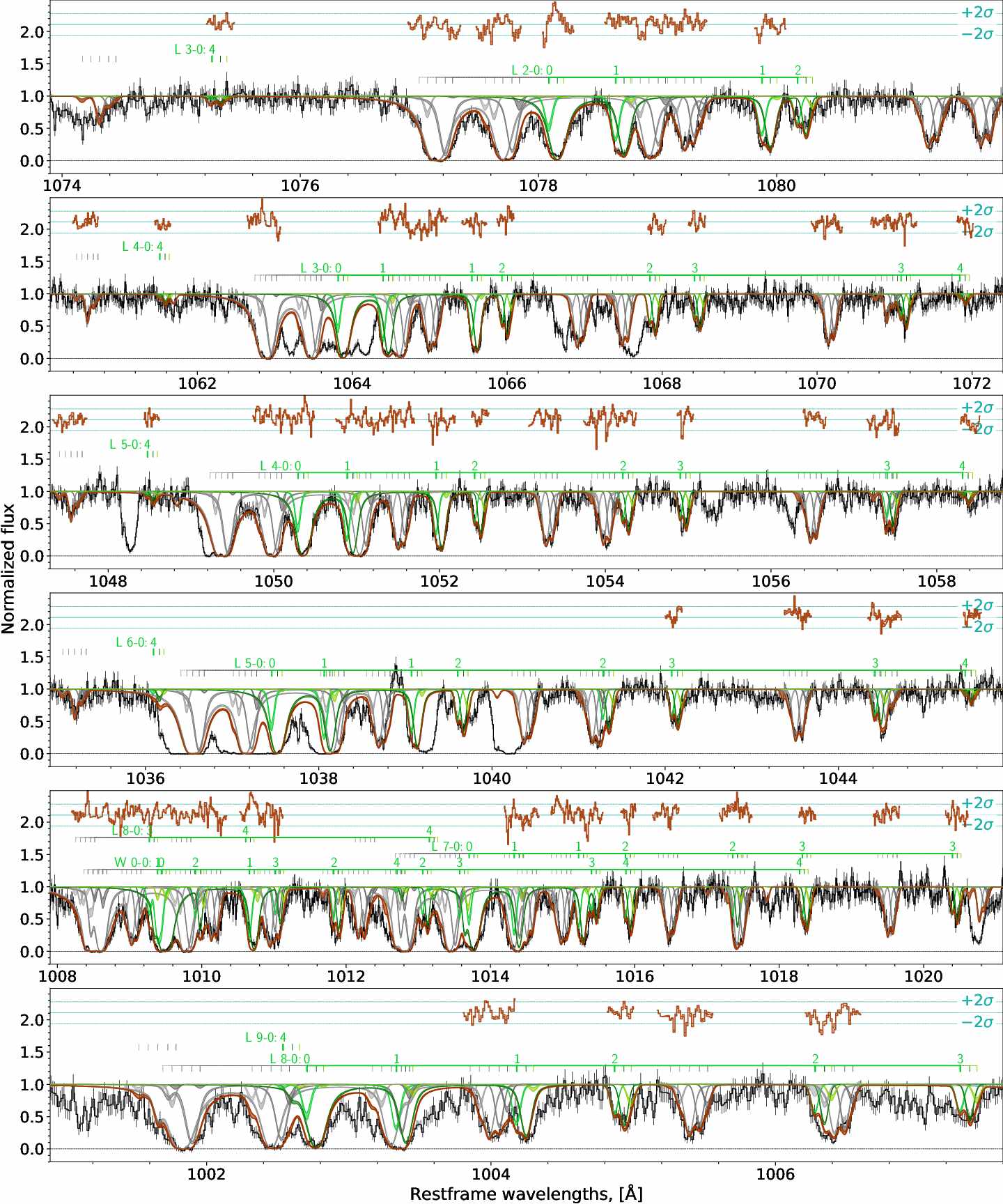}
    \caption{Fit to H2 absorption lines towards LH10 3102 in LMC. Lines are the same as for \ref{fig:lines_H2_Sk67_2}.
    }
    \label{fig:lines_H2_LH10_3102}
\end{figure*}

\begin{table*}
    \caption{Fit results of H$_2$ lines towards LH10 3120}
    \label{tab:LH10_3120}
    \begin{tabular}{cccccc}
    \hline
    \hline
    species & comp & 1 & 2 & 3 & 4  \\
            & z & $-0.0000008(^{+5}_{-7})$ & $0.0000846(^{+5}_{-7})$ & $0.0009079(^{+20}_{-8})$ & $0.0009645(^{+15}_{-9})$ \\
    \hline 
     ${\rm H_2\, J=0}$ & b\,km/s & $0.54^{+0.38}_{-0.04}$ & $2.3^{+0.5}_{-0.7}$ & $2.57^{+0.29}_{-0.77}$ & $0.58^{+0.40}_{-0.08}$ \\
                       & $\log N$ & $18.392^{+0.032}_{-0.031}$ & $18.350^{+0.032}_{-0.022}$ & $17.86^{+0.04}_{-0.07}$ & $17.40^{+0.05}_{-0.09}$\\
    ${\rm H_2\, J=1}$ & b\,km/s & $0.88^{+0.26}_{-0.34}$ &  $2.4^{+0.5}_{-0.6}$ & $3.12^{+0.15}_{-0.09}$ &  $2.10^{+0.19}_{-0.32}$\\
                      & $\log N$ &$18.562^{+0.028}_{-0.013}$ & $18.499^{+0.017}_{-0.032}$ & $17.97^{+0.05}_{-0.05}$ & $17.71^{+0.10}_{-0.06}$\\
    ${\rm H_2\, J=2}$ & b\,km/s & $2.91^{+0.16}_{-0.12}$ & $3.88^{+0.17}_{-0.18}$ & $3.10^{+0.13}_{-0.13}$ & $3.02^{+0.11}_{-0.13}$\\
                      & $\log N$ & $17.73^{+0.04}_{-0.04}$ & $17.25^{+0.11}_{-0.14}$ & $15.46^{+0.11}_{-0.23}$ & $16.38^{+0.39}_{-0.08}$ \\
    ${\rm H_2\, J=3}$ & b\,km/s & $3.29^{+0.16}_{-0.18}$ & $4.51^{+0.23}_{-0.23}$ & $3.15^{+0.13}_{-0.09}$ & $3.32^{+0.16}_{-0.08}$ \\
                      & $\log N$ & $17.36^{+0.08}_{-0.11}$ & $16.28^{+0.14}_{-0.17}$ & $14.73^{+0.07}_{-0.07}$ & $16.05^{+0.17}_{-0.12}$\\
    ${\rm H_2\, J=4}$ & b\,km/s &  $3.47^{+0.52}_{-0.25}$ & $5.06^{+0.26}_{-0.26}$ & -- & --\\
    				  & $\log N$ & $15.26^{+0.26}_{-0.16}$ & $14.03^{+0.19}_{-0.24}$ & $13.45^{+0.21}_{-0.20}$ & $14.21^{+0.11}_{-0.09}$\\
    ${\rm H_2\, J=5}$ & b\,km/s & -- & -- & -- & --\\
    				  & $\log N$ & $14.68^{+0.13}_{-0.09}$ &$13.85^{+0.19}_{-0.39}$ & $13.83^{+0.13}_{-0.14}$ & $12.9^{+0.4}_{-0.3}$\\
   \hline 
         & $\log N_{\rm tot}$ & $18.84^{+0.02}_{-0.01}$ & $18.75^{+0.02}_{-0.02}$ & $18.22^{+0.03}_{-0.04}$ & $17.90^{+0.07}_{-0.05}$ \\
    \hline 
    HD\, J=0 & b\,km/s & $0.518^{+0.374}_{-0.018}$ & $2.4^{+0.5}_{-0.9}$ & $2.66^{+0.31}_{-0.90}$ & $0.522^{+0.507}_{-0.022}$ \\
             & $\log N$ & $\lesssim 16.1$ & $\lesssim 15.5$ & $\lesssim 15.3$ & $\lesssim 15.4$ \\
                  
    \hline   
    \end{tabular}
    \begin{tablenotes}
     \item Doppler parameters of H$_2$ ${\rm J = 5}$ rotational level in component 1 and 2 and ${\rm J = 4, 5}$ in components 3 and 4 were tied to H$_2$ ${\rm J = 4}$ and ${\rm J = 3}$, respectively. 
    \end{tablenotes}
\end{table*}

\begin{figure*}
    \centering
    \includegraphics[width=\linewidth]{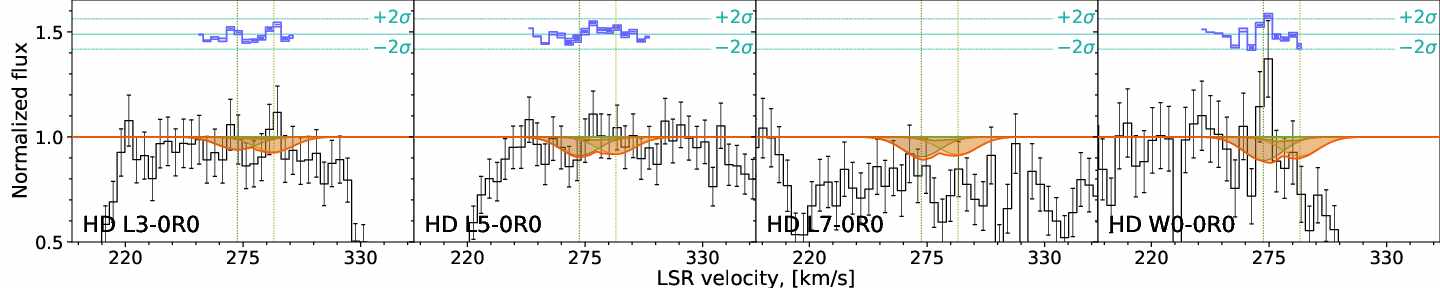}
    \caption{Fit to HD absorption lines towards LH10 3120 in LMC. Lines are the same as for \ref{fig:lines_HD_Sk67_2}.
    }
    \label{fig:lines_HD_LH10_3120}
\end{figure*}

\begin{figure*}
    \centering
    \includegraphics[width=\linewidth]{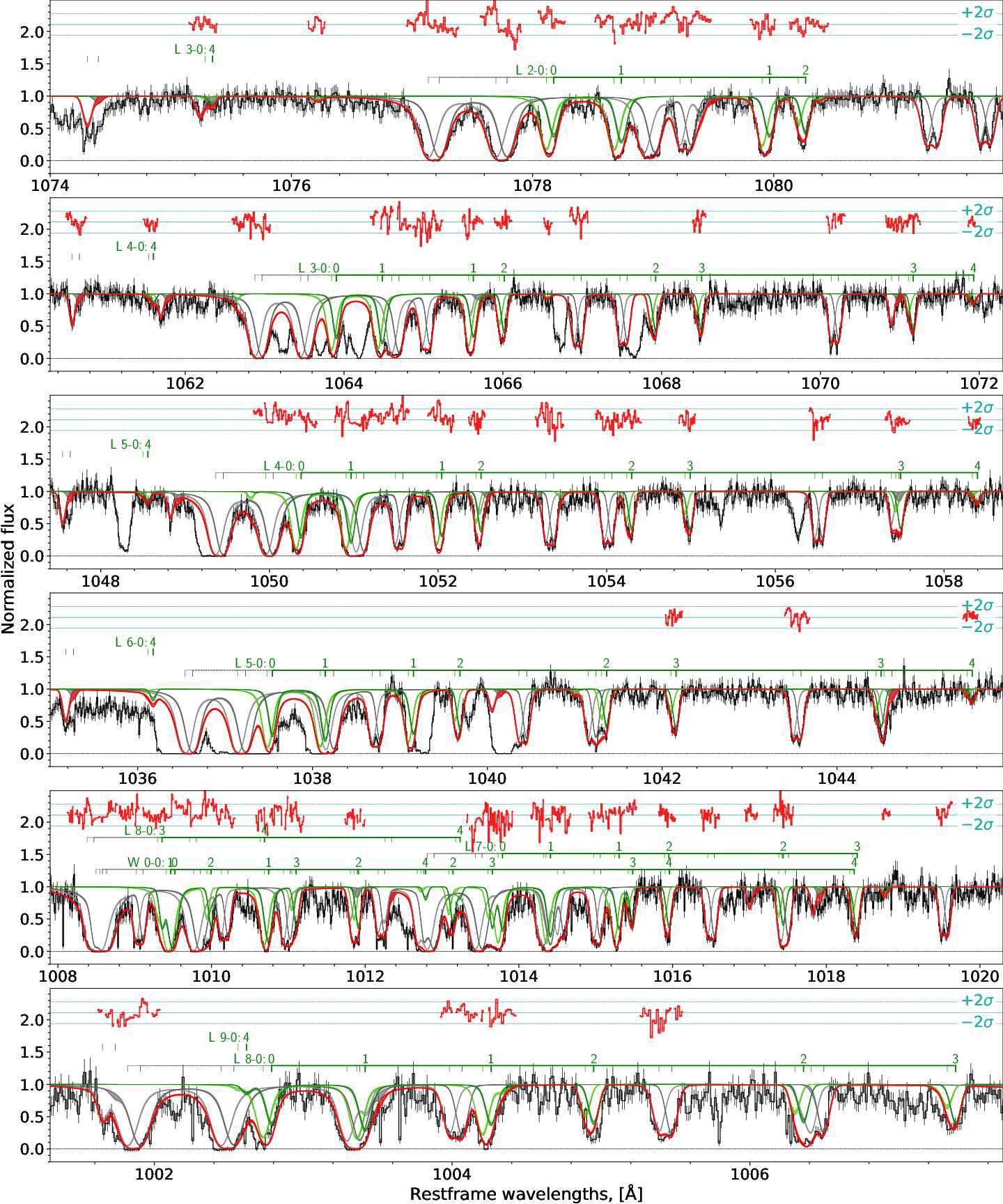}
    \caption{Fit to H2 absorption lines towards LH10 3120 in LMC. Lines are the same as for \ref{fig:lines_H2_Sk67_2}.
    }
    \label{fig:lines_H2_LH10_3120}
\end{figure*}

\begin{table*}
    \caption{Fit results of H$_2$ lines towards PGMW 3157}
    \label{tab:PGMW3157}
    \begin{tabular}{ccccccc}
    \hline
    \hline
    species & comp & 1 & 2 & 3 & 4 & 5\\
            & z & $-0.0000029(^{+11}_{-37})$ & $0.0000745(^{+40}_{-22})$ & $0.000122(^{+4}_{-5})$ & $0.0008904(^{+54}_{-29})$ & $0.0009579(^{+20}_{-40})$ \\
    \hline 
     ${\rm H_2\, J=0}$ & b\,km/s & $0.71^{+0.67}_{-0.12}$ & $1.0^{+0.6}_{-0.4}$ & $0.78^{+0.82}_{-0.26}$ &  $0.9^{+0.6}_{-0.4}$ & $3.2^{+1.5}_{-0.9}$ \\
                       & $\log N$ & $16.82^{+0.18}_{-0.46}$ & $18.38^{+0.08}_{-0.12}$ & $17.3^{+0.4}_{-0.6}$ & $16.9^{+0.4}_{-0.5}$ & $18.85^{+0.05}_{-0.03}$ \\
    ${\rm H_2\, J=1}$ & b\,km/s & $1.50^{+0.73}_{-0.21}$ & $2.3^{+0.6}_{-0.6}$ & $1.7^{+0.4}_{-0.8}$ & $2.6^{+0.7}_{-0.7}$ & $4.3^{+0.9}_{-0.7}$ \\
                      & $\log N$ &$17.10^{+0.20}_{-0.37}$ & $18.52^{+0.06}_{-0.07}$ & $15.7^{+0.7}_{-0.5}$ & $17.15^{+0.29}_{-0.82}$ & $18.76^{+0.04}_{-0.06}$ \\
    ${\rm H_2\, J=2}$ & b\,km/s & $3.8^{+0.5}_{-0.5}$ & $2.8^{+0.7}_{-0.4}$ & $4.0^{+0.5}_{-1.7}$ & $2.8^{+0.8}_{-0.5}$ & $5.3^{+0.6}_{-0.6}$ \\
                      & $\log N$ & $15.93^{+0.31}_{-0.35}$ & $17.16^{+0.29}_{-0.49}$ & $14.43^{+0.23}_{-0.16}$ &$15.64^{+0.22}_{-0.63}$ & $17.54^{+0.33}_{-0.14}$ \\
    ${\rm H_2\, J=3}$ & b\,km/s & $4.6^{+0.4}_{-0.7}$ & $4.4^{+0.5}_{-0.6}$ & $4.9^{+0.4}_{-2.0}$ & $3.5^{+1.1}_{-1.0}$ & $6.0^{+1.9}_{-0.6}$ \\
                      & $\log N$ & $15.40^{+0.30}_{-0.16}$ & $16.61^{+0.10}_{-0.61}$ & $14.2^{+0.3}_{-0.3}$ & $14.82^{+0.23}_{-0.25}$ & $17.09^{+0.21}_{-0.61}$ \\
    ${\rm H_2\, J=4}$ & b\,km/s & -- & --  & -- & -- & $10.5^{+2.2}_{-3.3}$ \\
    				  & $\log N$ & $14.30^{+0.09}_{-0.14}$ &  $14.53^{+0.09}_{-0.11}$ & $13.68^{+0.30}_{-2.11}$ &$14.05^{+0.16}_{-0.23}$ & $14.83^{+0.04}_{-0.08}$ \\
    ${\rm H_2\, J=5}$ & $\log N$ & $14.17^{+0.15}_{-0.24}$ & $13.6^{+0.5}_{-0.7}$ &  $13.9^{+0.3}_{-3.0}$ &$14.39^{+0.17}_{-0.18}$ & $14.54^{+0.09}_{-0.09}$ \\
    \hline 
         & $\log N_{\rm tot}$ & $17.30^{+0.15}_{-0.23}$ & $18.77^{+0.05}_{-0.06}$ & $17.31^{+0.39}_{-0.57}$ & $17.35^{+0.25}_{-0.38}$ & $19.13^{+0.03}_{-0.03}$  \\
    \hline
    HD\,J=0 & b\,km/s &$0.88^{+0.44}_{-0.25}$ & $1.0^{+0.5}_{-0.4}$ & $0.79^{+0.53}_{-0.29}$ & $1.1^{+0.4}_{-0.4}$ &  $2.6^{+1.9}_{-1.2}$ \\
            & $\log N$& $\lesssim 16.6$ & $\lesssim 16.4$ & $\lesssim 16.4$ & $\lesssim 16.5$ & $\lesssim 16.5$ \\
                  
    \hline   
    \end{tabular}
    \begin{tablenotes}
     \item Doppler parameters of H$_2$ ${\rm J = 4, 5}$ rotational levels in 1, 2, 3 and 4 components and H$_2$ ${\rm J = 5}$ in 5 components  were tied to H$_2$ ${\rm J = 3}$ and ${\rm J=4}$, respectively.
    \end{tablenotes}
\end{table*}

\begin{figure*}
    \centering
    \includegraphics[width=\linewidth]{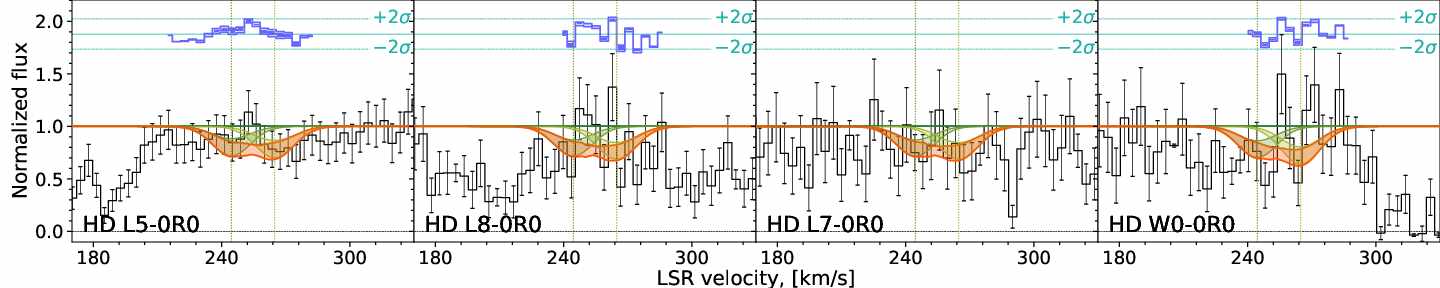}
    \caption{Fit to HD absorption lines towards PGMW 3157 in LMC. Lines are the same as for \ref{fig:lines_HD_Sk67_2}.
    }
    \label{fig:lines_HD_PGMW3157}
\end{figure*}

\begin{figure*}
    \centering
    \includegraphics[width=\linewidth]{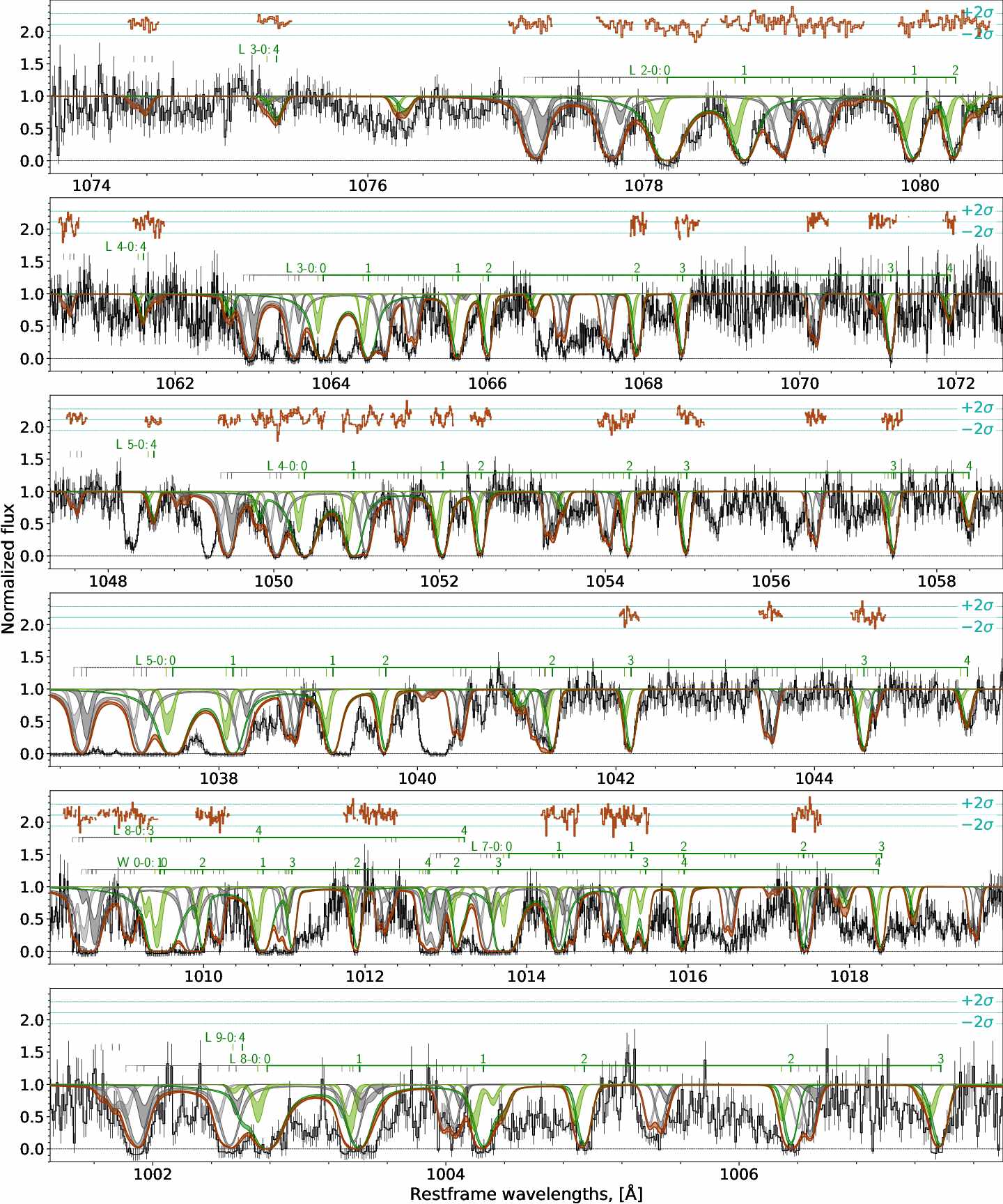}
    \caption{Fit to H2 absorption lines towards PGMW 3157 in LMC. Lines are the same as for \ref{fig:lines_H2_Sk67_2}.
    }
    \label{fig:lines_H2_PGMW3157}
\end{figure*}

\begin{table*}
    \caption{Fit results of H$_2$ lines towards PGMW 3223}
    \label{tab:PGMW3223}
    \begin{tabular}{cccccc}
    \hline
    \hline
    species & comp & 1 & 2 & 3 & 4\\
            & z & $-0.0000170(^{+6}_{-14})$ & $0.0000652(^{+9}_{-7})$ & $0.0008933(^{+8}_{-7})$ & $0.0009599(^{+18}_{-30})$ \\
    \hline 
     ${\rm H_2\, J=0}$ & b\,km/s &$1.5^{+1.4}_{-0.3}$ & $2.49^{+0.23}_{-0.62}$ & $0.68^{+0.26}_{-0.16}$ &  $0.53^{+0.21}_{-0.03}$\\
                       & $\log N$ & $16.71^{+0.20}_{-0.37}$ & $17.57^{+0.04}_{-0.07}$ & $18.561^{+0.021}_{-0.016}$ & $14.8^{+0.4}_{-0.8}$\\
    ${\rm H_2\, J=1}$ & b\,km/s & $2.29^{+1.84}_{-0.25}$ &$2.79^{+0.30}_{-0.34}$ & $1.2^{+0.8}_{-0.4}$ & $1.23^{+0.64}_{-0.24}$ \\
                      & $\log N$ & $15.60^{+0.58}_{-0.18}$ & $17.86^{+0.06}_{-0.06}$ & $18.378^{+0.024}_{-0.023}$ & $14.73^{+0.32}_{-0.14}$ \\
    ${\rm H_2\, J=2}$ & b\,km/s & $6.19^{+0.31}_{-0.62}$ &$5.3^{+0.3}_{-0.7}$ &$3.8^{+0.5}_{-0.5}$ & $2.7^{+0.3}_{-0.5}$ \\
                      & $\log N$ & $14.82^{+0.03}_{-0.04}$ & $15.13^{+0.14}_{-0.07}$ & $15.54^{+0.38}_{-0.18}$ & $14.24^{+0.08}_{-0.11}$ \\
    ${\rm H_2\, J=3}$ & b\,km/s &$6.3^{+0.4}_{-0.4}$ & $5.7^{+0.3}_{-0.3}$ & $7.9^{+0.8}_{-1.0}$ & $2.93^{+0.18}_{-0.26}$ \\
                      & $\log N$ &$14.718^{+0.026}_{-0.039}$ & $14.742^{+0.035}_{-0.031}$ & $15.03^{+0.03}_{-0.04}$ & $14.57^{+0.04}_{-0.12}$ \\
    ${\rm H_2\, J=4}$ & b\,km/s & -- & -- &  $8.7^{+0.5}_{-1.4}$ & --\\
    				  & $\log N$ & $13.91^{+0.10}_{-0.15}$ & $13.96^{+0.09}_{-0.13}$ & $14.39^{+0.05}_{-0.04}$ & $13.86^{+0.15}_{-0.11}$\\
    \hline 
         & $\log N_{\rm tot}$ & $16.75^{+0.19}_{-0.32}$ & $18.04^{+0.04}_{-0.05}$ & $18.78^{+0.02}_{-0.01}$ & $15.27^{+0.22}_{-0.17}$ \\
    \hline
    HD\, J=0 & b\,km/s &$1.5^{+1.8}_{-0.4}$ & $2.51^{+0.28}_{-0.72}$ & $0.514^{+0.336}_{-0.014}$ & $0.510^{+0.225}_{-0.010}$ \\
             & $\log N$ & $\lesssim 14.5$ & $\lesssim 15.2$ & $\lesssim 15.9$ & $\lesssim 15.7$ \\
    \hline   
    \end{tabular}
    \begin{tablenotes}
     \item Doppler parameters of H$_2$ ${\rm J = 4}$ rotational levels in 1, 2 and 4 components were tied to H$_2$ ${\rm J = 3}$.
     \item Resolution was fixed to be $R = 18000$
    \end{tablenotes}
\end{table*}

\begin{figure*}
    \centering
    \includegraphics[width=\linewidth]{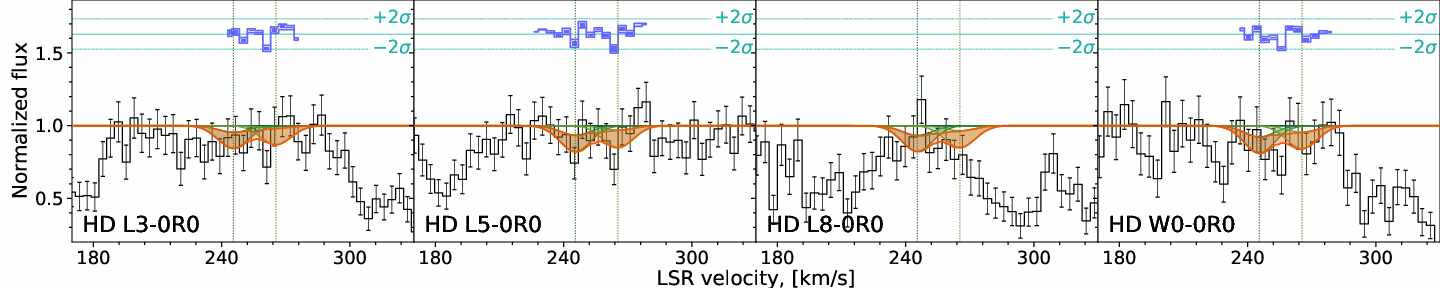}
    \caption{Fit to HD absorption lines towards PGMW 3223 in LMC. Lines are the same as for \ref{fig:lines_HD_Sk67_2}.
    }
    \label{fig:lines_HD_PGMW3223}
\end{figure*}

\begin{figure*}
    \centering
    \includegraphics[width=\linewidth]{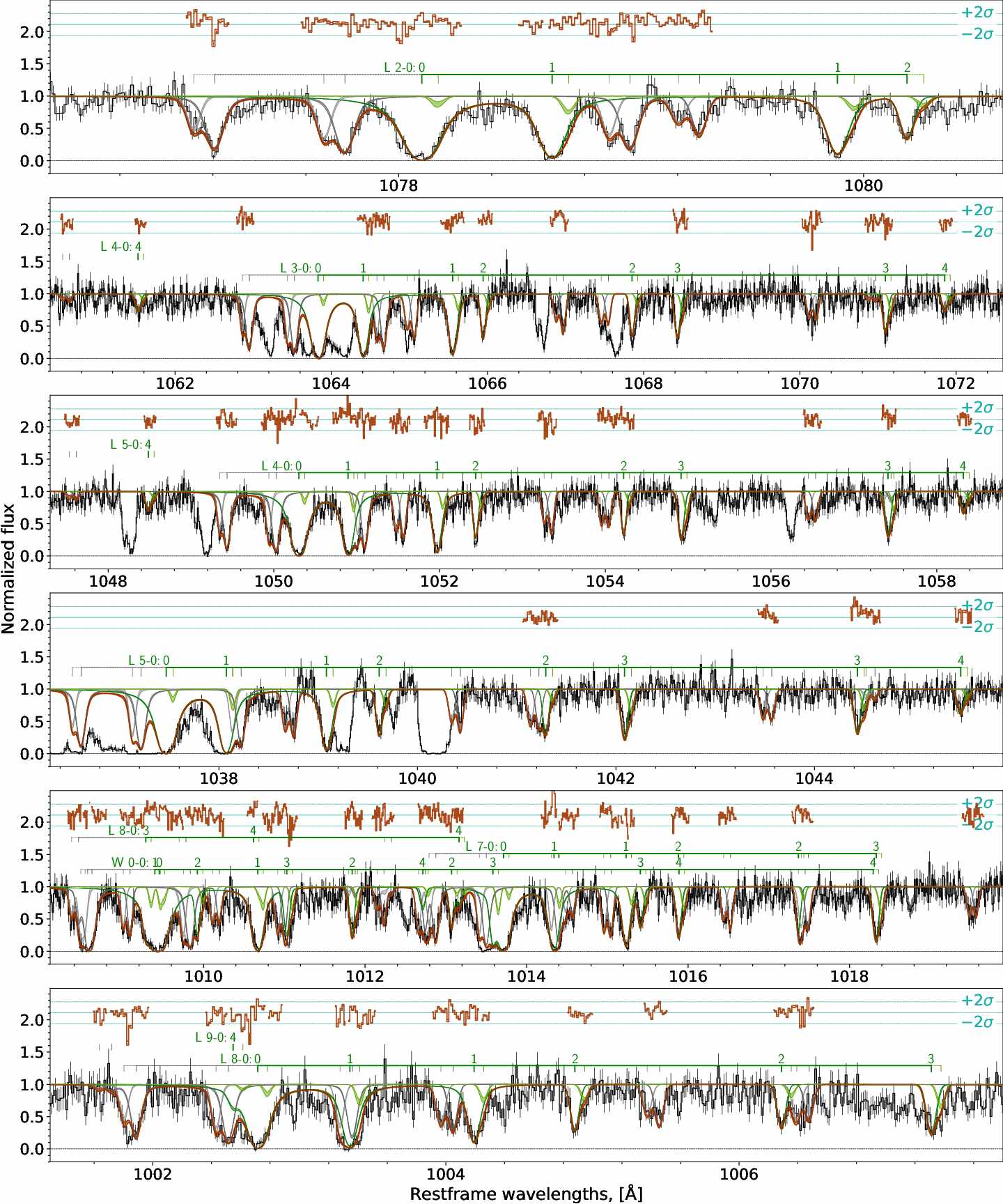}
    \caption{Fit to H2 absorption lines towards PGMW 3223 in LMC. Lines are the same as for \ref{fig:lines_H2_Sk67_2}.
    }
    \label{fig:lines_H2_PGMW3223}
\end{figure*}

\begin{table*}
    \caption{Fit results of H$_2$ lines towards Sk-66 35}
    \label{tab:Sk66_35}
    \begin{tabular}{cccccccc}
    \hline
    \hline
    species & comp & 1 & 2 & 3 & 4 & 5\\
            & z & $-0.0000192(^{+21}_{-14})$ &  $0.0000730(^{+8}_{-8})$ & $0.000882(^{+10}_{-10})$ & $0.0009186(^{+15}_{-14})$ & $0.000954(^{+9}_{-4})$ &  \\
    \hline 
     ${\rm H_2\, J=0}$ & b\,km/s & $1.84^{+0.29}_{-0.24}$ & $1.5^{+0.5}_{-0.9}$ &  $0.68^{+0.98}_{-0.18}$ & $0.73^{+0.76}_{-0.22}$ & $0.9^{+0.6}_{-0.3}$ \\
                       & $\log N$ & $15.24^{+0.34}_{-0.20}$ & $17.827^{+0.041}_{-0.031}$ & $18.22^{+0.27}_{-0.23}$ & $18.85^{+0.19}_{-0.06}$ & $18.61^{+0.19}_{-0.14}$ \\
    ${\rm H_2\, J=1}$ & b\,km/s &$1.99^{+0.26}_{-0.33}$ & $2.5^{+0.5}_{-0.8}$ &$0.68^{+1.50}_{-0.18}$ & $0.77^{+0.76}_{-0.18}$ & $1.3^{+0.6}_{-0.6}$ \\
                      & $\log N$ & $17.11^{+0.12}_{-0.29}$ & $18.17^{+0.04}_{-0.05}$ & $17.70^{+0.40}_{-0.25}$ & $19.124^{+0.019}_{-0.021}$ & $17.47^{+0.31}_{-0.89}$ \\
    ${\rm H_2\, J=2}$ & b\,km/s &$3.3^{+0.4}_{-0.8}$ & $3.53^{+0.46}_{-0.21}$ &$0.79^{+1.94}_{-0.29}$&  $1.6^{+0.4}_{-0.7}$ & $1.5^{+0.8}_{-0.5}$ \\
                      & $\log N$ & $15.10^{+0.25}_{-0.21}$ & $17.28^{+0.09}_{-0.35}$ & $14.4^{+1.4}_{-0.3}$ & $18.142^{+0.029}_{-0.081}$ & $14.8^{+0.9}_{-0.3}$ \\
    ${\rm H_2\, J=3}$ & b\,km/s & $4.0^{+1.1}_{-0.7}$ & $3.36^{+1.13}_{-0.20}$ &$0.90^{+2.71}_{-0.23}$ &  $1.7^{+0.5}_{-0.6}$ & $7.39^{+5.58}_{-5.94}$ \\
                      & $\log N$ & $14.56^{+0.08}_{-0.08}$ & $16.64^{+0.21}_{-0.65}$ & $13.1^{+1.2}_{-0.8}$ &$17.73^{+0.06}_{-0.13}$ & $14.20^{+0.30}_{-0.21}$ \\
    ${\rm H_2\, J=4}$ & b\,km/s & -- & -- &  $4.6^{+3.9}_{-2.1}$ & $3.05^{+0.40}_{-0.22}$ & -- \\
    				  & $\log N$ & $13.04^{+0.31}_{-0.58}$ & $14.12^{+0.09}_{-0.16}$ & $10.29^{+2.13}_{-0.27}$ & $16.01^{+0.10}_{-0.39}$ &  $12.9^{+0.7}_{-1.2}$\\
    ${\rm H_2\, J=5}$ & $\log N$ &-- & -- & $10.5^{+1.5}_{-0.5}$ & $14.59^{+0.10}_{-0.11}$ & -- \\
    \hline 
         & $\log N_{\rm tot}$ & $17.13^{+0.12}_{-0.28}$ & $18.38^{+0.03}_{-0.04}$ & $18.34^{+0.24}_{-0.18}$ & $19.35^{+0.07}_{-0.02}$ & $18.64^{+0.18}_{-0.13}$ \\
    \hline 
    HD\, J=0 & b\,km/s &  $1.80^{+0.37}_{-0.28}$ & $0.530^{+1.180}_{-0.030}$ & $0.55^{+1.07}_{-0.05}$ & $0.54^{+1.03}_{-0.04}$ & $0.80^{+0.70}_{-0.30}$ \\
             & $\log N$ & $\lesssim 14.5$ & $\lesssim 16.1$ & $\lesssim 16.2$ & $\lesssim 15.9$ & $\lesssim 15.7$ \\
    \hline   
    \end{tabular}
    \begin{tablenotes}
     \item Doppler parameters of H$_2$ ${\rm J = 4}$ rotational levels in 1, 2 and 5 components  were tied to H$_2$ ${\rm J = 3}$.
    \end{tablenotes}
\end{table*}

\begin{figure*}
    \centering
    \includegraphics[width=\linewidth]{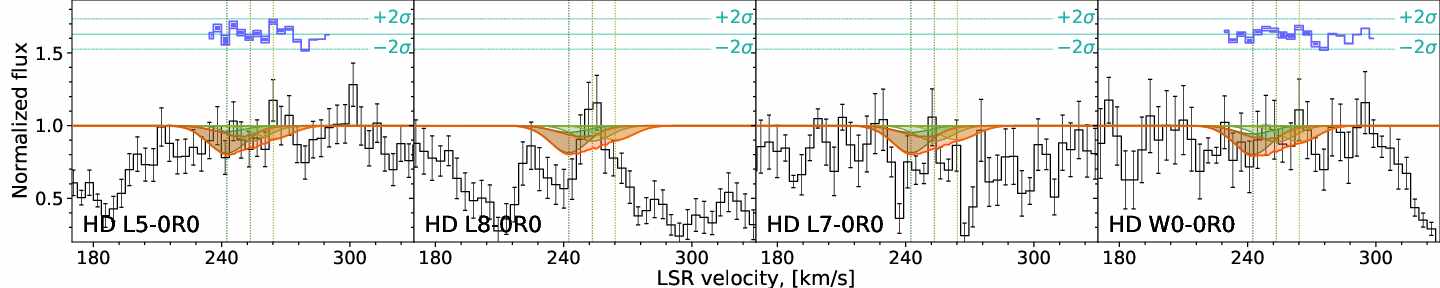}
    \caption{Fit to HD absorption lines towards Sk-66 35 in LMC. Lines are the same as for \ref{fig:lines_HD_Sk67_2}.
    }
    \label{fig:lines_HD_Sk66_35}
\end{figure*}

\begin{figure*}
    \centering
    \includegraphics[width=\linewidth]{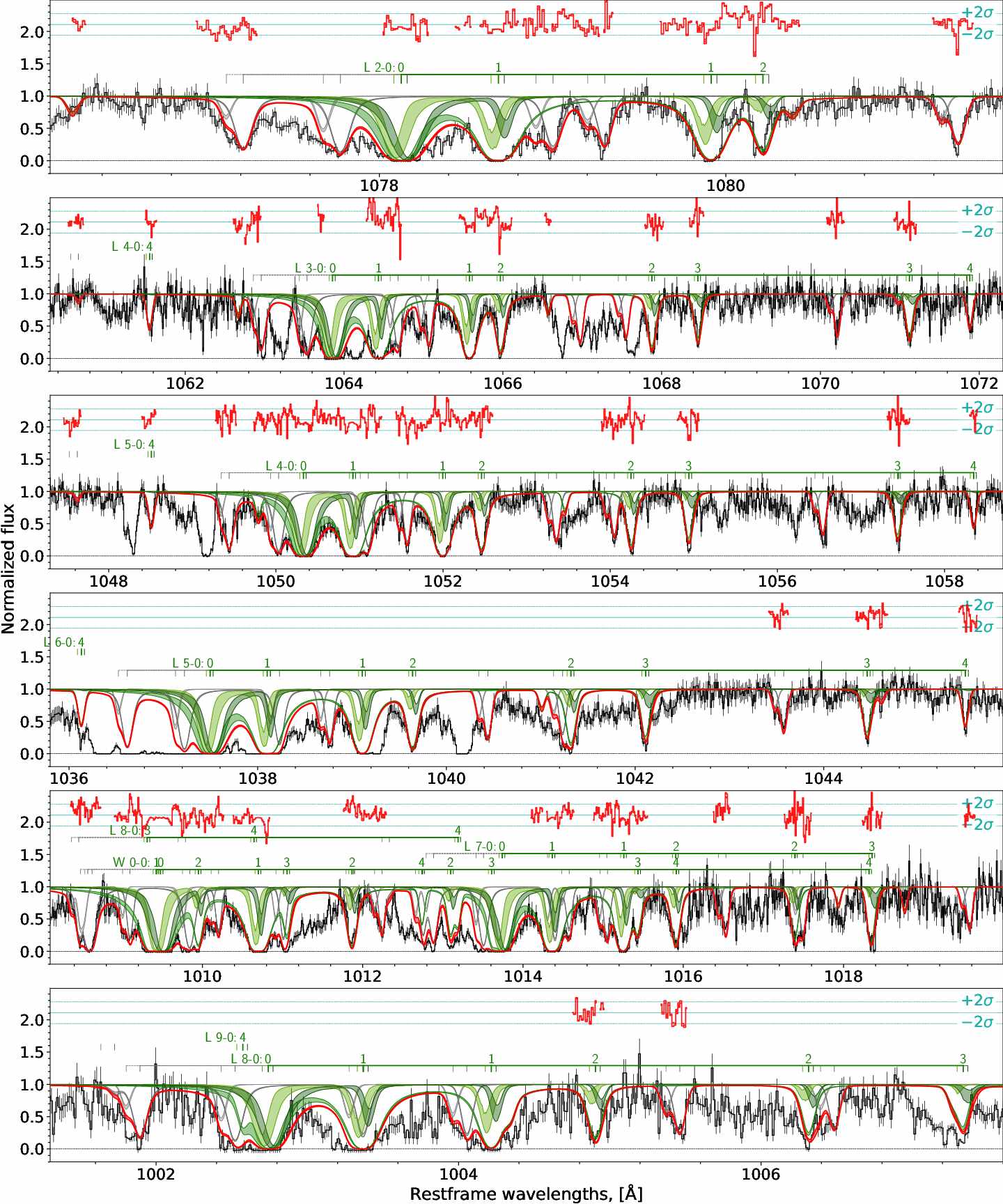}
    \caption{Fit to H2 absorption lines towards Sk-66 35 in LMC. Lines are the same as for \ref{fig:lines_H2_Sk67_2}.
    }
    \label{fig:lines_H2_Sk66_35}
\end{figure*}

\begin{table*}
    \caption{Fit results of H$_2$ lines towards Sk-69 52}
    \label{tab:Sk69_52}
    \begin{tabular}{cccc}
    \hline
    \hline
    species & comp & 1 & 2  \\
            & z & $0.0000808(^{+9}_{-8})$ & $0.0008653(^{+12}_{-10})$ \\
    \hline 
     ${\rm H_2\, J=0}$ & b\,km/s & $0.526^{+0.429}_{-0.026}$ & $0.506^{+0.171}_{-0.006}$ \\
                       & $\log N$ & $18.454^{+0.026}_{-0.020}$ & $18.452^{+0.038}_{-0.021}$\\
    ${\rm H_2\, J=1}$ & b\,km/s & $0.77^{+0.43}_{-0.25}$ & $0.63^{+0.13}_{-0.13}$\\
                      & $\log N$ & $18.487^{+0.023}_{-0.021}$ & $18.158^{+0.026}_{-0.026}$\\
    ${\rm H_2\, J=2}$ & b\,km/s & $1.04^{+0.61}_{-0.16}$ & $0.79^{+0.14}_{-0.17}$\\
                      & $\log N$ & $17.63^{+0.04}_{-0.07}$ & $16.70^{+0.08}_{-0.16}$\\
    ${\rm H_2\, J=3}$ & b\,km/s & $1.60^{+0.12}_{-0.53}$ & $0.99^{+0.08}_{-0.16}$\\
                      & $\log N$ & $17.09^{+0.14}_{-0.11}$ & $15.98^{+0.35}_{-0.27}$\\
    ${\rm H_2\, J=4}$ & $\log N$ & $14.9^{+0.6}_{-0.4}$ &  $14.7^{+0.6}_{-0.3}$ \\
    \hline 
         & $\log N_{\rm tot}$ & $18.81^{+0.02}_{-0.01}$ & $18.64^{+0.02}_{-0.02}$ \\
    \hline 
    HD\, J=0 & b\,km/s &$0.519^{+0.419}_{-0.019}$ & $0.508^{+0.171}_{-0.008}$ \\
             & $\log N$ & $\lesssim 16.3$ & $\lesssim 15.6$ \\
    \hline   
    \end{tabular}
    \begin{tablenotes}
     \item Doppler parameters of H$_2$ ${\rm J = 4}$ rotational levels in both components  were tied to H$_2$ ${\rm J = 3}$. 
    \end{tablenotes}
\end{table*}

\begin{figure*}
    \centering
    \includegraphics[width=\linewidth]{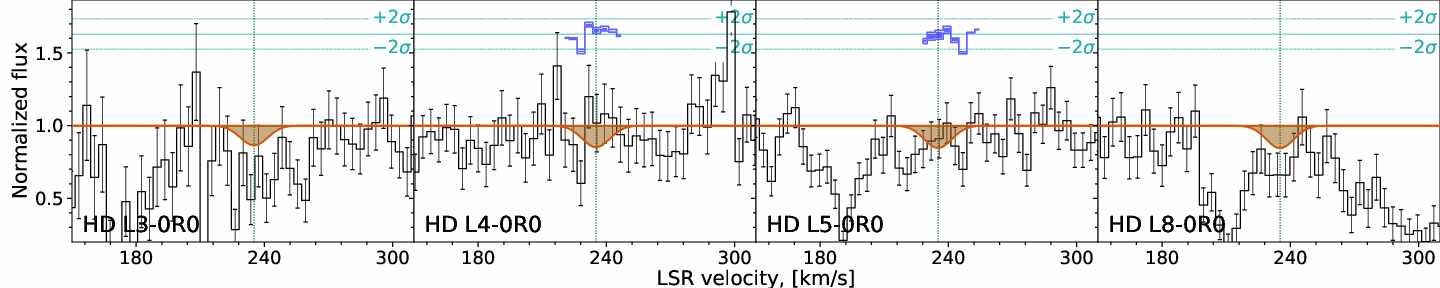}
    \caption{Fit to HD absorption lines towards Sk-69 52 in LMC. Lines are the same as for \ref{fig:lines_HD_Sk67_2}.
    }
    \label{fig:lines_HD_Sk69_52}
\end{figure*}

\begin{figure*}
    \centering
    \includegraphics[width=\linewidth]{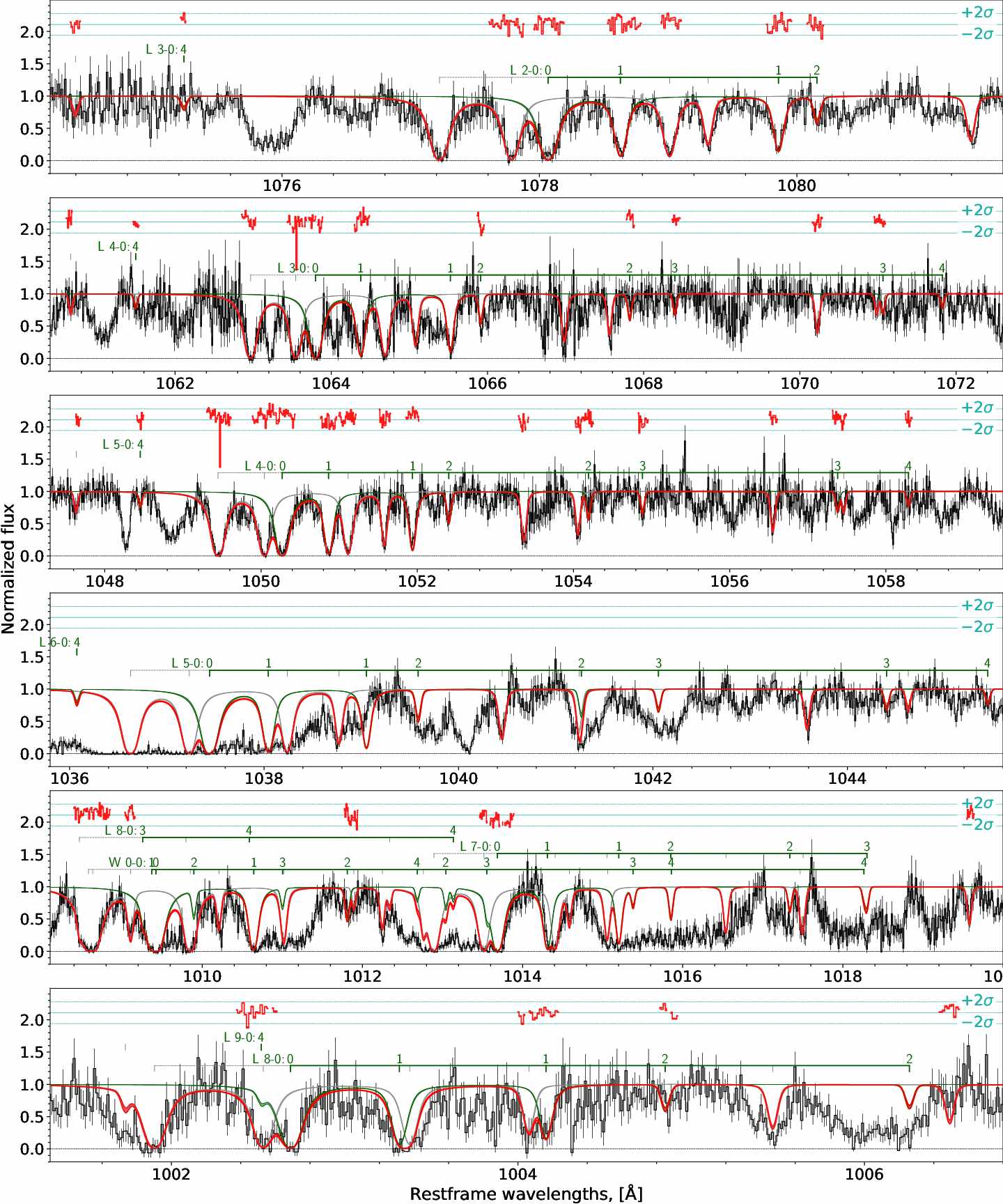}
    \caption{Fit to H2 absorption lines towards Sk-69 52 in LMC. Lines are the same as for \ref{fig:lines_H2_Sk67_2}.
    }
    \label{fig:lines_H2_Sk69_52}
\end{figure*}

\begin{table*}
    \caption{Fit results of H$_2$ lines towards Sk-65 21}
    \label{tab:Sk65_21}
    \begin{tabular}{cccc}
    \hline
    \hline
    species & comp & 1 & 2  \\
            & z & $0.00010030(^{+14}_{-27})$ &  $0.00082643(^{+26}_{-34})$ \\
    \hline 
     ${\rm H_2\, J=0}$ & b\,km/s &$2.63^{+0.12}_{-0.06}$ & $1.5^{+0.6}_{-0.5}$\\
                       & $\log N$ &$17.646^{+0.008}_{-0.020}$ & $17.969^{+0.014}_{-0.012}$\\
    ${\rm H_2\, J=1}$ & b\,km/s & $2.28^{+0.09}_{-0.09}$ & $2.00^{+0.21}_{-0.45}$\\
                      & $\log N$ & $17.978^{+0.007}_{-0.014}$ & $18.147^{+0.009}_{-0.015}$\\
    ${\rm H_2\, J=2}$ & b\,km/s &$2.34^{+0.27}_{-0.20}$ &  $2.48^{+0.06}_{-0.12}$\\
                      & $\log N$ &$17.23^{+0.12}_{-0.14}$ & $16.93^{+0.07}_{-0.04}$\\
    ${\rm H_2\, J=3}$ & b\,km/s &$4.99^{+0.32}_{-0.23}$ & $3.34^{+0.17}_{-0.06}$\\
                      & $\log N$ &  $15.26^{+0.10}_{-0.04}$ & $15.50^{+0.04}_{-0.04}$\\
    ${\rm H_2\, J=4}$ & b\,km/s & $6.2^{+0.8}_{-0.9}$ & $3.36^{+0.19}_{-0.18}$\\
    				  & $\log N$ & $14.342^{+0.020}_{-0.015}$ & $14.07^{+0.04}_{-0.05}$\\
    ${\rm H_2\, J=5}$ & b\,km/s & $6.6^{+1.7}_{-1.0}$ & -- \\
    				  & $\log N$ & $13.918^{+0.026}_{-0.038}$ & $13.71^{+0.08}_{-0.09}$\\
    \hline 
         & $\log N_{\rm tot}$ & $18.19^{+0.02}_{-0.02}$ & $18.38^{+0.01}_{-0.01}$ \\
    \hline
    HD J=0  & b\,km/s & $2.62^{+0.15}_{-0.06}$ & $0.530^{+0.413}_{-0.030}$ \\
            & $\log N$ & $\lesssim 13.6$ & $\lesssim 15.9$ \\
    \hline   
    \end{tabular}
    \begin{tablenotes}
     \item Doppler parameter of H$_2$ ${\rm J = 5}$ rotational levels in the 2 component  was tied to H$_2$ ${\rm J = 4}$. 
    \end{tablenotes}
\end{table*}

\begin{figure*}
    \centering
    \includegraphics[width=\linewidth]{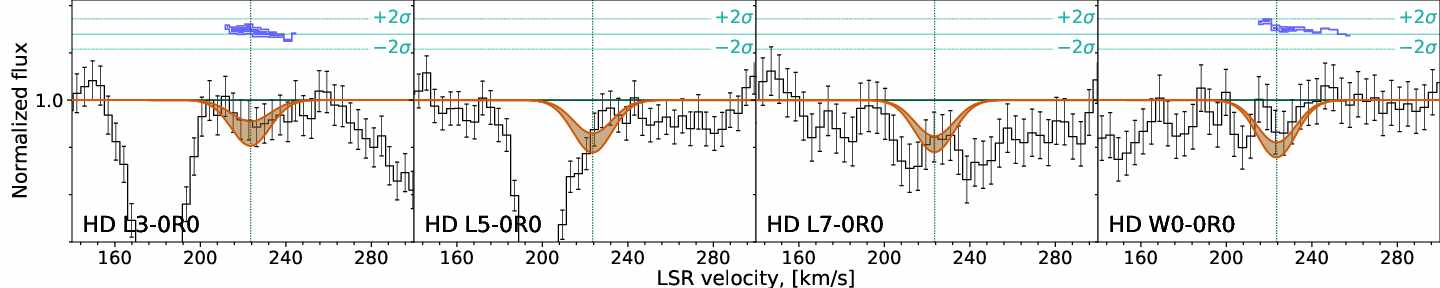}
    \caption{Fit to HD absorption lines towards Sk-65 21 in LMC. Lines are the same as for \ref{fig:lines_HD_Sk67_2}.
    }
    \label{fig:lines_HD_Sk65_21}
\end{figure*}

\begin{figure*}
    \centering
    \includegraphics[width=\linewidth]{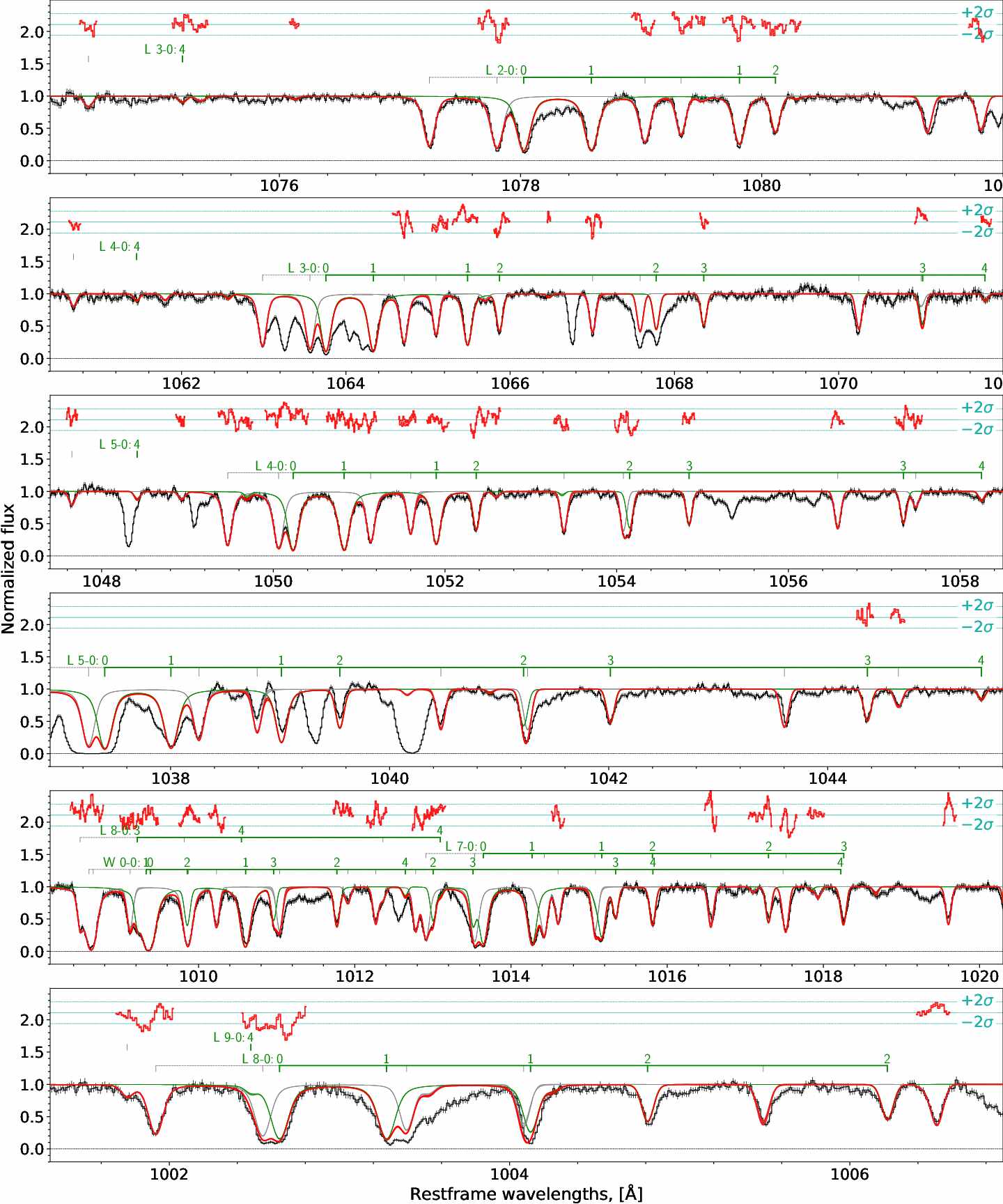}
    \caption{Fit to H2 absorption lines towards Sk-65 21 in LMC. Lines are the same as for \ref{fig:lines_H2_Sk67_2}.
    }
    \label{fig:lines_H2_Sk65_21}
\end{figure*}

\begin{table*}
    \caption{Fit results of H$_2$ lines towards Sk-66 51}
    \label{tab:Sk66_51}
    \begin{tabular}{cccc}
    \hline
    \hline
    species & comp & 1 & 2 \\
            & z & $0.0000861(^{+3}_{-10})$ & $0.0010389(^{+8}_{-7})$ \\
    \hline 
     ${\rm H_2\, J=0}$ & b\,km/s & $2.6^{+0.3}_{-0.3}$ & $0.70^{+0.22}_{-0.20}$ \\
                       & $\log N$ & $16.99^{+0.11}_{-0.18}$ &$17.854^{+0.026}_{-0.026}$  \\
    ${\rm H_2\, J=1}$ & b\,km/s & $2.89^{+0.22}_{-0.21}$ &$1.01^{+0.25}_{-0.34}$ \\
                      & $\log N$ &  $17.32^{+0.10}_{-0.09}$ &$17.656^{+0.026}_{-0.020}$ \\
    ${\rm H_2\, J=2}$ & b\,km/s & $2.84^{+0.37}_{-0.16}$ & $1.41^{+0.28}_{-0.15}$\\
                      & $\log N$ & $15.95^{+0.12}_{-0.32}$ &$16.46^{+0.14}_{-0.27}$ \\
    ${\rm H_2\, J=3}$ & b\,km/s & $8.3^{+0.6}_{-0.9}$ & $1.58^{+0.19}_{-0.15}$\\
                      & $\log N$ & $14.539^{+0.028}_{-0.028}$ &  $15.56^{+0.28}_{-0.24}$\\
    ${\rm H_2\, J=4}$ & $\log N$ & $13.62^{+0.20}_{-0.23}$ & $13.88^{+0.16}_{-0.22}$ \\
    \hline 
         & $\log N_{\rm tot}$ & $17.50^{+0.08}_{-0.08}$ & $18.08^{+0.02}_{-0.02}$ \\
    \hline
    HD J=0 & b\,km/s & $2.64^{+0.28}_{-0.37}$ & $0.511^{+0.345}_{-0.011}$ \\
           & $\log N$ & $\lesssim 14.2$ & $\lesssim 15.3$ \\
    \hline   
    \end{tabular}
    \begin{tablenotes}
     \item Doppler parameters of H$_2$ ${\rm J = 4}$ rotational levels  were tied to H$_2$ ${\rm J = 3}$. 
    \end{tablenotes}
\end{table*}

\begin{figure*}
    \centering
    \includegraphics[width=\linewidth]{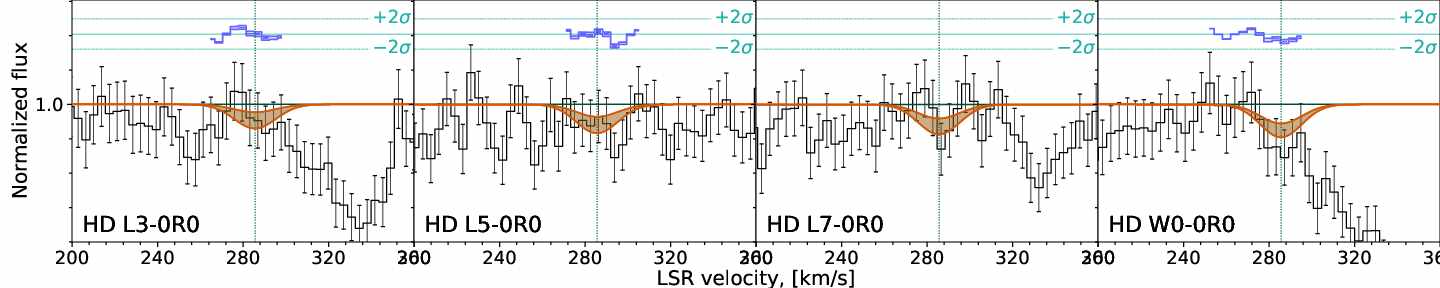}
    \caption{Fit to HD absorption lines towards Sk-66 51 in LMC. Lines are the same as for \ref{fig:lines_HD_Sk67_2}.
    }
    \label{fig:lines_HD_Sk66_51}
\end{figure*}

\begin{figure*}
    \centering
    \includegraphics[width=\linewidth]{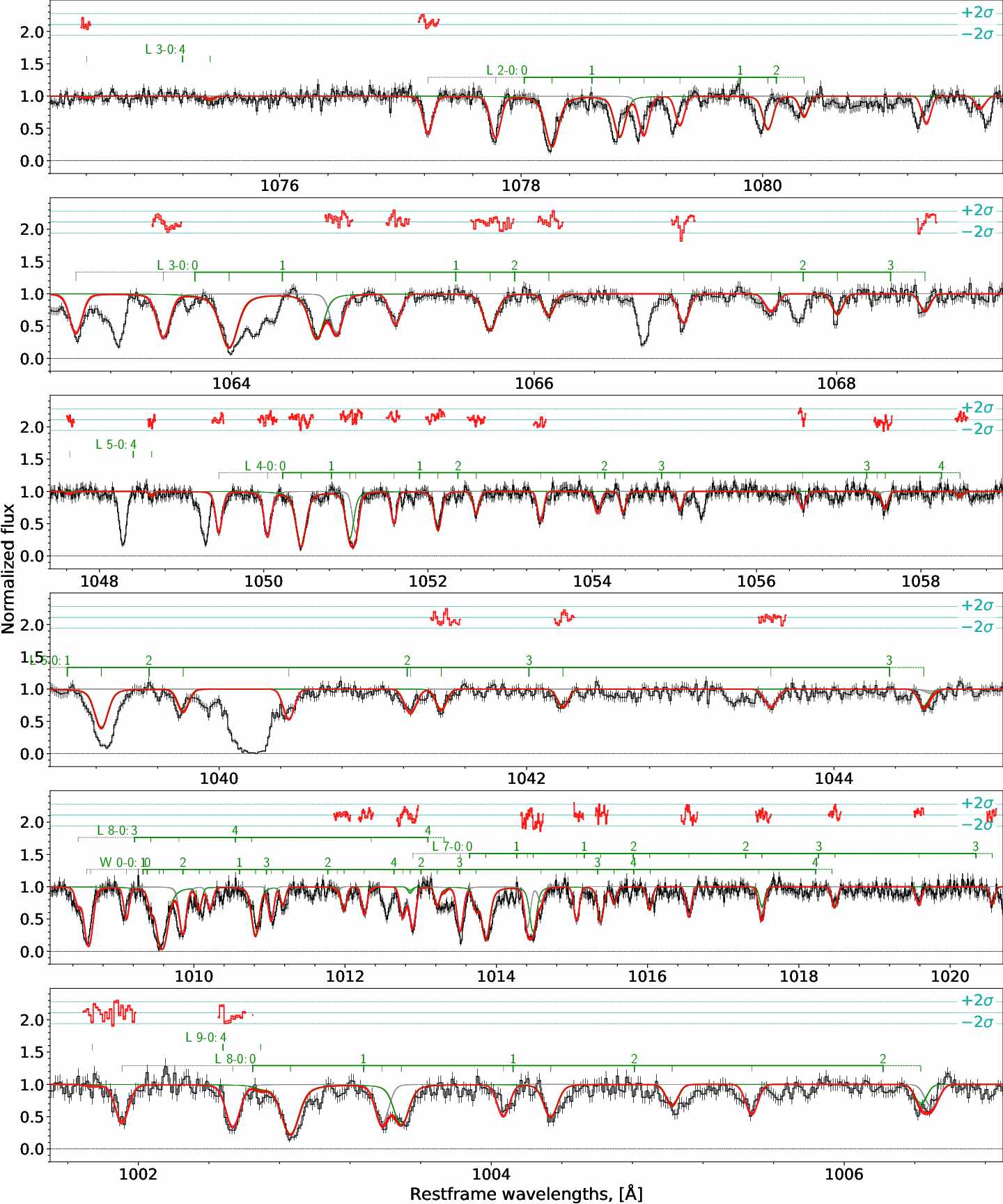}
    \caption{Fit to H2 absorption lines towards Sk-66 51 in LMC. Lines are the same as for \ref{fig:lines_H2_Sk67_2}.
    }
    \label{fig:lines_H2_Sk66_51}
\end{figure*}

\begin{table*}
    \caption{Fit results of H$_2$ lines towards Sk-70 79}
    \label{tab:Sk70_79}
    \begin{tabular}{cccc}
    \hline
    \hline
    species & comp & 1 & 2 \\
            & z &  $0.0000214(^{+11}_{-9})$ & $0.0007824(^{+4}_{-3})$\\
    \hline 
     ${\rm H_2\, J=0}$ & b\,km/s & $1.03^{+0.22}_{-0.32}$ & $0.59^{+0.51}_{-0.08}$ \\
                       & $\log N$ &  $17.792^{+0.025}_{-0.025}$ & $20.1386^{+0.0049}_{-0.0017}$\\
    ${\rm H_2\, J=1}$ & b\,km/s & $1.28^{+0.09}_{-0.26}$ & $0.95^{+0.31}_{-0.37}$ \\
                      & $\log N$ &  $16.10^{+0.17}_{-0.10}$ & $19.9712^{+0.0048}_{-0.0013}$\\
    ${\rm H_2\, J=2}$ & b\,km/s & $4.65^{+0.26}_{-0.25}$ & $1.0^{+0.4}_{-0.4}$\\
                      & $\log N$ &  $15.20^{+0.11}_{-0.07}$ & $18.768^{+0.005}_{-0.009}$ \\
    ${\rm H_2\, J=3}$ & b\,km/s & $4.50^{+0.19}_{-0.25}$ & $4.88^{+0.23}_{-0.35}$\\
                      & $\log N$ & $14.716^{+0.085}_{-0.021}$ & $17.53^{+0.11}_{-0.09}$\\
    ${\rm H_2\, J=4}$ & b\,km/s & -- &  $5.48^{+0.26}_{-0.33}$\\
                      & $\log N$ & $13.3^{+0.4}_{-0.3}$ & $16.04^{+0.16}_{-0.11}$ \\
    ${\rm H_2\, J=5}$ & b\,km/s & -- &  $5.1^{+0.4}_{-0.4}$  \\
                      & $\log N$ & -- & $15.12^{+0.16}_{-0.04}$\\
    ${\rm H_2\, J=6}$ & b\,km/s & -- &  $5.8^{+3.3}_{-1.0}$\\
                      & $\log N$ & -- &  $14.17^{+0.06}_{-0.07}$\\
    \hline 
         & $\log N_{\rm tot}$ & $17.80^{+0.02}_{-0.03}$ & $20.375^{+0.003}_{-0.002}$ \\
    \hline
    HD J=0 & b\,km/s & $1.00^{+0.17}_{-0.28}$ & $2.0^{+0.7}_{-0.6}$ \\
           & $\log N$ & $\lesssim 16.1$ & $15.6^{+0.7}_{-0.5}$ \\
    \hline   
    \end{tabular}
    \begin{tablenotes}
     \item Doppler parameter of H$_2$ ${\rm J = 4}$ rotational level in 1 component  was tied to H$_2$ ${\rm J = 3}$. 
    \end{tablenotes}
\end{table*}

\begin{figure*}
    \centering
    \includegraphics[width=\linewidth]{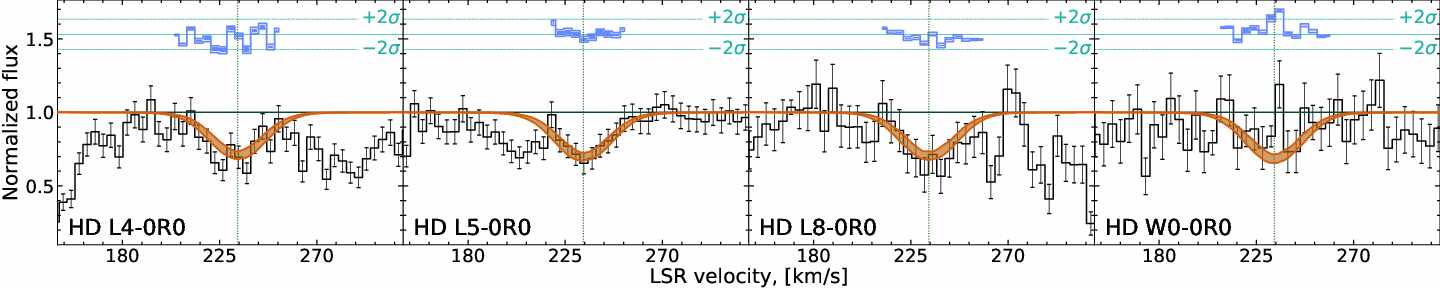}
    \caption{Fit to HD absorption lines towards Sk-70 79 in LMC. Lines are the same as for \ref{fig:lines_HD_Sk67_2}.
    }
    \label{fig:lines_HD_Sk70_79_appendix}
\end{figure*}

\begin{figure*}
    \centering
    \includegraphics[width=\linewidth]{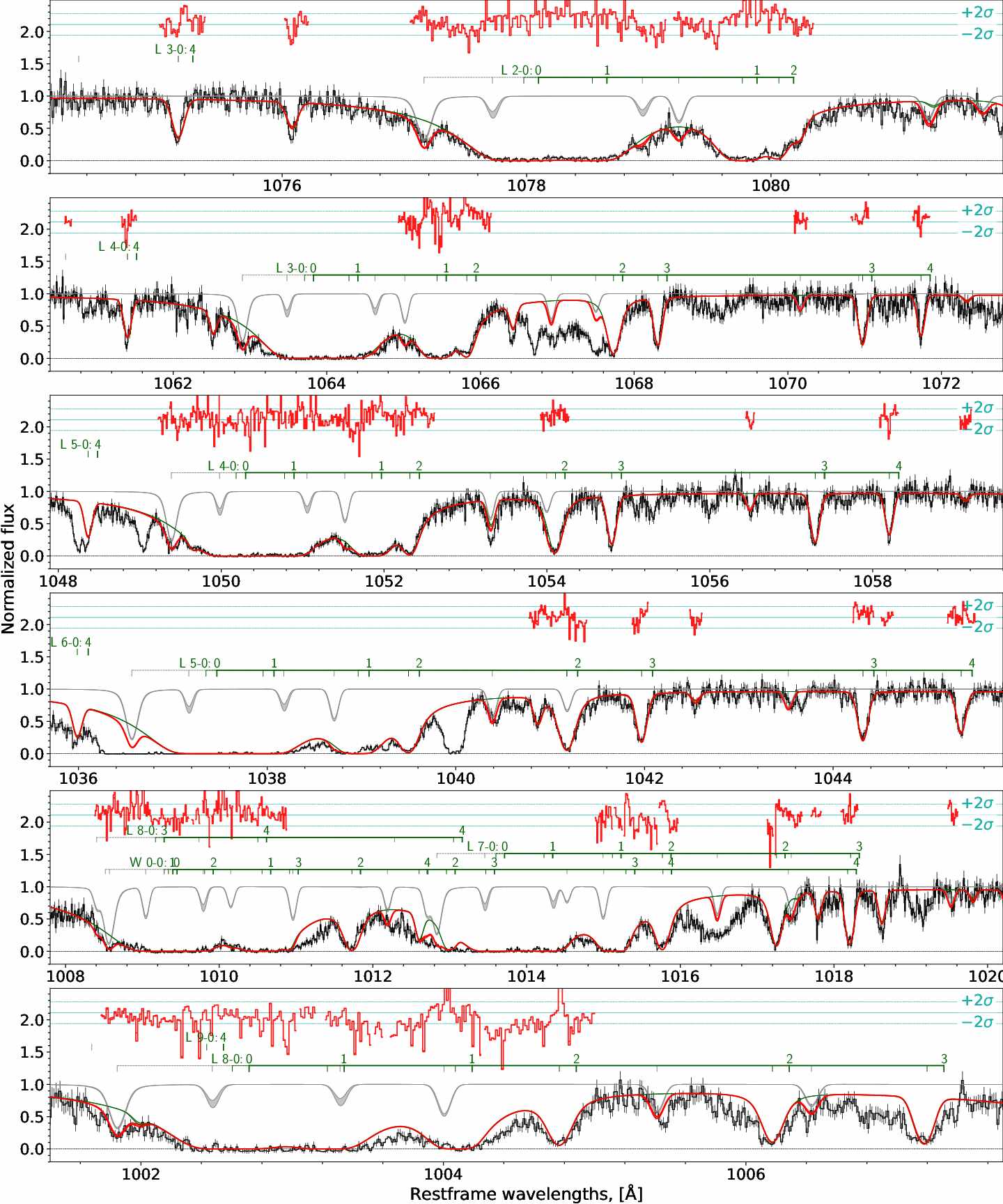}
    \caption{Fit to H2 absorption lines towards Sk-70 79 in LMC. Lines are the same as for \ref{fig:lines_H2_Sk67_2}.
    }
    \label{fig:lines_H2_Sk70_79}
\end{figure*}

\clearpage
\begin{table*}
    \caption{Fit results of H$_2$ lines towards Sk-68 52}
    \label{tab:Sk68_52}
    \begin{tabular}{cccc}
    \hline
    \hline
    species & comp & 1 & 2 \\
            & z & $0.0000709(^{+5}_{-9})$ & $0.0008299(^{+5}_{-5})$ \\
    \hline 
     ${\rm H_2\, J=0}$ & b\,km/s & $1.6^{+0.4}_{-0.4}$ & $0.72^{+0.37}_{-0.22}$ \\
                       & $\log N$ & $17.17^{+0.05}_{-0.04}$ & $19.335^{+0.007}_{-0.009}$ \\
    ${\rm H_2\, J=1}$ & b\,km/s & $2.00^{+0.15}_{-0.44}$ & $1.1^{+0.5}_{-0.4}$ \\
                      & $\log N$ & $17.903^{+0.034}_{-0.025}$ & $19.014^{+0.007}_{-0.008}$ \\
    ${\rm H_2\, J=2}$ & b\,km/s & $2.16^{+0.10}_{-0.09}$ & $2.55^{+0.11}_{-0.13}$ \\
                      & $\log N$ & $17.195^{+0.050}_{-0.029}$ & $17.26^{+0.04}_{-0.06}$ \\
    ${\rm H_2\, J=3}$ & b\,km/s & $2.12^{+0.08}_{-0.11}$ & $2.59^{+0.08}_{-0.12}$ \\
                      & $\log N$ & $16.87^{+0.08}_{-0.07}$ & $17.23^{+0.04}_{-0.06}$ \\
    ${\rm H_2\, J=4}$ & b\,km/s & $2.12^{+0.09}_{-0.10}$ & $2.54^{+0.10}_{-0.09}$ \\
    				  & $\log N$ & $14.85^{+0.16}_{-0.09}$ &$16.08^{+0.07}_{-0.10}$ \\
    ${\rm H_2\, J=5}$ & b\,km/s & -- & $2.67^{+0.10}_{-0.10}$ \\
    				  & $\log N$ & $14.16^{+0.12}_{-0.09}$ & $15.61^{+0.11}_{-0.11}$ \\
    ${\rm H_2\, J=6}$ & $\log N$ & -- & $13.95^{+0.17}_{-0.15}$ \\
    \hline 
         & $\log N_{\rm tot}$ & $18.07^{+0.03}_{-0.02}$ & $19.510^{+0.004}_{-0.007}$ \\
    \hline
    HD J=0 & b\,km/s & $1.7^{+0.4}_{-0.5}$ & $0.517^{+0.466}_{-0.017}$ \\
           & $\log N$ & $\lesssim 15.8$ & $\lesssim 15.6$ \\
    \hline   
    \end{tabular}
    \begin{tablenotes}
     \item Doppler parameters of H$_2$ ${\rm J = 5}$ rotational level in component 1 and H$_2$ ${\rm J = 6}$ rotational level in component 2   were tied to H$_2$ ${\rm J = 4}$ and H$_2$ ${\rm J = 5}$, respectively. 
    \end{tablenotes}
\end{table*}

\begin{figure*}
    \centering
    \includegraphics[width=\linewidth]{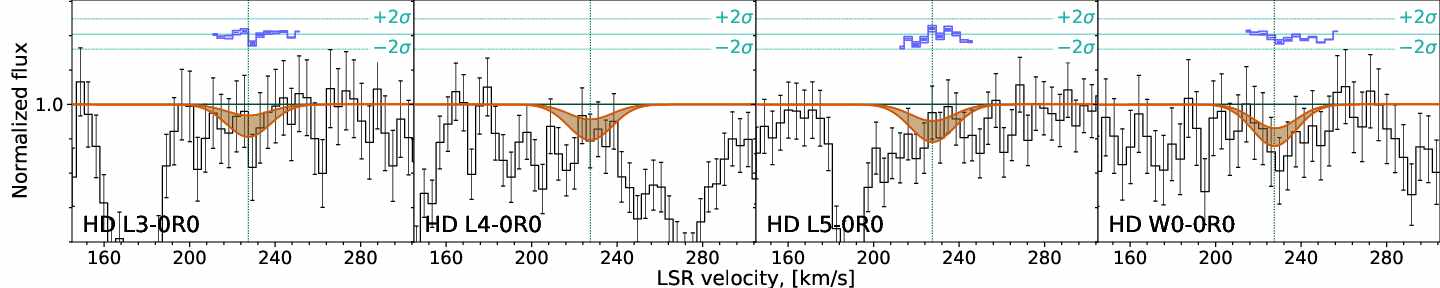}
    \caption{Fit to HD absorption lines towards Sk-68 52 in LMC. Lines are the same as for \ref{fig:lines_HD_Sk67_2}.
    }
    \label{fig:lines_HD_Sk68_52}
\end{figure*}

\begin{figure*}
    \centering
    \includegraphics[width=\linewidth]{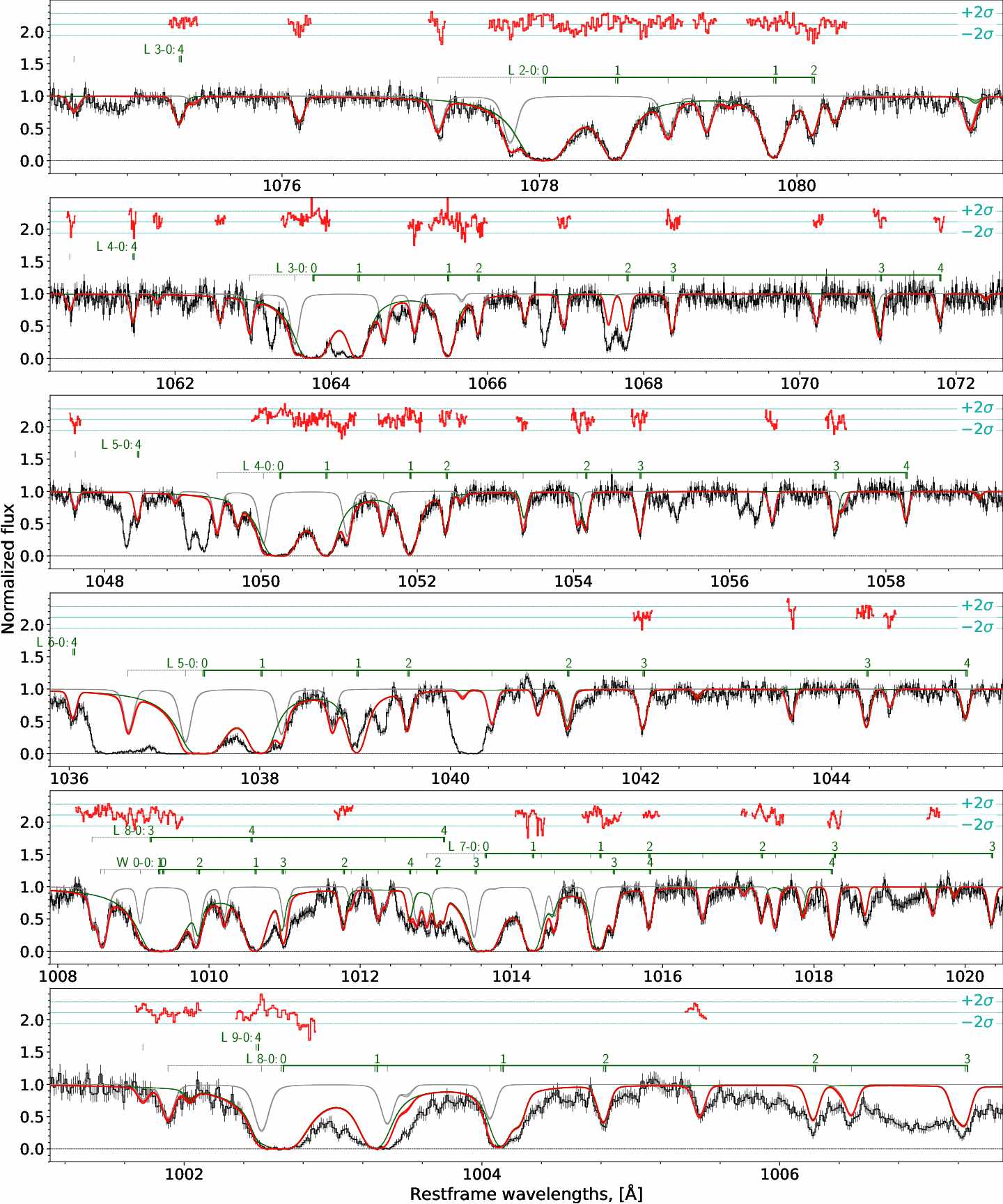}
    \caption{Fit to H2 absorption lines towards Sk-68 52 in LMC. Lines are the same as for \ref{fig:lines_H2_Sk67_2}.
    }
    \label{fig:lines_H2_Sk68_52}
\end{figure*}

\begin{table*}
    \caption{Fit results of H$_2$ lines towards Sk-71 8}
    \label{tab:Sk71_8}
    \begin{tabular}{ccccc}
    \hline
    \hline
    species & comp & 1 & 2 & 3 \\
            & z & $0.0000133(^{+10}_{-10})$ & $0.0006226(^{+15}_{-19})$ & $0.0007353(^{+11}_{-14})$ \\
    \hline 
     ${\rm H_2\, J=0}$ & b\,km/s &$1.8^{+0.6}_{-0.4}$ &  $0.85^{+0.35}_{-0.31}$ & $6.13^{+0.31}_{-0.41}$\\
                       & $\log N$ & $17.72^{+0.05}_{-0.04}$ & $17.57^{+0.05}_{-0.15}$ & $18.08^{+0.08}_{-0.05}$\\
    ${\rm H_2\, J=1}$ & b\,km/s & $3.0^{+0.4}_{-0.5}$ & $2.7^{+0.4}_{-0.6}$ & $5.6^{+0.4}_{-0.4}$\\
                      & $\log N$ &  $18.239^{+0.026}_{-0.042}$ & $18.02^{+0.06}_{-0.08}$ & $18.518^{+0.048}_{-0.030}$\\
    ${\rm H_2\, J=2}$ & b\,km/s & $3.03^{+0.44}_{-0.25}$ &$2.9^{+0.4}_{-0.4}$ & $5.9^{+1.7}_{-0.5}$\\
                      & $\log N$ & $17.29^{+0.09}_{-0.32}$ & $17.12^{+0.12}_{-0.35}$ & $15.93^{+0.88}_{-0.15}$\\
    ${\rm H_2\, J=3}$ & b\,km/s &  $3.5^{+0.3}_{-0.3}$ & $3.7^{+0.5}_{-0.4}$ & $7.72^{+1.02}_{-0.27}$\\
                      & $\log N$ & $16.52^{+0.33}_{-0.27}$ &  $16.72^{+0.19}_{-0.45}$ & $16.02^{+0.17}_{-0.12}$\\
    ${\rm H_2\, J=4}$ & b\,km/s & $3.57^{+0.54}_{-0.17}$ & $3.9^{+1.4}_{-0.7}$ & $9.3^{+0.4}_{-1.4}$\\
    				  & $\log N$ & $14.43^{+0.12}_{-0.10}$ & $14.92^{+0.20}_{-0.12}$ & $15.15^{+0.08}_{-0.05}$\\
    ${\rm H_2\, J=5}$ & b\,km/s & -- & $3.8^{+1.3}_{-0.6}$ &$8.9^{+1.4}_{-1.1}$ \\
    				  & $\log N$ & $14.19^{+0.09}_{-0.27}$ &  $14.74^{+0.18}_{-0.15}$ & $15.10^{+0.09}_{-0.05}$\\
    \hline 
         & $\log N_{\rm tot}$ & $18.40^{+0.02}_{-0.04}$ & $18.21^{+0.04}_{-0.07}$ & $18.66^{+0.04}_{-0.03}$ \\
    \hline
    HD J=0  & b\,km/s &   $1.7^{+0.6}_{-0.4}$ & $0.95^{+0.29}_{-0.31}$ & $6.1^{+0.4}_{-0.4}$ \\
            & $\log N$ & $\lesssim 16.0$ & $\lesssim 16.4$ & $\lesssim 14.3$ \\
    \hline   
    \end{tabular}
    \begin{tablenotes}
     \item Doppler parameter of H$_2$ ${\rm J = 5}$ rotational level in component 1 was tied to H$_2$ ${\rm J = 4}$. 
    \end{tablenotes}
\end{table*}

\begin{figure*}
    \centering
    \includegraphics[width=\linewidth]{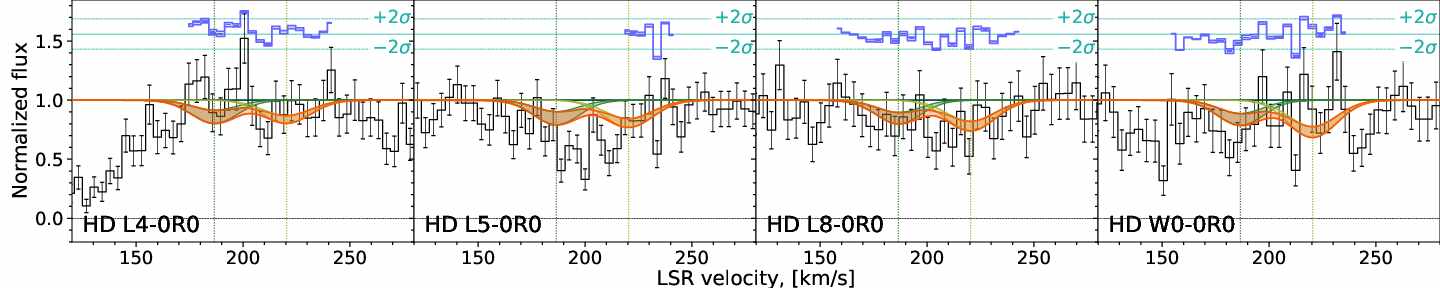}
    \caption{Fit to HD absorption lines towards Sk-71 8 in LMC. Lines are the same as for \ref{fig:lines_HD_Sk67_2}.
    }
    \label{fig:lines_HD_Sk71_8}
\end{figure*}

\begin{figure*}
    \centering
    \includegraphics[width=\linewidth]{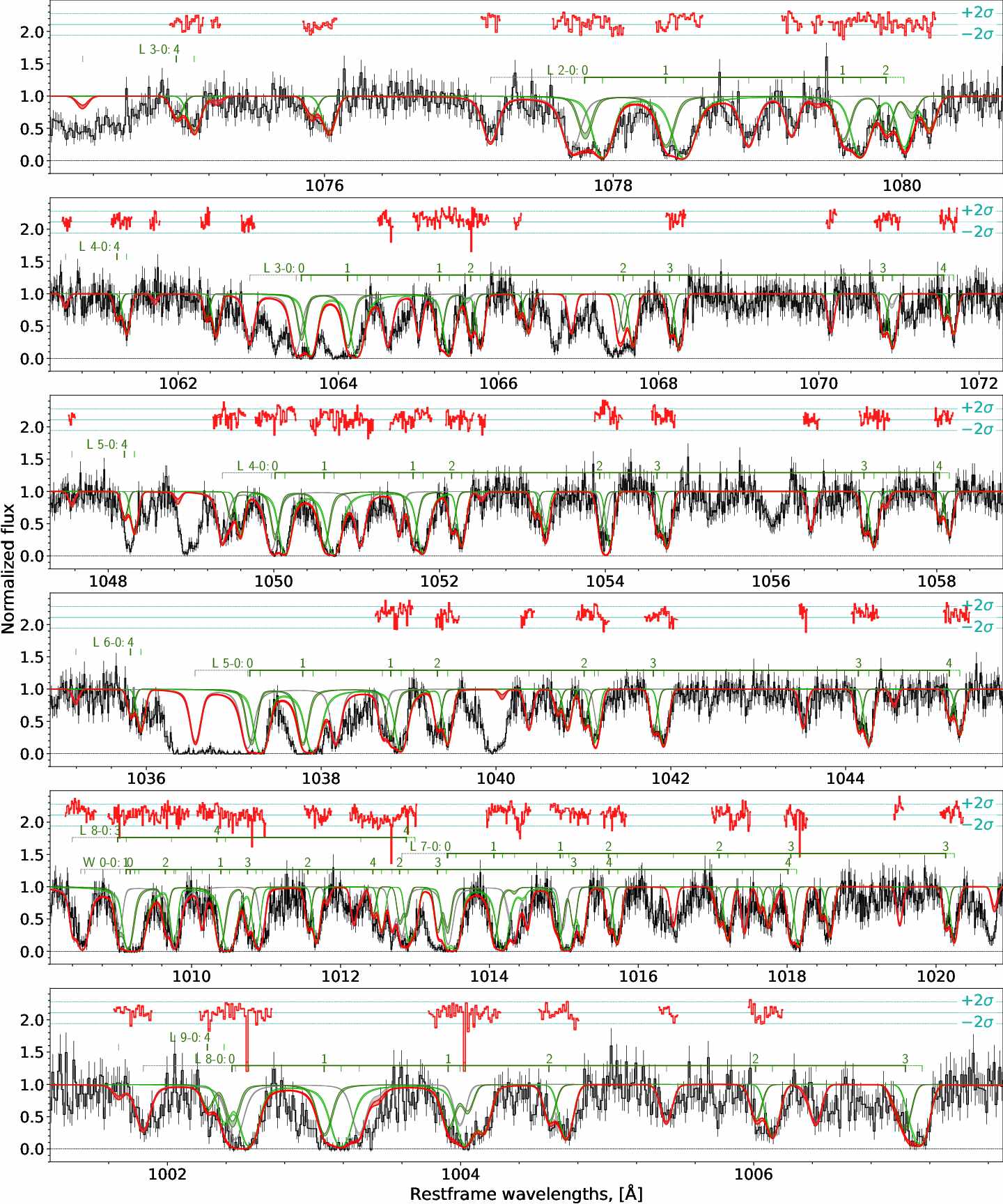}
    \caption{Fit to H2 absorption lines towards Sk-71 8 in LMC. Lines are the same as for \ref{fig:lines_H2_Sk67_2}.
    }
    \label{fig:lines_H2_Sk71_8}
\end{figure*}

\begin{table*}
    \caption{Fit results of H$_2$ lines towards Sk-70 85}
    \label{tab:Sk70_85}
    \begin{tabular}{ccccc}
    \hline
    \hline
    species & comp & 1 & 2 & 3 \\
            & z & $0.0000733(^{+3}_{-8})$ & $0.0008561(^{+6}_{-7})$ &  $0.0009449(^{+16}_{-13})$ \\
    \hline 
     ${\rm H_2\, J=0}$ & b\,km/s & $0.66^{+0.61}_{-0.16}$ & $1.0^{+0.6}_{-0.5}$ & $1.8^{+0.4}_{-0.7}$\\
                       & $\log N$ & $19.342^{+0.013}_{-0.008}$ & $18.37^{+0.07}_{-0.05}$ & $16.80^{+0.23}_{-0.46}$\\
    ${\rm H_2\, J=1}$ & b\,km/s & $1.4^{+0.9}_{-0.5}$ & $3.99^{+0.32}_{-0.22}$ & $2.13^{+0.31}_{-0.58}$\\
                      & $\log N$ & $18.788^{+0.015}_{-0.014}$ & $18.13^{+0.06}_{-0.05}$ & $16.93^{+0.25}_{-0.58}$ \\
    ${\rm H_2\, J=2}$ & b\,km/s & $3.52^{+0.30}_{-0.35}$ & $4.9^{+0.7}_{-0.7}$ & $2.3^{+0.7}_{-0.4}$\\
                      & $\log N$ & $17.40^{+0.19}_{-0.14}$ & $15.79^{+0.49}_{-0.24}$ & $15.03^{+0.47}_{-0.22}$\\
    ${\rm H_2\, J=3}$ & b\,km/s & $3.6^{+0.6}_{-0.4}$ & $6.2^{+0.7}_{-0.4}$ & $3.8^{+0.7}_{-0.7}$\\
                      & $\log N$ &  $16.5^{+0.4}_{-0.4}$ & $15.60^{+0.17}_{-0.13}$ & $14.84^{+0.18}_{-0.10}$\\
    ${\rm H_2\, J=4}$ & b\,km/s & $3.57^{+1.04}_{-0.31}$ & $6.7^{+2.0}_{-0.9}$ & -- \\
    				  & $\log N$ & $14.74^{+0.09}_{-0.10}$ & $14.42^{+0.05}_{-0.05}$ & $13.39^{+0.31}_{-0.40}$\\
    ${\rm H_2\, J=5}$ & b\,km/s & -- & $7.6^{+3.6}_{-1.5}$ & -- \\
    				  & $\log N$ & -- & $14.36^{+0.06}_{-0.06}$ & -- \\
   \hline 
         & $\log N_{\rm tot}$ & $19.45^{+0.01}_{-0.01}$ & $18.57^{+0.05}_{-0.04}$ & $17.18^{+0.18}_{-0.30}$ \\
    \hline 
    HD J=0 & b\,km/s & $0.526^{+0.533}_{-0.026}$ & $0.529^{+0.680}_{-0.029}$ & $0.530^{+1.456}_{-0.030}$ \\
           & $\log N$ & $\lesssim 15.9$ & $\lesssim 16.0$ & $\lesssim 15.9$ \\
    \hline   
    \end{tabular}
    \begin{tablenotes}
     \item Doppler parameters of H$_2$ ${\rm J = 4}$ rotational level in component 3 was tied to H$_2$ ${\rm J = 3}$. 
    \end{tablenotes}
\end{table*}

\begin{figure*}
    \centering
    \includegraphics[width=\linewidth]{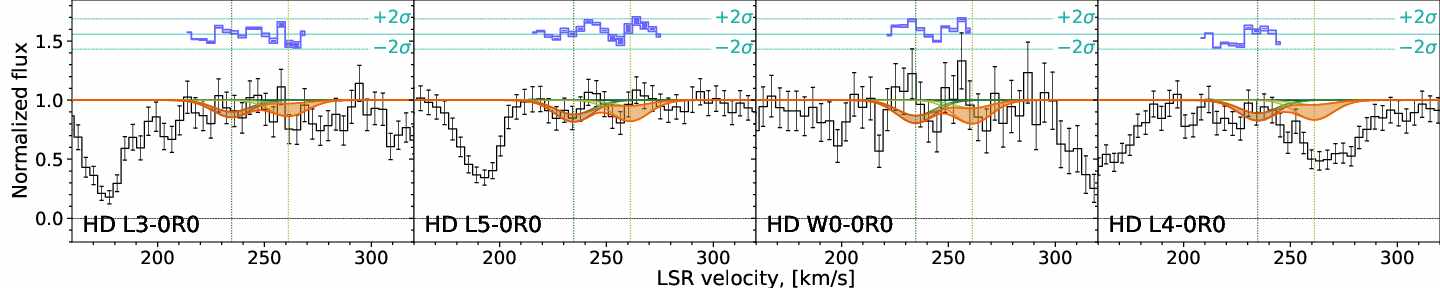}
    \caption{Fit to HD absorption lines towards Sk-70 85 in LMC. Lines are the same as for \ref{fig:lines_HD_Sk67_2}.
    }
    \label{fig:lines_HD_Sk70_85}
\end{figure*}

\begin{figure*}
    \centering
    \includegraphics[width=\linewidth]{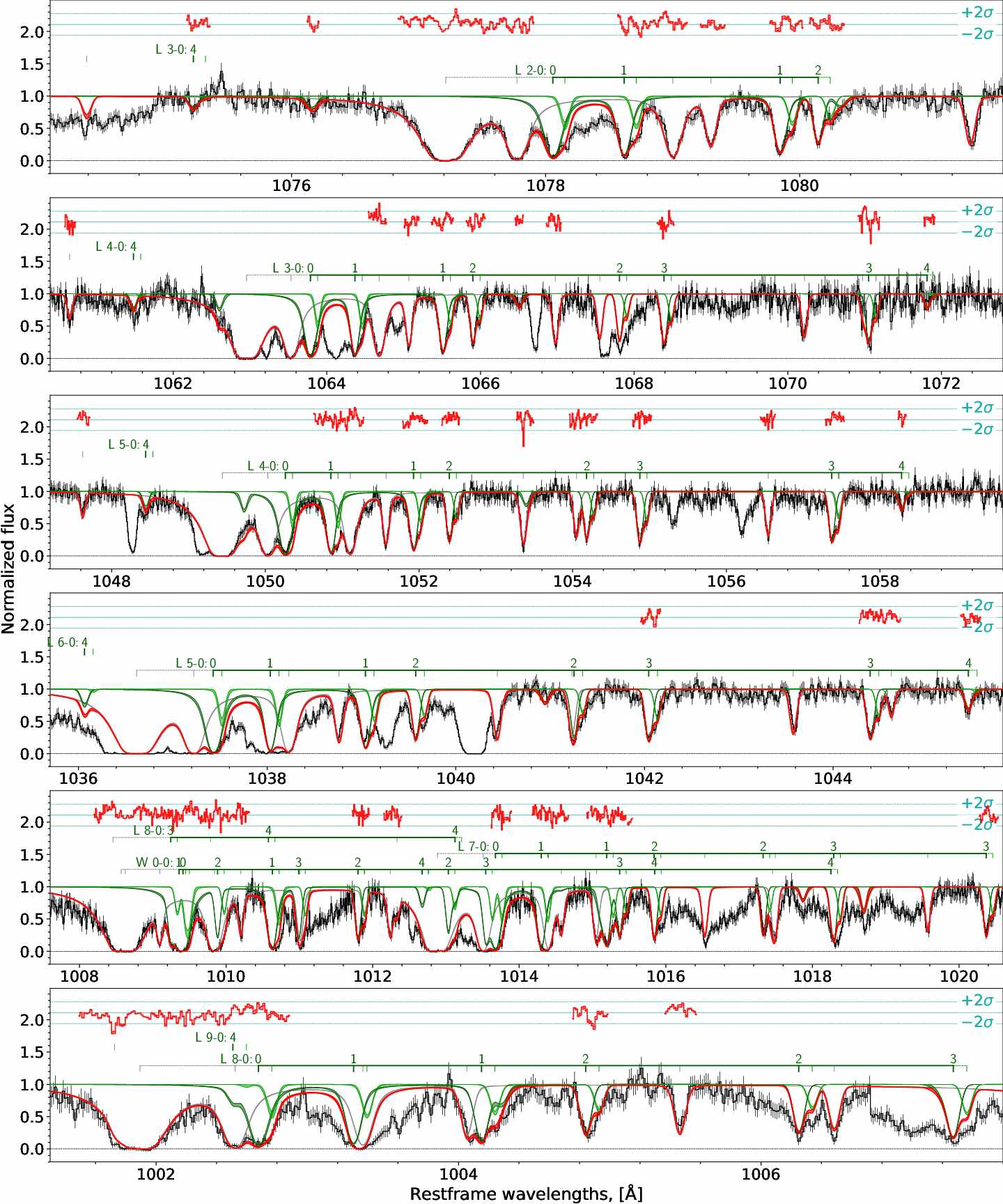}
    \caption{Fit to H2 absorption lines towards Sk-70 85 in LMC. Lines are the same as for \ref{fig:lines_H2_Sk67_2}.
    }
    \label{fig:lines_H2_Sk70_85}
\end{figure*}

\begin{table*}
    \caption{Fit results of H$_2$ lines towards Sk-69 106}
    \label{tab:Sk69_106}
    \begin{tabular}{cccc}
    \hline
    \hline
    species & comp & 1 & 2  \\
            & z & $0.0000700(^{+5}_{-6})$ & $0.0008451(^{+7}_{-6})$ \\
    \hline 
     ${\rm H_2\, J=0}$ & b\,km/s & $2.7^{+0.3}_{-1.4}$ & $0.87^{+0.95}_{-0.20}$\\
                       & $\log N$ & $18.350^{+0.017}_{-0.022}$ & $18.544^{+0.032}_{-0.009}$ \\
    ${\rm H_2\, J=1}$ & b\,km/s &$3.10^{+0.24}_{-0.48}$ & $1.8^{+0.7}_{-0.5}$\\
                      & $\log N$ & $18.257^{+0.021}_{-0.031}$ & $18.388^{+0.026}_{-0.018}$ \\
    ${\rm H_2\, J=2}$ & b\,km/s & $3.50^{+0.17}_{-0.24}$ & $2.4^{+0.4}_{-0.4}$ \\
                      & $\log N$ &  $17.40^{+0.07}_{-0.08}$ & $16.3^{+0.3}_{-0.5}$\\
    ${\rm H_2\, J=3}$ & b\,km/s & $3.84^{+0.26}_{-0.21}$ & $3.48^{+0.23}_{-0.31}$\\
                      & $\log N$ & $16.78^{+0.10}_{-0.20}$ & $15.88^{+0.26}_{-0.16}$\\
    ${\rm H_2\, J=4}$ & b\,km/s & $9.93^{+0.07}_{-1.79}$ & $3.76^{+0.13}_{-0.16}$ \\
    				  & $\log N$ & $14.23^{+0.07}_{-0.05}$ & $14.81^{+0.06}_{-0.08}$\\
    ${\rm H_2\, J=5}$ & $\log N$ & -- & $15.03^{+0.05}_{-0.07}$\\
    ${\rm H_2\, J=6}$ & $\log N$ & -- &$14.23^{+0.09}_{-0.09}$ \\ 
   \hline 
         & $\log N_{\rm tot}$ & $18.64^{+0.01}_{-0.02}$ & $18.78^{+0.02}_{-0.01}$ \\
    \hline
    HD J=0 & b\,km/s & $2.69^{+0.10}_{-2.19}$ & $0.9^{+1.0}_{-0.4}$ \\
           & $\log N$ & $\lesssim 14.3$ & $\lesssim 15.6$ \\ 
    \hline   
    \end{tabular}
    \begin{tablenotes}
     \item Doppler parameters of H$_2$ ${\rm J = 5, 6}$ rotational level in component 2 were tied to H$_2$ ${\rm J = 4}$. 
     \item covering factor was found to be cf = $0.974^{+0.008}_{-0.005}$
    \end{tablenotes}
\end{table*}

\begin{figure*}
    \centering
    \includegraphics[width=\linewidth]{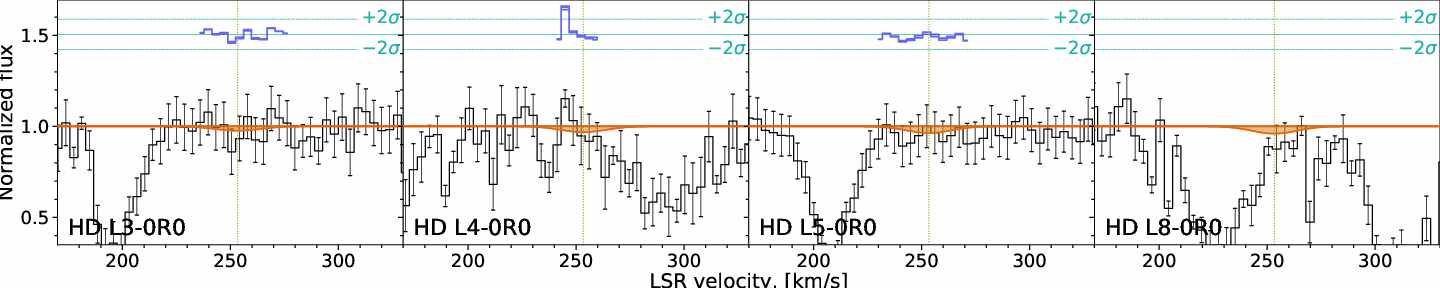}
    \caption{Fit to HD absorption lines towards Sk-60 106 in LMC. Lines are the same as for \ref{fig:lines_HD_Sk67_2}.
    }
    \label{fig:lines_HD_Sk69_106}
\end{figure*}

\begin{figure*}
    \centering
    \includegraphics[width=\linewidth]{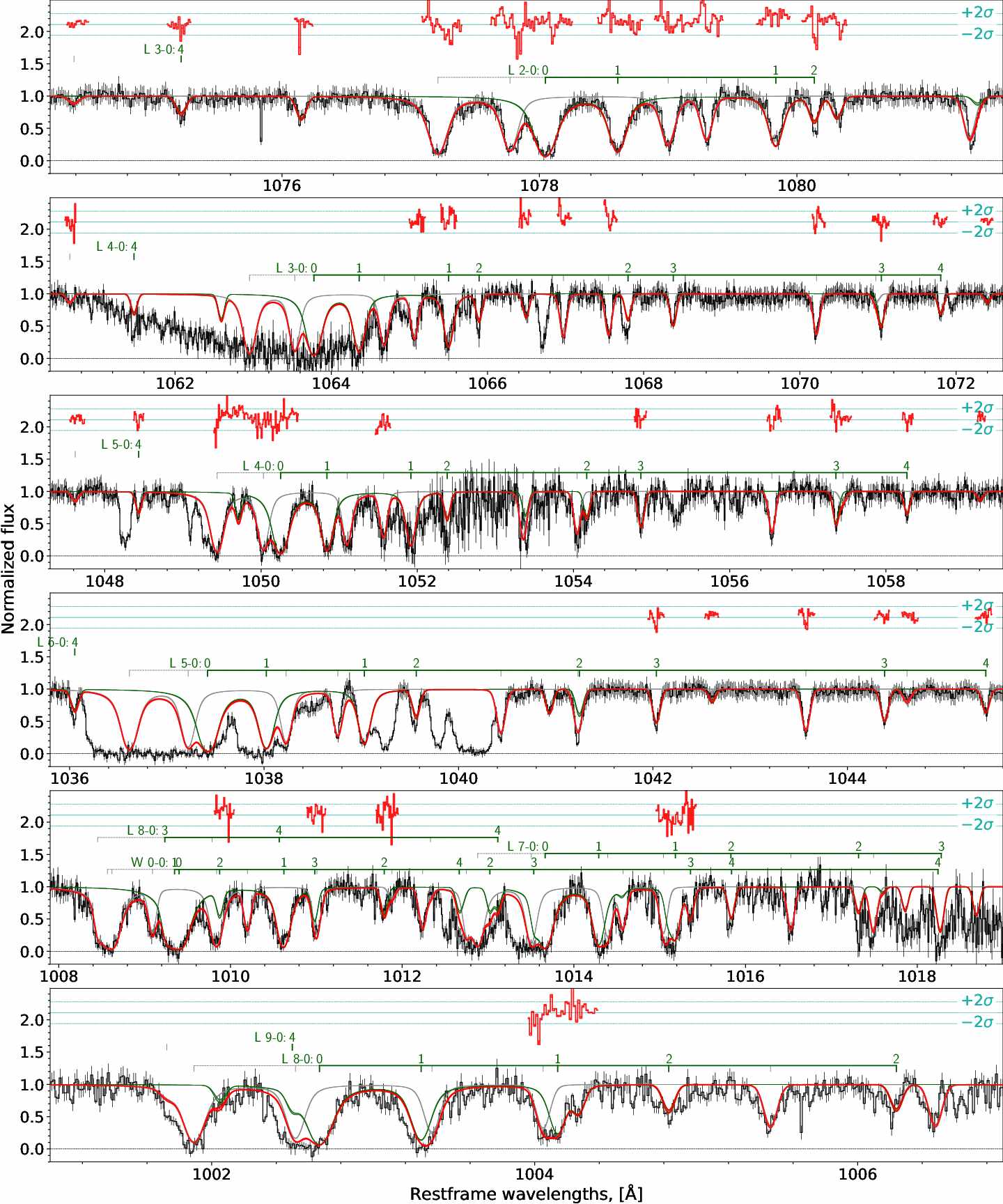}
    \caption{Fit to H2 absorption lines towards Sk-69 106 in LMC. Lines are the same as for \ref{fig:lines_H2_Sk67_2}.
    }
    \label{fig:lines_H2_Sk69_106}
\end{figure*}

\begin{table*}
    \caption{Fit results of H$_2$ lines towards Sk-68 73}
    \label{tab:Sk70_85}
    \begin{tabular}{ccccc}
    \hline
    \hline
    species & comp & 1 & 2 & 3 \\
            & z & $0.0000414(^{+17}_{-11})$ & $0.000078(^{+5}_{-9})$ & $0.00098129(^{+25}_{-97})$ \\
    \hline 
     ${\rm H_2\, J=0}$ & b\,km/s & $3.3^{+1.0}_{-1.4}$ & $0.99^{+1.26}_{-0.25}$ &$0.61^{+0.68}_{-0.11}$ \\
                       & $\log N$ & $17.96^{+0.06}_{-0.11}$ & $17.14^{+0.27}_{-0.16}$ & $20.092^{+0.008}_{-0.015}$\\
    ${\rm H_2\, J=1}$ & b\,km/s & $3.9^{+0.4}_{-1.4}$ & $2.6^{+0.5}_{-1.2}$ & $1.0^{+0.5}_{-0.4}$ \\
                      & $\log N$ & $18.03^{+0.07}_{-0.19}$ &$16.3^{+0.5}_{-0.5}$ & $19.672^{+0.009}_{-0.008}$ \\
    ${\rm H_2\, J=2}$ & b\,km/s & $4.48^{+0.18}_{-0.45}$ &$3.30^{+1.37}_{-0.20}$ & $2.3^{+0.4}_{-1.0}$ \\
                      & $\log N$ & $16.97^{+0.18}_{-0.20}$ &$14.62^{+0.60}_{-0.27}$ & $18.344^{+0.022}_{-0.016}$ \\
    ${\rm H_2\, J=3}$ & b\,km/s & $4.27^{+0.28}_{-0.41}$ & $4.75^{+0.20}_{-0.46}$ &$4.51^{+0.26}_{-0.35}$ \\
                      & $\log N$ & $15.96^{+0.35}_{-0.17}$ & $13.5^{+0.3}_{-0.8}$ &$17.89^{+0.04}_{-0.08}$ \\
    ${\rm H_2\, J=4}$ & b\,km/s & -- & -- & $4.27^{+0.24}_{-0.44}$\\
    				  & $\log N$ & $14.05^{+0.12}_{-0.09}$ & $10.72^{+0.25}_{-0.66}$ & $16.05^{+0.29}_{-0.17}$\\
    ${\rm H_2\, J=5}$ & b\,km/s & -- & -- & $5.8^{+1.3}_{-0.8}$\\
    				  & $\log N$ & $14.26^{+0.05}_{-0.15}$ & $10.6^{+0.6}_{-0.5}$ & $15.29^{+0.41}_{-0.08}$\\
    ${\rm H_2\, J=6}$ & b\,km/s & -- & -- & $5.4^{+2.0}_{-0.6}$ \\
                      & $\log N$ & -- & -- & $14.41^{+0.05}_{-0.14}$ \\
   \hline 
         & $\log N_{\rm tot}$ & $18.32^{+0.05}_{-0.10}$ & $17.20^{+0.25}_{-0.14}$ & $20.240^{+0.005}_{-0.012}$ \\
     \hline
     HD J=0 & b\,km/s & $3.7^{+0.7}_{-1.7}$ & $0.9^{+1.1}_{-0.4}$ & $8.64^{+7.31}_{-5.94}$ \\
            & $\log N$ & $\lesssim 15.6$ & $\lesssim 16.0$ & $14.16^{+0.08}_{-0.08}$ \\
    \hline   
    \end{tabular}
    \begin{tablenotes}
     \item Doppler parameters of H$_2$ ${\rm J = 4, 5}$ rotational level in components 1 and 2 were tied to H$_2$ ${\rm J = 3}$. 
    \end{tablenotes}
\end{table*}

\begin{figure*}
    \centering
    \includegraphics[width=\linewidth]{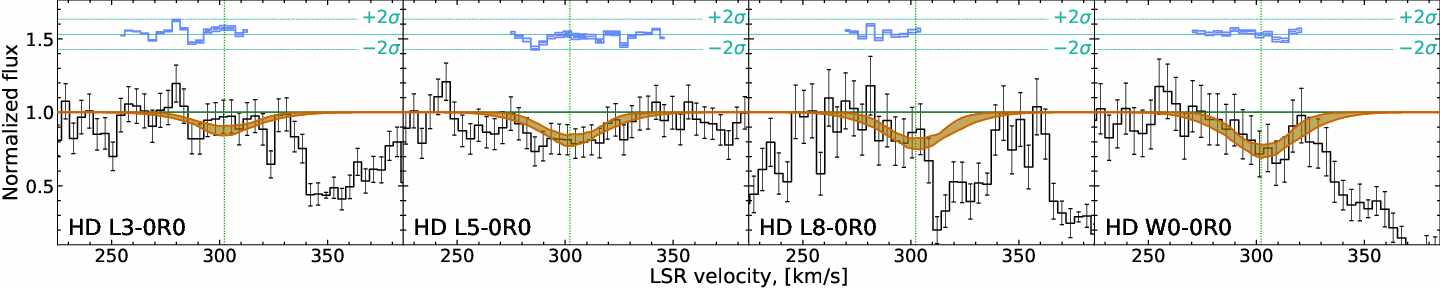}
    \caption{Fit to HD absorption lines towards Sk-68 73 in LMC. Lines are the same as for \ref{fig:lines_HD_Sk67_2}.
    }
    \label{fig:lines_HD_Sk68_73_appendix}
\end{figure*}

\begin{figure*}
    \centering
    \includegraphics[width=\linewidth]{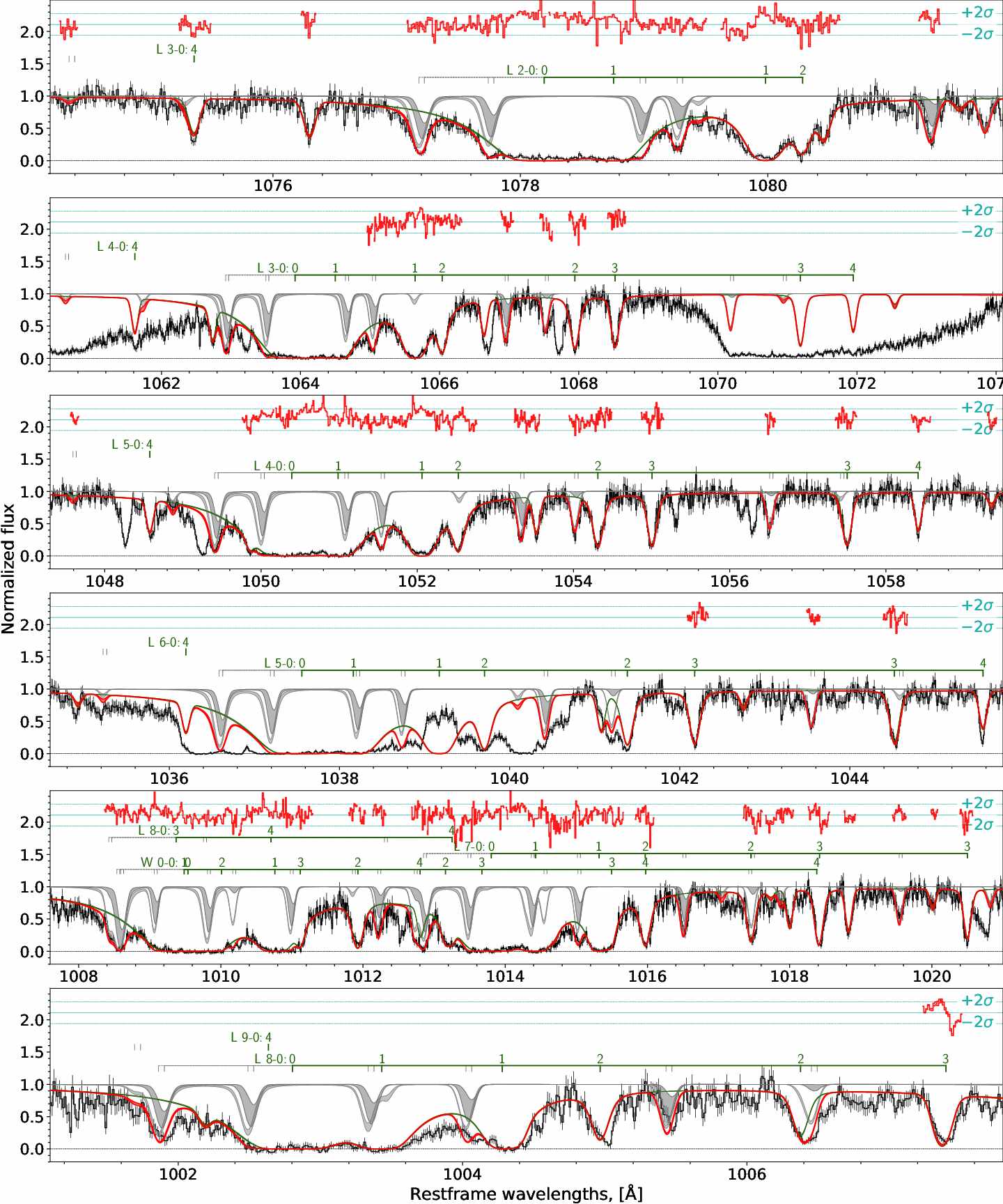}
    \caption{Fit to H2 absorption lines towards Sk-68 73 in LMC. Lines are the same as for \ref{fig:lines_H2_Sk67_2}.
    }
    \label{fig:lines_H2_Sk68_73}
\end{figure*}

\clearpage
\begin{table*}
    \caption{Fit results of H$_2$ lines towards Sk-67 105}
    \label{tab:Sk67_105}
    \begin{tabular}{cccccc}
    \hline
    \hline
    species & comp & 1 & 2 & 3 & 4 \\
            & z & $0.0000649(^{+5}_{-9})$ & $0.0001037(^{+28}_{-14})$ & $0.00100693(^{+39}_{-29})$ & $0.0010352(^{+28}_{-11})$ \\
    \hline 
     ${\rm H_2\, J=0}$ & b\,km/s &$2.14^{+0.16}_{-0.21}$ & $1.89^{+0.17}_{-0.29}$ & $2.3^{+0.6}_{-0.8}$ & $0.66^{+0.10}_{-0.12}$\\
                       & $\log N$ & $17.32^{+0.04}_{-0.05}$ & $14.81^{+0.10}_{-0.20}$ & $19.275^{+0.006}_{-0.010}$ & $17.46^{+0.29}_{-0.87}$\\
    ${\rm H_2\, J=1}$ & b\,km/s & $2.43^{+0.19}_{-0.12}$ & $4.3^{+0.5}_{-0.7}$ & $2.7^{+0.3}_{-0.4}$ & $1.03^{+0.13}_{-0.14}$\\
                      & $\log N$ & $17.82^{+0.04}_{-0.04}$ & $14.58^{+0.13}_{-0.04}$ & $18.878^{+0.006}_{-0.007}$ &$15.39^{+0.32}_{-0.30}$ \\
    ${\rm H_2\, J=2}$ & b\,km/s & $4.14^{+0.09}_{-0.34}$ & $5.08^{+0.23}_{-0.69}$ & $3.48^{+0.24}_{-0.18}$ & $1.19^{+0.24}_{-0.10}$\\
                      & $\log N$ & $15.386^{+0.027}_{-0.071}$ & $13.96^{+0.09}_{-0.06}$ & $15.82^{+0.15}_{-0.12}$ & $14.09^{+0.21}_{-0.09}$\\
    ${\rm H_2\, J=3}$ & b\,km/s & $5.9^{+0.4}_{-2.4}$ & $5.43^{+0.29}_{-0.32}$ & $3.29^{+0.16}_{-0.15}$ &$6.02^{+0.37}_{-0.31}$ \\
                      & $\log N$ & $14.63^{+0.13}_{-0.04}$ & $13.99^{+0.07}_{-0.05}$ & $15.88^{+0.09}_{-0.14}$ & $13.20^{+0.19}_{-0.06}$\\
    ${\rm H_2\, J=4}$ & b\,km/s & -- & -- & $3.22^{+0.31}_{-0.15}$ & --\\
    				  & $\log N$ &$13.32^{+0.13}_{-0.18}$ & $13.19^{+0.11}_{-0.19}$ & $14.72^{+0.05}_{-0.03}$ & $11.1^{+0.4}_{-0.8}$\\
    ${\rm H_2\, J=5}$ & b\,km/s & -- & -- & $3.35^{+0.30}_{-0.25}$ & -- \\
    				  & $\log N$ & -- & -- & $14.471^{+0.049}_{-0.017}$ & --  \\
    ${\rm H_2\, J=6}$ & b\,km/s & -- & -- & $22.0^{+4.1}_{-2.3}$ & -- \\
    				  & $\log N$ & -- & -- &$13.74^{+0.09}_{-0.09}$ & -- \\
   \hline 
         & $\log N_{\rm tot}$ & $17.94^{+0.03}_{-0.03}$ & $15.09^{+0.07}_{-0.10}$ & $19.422^{+0.043}_{-0.007}$ & $17.46^{+0.29}_{-0.85}$ \\
    \hline
    HD J=0 & b\,km/s &   $2.14^{+0.17}_{-0.24}$ & $1.87^{+0.23}_{-0.31}$ &$2.2^{+0.7}_{-0.8}$ & $0.65^{+0.11}_{-0.10}$ \\
           & $\log N$ & $\lesssim 13.6$ & $\lesssim 13.8$ & $\lesssim 14.8$ & $\lesssim 13.9$ \\
    \hline   
     & cf$^{\dagger}$ & \multicolumn{4}{c}{$0.8882^{+0.0022}_{-0.0023}$} \\
    \end{tabular}
    \begin{tablenotes}
     \item Doppler parameters of H$_2$ ${\rm J = 4}$ rotational level in component 1, 2 and 3 were tied to H$_2$ ${\rm J = 3}$.
     \item $\dagger$ Covering factor was used for all H$_2$ lines.
    \end{tablenotes}
\end{table*}

\begin{figure*}
    \centering
    \includegraphics[width=\linewidth]{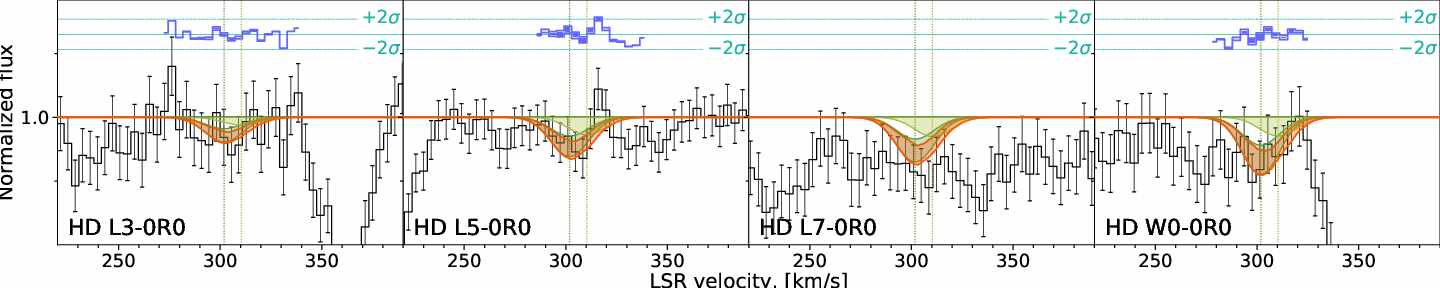}
    \caption{Fit to HD absorption lines towards Sk-67 105 in LMC. Lines are the same as for \ref{fig:lines_HD_Sk67_2}.
    }
    \label{fig:lines_HD_Sk67_105}
\end{figure*}

\begin{figure*}
    \centering
    \includegraphics[width=\linewidth]{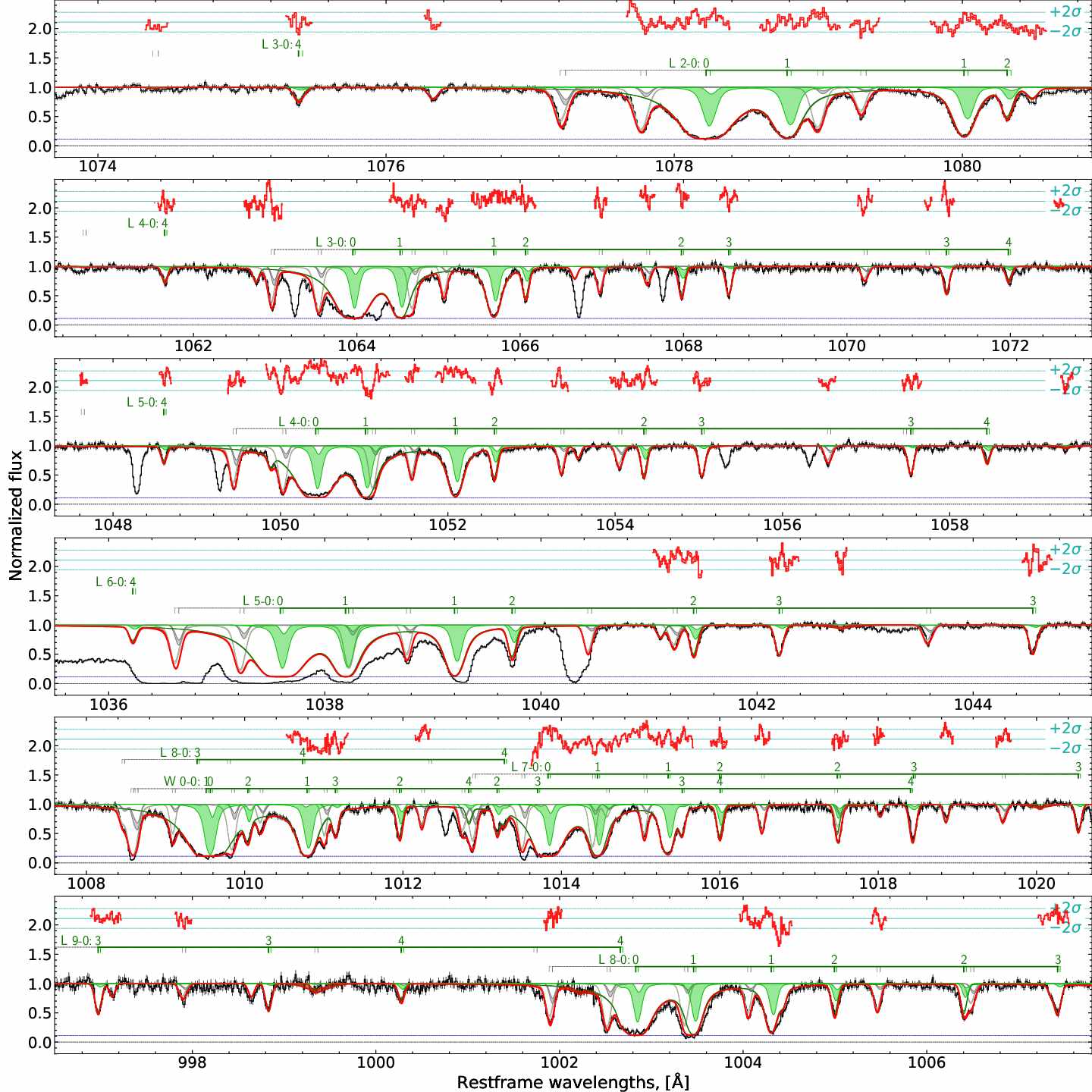}
    \caption{Fit to H2 absorption lines towards Sk-67 105 in LMC. Lines are the same as for \ref{fig:lines_H2_Sk67_2}. Blue dashed line shows covering factor.
    }
    \label{fig:lines_H2_Sk67_105}
\end{figure*}

\begin{table*}
    \caption{Fit results of H$_2$ lines towards BI 184}
    \label{tab:BI184}
    \begin{tabular}{ccccc}
    \hline
    \hline
    species & comp & 1 & 2 & 3 \\
            & z & $0.0000687(^{+18}_{-9})$ & $0.0008020(^{+9}_{-8})$ & $0.000938(^{+3}_{-4})$ \\
    \hline 
     ${\rm H_2\, J=0}$ & b\,km/s & $1.1^{+1.2}_{-0.6}$ & $1.7^{+3.3}_{-0.8}$ &$1.2^{+0.4}_{-0.4}$ \\
                       & $\log N$ & $19.134^{+0.022}_{-0.017}$ & $19.794^{+0.009}_{-0.019}$ &$14.0^{+0.4}_{-0.4}$ \\
    ${\rm H_2\, J=1}$ & b\,km/s & $3.0^{+0.6}_{-1.4}$ & $6.9^{+0.8}_{-1.3}$ &$1.82^{+0.09}_{-0.31}$ \\
                      & $\log N$ & $18.846^{+0.039}_{-0.020}$ & $19.204^{+0.028}_{-0.020}$ &  $15.40^{+0.29}_{-0.34}$\\
    ${\rm H_2\, J=2}$ & b\,km/s & $3.7^{+0.4}_{-0.5}$ & $7.0^{+0.7}_{-1.2}$ & $1.75^{+0.13}_{-0.20}$\\
                      & $\log N$ &$16.99^{+0.38}_{-0.30}$ & $16.32^{+0.62}_{-0.29}$ & $16.75^{+0.11}_{-0.21}$\\
    ${\rm H_2\, J=3}$ & b\,km/s & $4.0^{+0.5}_{-0.4}$ & $7.7^{+0.6}_{-0.8}$ & $1.85^{+0.12}_{-0.16}$\\
                      & $\log N$ & $15.93^{+0.35}_{-0.27}$ & $15.88^{+0.28}_{-0.17}$ & $15.01^{+0.33}_{-0.20}$ \\
    ${\rm H_2\, J=4}$ & b\,km/s & -- &  $7.6^{+0.8}_{-0.7}$ & -- \\
    				  & $\log N$ & $14.60^{+0.16}_{-0.16}$ & $15.10^{+0.12}_{-0.07}$ & $13.7^{+0.4}_{-1.5}$\\
    ${\rm H_2\, J=5}$ & b\,km/s & -- & $8.4^{+0.4}_{-0.4}$ & -- \\
    				  & $\log N$ & $14.04^{+0.15}_{-0.21}$ & $15.16^{+0.04}_{-0.06}$ & -- \\
    ${\rm H_2\, J=6}$ & $\log N$ & -- &$14.13^{+0.11}_{-0.05}$ & -- \\				 
    \hline 
         & $\log N_{\rm tot}$ & $19.32^{+0.02}_{-0.01}$ & $19.89^{+0.01}_{-0.02}$ & $16.78^{+0.11}_{-0.20}$ \\
    \hline
    HD J=0 & b\,km/s & $0.56^{+1.67}_{-0.06}$ & $1.4^{+2.1}_{-0.9}$ & $1.10^{+0.60}_{-0.30}$ \\
           & $\log N$ & $\lesssim 15.7$ & $\lesssim 16.7$ & $\lesssim 15.8$ \\
    \hline   
    \end{tabular}
    \begin{tablenotes}
     \item Doppler parameters of H$_2$ ${\rm J = 4, 5}$ in the 1 component and H$_2$ ${\rm J = 4}$ in the 3 component  rotational levels  were tied to H$_2$ ${\rm J = 3}$. 
    \end{tablenotes}
\end{table*}

\begin{figure*}
    \centering
    \includegraphics[width=\linewidth]{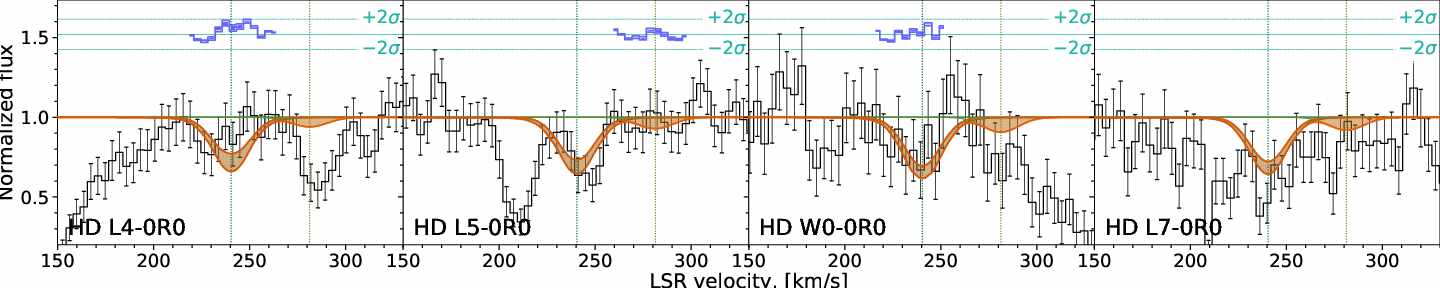}
    \caption{Fit to HD absorption lines towards BI 184 in LMC. Lines are the same as for \ref{fig:lines_HD_Sk67_2}.
    }
    \label{fig:lines_HD_BI184}
\end{figure*}

\begin{figure*}
    \centering
    \includegraphics[width=\linewidth]{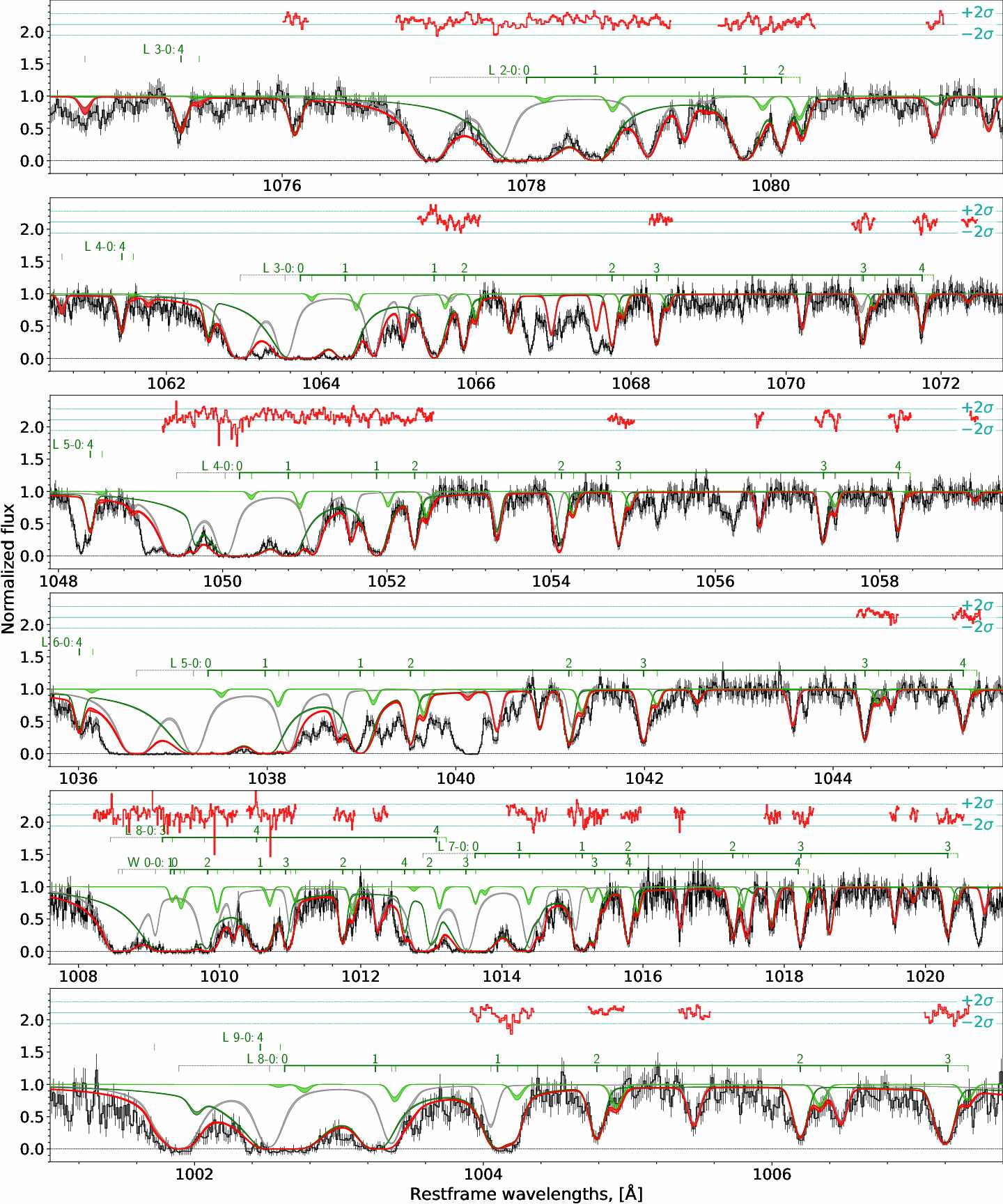}
    \caption{Fit to H2 absorption lines towards BI 184 in LMC. Lines are the same as for \ref{fig:lines_H2_Sk67_2}.
    }
    \label{fig:lines_H2_BI184}
\end{figure*}

\begin{table*}
    \caption{Fit results of H$_2$ lines towards Sk-71 38}
    \label{tab:Sk71_38}
    \begin{tabular}{cccccc}
    \hline
    \hline
    species & comp & 1 & 2 & 3 & 4\\
            & z & $0.0000338(^{+9}_{-7})$ & $0.0006714(^{+19}_{-20})$ & $0.0008106(^{+8}_{-5})$ & $0.0009343(^{+23}_{-14})$ \\
    \hline 
     ${\rm H_2\, J=0}$ & b\,km/s & $0.525^{+0.318}_{-0.025}$ & $0.530^{+0.126}_{-0.030}$ & $3.4^{+0.4}_{-1.6}$ & $0.68^{+0.08}_{-0.17}$\\
                       & $\log N$ & $19.052^{+0.015}_{-0.017}$ & $17.26^{+0.10}_{-0.11}$ &$18.10^{+0.11}_{-0.08}$ &$17.20^{+0.07}_{-0.07}$ \\
    ${\rm H_2\, J=1}$ & b\,km/s & $0.69^{+0.39}_{-0.14}$ & $0.63^{+0.18}_{-0.07}$ & $4.38^{+0.34}_{-0.21}$ & $0.83^{+0.14}_{-0.13}$ \\
                      & $\log N$ & $18.768^{+0.009}_{-0.018}$ & $17.01^{+0.09}_{-0.07}$ & $18.32^{+0.06}_{-0.04}$ & $17.08^{+0.09}_{-0.09}$\\
    ${\rm H_2\, J=2}$ & b\,km/s & $1.60^{+0.30}_{-0.40}$ & $0.76^{+0.22}_{-0.12}$ & $4.35^{+0.24}_{-0.25}$ &$0.97^{+0.06}_{-0.15}$ \\
                      & $\log N$ & $17.517^{+0.024}_{-0.068}$ & $15.46^{+0.39}_{-0.27}$ & $16.32^{+0.24}_{-0.14}$ & $15.45^{+0.30}_{-0.21}$\\
    ${\rm H_2\, J=3}$ & b\,km/s &  $2.07^{+0.15}_{-0.19}$ & $1.14^{+0.08}_{-0.17}$ &  $4.43^{+0.28}_{-0.21}$ & $1.01^{+0.04}_{-0.13}$\\
                      & $\log N$ & $17.11^{+0.10}_{-0.07}$ & $14.95^{+0.39}_{-0.25}$ & $16.07^{+0.22}_{-0.13}$ &  $15.10^{+0.17}_{-0.35}$\\
    ${\rm H_2\, J=4}$ & b\,km/s & $1.99^{+0.15}_{-0.23}$ & -- & $4.50^{+0.40}_{-0.30}$ & -- \\
    				  & $\log N$ &$14.73^{+0.21}_{-0.14}$ & $14.05^{+0.21}_{-0.27}$ &$14.90^{+0.06}_{-0.07}$ &  $14.09^{+0.14}_{-0.28}$\\
    ${\rm H_2\, J=5}$ & b\,km/s & -- & -- & $5.1^{+3.2}_{-0.8}$ & -- \\
    				  & $\log N$ & $12.8^{+0.6}_{-0.4}$ & -- & $14.47^{+0.06}_{-0.04}$ & -- \\
    \hline 
         & $\log N_{\rm tot}$ & $19.25^{+0.01}_{-0.01}$ & $17.46^{+0.07}_{-0.07}$ & $18.53^{+0.06}_{-0.04}$ & $17.45^{+0.06}_{-0.05}$ \\
     \hline
     HD J=0 & b\,km/s & $0.515^{+0.320}_{-0.015}$ & $0.506^{+0.160}_{-0.006}$ & $0.59^{+1.70}_{-0.09}$ & $0.67^{+0.10}_{-0.10}$ \\
            & $\log N$ & $\lesssim 15.5$ & $\lesssim 16.1$ & $\lesssim 16.0$ & $\lesssim 14.7$ \\
    \hline   
    \end{tabular}
    \begin{tablenotes}
     \item Doppler parameters of H$_2$ ${\rm J = 5}$ rotational level in the 1 component and  H$_2$ ${\rm J = 4}$ rotational level in the 2 and 4 components  were tied to H$_2$ ${\rm J = 4}$ and H$_2$ ${\rm J = 5}$, respectively. 
    \end{tablenotes}
\end{table*}

\begin{figure*}
    \centering
    \includegraphics[width=\linewidth]{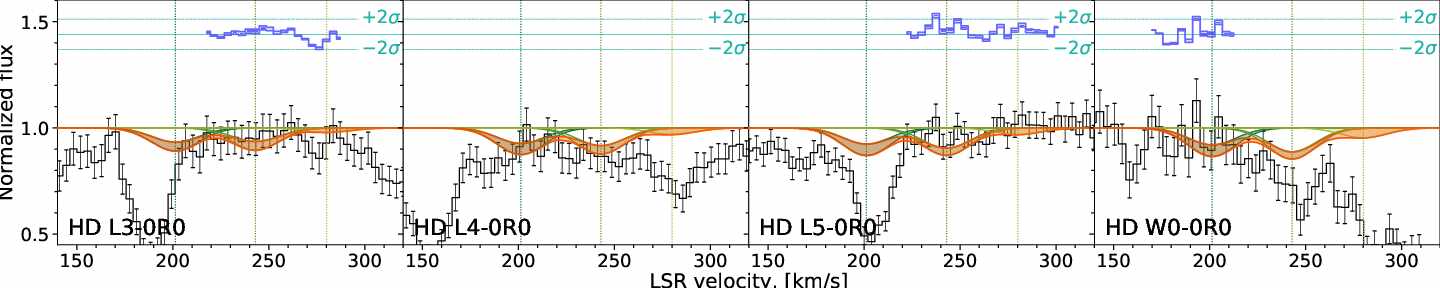}
    \caption{Fit to HD absorption lines towards Sk-71 38 in LMC. Lines are the same as for \ref{fig:lines_HD_Sk67_2}.
    }
    \label{fig:lines_HD_Sk71_38}
\end{figure*}

\begin{figure*}
    \centering
    \includegraphics[width=\linewidth]{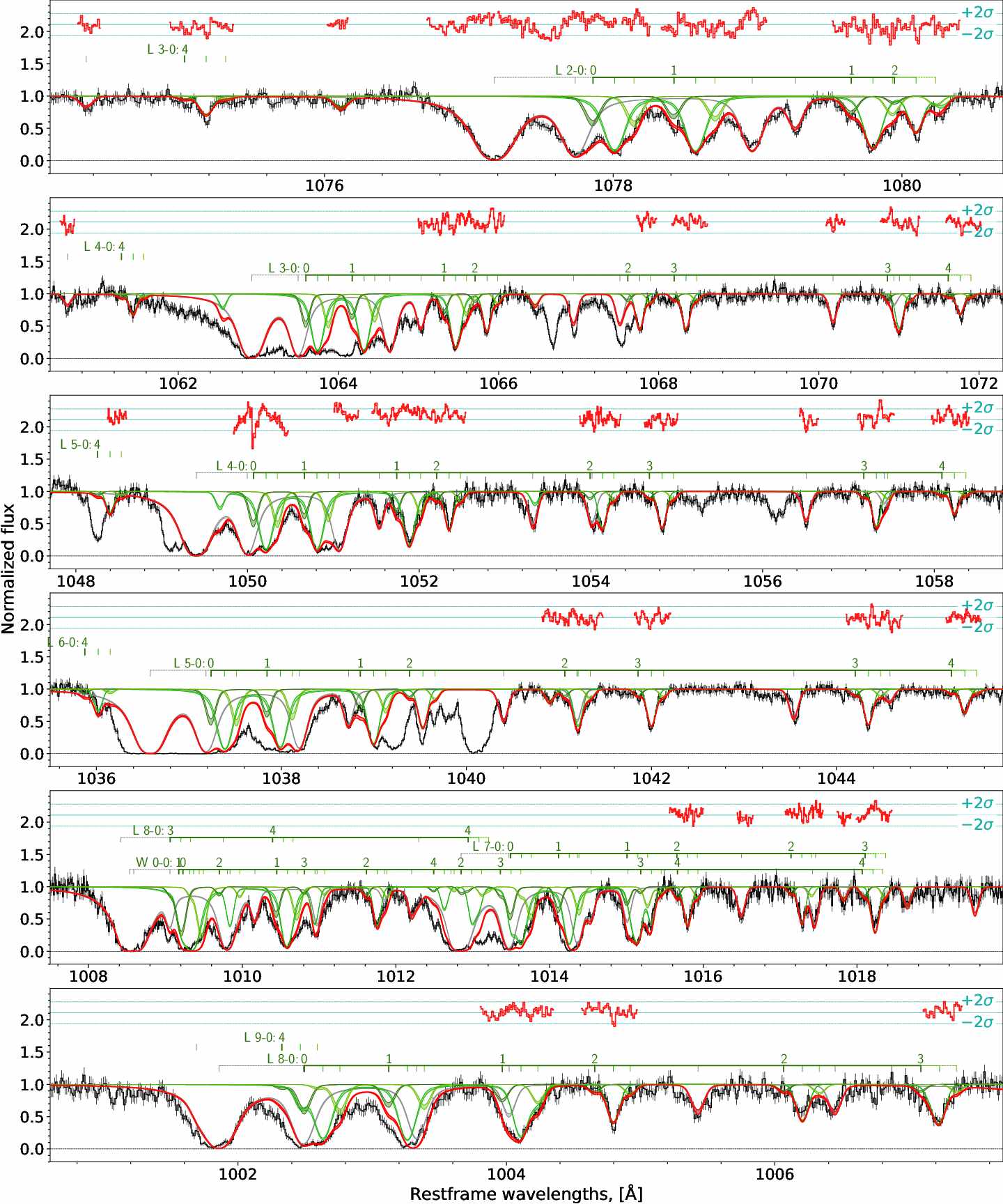}
    \caption{Fit to H2 absorption lines towards Sk-71 38 in LMC. Lines are the same as for \ref{fig:lines_H2_Sk67_2}.
    }
    \label{fig:lines_H2_Sk71_38}
\end{figure*}

\begin{table*}
    \caption{Fit results of H$_2$ lines towards Sk-71 45}
    \label{tab:Sk71_45}
    \begin{tabular}{cccc}
    \hline
    \hline
    species & comp & 1 & 2 \\
            & z & $0.00009854(^{+22}_{-30})$ & $0.00084775(^{+19}_{-16})$ \\
    \hline 
     ${\rm H_2\, J=0}$ & b\,km/s & $0.9^{+0.5}_{-0.4}$ & $3.90^{+0.20}_{-0.16}$\\
                       & $\log N$ & $18.877^{+0.004}_{-0.004}$ & $18.119^{+0.016}_{-0.011}$\\
    ${\rm H_2\, J=1}$ & b\,km/s &  $2.09^{+0.17}_{-0.85}$ & $4.21^{+0.11}_{-0.08}$\\
                      & $\log N$ &  $18.576^{+0.006}_{-0.007}$ &$18.208^{+0.009}_{-0.010}$ \\
    ${\rm H_2\, J=2}$ & b\,km/s & $2.27^{+0.13}_{-0.11}$ & $3.41^{+0.07}_{-0.10}$\\
                      & $\log N$ & $17.421^{+0.039}_{-0.019}$ & $17.498^{+0.023}_{-0.026}$\\
    ${\rm H_2\, J=3}$ & b\,km/s & $2.63^{+0.18}_{-0.15}$ & $3.45^{+0.13}_{-0.07}$\\
                      & $\log N$ & $16.74^{+0.07}_{-0.21}$ &  $17.455^{+0.032}_{-0.031}$\\
    ${\rm H_2\, J=4}$ & b\,km/s & $2.55^{+0.22}_{-0.20}$ &$4.02^{+0.22}_{-0.22}$ \\
    				  & $\log N$ &  $14.65^{+0.07}_{-0.04}$ & $14.679^{+0.026}_{-0.034}$\\
    ${\rm H_2\, J=5}$ & b\,km/s & -- &$8.5^{+0.4}_{-0.4}$	 \\		
                      & $\log N$ & $13.54^{+0.07}_{-0.10}$ & $14.464^{+0.010}_{-0.010}$\\
    ${\rm H_2\, J=6}$ & $\log N$ & -- & $13.51^{+0.08}_{-0.05}$ \\
    \hline 
         & $\log N_{\rm tot}$ & $19.065^{+0.003}_{-0.004}$ & $18.55^{+0.01}_{-0.01}$ \\
     \hline
     HD J=0 &  b\,km/s & $1.0^{+0.3}_{-0.5}$ &  $3.92^{+0.21}_{-0.22}$ \\
            & $\log N$ & $\lesssim 13.7$ & $\lesssim 13.3$ \\
    \hline   
    \end{tabular}
    \begin{tablenotes}
    \item Doppler parameters H$_2$ $\rm J=5$ in 1 component and $J=6$ in 2 component were tied to $\rm J=4$ and $\rm J=5$, respectively.
    \end{tablenotes}
\end{table*}

\begin{figure*}
    \centering
    \includegraphics[width=\linewidth]{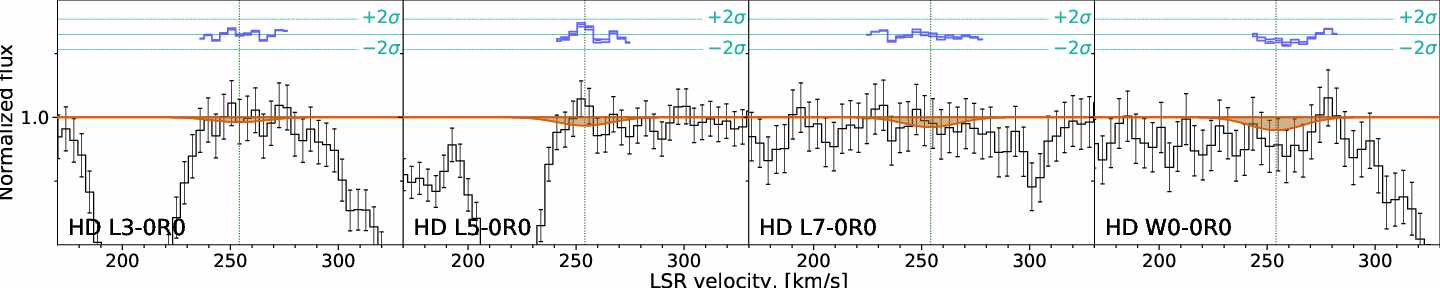}
    \caption{Fit to HD absorption lines towards Sk-71 45 in LMC. Lines are the same as for \ref{fig:lines_HD_Sk67_2}.
    }
    \label{fig:lines_HD_Sk71_45}
\end{figure*}

\begin{figure*}
    \centering
    \includegraphics[width=\linewidth]{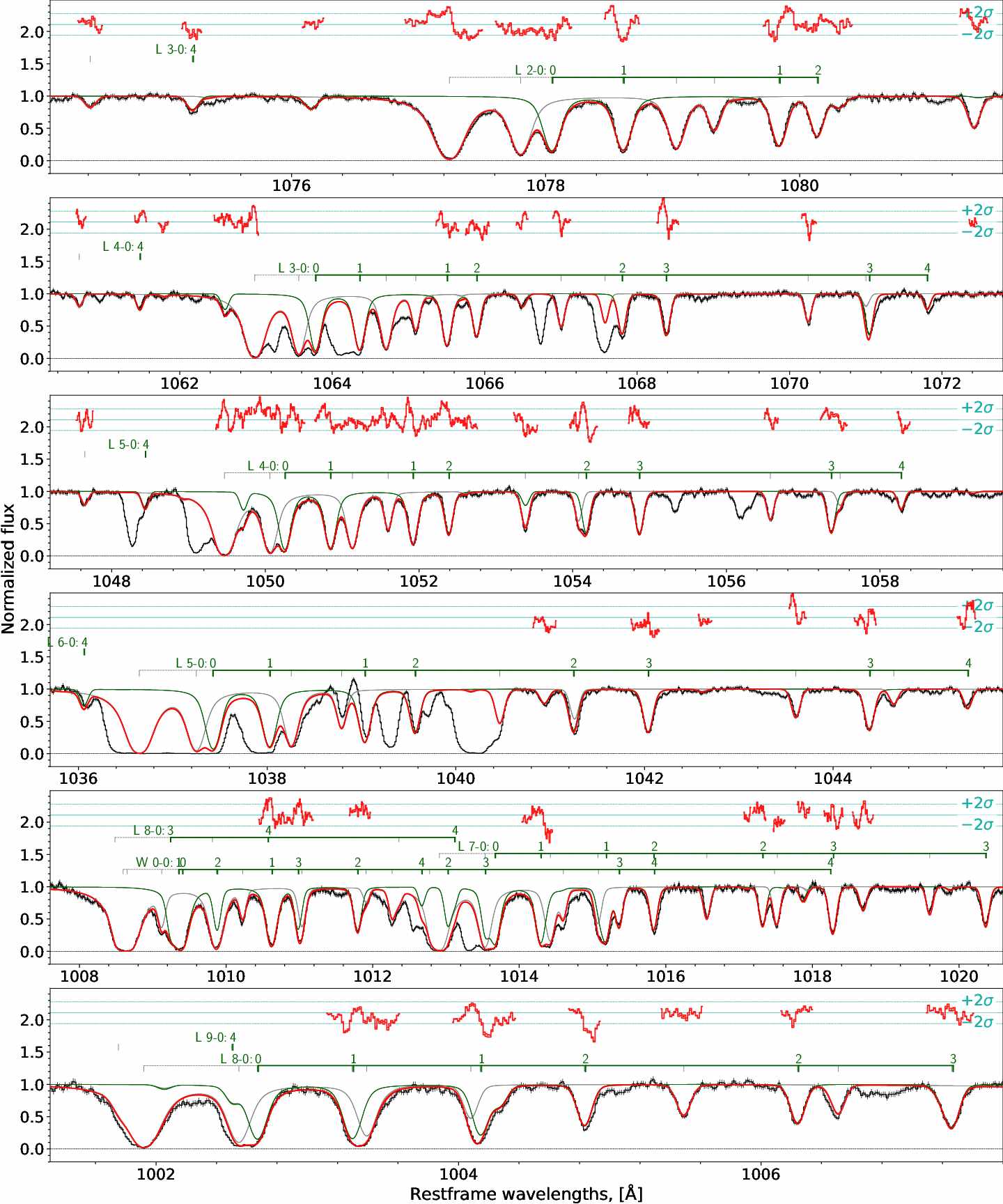}
    \caption{Fit to H2 absorption lines towards Sk-71 45 in LMC. Lines are the same as for \ref{fig:lines_H2_Sk67_2}.
    }
    \label{fig:lines_H2_Sk71_45}
\end{figure*}

\begin{table*}
    \caption{Fit results of H$_2$ lines towards Sk-71 46}
    \label{tab:Sk71_46}
    \begin{tabular}{cccc}
    \hline
    \hline
    species & comp & 1 & 2 \\
            & z & $0.0000675(^{+20}_{-22})$ & $0.0008119(^{+12}_{-14})$ \\
    \hline 
     ${\rm H_2\, J=0}$ & b\,km/s & $1.6^{+2.0}_{-1.1}$ &  $1.6^{+2.0}_{-1.1}$\\
                       & $\log N$ &$19.39^{+0.04}_{-0.05}$ &  $20.14^{+0.04}_{-0.06}$ \\
    ${\rm H_2\, J=1}$ & b\,km/s & $5.2^{+1.1}_{-2.3}$ & $5.0^{+1.0}_{-2.3}$ \\
                      & $\log N$ &  $18.96^{+0.08}_{-0.14}$ & $19.860^{+0.027}_{-0.008}$ \\
    ${\rm H_2\, J=2}$ & b\,km/s & $6.5^{+1.1}_{-0.7}$ & $5.6^{+0.9}_{-0.5}$ \\
                      & $\log N$ & $17.1^{+0.3}_{-0.6}$ & $18.11^{+0.16}_{-0.16}$\\
    ${\rm H_2\, J=3}$ & b\,km/s & $8.7^{+2.4}_{-1.0}$ &  $6.0^{+1.3}_{-0.7}$\\
                      & $\log N$ & $15.80^{+0.52}_{-0.20}$ & $17.4^{+0.4}_{-0.7}$ \\
    ${\rm H_2\, J=4}$ & b\,km/s & $15.17^{+4.12}_{-4.32}$ & $6.8^{+0.9}_{-1.2}$ \\
    				  & $\log N$ & $14.85^{+0.08}_{-0.09}$ & $15.51^{+0.40}_{-0.18}$ \\
    \hline 
         & $\log N_{\rm tot}$ & $19.53^{+0.04}_{-0.05}$ & $20.32^{+0.03}_{-0.04}$ \\
    \hline
    HD J=0 & b\,km/s & $1.7^{+1.0}_{-0.9}$ & $3.1^{+5.3}_{-2.5}$ \\
           & $\log N$ & $\lesssim 17.0$ &  $14.60^{+1.08}_{-0.30}$ \\ 
    \hline   
    \end{tabular}
    \begin{tablenotes}
    \item 
    \end{tablenotes}
\end{table*}

\begin{figure*}
    \centering
    \includegraphics[width=\linewidth]{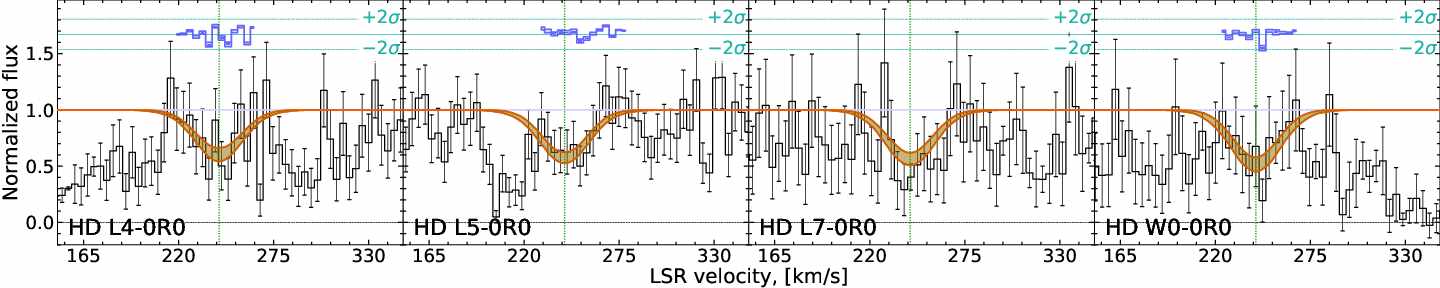}
    \caption{Fit to HD absorption lines towards Sk-71 46 in LMC. Lines are the same as for \ref{fig:lines_HD_Sk67_2}.
    }
    \label{fig:lines_HD_Sk71_46}
\end{figure*}

\begin{figure*}
    \centering
    \includegraphics[width=\linewidth]{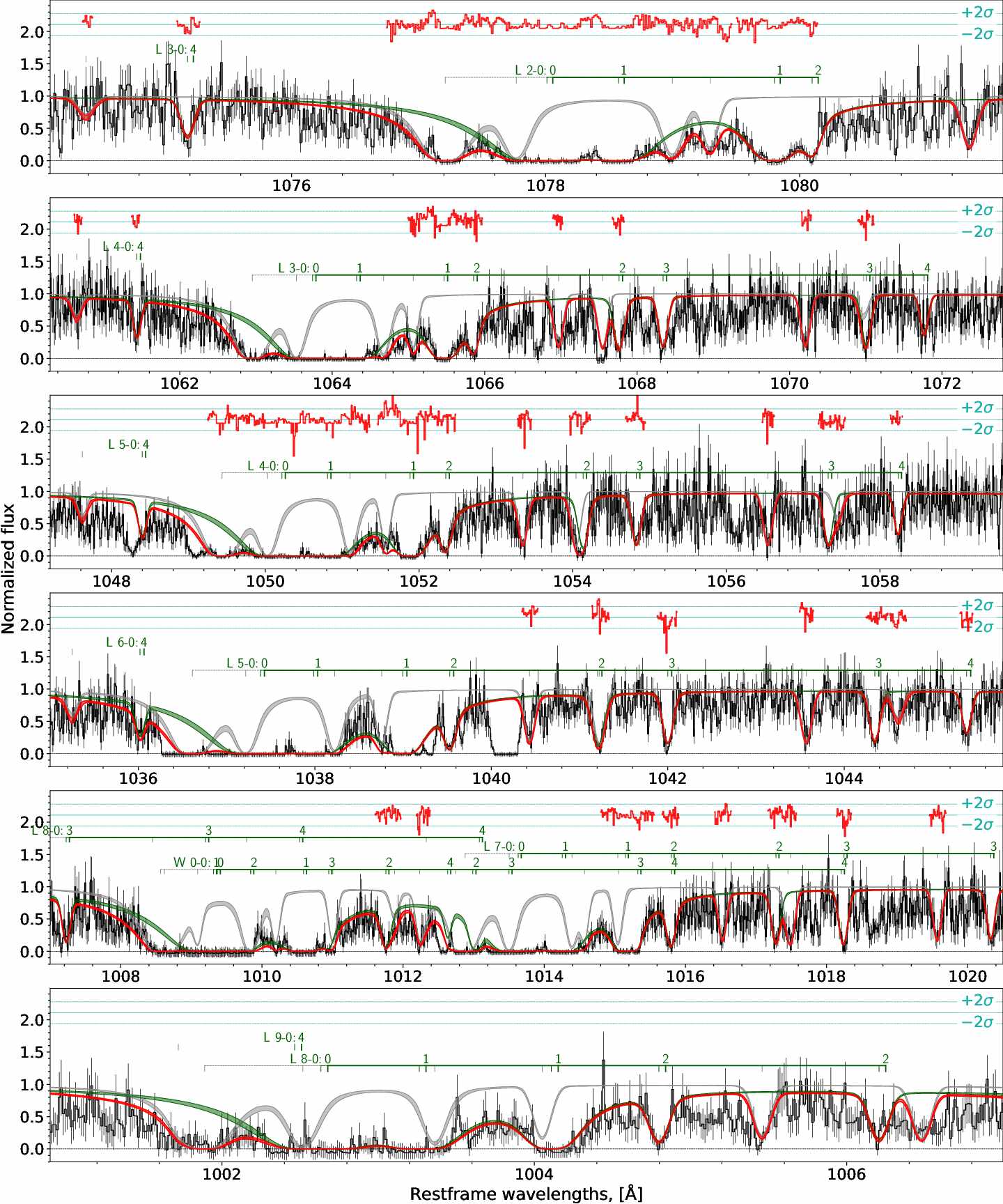}
    \caption{Fit to H2 absorption lines towards Sk-71 46 in LMC. Lines are the same as for \ref{fig:lines_H2_Sk67_2}.
    }
    \label{fig:lines_H2_Sk71_46}
\end{figure*}

\begin{table*}
    \caption{Fit results of H$_2$ lines towards Sk-69 191}
    \label{tab:Sk69_191}
    \begin{tabular}{ccccc}
    \hline
    \hline
    species & comp & 1 & 2 & 3 \\
            & z & $0.0000422(^{+8}_{-10})$ &  $0.0007644(^{+9}_{-11})$ & $0.0008597(^{+65}_{-22})$  \\
    \hline 
     ${\rm H_2\, J=0}$ & b\,km/s &  $3.7^{+0.8}_{-0.9}$ & $0.99^{+1.22}_{-0.22}$ & $1.0^{+0.5}_{-0.5}$ \\
                       & $\log N$ &  $17.88^{+0.05}_{-0.05}$ &$18.775^{+0.017}_{-0.034}$ & $15.0^{+1.2}_{-3.2}$ \\
    ${\rm H_2\, J=1}$ & b\,km/s & $3.7^{+0.4}_{-0.4}$ & $2.3^{+0.9}_{-0.7}$ & $1.9^{+0.4}_{-0.9}$ \\
                      & $\log N$ & $17.94^{+0.06}_{-0.10}$ & $18.817^{+0.023}_{-0.011}$ & $14.76^{+0.57}_{-0.19}$\\
    ${\rm H_2\, J=2}$ & b\,km/s & $4.0^{+0.7}_{-0.5}$ & $3.46^{+0.19}_{-1.08}$ & $2.7^{+1.0}_{-0.8}$ \\
                      & $\log N$ & $16.3^{+0.5}_{-0.4}$ & $17.45^{+0.10}_{-0.50}$ & $14.1^{+0.4}_{-0.4}$ \\
    ${\rm H_2\, J=3}$ & b\,km/s & $4.0^{+1.2}_{-0.4}$ & $6.34^{+0.23}_{-2.08}$ & $4.1^{+0.5}_{-0.6}$ \\
                      & $\log N$ & $14.98^{+0.19}_{-0.11}$ & $15.27^{+0.55}_{-0.14}$ & $14.50^{+0.10}_{-0.04}$\\
    ${\rm H_2\, J=4}$ & b\,km/s & $7.28^{+5.03}_{-3.97}$ & $6.2^{+1.2}_{-1.9}$ & -- \\
    				  & $\log N$ &$14.12^{+0.05}_{-0.11}$ & $14.58^{+0.10}_{-0.04}$ & $13.9^{+0.3}_{-0.5}$\\
   \hline 
         & $\log N_{\rm tot}$ & $18.22^{+0.04}_{-0.06}$ & $19.11^{+0.01}_{-0.02}$ & $15.34^{+0.92}_{-0.30}$  \\
    \hline
    HD J=0 & b\,km/s & $3.8^{+0.9}_{-1.2}$ & $0.94^{+0.37}_{-0.30}$ & $0.525^{+0.769}_{-0.025}$ \\
           & $\log N$ &  $\lesssim 15.9$ & $\lesssim 15.9$ & $\lesssim 15.1$ \\ 
    \hline   
    \end{tabular}
    \begin{tablenotes}
     \item Doppler parameter of H$_2$ ${\rm J = 4}$ rotational level in the 3 component was tied to H$_2$ ${\rm J = 3}$. 
    \end{tablenotes}
\end{table*}

\begin{figure*}
    \centering
    \includegraphics[width=\linewidth]{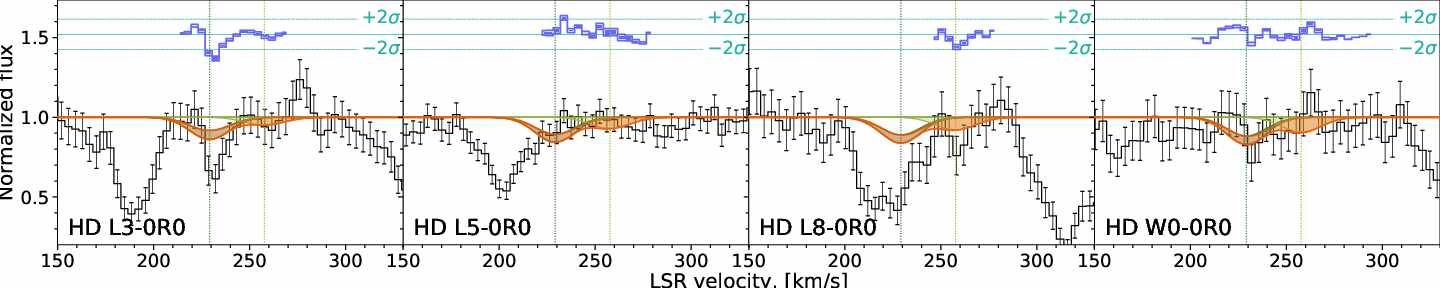}
    \caption{Fit to HD absorption lines towards Sk-69 191 in LMC. Lines are the same as for \ref{fig:lines_HD_Sk67_2}.
    }
    \label{fig:lines_HD_Sk69_191}
\end{figure*}

\begin{figure*}
    \centering
    \includegraphics[width=\linewidth]{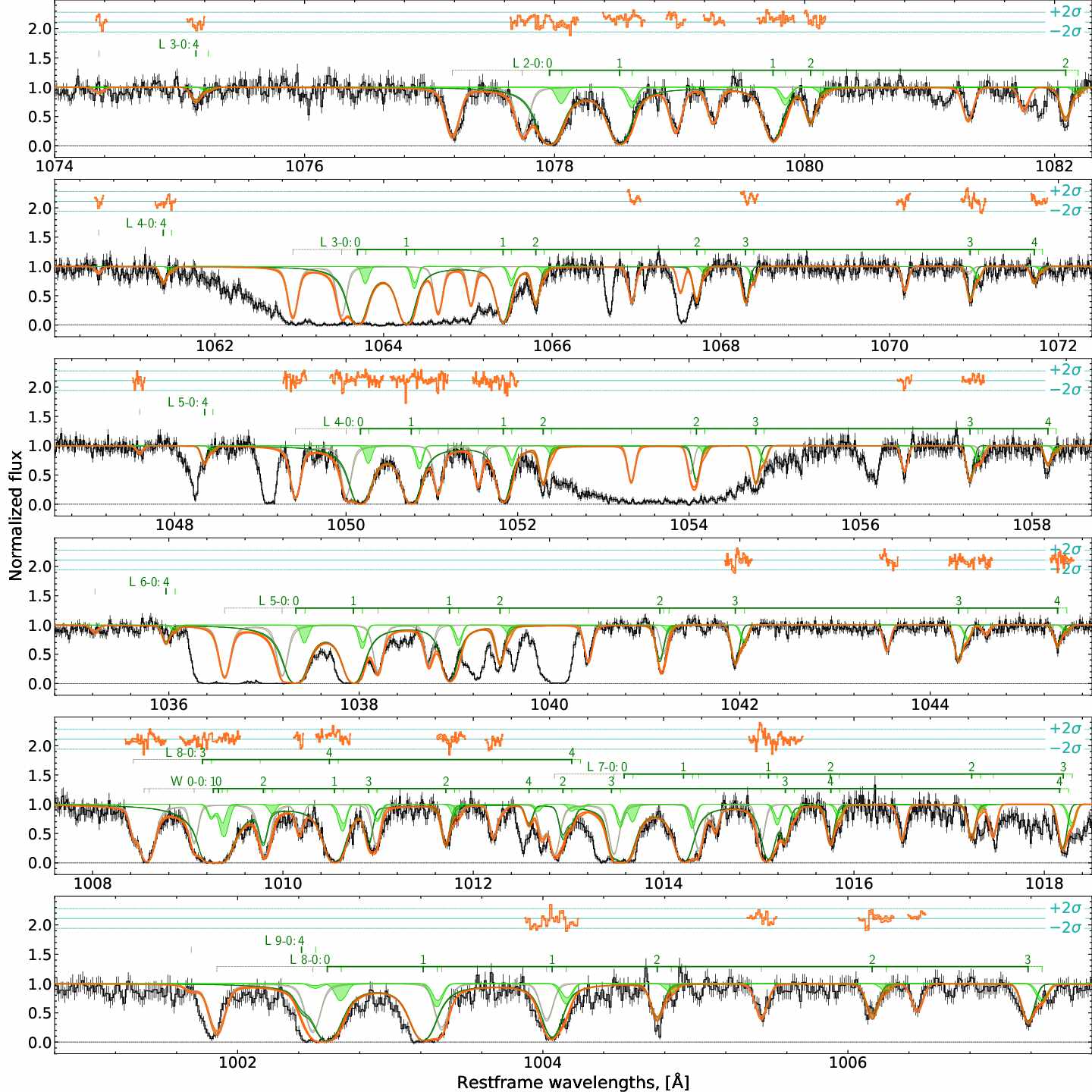}
    \caption{Fit to H2 absorption lines towards Sk-69 191 in LMC. Lines are the same as for \ref{fig:lines_H2_Sk67_2}.
    }
    \label{fig:lines_H2_Sk69_191}
\end{figure*}

\begin{table*}
    \caption{Fit results of H$_2$ lines towards J\,0534-6932}
    \label{tab:J0534}
    \begin{tabular}{ccccc}
    \hline
    \hline
    species & comp & 1 & 2 & 3 \\
            & z & $0.0000798(^{+12}_{-15})$ & $0.0009025(^{+8}_{-14})$ & $0.0009818(^{+26}_{-44})$ \\
    \hline 
     ${\rm H_2\, J=0}$ & b\,km/s & $0.516^{+0.020}_{-0.012}$ & $2.3^{+0.4}_{-1.5}$ & $0.55^{+0.05}_{-0.05}$\\
                       & $\log N$ & $18.46^{+0.04}_{-0.06}$ & $19.504^{+0.015}_{-0.013}$ & $17.07^{+0.31}_{-0.94}$ \\
    ${\rm H_2\, J=1}$ & b\,km/s & $0.512^{+0.032}_{-0.005}$ & $3.0^{+0.9}_{-1.2}$ & $0.60^{+0.22}_{-0.06}$ \\
                      & $\log N$ &$18.36^{+0.03}_{-0.03}$ & $19.607^{+0.007}_{-0.009}$ & $16.7^{+0.4}_{-0.4}$\\
    ${\rm H_2\, J=2}$ & b\,km/s &$0.527^{+0.025}_{-0.017}$ & $4.2^{+0.3}_{-0.5}$ & $0.78^{+0.21}_{-0.14}$\\
                      & $\log N$ & $17.588^{+0.034}_{-0.027}$ & $18.359^{+0.023}_{-0.039}$ &  $15.66^{+0.40}_{-0.18}$\\
    ${\rm H_2\, J=3}$ & b\,km/s & $0.547^{+0.012}_{-0.033}$ & $4.00^{+0.50}_{-0.18}$ & $2.18^{+0.09}_{-0.10}$\\
                      & $\log N$ & $17.20^{+0.05}_{-0.03}$ & $17.76^{+0.05}_{-0.14}$ & $15.77^{+0.28}_{-0.18}$\\
    ${\rm H_2\, J=4}$ & b\,km/s & -- & $5.19^{+0.41}_{-0.30}$ & -- \\
    				  & $\log N$ & $16.38^{+0.16}_{-0.17}$ & $16.26^{+0.19}_{-0.19}$ & $14.64^{+0.15}_{-0.24}$\\
    ${\rm H_2\, J=5}$ & b\,km/s & -- & $5.4^{+1.0}_{-0.5}$ & -- \\
    				  & $\log N$ & -- & $15.48^{+0.16}_{-0.26}$ & $14.74^{+0.27}_{-0.18}$\\
   \hline 
         & $\log N_{\rm tot}$ & $18.76^{+0.02}_{0.03}$ & $19.88^{+0.01}_{-0.01}$ & $17.24^{+0.26}_{-0.41}$ \\
    \hline
    HD J=0 & b\,km/s & $0.521^{+0.012}_{-0.021}$ & $2.4^{+0.7}_{-0.7}$ & $0.5029^{+0.0836}_{-0.0029}$ \\ 
           & $\log N$ & $\lesssim 16.1$ &  $14.5^{+0.9}_{-0.4}$ & $\lesssim 16.0$ \\
    \hline   
    \end{tabular}
    \begin{tablenotes}
     \item Doppler parameter of H$_2$ ${\rm J = 4}$ rotational level in the 1 and 3 component and ${\rm J = 5}$ in the 3 component was tied to H$_2$ ${\rm J = 3}$ and ${\rm J = 4}$, respectively. 
    \end{tablenotes}
\end{table*}

\begin{figure*}
    \centering
    \includegraphics[width=\linewidth]{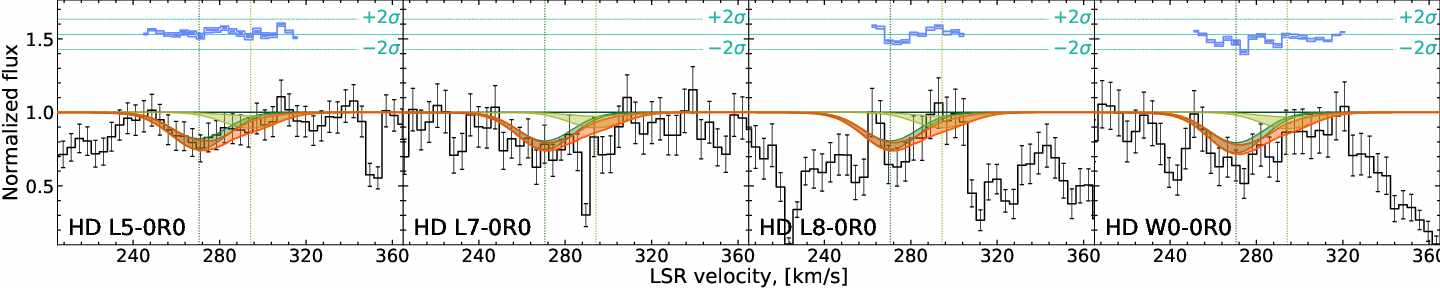}
    \caption{Fit to HD absorption lines towards J 0534-6932 in LMC. Lines are the same as for \ref{fig:lines_HD_Sk67_2}.
    }
    \label{fig:lines_HD_J0534_appendix}
\end{figure*}

\begin{figure*}
    \centering
    \includegraphics[width=\linewidth]{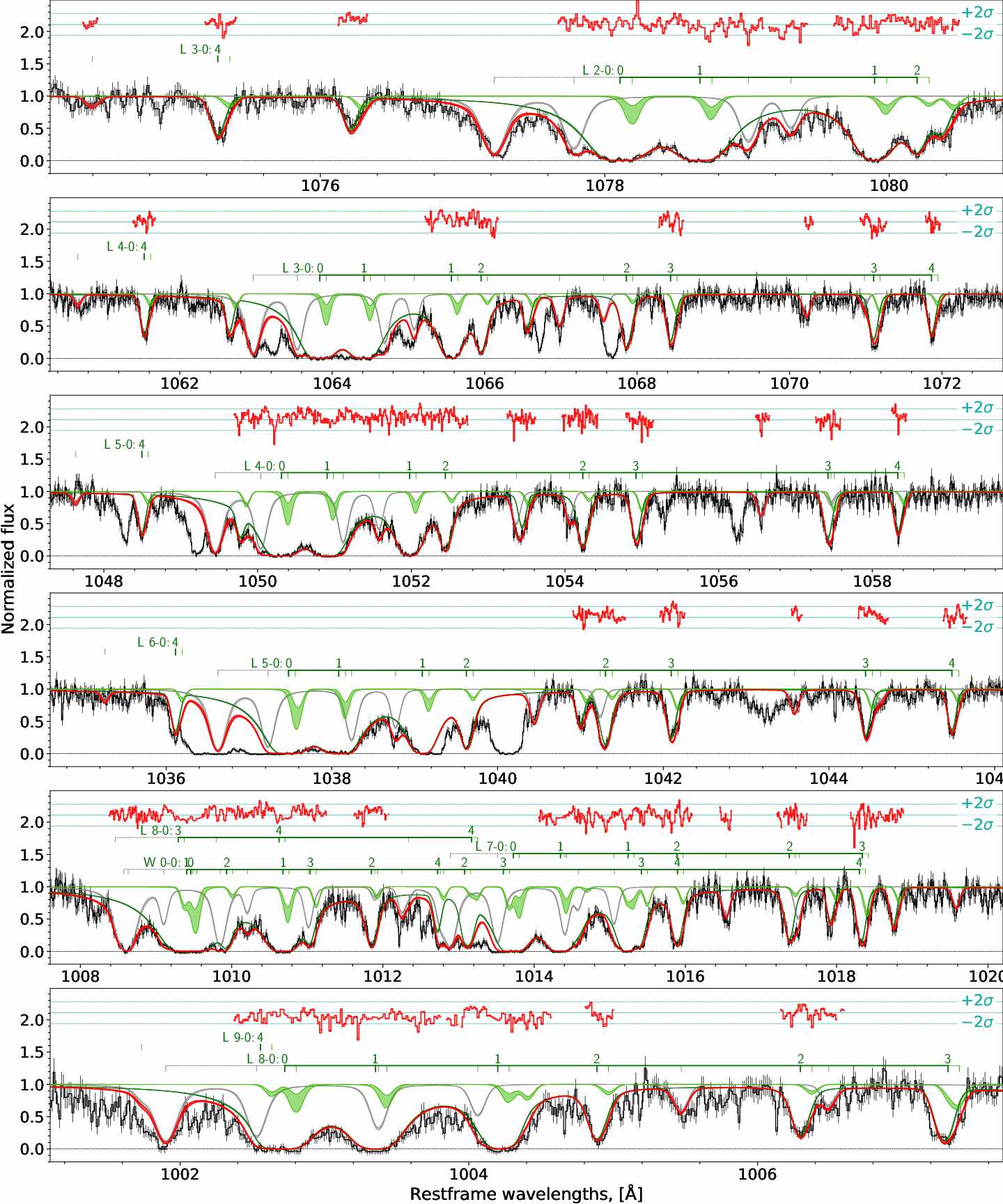}
    \caption{Fit to H2 absorption lines towards J 0534-6932 in LMC. Lines are the same as for \ref{fig:lines_H2_Sk67_2}.
    }
    \label{fig:lines_H2_J0534}
\end{figure*}

\begin{table*}
    \caption{Fit results of H$_2$ lines towards BI 237}
    \label{tab:BI237}
    \begin{tabular}{cccc}
    \hline
    \hline
    species & comp & 1 & 2 \\
            & z & $0.0000738(^{+8}_{-9})$ & $0.00098790(^{+30}_{-60})$ \\
    \hline 
     ${\rm H_2\, J=0}$ & b\,km/s & $1.0^{+0.7}_{-0.5}$ & $1.8^{+1.1}_{-1.2}$\\
                       & $\log N$ & $18.374^{+0.028}_{-0.032}$ & $19.892^{+0.023}_{-0.013}$\\
    ${\rm H_2\, J=1}$ & b\,km/s & $2.1^{+1.1}_{-0.8}$ & $3.9^{+0.7}_{-1.3}$\\
                      & $\log N$ & $18.510^{+0.030}_{-0.050}$ &$19.875^{+0.004}_{-0.009}$ \\
    ${\rm H_2\, J=2}$ & b\,km/s & $4.3^{+0.6}_{-0.5}$ & $4.18^{+0.64}_{-0.29}$\\
                      & $\log N$ & $16.32^{+0.53}_{-0.25}$ & $18.506^{+0.023}_{-0.025}$\\
    ${\rm H_2\, J=3}$ & b\,km/s & $4.4^{+0.7}_{-0.5}$ & $4.5^{+0.4}_{-0.4}$\\
                      & $\log N$ & $15.20^{+0.14}_{-0.18}$ & $18.190^{+0.030}_{-0.060}$\\
    ${\rm H_2\, J=4}$ & b\,km/s & $19.70^{+0.30}_{-3.80}$ & $4.5^{+0.4}_{-0.4}$\\
    				  & $\log N$ & $14.05^{+0.10}_{-0.12}$ & $16.25^{+0.29}_{-0.23}$\\
    ${\rm H_2\, J=5}$ & b\,km/s & -- & $4.6^{+0.5}_{-0.5}$ \\
    				  & $\log N$ & -- & $15.76^{+0.15}_{-0.27}$ \\
    ${\rm H_2\, J=6}$ & b\,km/s & -- & $4.7^{+0.7}_{-0.6}$ \\
                      & $\log N$ & -- & $14.46^{+0.06}_{-0.07}$ \\
    \hline 
         & $\log N_{\rm tot}$ & $18.75^{+0.02}_{-0.03}$ & $20.20^{+0.01}_{-0.01}$ \\
     \hline
     HD J=0 & b\,km/s & $0.53^{+0.91}_{-0.03}$ & $1.1^{+0.7}_{-0.6}$ \\
            & $\log N$ & $\lesssim 15.2$ & $14.35^{+1.07}_{-0.19}$ \\
    \hline   
    \end{tabular}
    \begin{tablenotes}
    \item 
    \end{tablenotes}
\end{table*}

\begin{figure*}
    \centering
    \includegraphics[width=\linewidth]{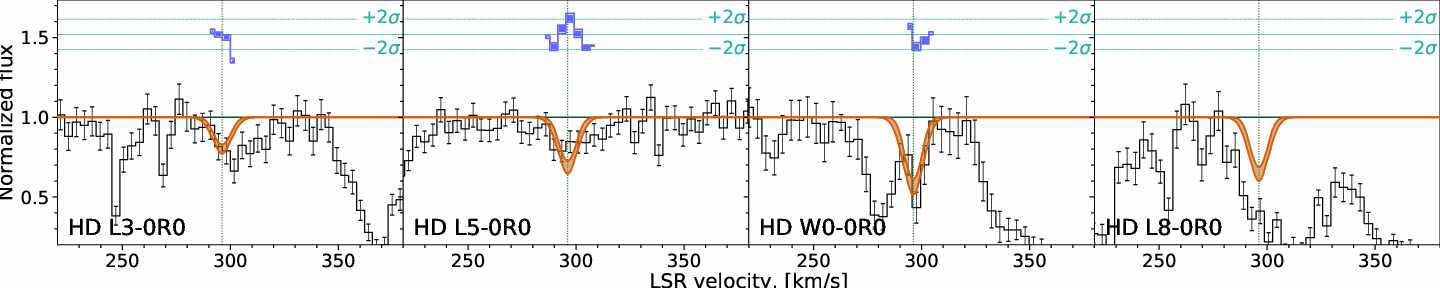}
    \caption{Fit to HD absorption lines towards BI 237 in LMC. Lines are the same as for \ref{fig:lines_HD_Sk67_2}.
    }
    \label{fig:lines_HD_BI237}
\end{figure*}

\begin{figure*}
    \centering
    \includegraphics[width=\linewidth]{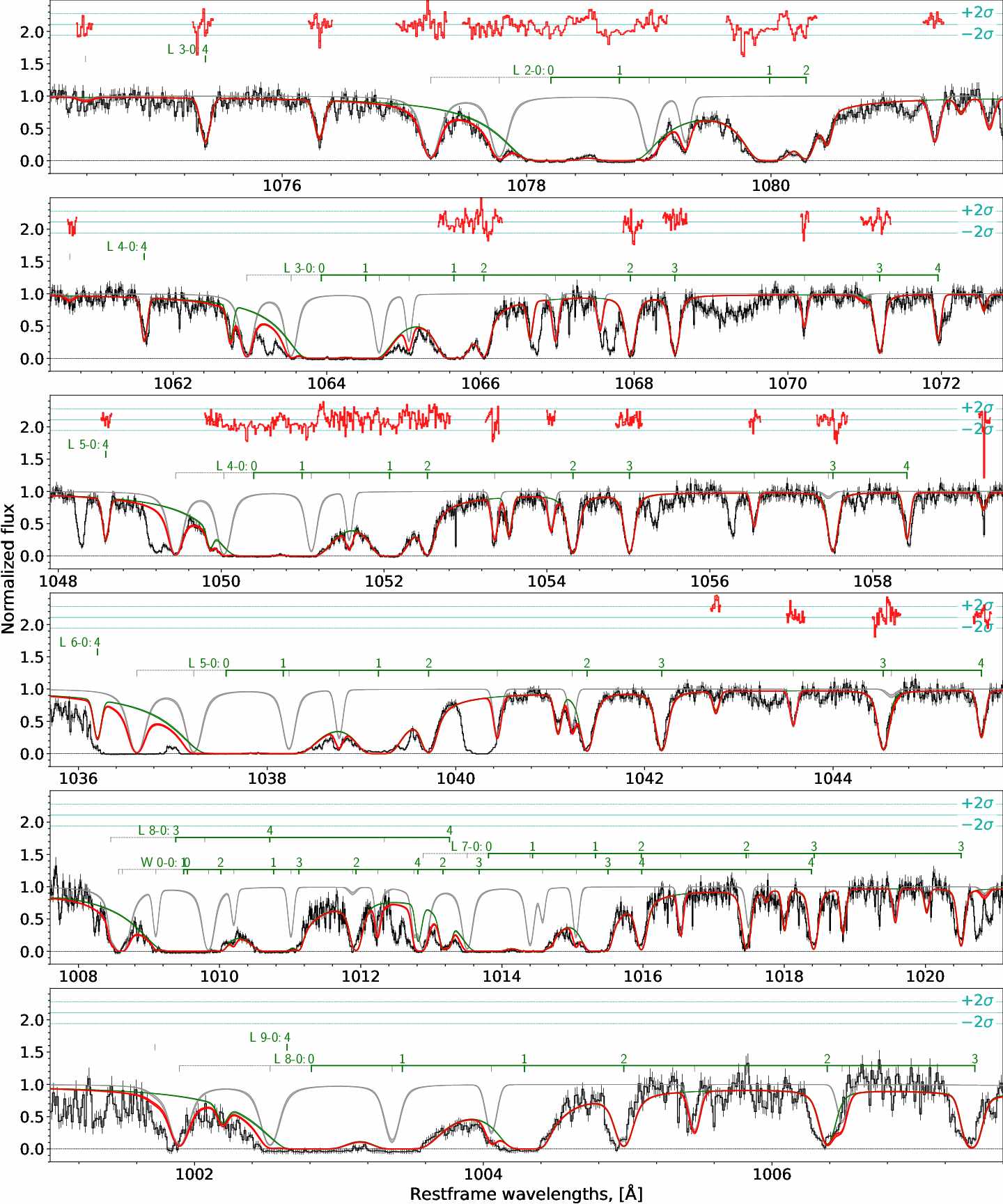}
    \caption{Fit to H2 absorption lines towards BI 237 in LMC. Lines are the same as for \ref{fig:lines_H2_Sk67_2}.
    }
    \label{fig:lines_H2_BI237}
\end{figure*}

\begin{table*}
    \caption{Fit results of H$_2$ lines towards Sk-68 129}
    \label{tab:Sk68_129}
    \begin{tabular}{cccc}
    \hline
    \hline
    species & comp & 1 & 2 \\
            & z & $0.0000688(^{+25}_{-28})$ & $0.0009205(^{+11}_{-18})$ \\
    \hline 
     ${\rm H_2\, J=0}$ & b\,km/s & $0.8^{+0.7}_{-0.3}$ &  $1.4^{+1.4}_{-0.9}$\\
                       & $\log N$ & $18.01^{+0.15}_{-0.21}$ &$20.26^{+0.13}_{-0.04}$ \\
    ${\rm H_2\, J=1}$ & b\,km/s & $1.0^{+0.9}_{-0.4}$ & $2.6^{+2.1}_{-1.1}$\\
                      & $\log N$ & $18.6^{+0.7}_{-0.7}$ & $20.163^{+0.023}_{-0.031}$\\
    ${\rm H_2\, J=2}$ & b\,km/s & $1.3^{+1.8}_{-0.5}$ & $5.1^{+0.9}_{-2.0}$\\
                      & $\log N$ & $17.88^{+0.16}_{-0.15}$ & $18.91^{+0.05}_{-0.06}$\\
    ${\rm H_2\, J=3}$ & b\,km/s &$3.02^{+0.18}_{-1.74}$ & $6.0^{+0.7}_{-1.1}$\\
                      & $\log N$ &  $17.64^{+0.16}_{-0.26}$ & $18.63^{+0.05}_{-0.07}$ \\
    ${\rm H_2\, J=4}$ & b\,km/s & -- & $5.7^{+0.9}_{-0.6}$\\
    				  & $\log N$ & $14.74^{+1.79}_{-0.19}$ &  $17.08^{+0.23}_{-0.55}$\\
    ${\rm H_2\, J=5}$ & b\,km/s &  -- & $7.6^{+1.7}_{-1.7}$\\
    				  & $\log N$ & -- & $16.4^{+0.5}_{-0.7}$ \\
    \hline 
         & $\log N_{\rm tot}$ & $18.81^{+0.56}_{-0.33}$ & $20.53^{+0.07}_{-0.03}$ \\
     \hline
     HD J=0 & b\,km/s &$1.0^{+0.4}_{-0.5}$ & $1.4^{+0.7}_{-0.9}$ \\
            & $\log N$ & $\lesssim 17.1$ & $\lesssim 17.2$ \\
    \hline   
    \end{tabular}
    \begin{tablenotes}
    \item H$_2$ $\rm J=4$ Doppler parameter in 1 component was tied to H$_2$ $\rm J=4$.
    \end{tablenotes}
\end{table*}

\begin{figure*}
    \centering
    \includegraphics[width=\linewidth]{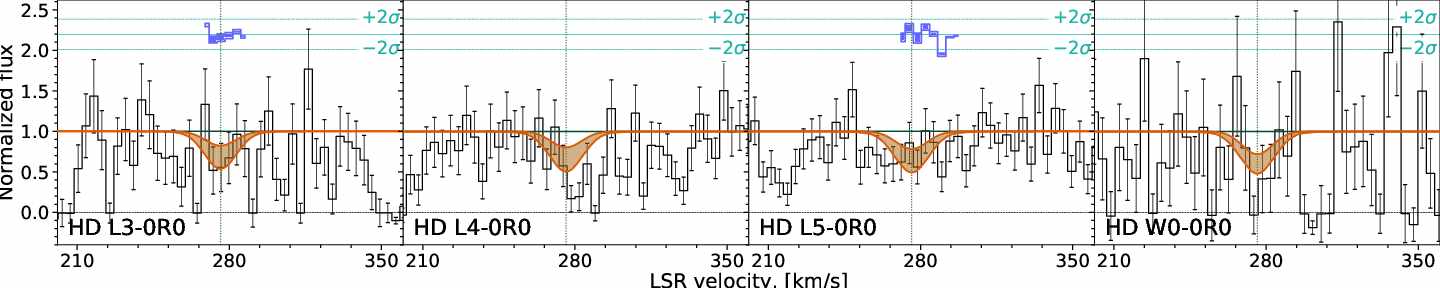}
    \caption{Fit to HD absorption lines towards Sk-68 129 in LMC. Lines are the same as for \ref{fig:lines_HD_Sk67_2}.
    }
    \label{fig:lines_HD_Sk68_129}
\end{figure*}

\begin{figure*}
    \centering
    \includegraphics[width=\linewidth]{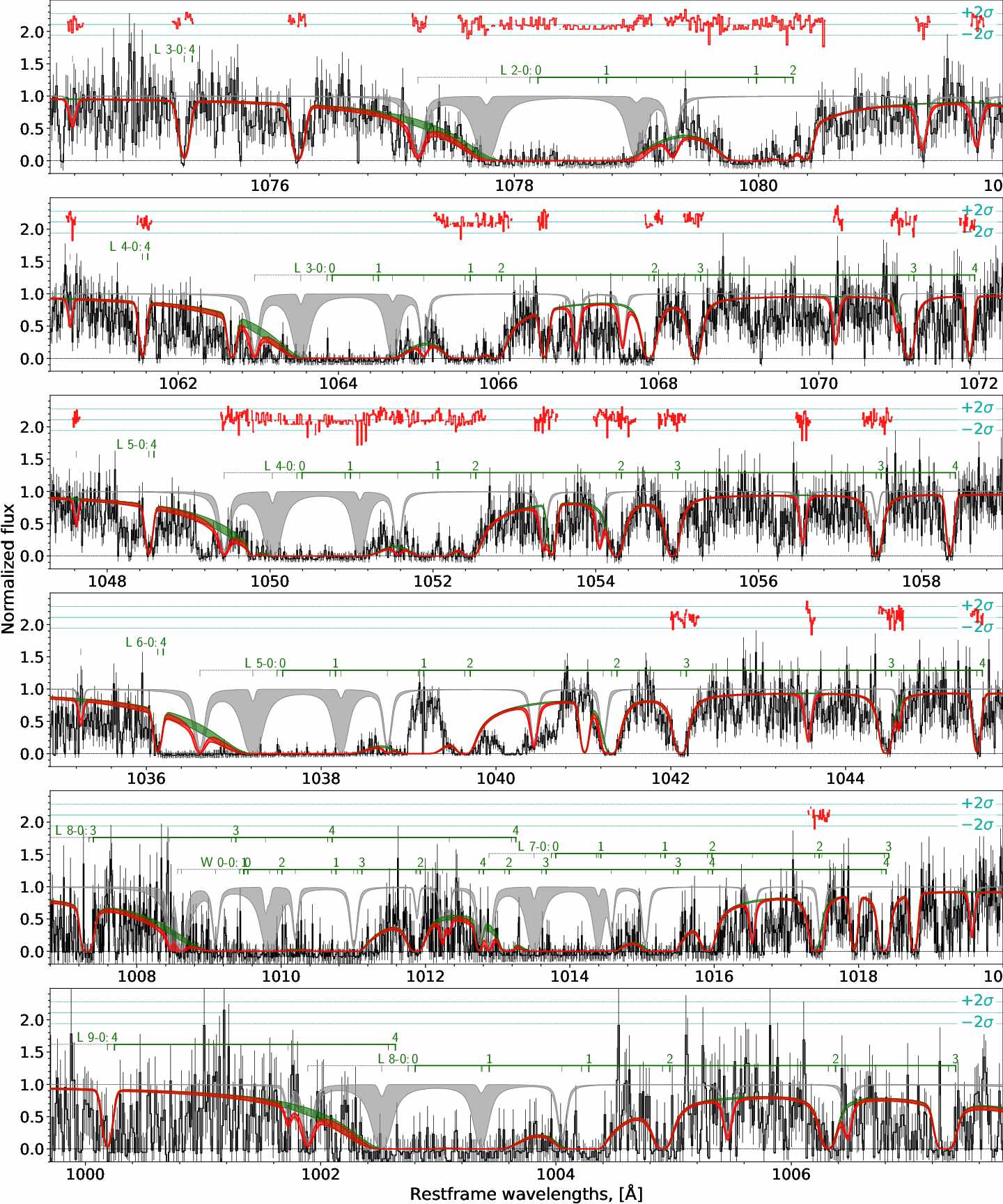}
    \caption{Fit to H2 absorption lines towards Sk-68 129 in LMC. Lines are the same as for \ref{fig:lines_H2_Sk67_2}.
    }
    \label{fig:lines_H2_Sk68_129}
\end{figure*}

\begin{table*}
    \caption{Fit results of H$_2$ lines towards Sk-69 220}
    \label{tab:Sk69_220}
    \begin{tabular}{cccc}
    \hline
    \hline
    species & comp & 1 & 2 \\
            & z & $0.0000472(^{+9}_{-15})$ & $0.0009449(^{+11}_{-10})$\\
    \hline 
     ${\rm H_2\, J=0}$ & b\,km/s &$1.6^{+0.4}_{-0.6}$ & $0.56^{+0.10}_{-0.06}$ \\
                       & $\log N$ & $17.741^{+0.040}_{-0.031}$ & $18.884^{+0.011}_{-0.019}$\\
    ${\rm H_2\, J=1}$ & b\,km/s & $2.03^{+0.15}_{-0.17}$ & $0.77^{+0.19}_{-0.18}$\\
                      & $\log N$ & $18.039^{+0.030}_{-0.027}$ & $18.976^{+0.017}_{-0.013}$ \\
    ${\rm H_2\, J=2}$ & b\,km/s &$2.21^{+0.06}_{-0.15}$ &$1.00^{+0.28}_{-0.23}$ \\
                      & $\log N$ & $17.17^{+0.07}_{-0.05}$ & $18.103^{+0.019}_{-0.024}$\\
    ${\rm H_2\, J=3}$ & b\,km/s & $2.10^{+0.07}_{-0.13}$ & $1.23^{+0.32}_{-0.20}$\\
                      & $\log N$ & $16.28^{+0.16}_{-0.30}$ &$17.997^{+0.015}_{-0.037}$ \\
    ${\rm H_2\, J=4}$ & b\,km/s & -- & $1.51^{+0.11}_{-0.24}$\\
    				  & $\log N$ &$13.65^{+0.30}_{-0.73}$ & $16.54^{+0.20}_{-0.13}$\\
    ${\rm H_2\, J=5}$ & $\log N$ & -- & $16.43^{+0.22}_{-0.31}$ \\
   \hline 
         & $\log N_{\rm tot}$ & $18.26^{+0.02}_{-0.02}$ & $19.29^{+0.01}_{-0.01}$ \\
     \hline
     HD J=0 & b\,km/s &$1.4^{+0.4}_{-0.4}$ & $0.58^{+0.08}_{-0.08}$ \\
            & $\log N$ & $\lesssim 16.4$ & $\lesssim 16.4$ \\
    \hline   
    \end{tabular}
    \begin{tablenotes}
    \item  Doppler parameters of H$_2$ $\rm J=4$ in 1 component and $\rm J=5$ in 2 component were tied to H$_2$ $\rm J=3$ and $\rm J=4$, respectively.
    \end{tablenotes}
\end{table*}

\begin{figure*}
    \centering
    \includegraphics[width=\linewidth]{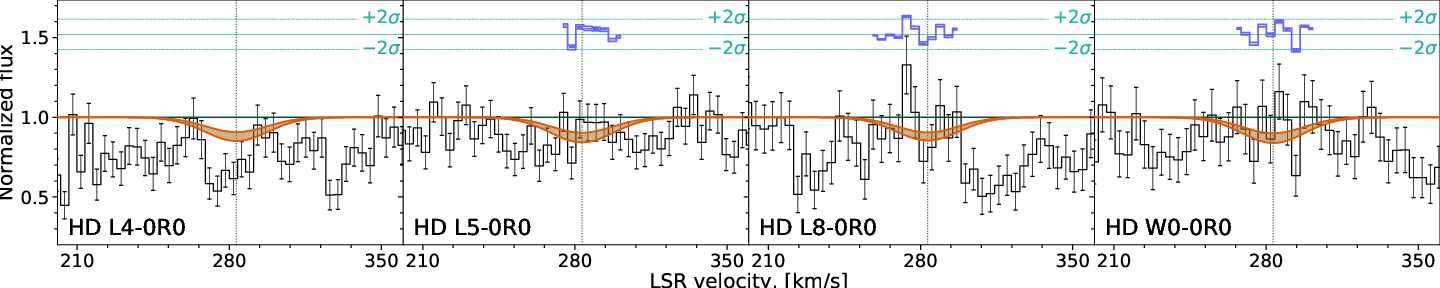}
    \caption{Fit to HD absorption lines towards Sk-69 220 in LMC. Lines are the same as for \ref{fig:lines_HD_Sk67_2}.
    }
    \label{fig:lines_HD_Sk69_106}
\end{figure*}

\begin{figure*}
    \centering
    \includegraphics[width=\linewidth]{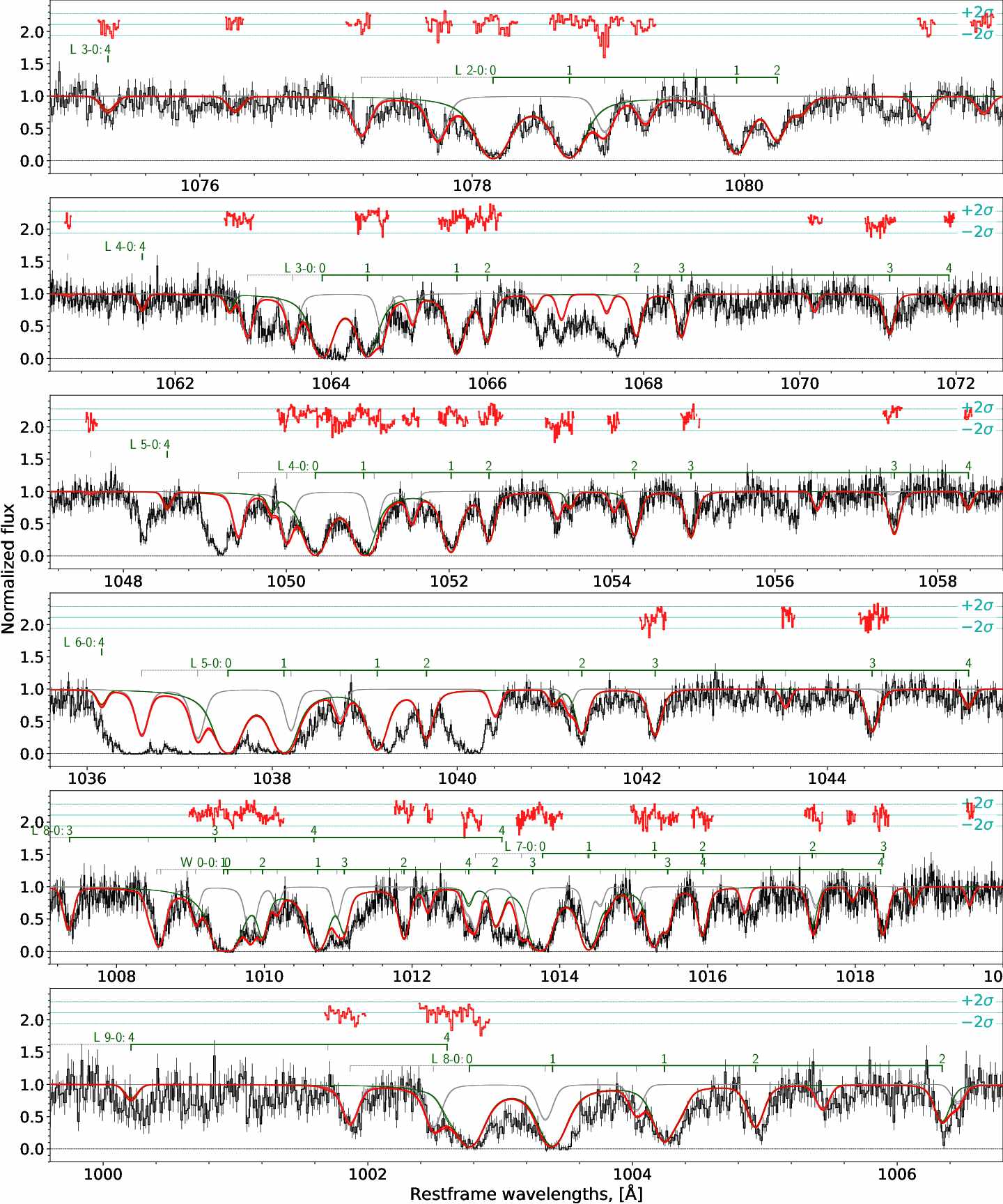}
    \caption{Fit to H2 absorption lines towards Sk-69 220 in LMC. Lines are the same as for \ref{fig:lines_H2_Sk67_2}.
    }
    \label{fig:lines_H2_Sk69_220}
\end{figure*}

\begin{table*}
    \caption{Fit results of H$_2$ lines towards Sk-69 223}
    \label{tab:Sk69_223}
    \begin{tabular}{ccccc}
    \hline
    \hline
    species & comp & 1 & 2 & 3 \\
            & z & $0.0001193(^{+15}_{-5})$ & $0.0009105(^{+20}_{-18})$ & $0.0010285(^{+7}_{-11})$\\
    \hline 
     ${\rm H_2\, J=0}$ & b\,km/s & $0.55^{+0.25}_{-0.05}$ & $0.88^{+0.60}_{-0.30}$ & $0.71^{+0.34}_{-0.21}$\\
                       & $\log N$ & $18.580^{+0.010}_{-0.032}$ & $19.05^{+0.10}_{-0.13}$ & $19.675^{+0.016}_{-0.038}$\\
    ${\rm H_2\, J=1}$ & b\,km/s & $0.66^{+0.23}_{-0.15}$ & $1.76^{+0.27}_{-0.46}$ & $1.63^{+0.31}_{-0.51}$\\
                      & $\log N$ & $18.800^{+0.034}_{-0.015}$ & $18.56^{+0.04}_{-0.09}$ & $19.545^{+0.016}_{-0.013}$\\
    ${\rm H_2\, J=2}$ & b\,km/s & $0.82^{+0.23}_{-0.24}$ & $2.02^{+0.06}_{-0.14}$ & $1.9^{+0.3}_{-0.3}$\\
                      & $\log N$ & $18.169^{+0.022}_{-0.016}$ & $16.63^{+0.06}_{-0.19}$ & $18.304^{+0.016}_{-0.017}$\\
    ${\rm H_2\, J=3}$ & b\,km/s & $0.82^{+0.32}_{-0.15}$ & $2.02^{+0.09}_{-0.09}$ & $2.08^{+0.14}_{-0.30}$\\
                      & $\log N$ & $17.900^{+0.015}_{-0.023}$ & $16.74^{+0.10}_{-0.15}$ &  $18.313^{+0.019}_{-0.017}$\\
    ${\rm H_2\, J=4}$ & b\,km/s & -- & -- & $2.15^{+0.13}_{-0.18}$\\
    				  & $\log N$ & $15.86^{+0.31}_{-0.29}$ & $15.82^{+0.25}_{-0.18}$ & $17.30^{+0.06}_{-0.06}$\\
    ${\rm H_2\, J=5}$ & $\log N$ & -- & $16.15^{+0.35}_{-0.23}$ & $16.92^{+0.14}_{-0.17}$\\
    \hline 
         & $\log N_{\rm tot}$ & $19.09^{+0.02}_{-0.01}$ & $19.18^{+0.08}_{-0.10}$ & $19.94^{+0.01}_{-0.02}$ \\
    \hline
    HD J=0 & b\,km/s &  $0.511^{+0.285}_{-0.011}$ &  $0.9^{+0.4}_{-0.4}$ & $0.516^{+0.462}_{-0.016}$ \\
           & $\log N$ & $\lesssim 15.7$ & $\lesssim 16.3$ & $\lesssim 16.2$ \\
    \hline   
    \end{tabular}
    \begin{tablenotes}
    \item Doppler parameters of H$_2$ $\rm J=4$ in 1 and 2 components and $\rm J=5$ in 2 and 3 components were tied to $\rm J=3$ and $\rm J=4$, respectively.  
    \end{tablenotes}
\end{table*}

\begin{figure*}
    \centering
    \includegraphics[width=\linewidth]{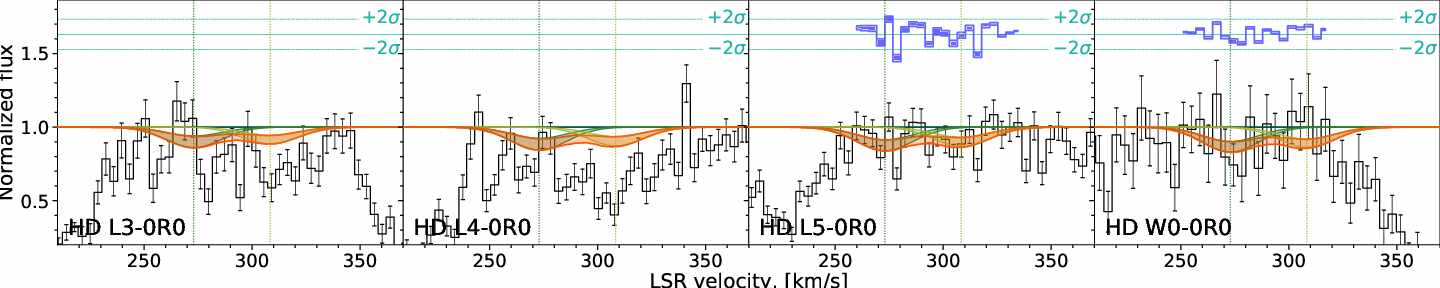}
    \caption{Fit to HD absorption lines towards Sk-69 223 in LMC. Lines are the same as for \ref{fig:lines_HD_Sk67_2}.
    }
    \label{fig:lines_HD_Sk69_223}
\end{figure*}

\begin{figure*}
    \centering
    \includegraphics[width=\linewidth]{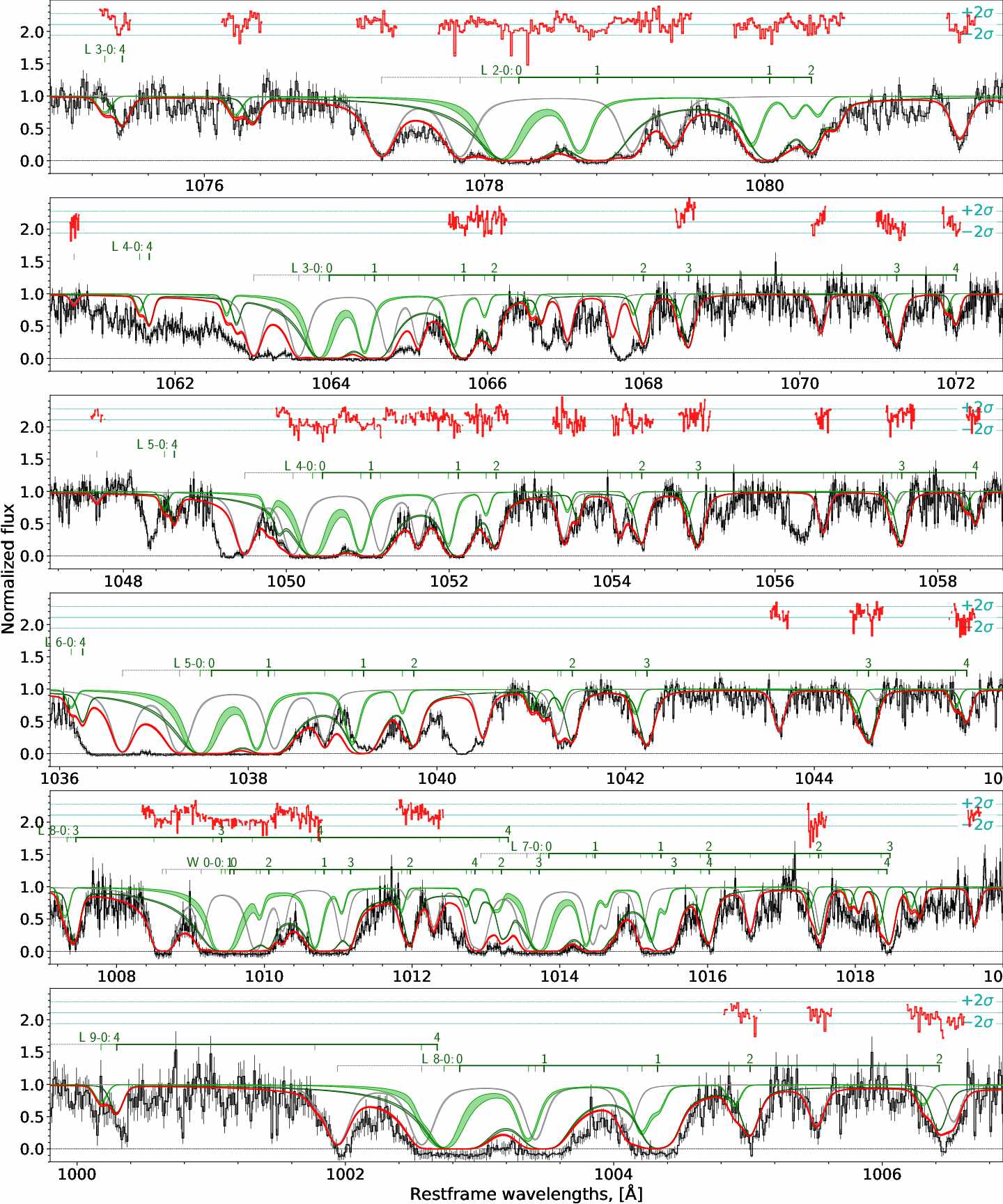}
    \caption{Fit to H2 absorption lines towards Sk-69 223 in LMC. Lines are the same as for \ref{fig:lines_H2_Sk67_2}.
    }
    \label{fig:lines_H2_Sk69_223}
\end{figure*}

\begin{table*}
    \caption{Fit results of H$_2$ lines towards Sk-66 172}
    \label{tab:Sk66_172}
    \begin{tabular}{ccccccc}
    \hline
    \hline
    species & comp & 1 & 2 & 3 & 4 & 5 \\
            & z & $0.0000844(^{+19}_{-13})$ & $0.000965(^{+3}_{-5})$ & $0.0010056(^{+17}_{-30})$ & $0.001104(^{+5}_{-4})$ & $0.001252(^{+4}_{-11})$ \\
    \hline 
     ${\rm H_2\, J=0}$ & b\,km/s & $4.4^{+0.7}_{-0.5}$ & $1.3^{+1.7}_{-0.6}$ & $2.2^{+0.8}_{-1.1}$ & $4.1^{+0.7}_{-0.9}$ & $0.95^{+0.61}_{-0.31}$ \\
                       & $\log N$ & $17.82^{+0.08}_{-0.07}$ & $18.01^{+0.07}_{-0.05}$ &  $15.8^{+0.5}_{-0.8}$ & $16.34^{+0.30}_{-0.83}$ & $15.9^{+0.4}_{-0.8}$ \\
    ${\rm H_2\, J=1}$ & b\,km/s & $3.9^{+0.4}_{-0.6}$ & $2.3^{+0.6}_{-0.9}$ & $5.9^{+1.0}_{-0.9}$ & $9.1^{+0.9}_{-2.8}$ & $0.55^{+0.64}_{-0.05}$ \\
                      & $\log N$ & $17.61^{+0.11}_{-0.21}$ & $17.68^{+0.18}_{-0.08}$ & $16.4^{+0.6}_{-0.3}$ & $14.30^{+0.16}_{-0.21}$ & $14.6^{+0.5}_{-0.5}$ \\
    ${\rm H_2\, J=2}$ & b\,km/s & $4.1^{+0.4}_{-0.7}$ & $0.69^{+4.65}_{-0.19}$ &  $6.1^{+1.0}_{-0.9}$ & $9.0^{+0.9}_{-5.1}$ &  $1.2^{+2.4}_{-0.7}$\\
                      & $\log N$ & $15.32^{+0.07}_{-0.25}$ &$13.4^{+1.0}_{-0.6}$ &  $15.50^{+0.49}_{-0.13}$ &  $12.8^{+0.6}_{-1.3}$ & $14.05^{+0.22}_{-0.17}$ \\
    ${\rm H_2\, J=3}$ & b\,km/s &$2.9^{+2.5}_{-0.5}$ & $3.9^{+2.9}_{-2.1}$ & $3.9^{+1.3}_{-0.6}$ & $0.65^{+14.35}_{-0.15}$ & $9.75^{+0.25}_{-1.25}$ \\
                      & $\log N$ &  $14.43^{+0.13}_{-0.07}$ & $14.57^{+0.19}_{-0.27}$ & $15.5^{+0.9}_{-0.3}$ & $14.1^{+0.4}_{-0.5}$ & $13.99^{+0.13}_{-0.09}$ \\
    ${\rm H_2\, J=4}$ & b\,km/s & -- & -- & $7.4^{+2.2}_{-1.0}$ & -- & -- \\
    				  & $\log N$ & $14.00^{+0.15}_{-0.15}$ & $12.2^{+1.1}_{-0.5}$ & $14.80^{+0.09}_{-0.06}$ &$13.3^{+0.5}_{-2.5}$ & $13.60^{+0.29}_{-0.69}$ \\
    ${\rm H_2\, J=5}$ & $\log N$ & -- & -- & $14.58^{+0.06}_{-0.05}$ & -- & -- \\
   \hline 
         & $\log N_{\rm tot}$ & $18.03^{+0.07}_{-0.08}$ & $18.18^{+0.08}_{-0.04}$ & $16.62^{+0.49}_{-0.19}$ & $16.35^{+0.30}_{-0.79}$ & $15.97^{+0.38}_{-0.67}$  \\
    \hline
    HD J=0 & b\,km/s & $4.5^{+0.7}_{-0.8}$ & $0.60^{+2.58}_{-0.10}$ & $2.5^{+0.4}_{-2.0}$ & $4.1^{+0.9}_{-1.1}$ & $0.9^{+0.6}_{-0.4}$ \\
            & $\log N$ & $\lesssim 14.0$ & $\lesssim 16.1$ & $\lesssim 16.1$ & $\lesssim 14.9$ & $\lesssim 15.5$ \\ 
    \hline   
    \end{tabular}
    \begin{tablenotes}
     \item Doppler parameters of H$_2$ ${\rm J = 4}$ rotational level in components 1, 2, 4 and 4 and ${\rm J = 5}$ in component 3 were tied to H$_2$ ${\rm J = 3}$ and ${\rm J = 4}$, respectively. 
    \end{tablenotes}
\end{table*}

\begin{figure*}
    \centering
    \includegraphics[width=\linewidth]{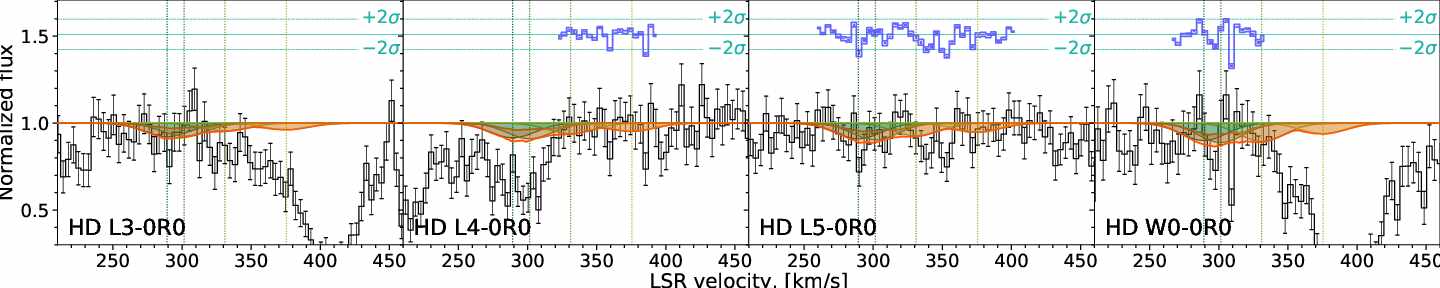}
    \caption{Fit to HD absorption lines towards Sk-66 172 in LMC. Lines are the same as for \ref{fig:lines_HD_Sk67_2}.
    }
    \label{fig:lines_HD_Sk66_172}
\end{figure*}

\begin{figure*}
    \centering
    \includegraphics[width=\linewidth]{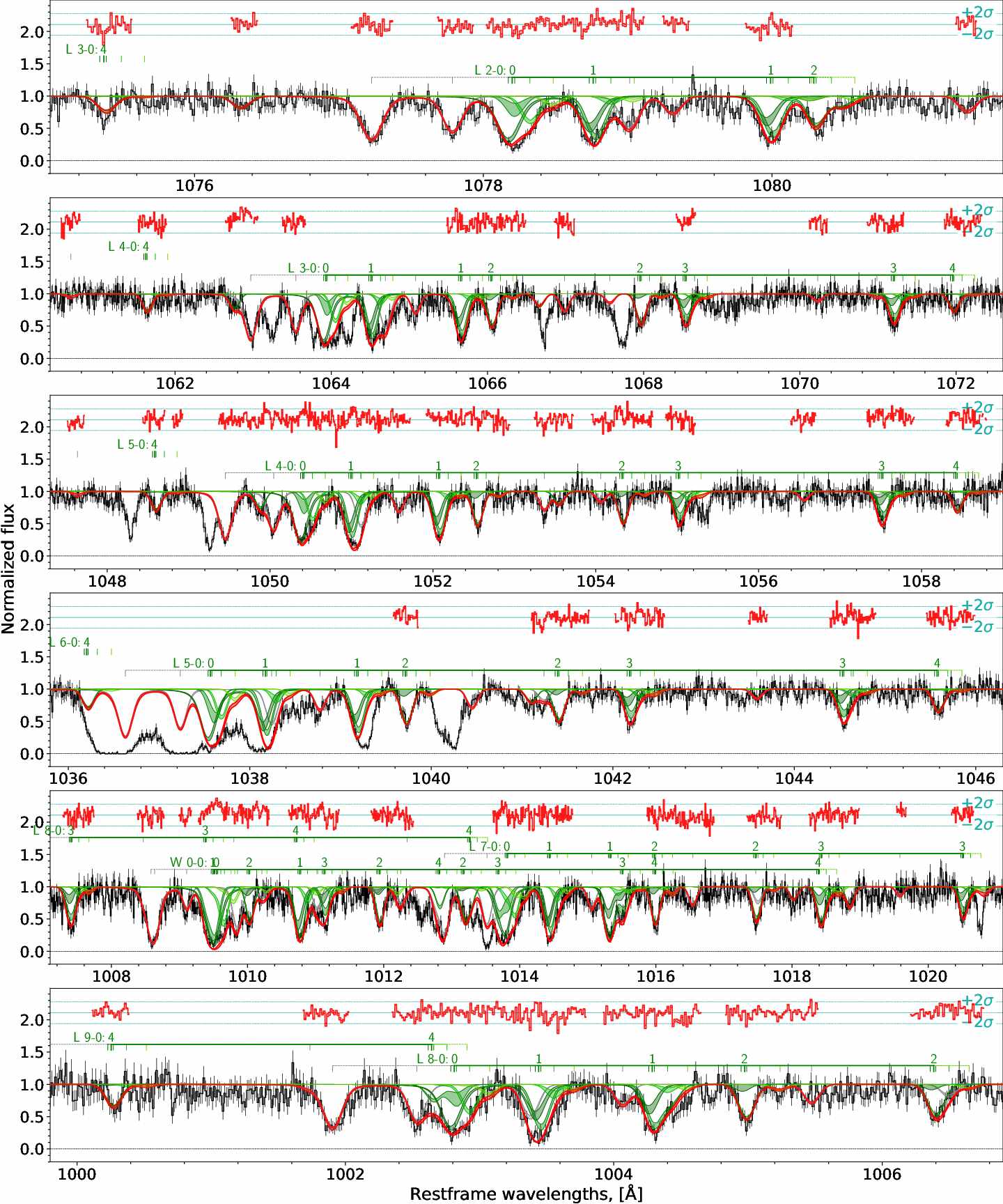}
    \caption{Fit to H2 absorption lines towards Sk-66 172 in LMC. Lines are the same as for \ref{fig:lines_H2_Sk67_2}.}
    \label{fig:lines_H2_Sk66_172}
\end{figure*}

\begin{table*}
    \caption{Fit results of H$_2$ lines towards Sk-69 228}
    \label{tab:Sk69_228}
    \begin{tabular}{ccccc}
    \hline
    \hline
    species & comp & 1 & 2 & 3 \\
            & z & $0.0000541(^{+12}_{-22})$ & $0.0008747(^{+23}_{-48})$ & $0.0009882(^{+27}_{-30})$ \\
    \hline 
     ${\rm H_2\, J=0}$ & b\,km/s & $0.66^{+1.49}_{-0.16}$ &$1.6^{+2.1}_{-0.4}$ & $2.3^{+1.0}_{-1.7}$\\
                       & $\log N$ & $17.91^{+0.09}_{-0.04}$ & $17.93^{+0.14}_{-0.17}$ &$18.657^{+0.068}_{-0.029}$ \\
    ${\rm H_2\, J=1}$ & b\,km/s & $2.9^{+0.8}_{-1.2}$ & $2.4^{+1.4}_{-0.6}$ & $4.4^{+0.5}_{-1.5}$\\
                      & $\log N$ & $18.439^{+0.027}_{-0.050}$ &$18.02^{+0.11}_{-0.21}$ & $18.44^{+0.10}_{-0.15}$\\
    ${\rm H_2\, J=2}$ & b\,km/s & $5.2^{+0.6}_{-1.0}$ & $2.6^{+1.5}_{-0.6}$ & $4.4^{+1.2}_{-0.5}$\\
                      & $\log N$ & $16.9^{+0.4}_{-0.4}$ & $15.16^{+0.85}_{-0.27}$ & $17.09^{+0.30}_{-0.78}$\\
    ${\rm H_2\, J=3}$ & b\,km/s & $5.7^{+1.9}_{-1.8}$ & $3.0^{+1.5}_{-0.4}$ & $6.1^{+0.6}_{-1.9}$\\
                      & $\log N$ & $15.14^{+0.59}_{-0.24}$ &$15.45^{+0.25}_{-0.51}$ & $15.54^{+0.78}_{-0.21}$\\
    ${\rm H_2\, J=4}$ & $\log N$ & $14.03^{+0.30}_{-0.46}$ & $14.62^{+0.43}_{-0.10}$ & $14.40^{+0.14}_{-0.19}$\\
    \hline 
         & $\log N_{\rm tot}$ & $ 18.56^{+0.03}_{-0.04}$ & $18.28^{+0.09}_{-0.13}$ & $18.87^{+0.06}_{-0.05}$ \\
    \hline
    HD J=0 & b\,km/s &$0.80^{+1.00}_{-0.30}$ & $1.9^{+0.8}_{-0.5}$ & $0.55^{+1.80}_{-0.05}$ \\
           & $\log N$ & $\lesssim 16.9$ & $\lesssim 16.8$ & $\lesssim 16.4$ \\ 
    \hline   
    \end{tabular}
    \begin{tablenotes}
    \item Doppler parameters of H$_2$ $\rm J=4$ in 1, 2 and 3 components were tied to $\rm J=3$.  
    \end{tablenotes}
\end{table*}

\begin{figure*}
    \centering
    \includegraphics[width=\linewidth]{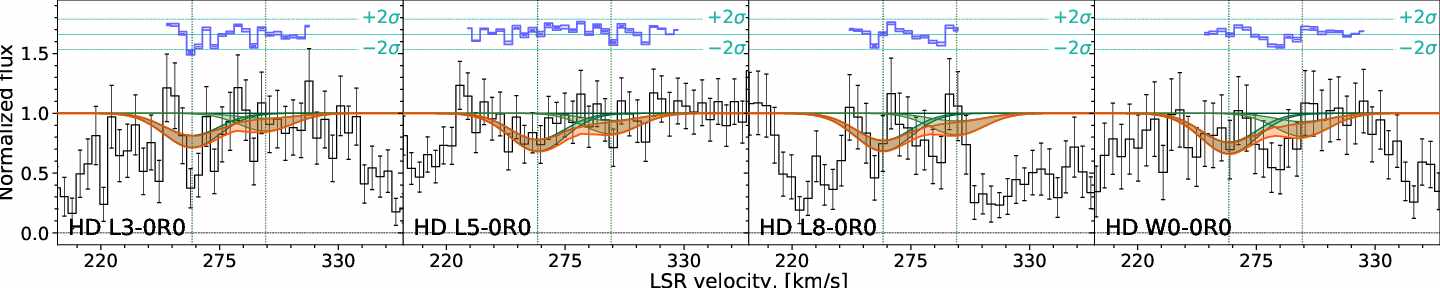}
    \caption{Fit to HD absorption lines towards Sk-69 228 in LMC. Lines are the same as for \ref{fig:lines_HD_Sk67_2}.
    }
    \label{fig:lines_HD_Sk69_228}
\end{figure*}

\begin{figure*}
    \centering
    \includegraphics[width=\linewidth]{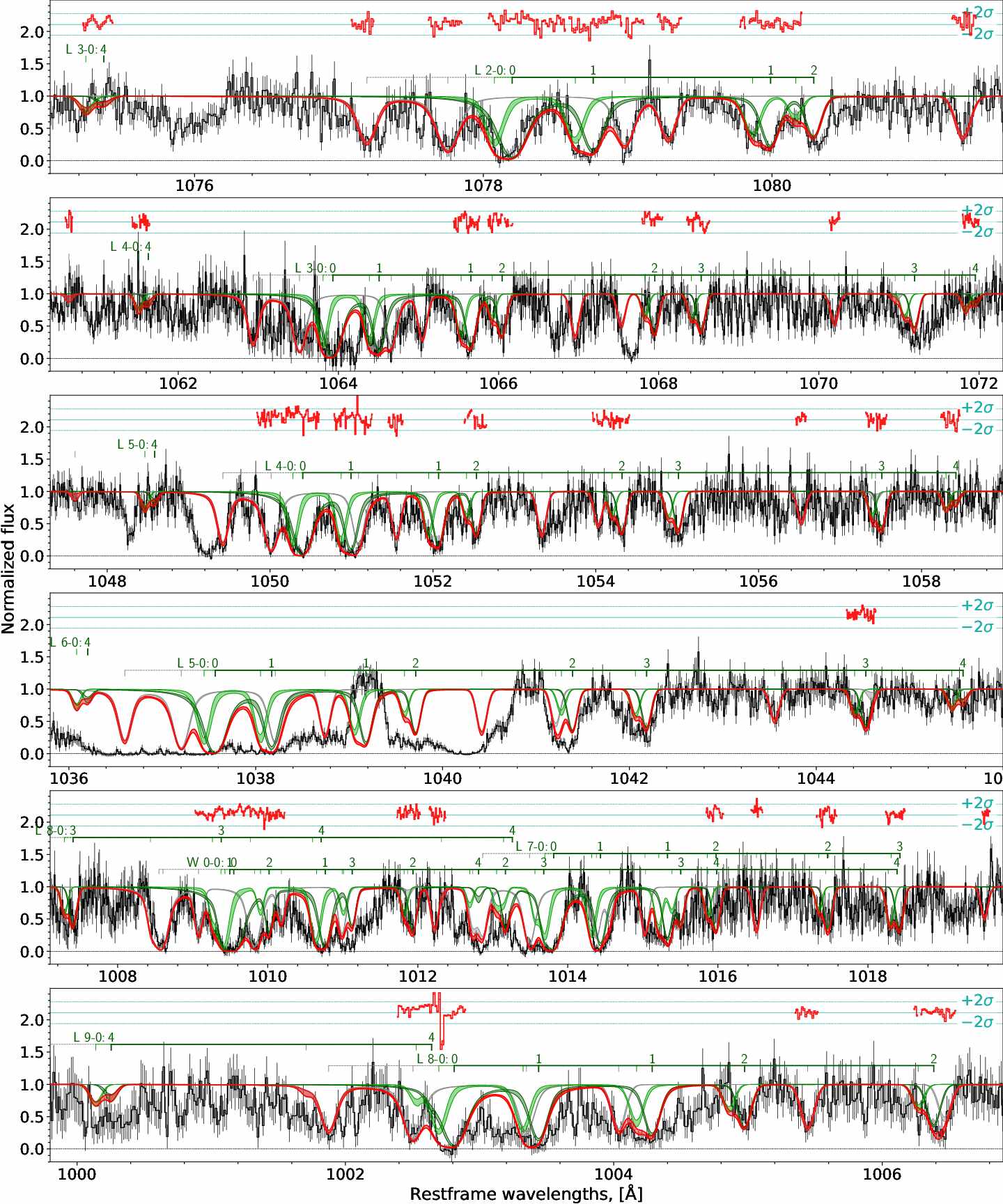}
    \caption{Fit to H2 absorption lines towards Sk-69 228 in LMC. Lines are the same as for \ref{fig:lines_H2_Sk67_2}.
    }
    \label{fig:lines_H2_Sk69_228}
\end{figure*}

\begin{table*}
    \caption{Fit results of H$_2$ lines towards BI 253}
    \label{tab:BI253}
    \begin{tabular}{ccccc}
    \hline
    \hline
    species & comp & 1 & 2 & 3 \\
            & z & $0.00005014(^{+117}_{-25})$ & $0.0008915(^{+14}_{-20})$ & $0.0009335(^{+67}_{-26})$ \\
    \hline 
     ${\rm H_2\, J=0}$ & b\,km/s & $2.64^{+0.35}_{-0.23}$ & $1.8^{+1.1}_{-1.2}$ & $7.1^{+0.8}_{-1.0}$ \\
                       & $\log N$ &$17.994^{+0.022}_{-0.030}$ & $19.7791^{+0.0016}_{-0.0177}$ & $17.98^{+0.19}_{-0.64}$\\
    ${\rm H_2\, J=1}$ & b\,km/s & $3.48^{+0.15}_{-0.28}$ & $4.9^{+0.7}_{-0.5}$ & $7.6^{+0.7}_{-0.7}$ \\
                      & $\log N$ & $18.08^{+0.04}_{-0.05}$ & $19.582^{+0.006}_{-0.010}$ & $17.48^{+0.43}_{-0.06}$\\
    ${\rm H_2\, J=2}$ & b\,km/s & $4.91^{+0.25}_{-0.43}$ & $4.90^{+0.26}_{-0.54}$ & $7.4^{+0.8}_{-0.5}$\\
                      & $\log N$ & $16.13^{+0.24}_{-0.07}$ & $18.429^{+0.020}_{-0.028}$ & $16.69^{+0.24}_{-0.10}$\\
    ${\rm H_2\, J=3}$ & b\,km/s & $5.4^{+1.2}_{-0.4}$ & $4.50^{+0.50}_{-0.30}$ &$7.9^{+0.8}_{-0.5}$ \\
                      & $\log N$ & $15.06^{+0.14}_{-0.07}$ & $18.230^{+0.033}_{-0.020}$ & $16.25^{+0.17}_{-0.10}$\\
    ${\rm H_2\, J=4}$ & b\,km/s & -- &$6.8^{+0.5}_{-0.9}$ & $10.3^{+1.0}_{-1.3}$\\
    				  & $\log N$ & $13.57^{+0.30}_{-0.78}$ & $15.45^{+0.07}_{-0.09}$ & $15.20^{+0.07}_{-0.07}$\\
    ${\rm H_2\, J=5}$ & b\,km/s & -- &$7.3^{+0.8}_{-0.7}$ & $10.3^{+1.1}_{-1.1}$\\
                      & $\log N$ & $13.5^{+0.3}_{-1.4}$ & $15.08^{+0.06}_{-0.06}$ & $14.82^{+0.09}_{-0.05}$\\
    ${\rm H_2\, J=6}$ & $\log N$ & -- & $14.44^{+0.05}_{-0.15}$ & $14.14^{+0.21}_{-0.17}$\\
    \hline 
         & $\log N_{\rm tot}$ & $18.35^{+0.02}_{-0.03}$ & $20.012^{+0.002}_{-0.012}$ & $18.12^{+0.12}_{-0.35}$ \\
       \hline         
    HD J=0 & b\,km/s &  $2.70^{+0.31}_{-0.34}$ & $0.59^{+0.46}_{-0.09}$ & $7.4^{+0.8}_{-1.2}$ \\
            & $\log N$ & $\lesssim 13.8$ & $15.3^{+0.4}_{-0.9}$ & $13.65^{+0.20}_{-0.23}$ \\
    \hline   
    \end{tabular}
    \begin{tablenotes}
    \item Doppler parameters of H$_2$ $\rm J=4, 5, 6$ in 1 component and  $\rm J = 6$ in 2 and 3 components were tied to $\rm J=3$ and $\rm J=5$, respectively.  
    \end{tablenotes}
\end{table*}

\begin{figure*}
    \centering
    \includegraphics[width=\linewidth]{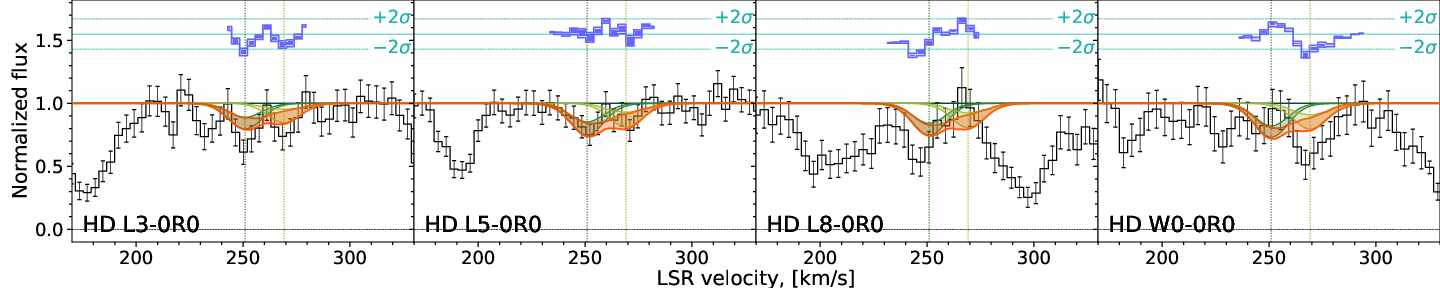}
    \caption{Fit to HD absorption lines towards BI 253 in LMC. Lines are the same as for \ref{fig:lines_HD_Sk67_2}.
    }
    \label{fig:lines_HD_BI253}
\end{figure*}

\begin{figure*}
    \centering
    \includegraphics[width=\linewidth]{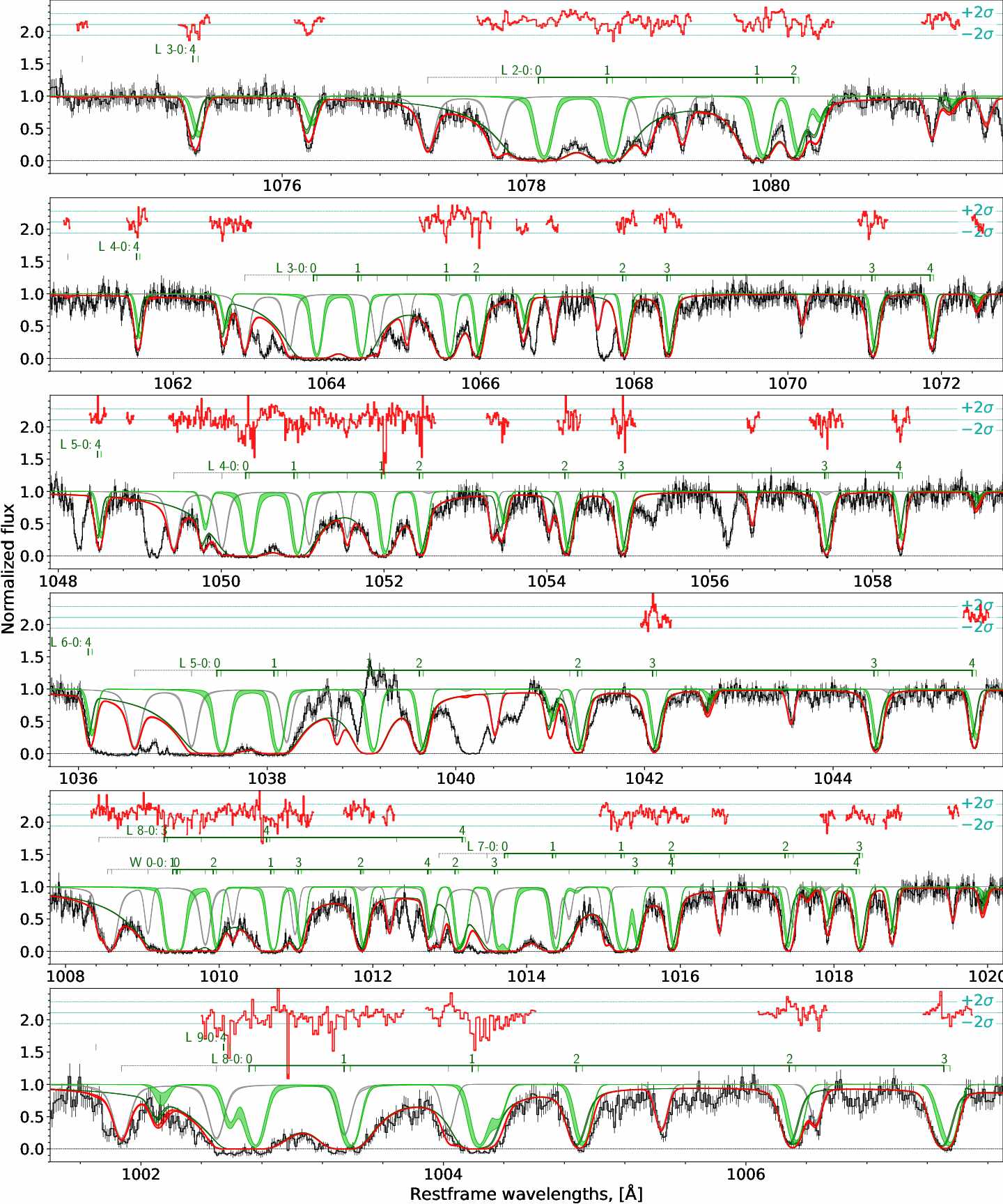}
    \caption{Fit to H2 absorption lines towards BI 253 in LMC. Lines are the same as for \ref{fig:lines_H2_Sk67_2}.
    }
    \label{fig:lines_H2_BI253}
\end{figure*}

\begin{table*}
    \caption{Fit results of H$_2$ lines towards Sk-68 135}
    \label{tab:Sk68_135}
    \begin{tabular}{ccccc}
    \hline
    \hline
    species & comp & 1 & 2 & 3 \\
            & z & $0.0000428(^{+9}_{-5})$ &  $0.0008364(^{+35}_{-23})$ & $0.00090870(^{+160}_{-30})$  \\
    \hline 
     ${\rm H_2\, J=0}$ & b\,km/s & $4.1^{+0.5}_{-0.5}$ & $0.530^{+0.330}_{-0.030}$ & $0.54^{+0.38}_{-0.04}$\\
                       & $\log N$ & $17.972^{+0.061}_{-0.021}$ & $15.8^{+1.2}_{-1.0}$ & $19.656^{+0.009}_{-0.017}$\\
    ${\rm H_2\, J=1}$ & b\,km/s & $5.08^{+0.18}_{-0.56}$ & $1.2^{+0.7}_{-0.4}$ & $1.3^{+1.4}_{-0.6}$\\
                      & $\log N$ & $18.11^{+0.04}_{-0.06}$ & $18.82^{+0.11}_{-0.06}$ & $19.681^{+0.016}_{-0.018}$\\
    ${\rm H_2\, J=2}$ & b\,km/s & $6.7^{+1.0}_{-0.8}$ & $1.93^{+0.45}_{-0.21}$ &$5.30^{+0.30}_{-0.30}$ \\
                      & $\log N$ & $15.41^{+0.11}_{-0.13}$ & $16.99^{+0.20}_{-0.42}$ & $18.489^{+0.033}_{-0.020}$\\
    ${\rm H_2\, J=3}$ & b\,km/s & $10.6^{+0.9}_{-1.1}$ & $1.92^{+0.54}_{-0.13}$ &  $5.83^{+0.22}_{-0.23}$\\
                      & $\log N$ & $15.011^{+0.024}_{-0.026}$ &  $15.70^{+0.22}_{-0.39}$ & $18.22^{+0.05}_{-0.04}$\\
    ${\rm H_2\, J=4}$ & b\,km/s & -- & -- & $5.53^{+0.34}_{-0.19}$ \\
    				  & $\log N$ & $13.04^{+0.44}_{-0.30}$ & $14.63^{+0.16}_{-0.21}$ & $17.09^{+0.07}_{-0.16}$\\
    ${\rm H_2\, J=5}$ & b\,km/s & -- & -- & $6.78^{+0.36}_{-0.20}$\\
    				  & $\log N$ & $12.7^{+0.4}_{-1.2}$ & $14.34^{+0.14}_{-0.16}$ & $16.54^{+0.07}_{-0.14}$\\
    ${\rm H_2\, J=6}$ & b\,km/s & -- &  -- & $9.9^{+1.7}_{-2.3}$\\
    				  & $\log N$ & -- & -- & $14.494^{+0.064}_{-0.024}$\\
    ${\rm H_2\, J=7}$ & $\log N$ & -- & -- & $14.46^{+0.04}_{-0.05}$\\ 				  
    \hline 
         & $\log N_{\rm tot}$ &  \\
     \hline    
     HD J=0 &  b\,km/s & $4.1^{+0.6}_{-0.6}$ & $0.515^{+0.342}_{-0.015}$ & $0.79^{+5.09}_{-0.29}$ \\
            & $\log N$ & $\lesssim 14.1$ & $\lesssim 15.9$ & $14.0^{+0.6}_{-0.3}$ \\
    \hline   
    \end{tabular}
    \begin{tablenotes}
     \item Doppler parameters of H$_2$ $\rm J=4$ and $\rm J=5$ in 1 and 2 components and $\rm J=7$ in 3 component were tied to $\rm J=3$, $\rm J=4$ and $\rm J=6$, respectively 
    \end{tablenotes}
\end{table*}

\begin{figure*}
    \centering
    \includegraphics[width=\linewidth]{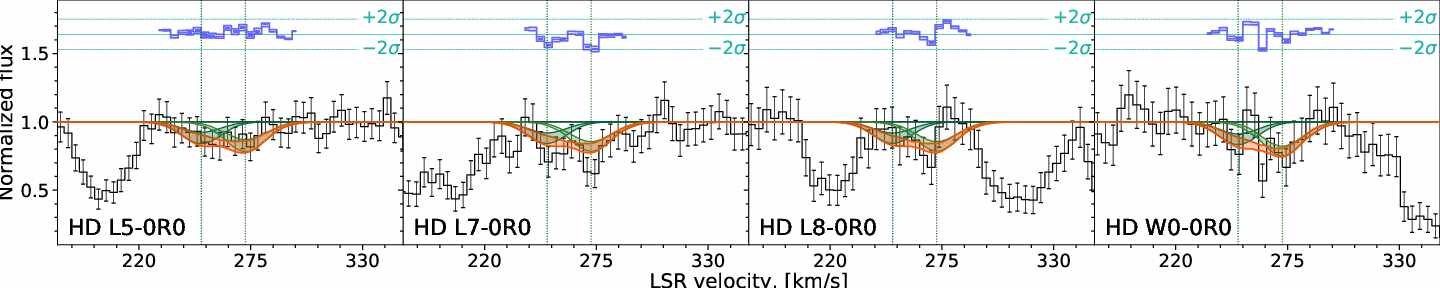}
    \caption{Fit to HD absorption lines towards Sk-68 135 in LMC. Lines are the same as for \ref{fig:lines_HD_Sk67_2}.
    }
    \label{fig:lines_HD_Sk68_135}
\end{figure*}

\begin{figure*}
    \centering
    \includegraphics[width=\linewidth]{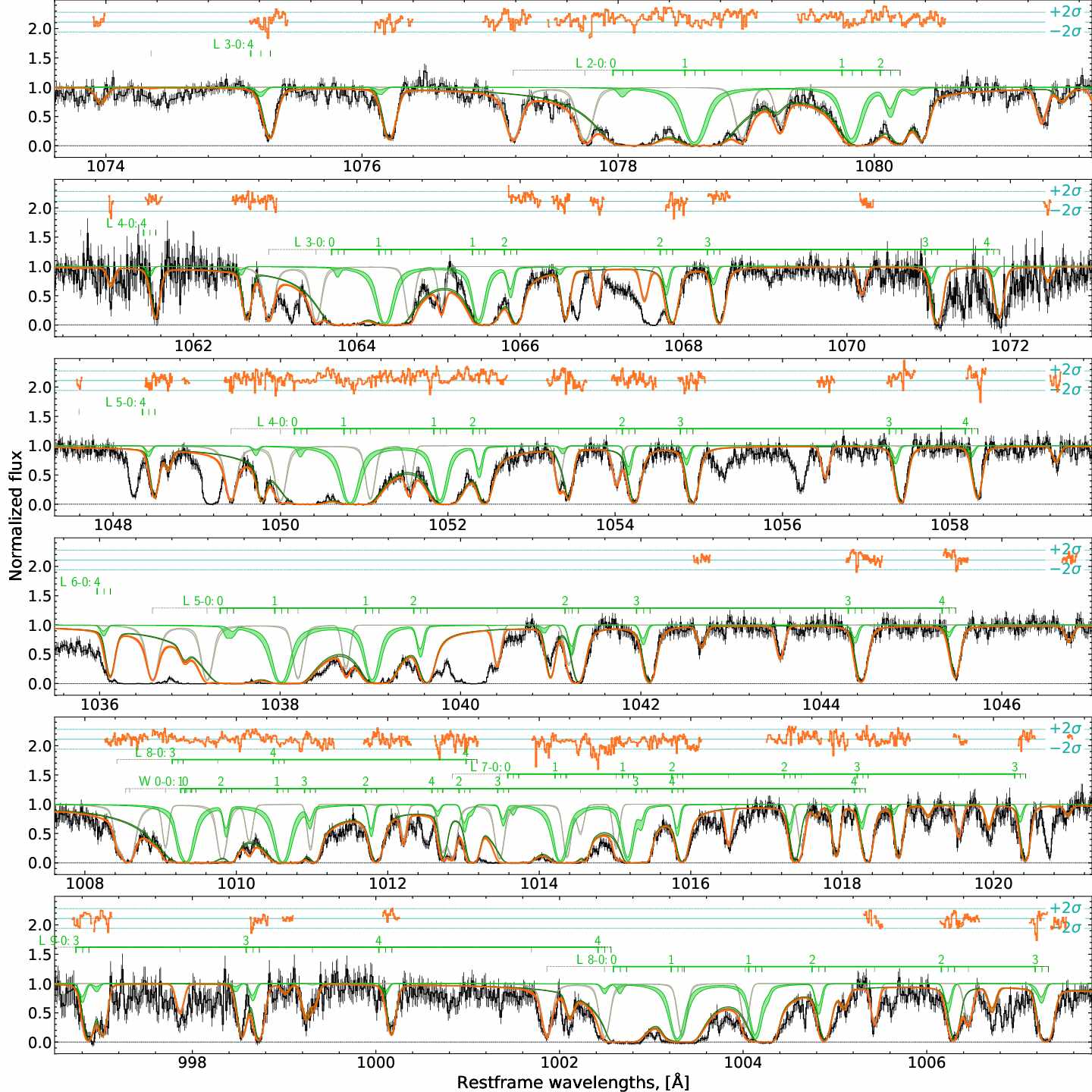}
    \caption{Fit to H2 absorption lines towards Sk-68 135 in LMC. Lines are the same as for \ref{fig:lines_H2_Sk67_2}.
    }
    \label{fig:lines_H2_Sk68_135}
\end{figure*}

\begin{table*}
    \caption{Fit results of H$_2$ lines towards Sk-68 137}
    \label{tab:Sk68_137}
    \begin{tabular}{ccccc}
    \hline
    \hline
    species & comp & 1 & 2 & 3 \\
            & z &  $0.0000866(^{+21}_{-9})$ &  $0.000929(^{+3}_{-3})$ &  $0.0010105(^{+42}_{-24})$ \\
    \hline 
     ${\rm H_2\, J=0}$ & b\,km/s & $0.67^{+0.45}_{-0.17}$ & $1.2^{+1.0}_{-0.5}$ & $2.5^{+1.6}_{-1.5}$\\
                       & $\log N$ &$18.29^{+0.07}_{-0.05}$ & $20.16^{+0.06}_{-0.55}$ & $20.36^{+0.07}_{-0.17}$\\
    ${\rm H_2\, J=1}$ & b\,km/s & $1.1^{+0.6}_{-0.3}$ & $2.8^{+0.9}_{-1.1}$ & $5.0^{+0.8}_{-1.8}$\\
                      & $\log N$ & $18.41^{+0.12}_{-0.14}$ & $19.74^{+0.11}_{-0.09}$ & $20.01^{+0.05}_{-0.05}$\\
    ${\rm H_2\, J=2}$ & b\,km/s & $1.57^{+0.57}_{-0.30}$  & $3.6^{+0.6}_{-0.7}$ & $5.8^{+0.5}_{-0.7}$\\
                      & $\log N$ & $17.50^{+0.10}_{-0.14}$ & $18.15^{+0.10}_{-0.12}$ & $18.67^{+0.06}_{-0.04}$\\
    ${\rm H_2\, J=3}$ & b\,km/s & $2.91^{+0.11}_{-0.16}$ & $4.0^{+0.5}_{-0.8}$ & $5.6^{+0.6}_{-0.6}$\\
                      & $\log N$ & $16.89^{+0.11}_{-0.12}$ & $18.08^{+0.10}_{-0.12}$ & $18.38^{+0.08}_{-0.09}$\\
    ${\rm H_2\, J=4}$ & b\,km/s & -- & $4.2^{+0.5}_{-0.5}$ & $5.4^{+0.8}_{-0.4}$\\
    				  & $\log N$ &  $14.29^{+0.08}_{-0.11}$ & $16.64^{+0.24}_{-0.31}$ & $16.46^{+0.18}_{-0.34}$ \\
    ${\rm H_2\, J=5}$ & b\,km/s & -- & -- & $8.4^{+4.3}_{-1.7}$\\
                      & $\log N$ & -- & $15.46^{+0.56}_{-0.25}$ & $15.06^{+0.08}_{-0.06}$\\
    ${\rm H_2\, J=6}$ & $\log N$ & -- & $14.54^{+0.06}_{-0.10}$ & -- \\
   		  
    \hline 
         & $\log N_{\rm tot}$ & $18.69^{+0.07}_{-0.07}$ & $20.31^{+0.05}_{-0.32}$ & $20.53^{+0.05}_{-0.11}$ \\
    \hline
    HD J=0 & b\,km/s & $0.54^{+0.55}_{-0.04}$ & $1.8^{+0.7}_{-0.7}$ & $1.3^{+1.5}_{-0.8}$ \\
           & $\log N$ & $\lesssim 15.2$ & $16.50^{+0.23}_{-1.63}$ & $14.48^{+1.07}_{-0.31}$ \\
    \hline   
    \end{tabular}
    \begin{tablenotes}
     \item Doppler parameters of H$_2$ $\rm J=4$ in 1 component and $\rm J=5, 6$ in  2 component were tied to $\rm J=3$ and $\rm J=4$, respectively.
    \end{tablenotes}
\end{table*}

\begin{figure*}
    \centering
    \includegraphics[width=\linewidth]{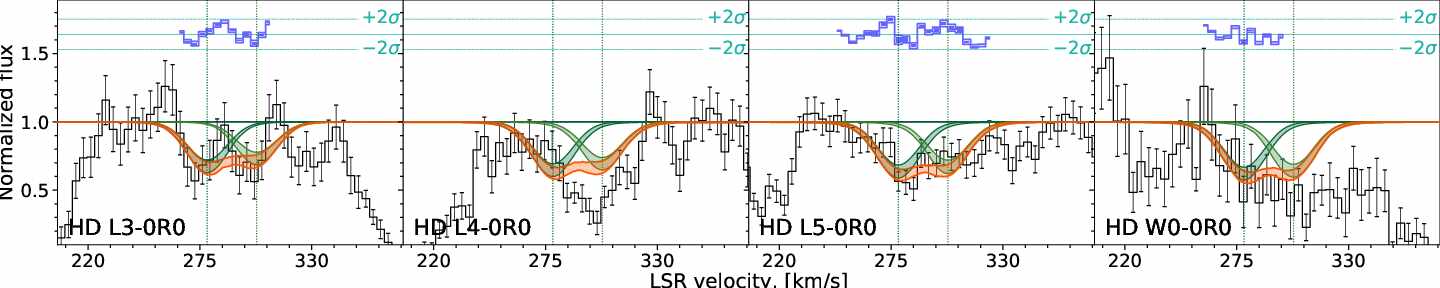}
    \caption{Fit to HD absorption lines towards Sk-68 137 in LMC. Lines are the same as for \ref{fig:lines_HD_Sk67_2}.
    }
    \label{fig:lines_HD_Sk68_137}
\end{figure*}

\begin{figure*}
    \centering
    \includegraphics[width=\linewidth]{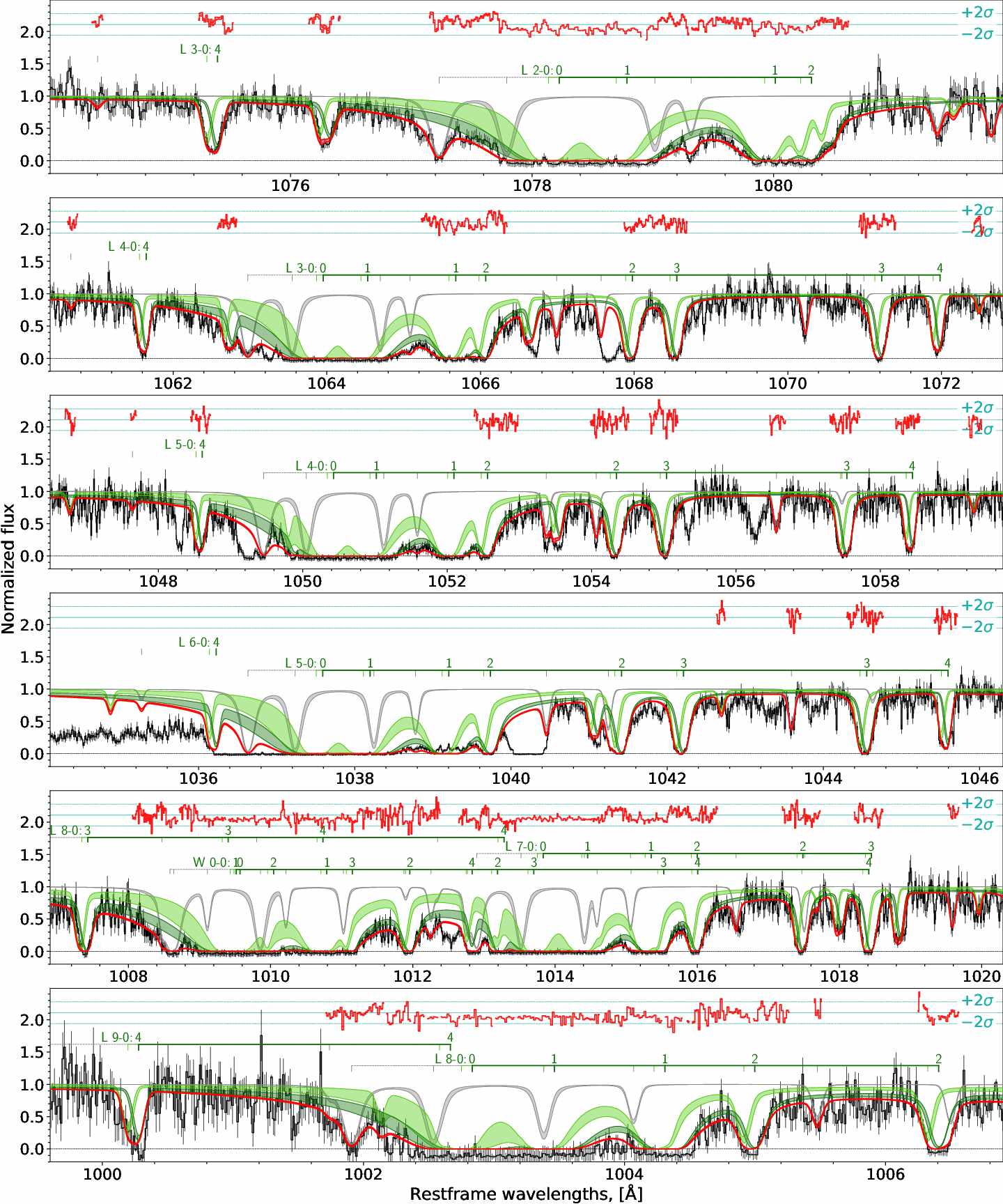}
    \caption{Fit to H2 absorption lines towards Sk-68 137 in LMC. Lines are the same as for \ref{fig:lines_H2_Sk67_2}.
    }
    \label{fig:lines_H2_Sk68_137}
\end{figure*}

\begin{table*}
    \caption{Fit results of H$_2$ lines towards Brey 77}
    \label{tab:Brey77}
    \begin{tabular}{cccccc}
    \hline
    \hline
    species & comp & 1 & 2 & 3 & 4 \\
            & z & $0.0000820(^{+8}_{-6})$ & $0.0009243(^{+41}_{-16})$ &  $0.0009878(^{+36}_{-19})$ & $0.0010553(^{+25}_{-16})$ \\
    \hline 
     ${\rm H_2\, J=0}$ & b\,km/s & $0.59^{+1.13}_{-0.09}$ & $1.6^{+0.5}_{-0.5}$ & $2.6^{+0.5}_{-0.6}$ &$0.69^{+0.08}_{-0.18}$ \\
                       & $\log N$ &$18.855^{+0.031}_{-0.040}$ &$18.83^{+0.08}_{-0.11}$ & $18.93^{+0.08}_{-0.07}$ & $18.04^{+0.12}_{-0.13}$\\
    ${\rm H_2\, J=1}$ & b\,km/s & $2.7^{+0.9}_{-0.4}$ &$1.8^{+0.4}_{-0.3}$ & $3.34^{+0.18}_{-0.34}$ & $1.78^{+0.66}_{-0.22}$ \\
                      & $\log N$ &$18.748^{+0.014}_{-0.035}$ & $19.05^{+0.04}_{-0.09}$ & $19.20^{+0.07}_{-0.05}$ &  $17.08^{+0.25}_{-0.15}$\\
    ${\rm H_2\, J=2}$ & b\,km/s & $4.12^{+0.38}_{-0.26}$ &$1.99^{+0.21}_{-0.32}$ &  $2.99^{+0.20}_{-0.23}$ &$2.21^{+0.31}_{-0.32}$ \\
                      & $\log N$ &$17.72^{+0.11}_{-0.14}$ & $17.94^{+0.10}_{-0.07}$ &  $18.16^{+0.08}_{-0.09}$ &$16.22^{+0.19}_{-0.34}$ \\
    ${\rm H_2\, J=3}$ & b\,km/s & $4.15^{+0.29}_{-0.24}$ &  $2.31^{+0.12}_{-0.51}$ &$3.37^{+0.27}_{-0.28}$ & $3.23^{+0.18}_{-0.39}$\\
                      & $\log N$ & $17.24^{+0.04}_{-0.24}$ & $18.06^{+0.11}_{-0.05}$ &$17.77^{+0.09}_{-0.24}$ & $15.98^{+0.25}_{-0.17}$\\
    ${\rm H_2\, J=4}$ & b\,km/s & -- & $2.47^{+0.19}_{-0.27}$ & $3.62^{+0.42}_{-0.29}$ & $3.13^{+0.56}_{-0.25}$\\
    				  & $\log N$ & $14.61^{+0.18}_{-0.05}$ & $16.83^{+0.05}_{-0.20}$ & $15.43^{+0.24}_{-0.11}$ & $14.73^{+0.14}_{-0.09}$\\
    ${\rm H_2\, J=5}$ & b\,km/s & -- & $2.40^{+0.18}_{-0.27}$ & $4.09^{+0.35}_{-0.14}$ & $3.2^{+0.5}_{-0.3}$\\
                      & $\log N$ & -- & $15.62^{+0.30}_{-0.22}$ & $14.79^{+0.10}_{-0.14}$ &$14.33^{+0.09}_{-0.10}$ \\
    ${\rm H_2\, J=6}$ & $\log N$ & -- &$14.01^{+0.17}_{-0.30}$ &$12.5^{+0.6}_{-0.5}$ &  $12.4^{+0.3}_{-0.3}$\\
    \hline 
         & $\log N_{\rm tot}$ & $19.13^{+0.02}_{-0.03}$ & $19.30^{+0.04}_{-0.06}$ & $19.42^{+0.05}_{-0.04}$ & $18.10^{+0.11}_{-0.11}$ \\
    \hline
    HD J=0 & b\,km/s & $0.84^{+0.35}_{-0.31}$ & $1.5^{+0.6}_{-0.6}$ & $2.6^{+0.5}_{-1.0}$ & $0.71^{+0.06}_{-0.13}$ \\
           & $\log N$ &  $\lesssim 16.2$ & $\lesssim 15.7$ $\lesssim 16.1$ & $\lesssim 16.4$ \\              
    \hline   
    \end{tabular}
    \begin{tablenotes}
     \item Doppler parameters of H$_2$ $\rm J=4$ in 1 component and $\rm J=6$ in  2, 3 and 4 components were tied to $\rm J=3$ and $\rm J=5$, respectively.
    \end{tablenotes}
\end{table*}

\begin{figure*}
    \centering
    \includegraphics[width=\linewidth]{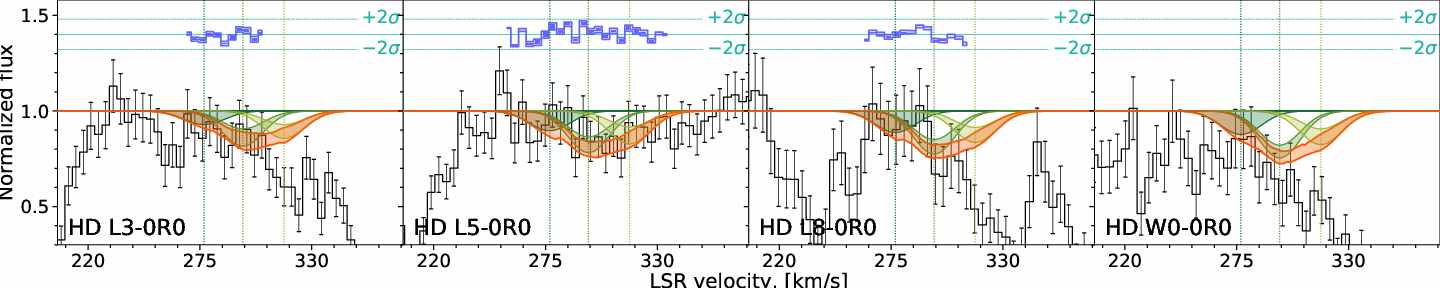}
    \caption{Fit to HD absorption lines towards Brey 77 in LMC. Lines are the same as for \ref{fig:lines_HD_Sk67_2}.
    }
    \label{fig:lines_HD_Brey77}
\end{figure*}

\begin{figure*}
    \centering
    \includegraphics[width=\linewidth]{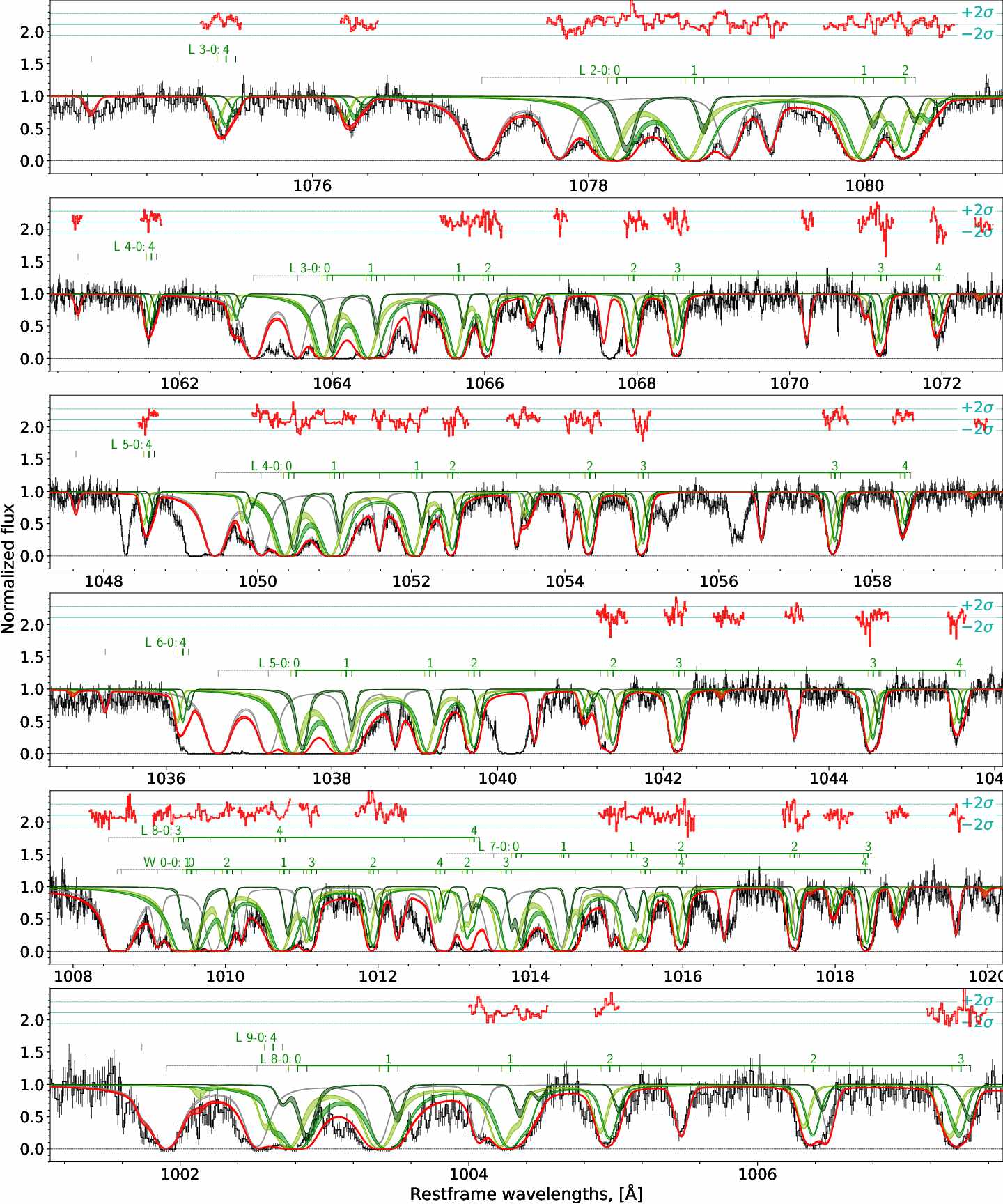}
    \caption{Fit to H2 absorption lines towards Brey 77  in LMC. Lines are the same as for \ref{fig:lines_H2_Sk67_2}.
    }
    \label{fig:lines_H2_Brey77}
\end{figure*}

\clearpage
\begin{table*}
    \caption{Fit results of H$_2$ lines towards Sk-69 243}
    \label{tab:Sk68_137}
    \begin{tabular}{ccccccc}
    \hline
    \hline
    species & comp & 1 & 2 & 3 & 4 & 5 \\
            & z & $0.0000878(^{+7}_{-4})$ & $0.0009040(^{+11}_{-34})$ & $0.000935(^{+3}_{-8})$ & $0.0009674(^{+19}_{-12})$ & $0.0010422(^{+7}_{-29})$   \\
    \hline 
     ${\rm H_2\, J=0}$ & b\,km/s & $1.5^{+0.8}_{-0.8}$ & $0.54^{+0.30}_{-0.04}$ & $1.4^{+1.6}_{-0.6}$ & $0.67^{+1.59}_{-0.17}$ & $1.7^{+0.8}_{-1.0}$\\
                       & $\log N$ &$19.012^{+0.031}_{-0.018}$ & $17.81^{+0.31}_{-0.25}$ & $18.91^{+0.12}_{-0.09}$ & $19.24^{+0.05}_{-0.06}$ & $16.4^{+0.6}_{-0.5}$\\
    ${\rm H_2\, J=1}$ & b\,km/s & $2.9^{+0.7}_{-1.2}$ & $0.84^{+0.46}_{-0.31}$ & $2.3^{+2.2}_{-1.1}$ & $3.05^{+0.21}_{-2.12}$ & $4.52^{+0.20}_{-0.85}$\\
                      & $\log N$ & $18.719^{+0.017}_{-0.028}$ & $18.11^{+0.34}_{-0.29}$ & $19.14^{+0.08}_{-0.06}$ & $19.330^{+0.014}_{-0.084}$ & $16.52^{+0.58}_{-0.19}$\\
    ${\rm H_2\, J=2}$ & b\,km/s & $3.49^{+0.39}_{-0.10}$ & $1.45^{+0.47}_{-0.22}$ & $5.4^{+0.7}_{-0.8}$ & $4.3^{+0.3}_{-0.6}$ & $4.2^{+0.4}_{-0.7}$\\
                      & $\log N$ & $17.45^{+0.13}_{-0.15}$ & $16.39^{+0.33}_{-0.15}$ & $16.98^{+0.24}_{-0.42}$ & $18.605^{+0.025}_{-0.024}$ & $15.40^{+0.18}_{-0.18}$\\
    ${\rm H_2\, J=3}$ & b\,km/s &  $3.56^{+0.36}_{-0.31}$ & $1.80^{+0.51}_{-0.24}$ &$6.3^{+0.9}_{-0.8}$ &$4.1^{+0.4}_{-0.4}$ & $4.7^{+0.6}_{-0.3}$\\
                      & $\log N$ & $16.72^{+0.24}_{-0.33}$ & $15.97^{+0.47}_{-0.17}$ & $15.15^{+0.96}_{-0.27}$ & $18.439^{+0.044}_{-0.024}$ & $15.55^{+0.12}_{-0.05}$\\
    ${\rm H_2\, J=4}$ & b\,km/s & $6.3^{+1.1}_{-1.5}$ & $2.2^{+0.5}_{-0.3}$ & $7.3^{+0.4}_{-1.3}$ & $3.8^{+0.7}_{-0.4}$ & $6.0^{+0.5}_{-0.8}$\\
    				  & $\log N$ & $14.334^{+0.060}_{-0.028}$ & $14.76^{+0.24}_{-0.10}$ & $13.84^{+0.36}_{-0.27}$ &$16.1^{+0.4}_{-0.4}$ & $14.683^{+0.040}_{-0.029}$\\
    ${\rm H_2\, J=5}$ & b\,km/s & -- & $3.1^{+1.2}_{-0.9}$ &$8.1^{+3.1}_{-1.0}$ & $5.6^{+0.6}_{-1.1}$ & $5.8^{+1.1}_{-0.6}$\\
                      & $\log N$ & $13.89^{+0.08}_{-0.16}$ & $14.46^{+0.13}_{-0.10}$ & $13.48^{+0.41}_{-0.29}$ &$14.80^{+0.05}_{-0.05}$ & $14.448^{+0.031}_{-0.066}$\\
    \hline 
         & $\log N_{\rm tot}$ & $19.20^{+0.02}_{-0.02}$ & $18.29^{+0.27}_{-0.19}$ & $19.34^{+0.07}_{-0.05}$ & $19.66^{+0.02}_{-0.04}$ & $16.80^{+0.45}_{-0.16}$ \\
     \hline
     HD J=0 & b\,km/s & $0.59^{+0.88}_{-0.09}$ & $0.55^{+0.36}_{-0.05}$ & $1.6^{+1.7}_{-0.9}$ & $0.57^{+1.10}_{-0.07}$ & $0.70^{+0.51}_{-0.20}$ \\
            & $\log N$ & $\lesssim 15.9$ & $\lesssim 15.9$ & $\lesssim 15.6$ & $\lesssim 16.0$ & $\lesssim 16.1$ \\
    \hline   
    \end{tabular}
    \begin{tablenotes}
     \item Doppler parameter of H$_2$ $\rm J=5$ in 1 component was tied to $\rm J=4$.
    \end{tablenotes}
\end{table*}

\begin{figure*}
    \centering
    \includegraphics[width=\linewidth]{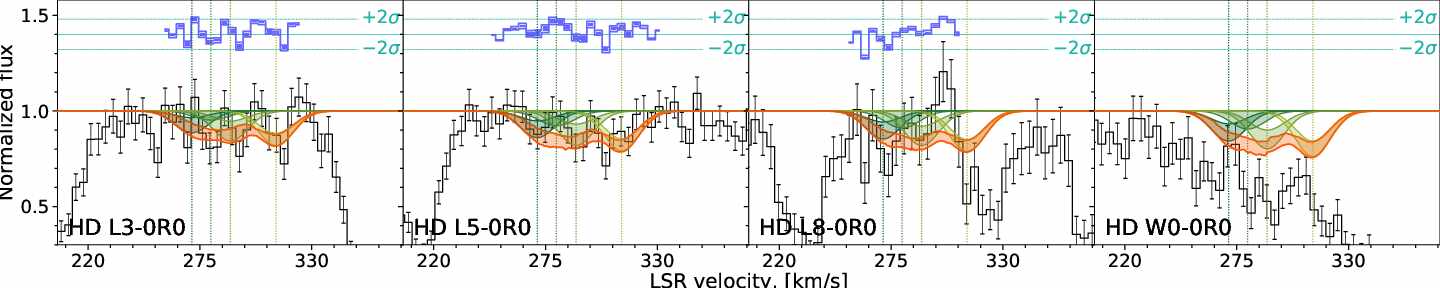}
    \caption{Fit to HD absorption lines towards Sk-69 243 in LMC. Lines are the same as for \ref{fig:lines_HD_Sk67_2}.
    }
    \label{fig:lines_HD_Sk69_243}
\end{figure*}

\begin{figure*}
    \centering
    \includegraphics[width=\linewidth]{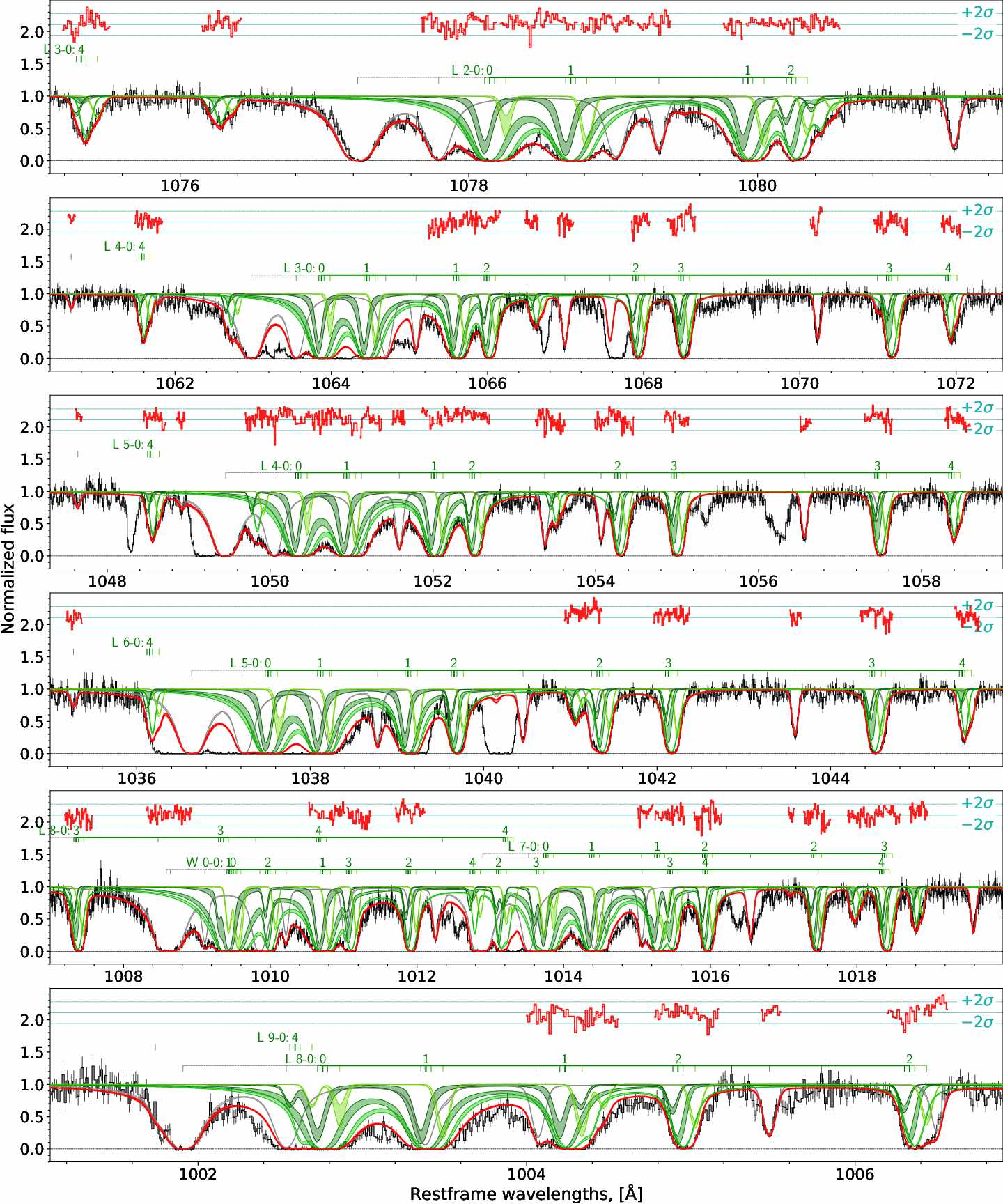}
    \caption{Fit to H2 absorption lines towards Sk-69 243 in LMC. Lines are the same as for \ref{fig:lines_H2_Sk67_2}.
    }
    \label{fig:lines_H2_Sk69_243}
\end{figure*}

\begin{table*}
    \caption{Fit results of H$_2$ lines towards Sk-69 246}
    \label{tab:Sk69_246}
    \begin{tabular}{cccc}
    \hline
    \hline
    species & comp & 1 & 2 \\
            & z & $0.00004578(^{+5}_{-4})$ &  $0.00091052(^{+14}_{-15})$ \\
    \hline 
     ${\rm H_2\, J=0}$ & b\,km/s & $2.72^{+0.24}_{-1.03}$ & $0.71^{+0.72}_{-0.20}$ \\
                       & $\log N$ & $18.553^{+0.004}_{-0.005}$ & $19.4874^{+0.0027}_{-0.0047}$ \\
    ${\rm H_2\, J=1}$ & b\,km/s & $2.3^{+0.5}_{-0.5}$ & $1.3^{+0.6}_{-0.5}$ \\
                      & $\log N$ & $18.725^{+0.008}_{-0.005}$ & $19.3945^{+0.0016}_{-0.0026}$ \\
    ${\rm H_2\, J=2}$ & b\,km/s & $3.22^{+0.10}_{-0.06}$ & $5.36^{+0.06}_{-0.04}$ \\
                      & $\log N$ & $17.431^{+0.022}_{-0.045}$ & $18.107^{+0.007}_{-0.016}$ \\
    ${\rm H_2\, J=3}$ & b\,km/s & $2.80^{+0.10}_{-0.06}$ & $5.945^{+0.116}_{-0.017}$ \\
                      & $\log N$ & $17.195^{+0.053}_{-0.018}$ & $17.811^{+0.029}_{-0.005}$ \\
    ${\rm H_2\, J=4}$ & b\,km/s & $3.863^{+0.011}_{-0.010}$ & $7.32^{+0.08}_{-0.18}$ \\
    				  & $\log N$ & $14.3891^{+0.0045}_{-0.0023}$ & $15.460^{+0.023}_{-0.023}$ \\
    ${\rm H_2\, J=5}$ & b\,km/s & -- & $6.77^{+0.19}_{-0.14}$ \\
    				  & $\log N$ & -- & $15.146^{+0.014}_{-0.016}$  \\
    ${\rm H_2\, J=6}$ & b\,km/s & -- & $7.4^{+2.8}_{-0.6}$ \\
    				  & $\log N$ & -- & $13.920^{+0.045}_{-0.021}$ \\
    \hline 
         & $\log N_{\rm tot}$ & $18.969^{+0.004}_{-0.004}$ & $19.7593^{+0.0002}_{-0.0037}$ \\
     \hline
     HD J=0 & b\,km/s & $2.60^{+0.30}_{-0.80}$ & $7.80^{+3.83}_{-3.65}$ \\
            & $\log N$ & $\lesssim 13.2$ & $13.89^{+0.06}_{-0.06}$ \\
    \hline   
    \end{tabular}
    \begin{tablenotes}
     \item 
    \end{tablenotes}
\end{table*}

\begin{figure*}
    \centering
    \includegraphics[width=\linewidth]{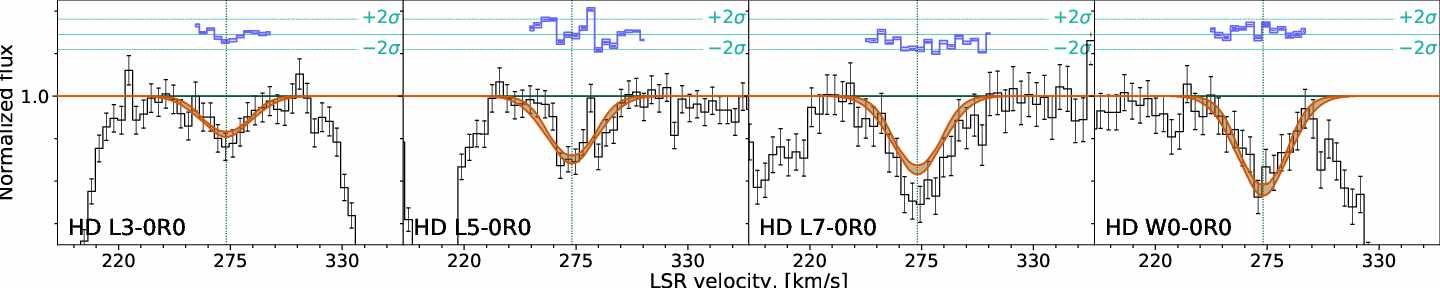}
    \caption{Fit to HD absorption lines towards Sk-69 246 in LMC. Lines are the same as for \ref{fig:lines_HD_Sk67_2}.
    }
    \label{fig:lines_HD_Sk69_246}
\end{figure*}

\begin{figure*}
    \centering
    \includegraphics[width=\linewidth]{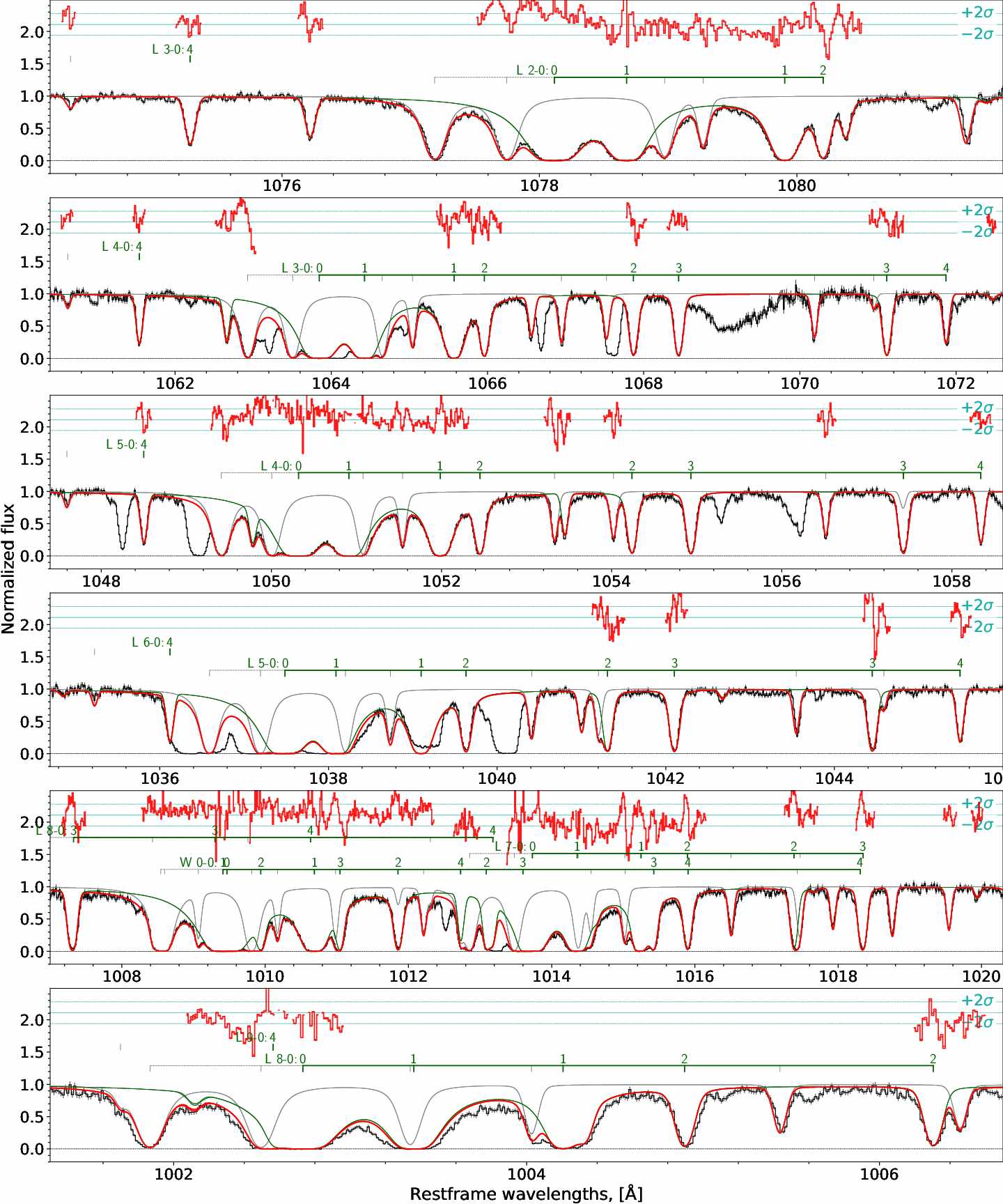}
    \caption{Fit to H2 absorption lines towards Sk-69 246 in LMC. Lines are the same as for \ref{fig:lines_H2_Sk67_2}.
    }
    \label{fig:lines_H2_Sk69_246}
\end{figure*}

\begin{table*}
    \caption{Fit results of H$_2$ lines towards Sk-68 140}
    \label{tab:Sk68_140}
    \begin{tabular}{cccc}
    \hline
    \hline
    species & comp & 1 & 2 \\
            & z &  $0.0000678(^{+29}_{-14})$ & $0.0009194(^{+18}_{-9})$ \\
    \hline 
     ${\rm H_2\, J=0}$ & b\,km/s & $0.529^{+0.730}_{-0.029}$ & $2.1^{+1.1}_{-1.3}$\\
                       & $\log N$ & $18.747^{+0.030}_{-0.062}$ & $20.21^{+0.04}_{-0.04}$\\
    ${\rm H_2\, J=1}$ & b\,km/s & $1.0^{+0.5}_{-0.5}$ & $4.5^{+0.9}_{-1.9}$\\
                      & $\log N$ & $18.65^{+0.17}_{-0.22}$ &  $19.916^{+0.020}_{-0.024}$\\
    ${\rm H_2\, J=2}$ & b\,km/s &  $1.5^{+0.6}_{-0.6}$ &  $5.4^{+0.6}_{-0.4}$\\
                      & $\log N$ & $17.53^{+0.12}_{-0.15}$ & $18.49^{+0.05}_{-0.09}$\\
    ${\rm H_2\, J=3}$ & b\,km/s & $4.06^{+0.21}_{-0.36}$ & $5.7^{+0.3}_{-0.5}$\\
                      & $\log N$ &  $16.92^{+0.20}_{-0.29}$ & $18.32^{+0.10}_{-0.05}$\\
    ${\rm H_2\, J=4}$ & b\,km/s & -- &  $5.6^{+0.3}_{-0.4}$\\
    				  & $\log N$ & $14.88^{+0.20}_{-0.16}$ &  $16.92^{+0.20}_{-0.25}$\\
    ${\rm H_2\, J=5}$ & b\,km/s & -- &  $6.07^{+0.20}_{-0.34}$\\
    				  & $\log N$ & -- & $16.40^{+0.18}_{-0.20}$ \\
    ${\rm H_2\, J=6}$ & $\log N$ & -- & $14.80^{+0.09}_{-0.16}$\\
    ${\rm H_2\, J=7}$ & $\log N$ & -- &  $14.44^{+0.15}_{-0.10}$\\
    ${\rm H_2\, J=8}$ & $\log N$ & -- & $13.84^{+0.20}_{-0.23}$\\
    \hline 
         & $\log N_{\rm tot}$ & $19.02^{+0.08}_{-0.09}$ & $20.40^{+0.03}_{-0.03}$\\
     \hline          
     HD J=0 &  b\,km/s & $0.530^{+0.590}_{-0.030}$ & $1.8^{+0.9}_{-0.9}$ \\
            & $\log N$ & $\lesssim 17.3$ & $\lesssim 17.1$ \\
    \hline   
    \end{tabular}
    \begin{tablenotes}
     \item Doppler parameters H$_2$ $\rm J=4$ in 1 component and $\rm J=6, 7, 8$ in 2 component were tied to $\rm J=3$ and $\rm J=5$, respectively.
     \end{tablenotes}
\end{table*}

\begin{figure*}
    \centering
    \includegraphics[width=\linewidth]{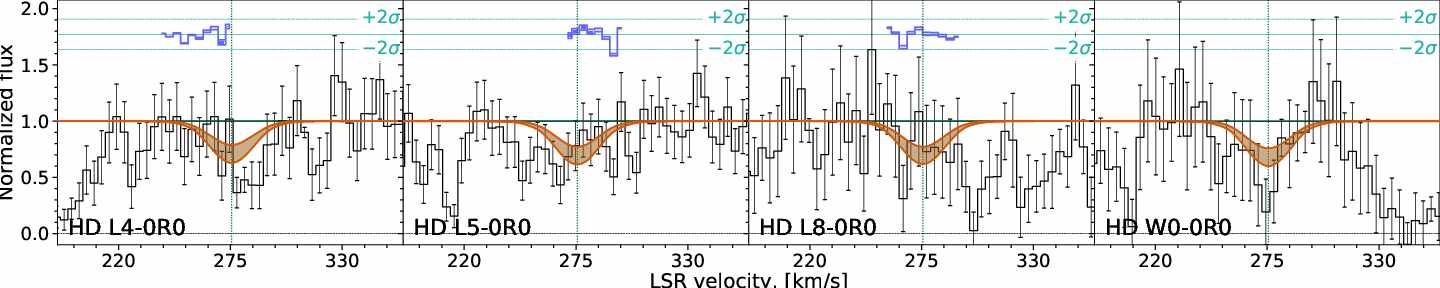}
    \caption{Fit to HD absorption lines towards Sk-68 140 in LMC. Lines are the same as for \ref{fig:lines_HD_Sk67_2}.
    }
    \label{fig:lines_HD_Sk68_140}
\end{figure*}

\begin{figure*}
    \centering
    \includegraphics[width=\linewidth]{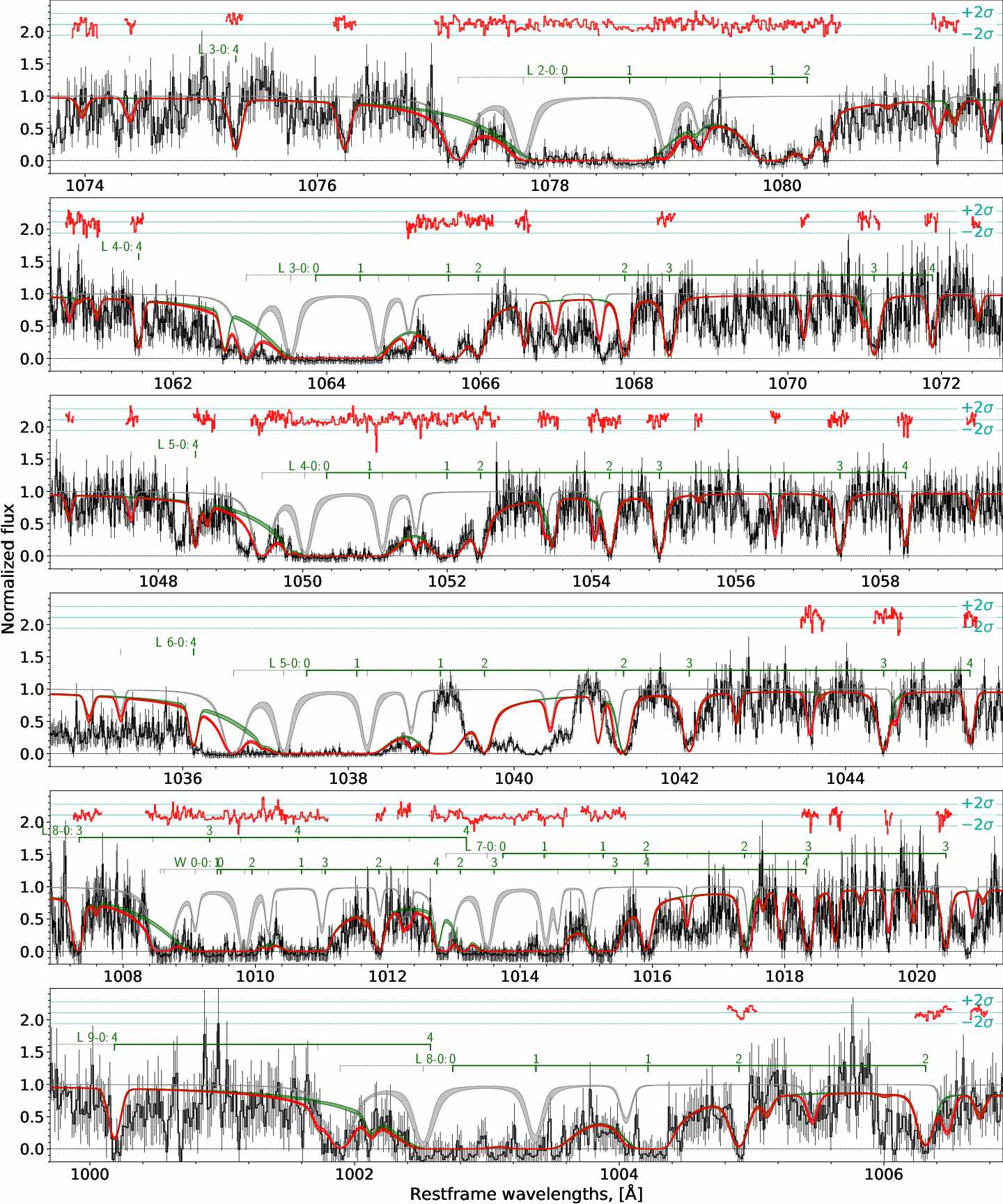}
    \caption{Fit to H2 absorption lines towards Sk-68 140 in LMC. Lines are the same as for \ref{fig:lines_H2_Sk67_2}.
    }
    \label{fig:lines_H2_Sk68_140}
\end{figure*}

\begin{table*}
    \caption{Fit results of H$_2$ lines towards Sk-71 50}
    \label{tab:Sk71_50}
    \begin{tabular}{cccccc}
    \hline
    \hline
    species & comp & 1 & 2 & 3 & 4\\
            & z &  $0.0000882(^{+9}_{-11})$ & $0.0007947(^{+31}_{-10})$ &  $0.0008923(^{+24}_{-46})$ & $0.0009745(^{+33}_{-28})$ \\
    \hline 
     ${\rm H_2\, J=0}$ & b\,km/s & $1.4^{+3.2}_{-0.5}$ &  $4.4^{+3.7}_{-1.0}$ & $1.6^{+1.8}_{-0.9}$ & $2.03^{+0.17}_{-1.37}$\\
                       & $\log N$ & $19.457^{+0.029}_{-0.012}$ & $19.47^{+0.10}_{-0.05}$ & $\lesssim 14.5$ &  $18.67^{+0.08}_{-0.09}$ \\
    ${\rm H_2\, J=1}$ & b\,km/s & $7.1^{+0.6}_{-1.7}$ & $8.9^{+0.7}_{-0.8}$ & $3.0^{+2.2}_{-0.8}$ & $1.9^{+2.1}_{-0.8}$\\
                      & $\log N$ & $19.09^{+0.05}_{-0.05}$ & $17.93^{+0.33}_{-0.28}$ & $19.43^{+0.05}_{-0.05}$ & $18.90^{+0.07}_{-0.11}$\\
    ${\rm H_2\, J=2}$ & b\,km/s & $6.9^{+0.8}_{-1.1}$ & $8.7^{+0.8}_{-0.6}$ & $5.6^{+0.4}_{-0.8}$ & $4.16^{+0.19}_{-0.36}$\\
                      & $\log N$ &  $16.28^{+0.62}_{-0.24}$ & $15.55^{+0.14}_{-0.09}$ & $16.28^{+0.50}_{-0.21}$ & $18.02^{+0.09}_{-0.13}$\\
    ${\rm H_2\, J=3}$ & b\,km/s & $7.2^{+0.9}_{-0.9}$ & $8.9^{+0.7}_{-0.4}$ &$5.4^{+0.4}_{-0.5}$ & $4.30^{+0.14}_{-0.21}$\\
                      & $\log N$ & $15.69^{+0.24}_{-0.22}$ & $15.59^{+0.08}_{-0.09}$ & $15.97^{+0.20}_{-0.24}$& $17.38^{+0.14}_{-0.16}$\\
    ${\rm H_2\, J=4}$ & b\,km/s & $7.7^{+1.6}_{-1.0}$& -- & -- & -- \\
    				  & $\log N$ & $14.90^{+0.08}_{-0.07}$& $14.94^{+0.04}_{-0.06}$ & $14.84^{+0.05}_{-0.17}$ & $14.73^{+0.31}_{-0.09}$ \\
    ${\rm H_2\, J=5}$ & $\log N$ & -- & $15.06^{+0.05}_{-0.05}$ & $14.18^{+0.14}_{-0.12}$ & $14.40^{+0.11}_{-0.10}$ \\
    
    \hline 
         & $\log N_{\rm tot}$ & $19.62^{+0.02}_{-0.02}$ & $19.47^{+0.09}_{-0.04}$ & $19.43^{+0.05}_{-0.05}$ & $19.15^{+0.05}_{-0.07}$ \\
     \hline
     HD J=0 & b\,km/s & $1.1^{+1.9}_{-0.6}$ & $3.6^{+2.9}_{-1.5}$ & $0.68^{+2.26}_{-0.18}$ & $1.5^{+0.6}_{-0.4}$ \\
            & $\log N$ & $\lesssim 16.6$ & $\lesssim 16.7$ & $\lesssim 16.5$ & $\lesssim 16.8$ \\ 
    \hline   
    \end{tabular}
    \begin{tablenotes}
    \item Doppler parameters H$_2$ $\rm J=4,5$ in 2, 3, 4 components were tied to $\rm J=3$. We could place only upper limit on H$_2$ $\rm J=0$ column density in 3 component, since all lines are blended.
    \end{tablenotes}
\end{table*}

\begin{figure*}
    \centering
    \includegraphics[width=\linewidth]{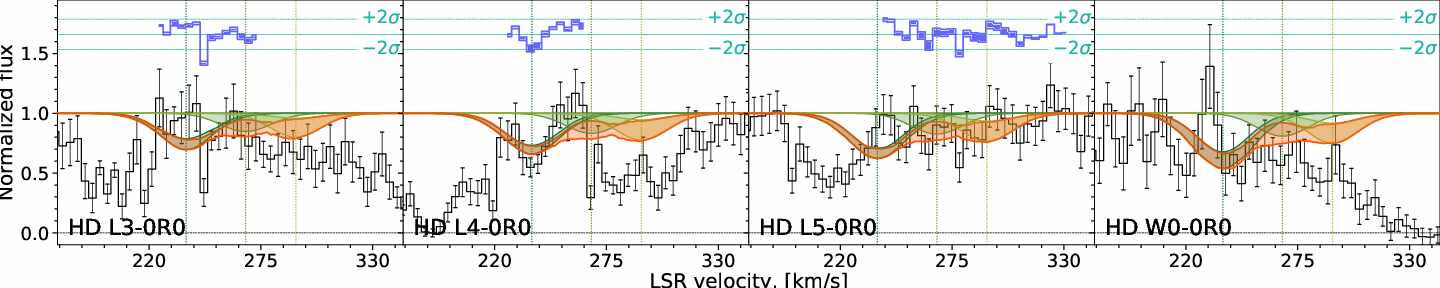}
    \caption{Fit to HD absorption lines towards Sk-71 50 in LMC. Lines are the same as for \ref{fig:lines_HD_Sk67_2}.
    }
    \label{fig:lines_HD_Sk71_50}
\end{figure*}

\begin{figure*}
    \centering
    \includegraphics[width=\linewidth]{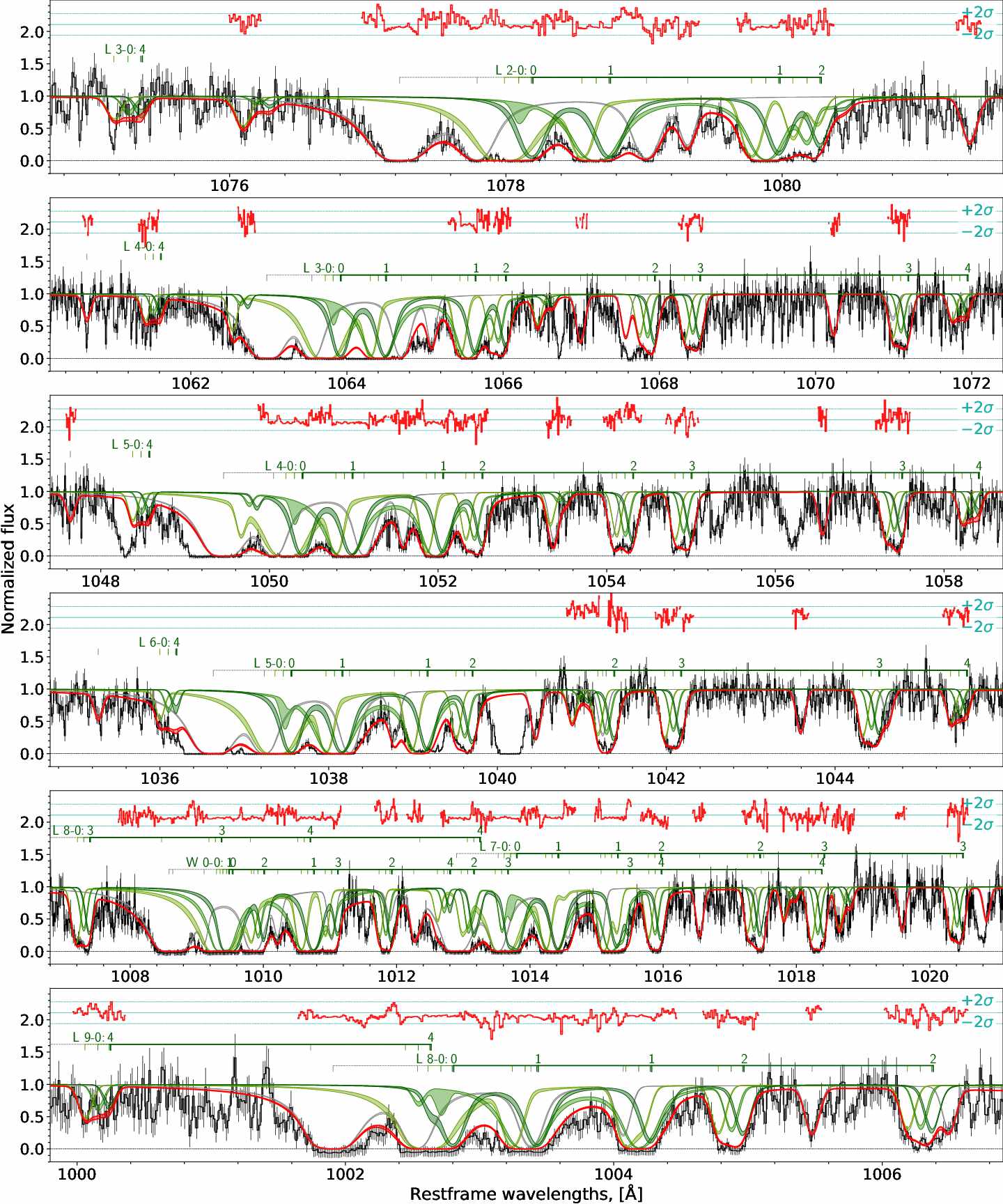}
    \caption{Fit to H2 absorption lines towards Sk-71 50 in LMC. Lines are the same as for \ref{fig:lines_H2_Sk67_2}.
    }
    \label{fig:lines_H2_Sk71_50}
\end{figure*}

\begin{table*}
    \caption{Fit results of H$_2$ lines towards Sk-69 279}
    \label{tab:Sk69_279}
    \begin{tabular}{ccccc}
    \hline
    \hline
    species & comp & 1 & 2 & 3 \\
            & z & $0.0000504(^{+38}_{-19})$ & $0.000832(^{+30}_{-18})$ & $0.0008871(^{+110}_{-21})$ \\
    \hline 
     ${\rm H_2\, J=0}$ & b\,km/s & $5.0^{+0.7}_{-1.5}$ & $1.3^{+0.3}_{-0.6}$ & $1.2^{+0.9}_{-0.7}$\\
                       & $\log N$ & $18.11^{+0.08}_{-0.18}$ & $19.10^{+0.23}_{-0.60}$ & $20.233^{+0.013}_{-0.055}$ \\
    ${\rm H_2\, J=1}$ & b\,km/s & $5.0^{+0.9}_{-0.5}$ & $1.7^{+2.7}_{-0.6}$ & $2.8^{+2.4}_{-0.8}$\\
                      & $\log N$ &$17.2^{+0.6}_{-0.3}$ & $18.5^{+0.5}_{-1.1}$ & $20.234^{+0.023}_{-0.017}$ \\
    ${\rm H_2\, J=2}$ & b\,km/s & $6.4^{+0.6}_{-0.5}$ & $6.1^{+3.3}_{-1.7}$ & $5.5^{+2.4}_{-0.7}$\\
                      & $\log N$ & $15.82^{+0.16}_{-0.28}$ & $16.0^{+1.1}_{-0.7}$ & $18.76^{+0.06}_{-0.10}$\\
    ${\rm H_2\, J=3}$ & b\,km/s & $6.8^{+0.6}_{-0.4}$ & $9.3^{+1.0}_{-1.2}$ & $8.6^{+1.5}_{-1.6}$\\
                      & $\log N$ & $14.76^{+0.12}_{-0.04}$ & $15.2^{+1.4}_{-0.7}$ & $18.35^{+0.14}_{-0.22}$\\
    ${\rm H_2\, J=4}$ & b\,km/s & -- & -- & $8.6^{+2.2}_{-2.8}$ \\
    				  & $\log N$ &$13.4^{+0.7}_{-1.2}$ & $14.69^{+0.27}_{-0.29}$ & $15.45^{+0.58}_{-0.18}$\\
    \hline 
         & $\log N_{\rm tot}$ & $18.16^{+0.14}_{-0.16}$ & $19.19^{+0.23}_{-0.44}$ & $20.54^{+0.01}_{-0.03}$\\
    \hline
    HD J=0 & b\,km/s & $5.3^{+0.6}_{-2.9}$ & $1.31^{+0.12}_{-0.81}$ & $0.9^{+3.6}_{-0.4}$ \\
           & $\log N$ & $\lesssim 16.7$ & $\lesssim 17.1$ & $\lesssim 17.0$ \\
    \hline   
    \end{tabular}
    \begin{tablenotes}
     \item Doppler parameters H$_2$ $\rm J=4$ in 1 and 2 components were tied to H$_2$ $\rm J=3$.
    \end{tablenotes}
\end{table*}

\begin{figure*}
    \centering
    \includegraphics[width=\linewidth]{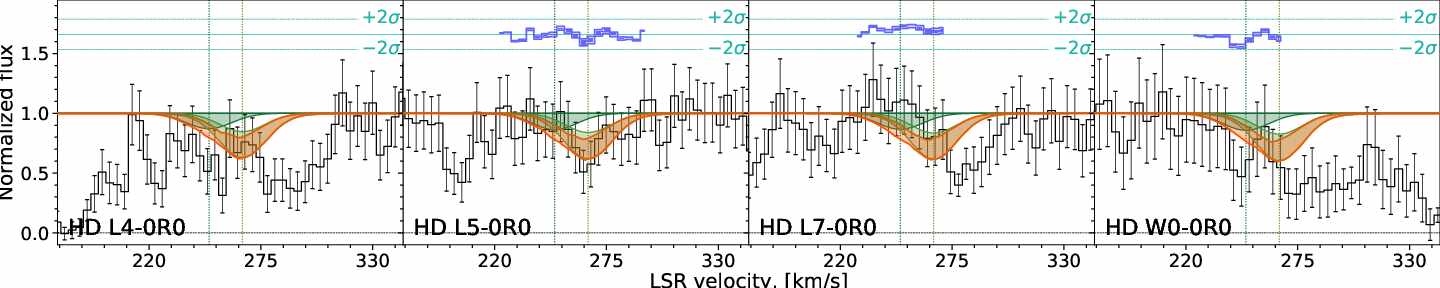}
    \caption{Fit to HD absorption lines towards Sk-69 279 in LMC. Lines are the same as for \ref{fig:lines_HD_Sk67_2}.
    }
    \label{fig:lines_HD_Sk69_279}
\end{figure*}

\begin{figure*}
    \centering
    \includegraphics[width=\linewidth]{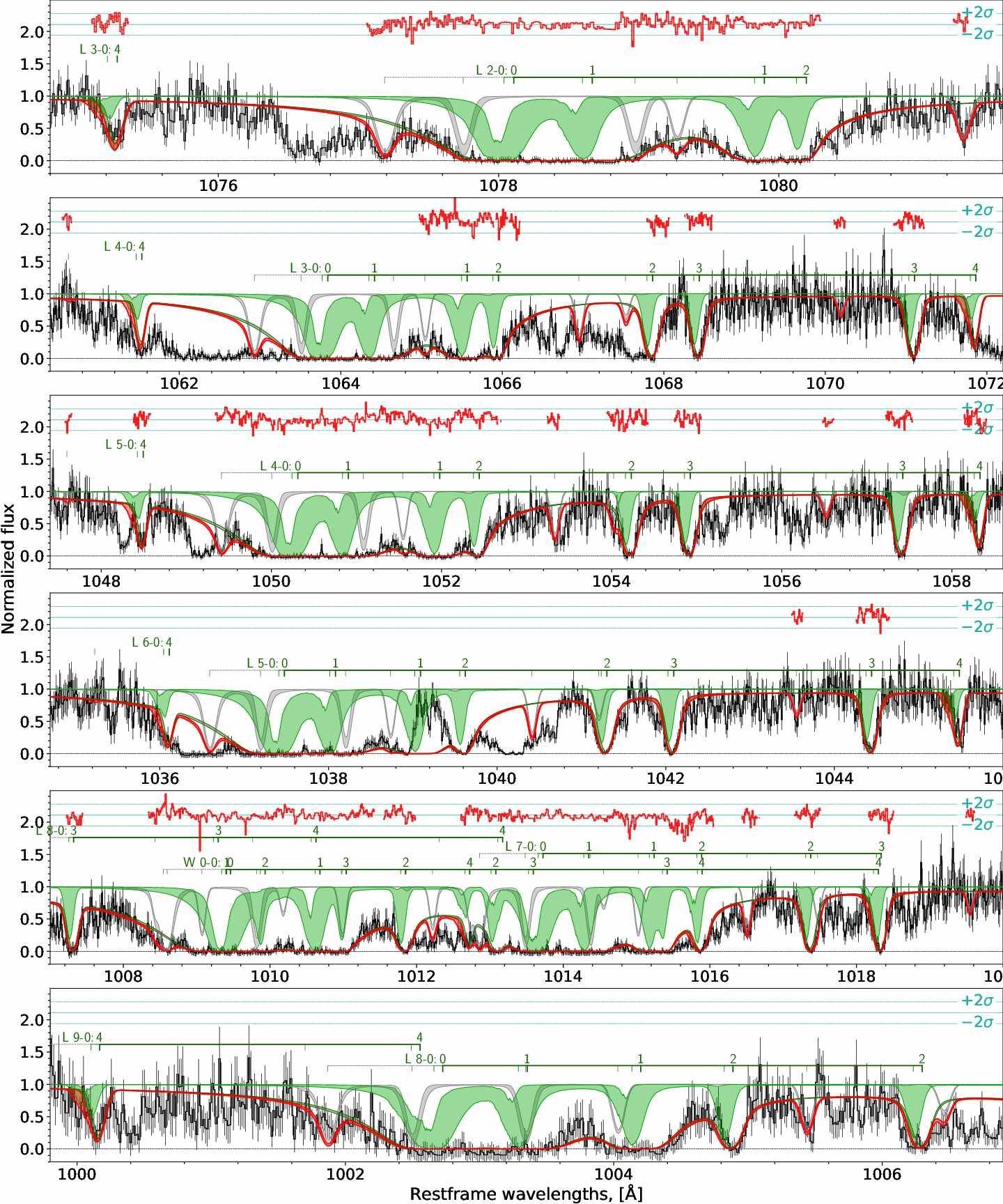}
    \caption{Fit to H2 absorption lines towards Sk-69 279 in LMC. Lines are the same as for \ref{fig:lines_H2_Sk67_2}.
    }
    \label{fig:lines_H2_Sk69_279}
\end{figure*}

\begin{table*}
    \caption{Fit results of H$_2$ lines towards Sk-68 155}
    \label{tab:Sk68_155}
    \begin{tabular}{cccccc}
    \hline
    \hline
    species & comp & 1 & 2 & 3 & 4 \\
            & z & $0.0000684(^{+17}_{-25})$ & $0.0009657(^{+27}_{-22})$ &$0.0010263(^{+29}_{-45})$ & $0.001113(^{+7}_{-4})$ \\
    \hline 
     ${\rm H_2\, J=0}$ & b\,km/s & $0.66^{+0.31}_{-0.16}$ &  $0.72^{+0.32}_{-0.20}$ & $0.9^{+0.5}_{-0.3}$ & $0.84^{+0.40}_{-0.22}$\\
                       & $\log N$ & $18.52^{+0.09}_{-0.09}$ & $19.57^{+0.07}_{-0.14}$ & $18.9^{+0.5}_{-0.7}$ & $18.98^{+0.23}_{-0.68}$\\
    ${\rm H_2\, J=1}$ & b\,km/s & $1.3^{+0.4}_{-0.5}$ & $0.81^{+0.76}_{-0.20}$ & $2.1^{+0.6}_{-0.9}$ & $1.3^{+0.7}_{-0.4}$\\
                      & $\log N$ &  $18.73^{+0.08}_{-0.09}$ & $19.781^{+0.022}_{-0.039}$ &$18.5^{+0.3}_{-1.1}$ & $17.6^{+0.3}_{-1.0}$\\
    ${\rm H_2\, J=2}$ & b\,km/s &  $2.73^{+0.18}_{-0.29}$ & $1.8^{+0.9}_{-0.4}$ & $3.1^{+0.7}_{-0.8}$ &  $1.8^{+0.7}_{-0.6}$\\
                      & $\log N$ & $17.78^{+0.09}_{-0.13}$ & $18.52^{+0.05}_{-0.09}$ &$16.3^{+0.7}_{-0.4}$ &  $14.6^{+0.6}_{-0.4}$\\
    ${\rm H_2\, J=3}$ & b\,km/s &$2.73^{+0.13}_{-0.25}$ & $2.84^{+0.18}_{-0.55}$ & $3.9^{+0.6}_{-1.0}$ &$4.03^{+0.30}_{-0.55}$ \\
                      & $\log N$ &$17.08^{+0.19}_{-0.32}$ &$18.46^{+0.05}_{-0.08}$ & $15.78^{+0.34}_{-0.24}$ & $14.49^{+0.17}_{-0.19}$ \\
    ${\rm H_2\, J=4}$ & b\,km/s & -- & $2.84^{+0.20}_{-0.18}$ & $5.02^{+0.19}_{-0.57}$ & $4.27^{+0.24}_{-0.43}$ \\
    				  & $\log N$ &$14.76^{+0.45}_{-0.21}$ &  $17.03^{+0.13}_{-0.17}$ &$14.75^{+0.14}_{-0.09}$  & $13.6^{+0.3}_{-0.3}$\\
    ${\rm H_2\, J=5}$ & $\log N$ & -- &$16.02^{+0.25}_{-0.64}$  &$14.68^{+0.13}_{-0.12}$ & $13.5^{+0.6}_{-0.4}$ \\
    ${\rm H_2\, J=6}$ & $\log N$ & -- & $14.18^{+0.09}_{-0.21}$ & $14.12^{+0.11}_{-0.21}$ & -- \\
    \hline 
         & $\log N_{\rm tot}$ & $18.97^{+0.06}_{-0.06}$ & $20.02^{+0.03}_{-0.05}$ & $19.03^{+0.40}_{-0.42}$ & $19.00^{+0.22}_{-0.62}$ \\
    \hline
    HD J=0 & b\,km/s & $0.70^{+0.23}_{-0.15}$ & $0.91^{+0.33}_{-0.19}$ & $0.94^{+0.40}_{-0.29}$ & $0.90^{+0.45}_{-0.21}$ \\
           & $\log N$ & $\lesssim 16.4$ & $\lesssim 16.8$ & $\lesssim 16.4$ & $16.8$ \\
    \hline   
    \end{tabular}
    \begin{tablenotes}
     \item Doppler parameters H$_2$ $\rm J=4$ in 1 component, $\rm J=5, 6$ in 2 and  3 components and $\rm J=5$ in 4 components were tied to H$_2$ $\rm J=3$ and $\rm J=4$, respectively.
    \end{tablenotes}
\end{table*}

\begin{figure*}
    \centering
    \includegraphics[width=\linewidth]{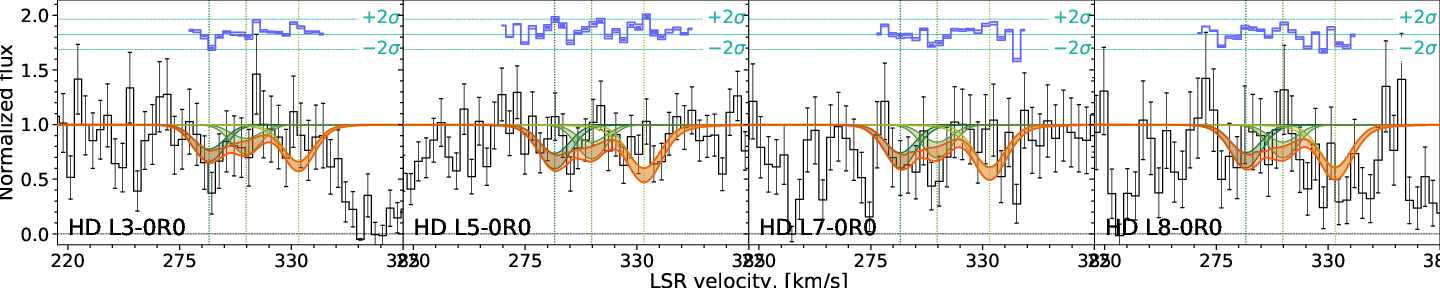}
    \caption{Fit to HD absorption lines towards Sk-68 155 in LMC. Lines are the same as for \ref{fig:lines_HD_Sk67_2}.
    }
    \label{fig:lines_HD_Sk68_155}
\end{figure*}

\begin{figure*}
    \centering
    \includegraphics[width=\linewidth]{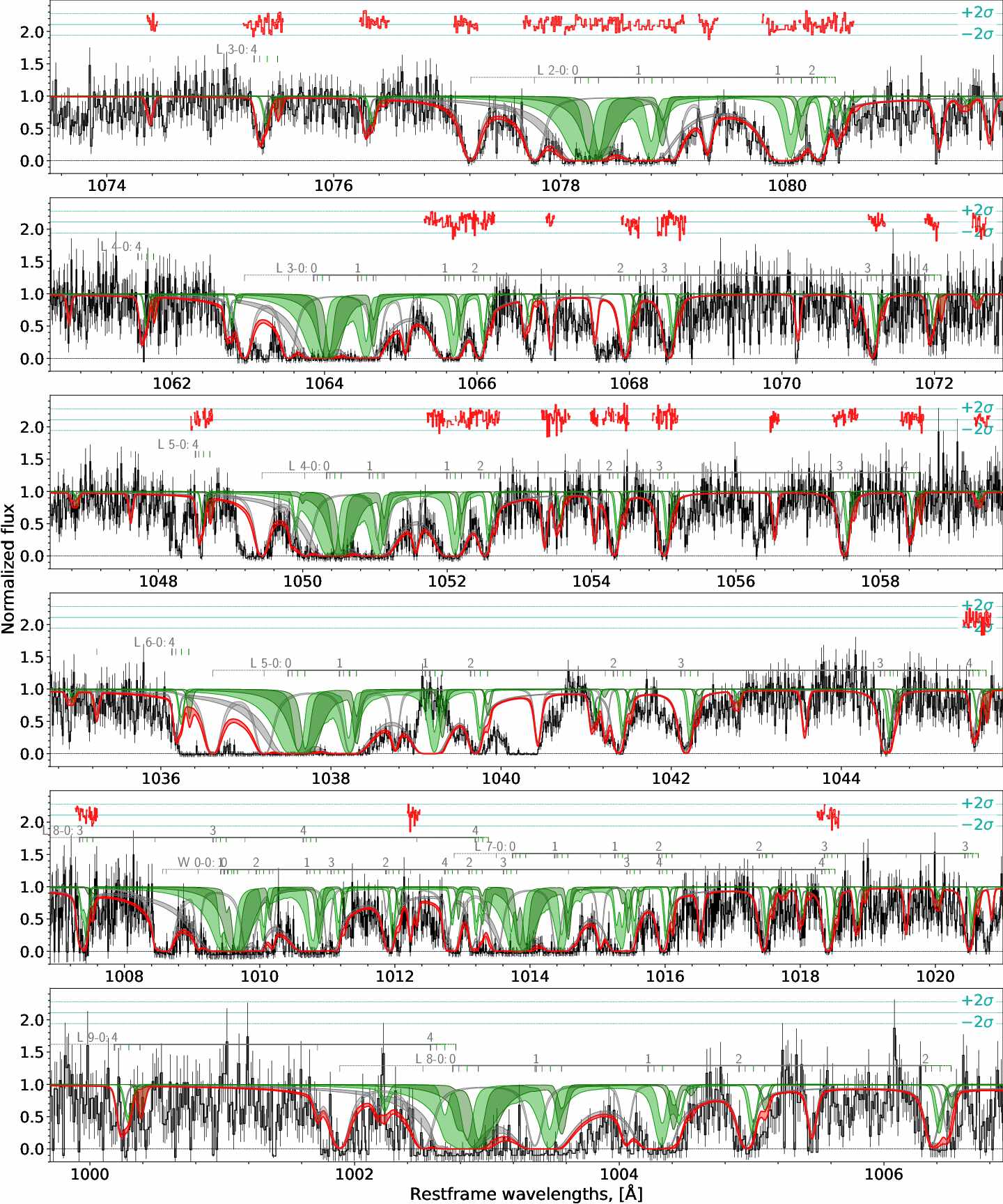}
    \caption{Fit to H2 absorption lines towards Sk-68 155 in LMC. Lines are the same as for \ref{fig:lines_H2_Sk67_2}.
    }
    \label{fig:lines_H2_Sk68_155}
\end{figure*}

\begin{table*}
    \caption{Fit results of H$_2$ lines towards Sk-69 297}
    \label{tab:Sk69_297}
    \begin{tabular}{ccccc}
    \hline
    \hline
    species & comp & 1 & 2 & 3 \\
            & z & $0.0000650(^{+20}_{-13})$ & $0.0008326(^{+8}_{-13})$ & $0.0009842(^{+12}_{-18})$ \\
    \hline 
     ${\rm H_2\, J=0}$ & b\,km/s & $0.509^{+0.084}_{-0.009}$ & $1.7^{+2.0}_{-1.2}$ & $0.518^{+0.245}_{-0.018}$\\
                       & $\log N$ & $18.25^{+0.06}_{-0.10}$ & $19.53^{+0.04}_{-0.04}$ & $18.44^{+0.11}_{-0.22}$\\
    ${\rm H_2\, J=1}$ & b\,km/s & $0.506^{+0.108}_{-0.006}$ & $5.9^{+1.3}_{-3.4}$ & $0.69^{+0.23}_{-0.17}$\\
                      & $\log N$ & $18.36^{+0.06}_{-0.04}$ & $19.383^{+0.024}_{-0.037}$ & $18.827^{+0.060}_{-0.028}$\\
    ${\rm H_2\, J=2}$ & b\,km/s & $0.546^{+0.152}_{-0.023}$ & $6.7^{+0.8}_{-0.6}$ & $1.01^{+0.18}_{-0.36}$\\
                      & $\log N$ & $17.769^{+0.046}_{-0.027}$ & $17.61^{+0.21}_{-0.50}$ & $18.057^{+0.048}_{-0.026}$\\
    ${\rm H_2\, J=3}$ & b\,km/s & $0.701^{+0.024}_{-0.148}$ & $7.2^{+1.0}_{-0.6}$ & $1.28^{+0.23}_{-0.36}$\\
                      & $\log N$ & $17.61^{+0.06}_{-0.05}$ & $16.8^{+0.4}_{-0.4}$ & $17.981^{+0.031}_{-0.046}$\\
    ${\rm H_2\, J=4}$ & b\,km/s & -- & $8.2^{+1.3}_{-1.5}$ & $1.41^{+0.15}_{-0.29}$\\
    				  & $\log N$ & $15.84^{+0.21}_{-0.13}$ & $15.32^{+0.22}_{-0.07}$ & $16.53^{+0.13}_{-0.23}$\\
    ${\rm H_2\, J=5}$ & b\,km/s & -- & $11.2^{+5.7}_{-2.3}$ & -- \\
    				  & $\log N$ & -- &  $14.72^{+0.06}_{-0.10}$ & $15.8^{+0.4}_{-0.5}$\\
    \hline 
         & $\log N_{\rm tot}$ & $18.70^{+0.04}_{-0.04}$ & $19.77^{+0.03}_{-0.03}$ & $19.07^{+0.05}_{-0.05}$\\
    \hline
    HD J=0 & b\,km/s & $0.504^{+0.109}_{-0.004}$ & $0.60^{+2.78}_{-0.10}$ & $0.511^{+0.284}_{-0.011}$ \\ 
           & $\log N$ & $\lesssim 16.2$ & $\lesssim 16.3$ & $\lesssim 16.4$ \\
    \hline   
    \end{tabular}
    \begin{tablenotes}
     \item Doppler parameters H$_2$ $\rm J=4$ in 1 component and $\rm J=5$ in 3 component were tied to H$_2$ $\rm J=3$ and $\rm J=4$, respectively.
    \end{tablenotes}
\end{table*}

\begin{figure*}
    \centering
    \includegraphics[width=\linewidth]{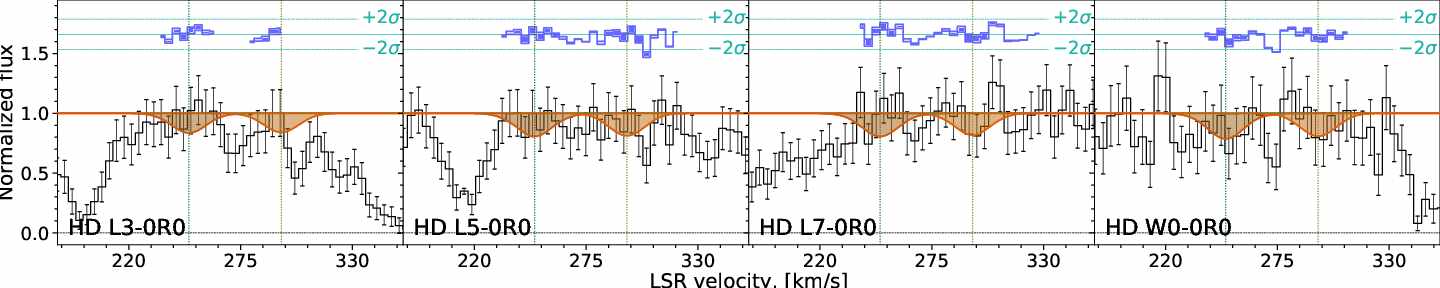}
    \caption{Fit to HD absorption lines towards Sk-69 297 in LMC. Lines are the same as for \ref{fig:lines_HD_Sk67_2}.
    }
    \label{fig:lines_HD_Sk69_297}
\end{figure*}

\begin{figure*}
    \centering
    \includegraphics[width=\linewidth]{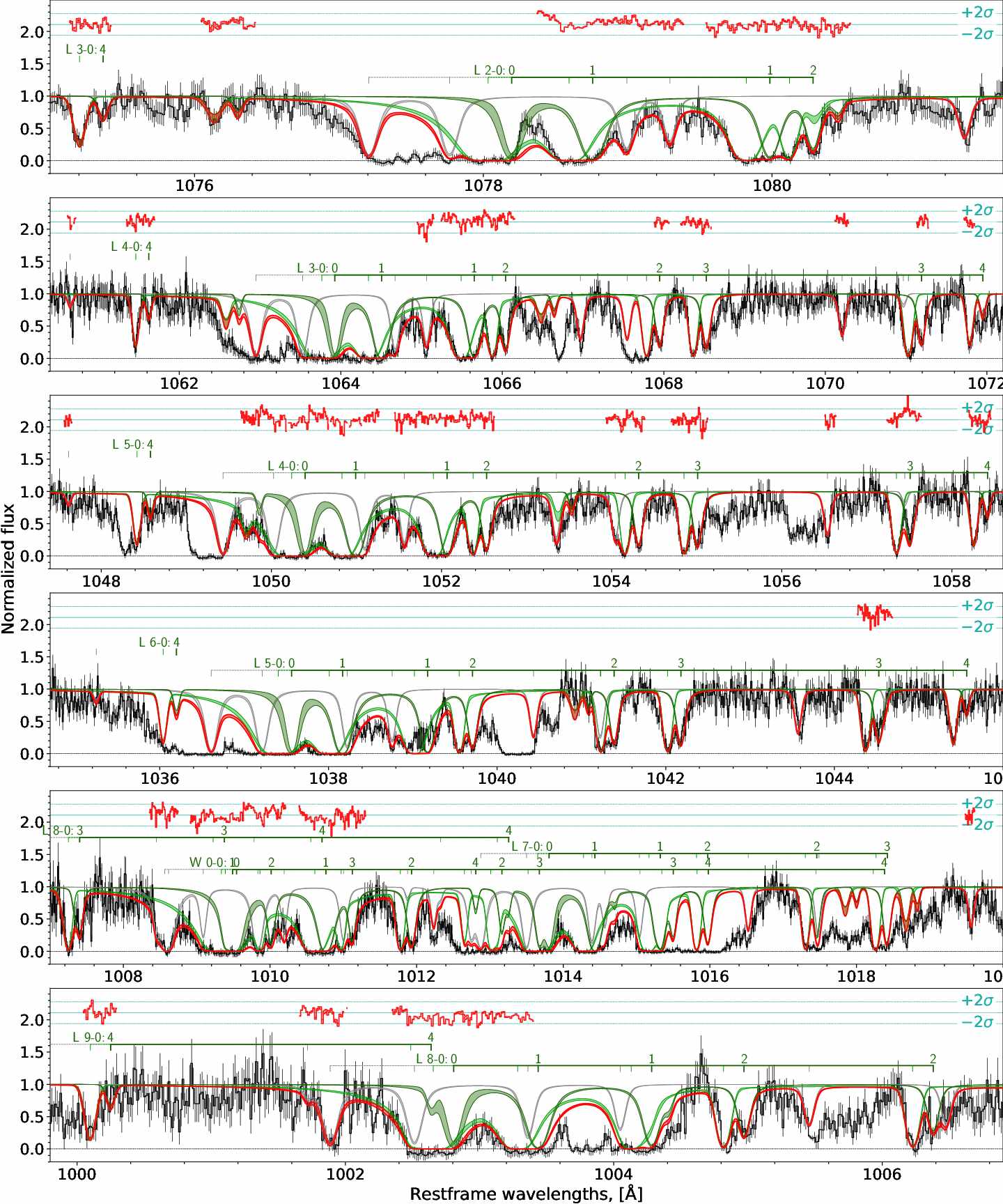}
    \caption{Fit to H2 absorption lines towards Sk-69 297 in LMC. Lines are the same as for \ref{fig:lines_H2_Sk67_2}.
    }
    \label{fig:lines_H2_Sk69_297}
\end{figure*}

\begin{table*}
    \caption{Fit results of H$_2$ lines towards Sk-70 115}
    \label{tab:Sk70_115}
    \begin{tabular}{ccccc}
    \hline
    \hline
    species & comp & 1 & 2 & 3 \\
            & z & $0.0000492(^{+9}_{-5})$ & $0.0007192(^{+3}_{-7})$ & $0.0009744(^{+10}_{-5})$ \\
    \hline 
     ${\rm H_2\, J=0}$ & b\,km/s &  $0.69^{+0.40}_{-0.18}$ & $0.9^{+0.6}_{-0.4}$ &$0.67^{+0.42}_{-0.17}$ \\
                       & $\log N$ & $18.071^{+0.020}_{-0.026}$ & $19.842^{+0.009}_{-0.008}$ & $17.50^{+0.14}_{-0.08}$\\
    ${\rm H_2\, J=1}$ & b\,km/s & $1.36^{+0.10}_{-0.28}$ & $1.2^{+0.6}_{-0.5}$ & $1.5^{+0.4}_{-0.4}$\\
                      & $\log N$ & $18.335^{+0.024}_{-0.018}$ &$19.364^{+0.006}_{-0.008}$ & $18.093^{+0.020}_{-0.022}$\\
    ${\rm H_2\, J=2}$ & b\,km/s & $1.47^{+0.06}_{-0.09}$ & $2.44^{+0.23}_{-0.25}$ & $1.54^{+0.48}_{-0.13}$\\
                      & $\log N$ & $17.587^{+0.024}_{-0.016}$ & $17.68^{+0.04}_{-0.04}$ & $17.12^{+0.06}_{-0.15}$\\
    ${\rm H_2\, J=3}$ & b\,km/s & $1.47^{+0.05}_{-0.09}$ & $2.56^{+0.15}_{-0.22}$ & $1.56^{+0.34}_{-0.16}$\\
                      & $\log N$ & $17.09^{+0.03}_{-0.06}$ & $17.11^{+0.10}_{-0.08}$ & $16.35^{+0.17}_{-0.49}$\\
    ${\rm H_2\, J=4}$ & b\,km/s & -- & $2.68^{+0.19}_{-0.15}$ &-- \\
    				  & $\log N$ &$14.01^{+0.07}_{-0.15}$ &$15.26^{+0.16}_{-0.15}$ & $13.76^{+0.10}_{-0.19}$\\
    ${\rm H_2\, J=5}$ & $\log N$ & $13.45^{+0.25}_{-0.36}$ & $14.62^{+0.09}_{-0.05}$ &  $13.1^{+0.3}_{-1.1}$\\
    \hline 
         & $\log N_{\rm tot}$ & $18.59^{+0.01}_{-0.01}$ & $19.97^{+0.01}_{-0.01}$ & $18.23^{+0.03}_{-0.02}$\\
    \hline
    HD J=0 & b\,km/s &$0.519^{+0.469}_{-0.019}$ & $1.10^{+0.30}_{-0.40}$ & $0.521^{+0.397}_{-0.021}$ \\
            & $\log N$ &  $\lesssim 15.4$ & $\lesssim 16.3$ & $\lesssim 15.5$ \\
    \hline   
    \end{tabular}
    \begin{tablenotes}
     \item Doppler parameters H$_2$ $\rm J=4, 5$ in 1 and 3 components and $\rm J=5$ in 2 component were tied to H$_2$ $\rm J=3$ and $\rm J=4$, respectively.
    \end{tablenotes}
\end{table*}

\begin{figure*}
    \centering
    \includegraphics[width=\linewidth]{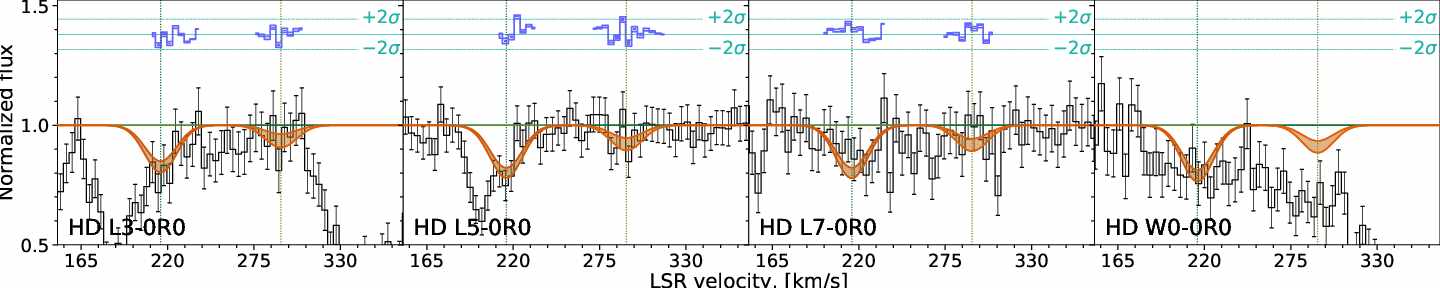}
    \caption{Fit to HD absorption lines towards Sk-70 115 in LMC. Lines are the same as for \ref{fig:lines_HD_Sk67_2}.
    }
    \label{fig:lines_HD_Sk70_115}
\end{figure*}

\begin{figure*}
    \centering
    \includegraphics[width=\linewidth]{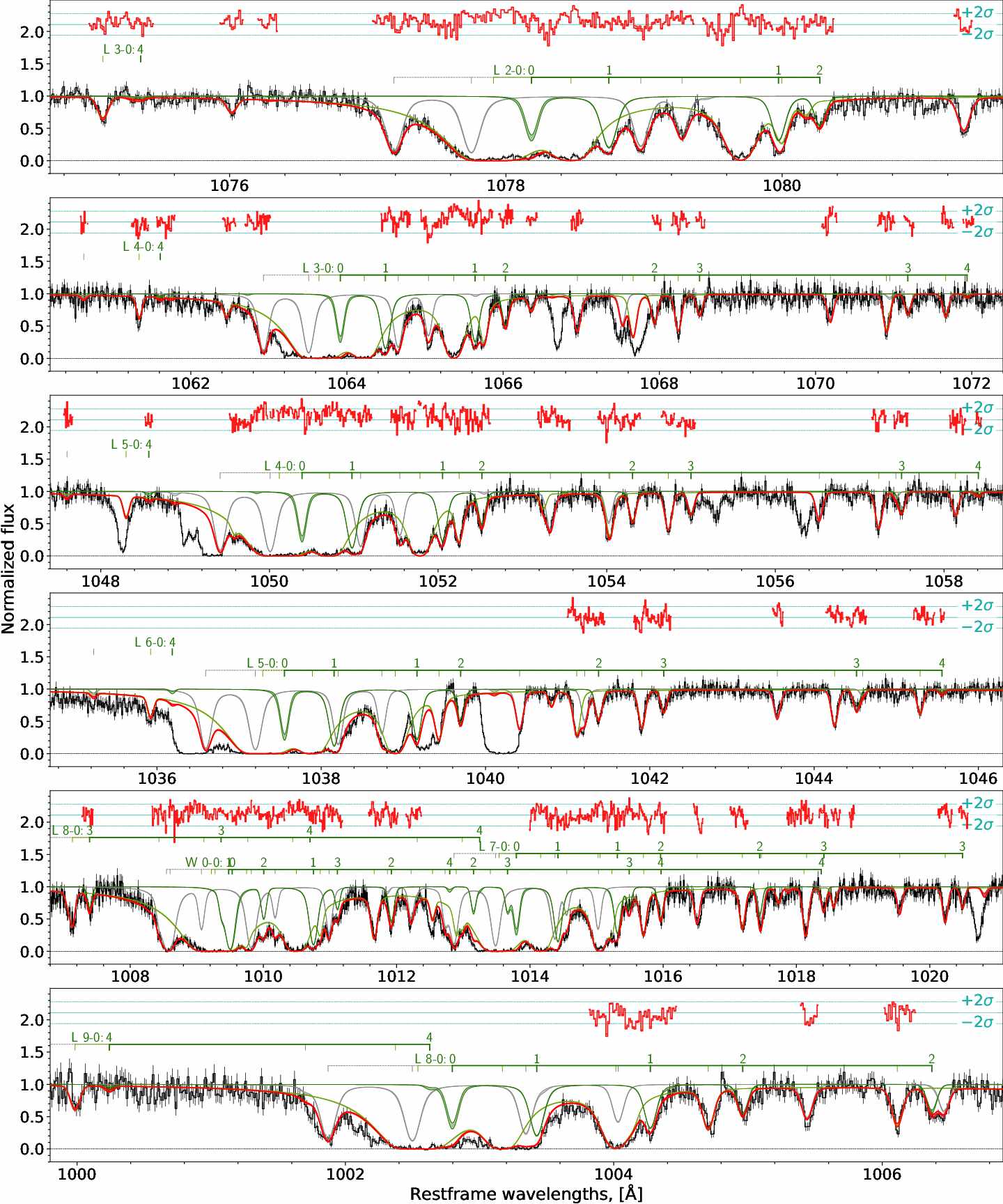}
    \caption{Fit to H2 absorption lines towards Sk-10 115 in LMC. Lines are the same as for \ref{fig:lines_H2_Sk67_2}.
    }
    \label{fig:lines_H2_Sk70_115}
\end{figure*}

\clearpage
\subsection{Small Magellanic Cloud}

\begin{table*}
    \caption{Fit results of H$_2$ lines towards AV 6}
    \label{tab:AV6}
    \begin{tabular}{cccccccc}
    \hline
    \hline
    species & comp & 1 & 2 & 3 &  4& 5 & 6\\
            & z &$-0.000021(^{+9}_{-5})$ & $0.0000533(^{+29}_{-14})$ &  $0.000090(^{+7}_{-4})$ & $0.0003581(^{+31}_{-37})$ & $0.0004402(^{+22}_{-34})$ & $0.000502(^{+6}_{-8})$ \\
    \hline 
     ${\rm H_2\, J=0}$ & b\,km/s & $0.53^{+0.22}_{-0.03}$ & $0.55^{+0.36}_{-0.05}$ & $0.71^{+0.16}_{-0.16}$ & $0.75^{+0.48}_{-0.13}$ & $1.5^{+0.9}_{-0.7}$ & $1.5^{+0.9}_{-0.8}$\\
                       & $\log N$ &  $15.2^{+0.4}_{-0.5}$ & $17.38^{+0.08}_{-0.09}$ & $11.3^{+2.4}_{-1.0}$ & $16.73^{+0.30}_{-0.69}$ & $18.61^{+0.06}_{-0.05}$ & $17.42^{+0.15}_{-0.68}$\\
    ${\rm H_2\, J=1}$ & b\,km/s & $0.55^{+0.27}_{-0.05}$ & $2.8^{+0.4}_{-0.4}$ & $0.73^{+0.27}_{-0.14}$ & $1.2^{+0.4}_{-0.4}$ & $5.3^{+0.7}_{-1.4}$ & $2.3^{+1.0}_{-1.1}$\\
                      & $\log N$ & $14.9^{+0.5}_{-0.4}$ & $17.49^{+0.09}_{-0.19}$ & $16.01^{+0.21}_{-0.55}$ & $16.08^{+0.31}_{-0.46}$ &$18.896^{+0.030}_{-0.022}$ &  $15.1^{+0.4}_{-0.5}$\\
    ${\rm H_2\, J=2}$ & b\,km/s & $0.85^{+0.41}_{-0.22}$ & $2.9^{+0.4}_{-0.4}$ & $1.7^{+0.5}_{-0.4}$ & $2.1^{+0.4}_{-0.6}$ &$4.9^{+0.3}_{-0.8}$ & $2.7^{+1.2}_{-0.8}$\\
                      & $\log N$ & $14.32^{+0.21}_{-0.47}$ & $15.68^{+0.45}_{-0.21}$ & $14.89^{+0.21}_{-0.36}$ & $14.69^{+0.18}_{-0.26}$ & $17.52^{+0.24}_{-0.09}$ & $14.2^{+0.4}_{-0.3}$\\
    ${\rm H_2\, J=3}$ & b\,km/s & $1.1^{+1.2}_{-0.3}$ & $2.8^{+0.5}_{-0.4}$ & $4.7^{+3.8}_{-1.4}$ & $8.0^{+2.4}_{-2.0}$ & $5.2^{+0.5}_{-0.9}$ & $2.9^{+1.5}_{-0.7}$\\
                      & $\log N$ &  $14.07^{+0.34}_{-0.17}$ & $14.89^{+0.17}_{-0.18}$ & $14.35^{+0.12}_{-0.12}$ & $14.52^{+0.06}_{-0.06}$ &  $16.10^{+0.65}_{-0.19}$ & $14.20^{+0.23}_{-0.17}$\\
    ${\rm H_2\, J=4}$ & b\,km/s & -- & -- & -- & -- &$5.2^{+1.1}_{-0.8}$ & -- \\
    				  & $\log N$ &$13.6^{+0.4}_{-0.4}$ & $14.29^{+0.09}_{-0.16}$ & $13.5^{+0.4}_{-0.4}$ & $13.93^{+0.13}_{-0.13}$ & $14.65^{+0.08}_{-0.07}$ & $13.77^{+0.20}_{-0.49}$ \\
    
    \hline 
         & $\log N_{\rm tot}$ & $15.43^{+0.32}_{-0.24}$ & $17.74^{+0.06}_{-0.11}$ & $16.05^{+0.19}_{-0.46}$ & $16.82^{+0.26}_{-0.46}$ & $19.09^{+0.03}_{-0.02}$ & $17.42^{+0.15}_{-0.67}$ \\
    \hline
    HD J=0 & b\,km/s &$0.54^{+0.20}_{-0.04}$ & $0.57^{+0.47}_{-0.07}$ & $0.77^{+0.09}_{-0.21}$ & $0.85^{+0.27}_{-0.26}$ & $1.1^{+1.0}_{-0.6}$ & $0.72^{+1.48}_{-0.22}$ \\
            & $\log N$ & $\lesssim 15.9$ & $\lesssim 16.1$ & $\lesssim 15.8$ & $\lesssim 15.9$ &  $\lesssim 16.0$ & $\lesssim 16.2$ \\
    \hline   
    \end{tabular}
    \begin{tablenotes}
    \item Doppler parameters H$_2$ $\rm J=4$ in 1, 2, 3, 4 and 6 components were tied to H$_2$ $\rm J=3$, respectively.
    \end{tablenotes}
\end{table*}

\begin{figure*}
    \centering
    \includegraphics[width=\linewidth]{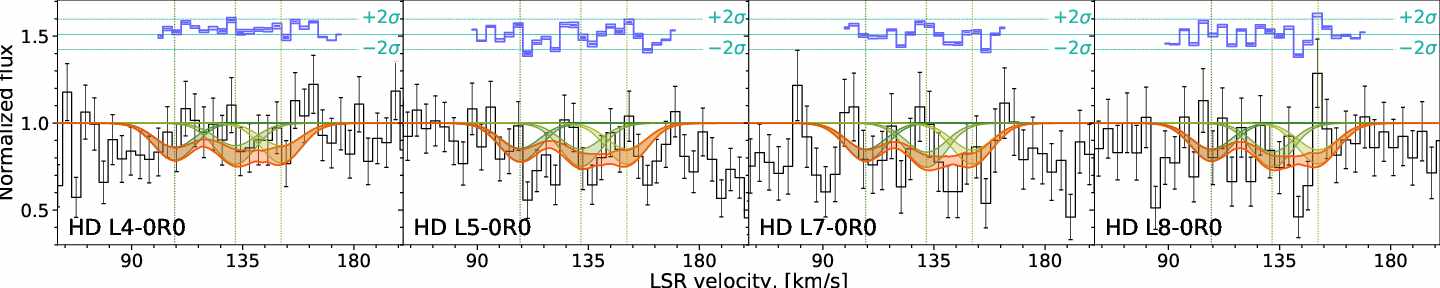}
    \caption{Fit to HD absorption lines towards AV 6 in SMC. Lines are the same as for \ref{fig:lines_HD_Sk67_2}.
    }
    \label{fig:lines_HD_AV6}
\end{figure*}

\begin{figure*}
    \centering
    \includegraphics[width=\linewidth]{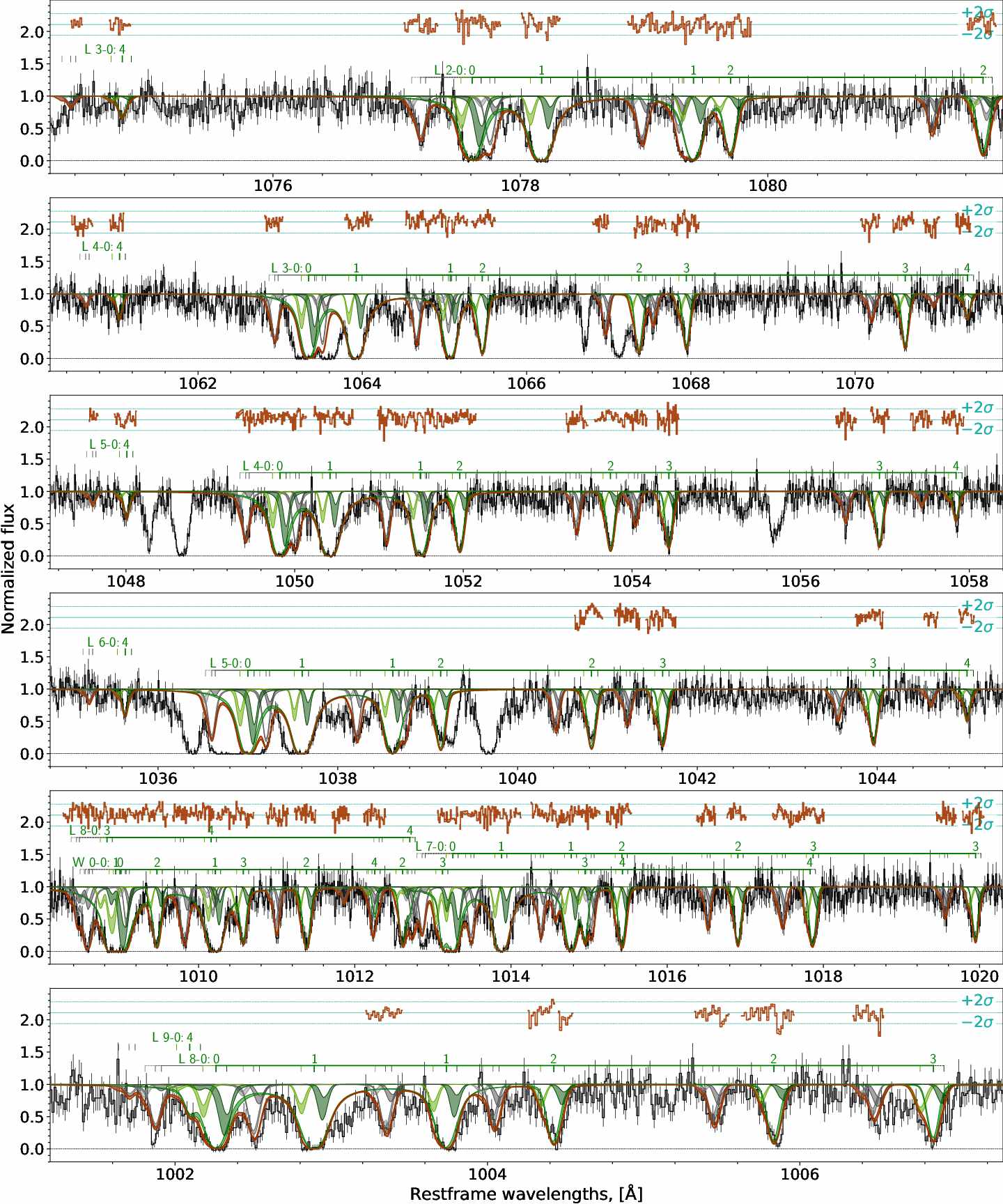}
    \caption{Fit to H2 absorption lines towards AV 6 in SMC. Lines are the same as for \ref{fig:lines_H2_Sk67_2}.
    }
    \label{fig:lines_H2_AV6}
\end{figure*}

\clearpage
\begin{table*}
    \caption{Fit results of H$_2$ lines towards AV 14}
    \label{tab:AV14}
    \begin{tabular}{cccccccc}
    \hline
    \hline
    species & comp & 1 & 2 & 3 &  4& 5 & 6\\
            & z &  $0.00004100(^{+118}_{-28})$ & $0.0001037(^{+23}_{-24})$ & $0.0003052(^{+32}_{-17})$ & $0.0003724(^{+27}_{-7})$ & $0.0004377(^{+20}_{-23})$ & $0.0004959(^{+29}_{-15})$ \\
    \hline 
     ${\rm H_2\, J=0}$ & b\,km/s &  $1.12^{+0.23}_{-0.50}$ & $3.2^{+0.5}_{-1.6}$ & $1.35^{+0.48}_{-0.19}$ & $1.3^{+0.5}_{-0.3}$ & $0.71^{+0.41}_{-0.21}$ & $1.94^{+0.29}_{-0.44}$\\
                       & $\log N$ &  $17.621^{+0.023}_{-0.041}$ & $10.09^{+0.64}_{-0.09}$ $^{\ast}$ & $15.6^{+0.5}_{-0.4}$ & $17.61^{+0.12}_{-0.11}$ &  $18.08^{+0.12}_{-0.11}$ & $16.58^{+0.26}_{-0.99}$\\
    ${\rm H_2\, J=1}$ & b\,km/s & $1.62^{+0.09}_{-0.51}$ & $3.9^{+1.5}_{-0.6}$ & $1.8^{+0.5}_{-0.4}$ & $1.81^{+0.30}_{-0.43}$ & $1.20^{+0.22}_{-0.51}$ & $2.0^{+0.3}_{-0.4}$\\
                      & $\log N$ & $17.773^{+0.042}_{-0.015}$ & $11.8^{+0.8}_{-0.3}$ $^{\ast}$ & $14.78^{+0.36}_{-0.17}$ &  $17.98^{+0.10}_{-0.05}$ & $17.968^{+0.029}_{-0.098}$ & $14.97^{+0.33}_{-0.14}$\\
    ${\rm H_2\, J=2}$ & b\,km/s & $1.85^{+0.41}_{-0.12}$ & $4.6^{+0.8}_{-0.9}$ & $3.2^{+1.0}_{-0.5}$ & $1.95^{+0.33}_{-0.15}$ & $2.47^{+0.17}_{-0.90}$ & $2.3^{+0.5}_{-0.3}$\\
                      & $\log N$ & $16.98^{+0.10}_{-0.22}$ & $14.00^{+0.09}_{-0.07}$ & $14.05^{+0.09}_{-0.10}$ & $17.12^{+0.07}_{-0.30}$ & $15.07^{+0.44}_{-0.17}$ & $14.33^{+0.07}_{-0.10}$\\
    ${\rm H_2\, J=3}$ & b\,km/s &  $1.75^{+0.39}_{-0.17}$ & $6.3^{+0.9}_{-1.6}$ & $5.1^{+1.1}_{-1.2}$ & $2.53^{+0.51}_{-0.22}$ & $3.7^{+0.4}_{-0.5}$ & $2.8^{+0.5}_{-0.5}$\\
                      & $\log N$ &  $16.39^{+0.17}_{-0.56}$ & $14.117^{+0.082}_{-0.019}$ &$14.33^{+0.03}_{-0.06}$ & $16.53^{+0.16}_{-0.65}$ & $15.05^{+0.15}_{-0.10}$ & $14.44^{+0.06}_{-0.10}$\\
    ${\rm H_2\, J=4}$ & b\,km/s & -- & -- & -- & $5.5^{+1.1}_{-1.1}$ & $4.8^{+1.3}_{-0.7}$ & --  \\
    				  & $\log N$ & $13.81^{+0.12}_{-0.12}$ & $13.05^{+0.57}_{-0.07}$ & $13.64^{+0.09}_{-0.16}$ & $14.15^{+0.07}_{-0.05}$ & $14.17^{+0.04}_{-0.05}$ & $13.90^{+0.08}_{-0.15}$\\
    ${\rm H_2\, J=5}$ & $\log N$ & -- & -- & -- & $13.89^{+0.09}_{-0.11}$ & $14.01^{+0.07}_{-0.12}$ & -- \\
    
    \hline 
         & $\log N_{\rm tot}$ & $18.05^{+0.03}_{-0.02}$ & $14.38^{0.08}_{-0.03}$ & $15.69^{+0.44}_{-0.29}$ & $18.18^{+0.07}_{-0.05}$ & $18.33^{+0.07}_{-0.07}$ & $16.60^{+0.25}_{-0.87}$ \\
    \hline
    HD J=0 & b\,km/s & $0.517^{+0.800}_{-0.017}$ & $2.5^{+1.2}_{-0.9}$ & $1.37^{+0.59}_{-0.28}$ & $1.20^{+0.60}_{-0.30}$ & $0.521^{+0.464}_{-0.021}$ & $1.8^{+0.4}_{-0.5}$ \\
           & $\log N$ & $\lesssim 15.2$ & $\lesssim 14.2$ & $\lesssim 14.4$ & $\lesssim 14.1$ & $\lesssim 15.3$ & $\lesssim 15.4$ \\
    \hline   
    \end{tabular}
    \begin{tablenotes}
    \item Doppler parameters H$_2$ $\rm J=4$ in 1, 2, 3 and 6 components and $\rm J = 5$ in 4 and 5 components were tied to H$_2$ $\rm J=3$ and $\rm J=4$, respectively.
    \item $\ast$ Upper limits?
    \end{tablenotes}
\end{table*}

\begin{figure*}
    \centering
    \includegraphics[width=\linewidth]{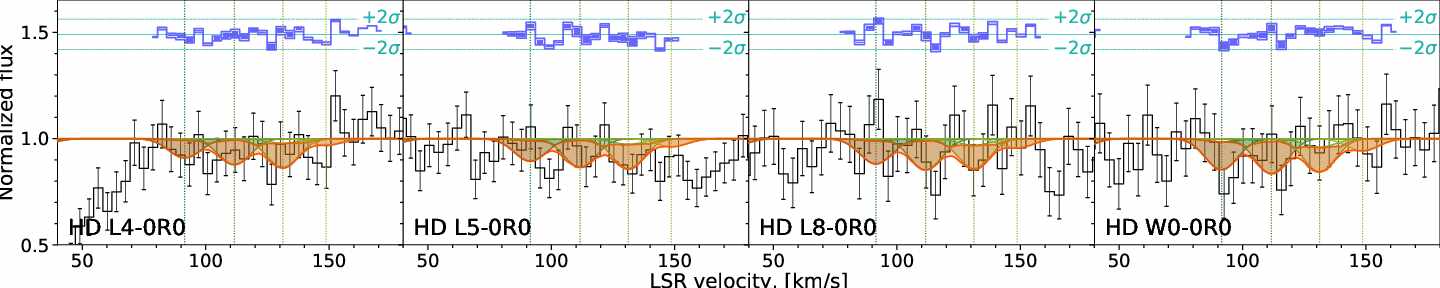}
    \caption{Fit to HD absorption lines towards AV 14 in SMC. Lines are the same as for \ref{fig:lines_HD_Sk67_2}.
    }
    \label{fig:lines_HD_AV14}
\end{figure*}

\begin{figure*}
    \centering
    \includegraphics[width=\linewidth]{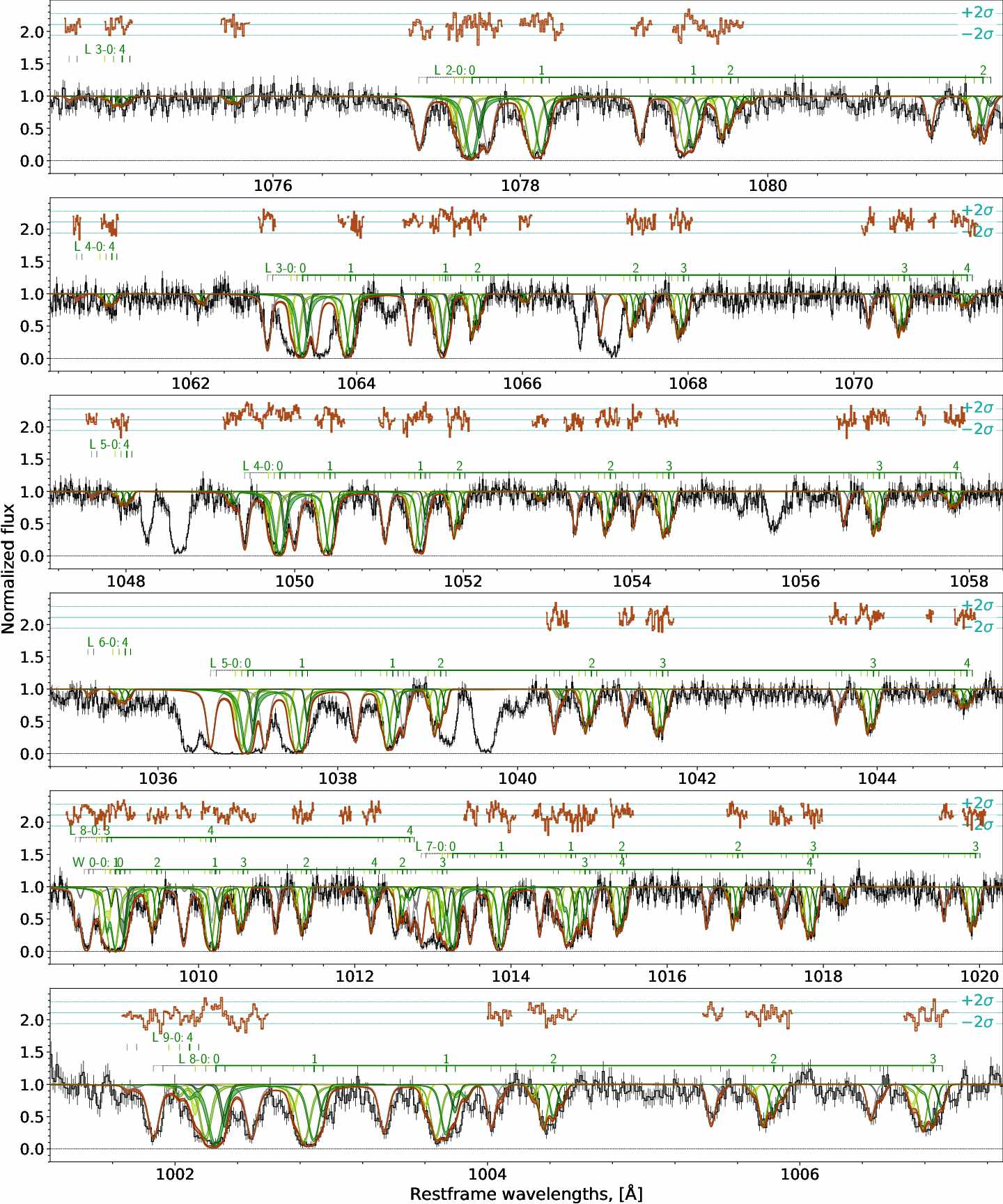}
    \caption{Fit to H2 absorption lines towards AV 14 in SMC. Lines are the same as for \ref{fig:lines_H2_Sk67_2}.
    }
    \label{fig:lines_H2_AV14}
\end{figure*}

\begin{table*}
    \caption{Fit results of H$_2$ lines towards AV 15}
    \label{tab:AV15}
    \begin{tabular}{ccccc}
    \hline
    \hline
    species & comp & 1 & 2 & 3 \\
            & z & $0.0000450(^{+4}_{-6})$ & $0.0004001(^{+16}_{-28})$ & $0.0004592(^{+9}_{-6})$ \\
    \hline 
     ${\rm H_2\, J=0}$ & b\,km/s &  $1.59^{+0.55}_{-0.32}$ & $0.509^{+0.042}_{-0.009}$ & $3.1^{+0.3}_{-1.0}$ \\
                       & $\log N$ & $17.11^{+0.05}_{-0.08}$ &  $16.08^{+0.07}_{-0.25}$ & $18.152^{+0.010}_{-0.031}$ \\
    ${\rm H_2\, J=1}$ & b\,km/s & $2.23^{+0.12}_{-0.08}$ & $0.514^{+0.075}_{-0.014}$ & $3.22^{+0.19}_{-0.14}$ \\
                      & $\log N$ & $17.647^{+0.025}_{-0.026}$ & $16.55^{+0.17}_{-0.16}$ & $18.135^{+0.016}_{-0.039}$ \\
    ${\rm H_2\, J=2}$ & b\,km/s & $2.34^{+0.05}_{-0.12}$ & $0.71^{+0.18}_{-0.12}$ & $4.8^{+0.6}_{-0.4}$ \\
                      & $\log N$ & $16.40^{+0.14}_{-0.07}$ & $15.30^{+0.23}_{-0.42}$ & $15.39^{+0.25}_{-0.11}$ \\
    ${\rm H_2\, J=3}$ & b\,km/s & $2.26^{+0.12}_{-0.08}$ & $3.03^{+0.15}_{-0.23}$ & $6.1^{+0.3}_{-0.4}$ \\
                      & $\log N$ & $15.15^{+0.24}_{-0.28}$ & $14.24^{+0.07}_{-0.05}$ & $15.41^{+0.05}_{-0.08}$ \\
    ${\rm H_2\, J=4}$ & b\,km/s & -- & -- & $9.8^{+1.6}_{-0.8}$ \\
    				  & $\log N$ & $13.66^{+0.18}_{-0.16}$ & $13.10^{+0.29}_{-0.41}$ & $14.497^{+0.029}_{-0.027}$\\
    ${\rm H_2\, J=5}$ & b\,km/s & -- & -- & $14.2^{+1.8}_{-2.2}$ \\
    				  & $\log N$ & -- & $12.8^{+0.5}_{-0.5}$ & $14.246^{+0.035}_{-0.031}$\\
    
    \hline 
         & $\log N_{\rm tot}$ & $17.78^{+0.02}_{-0.03}$ & $19.69^{+0.13}_{-0.12}$ & $18.45^{+0.01}_{-0.03}$ \\
     \hline
     HD J=0 & b\,km/s & $1.6^{+0.6}_{-0.5}$ & $0.5018^{+0.0484}_{-0.0018}$ & $0.50^{+3.69}_{-0.00}$ \\
            & $\log N$ & $\lesssim 14.5$ & $\lesssim 15.0$ & $\lesssim 15.3$ \\ 
    \hline   
    \end{tabular}
    \begin{tablenotes}
    \item Doppler parameters H$_2$ $\rm J=4$ and $\rm J = 5$ in 2 component were tied to H$_2$ $\rm J=3$.
    \end{tablenotes}
\end{table*}

\begin{figure*}
    \centering
    \includegraphics[width=\linewidth]{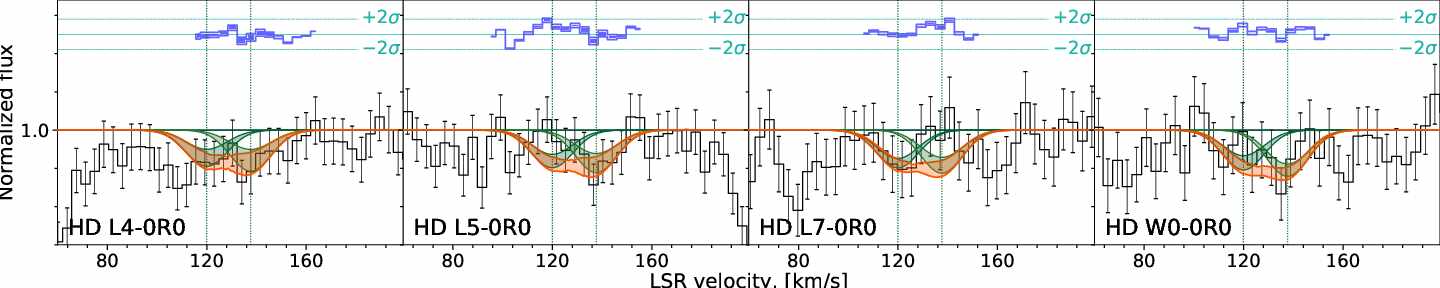}
    \caption{Fit to HD absorption lines towards AV 15 in SMC. Lines are the same as for \ref{fig:lines_HD_Sk67_2}.
    }
    \label{fig:lines_HD_AV15}
\end{figure*}

\begin{figure*}
    \centering
    \includegraphics[width=\linewidth]{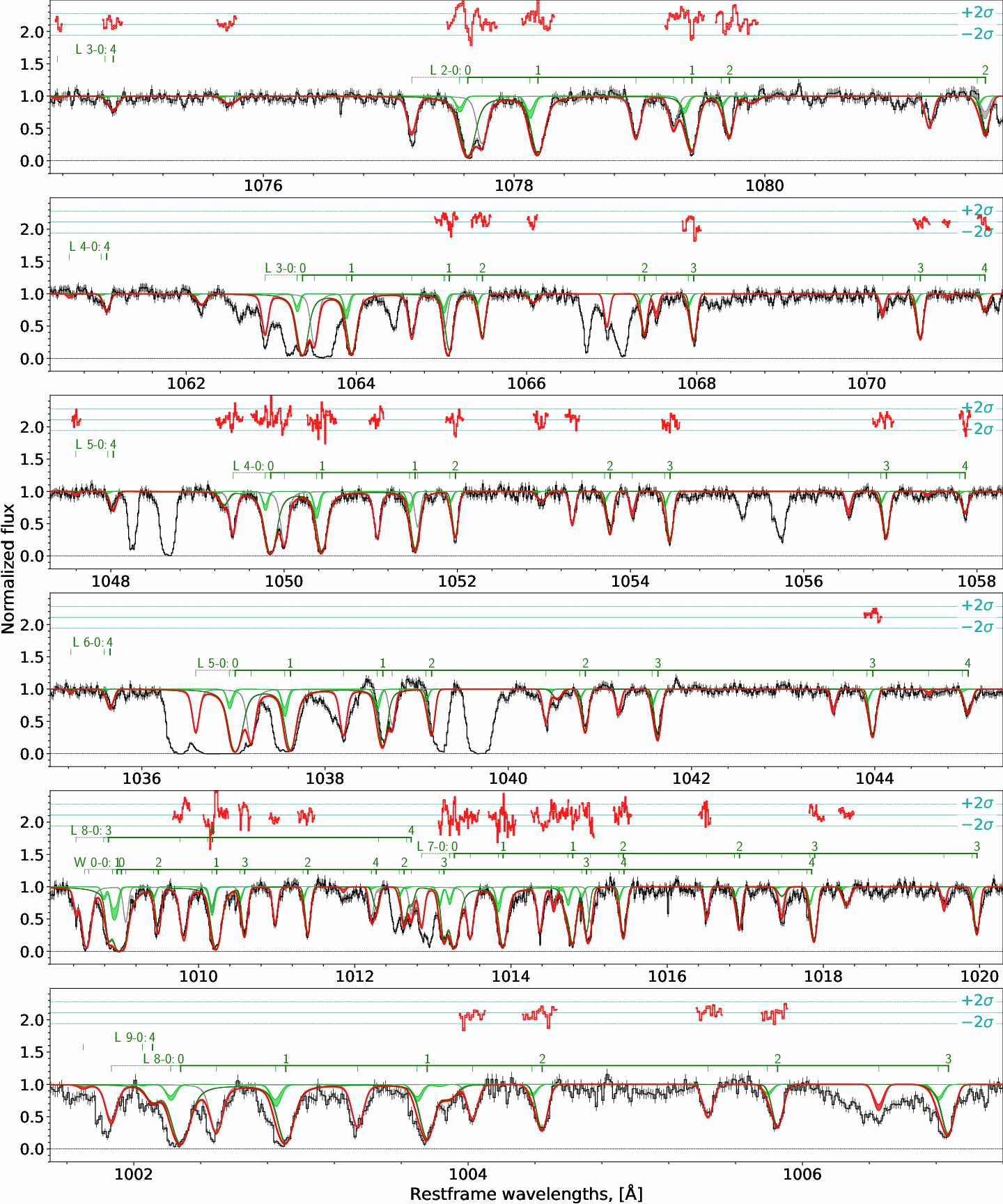}
    \caption{Fit to H2 absorption lines towards AV 15 in SMC. Lines are the same as for \ref{fig:lines_H2_Sk67_2}.
    }
    \label{fig:lines_H2_AV15}
\end{figure*}

\begin{table*}
    \caption{Fit results of H$_2$ lines towards AV 16}
    \label{tab:AV16}
    \begin{tabular}{ccccc}
    \hline
    \hline
    species & comp & 1 & 2 & 3 \\
            & z & $0.0000915(^{+23}_{-24})$ & $0.0004268(^{+17}_{-23})$ & $0.0005072(^{+18}_{-20})$ \\
    \hline 
     ${\rm H_2\, J=0}$ & b\,km/s & $0.84^{+0.82}_{-0.29}$ &  $1.2^{+0.6}_{-0.7}$ & $0.9^{+0.9}_{-0.3}$ \\
                       & $\log N$ & $18.61^{+0.13}_{-0.14}$ & $19.80^{+0.17}_{-0.14}$ & $19.86^{+0.13}_{-0.15}$ \\
    ${\rm H_2\, J=1}$ & b\,km/s & $1.8^{+1.3}_{-0.3}$ & $2.1^{+1.2}_{-0.5}$ & $2.3^{+0.7}_{-0.8}$ \\
                      & $\log N$ & $17.76^{+0.14}_{-0.37}$ & $19.711^{+0.022}_{-0.042}$ & $19.40^{+0.04}_{-0.06}$ \\
    ${\rm H_2\, J=2}$ & b\,km/s & $3.32^{+0.16}_{-1.39}$ & $3.3^{+0.5}_{-0.6}$ & $3.0^{+0.3}_{-0.5}$ \\
                      & $\log N$ & $15.2^{+1.0}_{-1.0}$ & $17.29^{+0.31}_{-0.35}$ &  $16.17^{+0.30}_{-0.48}$\\
    ${\rm H_2\, J=3}$ & b\,km/s & $3.57^{+0.26}_{-0.72}$ & $4.51^{+0.19}_{-0.33}$ & $4.64^{+0.25}_{-0.29}$ \\
                      & $\log N$ & $15.28^{+0.40}_{-0.27}$ & $15.71^{+0.26}_{-0.11}$ & $15.72^{+0.24}_{-0.09}$ \\
    ${\rm H_2\, J=4}$ & $\log N$ &$14.31^{+0.11}_{-0.17}$ & $14.90^{+0.08}_{-0.06}$ & $14.64^{+0.05}_{-0.04}$ \\
    
    \hline 
         & $\log N_{\rm tot}$ & $18.67^{+0.12}_{-0.13}$ & $20.06^{+0.10}_{-0.07}$ & $19.99^{+0.10}_{-0.11}$ \\
    \hline
    HD J=0 & b\,km/s &$0.54^{+0.92}_{-0.04}$ & $1.4^{+0.5}_{-0.5}$ & $0.80^{+0.60}_{-0.30}$ \\
            & $\log N$ & $\lesssim 15.8$ & $\lesssim 16.7$ & $\lesssim 16.4$ \\
    \hline   
    \end{tabular}
    \begin{tablenotes}
    \item Doppler parameters H$_2$ $\rm J=4$ were tied to H$_2$ $\rm J=3$.
    \end{tablenotes}
\end{table*}

\begin{figure*}
    \centering
    \includegraphics[width=\linewidth]{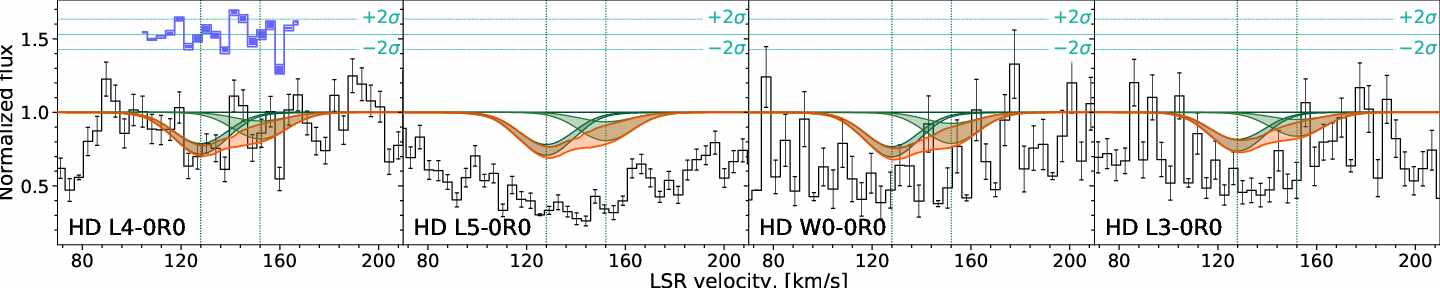}
    \caption{Fit to HD absorption lines towards AV 16 in SMC. Lines are the same as for \ref{fig:lines_HD_Sk67_2}.
    }
    \label{fig:lines_HD_AV16}
\end{figure*}

\begin{figure*}
    \centering
    \includegraphics[width=\linewidth]{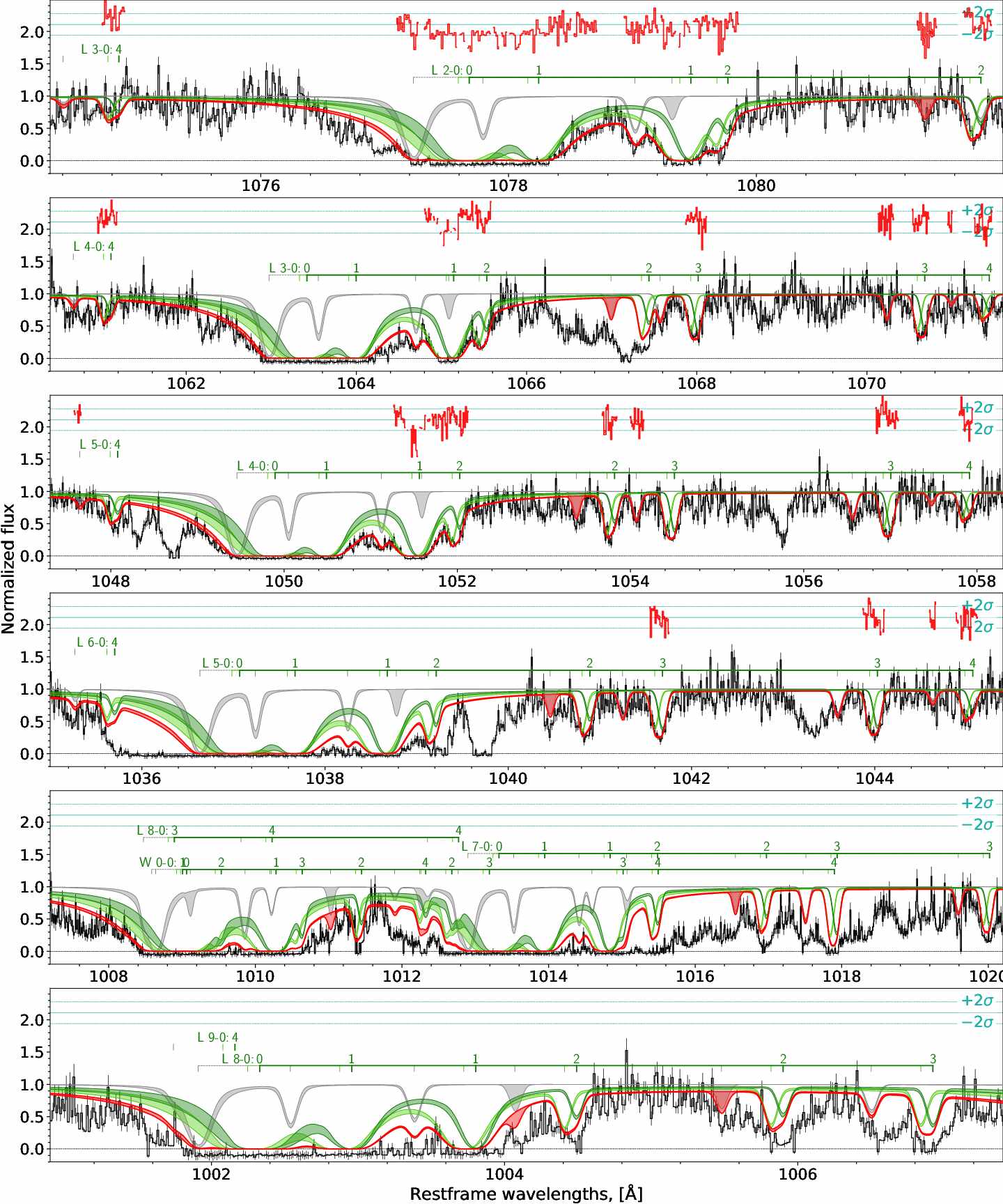}
    \caption{Fit to H2 absorption lines towards AV 16 in SMC. Lines are the same as for \ref{fig:lines_H2_Sk67_2}.
    }
    \label{fig:lines_H2_AV16}
\end{figure*}

\begin{table*}
    \caption{Fit results of H$_2$ lines towards AV 18}
    \label{tab:AV18}
    \begin{tabular}{ccccccc}
    \hline
    \hline
    species & comp & 1 & 2 & 3 & 4 & 5 \\
            & z & $0.0000328(^{+34}_{-22})$ & $0.000345(^{+6}_{-3})$ & $0.0004358(^{+13}_{-16})$ & $0.0005143(^{+5}_{-26})$ & $0.0005870(^{+15}_{-46})$  \\
    \hline 
     ${\rm H_2\, J=0}$ & b\,km/s & $2.3^{+0.7}_{-0.8}$ & $1.2^{+2.0}_{-0.3}$ & $1.1^{+1.1}_{-0.4}$ & $2.5^{+1.3}_{-0.8}$ & $1.8^{+0.6}_{-0.6}$\\
                       & $\log N$ & $18.84^{+0.07}_{-0.07}$ & $15.80^{+0.23}_{-0.28}$ &$20.177^{+0.048}_{-0.021}$ & $18.74^{+0.29}_{-0.41}$ & $19.45^{+0.07}_{-0.13}$\\
    ${\rm H_2\, J=1}$ & b\,km/s & $3.1^{+0.3}_{-0.9}$ &  $2.9^{+1.8}_{-1.1}$ &$2.2^{+0.6}_{-0.9}$ & $4.0^{+0.5}_{-1.1}$ & $2.1^{+0.5}_{-0.4}$ \\
                      & $\log N$ & $17.08^{+0.79}_{-0.14}$ & $15.56^{+0.42}_{-0.18}$ & $20.141^{+0.026}_{-0.038}$ & $18.99^{+0.19}_{-0.41}$ & $19.59^{+0.06}_{-0.05}$\\
    ${\rm H_2\, J=2}$ & b\,km/s & $4.8^{+0.4}_{-1.3}$ &  $6.4^{+1.9}_{-1.2}$ & $2.22^{+0.82}_{-0.27}$ &$4.0^{+0.8}_{-0.4}$ & $2.34^{+0.24}_{-0.32}$ \\
                      & $\log N$ & $15.52^{+0.27}_{-0.50}$ &$14.78^{+0.08}_{-0.12}$  & $18.25^{+0.06}_{-0.12}$ & $17.10^{+0.29}_{-0.60}$ &  $17.12^{+0.19}_{-0.19}$\\
    ${\rm H_2\, J=3}$ & b\,km/s & $5.4^{+0.9}_{-0.4}$ & $15.58^{+3.63}_{-6.40}$ &$2.8^{+0.6}_{-0.4}$ &  $5.1^{+0.4}_{-0.7}$ & $12.3^{+2.0}_{-1.2}$\\
                      & $\log N$ & $14.88^{+0.12}_{-0.05}$ & $14.52^{+0.11}_{-0.08}$ & $17.62^{+0.16}_{-0.27}$ & $15.65^{+0.22}_{-0.20}$ & $14.859^{+0.031}_{-0.041}$ \\
    ${\rm H_2\, J=4}$ & b\,km/s & -- & -- & $3.1^{+0.7}_{-0.4}$ & $5.1^{+0.9}_{-0.5}$ & -- \\
                        & $\log N$ & $13.4^{+0.4}_{-0.6}$ & $14.01^{+0.13}_{-0.16}$ & $15.4^{+0.3}_{-0.3}$ & $15.08^{+0.09}_{-0.11}$ & $10.5^{+1.3}_{-0.5}$ \\
    
    \hline 
         & $\log N_{\rm tot}$ & $18.84^{+0.08}_{-0.07}$ & $16.04^{+0.22}_{-0.15}$ & $20.46^{+0.03}_{-0.02}$ & $19.19^{0.17}_{-0.26}$ & $19.83^{0.05}_{-0.06}$  \\
    \hline
    HD J=0 & b\,km/s & $2.2^{+0.8}_{-0.8}$ & $1.4^{+2.5}_{-0.4}$ & $1.4^{+14.5}_{-0.9}$ &$2.0^{+1.0}_{-0.6}$ & $1.5^{+0.6}_{-0.5}$ \\
            & $\log N$ & $\lesssim 15.9$ & $\lesssim 15.6$ & $14.63^{+0.28}_{-0.20}$ & $14.59^{+1.02}_{-0.17}$ & $\lesssim 16.2$ \\
    \hline   
    \end{tabular}
    \begin{tablenotes}
    \item Doppler parameters H$_2$ $\rm J=4$ in components 1, 2 and 5 were tied to H$_2$ $\rm J=3$.
    \end{tablenotes}
\end{table*}

\begin{figure*}
    \centering
    \includegraphics[width=\linewidth]{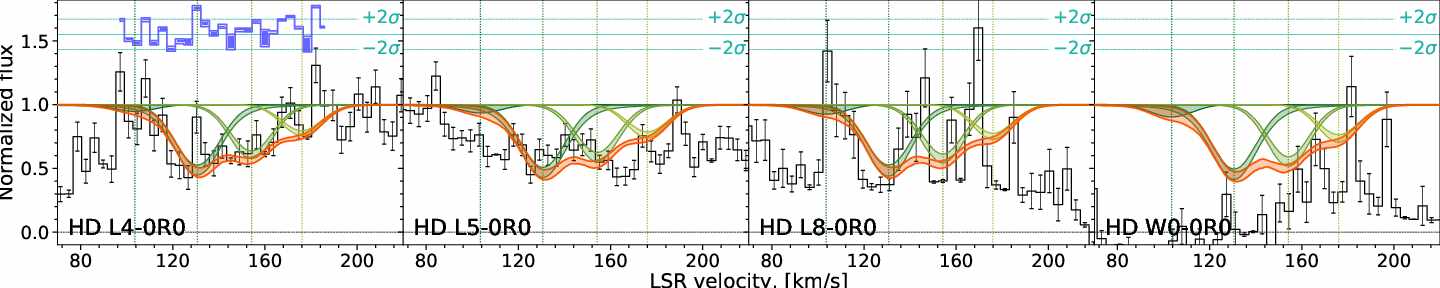}
    \caption{Fit to HD absorption lines towards AV 18 in SMC. Lines are the same as for \ref{fig:lines_HD_Sk67_2}.
    }
    \label{fig:lines_HD_AV18}
\end{figure*}

\begin{figure*}
    \centering
    \includegraphics[width=\linewidth]{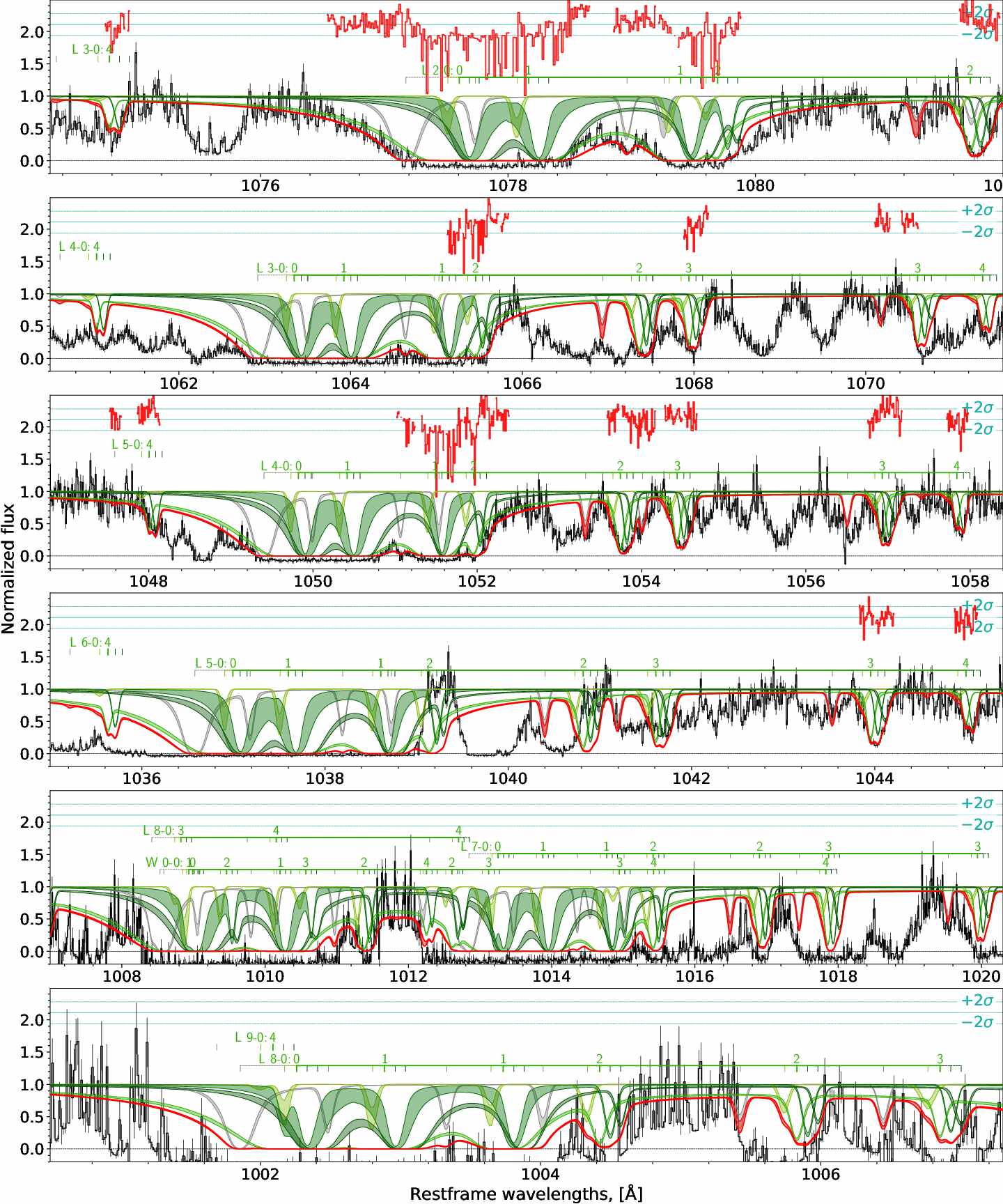}
    \caption{Fit to H2 absorption lines towards AV 18 in SMC. Lines are the same as for \ref{fig:lines_H2_Sk67_2}.
    }
    \label{fig:lines_H2_AV18}
\end{figure*}

\begin{table*}
    \caption{Fit results of H$_2$ lines towards AV 22}
    \label{tab:AV22}
    \begin{tabular}{cccccc}
    \hline
    \hline
    species & comp & 1 & 2 & 3 & 4 \\
            & z & $0.0000570(^{+27}_{-13})$ & $0.000375(^{+3}_{-4})$ & $0.000445(^{+4}_{-3})$ & $0.0005130(^{+12}_{-45})$ \\
    \hline 
     ${\rm H_2\, J=0}$ & b\,km/s &$0.61^{+0.16}_{-0.09}$ & $0.65^{+0.38}_{-0.14}$ &$1.8^{+0.4}_{-0.3}$ & $0.68^{+0.33}_{-0.16}$ \\
                       & $\log N$ &$17.38^{+0.07}_{-0.06}$ & $17.83^{+0.07}_{-0.11}$  & $16.87^{+0.35}_{-0.08}$ &  $17.12^{+0.22}_{-0.06}$\\
    ${\rm H_2\, J=1}$ & b\,km/s & $0.72^{+0.26}_{-0.15}$ & $1.9^{+0.5}_{-0.7}$ & $2.1^{+0.3}_{-0.4}$ & $0.71^{+0.48}_{-0.13}$\\
                      & $\log N$ & $17.77^{+0.07}_{-0.07}$ & $17.66^{+0.11}_{-0.12}$ & $17.33^{+0.21}_{-0.04}$ & $17.20^{+0.16}_{-0.16}$\\
    ${\rm H_2\, J=2}$ & b\,km/s & $1.5^{+0.6}_{-0.6}$ & $2.2^{+0.4}_{-0.7}$ & $3.05^{+0.18}_{-0.11}$ & $1.31^{+0.52}_{-0.27}$\\
                      & $\log N$ & $16.21^{+0.25}_{-0.36}$ & $14.8^{+0.8}_{-0.4}$ &$16.42^{+0.22}_{-0.10}$ & $16.07^{+0.20}_{-1.10}$\\
    ${\rm H_2\, J=3}$ & b\,km/s & $6.8^{+2.1}_{-1.5}$ & $2.67^{+0.14}_{-0.89}$ & $2.92^{+0.19}_{-0.14}$ &$2.64^{+0.13}_{-0.29}$ \\
                      & $\log N$ & $14.43^{+0.12}_{-0.14}$ & $14.74^{+0.39}_{-0.17}$  &$15.58^{+0.39}_{-0.17}$ &$15.04^{+0.20}_{-0.22}$ \\
    ${\rm H_2\, J=4}$ & b\,km/s & $10.4^{+1.9}_{-2.3}$ & $13.82^{+0.29}_{-0.08}$ & $14.01^{+0.11}_{-0.28}$ & $14.17^{+0.12}_{-0.14}$\\
                      & $\log N$ & $13.93^{+0.16}_{-0.17}$ & $13.91^{+0.20}_{-0.21}$ &$14.00^{+0.26}_{-0.08}$ &  $14.40^{+0.20}_{-0.07}$\\
    
    \hline 
         & $\log N_{\rm tot}$ & $17.93^{+0.05}_{-0.05}$ & $18.06^{+0.06}_{-0.08}$ & $17.50^{+0.18}_{-0.03}$ & $17.48^{+0.14}_{-0.08}$ \\
    \hline
    HD J=0 & b\,km/s & $0.60^{+0.15}_{-0.10}$ & $0.68^{+0.27}_{-0.18}$ & $1.8^{+0.4}_{-0.4}$ & $0.72^{+0.21}_{-0.22}$ \\
           & $\log N$ & $\lesssim 16.1$ & $\lesssim 16.0$ & $\lesssim 15.8$ & $16.2$ \\     
                  
    \hline   
    \end{tabular}
    \begin{tablenotes}
    \item Doppler parameters H$_2$ $\rm J=4$ were tied to H$_2$ $\rm J=3$.
    \end{tablenotes}
\end{table*}

\begin{figure*}
    \centering
    \includegraphics[width=\linewidth]{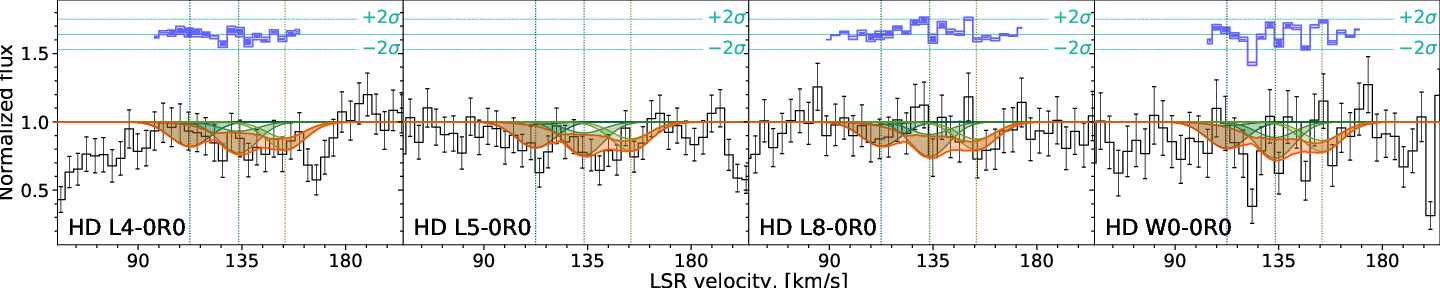}
    \caption{Fit to HD absorption lines towards AV 22 in SMC. Lines are the same as for \ref{fig:lines_HD_Sk67_2}.
    }
    \label{fig:lines_HD_AV22}
\end{figure*}

\begin{figure*}
    \centering
    \includegraphics[width=\linewidth]{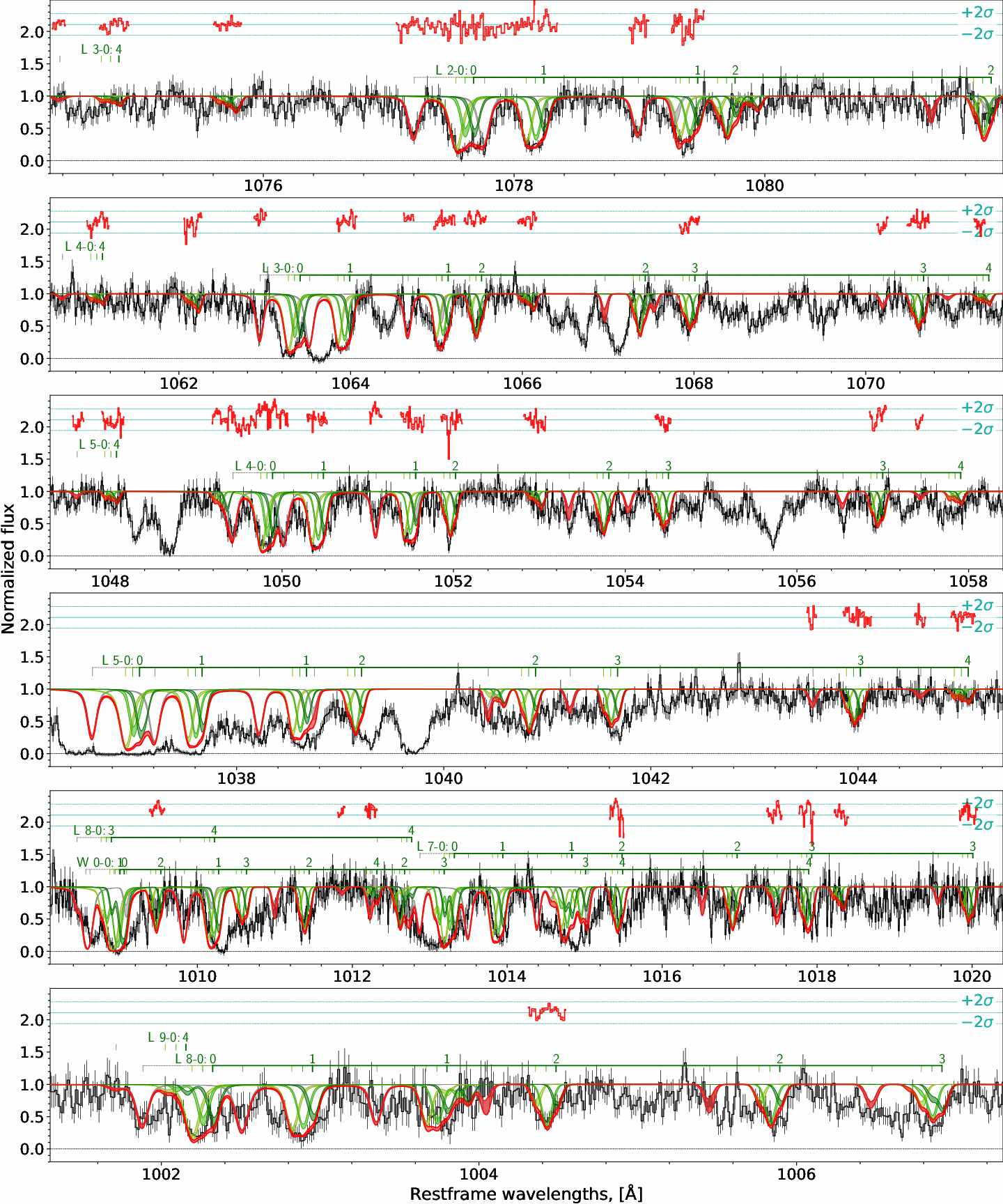}
    \caption{Fit to H2 absorption lines towards AV 22 in SMC. Lines are the same as for \ref{fig:lines_H2_Sk67_2}.
    }
    \label{fig:lines_H2_AV22}
\end{figure*}

\begin{table*}
    \caption{Fit results of H$_2$ lines towards AV 26}
    \label{tab:AV26}
    \begin{tabular}{ccccccc}
    \hline
    \hline
    species & comp & 1 & 2 & 3 & 4 & 5 \\
            & z & $0.0000587(^{+24}_{-62})$ & $0.000117(^{+48}_{-9})$ & $0.000398(^{+7}_{-9})$ &  $0.0004420(^{+50}_{-30})$ & $0.000507(^{+8}_{-10})$ \\
    \hline 
     ${\rm H_2\, J=0}$ & b\,km/s & $4.0^{+1.9}_{-1.9}$ &$1.0^{+0.4}_{-0.5}$ & $0.70^{+0.42}_{-0.20}$ & $2.02^{+0.22}_{-1.39}$ & $0.78^{+0.84}_{-0.28}$\\
                       & $\log N$ & $17.0^{+0.6}_{-0.8}$ &$16.5^{+1.5}_{-1.4}$ & $18.94^{+0.21}_{-0.44}$ &$20.610^{+0.022}_{-0.007}$ & $18.6^{+0.4}_{-0.8}$\\
    ${\rm H_2\, J=1}$ & b\,km/s & $4.2^{+2.8}_{-0.9}$ &$1.4^{+2.0}_{-0.5}$ & $0.93^{+0.88}_{-0.22}$ &$2.8^{+2.8}_{-1.2}$ & $1.6^{+1.0}_{-0.8}$\\
                      & $\log N$ & $15.76^{+1.18}_{-0.17}$ & $14.7^{+1.1}_{-0.5}$ & $18.81^{+0.40}_{-0.18}$ & $19.992^{+0.031}_{-0.037}$ &$16.8^{+0.8}_{-0.6}$ \\
    ${\rm H_2\, J=2}$ & b\,km/s &  $7.0^{+2.3}_{-1.8}$ &$4.0^{+1.7}_{-2.2}$ & $2.3^{+0.7}_{-1.0}$ & $5.9^{+1.4}_{-2.6}$ &$4.1^{+2.2}_{-0.7}$ \\
                      & $\log N$ & $15.03^{+0.14}_{-0.11}$ & $13.4^{+0.9}_{-0.6}$ & $17.23^{+0.33}_{-0.30}$ & $15.32^{+0.36}_{-0.24}$ & $14.75^{+0.13}_{-0.17}$\\
    ${\rm H_2\, J=3}$ & b\,km/s &$9.0^{+2.0}_{-2.2}$ &$4.6^{+2.5}_{-2.1}$ & $2.2^{+1.5}_{-0.5}$ & $11.2^{+1.3}_{-5.0}$ & $4.4^{+2.3}_{-0.6}$ \\
                      & $\log N$ & $14.74^{+0.05}_{-0.07}$ & $13.0^{+1.1}_{-0.5}$ & $15.81^{+0.27}_{-1.01}$ & $15.18^{+0.10}_{-0.15}$ & $14.33^{+0.35}_{-0.19}$\\
    ${\rm H_2\, J=4}$ & b\,km/s & -- & $9.82^{+9.83}_{-4.23}$ & -- & -- & --\\
    				  & $\log N$ & $13.98^{+0.13}_{-0.21}$ &$12.3^{+0.9}_{-1.1}$ & $14.60^{+0.26}_{-0.34}$ & $14.55^{+0.08}_{-0.14}$ & $13.4^{+0.5}_{-0.7}$\\
    \hline 
         & $\log N_{\rm tot}$ & $17.01^{+0.58}_{-0.66}$ & $16.53^{+1.21}_{-1.28}$ & $19.19^{+0.24}_{-0.22}$ & $20.70^{+0.02}_{-0.01}$ & $18.66^{+0.40}_{-0.77}$ \\
    \hline
    HD J=0 & b\,km/s &  $4.5^{+1.9}_{-2.2}$ & $1.00^{+0.40}_{-0.30}$ & $0.80^{+0.46}_{-0.19}$ & $2.03^{+0.16}_{-1.02}$ & $0.72^{+0.79}_{-0.22}$ \\
            & $\log N$ & $\lesssim 16.0$ & $\lesssim 16.4$ & $\lesssim 16.7$ &  $14.4^{+1.7}_{-0.4}$ & $\lesssim 16.4$ \\
    \hline   
    \end{tabular}
    \begin{tablenotes}
    \item Doppler parameters H$_2$ $\rm J=4$ in 1, 3 and 4 components were tied to H$_2$ $\rm J=3$.
    \end{tablenotes}
\end{table*}

\begin{figure*}
    \centering
    \includegraphics[width=\linewidth]{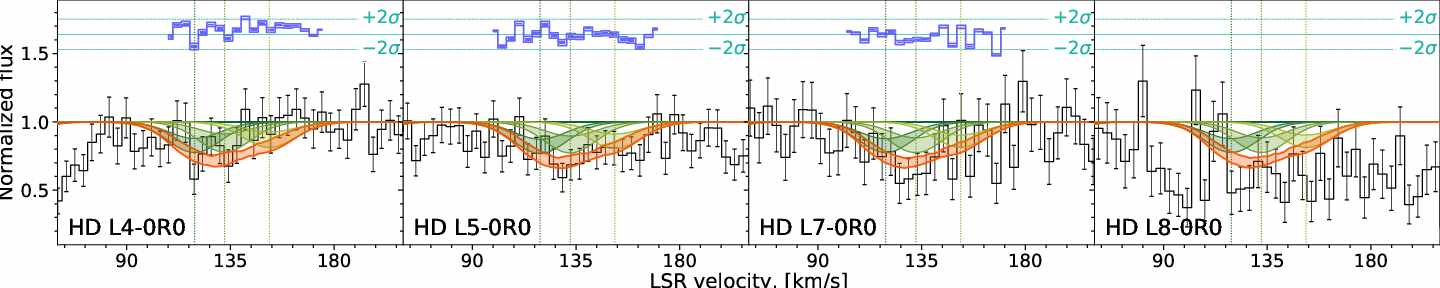}
    \caption{Fit to HD absorption lines towards AV 26 in SMC. Lines are the same as for \ref{fig:lines_HD_Sk67_2}.
    }
    \label{fig:lines_HD_AV26}
\end{figure*}

\begin{figure*}
    \centering
    \includegraphics[width=\linewidth]{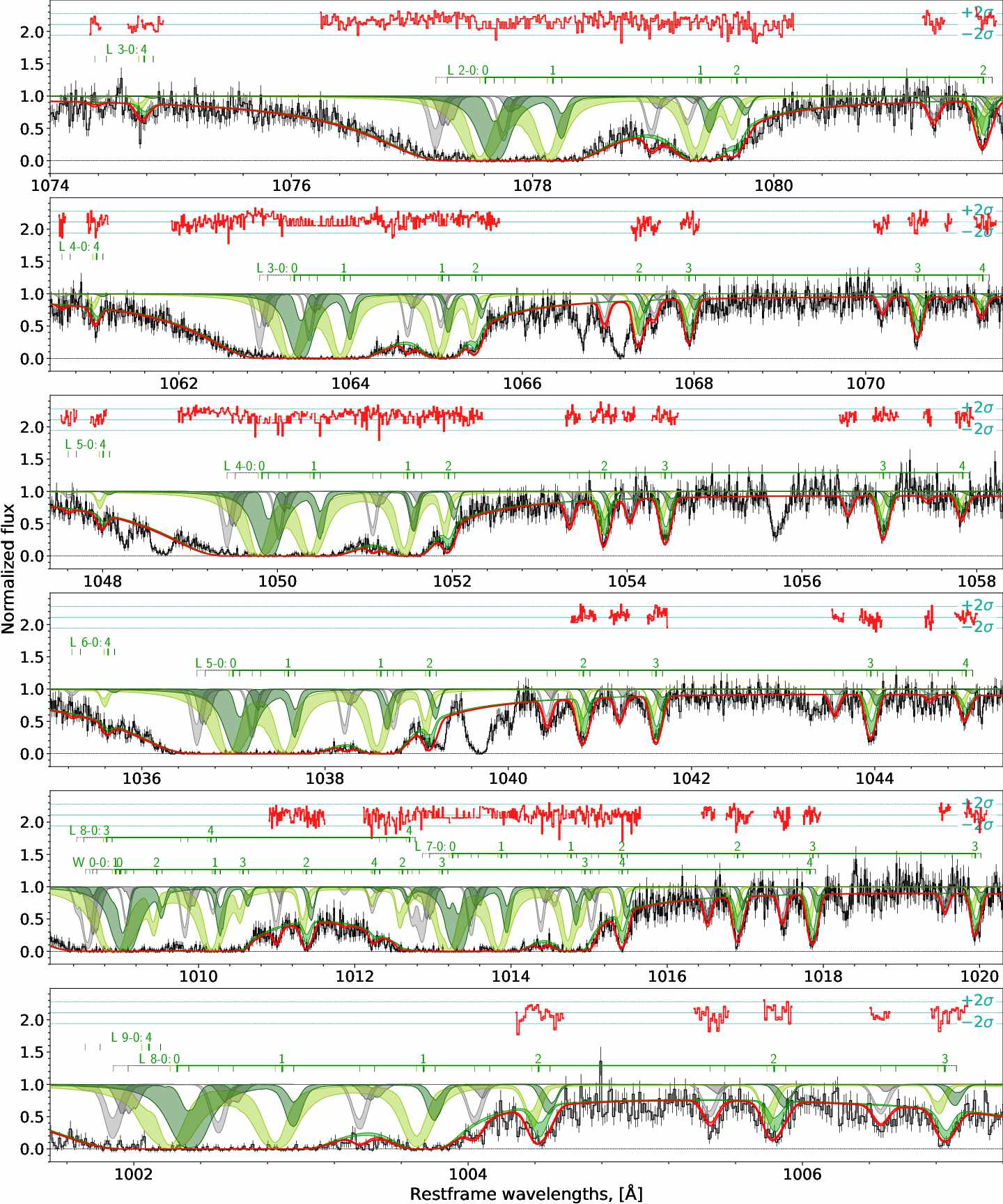}
    \caption{Fit to H2 absorption lines towards AV 26 in SMC. Lines are the same as for \ref{fig:lines_H2_Sk67_2}.
    }
    \label{fig:lines_H2_AV26}
\end{figure*}

\begin{table*}
    \caption{Fit results of H$_2$ lines towards AV 39a}
    \label{tab:AV39a}
    \begin{tabular}{cccccc}
    \hline
    \hline
    species & comp & 1 & 2 & 3 & 4  \\
            & z & $0.0000054(^{+10}_{-4})$ & $0.0003459(^{+28}_{-20})$ & $0.0003907(^{+33}_{-24})$ & $0.0004330(^{+23}_{-53})$ \\
    \hline 
     ${\rm H_2\, J=0}$ & b\,km/s &$3.4^{+0.4}_{-0.9}$ & $0.67^{+0.69}_{-0.11}$ & $1.1^{+0.6}_{-0.4}$ &  $0.68^{+0.17}_{-0.17}$\\
                       & $\log N$ &  $17.91^{+0.05}_{-0.06}$ & $16.75^{+0.22}_{-0.42}$ & $18.19^{+0.08}_{-0.06}$ & $16.95^{+0.49}_{-0.27}$\\
    ${\rm H_2\, J=1}$ & b\,km/s &$3.62^{+0.17}_{-0.41}$ & $1.2^{+0.7}_{-0.4}$ & $1.5^{+0.7}_{-0.3}$ & $0.88^{+0.22}_{-0.25}$ \\
                      & $\log N$ & $17.95^{+0.05}_{-0.04}$ & $17.03^{+0.28}_{-0.40}$ & $18.46^{+0.05}_{-0.06}$ &$16.3^{+0.5}_{-0.5}$ \\
    ${\rm H_2\, J=2}$ & b\,km/s & $3.36^{+0.23}_{-0.29}$ &$1.9^{+0.5}_{-0.8}$ &$2.46^{+0.57}_{-0.25}$ & $1.9^{+0.8}_{-0.5}$ \\
                      & $\log N$ & $16.58^{+0.24}_{-0.17}$ & $14.65^{+1.21}_{-0.24}$ & $17.01^{+0.14}_{-0.77}$ & $14.58^{+0.46}_{-0.17}$\\
    ${\rm H_2\, J=3}$ & b\,km/s & $4.06^{+0.23}_{-0.13}$ &$4.56^{+0.28}_{-0.19}$ & $3.07^{+0.13}_{-0.22}$ &$4.45^{+0.20}_{-0.68}$ \\
                      & $\log N$ & $16.20^{+0.15}_{-0.14}$ & $14.76^{+0.06}_{-0.10}$ & $15.88^{+0.28}_{-0.26}$ & $14.56^{+0.12}_{-0.12}$\\
    ${\rm H_2\, J=4}$ & $\log N$ & $14.31^{+0.06}_{-0.20}$ & $13.79^{+0.15}_{-0.34}$ & $14.78^{+0.12}_{-0.14}$ & $13.58^{+0.25}_{-0.41}$\\
    ${\rm H_2\, J=4}$ & $\log N$ & -- & $13.75^{+0.25}_{-0.38}$ &$14.33^{+0.17}_{-0.13}$ & $13.7^{+0.4}_{-0.6}$ \\
    \hline 
         & $\log N_{\rm tot}$ & $18.25^{+0.03}_{-0.03}$ & $17.22^{+0.21}_{-0.26}$ & $18.66^{+0.04}_{-0.04}$ & $17.03^{+0.44}_{-0.22}$   \\
     \hline
     HD J=0 & b\,km/s & $3.2^{+0.6}_{-1.1}$ & $0.71^{+0.47}_{-0.21}$ & $1.1^{+0.7}_{-0.4}$ & $0.70^{+0.21}_{-0.10}$ \\
            & $\log N$ & $\lesssim 16.4$ & $\lesssim 16.1$ & $\lesssim 15.5$ & $\lesssim 16.3$ \\
    \hline   
    \end{tabular}
    \begin{tablenotes}
    \item Doppler parameters H$_2$ $\rm J=4, 5$  were tied to H$_2$ $\rm J=3$.
    \end{tablenotes}
\end{table*}

\begin{figure*}
    \centering
    \includegraphics[width=\linewidth]{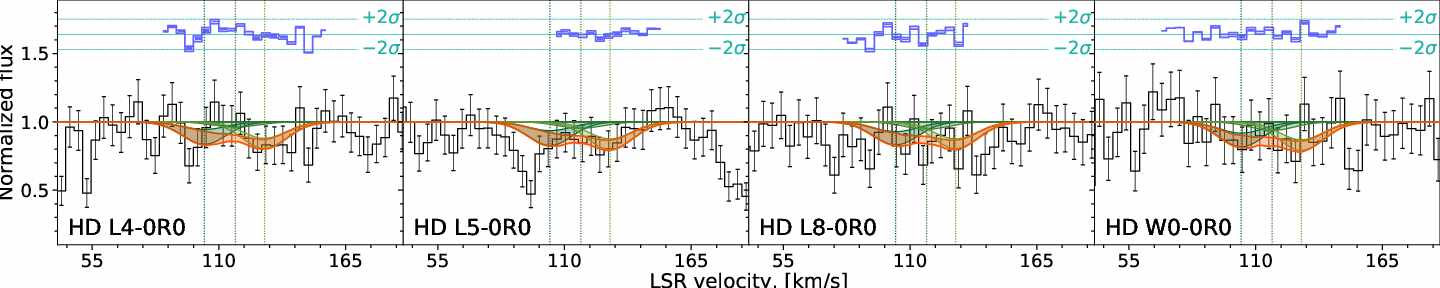}
    \caption{Fit to HD absorption lines towards AV 39a in SMC. Lines are the same as for \ref{fig:lines_HD_Sk67_2}.
    }
    \label{fig:lines_HD_AV39a}
\end{figure*}

\begin{figure*}
    \centering
    \includegraphics[width=\linewidth]{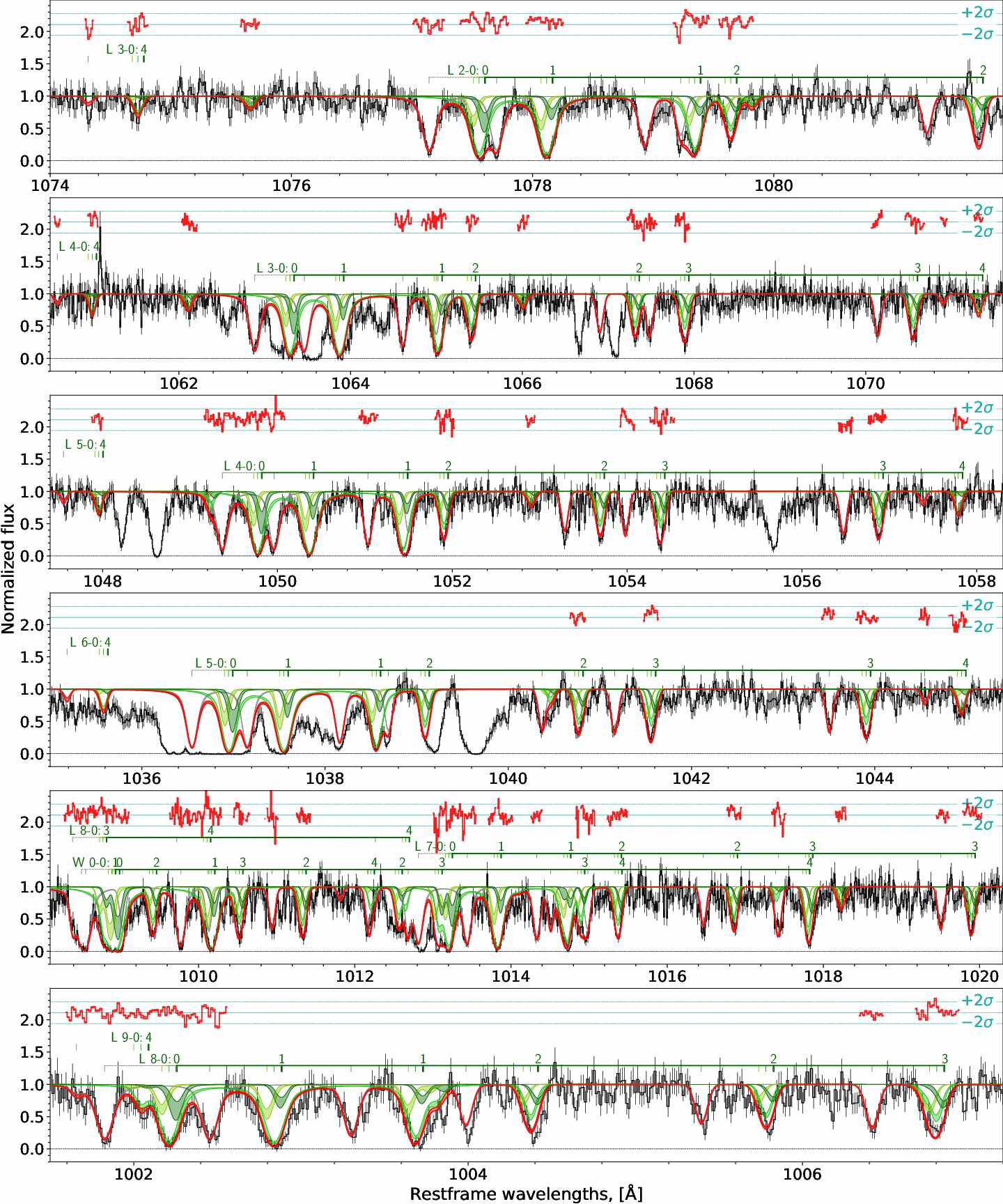}
    \caption{Fit to H2 absorption lines towards AV 39a in SMC. Lines are the same as for \ref{fig:lines_H2_Sk67_2}.
    }
    \label{fig:lines_H2_AV39a}
\end{figure*}

\begin{table*}
    \caption{Fit results of H$_2$ lines towards AV 47}
    \label{tab:AV47}
    \begin{tabular}{ccccc}
    \hline
    \hline
    species & comp & 1 & 2 & 3  \\
            & z & $0.00004510(^{+30}_{-45})$ & $0.0003887(^{+10}_{-29})$ &  $0.0004306(^{+12}_{-13})$\\
    \hline 
     ${\rm H_2\, J=0}$ & b\,km/s &$3.22^{+0.26}_{-0.06}$ & $0.75^{+0.20}_{-0.24}$ &$0.96^{+0.79}_{-0.25}$ \\
                       & $\log N$ & $17.23^{+0.07}_{-0.04}$ & $17.32^{+0.13}_{-0.05}$ & $17.976^{+0.040}_{-0.028}$\\
    ${\rm H_2\, J=1}$ & b\,km/s &$3.33^{+0.10}_{-0.11}$ & $1.05^{+0.33}_{-0.28}$ & $1.8^{+0.5}_{-0.5}$\\
                      & $\log N$ & $17.491^{+0.047}_{-0.028}$ & $17.50^{+0.08}_{-0.13}$ & $18.198^{+0.025}_{-0.035}$\\
    ${\rm H_2\, J=2}$ & b\,km/s & $3.29^{+0.17}_{-0.15}$ &$1.35^{+0.40}_{-0.27}$ & $2.18^{+0.19}_{-0.25}$ \\
                      & $\log N$ &$16.01^{+0.13}_{-0.20}$ & $14.84^{+0.51}_{-0.25}$ & $17.34^{+0.06}_{-0.08}$\\
    ${\rm H_2\, J=3}$ & b\,km/s & $3.48^{+0.16}_{-0.18}$ &$3.04^{+0.15}_{-0.21}$ &$2.37^{+0.19}_{-0.29}$ \\
                      & $\log N$ & $14.749^{+0.049}_{-0.028}$ & $14.36^{+0.09}_{-0.09}$ & $17.03^{+0.12}_{-0.19}$\\
    ${\rm H_2\, J=4}$ & b\,km/s & -- & -- &$2.32^{+0.57}_{-0.27}$ \\
    				  & $\log N$ & $13.83^{+0.08}_{-0.07}$ &$13.38^{+0.20}_{-0.29}$ &  $14.41^{+0.10}_{-0.08}$\\
    \hline 
         & $\log N_{\rm tot}$ & $17.69^{+0.04}_{-0.02}$ & $17.72^{+0.07}_{-0.08}$ & $18.45^{+0.02}_{-0.02}$ \\
    \hline
    HD J=0 & b\,km/s & $3.22^{+0.32}_{-0.07}$ & $0.510^{+0.284}_{-0.010}$ & $1.2^{+0.6}_{-0.5}$ \\
             & $\log N$ &  $\lesssim 13.6$ & $\lesssim 15.4$ & $\lesssim 13.5$ \\
    \hline   
    \end{tabular}
    \begin{tablenotes}
    \item Doppler parameters H$_2$ $\rm J=4$ in 1and 2 components were tied to H$_2$ $\rm J=3$.
    \end{tablenotes}
\end{table*}

\begin{figure*}
    \centering
    \includegraphics[width=\linewidth]{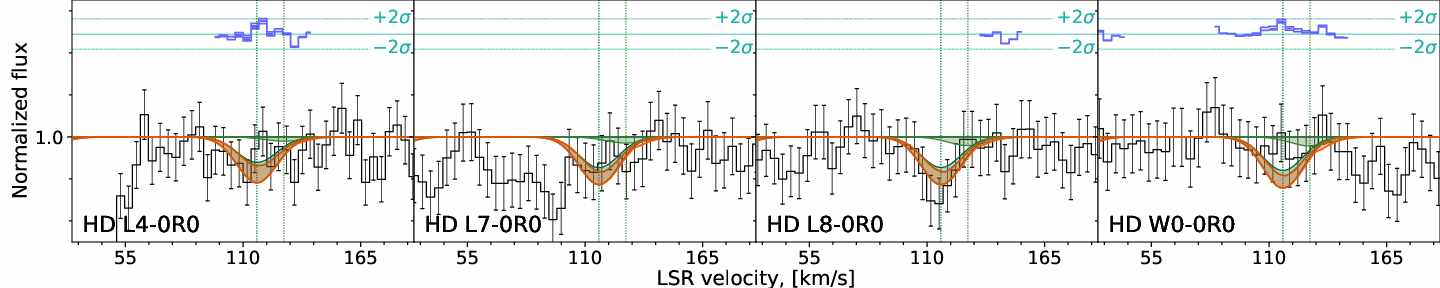}
    \caption{Fit to HD absorption lines towards AV 47 in SMC. Lines are the same as for \ref{fig:lines_HD_Sk67_2}.
    }
    \label{fig:lines_HD_AV47}
\end{figure*}

\begin{figure*}
    \centering
    \includegraphics[width=\linewidth]{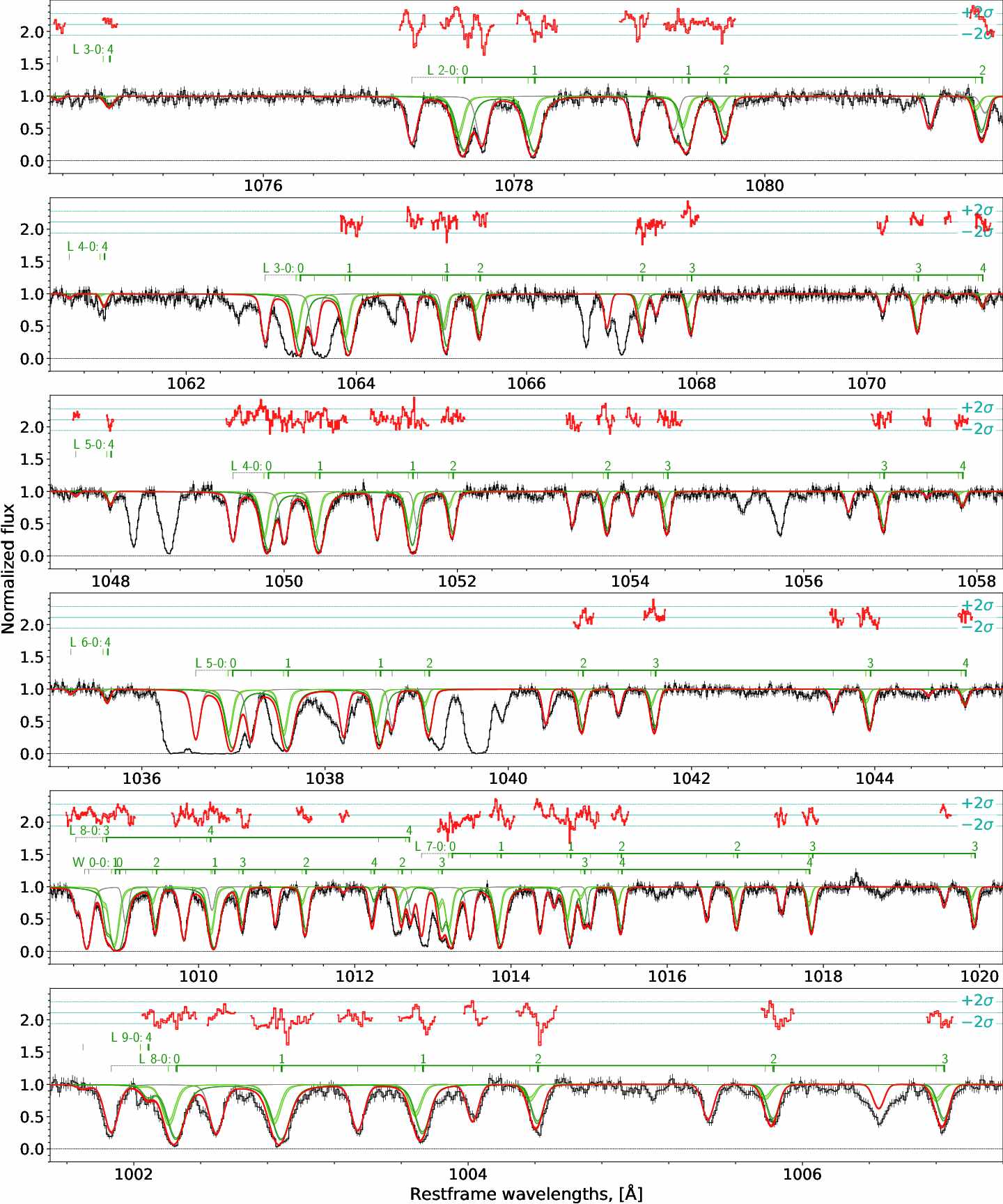}
    \caption{Fit to H2 absorption lines towards AV 47 in SMC. Lines are the same as for \ref{fig:lines_H2_Sk67_2}.
    }
    \label{fig:lines_H2_AV47}
\end{figure*}

\begin{table*}
    \caption{Fit results of H$_2$ lines towards AV 60a}
    \label{tab:AV60a}
    \begin{tabular}{cccccc}
    \hline
    \hline
    species & comp & 1 & 2 & 3 & 4 \\
            & z & $0.0000206(^{+15}_{-14})$ & $0.0004171(^{+31}_{-55})$ & $0.0004789(^{+38}_{-24})$ & $0.0007262(^{+20}_{-20})$ \\
    \hline 
     ${\rm H_2\, J=0}$ & b\,km/s &$4.9^{+0.7}_{-0.5}$ & $1.0^{+1.0}_{-0.4}$ & $2.2^{+0.9}_{-1.4}$ & $1.3^{+0.7}_{-0.6}$\\
                       & $\log N$ & $16.9^{+0.4}_{-0.4}$ & $18.46^{+0.22}_{-0.26}$ & $19.585^{+0.019}_{-0.042}$ & $18.04^{+0.17}_{-0.04}$\\
    ${\rm H_2\, J=1}$ & b\,km/s &$5.2^{+0.6}_{-0.6}$ &  $2.2^{+0.7}_{-0.9}$ & $3.6^{+0.9}_{-1.1}$ & $2.3^{+0.5}_{-0.5}$\\
                      & $\log N$ & $16.3^{+0.4}_{-0.3}$ & $18.46^{+0.12}_{-0.13}$ & $19.13^{+0.04}_{-0.05}$& $17.38^{+0.16}_{-0.27}$\\
    ${\rm H_2\, J=2}$ & b\,km/s & $6.0^{+0.8}_{-0.8}$ & $2.7^{+0.5}_{-0.7}$ & $4.5^{+0.6}_{-0.4}$ & $3.0^{+1.5}_{-0.8}$\\
                      & $\log N$ & $15.09^{+0.11}_{-0.08}$ & $15.9^{+0.5}_{-0.4}$ & $17.03^{+0.17}_{-0.41}$ & $14.74^{+0.36}_{-0.22}$\\
    ${\rm H_2\, J=3}$ & b\,km/s & $6.6^{+1.5}_{-1.1}$ & $4.0^{+1.2}_{-0.7}$ & $4.7^{+0.8}_{-0.5}$ &  $4.4^{+3.3}_{-1.1}$ \\
                      & $\log N$ & $14.60^{+0.09}_{-0.06}$ & $15.14^{+0.35}_{-0.17}$ &$16.41^{+0.20}_{-0.46}$ & $14.43^{+0.11}_{-0.08}$\\
    ${\rm H_2\, J=4}$ & b\,km/s & -- &  $5.2^{+4.1}_{-1.7}$ & $5.6^{+2.1}_{-1.0}$ & -- \\
    				  & $\log N$ & $12.9^{+0.4}_{-0.6}$ & $14.37^{+0.12}_{-0.08}$ & $14.55^{+0.11}_{-0.07}$ & $13.63^{+0.25}_{-0.99}$\\
     
    \hline 
         & $\log N_{\rm tot}$ & $16.97^{+0.34}_{-0.28}$ & $18.76^{+0.13}_{-0.13}$ & $19.72^{+0.02}_{-0.03}$ & $18.12^{+0.15}_{-0.05}$  \\
    \hline
    HD J=0 & b\,km/s &$4.9^{+0.8}_{-0.6}$ & $0.55^{+1.44}_{-0.05}$ & $1.1^{+1.4}_{-0.6}$ & $1.0^{+0.6}_{-0.5}$ \\
           & $\log N$ & $\lesssim 14.1$ & $\lesssim 15.9$ & $\lesssim 16.6$ & $\lesssim 16.1$ \\
    \hline   
    \end{tabular}
    \begin{tablenotes}
    \item Doppler parameters H$_2$ $\rm J=4$ in 1 and 4 components were tied to H$_2$ $\rm J=3$.
    \end{tablenotes}
\end{table*}

\begin{figure*}
    \centering
    \includegraphics[width=\linewidth]{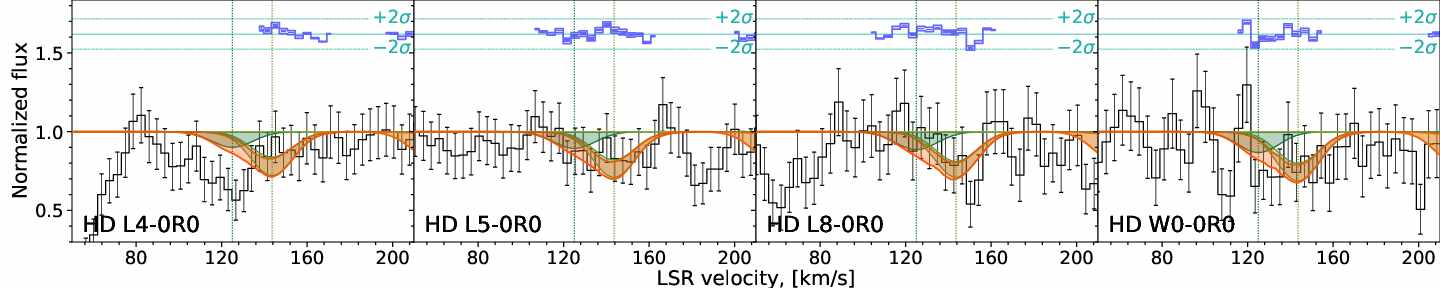}
    \caption{Fit to HD absorption lines towards AV 60a in SMC. Lines are the same as for \ref{fig:lines_HD_Sk67_2}.
    }
    \label{fig:lines_H2_AV60a}
\end{figure*}

\begin{figure*}
    \centering
    \includegraphics[width=\linewidth]{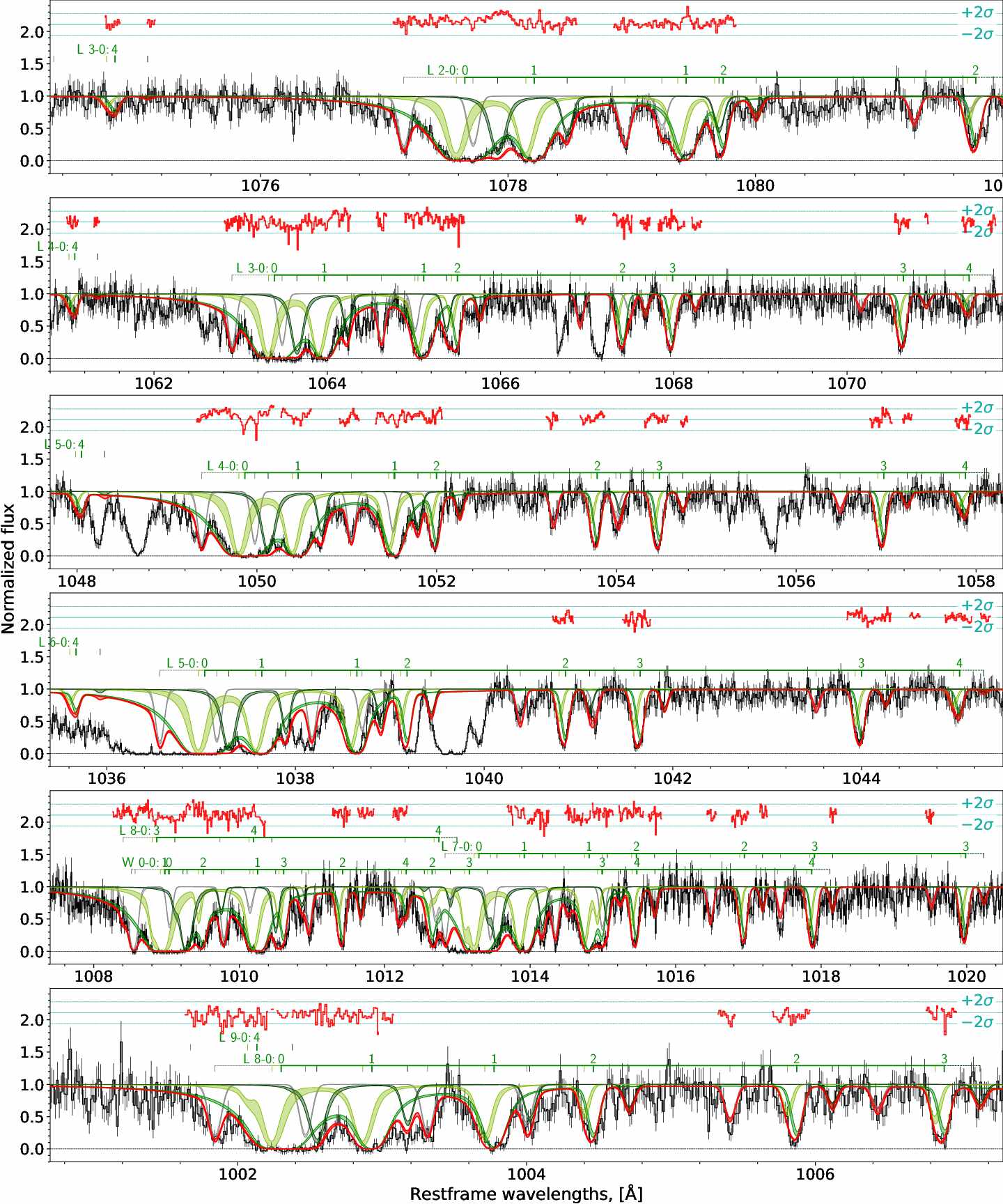}
    \caption{Fit to H2 absorption lines towards AV 60a in SMC. Lines are the same as for \ref{fig:lines_H2_Sk67_2}.
    }
    \label{fig:lines_HD_AV60a}
\end{figure*}

\begin{table*}
    \caption{Fit results of H$_2$ lines towards AV 69}
    \label{tab:AV69}
    \begin{tabular}{ccccc}
    \hline
    \hline
    species & comp & 1 & 2 & 3  \\
            & z & $0.0000469(^{+3}_{-4})$ & $0.0003763(^{+21}_{-24})$ & $0.0004283(^{+12}_{-16})$ \\
    \hline 
     ${\rm H_2\, J=0}$ & b\,km/s & $3.65^{+0.14}_{-0.12}$ & $1.1^{+1.0}_{-0.4}$ & $1.6^{+0.5}_{-0.8}$\\
                       & $\log N$ & $17.971^{+0.015}_{-0.023}$ & $17.96^{+0.07}_{-0.06}$ & $18.454^{+0.020}_{-0.036}$\\
    ${\rm H_2\, J=1}$ & b\,km/s & $3.50^{+0.10}_{-0.15}$ & $1.9^{+0.8}_{-0.4}$ & $2.68^{+0.40}_{-0.26}$\\
                      & $\log N$ & $17.900^{+0.027}_{-0.021}$ & $18.09^{+0.06}_{-0.08}$ & $18.640^{+0.028}_{-0.021}$\\
    ${\rm H_2\, J=2}$ & b\,km/s & $3.62^{+0.16}_{-0.16}$ &$2.25^{+0.37}_{-0.30}$ & $2.92^{+0.27}_{-0.26}$\\
                      & $\log N$ & $15.99^{+0.13}_{-0.16}$ & $15.94^{+0.23}_{-0.21}$ & $17.52^{+0.05}_{-0.05}$\\
    ${\rm H_2\, J=3}$ & b\,km/s & $3.79^{+0.20}_{-0.19}$ & $3.76^{+0.22}_{-0.67}$ &$2.71^{+0.28}_{-0.24}$ \\
                      & $\log N$ &$15.19^{+0.10}_{-0.07}$ &$14.92^{+0.13}_{-0.08}$ &$17.48^{+0.07}_{-0.06}$ \\
    ${\rm H_2\, J=4}$ & b\,km/s & -- &$4.31^{+0.26}_{-0.51}$ & $3.6^{+0.7}_{-0.5}$\\
    				  & $\log N$ &$13.84^{+0.08}_{-0.06}$ &  $13.92^{+0.08}_{-0.10}$ & $14.62^{+0.06}_{-0.05}$\\
    ${\rm H_2\, J=4}$ & $\log N$ & -- & $13.82^{+0.10}_{-0.04}$ & $14.19^{+0.05}_{-0.06}$  \\
    \hline 
         & $\log N_{\rm tot}$ & $18.24^{+0.01}_{-0.02}$ & $18.34^{+0.05}_{-0.05}$ & $18.89^{+0.02}_{-0.02}$  \\
     \hline
     HD J=0 & b\,km/s & $3.62^{+0.19}_{-0.12}$ & $0.56^{+1.41}_{-0.06}$ & $0.530^{+1.110}_{-0.030}$ \\
            & $\log N$ & $\lesssim 13.6$ & $\lesssim 15.1$ & $\lesssim 15.3$ \\
    \hline   
    \end{tabular}
    \begin{tablenotes}
    \item Doppler parameters H$_2$ $\rm J=4$ in 1 component and $\rm J=5$ in 2 and 3 components were tied to H$_2$ $\rm J=3$ and $\rm J=4$, respectively.
    \end{tablenotes}
\end{table*}

\begin{figure*}
    \centering
    \includegraphics[width=\linewidth]{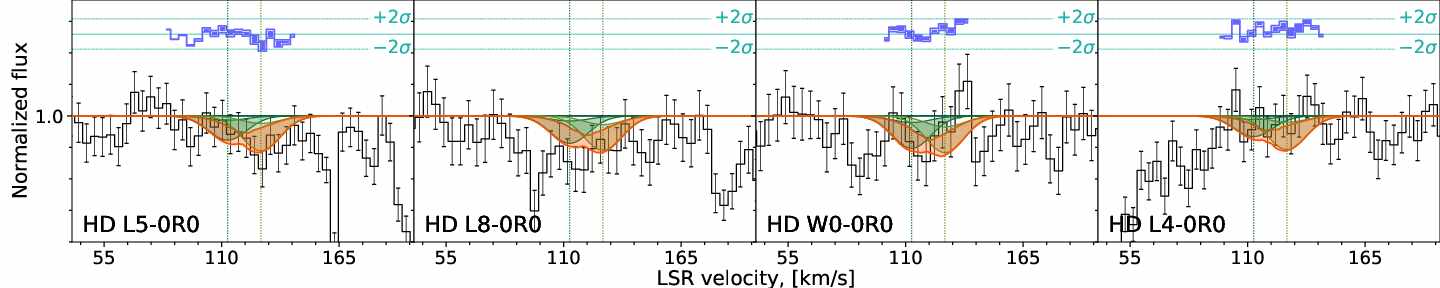}
    \caption{Fit to HD absorption lines towards AV 69 in SMC. Lines are the same as for \ref{fig:lines_HD_Sk67_2}.
    }
    \label{fig:lines_HD_AV69}
\end{figure*}

\begin{figure*}
    \centering
    \includegraphics[width=\linewidth]{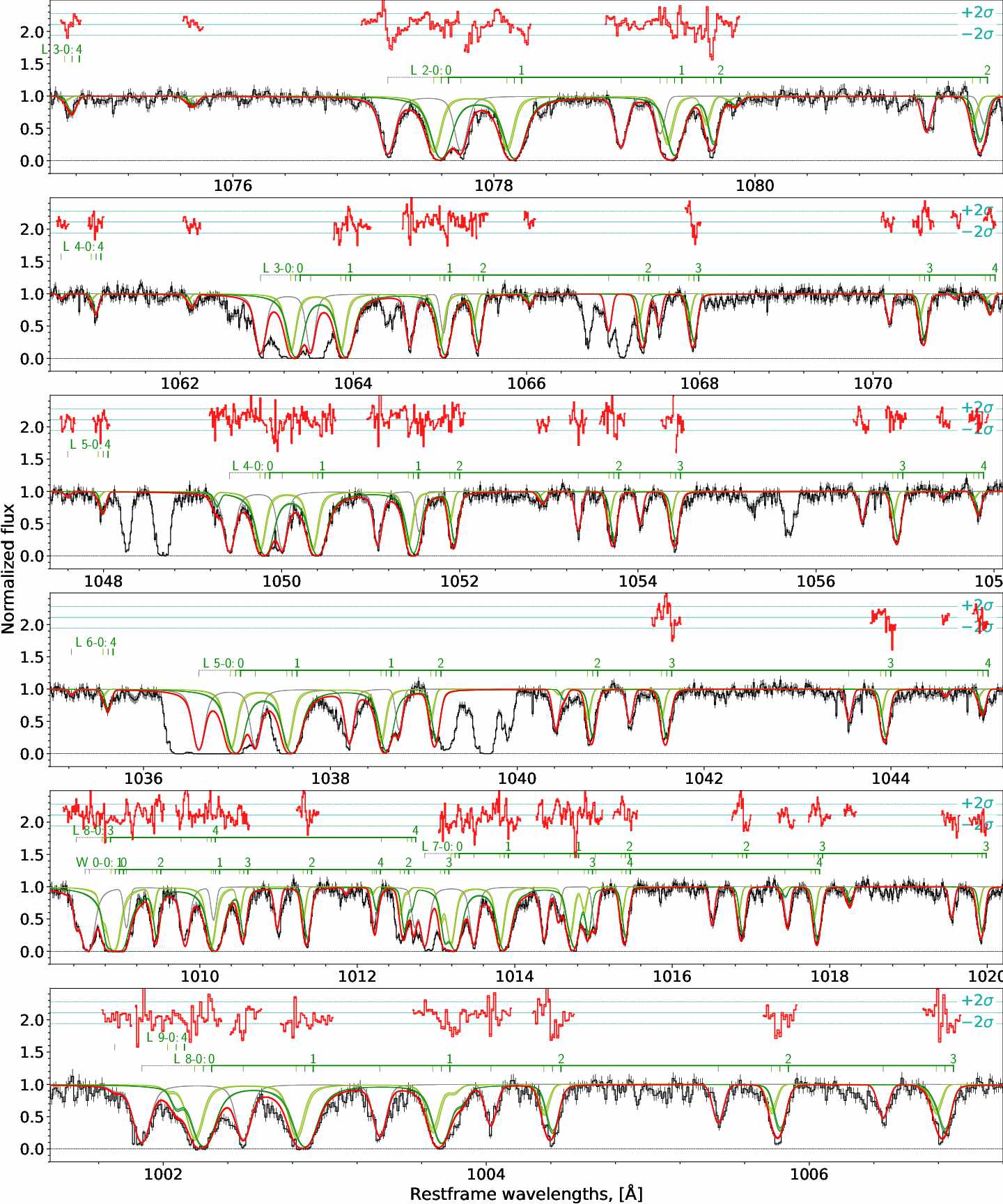}
    \caption{Fit to H2 absorption lines towards AV 69 in SMC. Lines are the same as for \ref{fig:lines_H2_Sk67_2}.
    }
    \label{fig:lines_H2_AV69}
\end{figure*}

\begin{table*}
    \caption{Fit results of H$_2$ lines towards AV 75}
    \label{tab:AV75}
    \begin{tabular}{ccccc}
    \hline
    \hline
    species & comp & 1 & 2 & 3  \\
            & z & $0.0000425(^{+4}_{-4})$ & $0.0003781(^{+16}_{-8})$ & $0.0004225(^{+53}_{-13})$ \\
    \hline 
     ${\rm H_2\, J=0}$ & b\,km/s & $1.5^{+0.5}_{-0.7}$ & $0.78^{+0.39}_{-0.27}$ & $3.7^{+0.4}_{-0.6}$\\
                       & $\log N$ &$17.521^{+0.025}_{-0.012}$ & $18.501^{+0.015}_{-0.018}$ & $16.29^{+0.10}_{-0.47}$ \\
    ${\rm H_2\, J=1}$ & b\,km/s & $2.54^{+0.13}_{-0.08}$ & $1.3^{+0.6}_{-0.4}$ & $3.6^{+0.5}_{-0.4}$\\
                      & $\log N$ &$18.119^{+0.015}_{-0.011}$ & $18.569^{+0.014}_{-0.030}$ & $17.71^{+0.12}_{-0.09}$ \\
    ${\rm H_2\, J=2}$ & b\,km/s &$2.52^{+0.09}_{-0.11}$ & $2.1^{+0.4}_{-0.4}$ & $3.7^{+0.3}_{-0.5}$\\
                      & $\log N$ & $17.23^{+0.04}_{-0.06}$ & $17.52^{+0.04}_{-0.06}$ & $16.21^{+0.21}_{-0.20}$ \\
    ${\rm H_2\, J=3}$ & b\,km/s &$2.59^{+0.06}_{-0.14}$ & $2.33^{+0.18}_{-0.55}$ & $4.6^{+0.6}_{-0.7}$ \\
                      & $\log N$ & $16.56^{+0.08}_{-0.13}$ & $17.32^{+0.10}_{-0.05}$ & $15.73^{+0.23}_{-0.15}$ \\
    ${\rm H_2\, J=4}$ & b\,km/s & -- & $2.33^{+0.30}_{-0.65}$ & $5.2^{+1.3}_{-0.8}$\\
    				  & $\log N$ & $13.80^{+0.06}_{-0.11}$ & $15.12^{+0.15}_{-0.44}$ & $14.68^{+0.04}_{-0.05}$\\
    ${\rm H_2\, J=4}$ & b\,km/s & -- & $3.1^{+1.2}_{-0.8}$ &$11.1^{+2.5}_{-3.1}$ \\
                      & $\log N$ & -- & $14.38^{+0.12}_{-0.12}$ & $14.53^{+0.05}_{-0.06}$\\
    \hline 
         & $\log N_{\rm tot}$ & $18.27^{+0.01}_{-0.01}$ & $18.87^{+0.01}_{-0.02}$ & $17.74^{+0.11}_{-0.08}$  \\
    \hline
    HD J=0 & b\,km/s & $0.531^{+0.976}_{-0.031}$ & $0.521^{+0.530}_{-0.021}$ & $3.5^{+0.7}_{-0.5}$ \\
           & $\log N$ & $\lesssim 15.2$ & $\lesssim 15.3$ & $\lesssim 13.7$ \\
    \hline   
    \end{tabular}
    \begin{tablenotes}
    \item Doppler parameter H$_2$ $\rm J=4$ in 1 component was tied to H$_2$ $\rm J=3$.
    \end{tablenotes}
\end{table*}

\begin{figure*}
    \centering
    \includegraphics[width=\linewidth]{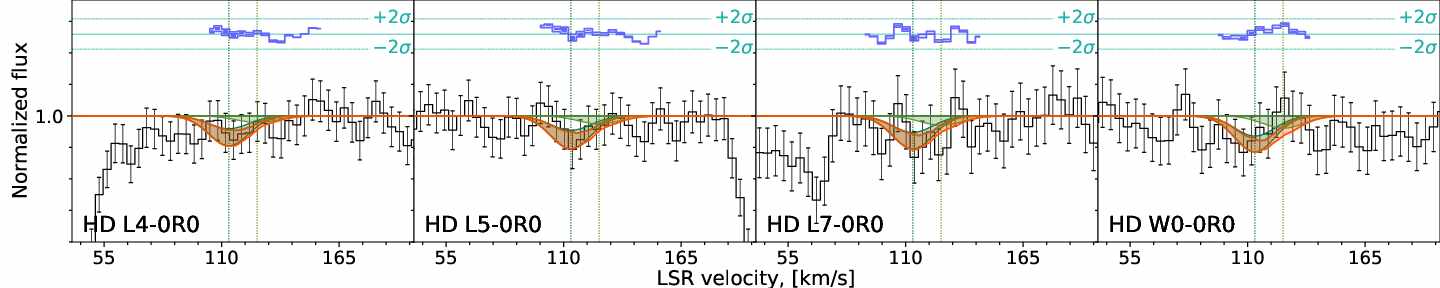}
    \caption{Fit to HD absorption lines towards AV 75 in SMC. Lines are the same as for \ref{fig:lines_HD_Sk67_2}.
    }
    \label{fig:lines_HD_AV75}
\end{figure*}

\begin{figure*}
    \centering
    \includegraphics[width=\linewidth]{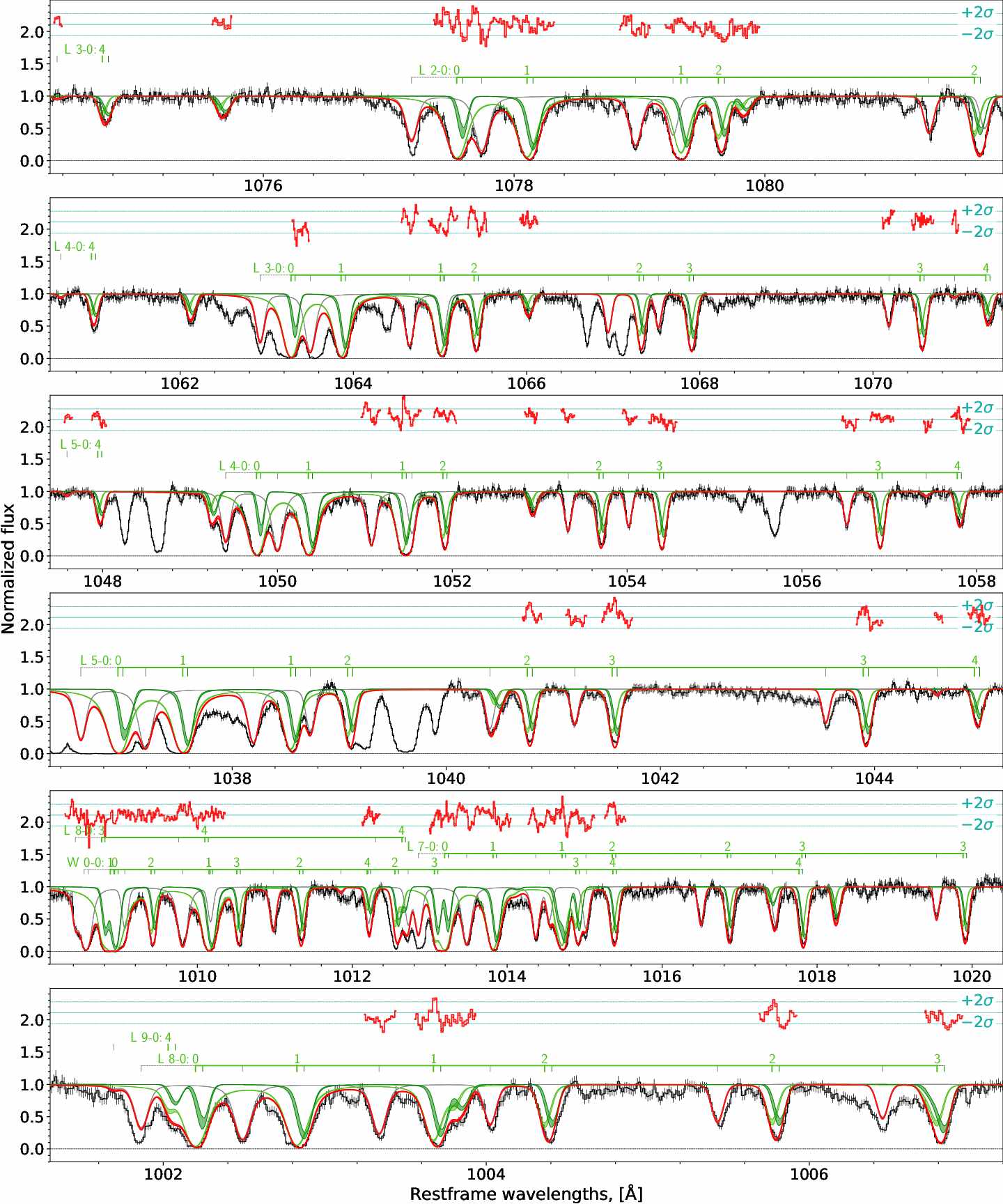}
    \caption{Fit to H2 absorption lines towards AV 75 in SMC. Lines are the same as for \ref{fig:lines_H2_Sk67_2}.
    }
    \label{fig:lines_H2_AV75}
\end{figure*}

\clearpage 
\begin{table*}
    \caption{Fit results of H$_2$ lines towards AV 80}
    \label{tab:AV81}
    \begin{tabular}{ccccc}
    \hline
    \hline
    species & comp & 1 & 2 & 3  \\
            & z & $0.0000672(^{+9}_{-8})$ & $0.0003962(^{+8}_{-13})$ & $0.0004504(^{+29}_{-23})$ \\
    \hline 
     ${\rm H_2\, J=0}$ & b\,km/s & $1.05^{+1.22}_{-0.31}$ & $0.67^{+0.64}_{-0.17}$ & $1.03^{+0.31}_{-0.45}$\\
                       & $\log N$ & $18.36^{+0.04}_{-0.05}$&$19.912^{+0.021}_{-0.022}$ & $18.72^{+0.25}_{-1.14}$\\
    ${\rm H_2\, J=1}$ & b\,km/s & $2.8^{+0.6}_{-0.9}$ & $1.1^{+0.8}_{-0.4}$ & $1.3^{+0.5}_{-0.4}$\\
                      & $\log N$ & $18.19^{+0.04}_{-0.08}$ & $19.957^{+0.009}_{-0.006}$ & $17.6^{+0.4}_{-0.5}$\\
    ${\rm H_2\, J=2}$ & b\,km/s & $3.30^{+0.24}_{-0.14}$ & $2.09^{+0.29}_{-0.99}$ &$1.72^{+0.35}_{-0.25}$\\
                      & $\log N$ & $16.83^{+0.10}_{-0.21}$ & $18.243^{+0.023}_{-0.023}$ & $15.81^{+0.27}_{-0.44}$\\
    ${\rm H_2\, J=3}$ & b\,km/s & $4.05^{+0.39}_{-0.26}$ & $3.0^{+0.5}_{-0.4}$ & $3.48^{+0.14}_{-0.41}$\\
                      & $\log N$ & $16.20^{+0.17}_{-0.30}$ & $17.78^{+0.06}_{-0.08}$ &$15.41^{+0.19}_{-0.23}$ \\
    ${\rm H_2\, J=4}$ & b\,km/s &  $4.1^{+0.6}_{-0.4}$ & $3.7^{+0.5}_{-0.4}$ & --\\
    				  & $\log N$ & $14.14^{+0.05}_{-0.09}$ &$15.81^{+0.20}_{-0.30}$ & $14.28^{+0.08}_{-0.10}$\\
    ${\rm H_2\, J=5}$ & b\,km/s & -- & $5.9^{+1.8}_{-1.6}$ & --\\
                      & $\log N$ & -- & $14.67^{+0.06}_{-0.06}$ & $13.95^{+0.15}_{-0.16}$\\
    \hline 
         & $\log N_{\rm tot}$ & $18.59^{+0.03}_{-0.04}$ & $20.24^{+0.01}_{-0.01}$ & $18.75^{+0.24}_{-0.86}$ \\
     \hline
     HD J=0 & b\,km/s & $1.2^{+1.1}_{-0.7}$ & $17.3^{+3.1}_{-2.2}$ &  $0.98^{+0.31}_{-0.48}$ \\
            & $\log N$ & $\lesssim 14.6$ & $14.504^{+0.030}_{-0.040}$ & $\lesssim 15.4$ \\
    \hline   
    \end{tabular}
    \begin{tablenotes}
    \item Doppler parameters H$_2$ $\rm J=4, 5$ in 3 component were tied to H$_2$ $\rm J=3$.
    \end{tablenotes}
\end{table*}

\begin{figure*}
    \centering
    \includegraphics[width=\linewidth]{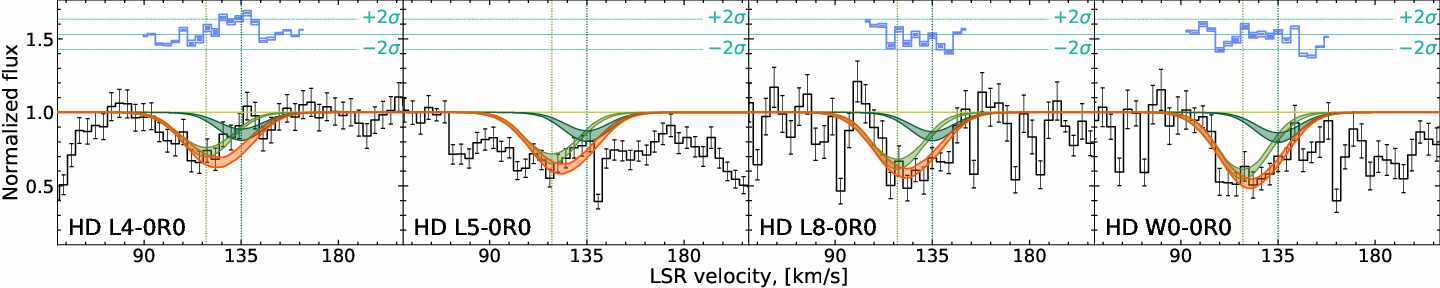}
    \caption{Fit to HD absorption lines towards AV 80 in SMC. Lines are the same as for \ref{fig:lines_HD_Sk67_2}.
    }
    \label{fig:lines_HD_AV80_appendix}
\end{figure*}

\begin{figure*}
    \centering
    \includegraphics[width=\linewidth]{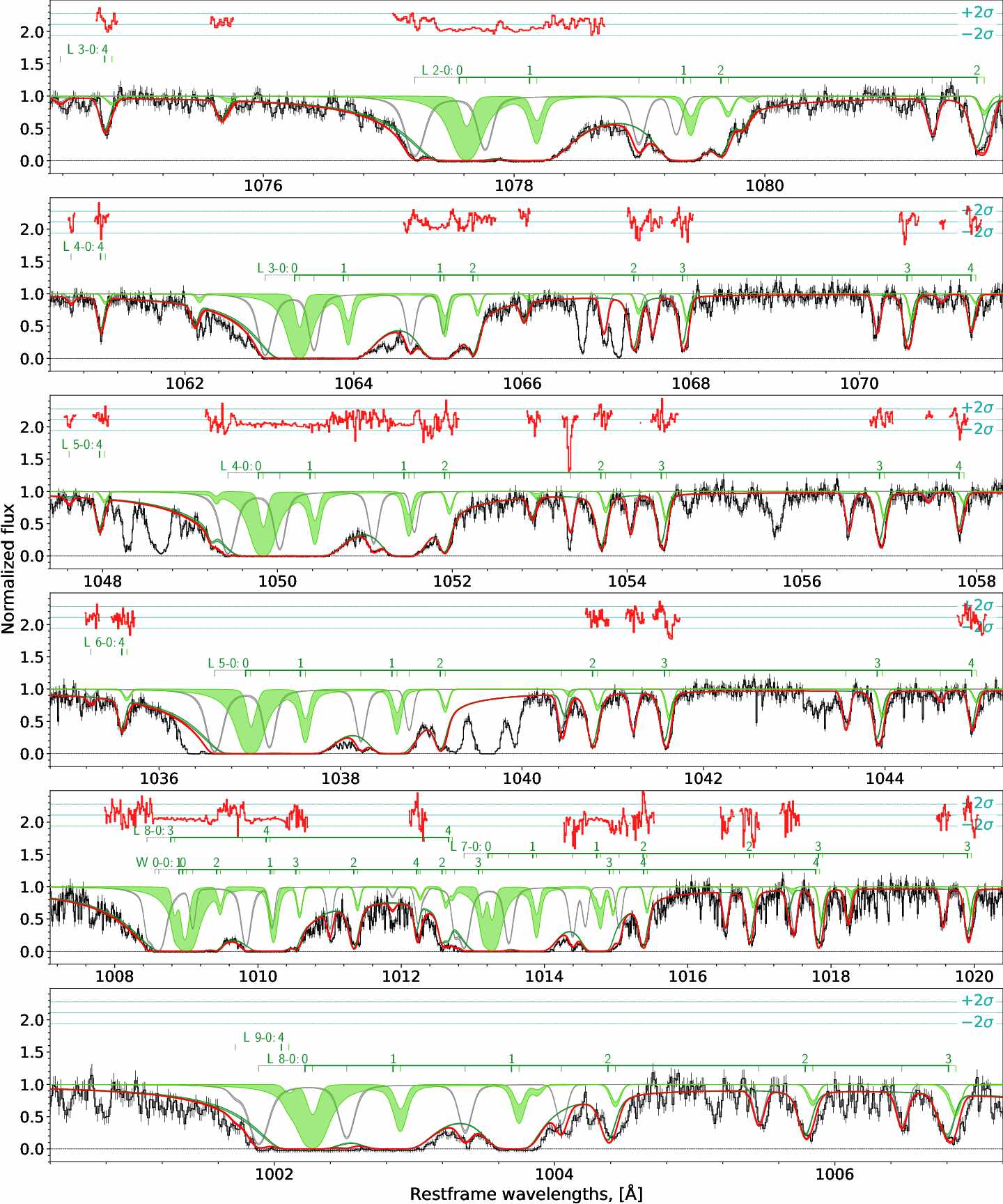}
    \caption{Fit to H2 absorption lines towards AV 80 in SMC. Lines are the same as for \ref{fig:lines_H2_Sk67_2}.
    }
    \label{fig:lines_H2_AV80}
\end{figure*}

\begin{table*}
    \caption{Fit results of H$_2$ lines towards AV 81}
    \label{tab:AV81}
    \begin{tabular}{cccccc}
    \hline
    \hline
    species & comp & 1 & 2 & 3 & 4 \\
            & z & $0.0000438(^{+5}_{-4})$ & $0.0003732(^{+5}_{-18})$ & $0.0004663(^{+8}_{-4})$ & $0.0005915(^{+5}_{-5})$\\
    \hline 
     ${\rm H_2\, J=0}$ & b\,km/s & $4.36^{+0.28}_{-0.11}$ & $0.55^{+0.11}_{-0.05}$ & $2.3^{+0.8}_{-1.2}$ &$3.90^{+0.24}_{-0.25}$ \\
                       & $\log N$ & $17.30^{+0.07}_{-0.10}$ & $15.35^{+0.27}_{-0.20}$ & $18.337^{+0.020}_{-0.019}$ & $17.23^{+0.07}_{-0.10}$\\
    ${\rm H_2\, J=1}$ & b\,km/s &$4.25^{+0.16}_{-0.16}$ & $2.62^{+0.24}_{-0.18}$ & $5.79^{+0.17}_{-0.15}$ &  $3.91^{+0.18}_{-0.17}$\\
                      & $\log N$ & $17.47^{+0.06}_{-0.10}$ & $15.82^{+0.23}_{-0.14}$ & $18.21^{+0.07}_{-0.04}$ & $17.75^{+0.05}_{-0.10}$\\
    ${\rm H_2\, J=2}$ & b\,km/s & $5.2^{+0.5}_{-0.5}$ & $3.09^{+0.42}_{-0.20}$ & $9.3^{+0.3}_{-0.4}$ & $4.32^{+0.61}_{-0.21}$\\
                      & $\log N$ &$15.14^{+0.10}_{-0.05}$ & $14.82^{+0.08}_{-0.07}$ & $15.49^{+0.05}_{-0.04}$ &  $16.42^{+0.11}_{-0.44}$\\
    ${\rm H_2\, J=3}$ & b\,km/s & $8.7^{+1.8}_{-1.4}$ & $3.3^{+0.8}_{-0.4}$ & $10.6^{+0.3}_{-0.6}$ & $5.9^{+0.8}_{-0.5}$\\
                      & $\log N$ & $14.456^{+0.024}_{-0.032}$ &$14.42^{+0.05}_{-0.04}$ &  $15.232^{+0.017}_{-0.023}$ &$14.93^{+0.05}_{-0.05}$ \\
    ${\rm H_2\, J=4}$ & b\,km/s & -- & -- & $12.1^{+1.9}_{-1.3}$ & -- \\
    				  & $\log N$ & $13.3^{+0.3}_{-1.0}$ &$12.59^{+0.37}_{-0.28}$ & $14.298^{+0.023}_{-0.025}$ & $13.4^{+0.3}_{-0.9}$\\
    ${\rm H_2\, J=5}$ & $\log N$ & -- & -- & $13.99^{+0.04}_{-0.05}$ & -- \\
    \hline 
         & $\log N_{\rm tot}$ & $17.69^{0.05}_{-0.07}$ & $15.99^{+0.18}_{-0.10}$ & $18.58^{+0.03}_{-0.02}$ & $17.88^{+0.04}_{-0.08}$ \\
    \hline
    HD J=0 & b\,km/s & $4.38^{+0.36}_{-0.13}$ & $0.59^{+0.05}_{-0.09}$ & $0.55^{+4.99}_{-0.05}$ & $3.88^{+0.27}_{-0.29}$ \\
            & $\log N$ & $\lesssim 13.5$ & $\lesssim 15.1$ & $\lesssim 15.5$ & $\lesssim 13.8$ \\
    \hline   
    \end{tabular}
    \begin{tablenotes}
    \item Doppler parameters H$_2$ $\rm J=4$ in 1, 2 and 4 components and $\rm J=5$  were tied to H$_2$ $\rm J=3$ and $\rm J=4$, respectively.
    \end{tablenotes}
\end{table*}

\begin{figure*}
    \centering
    \includegraphics[width=\linewidth]{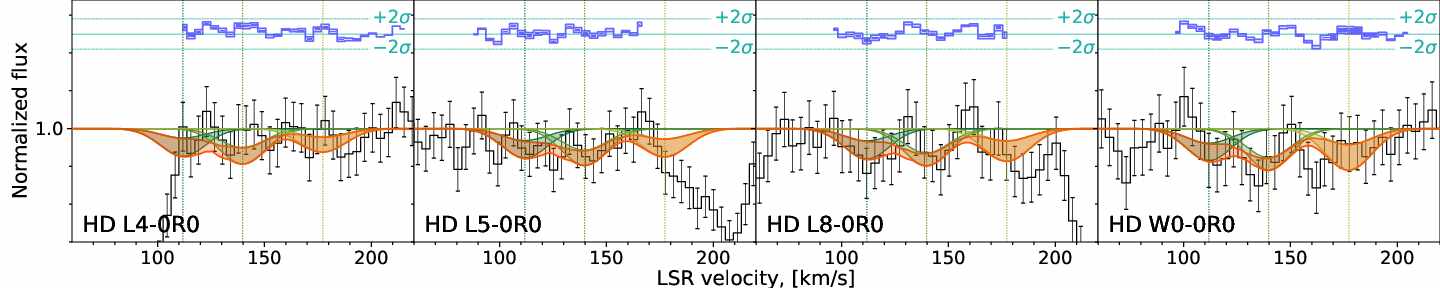}
    \caption{Fit to HD absorption lines towards AV 81 in SMC. Lines are the same as for \ref{fig:lines_HD_Sk67_2}.
    }
    \label{fig:lines_HD_AV81}
\end{figure*}

\begin{figure*}
    \centering
    \includegraphics[width=\linewidth]{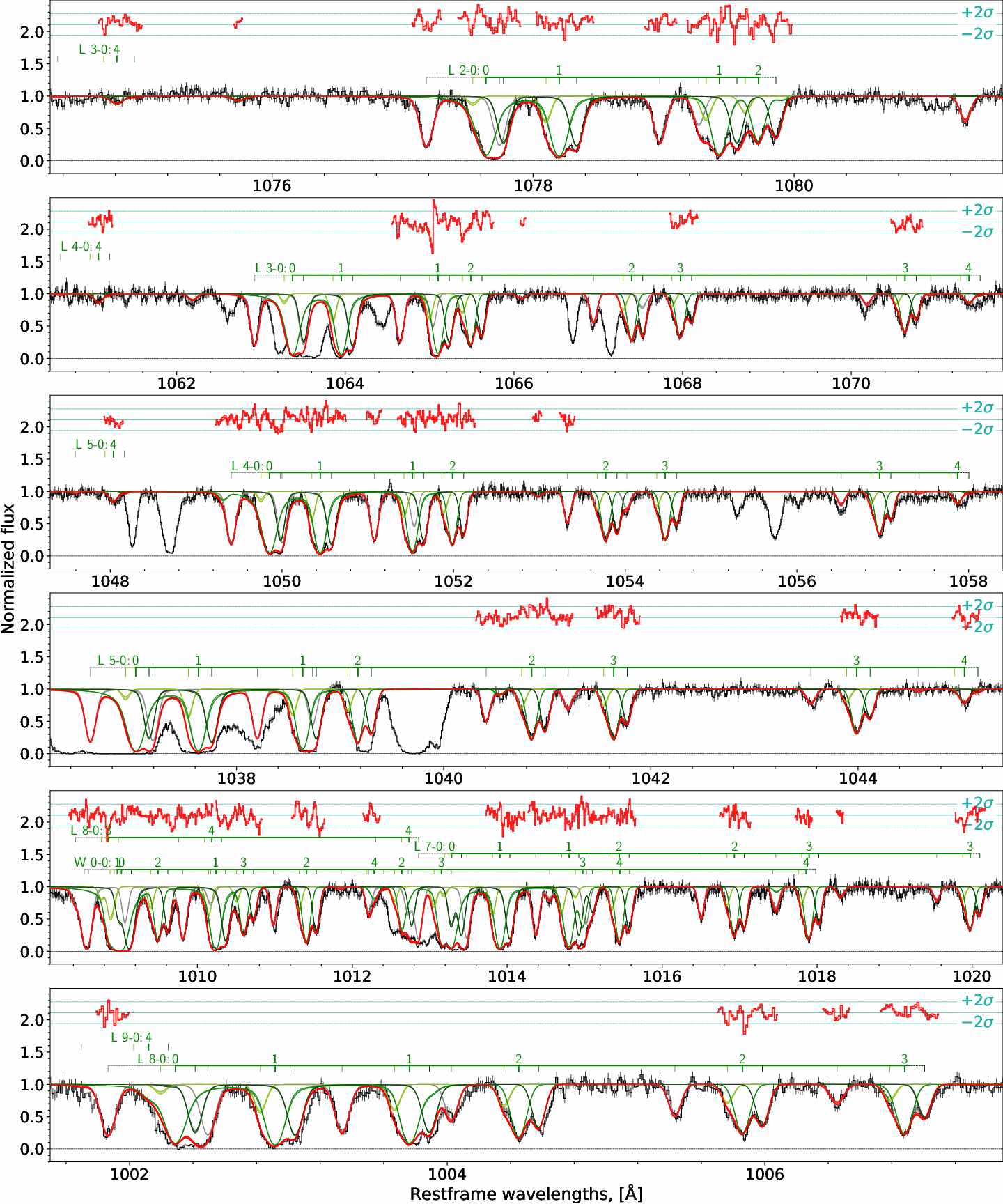}
    \caption{Fit to H2 absorption lines towards AV 81 in SMC. Lines are the same as for \ref{fig:lines_H2_Sk67_2}.
    }
    \label{fig:lines_H2_AV81}
\end{figure*}

\begin{table*}
    \caption{Fit results of H$_2$ lines towards AV 95}
    \label{tab:AV95}
    \begin{tabular}{cccccc}
    \hline
    \hline
    species & comp & 1 & 2 & 3 & 4\\
            & z & $0.00004786(^{+59}_{-24})$ & $0.0003124(^{+23}_{-15})$ & $0.0004152(^{+9}_{-32})$ & $0.0004637(^{+65}_{-25})$ \\
    \hline 
     ${\rm H_2\, J=0}$ & b\,km/s &$4.0^{+0.4}_{-0.7}$ & $9.1^{+0.7}_{-1.3}$ &$1.9^{+0.4}_{-1.2}$ & $0.59^{+0.21}_{-0.09}$ \\
                       & $\log N$ & $18.283^{+0.014}_{-0.031}$ & $15.10^{+0.09}_{-0.15}$ & $18.75^{+0.08}_{-0.12}$ &  $19.06^{+0.05}_{-0.05}$ \\
    ${\rm H_2\, J=1}$ & b\,km/s & $4.07^{+0.29}_{-0.23}$ & $9.85^{+0.15}_{-0.44}$ & $2.2^{+1.0}_{-0.9}$ & $0.67^{+0.29}_{-0.13}$ \\
                      & $\log N$ & $17.88^{+0.06}_{-0.11}$ & $15.07^{+0.06}_{-0.04}$ & $19.032^{+0.026}_{-0.028}$ & $18.16^{+0.13}_{-0.19}$ \\
    ${\rm H_2\, J=2}$ & b\,km/s & $4.12^{+0.30}_{-0.23}$ & $9.74^{+0.25}_{-0.41}$ & $3.91^{+0.23}_{-0.57}$ & $0.80^{+0.41}_{-0.17}$\\
                      & $\log N$ & $16.48^{+0.21}_{-0.20}$ & $14.324^{+0.030}_{-0.053}$ & $17.53^{+0.11}_{-0.09}$ & $15.27^{+0.41}_{-0.28}$ \\
    ${\rm H_2\, J=3}$ & b\,km/s & $4.22^{+0.31}_{-0.31}$ & $9.64^{+0.36}_{-0.18}$ & $4.0^{+0.5}_{-0.4}$ &  $0.93^{+0.33}_{-0.21}$\\
                      & $\log N$ & $15.51^{+0.17}_{-0.13}$ &  $14.302^{+0.030}_{-0.023}$ &  $16.94^{+0.18}_{-0.26}$ & $15.20^{+0.25}_{-0.14}$ \\
    ${\rm H_2\, J=4}$ & b\,km/s &$5.8^{+2.7}_{-0.8}$ & $9.83^{+0.17}_{-0.31}$ & $9.6^{+0.4}_{-1.5}$ & $5.5^{+2.7}_{-1.5}$ \\
    				  & $\log N$ &$14.135^{+0.039}_{-0.015}$ & $13.42^{+0.10}_{-0.15}$ & $14.438^{+0.031}_{-0.058}$ & $14.17^{+0.06}_{-0.09}$ \\
    ${\rm H_2\, J=5}$ & b\,km/s & -- & -- & -- & $8.5^{+3.0}_{-2.7}$\\
                      & $\log N$ & -- & $13.27^{+0.20}_{-0.24}$ &$13.86^{+0.09}_{-0.09}$ & $13.73^{+0.10}_{-0.16}$\\
    
    \hline 
         & $\log N_{\rm tot}$ & $18.43^{+0.02}_{-0.04}$ & $15.46^{+0.05}_{-0.06}$ & $19.23^{+0.03}_{-0.04}$ & $19.11^{+0.05}_{-0.05}$ \\
    \hline
    HD J=0 &  b\,km/s &$4.0^{+0.5}_{-0.9}$ &$9.1^{+0.9}_{-1.5}$ &$0.527^{+2.647}_{-0.027}$ & $0.74^{+0.08}_{-0.24}$ \\
            & $\log N$ & $\lesssim 13.1$ & $\lesssim 13.4$ & $13.75^{+0.39}_{-0.29}$ & $\lesssim 13.6$ \\
    \hline   
    \end{tabular}
    \begin{tablenotes}
    \item Doppler parameters H$_2$ $\rm J=5$ in 2 and 3 components were tied to H$_2$ $\rm J=4$.
    \end{tablenotes}
\end{table*}

\begin{figure*}
    \centering
    \includegraphics[width=\linewidth]{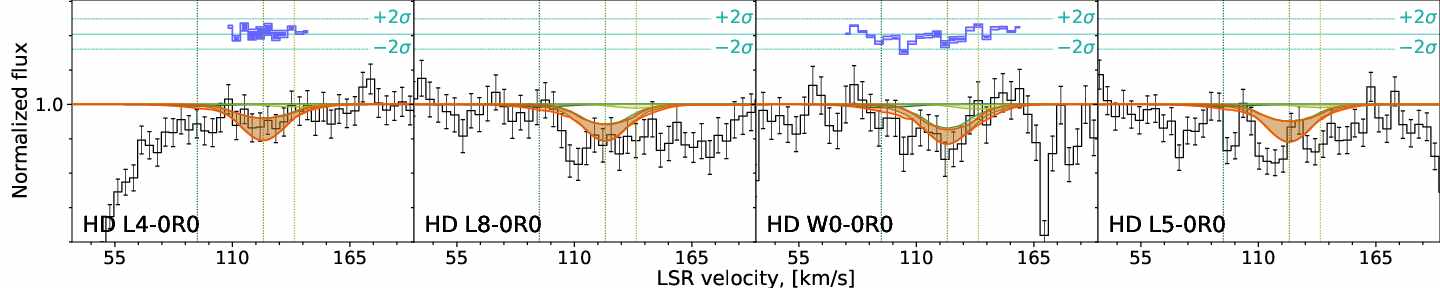}
    \caption{Fit to HD absorption lines towards AV 95 in SMC. Lines are the same as for \ref{fig:lines_HD_Sk67_2}.
    }
    \label{fig:lines_HD_AV95}
\end{figure*}

\begin{figure*}
    \centering
    \includegraphics[width=\linewidth]{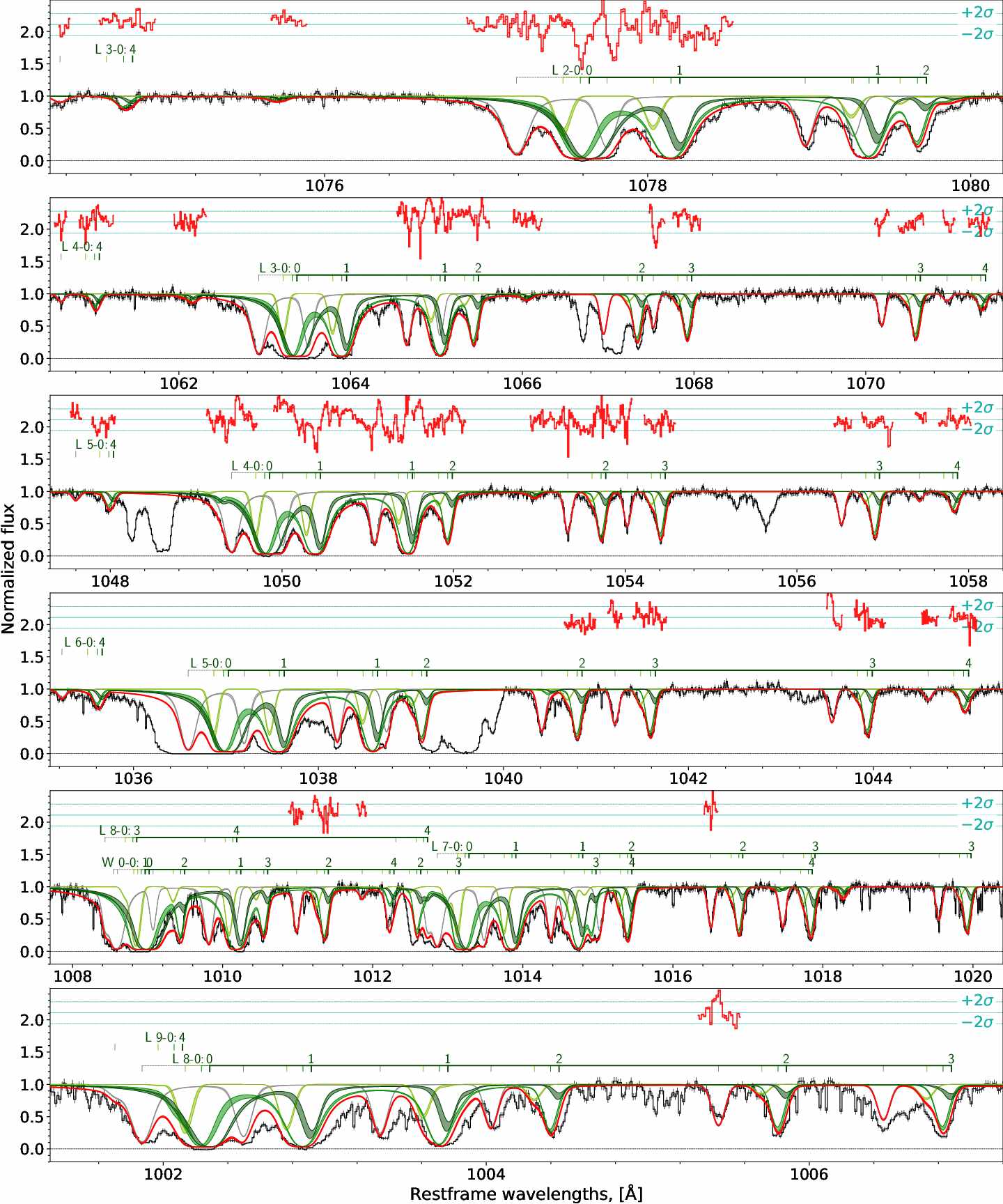}
    \caption{Fit to H2 absorption lines towards AV 95 in SMC. Lines are the same as for \ref{fig:lines_H2_Sk67_2}.
    }
    \label{fig:lines_H2_AV95}
\end{figure*}

\begin{table*}
    \caption{Fit results of H$_2$ lines towards AV 104}
    \label{tab:AV81}
    \begin{tabular}{ccccc}
    \hline
    \hline
    species & comp & 1 & 2 & 3 \\
            & z &$0.0000554(^{+7}_{-9})$ & $0.0003983(^{+11}_{-8})$ & $0.000492(^{+7}_{-3})$ \\
    \hline 
     ${\rm H_2\, J=0}$ & b\,km/s & $0.79^{+0.75}_{-0.29}$ & $1.0^{+0.6}_{-0.5}$ & $0.54^{+0.37}_{-0.04}$\\
                       & $\log N$ &$18.294^{+0.018}_{-0.036}$ &  $19.200^{+0.012}_{-0.015}$ &$16.5^{+0.4}_{-0.9}$ \\
    ${\rm H_2\, J=1}$ & b\,km/s & $2.7^{+0.6}_{-0.8}$ &$1.6^{+1.2}_{-0.3}$ & $0.79^{+0.66}_{-0.24}$\\
                      & $\log N$ &$18.08^{+0.06}_{-0.03}$ &$18.823^{+0.027}_{-0.009}$ & $16.42^{+0.26}_{-0.89}$\\
    ${\rm H_2\, J=2}$ & b\,km/s & $6.2^{+0.4}_{-1.0}$ & $3.8^{+0.5}_{-0.4}$ & $0.88^{+0.86}_{-0.31}$\\
                      & $\log N$ & $15.21^{+0.17}_{-0.09}$ & $16.84^{+0.24}_{-0.49}$ & $13.98^{+0.56}_{-0.27}$\\
    ${\rm H_2\, J=3}$ & b\,km/s & $6.5^{+1.4}_{-0.8}$ &$5.3^{+0.8}_{-0.9}$ & $1.2^{+3.3}_{-0.7}$\\
                      & $\log N$ & $14.99^{+0.08}_{-0.08}$ & $15.18^{+0.22}_{-0.13}$ & $14.08^{+0.15}_{-0.15}$\\
    ${\rm H_2\, J=4}$ & b\,km/s & $7.9^{+2.8}_{-1.6}$ & $5.9^{+2.8}_{-0.9}$ & $11.01^{+8.24}_{-6.22}$\\
    				  & $\log N$ &$14.226^{+0.029}_{-0.103}$ & $14.37^{+0.06}_{-0.07}$ & $13.61^{+0.17}_{-0.23}$\\
    \hline 
         & $\log N_{\rm tot}$ & $18.50^{+0.03}_{-0.03}$ & $19.35^{+0.01}_{-0.01}$ & $16.77^{+0.28}_{-0.42}$ \\
    \hline
    HD J=0 & b\,km/s &$0.54^{+0.96}_{-0.04}$ & $0.82^{+0.47}_{-0.32}$ & $0.518^{+0.335}_{-0.018}$ \\
           & $\log N$ & $\lesssim 15.3$ & $\lesssim 16.1$ & $\lesssim 15.9$ \\ 
    \hline   
    \end{tabular}
    \begin{tablenotes}
    \item Doppler parameters H$_2$ $\rm J=4$ in 1, 2 and 4 components and $\rm J=5$  were tied to H$_2$ $\rm J=3$ and $\rm J=4$, respectively.
    \end{tablenotes}
\end{table*}

\begin{figure*}
    \centering
    \includegraphics[width=\linewidth]{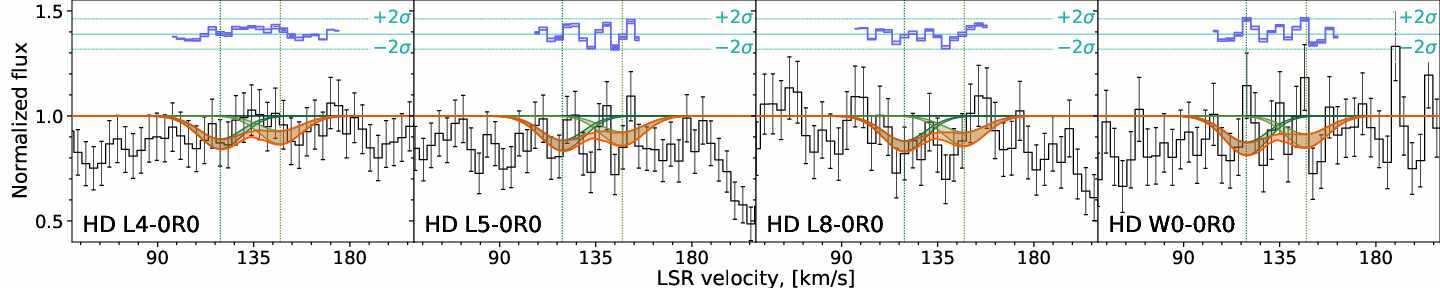}
    \caption{Fit to HD absorption lines towards AV 104 in SMC. Lines are the same as for \ref{fig:lines_HD_Sk67_2}.
    }
    \label{fig:lines_HD_AV104}
\end{figure*}

\begin{figure*}
    \centering
    \includegraphics[width=\linewidth]{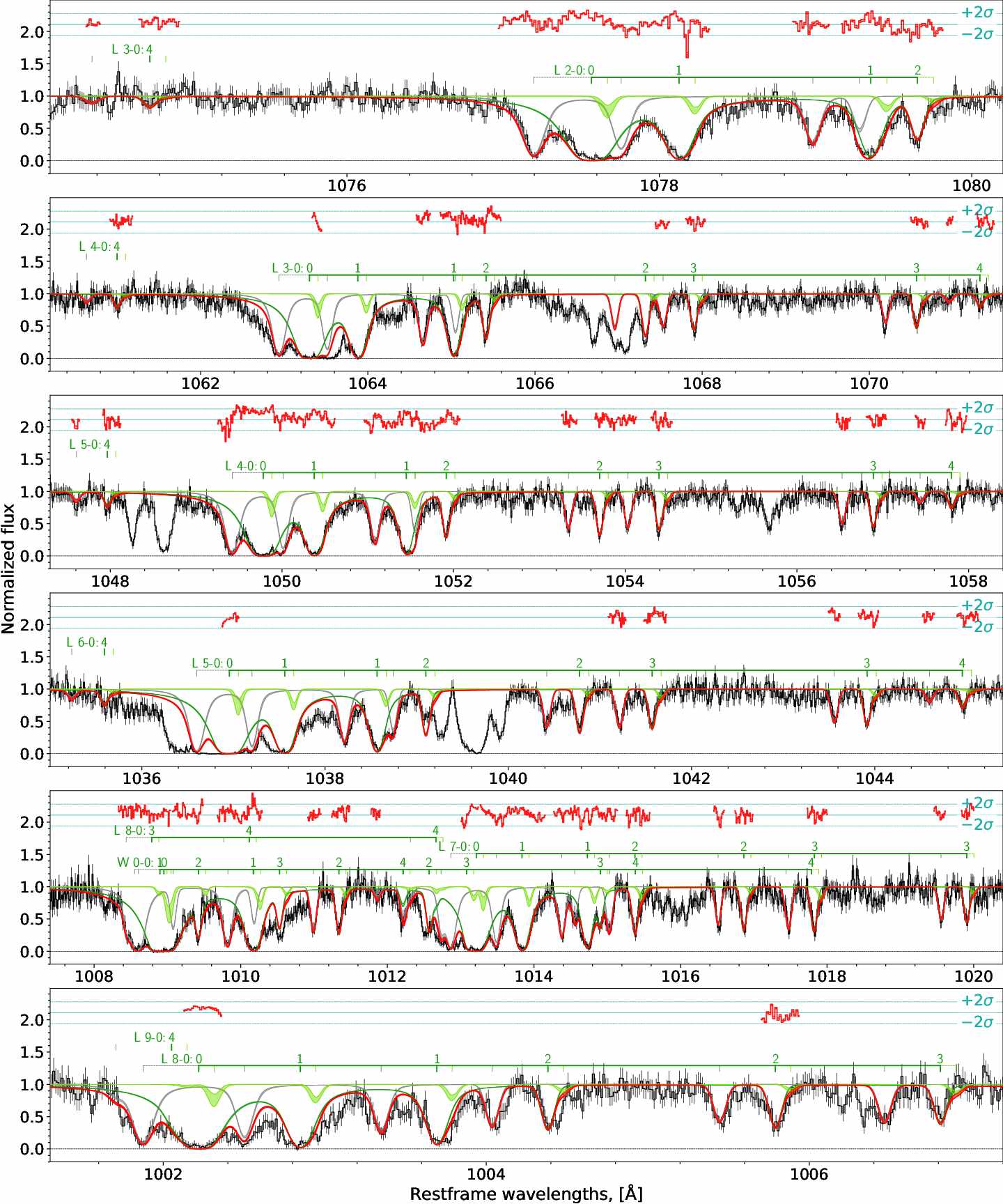}
    \caption{Fit to H2 absorption lines towards AV 104 in SMC. Lines are the same as for \ref{fig:lines_H2_Sk67_2}.
    }
    \label{fig:lines_H2_AV104}
\end{figure*}

\begin{table*}
    \caption{Fit results of H$_2$ lines towards WR 9}
    \label{tab:WR9}
    \begin{tabular}{ccccccccc}
    \hline
    \hline
    species & comp & 1 & 2 & 3 & 4 & 5 & 6 & 7 \\
            & z &$0.0000567(^{+27}_{-21})$ & $0.0001262(^{+12}_{-36})$ & $0.0002925(^{+28}_{-45})$ & $0.0003931(^{+20}_{-36})$ & $0.0004521(^{+20}_{-47})$ &  $0.0005001(^{+42}_{-19})$ & $0.0005601(^{+102}_{-26})$ \\
    \hline 
     ${\rm H_2\, J=0}$ & b\,km/s & $0.78^{+0.57}_{-0.22}$ & $0.75^{+0.41}_{-0.25}$ & $0.63^{+0.25}_{-0.11}$ & $2.2^{+0.6}_{-1.0}$ & $1.9^{+0.4}_{-0.8}$ & $1.5^{+0.6}_{-0.6}$ &  $0.63^{+0.62}_{-0.12}$\\
                       & $\log N$ &$17.27^{+0.15}_{-0.18}$ & $16.73^{+0.28}_{-0.16}$ & $15.9^{+0.4}_{-0.4}$ & $18.69^{+0.08}_{-0.05}$ & $18.02^{+0.07}_{-0.11}$ & $15.8^{+0.7}_{-0.4}$ & $17.36^{+0.11}_{-0.45}$\\
    ${\rm H_2\, J=1}$ & b\,km/s & $1.18^{+0.46}_{-0.20}$ & $2.90^{+0.20}_{-0.49}$ & $0.62^{+0.36}_{-0.09}$ & $3.29^{+0.30}_{-0.78}$ & $2.0^{+0.5}_{-0.4}$ & $2.4^{+0.4}_{-0.7}$ & $1.4^{+0.6}_{-0.4}$\\
                      & $\log N$ & $17.34^{+0.12}_{-0.08}$ & $16.43^{+0.30}_{-0.33}$ & $15.46^{+0.45}_{-0.22}$ & $16.14^{+0.48}_{-0.29}$ & $19.342^{+0.013}_{-0.015}$ &$16.23^{+0.49}_{-0.31}$ & $15.2^{+0.6}_{-0.3}$\\
    ${\rm H_2\, J=2}$ & b\,km/s & $2.32^{+0.24}_{-0.18}$ & $2.89^{+0.46}_{-0.26}$ & $1.8^{+0.5}_{-0.7}$ & $3.2^{+0.5}_{-0.3}$ &  $2.31^{+0.29}_{-0.36}$ & $2.52^{+0.32}_{-0.28}$ & $1.4^{+0.6}_{-0.3}$\\
                      & $\log N$ &$15.96^{+0.15}_{-0.10}$ &  $15.12^{+0.12}_{-0.26}$ & $14.24^{+0.26}_{-0.13}$ & $15.24^{+0.09}_{-0.20}$ & $17.33^{+0.15}_{-0.27}$ & $15.29^{+0.28}_{-0.32}$ & $14.33^{+0.25}_{-0.17}$ \\
    ${\rm H_2\, J=3}$ & b\,km/s & $9.1^{+0.9}_{-1.4}$ & $3.2^{+0.3}_{-0.6}$ & $3.5^{+0.7}_{-0.4}$ & $3.9^{+0.6}_{-0.4}$ & $2.93^{+0.33}_{-0.23}$ & $2.78^{+0.68}_{-0.24}$ & $2.29^{+0.27}_{-0.61}$\\
                      & $\log N$ & $14.42^{+0.05}_{-0.05}$ & $14.46^{+0.06}_{-0.10}$ & $14.23^{+0.07}_{-0.15}$ & $14.75^{+0.13}_{-0.11}$ & $16.30^{+0.20}_{-0.40}$ & $15.03^{+0.22}_{-0.19}$ &$14.63^{+0.05}_{-0.20}$ \\
    ${\rm H_2\, J=4}$ & b\,km/s & -- & -- & $3.8^{+0.3}_{-0.5}$ &$6.0^{+0.8}_{-0.4}$ &  $4.0^{+0.6}_{-0.9}$ & $3.00^{+1.01}_{-0.23}$ & $9.6^{+1.8}_{-0.7}$\\
    				  & $\log N$ &$11.2^{+0.4}_{-1.1}$ & $13.69^{+0.23}_{-0.36}$ & $13.64^{+0.12}_{-0.34}$ & $14.01^{+0.06}_{-0.17}$ & $14.52^{+0.11}_{-0.09}$ & $14.29^{+0.08}_{-0.19}$ & $13.76^{+0.09}_{-0.21}$\\
    ${\rm H_2\, J=5}$ & $\log N$ & $10.6^{+0.7}_{-0.5}$ & $13.7^{+0.4}_{-0.6}$ & -- & $14.14^{+0.05}_{-0.09}$ & $13.70^{+0.11}_{-0.31}$ & $14.18^{+0.06}_{-0.16}$ &  $13.75^{+0.20}_{-0.21}$\\
    \hline 
         & $\log N_{\rm tot}$ & $17.62^{+0.10}_{-0.09}$ & $16.91^{+0.22}_{-0.14}$ & $16.06^{+0.34}_{-0.25}$ & $18.69^{+0.08}_{-0.05}$ & $19.37^{+0.01}_{-0.02}$ & $16.43^{+0.42}_{-0.19}$ & $17.37^{+0.11}_{-0.44}$ \\
     \hline
     HD J=0 & b\,km/s & $0.9^{+0.6}_{-0.4}$ & $0.524^{+0.668}_{-0.024}$ & $0.62^{+0.30}_{-0.12}$ & $2.3^{+0.4}_{-1.8}$ & $2.1^{+0.4}_{-0.9}$ & $1.68^{+0.31}_{-1.18}$ & $0.57^{+0.76}_{-0.07}$ \\
            & $\log N_{\rm tot}$ & $\lesssim 14.9$ & $\lesssim 15.6$ & $\lesssim 15.7$ & $\lesssim 15.9$ & $\lesssim 15.7$ & $\lesssim 15.7$ & $\lesssim 15.9$ \\
    \hline   
    \end{tabular}
    \begin{tablenotes}
    \item Doppler parameters H$_2$ $\rm J=4, 5$ in 1 and  2 components and $\rm J=5$ in 4, 5, 6 and 7 components were tied to H$_2$ $\rm J=3$ and $\rm J=4$, respectively.
    \end{tablenotes}
\end{table*}

\begin{figure*}
    \centering
    \includegraphics[width=\linewidth]{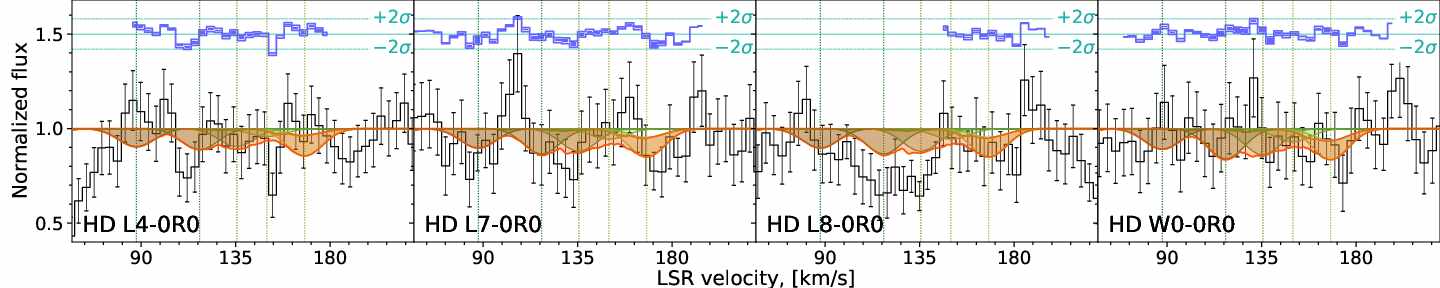}
    \caption{Fit to HD absorption lines towards WR 9 in SMC. Lines are the same as for \ref{fig:lines_HD_Sk67_2}.
    }
    \label{fig:lines_HD_WR9}
\end{figure*}

\begin{figure*}
    \centering
    \includegraphics[width=\linewidth]{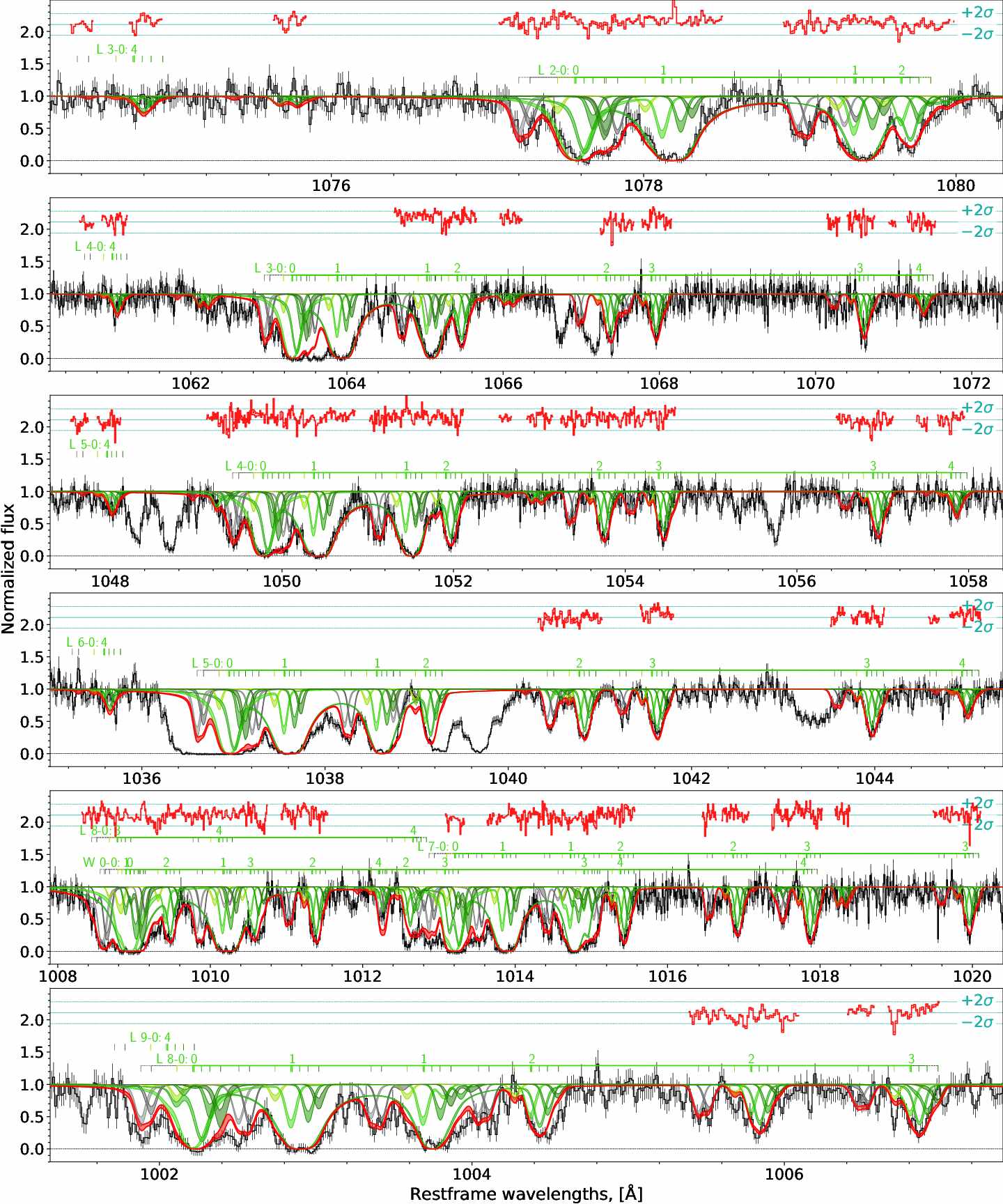}
    \caption{Fit to H2 absorption lines towards WR 9 in SMC. Lines are the same as for \ref{fig:lines_H2_Sk67_2}.
    }
    \label{fig:lines_H2_WR9}
\end{figure*}

\clearpage
\begin{table*}
    \caption{Fit results of H$_2$ lines towards AV 170}
    \label{tab:AV170}
    \begin{tabular}{cccc}
    \hline
    \hline
    species & comp & 1 & 2  \\
            & z & $0.0000632(^{+7}_{-9})$ & $0.0004605(^{+11}_{-6})$ \\
    \hline 
     ${\rm H_2\, J=0}$ & b\,km/s & $1.77^{+0.26}_{-0.63}$ & $0.73^{+0.67}_{-0.23}$\\
                       & $\log N$ & $17.59^{+0.04}_{-0.07}$ & $19.614^{+0.012}_{-0.009}$\\
    ${\rm H_2\, J=1}$ & b\,km/s & $2.02^{+0.17}_{-0.46}$ & $1.85^{+0.18}_{-0.71}$\\
                      & $\log N$ &  $17.85^{+0.05}_{-0.05}$ & $19.111^{+0.012}_{-0.012}$\\
    ${\rm H_2\, J=2}$ & b\,km/s & $2.27^{+0.12}_{-0.14}$ & $1.98^{+0.11}_{-0.37}$\\
                      & $\log N$ & $17.04^{+0.09}_{-0.09}$ & $17.15^{+0.08}_{-0.13}$\\
    ${\rm H_2\, J=3}$ & b\,km/s & $2.09^{+0.08}_{-0.11}$ & $2.05^{+0.09}_{-0.12}$\\
                      & $\log N$ & $16.19^{+0.18}_{-0.14}$ & $16.37^{+0.19}_{-0.11}$\\
    ${\rm H_2\, J=4}$ & b\,km/s & -- & --\\
    				  & $\log N$ &$13.82^{+0.13}_{-0.17}$ & $14.27^{+0.21}_{-0.16}$\\
    
    \hline 
         & $\log N_{\rm tot}$ & $18.09^{+0.03}_{-0.04}$ & $19.73^{+0.01}_{-0.01}$\\
     \hline
     HD J=0 & b\,km/s & $1.68^{+0.10}_{-1.18}$ &  $0.529^{+0.640}_{-0.029}$ \\
            & $\log N$ & $\lesssim 15.4$ & $\lesssim 16.1$ \\
    \hline   
    \end{tabular}
    \begin{tablenotes}
    \item Doppler parameters H$_2$ $\rm J=4$ in 1 and 2 components were tied to H$_2$ $\rm J=3$.
    \end{tablenotes}
\end{table*}

\begin{figure*}
    \centering
    \includegraphics[width=\linewidth]{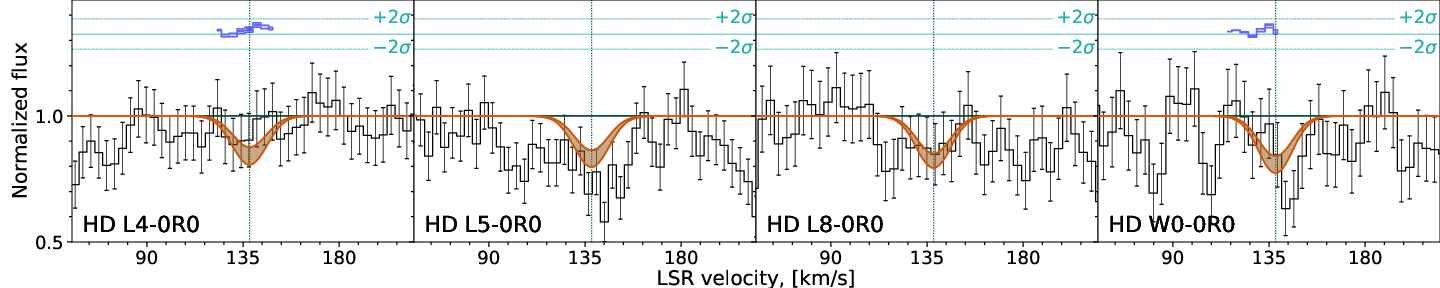}
    \caption{Fit to HD absorption lines towards AV 170 in SMC. Lines are the same as for \ref{fig:lines_HD_Sk67_2}.
    }
    \label{fig:lines_HD_AV170}
\end{figure*}

\begin{figure*}
    \centering
    \includegraphics[width=\linewidth]{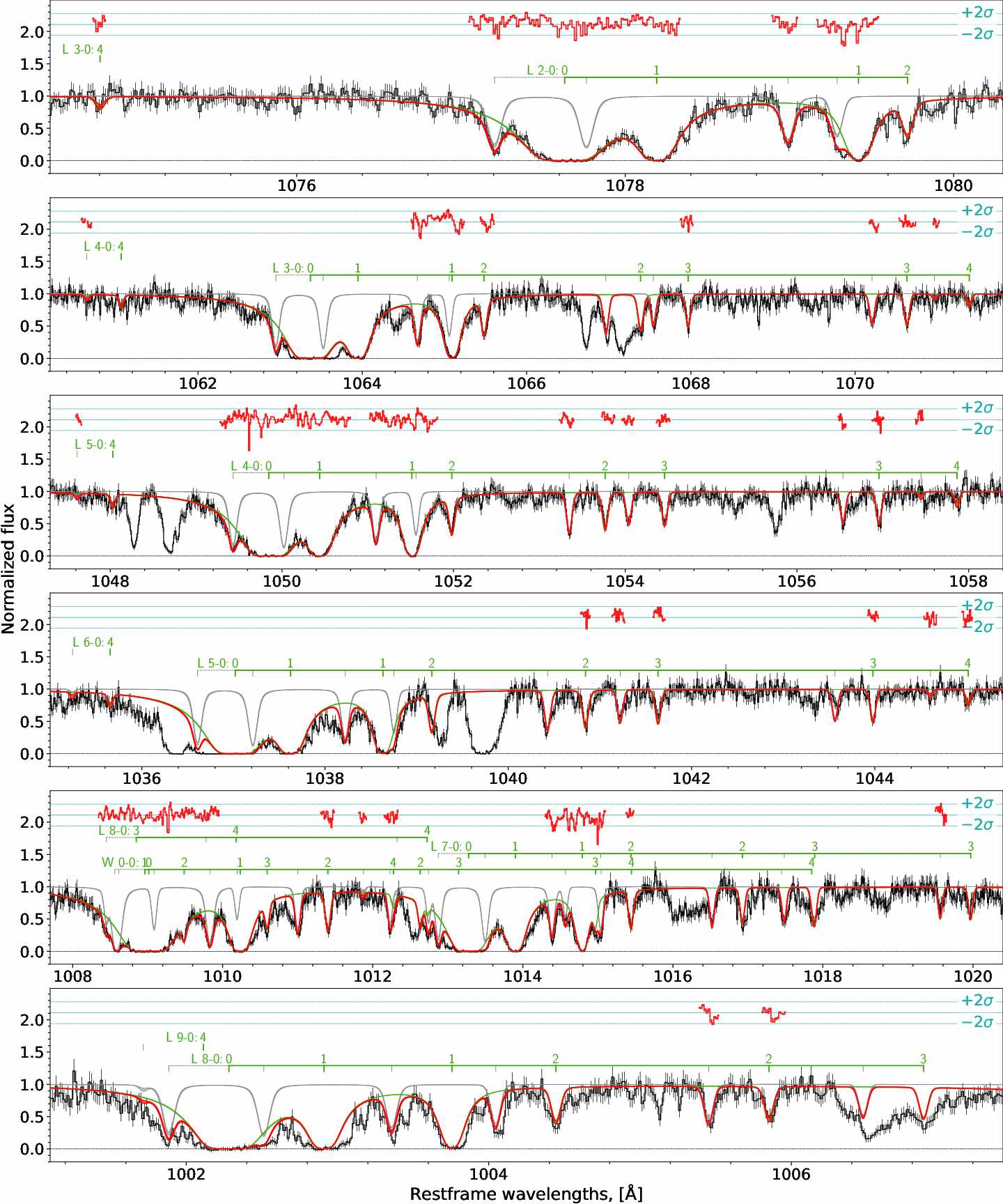}
    \caption{Fit to H2 absorption lines towards AV 170 in SMC. Lines are the same as for \ref{fig:lines_H2_Sk67_2}.
    }
    \label{fig:lines_H2_AV170}
\end{figure*}

\begin{table*}
    \caption{Fit results of H$_2$ lines towards AV 175}
    \label{tab:AV175}
    \begin{tabular}{ccccc}
    \hline
    \hline
    species & comp & 1 & 2 & 3  \\
            & z & $0.000048(^{+5}_{-5})$ & $0.000406(^{+4}_{-4})$ & $0.000462(^{+4}_{-4})$ \\
    \hline 
     ${\rm H_2\, J=0}$ & b\,km/s & $1.7^{+0.6}_{-0.8}$ & $0.9^{+0.5}_{-0.4}$ & $0.68^{+0.42}_{-0.18}$\\
                       & $\log N$ &$18.0^{+2.0}_{-1.8}$ & $19.02^{+0.34}_{-0.21}$ &$19.92^{+0.13}_{-0.12}$ \\
    ${\rm H_2\, J=1}$ & b\,km/s & $2.0^{+0.7}_{-0.3}$ & $1.3^{+0.7}_{-0.6}$ & $1.1^{+0.6}_{-0.4}$\\
                      & $\log N$ & $17.2^{+0.4}_{-0.7}$ & $19.31^{+0.16}_{-0.16}$ & $19.46^{+0.13}_{-0.08}$\\
    ${\rm H_2\, J=2}$ & b\,km/s & $2.7^{+0.7}_{-0.8}$ &$2.7^{+0.4}_{-0.6}$ & $1.7^{+0.5}_{-0.5}$ \\
                      & $\log N$ & $15.5^{+0.6}_{-0.5}$ & $17.3^{+0.3}_{-0.6}$ &$17.50^{+0.19}_{-0.29}$ \\
    ${\rm H_2\, J=3}$ & b\,km/s & $3.3^{+0.6}_{-0.7}$ &$2.84^{+0.78}_{-0.23}$ & $1.85^{+0.48}_{-0.27}$ \\
                      & $\log N$ & $15.12^{+0.37}_{-0.17}$ & $16.0^{+0.5}_{-0.5}$ &$17.19^{+0.29}_{-0.29}$ \\
    ${\rm H_2\, J=4}$ & b\,km/s & $3.8^{+0.5}_{-0.4}$ & $4.1^{+0.3}_{-0.3}$ & $4.16^{+0.23}_{-0.26}$ \\
    				  & $\log N$ & $14.03^{+0.26}_{-0.38}$ & $14.74^{+0.20}_{-0.20}$ &$15.49^{+0.23}_{-0.23}$ \\
    ${\rm H_2\, J=5}$ & $\log N$ & $14.10^{+0.31}_{-0.46}$ & $14.42^{+0.21}_{-0.22}$ &  $14.43^{+0.17}_{-0.24}$\\	 			  
    
    \hline 
         & $\log N_{\rm tot}$ & $18.10^{+1.07}_{-0.87}$ & $19.49^{+0.17}_{-0.12}$ & $20.05^{+0.10}_{-0.09}$\\
    \hline
    HD J=0 & b\,km/s &$1.7^{+0.7}_{-0.7}$ & $1.0^{+0.4}_{-0.4}$ & $0.77^{+0.32}_{-0.27}$ \\
           & $\log N$ & $\lesssim 14.8$ & $\lesssim 16.6$ & $\lesssim 16.4$ \\ 
    \hline   
    \end{tabular}
    \begin{tablenotes}
    \item Doppler parameters H$_2$ $\rm J=4$ in 1 and 2 components were tied to H$_2$ $\rm J=3$.
    \end{tablenotes}
\end{table*}

\begin{figure*}
    \centering
    \includegraphics[width=\linewidth]{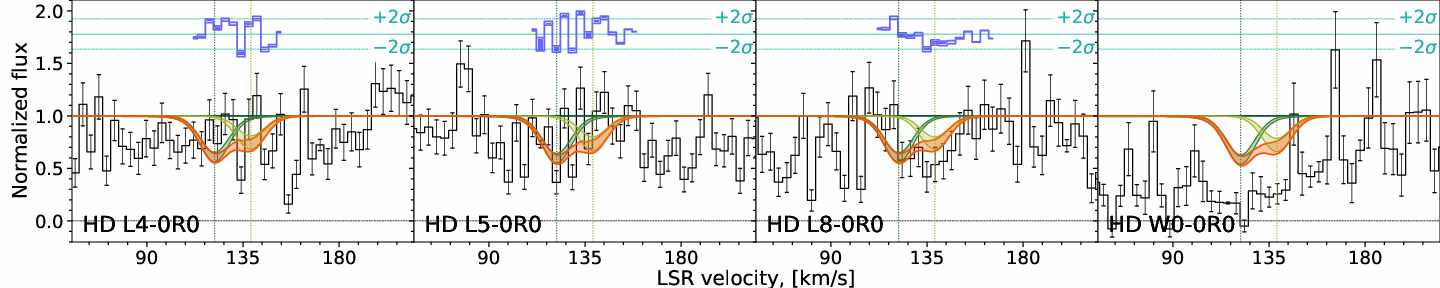}
    \caption{Fit to HD absorption lines towards AV 175 in SMC. Lines are the same as for \ref{fig:lines_HD_Sk67_2}.
    }
    \label{fig:lines_HD_AV175}
\end{figure*}

\begin{figure*}
    \centering
    \includegraphics[width=\linewidth]{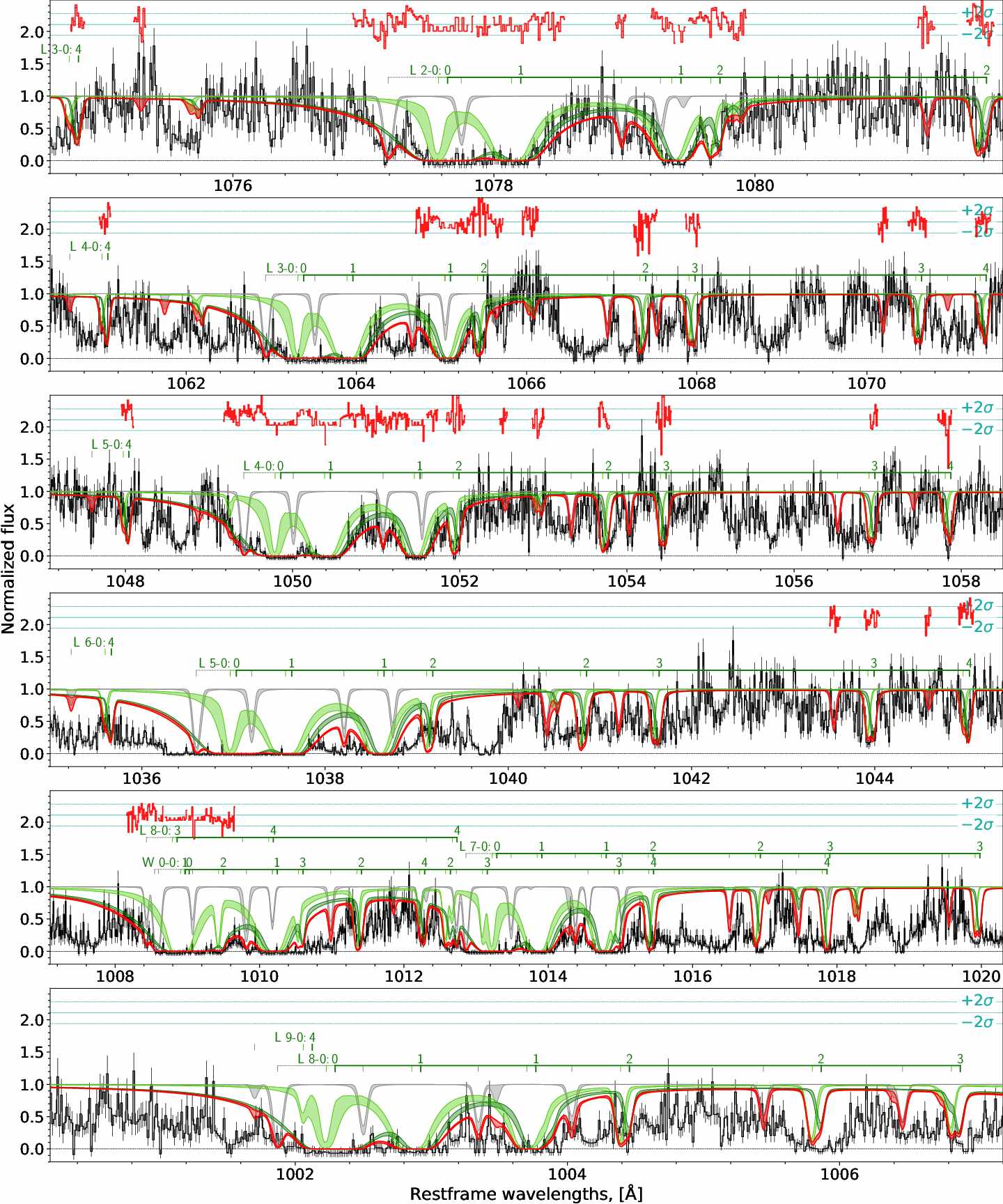}
    \caption{Fit to H2 absorption lines towards AV 175 in SMC. Lines are the same as for \ref{fig:lines_H2_Sk67_2}.
    }
    \label{fig:lines_H2_AV175}
\end{figure*}

\begin{table*}
    \caption{Fit results of H$_2$ lines towards NGC 346 -12}
    \label{tab:NGC346_12}
    \begin{tabular}{cccccc}
    \hline
    \hline
    species & comp & 1 & 2 & 3 & 4 \\
            & z & $0.0000270(^{+23}_{-53})$ & $0.0004634(^{+7}_{-21})$ & $0.0005553(^{+24}_{-27})$ & $0.0006113(^{+19}_{-33})$ \\
    \hline 
     ${\rm H_2\, J=0}$ & b\,km/s & $0.79^{+0.41}_{-0.29}$ & $1.5^{+0.5}_{-0.6}$ & $0.78^{+0.37}_{-0.21}$ & $0.66^{+0.44}_{-0.16}$ \\
                       & $\log N$ & $16.88^{+0.31}_{-0.34}$ & $19.39^{+0.27}_{-0.13}$ &  $18.8^{+0.5}_{-0.7}$ & $20.35^{+0.04}_{-0.07}$ \\
    ${\rm H_2\, J=1}$ & b\,km/s &$1.1^{+0.5}_{-0.5}$ & $1.8^{+0.8}_{-0.5}$ & $1.1^{+0.6}_{-0.4}$ & $0.89^{+1.08}_{-0.23}$ \\
                      & $\log N$ & $14.9^{+0.5}_{-4.8}$ & $19.27^{+0.05}_{-0.07}$ &  $18.0^{+0.5}_{-0.7}$ & $19.43^{+0.05}_{-0.04}$\\
    ${\rm H_2\, J=2}$ & b\,km/s &  $1.4^{+0.4}_{-0.4}$ & $2.5^{+0.6}_{-0.5}$ & $1.99^{+0.26}_{-0.68}$ & $2.8^{+0.5}_{-0.4}$ \\
                      & $\log N$ & $17.43^{+0.11}_{-0.12}$ & $17.88^{+0.07}_{-0.10}$ & $16.63^{+0.25}_{-0.51}$ & $17.2^{+0.4}_{-0.4}$ \\
    ${\rm H_2\, J=3}$ & b\,km/s & $1.93^{+0.24}_{-0.47}$ & $3.2^{+0.5}_{-0.6}$ & $3.17^{+0.28}_{-0.64}$ & $2.8^{+0.4}_{-0.4}$ \\
                      & $\log N$ & $17.10^{+0.20}_{-0.33}$ &  $17.33^{+0.27}_{-0.33}$ & $16.09^{+0.51}_{-0.14}$ & $16.35^{+0.44}_{-0.24}$ \\
    ${\rm H_2\, J=4}$ & b\,km/s & $2.3^{+1.6}_{-0.6}$ & $11.6^{+1.9}_{-1.5}$ &$3.3^{+0.5}_{-0.8}$ & $4.5^{+1.1}_{-0.9}$ \\
    				  & $\log N$ & $14.57^{+0.39}_{-0.19}$ &  $14.80^{+0.06}_{-0.03}$ & $14.84^{+0.24}_{-0.12}$ & $14.90^{+0.09}_{-0.09}$ \\
    ${\rm H_2\, J=5}$ & b\,km/s & $5.6^{+3.4}_{-2.4}$ & $12.7^{+1.9}_{-2.4}$ & $6.9^{+5.8}_{-2.8}$ &  $5.6^{+5.6}_{-1.0}$ \\
                      & $\log N$ & $14.0^{+0.4}_{-0.7}$ & $14.30^{+0.27}_{-0.49}$ &  $14.50^{+0.29}_{-0.10}$ & $14.53^{+0.22}_{-0.49}$\\	 			  
    
    \hline 
         & $\log N_{\rm tot}$ & $17.67^{+0.11}_{-0.11}$ & $19.64^{+0.17}_{-0.07}$ & $18.87^{+0.46}_{-0.51}$ & $20.40^{+0.04}_{[-0.06}$ \\
    \hline
    HD J=0 & b\,km/s &$0.98^{+0.47}_{-0.30}$ & $1.4^{+0.5}_{-0.5}$ & $0.71^{+0.36}_{-0.20}$ & $1.3^{+0.5}_{-0.4}$ \\
           & $\log N$ & $\lesssim 17.4$ & $\lesssim 13.4$ & $\lesssim 15.6$ & $\lesssim 13.6$ \\
    \hline   
    \end{tabular}
    \begin{tablenotes}
    \item 
    \end{tablenotes}
\end{table*}

\begin{figure*}
    \centering
    \includegraphics[width=\linewidth]{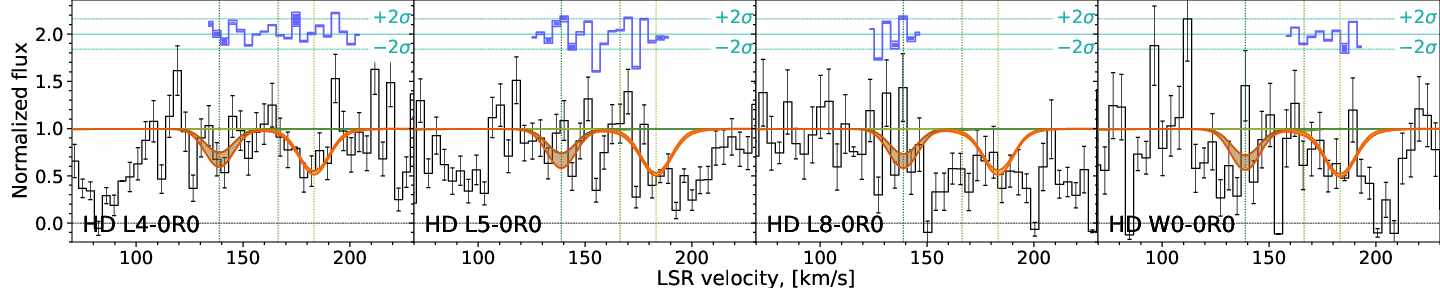}
    \caption{Fit to HD absorption lines towards NGC 346-12 in SMC. Lines are the same as for \ref{fig:lines_HD_Sk67_2}.
    }
    \label{fig:lines_HD_NGC346_12}
\end{figure*}

\begin{figure*}
    \centering
    \includegraphics[width=\linewidth]{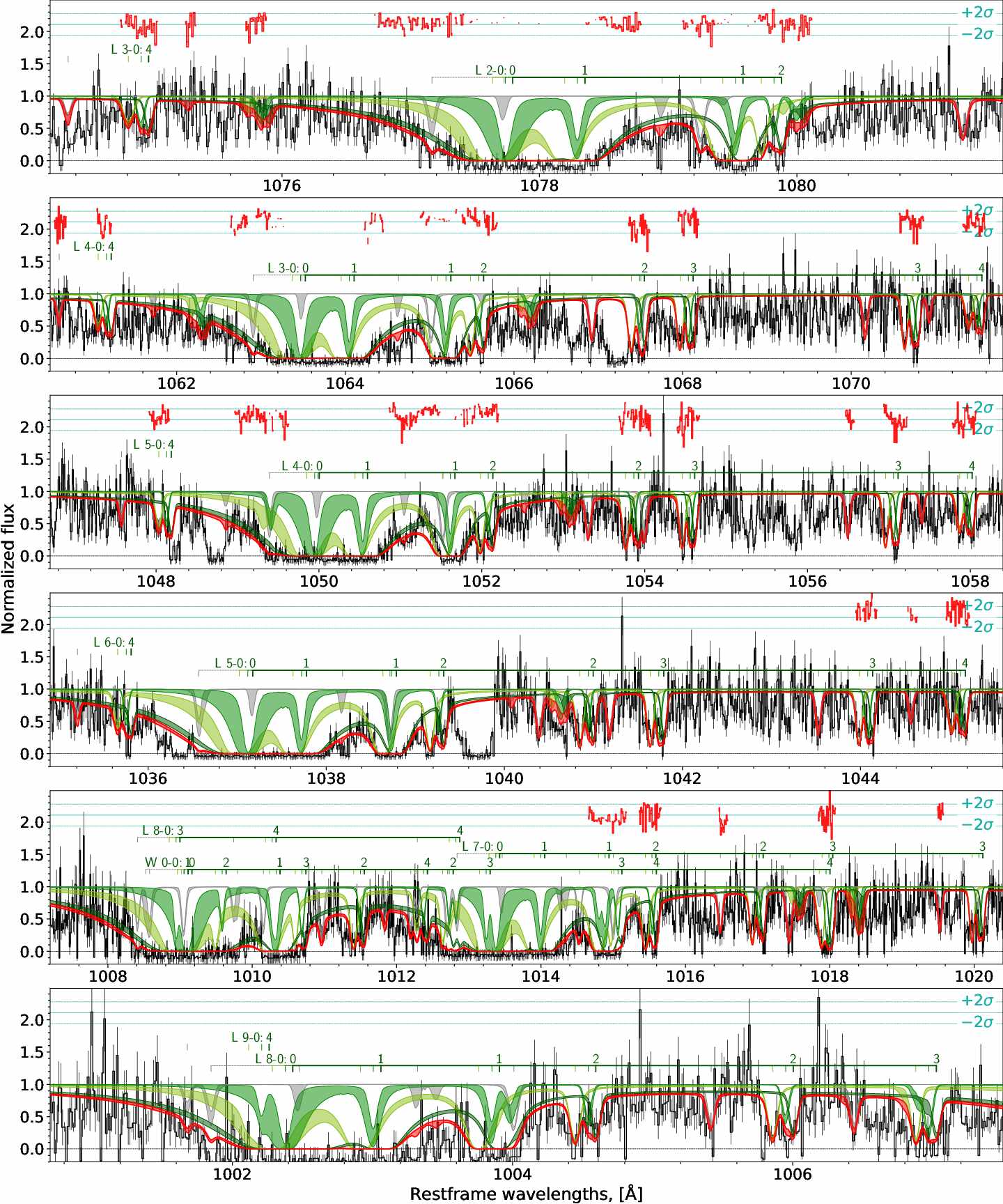}
    \caption{Fit to H2 absorption lines towards NGC 346-12 in SMC. Lines are the same as for \ref{fig:lines_H2_Sk67_2}.
    }
    \label{fig:lines_H2_NGC346_12}
\end{figure*}

\begin{table*}
    \caption{Fit results of H$_2$ lines towards AV 207}
    \label{tab:AV207}
    \begin{tabular}{cccccc}
    \hline
    \hline
    species & comp & 1 & 2 & 3 & 4  \\
            & z & $0.0000549(^{+27}_{-19})$ & $0.0001511(^{+20}_{-142})$ & $0.0005342(^{+31}_{-42})$ & $0.0006044(^{+59}_{-29})$ \\
    \hline 
     ${\rm H_2\, J=0}$ & b\,km/s & $1.13^{+0.88}_{-0.27}$ & $0.56^{+0.19}_{-0.06}$ & $0.62^{+0.46}_{-0.12}$ & $1.0^{+0.6}_{-0.5}$\\
                       & $\log N$ & $17.24^{+0.13}_{-0.08}$ &$15.31^{+0.30}_{-0.45}$ & $19.26^{+0.04}_{-0.04}$ & $18.53^{+0.14}_{-0.21}$\\
    ${\rm H_2\, J=1}$ & b\,km/s &$2.90^{+0.27}_{-0.33}$ & $0.77^{+0.22}_{-0.22}$ & $1.7^{+0.7}_{-0.6}$ &$1.6^{+0.8}_{-0.5}$ \\
                      & $\log N$ & $17.71^{+0.12}_{-0.08}$ & $15.59^{+0.28}_{-0.25}$ & $19.164^{+0.021}_{-0.084}$ & $18.70^{+0.09}_{-0.16}$\\
    ${\rm H_2\, J=2}$ & b\,km/s &$2.82^{+0.36}_{-0.13}$ & $1.80^{+0.65}_{-0.19}$ & $2.80^{+0.22}_{-0.42}$ & $2.76^{+0.15}_{-0.58}$\\
                      & $\log N$ & $16.49^{+0.16}_{-0.27}$ & $14.65^{+0.21}_{-0.23}$ & $17.29^{+0.09}_{-0.25}$ & $16.59^{+0.37}_{-0.28}$\\
    ${\rm H_2\, J=3}$ & b\,km/s &$3.07^{+0.24}_{-0.23}$ & $2.38^{+0.23}_{-0.41}$ & $2.76^{+0.30}_{-0.22}$ & $2.97^{+0.16}_{-0.52}$\\
                      & $\log N$ & $15.48^{+0.19}_{-0.35}$ & $14.25^{+0.32}_{-0.24}$ &$16.65^{+0.36}_{-0.14}$ &$16.09^{+0.26}_{-0.50}$ \\
    ${\rm H_2\, J=4}$ & b\,km/s & -- & -- &  $3.07^{+0.15}_{-0.29}$ & -- \\
    				  & $\log N$ & $14.38^{+0.17}_{-0.12}$ &  $13.90^{+0.30}_{-0.57}$ & $14.92^{+0.24}_{-0.25}$ & $14.54^{+0.16}_{-0.26}$\\
    ${\rm H_2\, J=4}$ & $\log N$ & -- & -- & $14.15^{+0.29}_{-0.28}$ & $14.07^{+0.32}_{-0.28}$\\
    
    \hline 
         & $\log N_{\rm tot}$ & $17.86^{+0.09}_{-0.06}$ & $15.82^{+0.21}_{-0.17}$ & $19.52^{+0.02}_{-0.04}$ & $18.93^{+0.08}_{-0.12}$ \\
    \hline
    HD J=0 & b\,km/s &$1.10^{+1.00}_{-0.30}$ & $0.510^{+0.247}_{-0.010}$ & $0.90^{+0.30}_{-0.40}$ &$0.54^{+0.89}_{-0.04}$ \\
           & $\log N$ & $\lesssim 16.0$ & $\lesssim 16.2$ & $\lesssim 16.8$ & $\lesssim 16.5$ \\
    \hline   
    \end{tabular}
    \begin{tablenotes}
    \item Doppler parameters H$_2$ $\rm J=4$ in 1 and 2 components, $\rm J=4, 5$ in 4 component and $\rm J=5$ in 3 component  were tied to H$_2$ $\rm J=3$ and $\rm J=4$, respectively .
    \end{tablenotes}
\end{table*}

\begin{figure*}
    \centering
    \includegraphics[width=\linewidth]{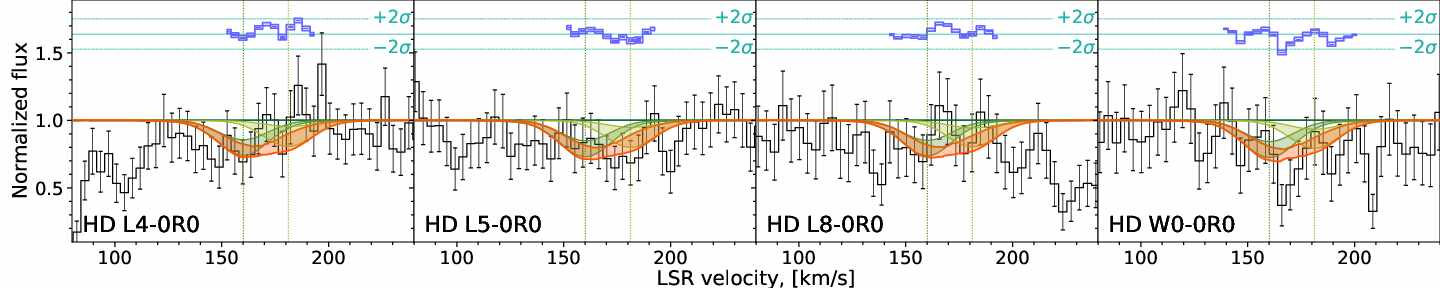}
    \caption{Fit to HD absorption lines towards AV 207 in SMC. Lines are the same as for \ref{fig:lines_HD_Sk67_2}.
    }
    \label{fig:lines_HD_AV207}
\end{figure*}

\begin{figure*}
    \centering
    \includegraphics[width=\linewidth]{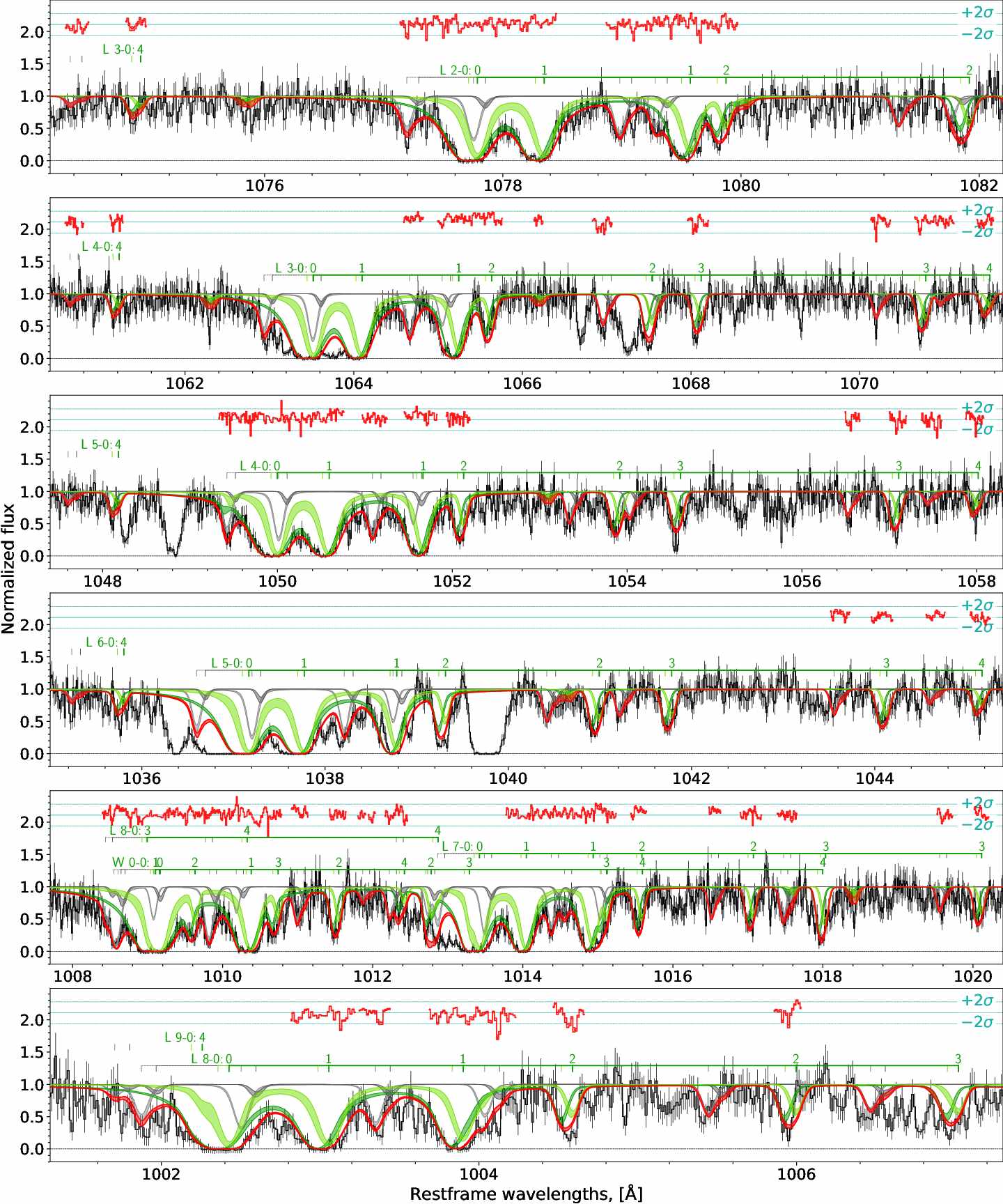}
    \caption{Fit to H2 absorption lines towards AV 207 in SMC. Lines are the same as for \ref{fig:lines_H2_Sk67_2}.
    }
    \label{fig:lines_H2_AV207}
\end{figure*}

\begin{table*}
    \caption{Fit results of H$_2$ lines towards AV 208}
    \label{tab:AV208}
    \begin{tabular}{ccccc}
    \hline
    \hline
    species & comp & 1 & 2 & 3 \\
            & z & $0.0000519(^{+25}_{-14})$ & $0.000412(^{+4}_{-5})$ & $0.0004794(^{+27}_{-136})$\\
    \hline 
     ${\rm H_2\, J=0}$ & b\,km/s & $0.69^{+0.46}_{-0.19}$ &  $0.57^{+0.42}_{-0.07}$ &$0.72^{+0.54}_{-0.16}$ \\
                       & $\log N$ & $17.57^{+0.08}_{-0.18}$ & $19.43^{+0.18}_{-0.33}$ &  $19.79^{+0.07}_{-0.20}$\\
    ${\rm H_2\, J=1}$ & b\,km/s & $1.03^{+0.66}_{-0.28}$ & $0.71^{+0.50}_{-0.21}$ & $1.17^{+0.60}_{-0.25}$ \\
                      & $\log N$ & $17.95^{+0.07}_{-0.08}$ & $19.53^{+0.07}_{-0.11}$ & $19.12^{+0.22}_{-0.06}$ \\
    ${\rm H_2\, J=2}$ & b\,km/s &  $2.01^{+0.08}_{-0.66}$ & $0.92^{+0.71}_{-0.24}$ & $1.98^{+0.14}_{-0.52}$ \\
                      & $\log N$ &  $17.16^{+0.14}_{-0.12}$ & $18.37^{+0.08}_{-0.17}$ & $17.69^{+0.51}_{-0.17}$ \\
    ${\rm H_2\, J=3}$ & b\,km/s & $2.01^{+0.13}_{-0.50}$ &$2.03^{+0.08}_{-1.04}$ & $2.02^{+0.13}_{-0.24}$ \\
                      & $\log N$ & $16.72^{+0.35}_{-0.13}$ & $18.32^{+0.09}_{-0.15}$ & $17.45^{+0.67}_{-0.18}$ \\
    ${\rm H_2\, J=4}$ & b\,km/s & -- &$2.01^{+0.14}_{-0.89}$ & $2.17^{+0.12}_{-0.12}$\\
    				  & $\log N$ & $14.72^{+0.59}_{-0.25}$ & $17.27^{+0.15}_{-0.83}$ & $15.9^{+0.4}_{-0.3}$ \\
    ${\rm H_2\, J=5}$ & $\log N$ & -- & $16.82^{+0.28}_{-0.61}$ &$14.6^{+0.9}_{-0.4}$ \\
    
    \hline 
         & $\log N_{\rm tot}$ & $18.17^{+0.05}_{-0.06}$ & $19.82^{+0.09}_{-0.12}$ & $19.88^{+0.07}_{-0.16}$ \\
     \hline
     HD J=0 & b\,km/s &$0.522^{+0.601}_{-0.022}$ & $2.5^{+5.3}_{-2.0}$ &  $0.70^{+0.51}_{-0.20}$ \\
            & $\log N$ & $\lesssim 16.6$ & $14.46^{+1.08}_{-0.29}$ & $\lesssim 16.8$ \\
    \hline   
    \end{tabular}
    \begin{tablenotes}
    \item Doppler parameters H$_2$ $\rm J=4$ in 1 component and $\rm J = 5$ in 2 and 3 components were tied to H$_2$ $\rm J=3$ and $\rm J=4$, respectively.
    \end{tablenotes}
\end{table*}

\begin{figure*}
    \centering
    \includegraphics[width=\linewidth]{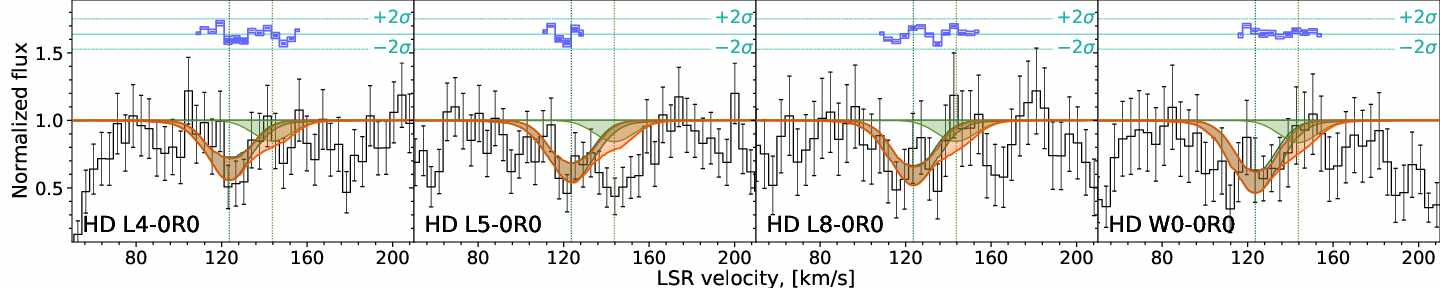}
    \caption{Fit to HD absorption lines towards AV 208 in SMC. Lines are the same as for \ref{fig:lines_HD_Sk67_2}.
    }
    \label{fig:lines_H2_AV208}
\end{figure*}

\begin{figure*}
    \centering
    \includegraphics[width=\linewidth]{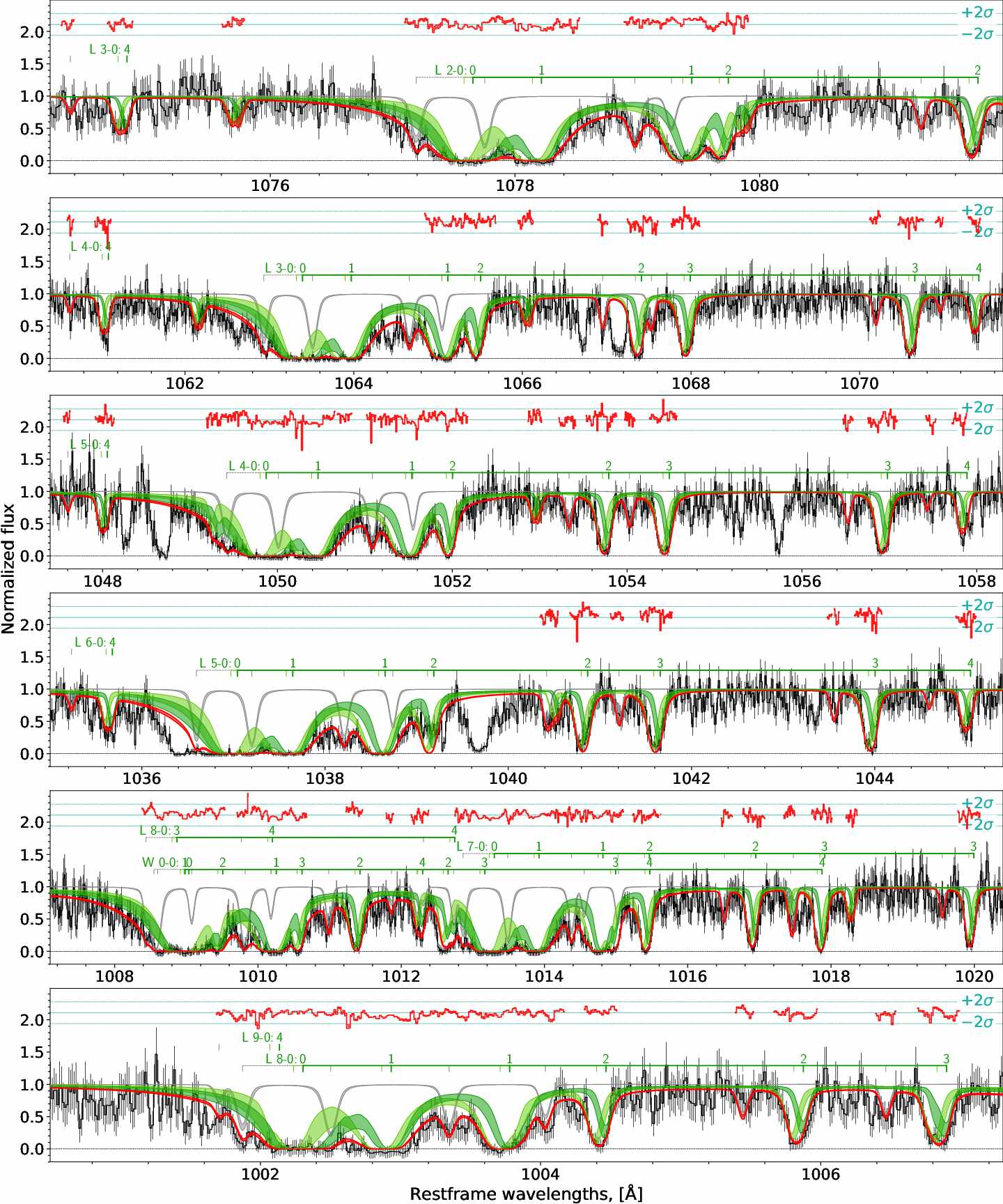}
    \caption{Fit to H2 absorption lines towards AV 208 in SMC. Lines are the same as for \ref{fig:lines_H2_Sk67_2}.
    }
    \label{fig:lines_HD_AV208}
\end{figure*}

\begin{table*}
    \caption{Fit results of H$_2$ lines towards AV 210}
    \label{tab:AV210}
    \begin{tabular}{ccccccc}
    \hline
    \hline
    species & comp & 1 & 2 & 3 & 4 & 5 \\
            & z &$-0.000082(^{+3}_{-18})$ & $0.0000308(^{+9}_{-14})$ & $0.0004312(^{+37}_{-14})$ & $0.0005496(^{+28}_{-12})$ & $0.0006074(^{+26}_{-73})$ \\
    \hline 
     ${\rm H_2\, J=0}$ & b\,km/s & $0.74^{+0.19}_{-0.11}$ & $1.8^{+0.9}_{-0.4}$ & $0.77^{+0.20}_{-0.16}$ & $1.33^{+0.13}_{-0.64}$ & $0.57^{+0.15}_{-0.05}$\\
                       & $\log N$ & $15.95^{+0.33}_{-0.27}$ & $17.50^{+0.07}_{-0.08}$ & $19.174^{+0.014}_{-0.031}$ & $17.51^{+0.31}_{-0.51}$ & $17.19^{+0.21}_{-0.31}$\\
    ${\rm H_2\, J=1}$ & b\,km/s & $0.81^{+0.36}_{-0.10}$ & $2.51^{+0.84}_{-0.24}$ &$1.4^{+1.2}_{-0.3}$ & $3.40^{+0.28}_{-0.22}$ &$0.72^{+0.25}_{-0.11}$ \\
                      & $\log N$ & $15.59^{+0.60}_{-0.06}$ & $17.76^{+0.04}_{-0.21}$ & $18.783^{+0.031}_{-0.031}$ & $16.91^{+0.33}_{-0.15}$ & $16.81^{+0.11}_{-0.34}$\\
    ${\rm H_2\, J=2}$ & b\,km/s & $1.8^{+1.5}_{-0.4}$ & $3.56^{+0.42}_{-0.19}$ & $3.57^{+0.25}_{-0.17}$ & $3.48^{+0.41}_{-0.29}$ & $0.78^{+0.40}_{-0.11}$\\
                      & $\log N$ & $15.0^{+0.4}_{-0.4}$  & $16.48^{+0.09}_{-0.74}$ & $17.60^{+0.08}_{-0.10}$ &$15.82^{+0.39}_{-0.12}$ & $13.1^{+0.8}_{-0.7}$ \\
    ${\rm H_2\, J=3}$ & b\,km/s & $3.2^{+0.5}_{-1.3}$  & $3.40^{+0.24}_{-0.20}$ &$3.59^{+0.19}_{-0.19}$ &$4.66^{+0.30}_{-1.10}$ &$1.66^{+0.07}_{-0.41}$ \\
                      & $\log N$ & $14.18^{+0.25}_{-0.26}$ & $15.29^{+0.42}_{-0.09}$ & $16.84^{+0.14}_{-0.23}$ & $15.46^{+0.43}_{-0.07}$ & $12.3^{+1.2}_{-1.5}$\\
    ${\rm H_2\, J=4}$ & b\,km/s &$3.7^{+0.6}_{-0.8}$ & -- & -- & $15.1^{+1.5}_{-2.5}$ &-- \\
    				  & $\log N$ & $13.5^{+0.4}_{-0.5}$ &$14.71^{+0.13}_{-0.05}$ & $14.35^{+0.13}_{-0.06}$ & $14.527^{+0.153}_{-0.020}$ & $11.2^{+0.9}_{-0.9}$\\
    ${\rm H_2\, J=5}$ & $\log N$ & $13.5^{+0.6}_{-0.3}$ & -- & -- &$14.63^{+0.04}_{-0.10}$ &$11.74^{+0.24}_{-1.26}$ \\
    
    \hline 
         & $\log N_{\rm tot}$ & $16.15^{+0.32}_{-0.16}$ & $17.96^{+0.03}_{-0.12}$ & $19.33^{+0.01}_{-0.02}$ & $17.62^{+0.26}_{-0.34}$ & $17.34^{+0.16}_{-0.22}$ \\
     \hline
     HD J=0 & b\,km/s & $0.72^{+0.20}_{-0.10}$ & $1.8^{+1.0}_{-0.4}$ & $0.77^{+0.19}_{-0.15}$ &  $1.12^{+0.23}_{-0.36}$ & $0.59^{+0.16}_{-0.06}$ \\
            & $\log N$ & $\lesssim 15.6$ & $\lesssim 14.3$ & $\lesssim 16.1$ & $\lesssim 16.6$ & $\lesssim 16.5$ \\
    \hline   
    \end{tabular}
    \begin{tablenotes}
    \item Doppler parameters H$_2$ $\rm J=4$ in 2, 3 and 5 components and $\rm J = 5$ in 1, 4 and 5 components were tied to H$_2$ $\rm J=3$ and $\rm J=4$, respectively.
    \end{tablenotes}
\end{table*}

\begin{figure*}
    \centering
    \includegraphics[width=\linewidth]{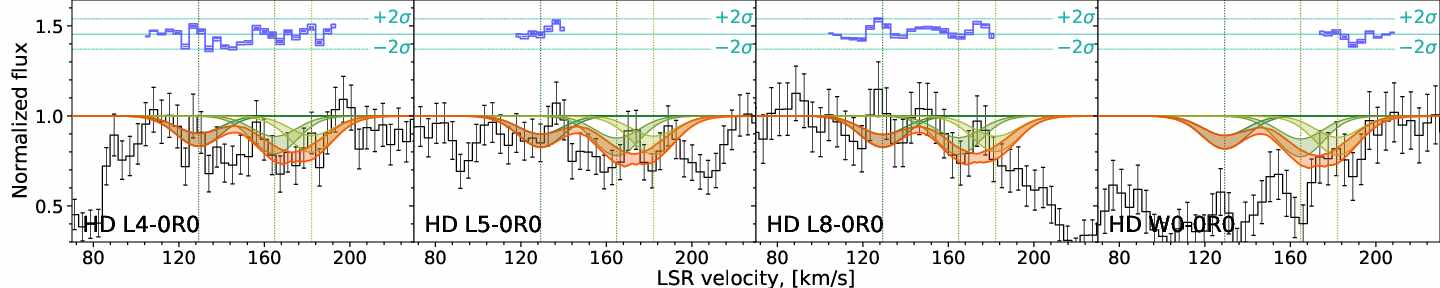}
    \caption{Fit to HD absorption lines towards AV 210 in SMC. Lines are the same as for \ref{fig:lines_HD_Sk67_2}.
    }
    \label{fig:lines_HD_AV210}
\end{figure*}

\begin{figure*}
    \centering
    \includegraphics[width=\linewidth]{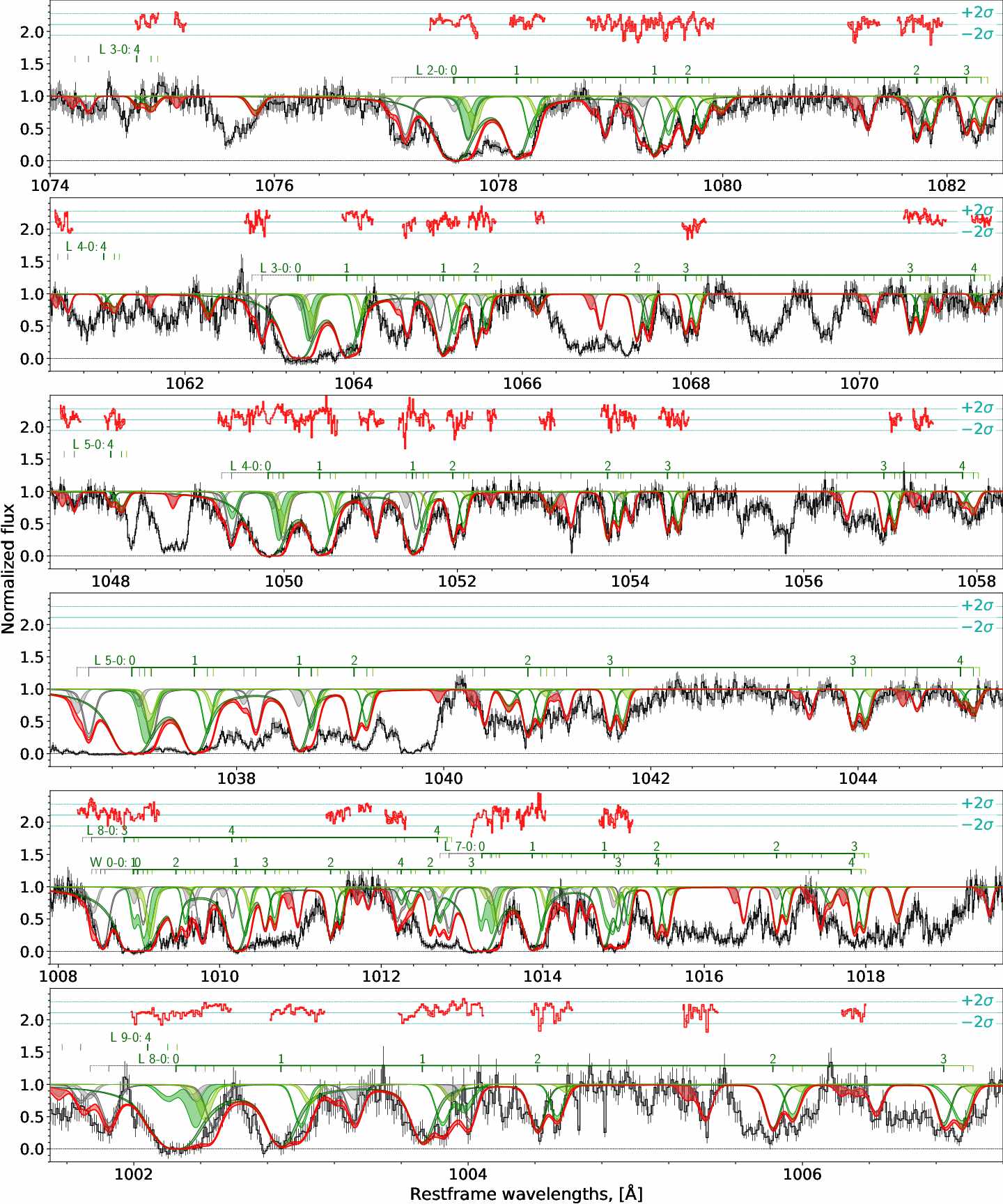}
    \caption{Fit to H2 absorption lines towards AV 210 in SMC. Lines are the same as for \ref{fig:lines_H2_Sk67_2}.
    }
    \label{fig:lines_H2_AV210}
\end{figure*}

\begin{table*}
    \caption{Fit results of H$_2$ lines towards AV 215}
    \label{tab:AV215}
    \begin{tabular}{cccccc}
    \hline
    \hline
    species & comp & 1 & 2 & 3 & 4 \\
            & z & $-0.000024(^{+9}_{-10})$ & $0.0000459(^{+17}_{-18})$ & $0.0004210(^{+64}_{-23})$ & $0.000486(^{+4}_{-4})$ \\
    \hline 
     ${\rm H_2\, J=0}$ & b\,km/s & $0.85^{+0.64}_{-0.28}$ & $3.0^{+0.8}_{-1.0}$ &$1.9^{+1.0}_{-1.3}$ &  $2.5^{+1.0}_{-1.6}$\\
                       & $\log N$ & $17.6^{+0.3}_{-0.6}$ & $18.65^{+0.07}_{-0.08}$ & $18.85^{+0.08}_{-0.21}$ & $19.24^{+0.07}_{-0.08}$\\
    ${\rm H_2\, J=1}$ & b\,km/s & $1.5^{+0.5}_{-0.8}$ & $3.6^{+0.4}_{-0.5}$ & $4.1^{+0.7}_{-1.7}$ & $2.0^{+2.4}_{-0.3}$\\
                      & $\log N$ & $14.86^{+1.47}_{-0.22}$ & $18.15^{+0.09}_{-0.11}$ & $19.20^{+0.06}_{-0.13}$ & $19.18^{+0.09}_{-0.06}$\\
    ${\rm H_2\, J=2}$ & b\,km/s &$6.0^{+1.1}_{-1.9}$ & $3.6^{+0.5}_{-0.4}$ & $4.6^{+0.9}_{-0.6}$ & $5.4^{+0.7}_{-0.5}$\\
                      & $\log N$ & $14.01^{+0.27}_{-0.26}$ & $16.52^{+0.48}_{-0.32}$ & $17.90^{+0.15}_{-0.16}$ & $17.52^{+0.23}_{-0.13}$\\
    ${\rm H_2\, J=3}$ & b\,km/s & $6.8^{+1.8}_{-1.1}$ & $3.78^{+0.67}_{-0.30}$ & $6.0^{+1.1}_{-0.7}$ & $5.5^{+0.5}_{-0.6}$\\
                      & $\log N$ & $13.79^{+0.24}_{-0.34}$ & $15.94^{+0.30}_{-0.32}$ & $17.33^{+0.18}_{-0.83}$ & $16.99^{+0.18}_{-0.24}$\\
    ${\rm H_2\, J=4}$ & b\,km/s & $8.2^{+1.6}_{-1.7}$ & $4.1^{+0.6}_{-0.6}$ & $7.1^{+0.5}_{-1.3}$ & $5.6^{+0.6}_{-0.6}$\\
                      & $\log N$ & $12.89^{+0.21}_{-0.95}$ & $14.44^{+0.09}_{-0.14}$ & $14.97^{+0.10}_{-0.10}$ & $14.88^{+0.07}_{-0.13}$\\
    ${\rm H_2\, J=4}$ & b\,km/s & -- & -- & -- & $5.9^{+1.5}_{-0.4}$\\ 
                      & $\log N$ &$12.2^{+1.1}_{-0.4}$ & $13.94^{+0.20}_{-0.25}$ & $14.60^{+0.08}_{-0.11}$ &$14.61^{+0.07}_{-0.14}$ \\
    ${\rm H_2\, J=4}$ & $\log N$ & -- & -- & $13.86^{+0.11}_{-0.10}$ & $13.89^{+0.15}_{-0.64}$\\ 
     \hline 
         & $\log N_{\rm tot}$ & $17.60^{+0.30}_{-0.60}$ & $18.77^{+0.06}_{-0.07}$ & $19.38^{+0.05}_{-0.10}$ & $19.52^{+0.06}_{-0.05}$ \\
    \hline
    HD J=0 & b\,km/s & $0.90^{+0.50}_{-0.30}$ & $2.8^{+1.3}_{-0.8}$ &  $1.2^{+0.7}_{-0.7}$ &  $0.62^{+2.33}_{-0.12}$ \\
            & $\log N$ & $\lesssim 16.4$ & $\lesssim 15.5$ & $\lesssim 16.6$ & $\lesssim 16.5$ \\
    \hline   
    \end{tabular}
    \begin{tablenotes}
    \item Doppler parameters H$_2$ $\rm J=4$ in 1 and 2 components, $\rm J=5, 6$ in 3 component and $\rm J=6$ in 4 component were tied to H$_2$ $\rm J=3$, $\rm J=4$ and $\rm J=6$, respectively.
    \end{tablenotes}
\end{table*}

\begin{figure*}
    \centering
    \includegraphics[width=\linewidth]{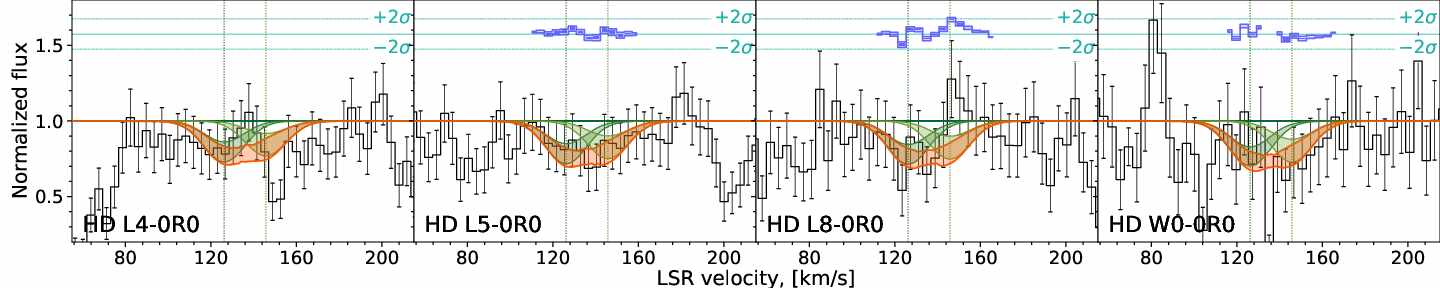}
    \caption{Fit to HD absorption lines towards AV 215 in SMC. Lines are the same as for \ref{fig:lines_HD_Sk67_2}.
    }
    \label{fig:lines_HD_AV215}
\end{figure*}

\begin{figure*}
    \centering
    \includegraphics[width=\linewidth]{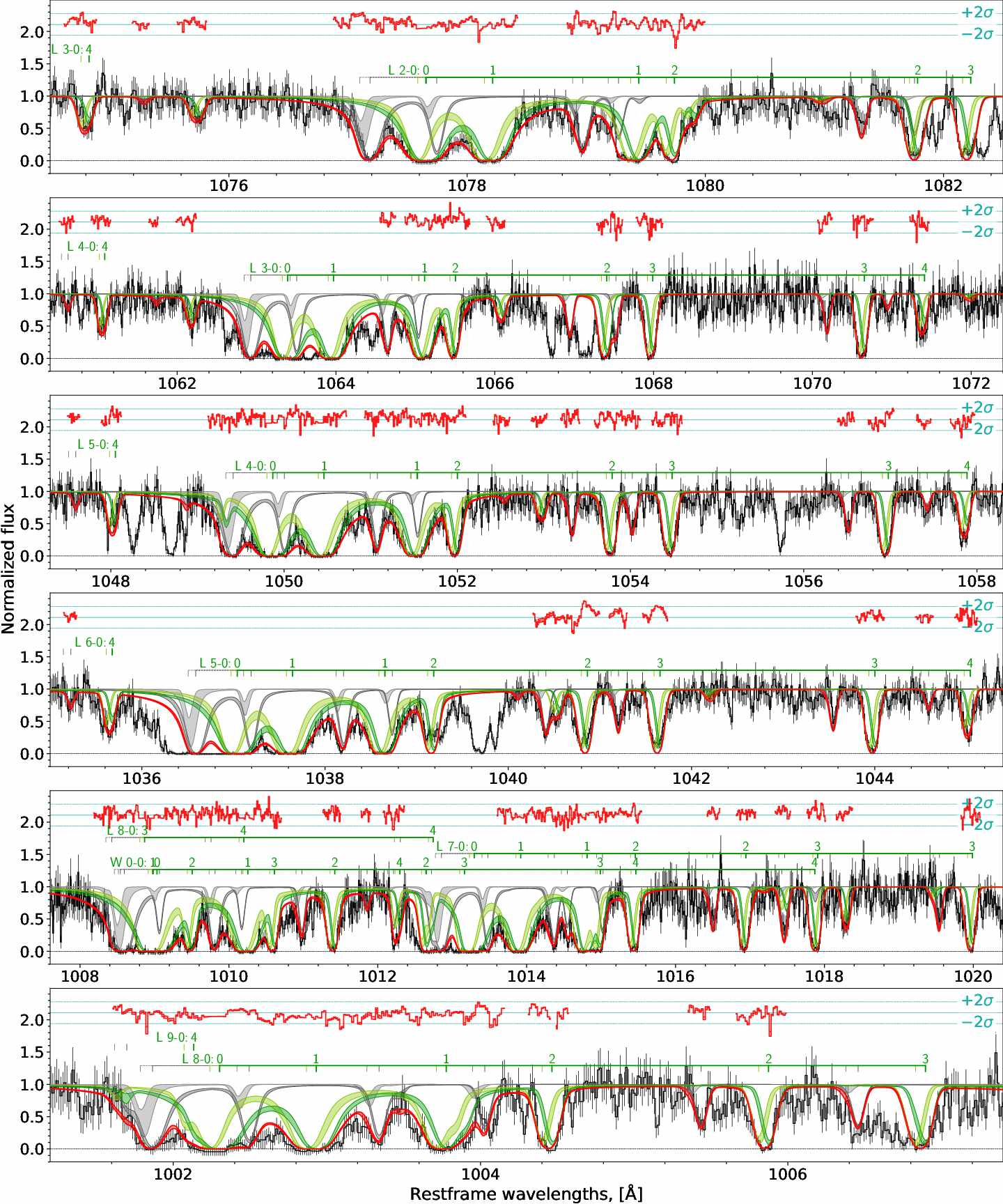}
    \caption{Fit to H2 absorption lines towards AV 215 in SMC. Lines are the same as for \ref{fig:lines_H2_Sk67_2}.
    }
    \label{fig:lines_H2_AV215}
\end{figure*}

\begin{table*}
    \caption{Fit results of H$_2$ lines towards AV 216}
    \label{tab:AV216}
    \begin{tabular}{cccccc}
    \hline
    \hline
    species & comp & 1 & 2 & 3 & 4 \\
            & z & $0.0000450(^{+13}_{-10})$ &$0.0004160(^{+44}_{-28})$ & $0.0004641(^{+23}_{-17})$ & $0.000542(^{+9}_{-5})$\\
    \hline 
     ${\rm H_2\, J=0}$ & b\,km/s & $2.06^{+0.18}_{-0.50}$ & $0.95^{+0.43}_{-0.26}$& $2.1^{+0.5}_{-0.7}$  & $0.519^{+0.220}_{-0.019}$\\
                       & $\log N$ & $17.46^{+0.06}_{-0.08}$ & $14.7^{+1.5}_{-1.9}$& $18.451^{+0.051}_{-0.030}$ & $15.5^{+0.5}_{-0.7}$\\
    ${\rm H_2\, J=1}$ & b\,km/s & $2.10^{+0.15}_{-0.26}$ & $1.72^{+0.26}_{-0.43}$& $2.76^{+0.13}_{-0.49}$ &$0.74^{+0.18}_{-0.15}$ \\
                      & $\log N$ & $17.64^{+0.07}_{-0.04}$ & $17.62^{+0.17}_{-0.27}$& $18.33^{+0.10}_{-0.04}$ & $15.7^{+0.4}_{-0.4}$\\
    ${\rm H_2\, J=2}$ & b\,km/s & $2.31^{+0.10}_{-0.10}$  & $2.03^{+0.10}_{-0.42}$& $2.69^{+0.21}_{-0.14}$ &  $1.07^{+0.14}_{-0.39}$\\
                      & $\log N$ & $16.73^{+0.12}_{-0.06}$ & $16.60^{+0.38}_{-0.22}$& $17.23^{+0.14}_{-0.18}$ & $14.89^{+0.26}_{-0.41}$\\
    ${\rm H_2\, J=3}$ & b\,km/s & $2.12^{+0.13}_{-0.07}$ & $2.16^{+0.12}_{-0.18}$ & $3.01^{+0.13}_{-0.14}$ & $1.56^{+0.36}_{-0.23}$\\
                      & $\log N$ &  $16.11^{+0.10}_{-0.32}$ & $16.40^{+0.21}_{-0.36}$ &$16.59^{+0.15}_{-0.21}$ & $14.92^{+0.36}_{-0.16}$\\
    ${\rm H_2\, J=4}$ & $\log N$ & $13.96^{+0.17}_{-0.35}$ &$14.62^{+0.24}_{-0.23}$ & $14.07^{+0.18}_{-0.20}$ & $14.26^{+0.33}_{-0.14}$\\
     \hline 
         & $\log N_{\rm tot}$ & $17.90^{+0.05}_{-0.04}$ & $17.68^{+0.16}_{+0.22}$ & $18.71^{+0.05}_{-0.02}$ & $15.99^{+0.31}_{-0.22}$ \\
    \hline
    HD J=0 & b\,km/s & $2.03^{+0.17}_{-0.76}$ & $0.88^{+0.50}_{-0.23}$ & $2.2^{+0.4}_{-1.1}$ & $0.59^{+0.17}_{-0.09}$ \\
           & $\log N$ & $\lesssim 16.0$ & $\lesssim 15.7$ & $\lesssim 16.0$ & $\lesssim 16.2$ \\     
    \hline   
    \end{tabular}
    \begin{tablenotes}
    \item Doppler parameters H$_2$ $\rm J=4$ in all of the components were tied to H$_2$ $\rm J=3$.
    \end{tablenotes}
\end{table*}

\begin{figure*}
    \centering
    \includegraphics[width=\linewidth]{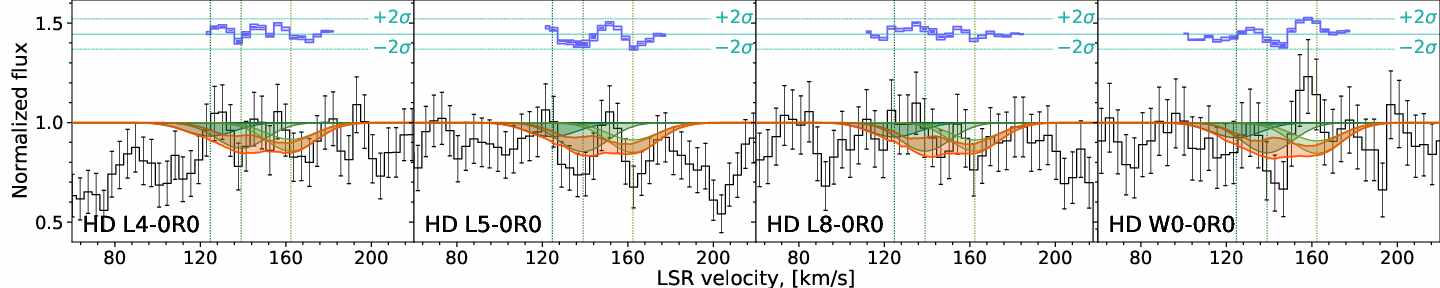}
    \caption{Fit to HD absorption lines towards AV 216 in SMC. Lines are the same as for \ref{fig:lines_HD_Sk67_2}.
    }
    \label{fig:lines_HD_AV216}
\end{figure*}

\begin{figure*}
    \centering
    \includegraphics[width=\linewidth]{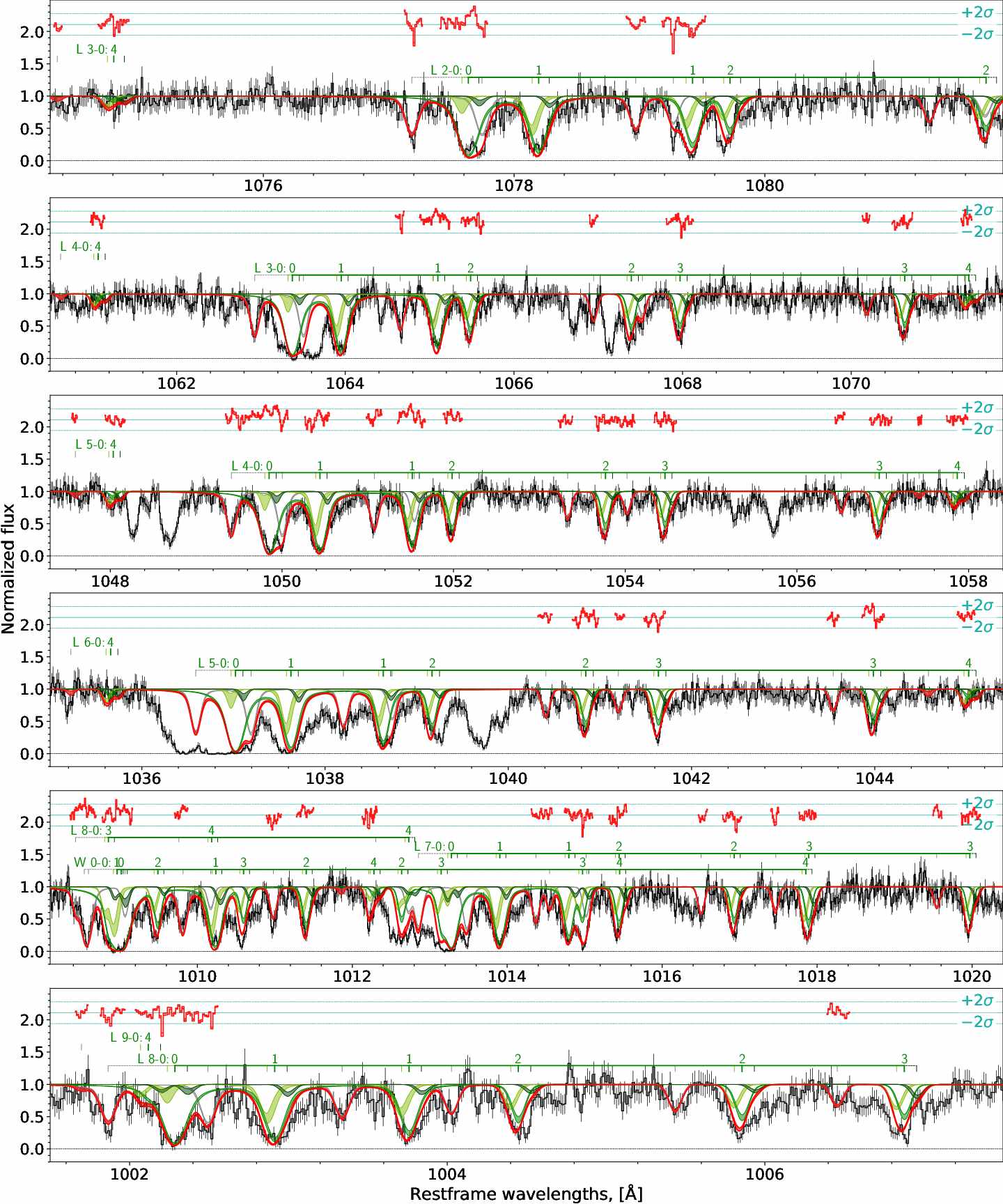}
    \caption{Fit to H2 absorption lines towards AV 216 in SMC. Lines are the same as for \ref{fig:lines_H2_Sk67_2}.
    }
    \label{fig:lines_H2_AV216}
\end{figure*}

\begin{table*}
    \caption{Fit results of H$_2$ lines towards NGC 346 -637}
    \label{tab:NGC346_637}
    \begin{tabular}{cccc}
    \hline
    \hline
    species & comp & 1 & 2   \\
            & z & $0.0000727(^{+11}_{-11})$ & $0.0005417(^{+19}_{-22})$ \\
    \hline 
     ${\rm H_2\, J=0}$ & b\,km/s &$0.81^{+0.56}_{-0.31}$ & $0.522^{+0.559}_{-0.022}$ \\
                       & $\log N$ & $17.83^{+0.05}_{-0.06}$ & $18.963^{+0.042}_{-0.025}$ \\
    ${\rm H_2\, J=1}$ & b\,km/s & $1.6^{+0.4}_{-0.7}$ &  $1.1^{+0.4}_{-0.5}$ \\
                      & $\log N$ & $18.14^{+0.04}_{-0.04}$ & $19.226^{+0.015}_{-0.021}$ \\
    ${\rm H_2\, J=2}$ & b\,km/s & $3.04^{+0.12}_{-0.24}$ & $1.18^{+0.63}_{-0.31}$ \\
                      & $\log N$ & $17.50^{+0.05}_{-0.04}$ & $17.92^{+0.06}_{-0.04}$ \\
    ${\rm H_2\, J=3}$ & b\,km/s & $2.93^{+0.15}_{-0.24}$ &$2.54^{+0.31}_{-0.38}$  \\
                      & $\log N$ & $17.47^{+0.07}_{-0.07}$ & $17.70^{+0.10}_{-0.11}$ \\
    ${\rm H_2\, J=4}$ & $\log N$ &  $15.75^{+0.19}_{-0.22}$ & $16.30^{+0.31}_{-0.37}$\\
    ${\rm H_2\, J=5}$ & $\log N$ & $13.9^{+0.4}_{-0.6}$ & $15.9^{+0.7}_{-0.5}$\\	 			  
    
    \hline 
         & $\log N_{\rm tot}$ & $18.43^{+0.03}_{-0.03}$ & $19.44^{+0.02}_{-0.02}$ \\
     \hline 
     HD J=0 & b\,km/s & $1.20^{+0.30}_{-0.40}$ & $0.522^{+0.519}_{-0.022}$ \\
            & $\log N$ & $\lesssim 16.6$ & $\lesssim 17.0$ \\
    \hline   
    \end{tabular}
    \begin{tablenotes}
    \item Doppler parameters H$_2$ $\rm J=4, 5$ were tied to H$_2$ $\rm J=3$.
    \end{tablenotes}
\end{table*}

\begin{figure*}
    \centering
    \includegraphics[width=\linewidth]{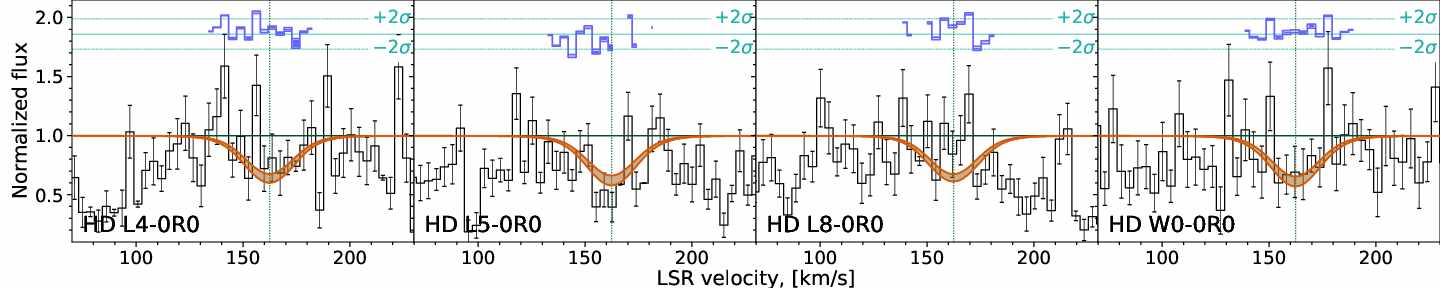}
    \caption{Fit to HD absorption lines towards NGC 346-637 in SMC. Lines are the same as for \ref{fig:lines_HD_Sk67_2}.
    }
    \label{fig:lines_HD_NGC346_637}
\end{figure*}

\begin{figure*}
    \centering
    \includegraphics[width=\linewidth]{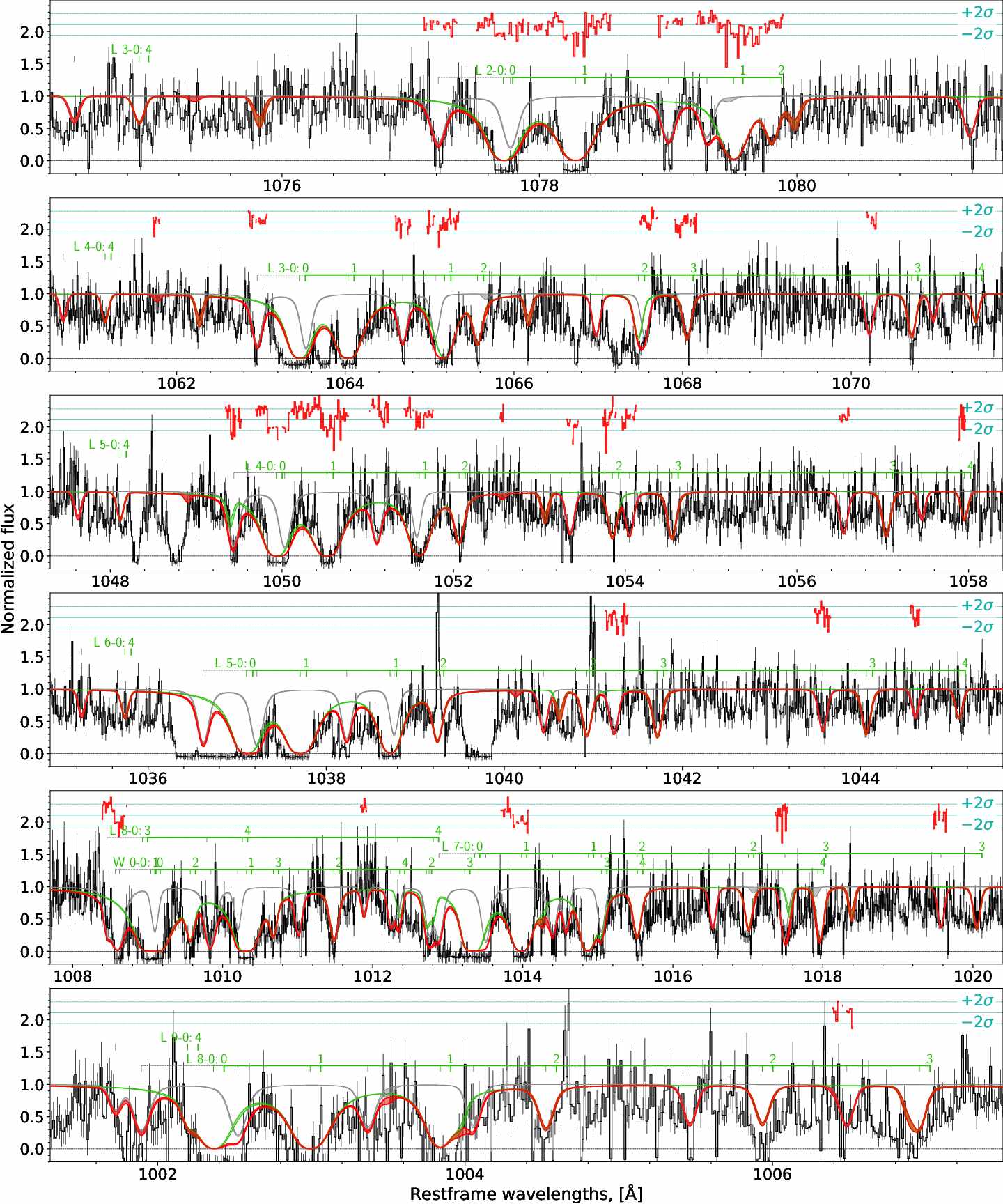}
    \caption{Fit to H2 absorption lines towards NGC 346-637 in SMC. Lines are the same as for \ref{fig:lines_H2_Sk67_2}.
    }
    \label{fig:lines_H2_NGC346_637}
\end{figure*}

\begin{table*}
    \caption{Fit results of H$_2$ lines towards AV 243}
    \label{tab:AV243}
    \begin{tabular}{ccccc}
    \hline
    \hline
    species & comp & 1 & 2 & 3  \\
            & z & $0.0000479(^{+9}_{-12})$ & $0.0004284(^{+25}_{-13})$ & $0.0004932(^{+12}_{-24})$ \\
    \hline 
     ${\rm H_2\, J=0}$ & b\,km/s & $2.2^{+0.5}_{-0.6}$ & $0.9^{+0.4}_{-0.4}$ & $1.2^{+0.6}_{-0.5}$ \\
                       & $\log N$ & $17.17^{+0.11}_{-0.13}$ & $18.702^{+0.030}_{-0.020}$ & $13.4^{+1.7}_{-1.7}$\\
    ${\rm H_2\, J=1}$ & b\,km/s & $2.63^{+0.41}_{-0.29}$ &$1.3^{+0.6}_{-0.4}$ & $1.95^{+0.43}_{-0.29}$\\
                      & $\log N$ & $17.55^{+0.11}_{-0.08}$ & $18.591^{+0.027}_{-0.049}$ & $17.22^{+0.19}_{-0.34}$\\
    ${\rm H_2\, J=2}$ & b\,km/s & $3.00^{+0.06}_{-0.39}$ & $1.7^{+0.7}_{-0.3}$ & $2.40^{+0.29}_{-0.27}$\\
                      & $\log N$ &  $16.10^{+0.20}_{-0.18}$ &$16.82^{+0.23}_{-0.43}$ & $15.97^{+0.26}_{-0.39}$\\
    ${\rm H_2\, J=3}$ & b\,km/s & $3.03^{+0.18}_{-0.28}$ & $2.5^{+0.6}_{-0.3}$ & $3.01^{+0.13}_{-0.23}$\\
                      & $\log N$ & $14.59^{+0.14}_{-0.09}$ & $15.30^{+0.39}_{-0.26}$ & $15.61^{+0.18}_{-0.20}$\\
    ${\rm H_2\, J=4}$ & $\log N$ &$14.01^{+0.09}_{-0.25}$ & $14.11^{+0.14}_{-0.15}$ &$14.28^{+0.08}_{-0.16}$ \\
    ${\rm H_2\, J=5}$ & $\log N$ & -- & $13.64^{+0.25}_{-0.22}$ &  $14.17^{+0.09}_{-0.20}$\\
     \hline 
         & $\log N_{\rm tot}$ & $17.71^{+0.08}_{-0.06}$ & $18.95^{+0.02}_{-0.02}$ & $17.25^{+0.18}_{-0.30}$ \\
     \hline
     HD J=0 & b\,km/s &$1.9^{+0.4}_{-1.0}$ & $0.523^{+0.632}_{-0.023}$ & $1.0^{+0.6}_{-0.5}$ \\
            & $\log N$ & $\lesssim 16.0$ & $\lesssim 15.7$ & $\lesssim 16.1$ \\
    \hline   
    \end{tabular}
    \begin{tablenotes}
    \item Doppler parameters H$_2$ $\rm J=4$ in all of the components and $\rm J=5$ in 2 and 3 components were tied to H$_2$ $\rm J=3$.
    \end{tablenotes}
\end{table*}

\begin{figure*}
    \centering
    \includegraphics[width=\linewidth]{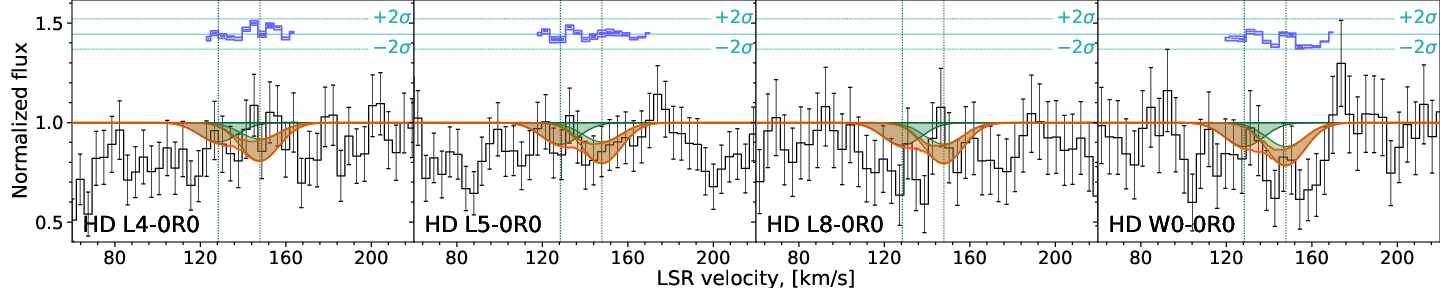}
    \caption{Fit to HD absorption lines towards AV 243 in SMC. Lines are the same as for \ref{fig:lines_HD_Sk67_2}.
    }
    \label{fig:lines_HD_AV243}
\end{figure*}

\begin{figure*}
    \centering
    \includegraphics[width=\linewidth]{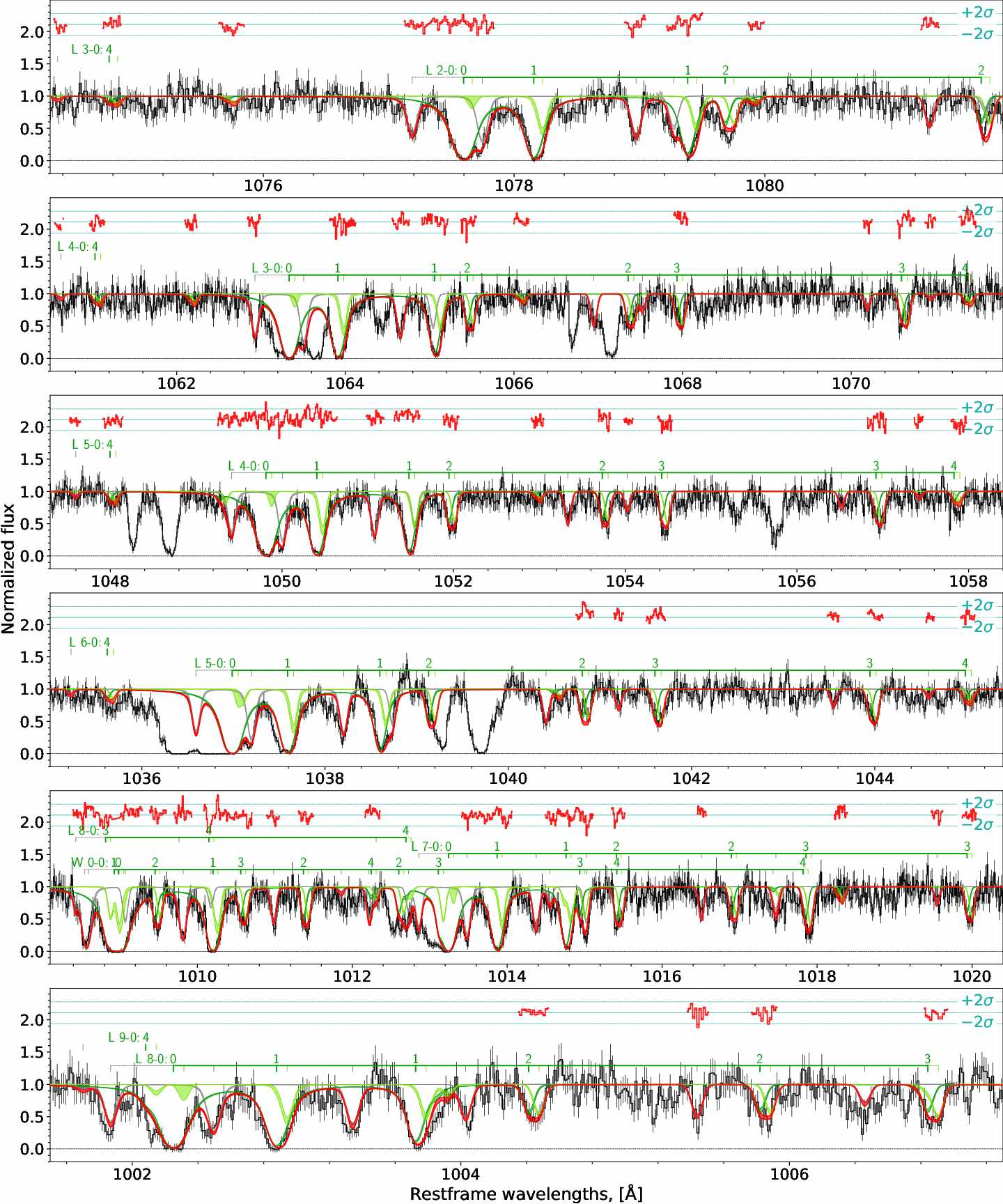}
    \caption{Fit to H2 absorption lines towards AV 243 in SMC. Lines are the same as for \ref{fig:lines_H2_Sk67_2}.
    }
    \label{fig:lines_H2_AV243}
\end{figure*}

\begin{table*}
    \caption{Fit results of H$_2$ lines towards AV 242}
    \label{tab:AV242}
    \begin{tabular}{ccccccc}
    \hline
    \hline
    species & comp & 1 & 2 & 3 & 4 & 5  \\
            & z & $0.0000531(^{+9}_{-7})$ & $0.0003297(^{+13}_{-9})$ & $0.0004411(^{+23}_{-26})$ & $0.0005468(^{+9}_{-25})$ & $0.0006079(^{+25}_{-46})$ \\
    \hline 
     ${\rm H_2\, J=0}$ & b\,km/s &$3.0^{+0.4}_{-0.4}$ & $1.75^{+0.49}_{-0.22}$ & $0.68^{+0.47}_{-0.17}$ & $1.7^{+0.5}_{-0.5}$ & $0.76^{+0.25}_{-0.15}$\\
                       & $\log N$ &$17.08^{+0.08}_{-0.24}$ & $16.82^{+0.15}_{-0.17}$ & $14.02^{+0.73}_{-0.24}$ & $17.08^{+0.08}_{-0.13}$ & $16.47^{+0.18}_{-0.31}$\\
    ${\rm H_2\, J=1}$ & b\,km/s &$3.14^{+0.44}_{-0.26}$ &  $2.59^{+0.41}_{-0.20}$ & $7.5^{+2.2}_{-1.5}$ & $2.55^{+0.25}_{-0.50}$ & $0.94^{+0.33}_{-0.13}$\\
                      & $\log N$ &$17.18^{+0.20}_{-0.21}$ & $16.92^{+0.21}_{-0.30}$ & $14.652^{+0.025}_{-0.046}$ & $17.60^{+0.10}_{-0.15}$ & $16.41^{+0.21}_{-0.23}$\\
    ${\rm H_2\, J=2}$ & b\,km/s &$4.0^{+1.0}_{-0.5}$ & $2.71^{+0.62}_{-0.28}$ &$10.2^{+1.7}_{-2.9}$ & $2.40^{+0.54}_{-0.30}$ & $1.23^{+0.41}_{-0.21}$\\
                      & $\log N$ &  $15.22^{+0.34}_{-0.14}$ & $15.62^{+0.17}_{-0.46}$ & $14.13^{+0.03}_{-0.11}$ & $16.2^{+0.4}_{-0.4}$ & $14.8^{+0.3}_{-0.4}$ \\
    ${\rm H_2\, J=3}$ & b\,km/s & $8.4^{+2.5}_{-1.2}$ &$3.1^{+0.7}_{-0.6}$ & $11.0^{+3.8}_{-1.9}$ & $2.89^{+0.82}_{-0.30}$ & $1.9^{+1.3}_{-0.7}$ \\
                      & $\log N$ & $14.590^{+0.032}_{-0.031}$ & $14.61^{+0.15}_{-0.08}$ &$14.369^{+0.027}_{-0.067}$ & $15.26^{+0.65}_{-0.06}$ & $14.27^{+0.09}_{-0.21}$ \\
    ${\rm H_2\, J=4}$ & b\,km/s & -- & -- & -- & $4.7^{+2.9}_{-1.6}$ & $17.15^{+5.59}_{-9.01}$\\
                      & $\log N$ & $13.85^{+0.10}_{-0.17}$ & $13.65^{+0.19}_{-0.22}$ & $13.86^{+0.06}_{-0.18}$ & $14.16^{+0.07}_{-0.05}$ & $12.4^{+0.6}_{-0.5}$\\
    ${\rm H_2\, J=5}$ & $\log N$ & -- & -- & -- & $13.93^{+0.09}_{-0.14}$ & $12.0^{+1.1}_{-0.4}$\\
     \hline 
         & $\log N_{\rm tot}$ & $17.44^{+0.13}_{-0.14}$ & $17.18^{+0.14}_{-0.16}$ & $15.00^{+0.16}_{-0.04}$ & $17.73^{+0.08}_{-0.11}$ & $16.75^{+0.14}_{-0.17}$ \\
    \hline
    HD J=0 & b\,km/s & $2.9^{+0.6}_{-0.4}$ & $1.65^{+0.64}_{-0.18}$ & $0.54^{+0.44}_{-0.04}$ & $1.7^{+0.7}_{-0.5}$ & $0.72^{+0.29}_{-0.13}$ \\
           & $\log N$ & $\lesssim 13.6$ & $\lesssim 14.0$ & $\lesssim 15.7$ & $\lesssim 15.9$ & $\lesssim 14.9$ \\  
    \hline   
    \end{tabular}
    \begin{tablenotes}
    \item Doppler parameters H$_2$ $\rm J=4$ in the 1, 2 and 3 components were tied to H$_2$ $\rm J=3$.
    \end{tablenotes}
\end{table*}

\begin{figure*}
    \centering
    \includegraphics[width=\linewidth]{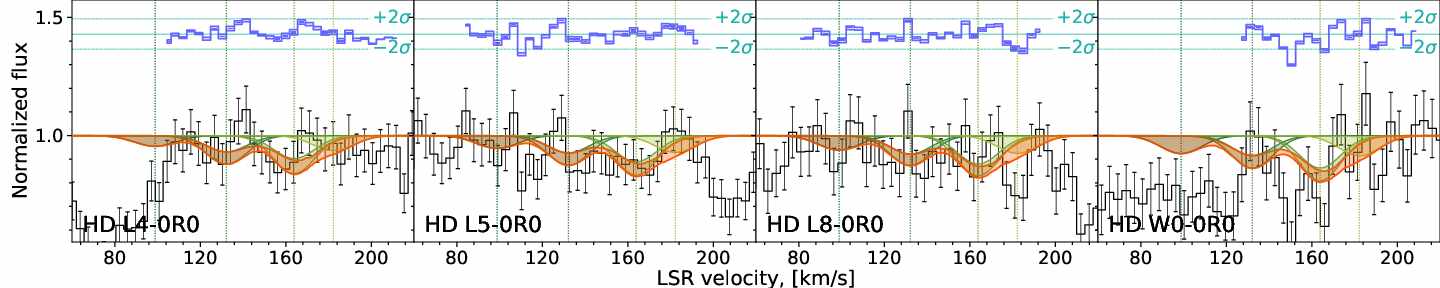}
    \caption{Fit to HD absorption lines towards AV 242 in SMC. Lines are the same as for \ref{fig:lines_HD_Sk67_2}.
    }
    \label{fig:lines_HD_AV242} 
\end{figure*}

\begin{figure*}
    \centering
    \includegraphics[width=\linewidth]{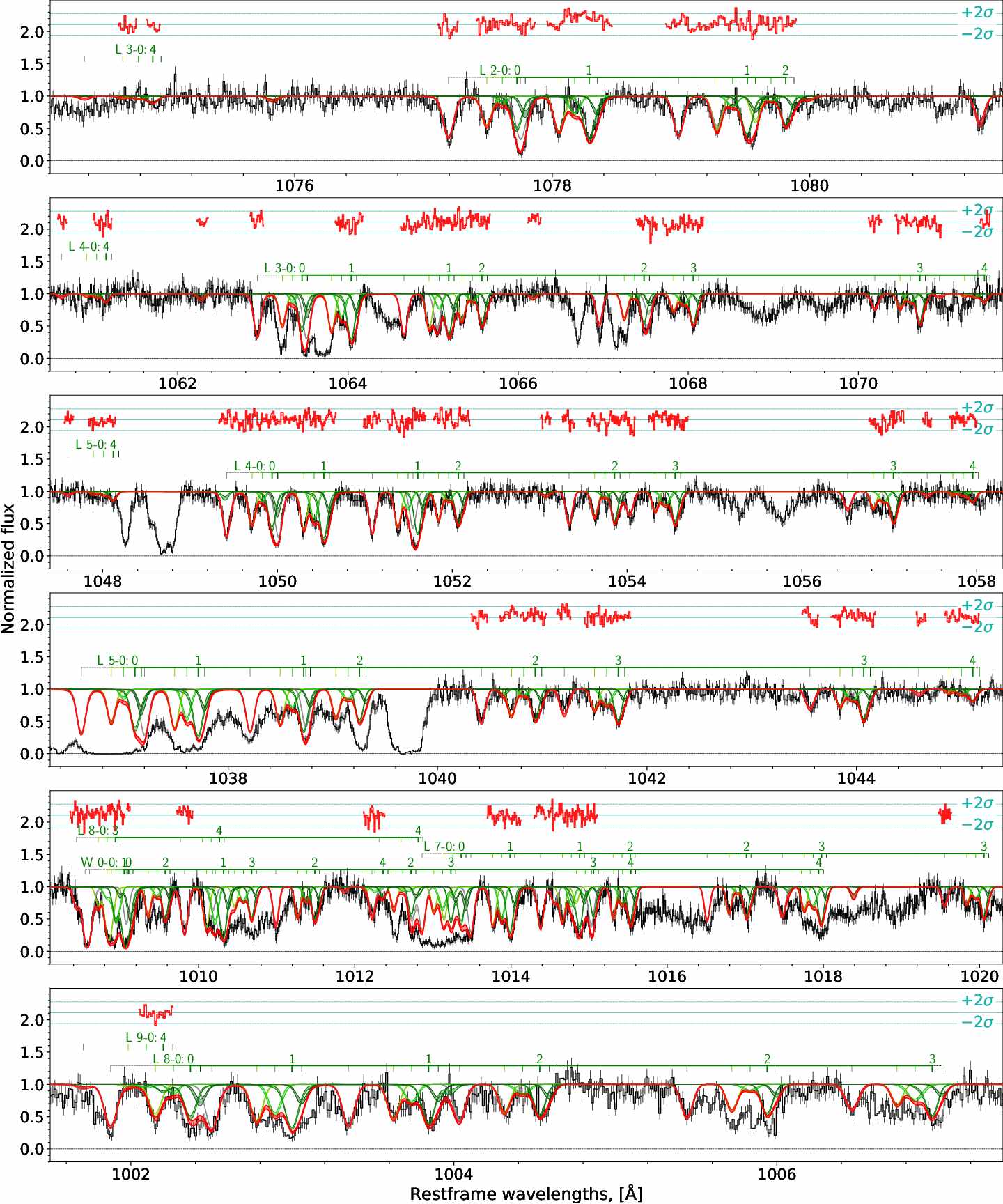}
    \caption{Fit to H2 absorption lines towards AV 242 in SMC. Lines are the same as for \ref{fig:lines_H2_Sk67_2}.
    }
    \label{fig:lines_H2_AV242}
\end{figure*}

\begin{table*}
    \caption{Fit results of H$_2$ lines towards AV 261}
    \label{tab:AV261}
    \begin{tabular}{cccccc}
    \hline
    \hline
    species & comp & 1 & 2 & 3 & 4   \\
            & z & $0.000008(^{+5}_{-16})$ &  $0.000057(^{+3}_{-4})$ & $0.000311(^{+9}_{-5})$ & $0.0004179(^{+30}_{-31})$ \\
    \hline 
     ${\rm H_2\, J=0}$ & b\,km/s & $1.7^{+0.5}_{-0.7}$ &$2.5^{+0.9}_{-0.9}$ & $1.5^{+0.8}_{-1.0}$ & $1.6^{+0.9}_{-0.9}$ \\
                       & $\log N$ & $17.54^{+0.22}_{-0.49}$ &  $17.79^{+0.11}_{-0.28}$ &$16.2^{+1.4}_{-0.7}$ & $19.33^{+0.04}_{-0.04}$\\
    ${\rm H_2\, J=1}$ & b\,km/s & $1.7^{+0.7}_{-0.6}$ & $3.5^{+0.4}_{-0.9}$ & $3.7^{+0.3}_{-1.1}$ & $2.4^{+1.5}_{-0.7}$\\
                      & $\log N$ & $17.47^{+0.28}_{-0.97}$ & $17.68^{+0.27}_{-0.23}$ & $16.2^{+0.6}_{-0.8}$ & $18.93^{+0.05}_{-0.06}$\\
    ${\rm H_2\, J=2}$ & b\,km/s &$2.0^{+0.8}_{-0.5}$ & $3.7^{+0.5}_{-0.6}$ & $4.5^{+0.8}_{-0.8}$ & $4.6^{+0.5}_{-0.5}$\\
                      & $\log N$ & $15.05^{+0.81}_{-0.23}$ & $16.3^{+0.4}_{-0.3}$ &$14.20^{+0.21}_{-0.18}$ & $16.8^{+0.5}_{-0.3}$\\
    ${\rm H_2\, J=3}$ & b\,km/s & $4.1^{+0.4}_{-1.0}$ & $4.20^{+0.27}_{-0.31}$ &$5.71^{+0.27}_{-0.96}$ & $5.19^{+0.21}_{-0.40}$\\
                      & $\log N$ & $14.05^{+0.47}_{-0.27}$ & $15.26^{+0.50}_{-0.17}$ &$14.44^{+0.18}_{-0.17}$ & $15.94^{+0.11}_{-0.42}$ \\
    ${\rm H_2\, J=4}$ & $\log N$ & $12.9^{+0.5}_{-1.8}$ & $14.1^{+0.4}_{-0.7}$ &$13.67^{+0.22}_{-0.27}$ & $14.64^{+0.10}_{-0.15}$ \\
    ${\rm H_2\, J=5}$ & $\log N$ & -- & -- & $13.95^{+0.25}_{-0.34}$ & $14.58^{+0.11}_{-0.12}$ \\
     \hline 
         & $\log N_{\rm tot}$ & $17.81^{+0.19}_{-0.35}$ & $18.05^{+0.15}_{-0.16}$ & $16.54^{+1.12}_{-0.37}$ & $19.38^{+0.03}_{}/-0.03$ \\
    \hline
    HD J=0 & b\,km/s &$1.6^{+0.5}_{-0.7}$ & $2.1^{+1.2}_{-0.7}$ & $1.1^{+0.7}_{-0.6}$ & $1.4^{+0.8}_{-0.6}$ \\
           & $\log N$ & $\lesssim 16.4$ &  $\lesssim 16.8$ & $\lesssim 16.5$ & $\lesssim 16.7$ \\
    \hline   
    \end{tabular}
    \begin{tablenotes}
    \item Doppler parameters H$_2$ $\rm J=4, 5$  were tied to H$_2$ $\rm J=3$.
    \end{tablenotes}
\end{table*}

\begin{figure*}
    \centering
    \includegraphics[width=\linewidth]{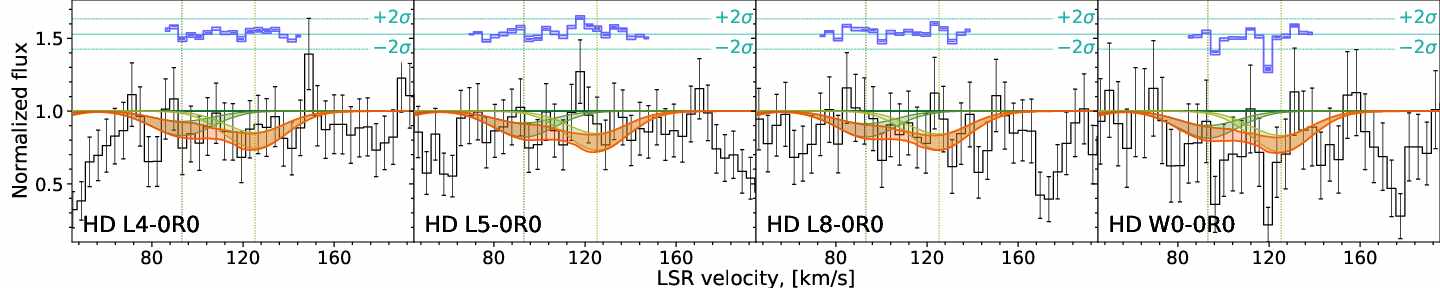}
    \caption{Fit to HD absorption lines towards AV 261 in SMC. Lines are the same as for \ref{fig:lines_HD_Sk67_2}.
    }
    \label{fig:lines_HD_AV261}
\end{figure*}

\begin{figure*}
    \centering
    \includegraphics[width=\linewidth]{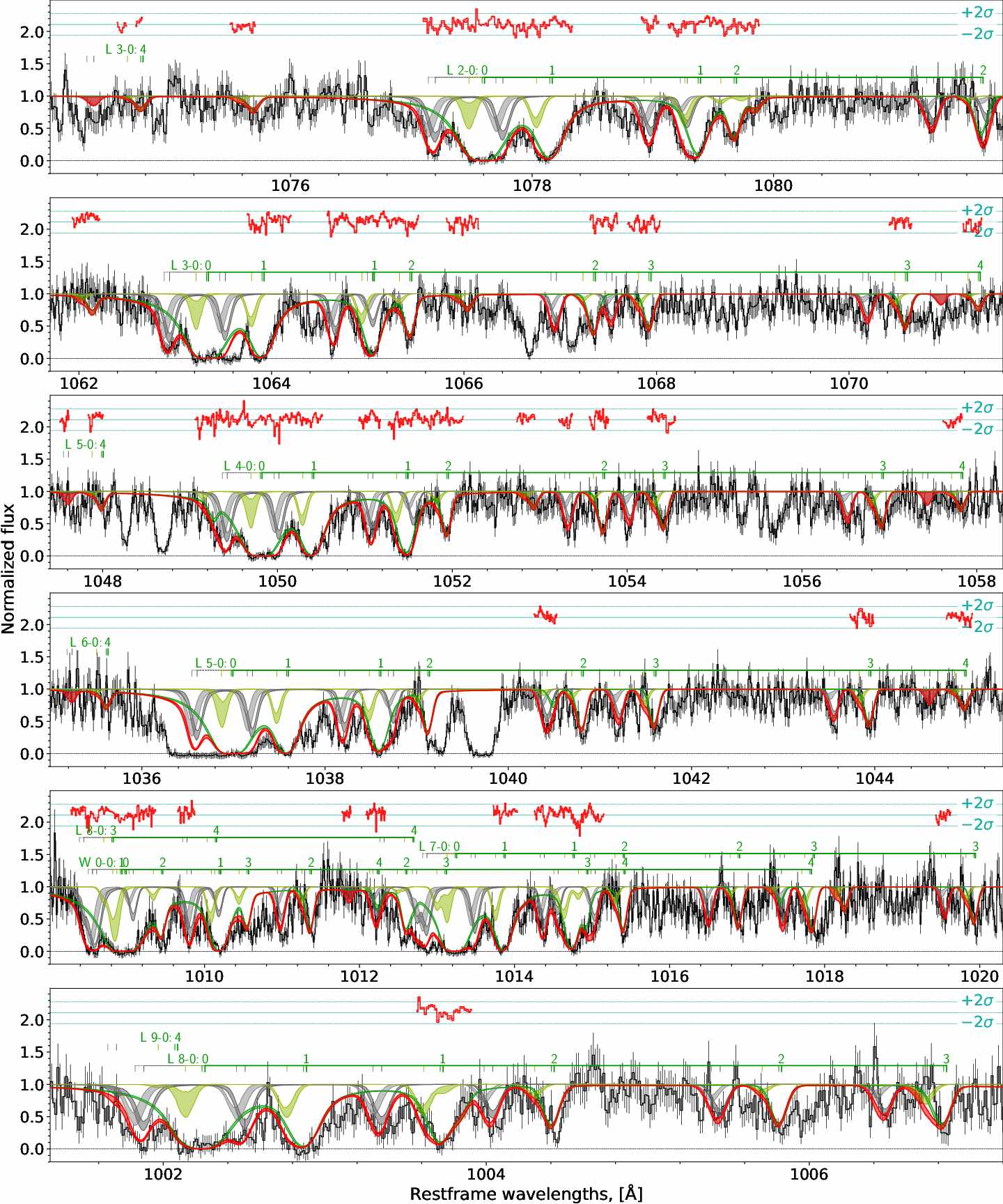}
    \caption{Fit to H2 absorption lines towards AV 161 in SMC. Lines are the same as for \ref{fig:lines_H2_Sk67_2}.
    }
    \label{fig:lines_H2_AV261}
\end{figure*}

\begin{table*}
    \caption{Fit results of H$_2$ lines towards AV 266}
    \label{tab:AV266}
    \begin{tabular}{ccccccc}
    \hline
    \hline
    species & comp & 1 & 2 & 3 & 4 & 5  \\
            & z & $0.0000329(^{+23}_{-31})$ & $0.000068(^{+7}_{-6})$ & $0.000355(^{+7}_{-4})$ & $0.0004245(^{+13}_{-20})$ & $0.000442(^{+6}_{-3})$ \\
    \hline 
     ${\rm H_2\, J=0}$ & b\,km/s & $0.70^{+1.06}_{-0.20}$ & $2.7^{+0.9}_{-1.6}$ 7 $4.6^{+0.8}_{-1.5}$ & $0.9^{+0.6}_{-0.3}$ & $1.0^{+1.0}_{-0.4}$ \\
                       & $\log N$ & $18.37^{+0.04}_{-0.07}$ & $17.48^{+0.21}_{-0.40}$ & $17.00^{+0.28}_{-0.72}$ & $18.891^{+0.025}_{-0.024}$ & $17.1^{+0.7}_{-0.5}$ \\
    ${\rm H_2\, J=1}$ & b\,km/s &$2.6^{+0.4}_{-1.2}$ & $2.7^{+1.0}_{-1.1}$ & $5.0^{+1.3}_{-1.4}$ & $2.5^{+0.6}_{-1.3}$ & $1.9^{+0.6}_{-1.2}$ \\
                      & $\log N$ & $18.10^{+0.05}_{-0.05}$ & $15.8^{+0.6}_{-0.5}$ & $15.28^{+0.41}_{-0.11}$ & $18.860^{+0.014}_{-0.022}$ &  $17.20^{+0.28}_{-0.55}$\\
    ${\rm H_2\, J=2}$ & b\,km/s & $2.4^{+0.6}_{-0.8}$ & $3.4^{+0.5}_{-1.4}$ & $5.1^{+1.7}_{-0.9}$ & $4.6^{+1.0}_{-0.9}$ & $1.9^{+1.3}_{-0.7}$ \\
                      & $\log N$ &  $16.4^{+0.3}_{-0.4}$ &  $14.81^{+0.26}_{-0.26}$ & $14.52^{+0.08}_{-0.10}$ & $17.11^{+0.19}_{-0.61}$ & $15.9^{+0.5}_{-0.4}$ \\
    ${\rm H_2\, J=3}$ & b\,km/s & $2.9^{+0.4}_{-0.8}$ &$3.3^{+0.9}_{-1.0}$ & $8.5^{+2.7}_{-2.7}$ & $5.6^{+0.9}_{-1.7}$ & $2.0^{+1.6}_{-0.5}$ \\
                      & $\log N$ & $15.65^{+0.29}_{-0.39}$ & $14.55^{+0.12}_{-0.27}$ &$14.46^{+0.04}_{-0.12}$ & $15.51^{+0.34}_{-0.27}$ & $15.35^{+0.63}_{-0.22}$ \\
    ${\rm H_2\, J=4}$ & $\log N$ & $14.03^{+0.12}_{-0.17}$ &$13.37^{+0.31}_{-0.55}$ &$13.35^{+0.34}_{-0.18}$  & $13.0^{+0.6}_{-0.9}$ & $14.32^{+0.23}_{-0.18}$\\
    ${\rm H_2\, J=5}$ & $\log N$ & $13.74^{+0.21}_{-0.19}$ & $13.1^{+0.6}_{-0.6}$ & $13.61^{+0.19}_{-0.29}$ & $11.2^{+0.8}_{-0.6}$ & $13.97^{+0.16}_{-0.14}$\\
     \hline 
         & $\log N_{\rm tot}$ & $18.56^{+0.03}_{-0.05}$ & $17.49^{+0.21}_{-0.39}$ & $17.01^{+0.27}_{-0.68}$ & $19.18^{+0.01}_{-0.02}$ & $17.49^{+0.44}_{-0.29}$  \\
     \hline
     HD J=0 & b\,km/s &$0.56^{+1.60}_{-0.06}$ & $2.9^{+0.9}_{-1.2}$ & $4.8^{+0.9}_{-1.9}$ & $0.53^{+1.02}_{-0.03}$ & $0.56^{+1.78}_{-0.06}$ \\
            & $\log N$ & $\lesssim 15.2$ & $\lesssim 15.0$ & $\lesssim 15.2$ & $\lesssim 15.6$ & $\lesssim 15.5$ \\
    \hline   
    \end{tabular}
    \begin{tablenotes}
    \item Doppler parameters H$_2$ $\rm J=4, 5$ were tied to H$_2$ $\rm J=3$.
    \end{tablenotes}
\end{table*}

\begin{figure*}
    \centering
    \includegraphics[width=\linewidth]{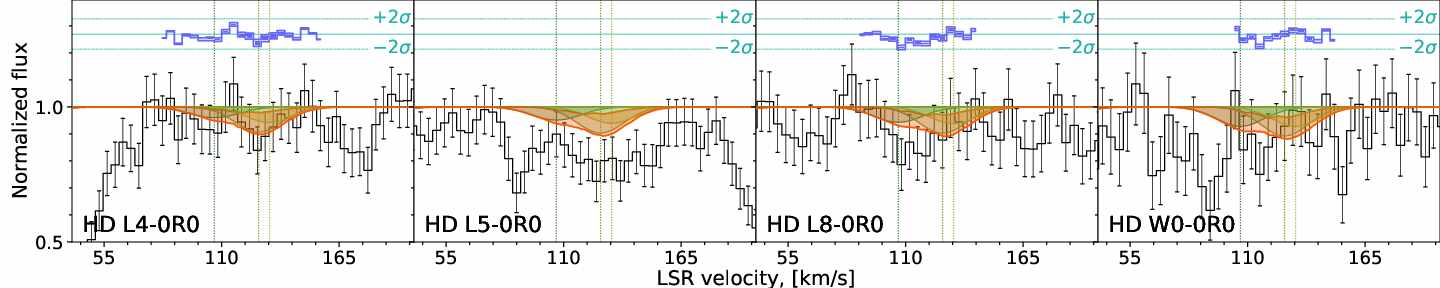}
    \caption{Fit to HD absorption lines towards AV 266 in SMC. Lines are the same as for \ref{fig:lines_HD_Sk67_2}.
    }
    \label{fig:lines_HD_AV266}
\end{figure*}

\begin{figure*}
    \centering
    \includegraphics[width=\linewidth]{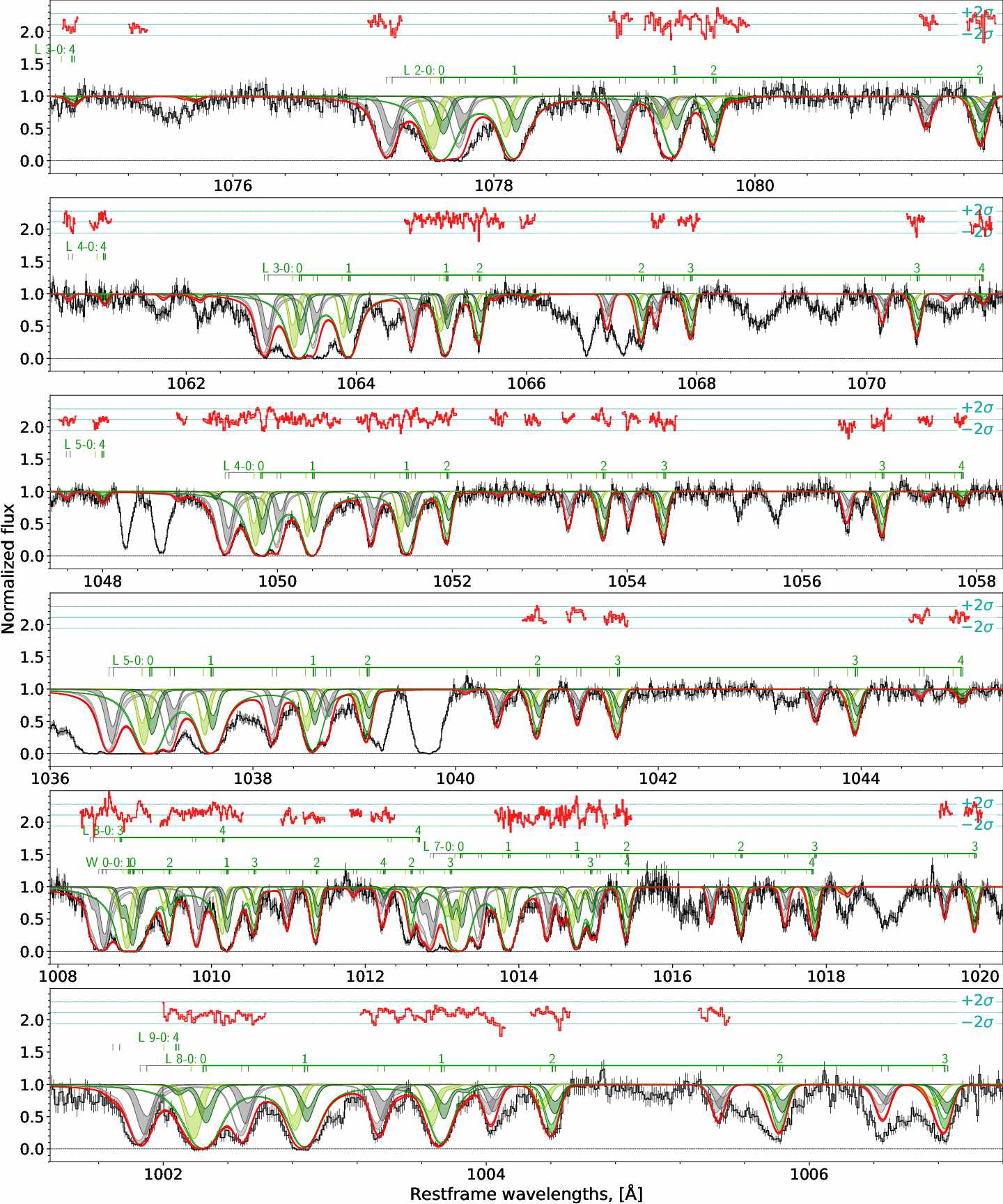}
    \caption{Fit to H2 absorption lines towards AV 266 in SMC. Lines are the same as for \ref{fig:lines_H2_Sk67_2}.
    }
    \label{fig:lines_H2_AV266}
\end{figure*}

\begin{table*}
    \caption{Fit results of H$_2$ lines towards AV 304}
    \label{tab:AV304}
    \begin{tabular}{cccccc}
    \hline
    \hline
    species & comp & 1 & 2 & 3 & 4  \\
            & z & $-0.000008(^{+5}_{-5})$ & $0.000051(^{+3}_{-7})$ & $0.0004037(^{+12}_{-21})$ & $0.000459(^{+7}_{-6})$ \\
    \hline 
     ${\rm H_2\, J=0}$ & b\,km/s & $0.73^{+0.21}_{-0.23}$ & $0.61^{+0.42}_{-0.11}$ & $0.59^{+0.38}_{-0.09}$ & $0.8^{+0.4}_{-0.3}$ \\
                       & $\log N$ & $18.43^{+0.12}_{-0.15}$ &$18.73^{+0.07}_{-0.11}$ & $19.13^{+0.07}_{-0.08}$ & $18.76^{+0.19}_{-0.27}$ \\
    ${\rm H_2\, J=1}$ & b\,km/s & $1.2^{+0.4}_{-0.4}$ & $0.86^{+0.42}_{-0.25}$ &$1.3^{+0.8}_{-0.4}$ &  $1.2^{+0.3}_{-0.4}$\\
                      & $\log N$ &$17.47^{+0.15}_{-0.14}$ & $17.11^{+0.20}_{-0.25}$ & $19.334^{+0.017}_{-0.010}$ & $16.8^{+0.6}_{-0.3}$\\
    ${\rm H_2\, J=2}$ & b\,km/s & $1.9^{+0.4}_{-0.5}$ & $2.65^{+0.25}_{-0.66}$ &$2.7^{+0.4}_{-0.5}$ & $1.56^{+0.31}_{-0.13}$ \\
                      & $\log N$ & $16.02^{+0.25}_{-0.57}$ & $15.84^{+0.26}_{-0.49}$ & $17.40^{+0.14}_{-0.19}$ & $15.57^{+0.52}_{-0.26}$\\
    ${\rm H_2\, J=3}$ & b\,km/s &$3.06^{+0.20}_{-0.68}$ & $2.96^{+0.30}_{-0.35}$ & $4.13^{+0.44}_{-0.28}$ &$1.78^{+0.16}_{-0.26}$ \\
                      & $\log N$ & $14.61^{+0.38}_{-0.13}$ & $15.01^{+0.37}_{-0.17}$ &$16.80^{+0.13}_{-0.30}$ & $14.96^{+0.38}_{-0.28}$\\
    ${\rm H_2\, J=4}$ & b\,km/s & -- & -- & $4.6^{+2.7}_{-0.6}$ &-- \\
                      & $\log N$ & $14.09^{+0.13}_{-0.19}$ & $13.70^{+0.23}_{-0.43}$ & $14.42^{+0.06}_{-0.09}$ & $13.73^{+0.30}_{-0.48}$\\
    \hline 
         & $\log N_{\rm tot}$ & $18.48^{+0.11}_{-0.13}$ & $18.74^{+0.07}_{-0.11}$ & $19.55^{+0.03}_{-0.03}$ & $18.77^{+0.19}_{-0.27}$ \\
     \hline
     HD J=0 & b\,km/s &$0.73^{+0.22}_{-0.15}$ & $0.61^{+0.38}_{-0.11}$ & $0.54^{+0.40}_{-0.04}$ & $0.519^{+0.523}_{-0.019}$ \\
            & $\log N$ & $\lesssim 16.2$ & $\lesssim 16.1$ & $\lesssim 15.8$ & $\lesssim 15.7$ \\ 
    \hline   
    \end{tabular}
    \begin{tablenotes}
    \item Doppler parameters H$_2$ $\rm J=4$ in 1, 2 and 4 components were tied to H$_2$ $\rm J=3$.
    \end{tablenotes}
\end{table*}

\begin{figure*}
    \centering
    \includegraphics[width=\linewidth]{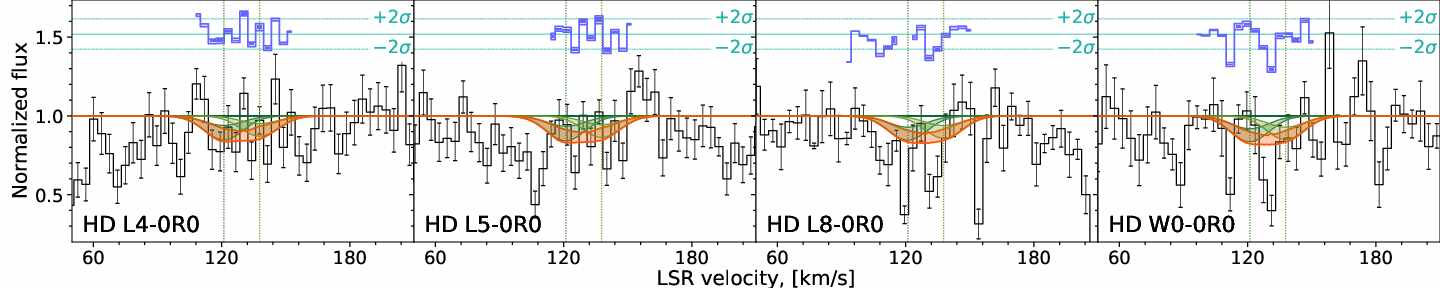}
    \caption{Fit to HD absorption lines towards AV 304 in SMC. Lines are the same as for \ref{fig:lines_HD_Sk67_2}.
    }
    \label{fig:lines_HD_AV304}
\end{figure*}

\begin{figure*}
    \centering
    \includegraphics[width=\linewidth]{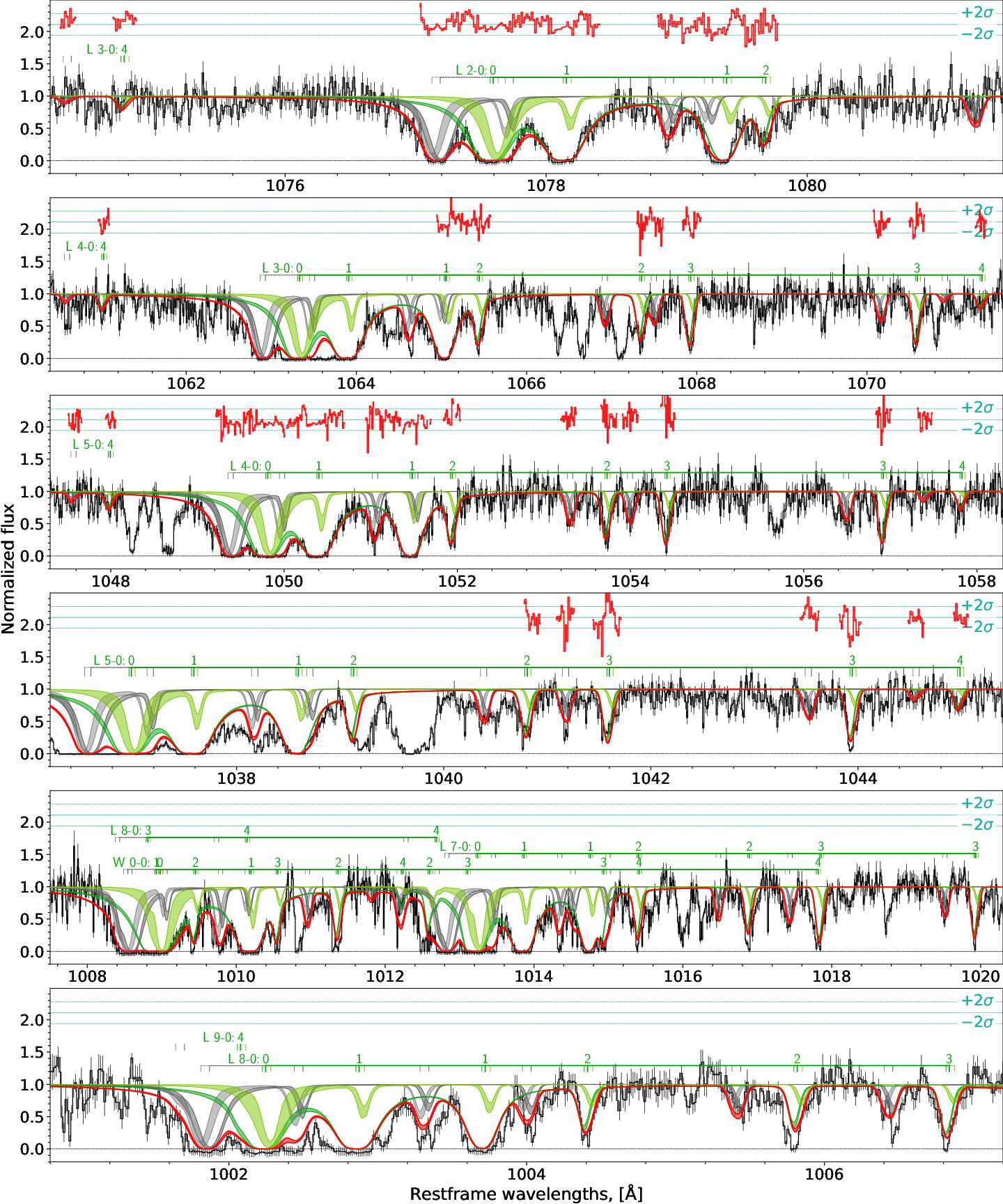}
    \caption{Fit to H2 absorption lines towards AV 304 in SMC. Lines are the same as for \ref{fig:lines_H2_Sk67_2}.
    }
    \label{fig:lines_H2_AV304}
\end{figure*}

\begin{table*}
    \caption{Fit results of H$_2$ lines towards AV 372}
    \label{tab:AV372}
    \begin{tabular}{ccccccc}
    \hline
    \hline
    species & comp & 1 & 2 & 3 & 4 & 5  \\
            & z & $0.0000443(^{+29}_{-11})$ & $0.000111(^{+4}_{-3})$ & $0.0004090(^{+21}_{-27})$ & $0.000470(^{+5}_{-6})$ & $0.0004907(^{+11}_{-32})$ \\
    \hline 
     ${\rm H_2\, J=0}$ & b\,km/s & $2.75^{+0.22}_{-0.86}$ & $0.63^{+0.39}_{-0.13}$ & $2.2^{+1.0}_{-1.3}$ & $0.8^{+0.4}_{-0.3}$ & $4.3^{+0.7}_{-0.9}$\\
                       & $\log N$ & $16.7^{+0.4}_{-0.5}$ & $13.99^{+1.15}_{-0.14}$ & $14.53^{+0.67}_{-0.14}$ & $18.23^{+0.08}_{-0.16}$ & $17.84^{+0.27}_{-0.23}$\\
    ${\rm H_2\, J=1}$ & b\,km/s &$2.8^{+0.5}_{-0.4}$ & $8.9^{+0.9}_{-2.0}$ & $7.0^{+1.0}_{-1.0}$ & $0.9^{+0.7}_{-0.3}$ &$4.6^{+0.6}_{-0.4}$ \\
                      & $\log N$ &$15.71^{+0.30}_{-0.35}$ & $14.57^{+0.07}_{-0.04}$ & $15.28^{+0.10}_{-0.09}$ & $18.39^{+0.05}_{-0.41}$ & $18.29^{+0.09}_{-0.27}$\\
    ${\rm H_2\, J=2}$ & b\,km/s & $4.2^{+0.8}_{-1.2}$ & $9.2^{+0.7}_{-1.7}$ & $9.3^{+0.6}_{-1.1}$ & $4.1^{+2.0}_{-1.4}$ &$4.46^{+0.87}_{-0.18}$ \\
                      & $\log N$ & $14.64^{+0.08}_{-0.09}$ & $13.90^{+0.11}_{-0.15}$ &$14.652^{+0.025}_{-0.076}$ & $15.2^{+0.8}_{-0.4}$ &$17.15^{+0.17}_{-0.39}$ \\
    ${\rm H_2\, J=3}$ & b\,km/s & $9.5^{+0.5}_{-1.0}$ & $9.90^{+0.10}_{-1.00}$ &$9.88^{+0.12}_{-0.39}$ & $6.6^{+1.3}_{-2.1}$ &$5.3^{+0.9}_{-0.3}$ \\
                      & $\log N$ & $14.10^{+0.07}_{-0.12}$ & $14.09^{+0.06}_{-0.12}$ & $14.70^{+0.05}_{-0.04}$ & $15.0^{+0.5}_{-0.5}$ & $16.68^{+0.28}_{-0.49}$\\
    ${\rm H_2\, J=4}$ & b\,km/s & -- & -- & -- & $8.2^{+1.5}_{-1.0}$ & $5.2^{+1.2}_{-0.4}$\\
                      & $\log N$ &$13.44^{+0.09}_{-0.17}$ & $13.19^{+0.15}_{-0.35}$ & $13.81^{+0.12}_{-0.14}$ & $13.4^{+0.7}_{-0.8}$ & $14.94^{+0.06}_{-0.11}$\\
    ${\rm H_2\, J=5}$ & $\log N$ & $13.66^{+0.15}_{-0.16}$ & $13.36^{+0.12}_{-0.72}$ & $13.81^{+0.14}_{-0.10}$ & $13.6^{+0.5}_{-0.9}$& $14.33^{+0.10}_{-0.05}$\\
     \hline 
         & $\log N_{\rm tot}$ & $16.74^{+0.37}_{-0.42}$ & $14.85^{+0.45}_{-0.04}$ & $15.52^{+0.15}_{-0.05}$ & $18.62^{+0.04}_{-0.21}$ & $18.45^{+0.10}_{-0.18}$ \\
     \hline
     HD J=0 & b\,km/s & $2.75^{+0.30}_{-0.88}$ & $0.60^{+0.46}_{-0.10}$ & $0.56^{+2.26}_{-0.06}$ & $0.88^{+0.25}_{-0.38}$ & $4.3^{+0.8}_{-1.0}$ \\
            & $\log N$ & $\lesssim 15.2$ & $\lesssim 15.8$ & $\lesssim 15.6$ & $\lesssim 16.2$ & $13.97^{+0.22}_{-0.23}$ \\
    \hline   
    \end{tabular}
    \begin{tablenotes}
    \item Doppler parameters H$_2$ $\rm J=4, 5$ in 1, 2 and 3 components and $\rm J=5$ in 4 and 5 components were tied to H$_2$ $\rm J=3$ and $\rm J=4$, respectively.
    \end{tablenotes}
\end{table*}

\begin{figure*}
    \centering
    \includegraphics[width=\linewidth]{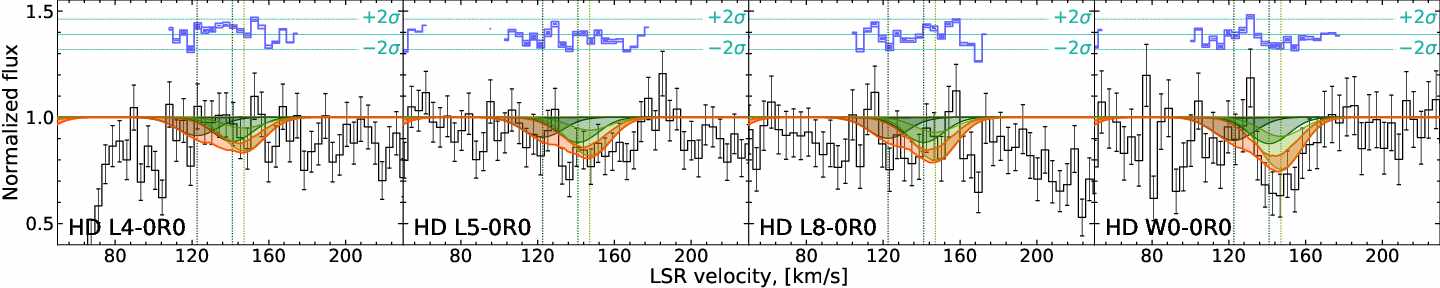}
    \caption{Fit to HD absorption lines towards AV 372 in SMC. Lines are the same as for \ref{fig:lines_HD_Sk67_2}.
    }
    \label{fig:lines_H2_AV372}
\end{figure*}

\begin{figure*}
    \centering
    \includegraphics[width=\linewidth]{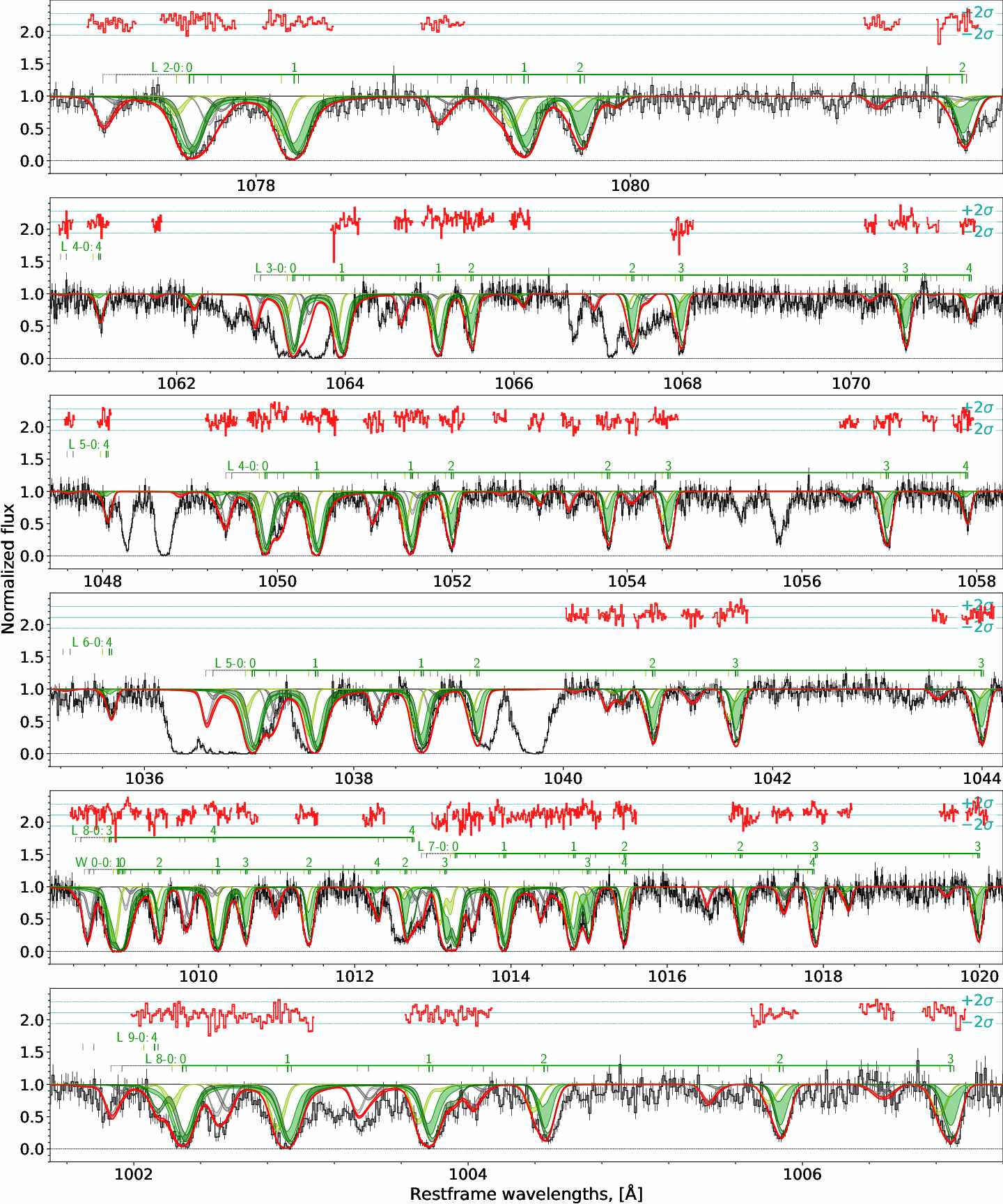}
    \caption{Fit to H2 absorption lines towards AV 372 in SMC. Lines are the same as for \ref{fig:lines_H2_Sk67_2}.
    }
    \label{fig:lines_HD_AV372}
\end{figure*}

\begin{table*}
    \caption{Fit results of H$_2$ lines towards AV 374}
    \label{tab:AV374}
    \begin{tabular}{ccccc}
    \hline
    \hline
    species & comp & 1 & 2 & 3   \\
            & z & $0.0000444(^{+11}_{-9})$ & $0.0002885(^{+12}_{-14})$ & $0.0004441(^{+12}_{-9})$  \\
    \hline 
     ${\rm H_2\, J=0}$ & b\,km/s & $0.57^{+0.15}_{-0.07}$ & $0.525^{+0.313}_{-0.025}$ & $0.8^{+0.4}_{-0.3}$ \\
                       & $\log N$ & $18.094^{+0.029}_{-0.036}$ & $17.66^{+0.05}_{-0.08}$ & $18.367^{+0.036}_{-0.031}$ \\
    ${\rm H_2\, J=1}$ & b\,km/s & $0.70^{+0.09}_{-0.18}$ & $0.68^{+0.38}_{-0.17}$ & $1.37^{+0.26}_{-0.62}$ \\
                      & $\log N$ & $18.190^{+0.026}_{-0.031}$ & $17.831^{+0.028}_{-0.034}$ & $18.317^{+0.022}_{-0.029}$ \\
    ${\rm H_2\, J=2}$ & b\,km/s & $0.92^{+0.05}_{-0.25}$ & $0.82^{+0.36}_{-0.26}$ & $1.82^{+0.23}_{-0.42}$ \\
                      & $\log N$ & $17.39^{+0.05}_{-0.05}$ & $16.94^{+0.11}_{-0.29}$ & $16.78^{+0.24}_{-0.22}$ \\
    ${\rm H_2\, J=3}$ & b\,km/s &$0.92^{+0.08}_{-0.14}$ & $0.87^{+0.41}_{-0.29}$ & $2.06^{+0.12}_{-0.31}$\\
                      & $\log N$ & $16.90^{+0.10}_{-0.10}$ & $16.30^{+0.25}_{-0.48}$ & $15.71^{+0.55}_{-0.21}$\\
    ${\rm H_2\, J=4}$ & $\log N$ & $14.9^{+0.6}_{-0.3}$ & $13.9^{+0.9}_{-0.8}$ & $14.30^{+0.29}_{-0.24}$ \\
    \hline 
         & $\log N_{\rm tot}$ & $18.49^{+0.02}_{-0.02}$ & $18.09^{+0.03}_{-0.04}$ & $18.65^{+0.02}_{-0.02}$ \\
     \hline
     HD J=0 & b\,km/s & $0.61^{+0.11}_{-0.11}$ & $0.515^{+0.339}_{-0.015}$ &  $0.520^{+0.465}_{-0.020}$ \\
            & $\log N$ & $\lesssim 16.1$ & $\lesssim 16.2$ & $\lesssim 15.9$ \\
    \hline   
    \end{tabular}
    \begin{tablenotes}
    \item Doppler parameters H$_2$ $\rm J=4$  were tied to H$_2$ $\rm J=3$.
    \end{tablenotes}
\end{table*}

\begin{figure*}
    \centering
    \includegraphics[width=\linewidth]{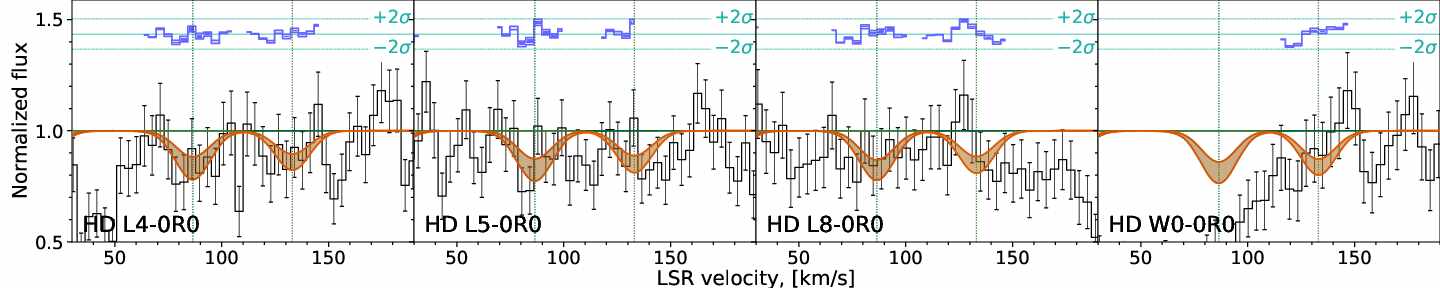}
    \caption{Fit to HD absorption lines towards AV 374 in SMC. Lines are the same as for \ref{fig:lines_HD_Sk67_2}.
    }
    \label{fig:lines_HD_AV374}
\end{figure*}

\begin{figure*}
    \centering
    \includegraphics[width=\linewidth]{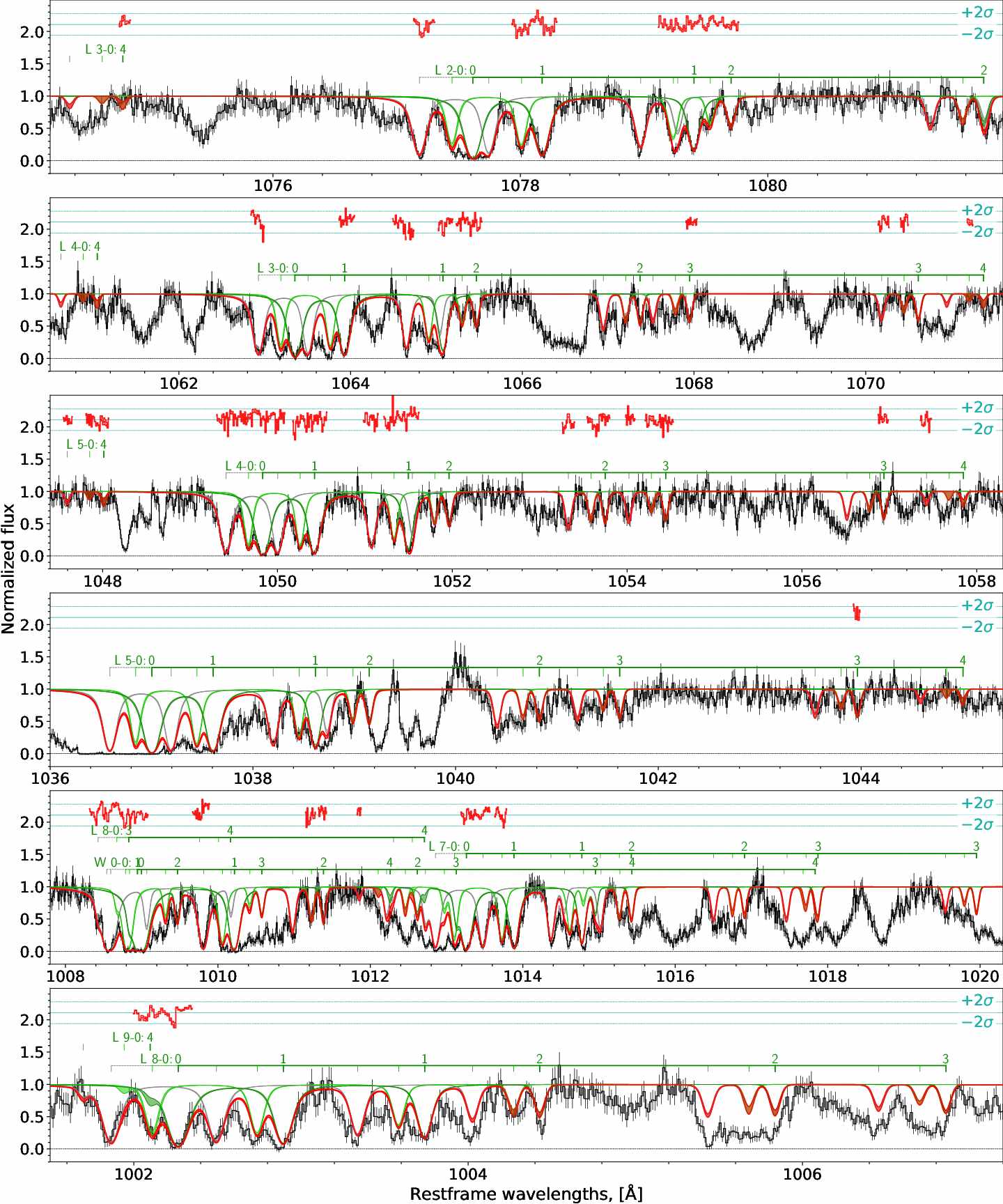}
    \caption{Fit to H2 absorption lines towards AV 374 in SMC. Lines are the same as for \ref{fig:lines_H2_Sk67_2}.
    }
    \label{fig:lines_H2_AV374}
\end{figure*}

\begin{table*}
    \caption{Fit results of H$_2$ lines towards AV 423}
    \label{tab:AV423}
    \begin{tabular}{cccccc}
    \hline
    \hline
    species & comp & 1 & 2 & 3 & 4  \\
            & z &  $0.0000593(^{+17}_{-21})$ & $0.000473(^{+5}_{-16})$ & $0.0004858(^{+48}_{-25})$ & $0.0006142(^{+30}_{-27})$ \\
    \hline 
     ${\rm H_2\, J=0}$ & b\,km/s & $0.60^{+0.75}_{-0.10}$ & $4.2^{+2.3}_{-0.9}$ &$5.0^{+0.8}_{-0.8}$ & $1.0^{+0.7}_{-0.3}$ \\
                       & $\log N$ & $17.32^{+0.10}_{-0.10}$ & $17.6^{+0.3}_{-0.4}$ & $17.87^{+0.23}_{-0.13}$ & $16.82^{+0.18}_{-0.32}$\\
    ${\rm H_2\, J=1}$ & b\,km/s & $1.7^{+0.3}_{-0.4}$ &$6.9^{+0.4}_{-0.7}$ & $5.4^{+0.7}_{-1.2}$ & $1.61^{+0.62}_{-0.27}$ \\
                      & $\log N$ & $17.59^{+0.05}_{-0.11}$ & $17.8^{+0.5}_{-0.8}$ & $18.35^{+0.10}_{-0.15}$ & $16.59^{+0.53}_{-0.26}$\\
    ${\rm H_2\, J=2}$ & b\,km/s & $1.9^{+0.5}_{-0.3}$ & $6.4^{+0.8}_{-0.4}$ & $5.62^{+0.25}_{-1.10}$ & $2.04^{+0.60}_{-0.24}$\\
                      & $\log N$ & $15.9^{+0.4}_{-0.6}$ & $15.48^{+0.25}_{-1.19}$ & $16.86^{+0.32}_{-0.31}$ & $14.96^{+0.40}_{-0.23}$ \\
    ${\rm H_2\, J=3}$ & b\,km/s & $2.53^{+0.18}_{-0.68}$ & $7.26^{+0.24}_{-0.52}$ & $6.4^{+1.0}_{-0.7}$ & $2.95^{+0.21}_{-0.48}$\\
                      & $\log N$ & $14.57^{+0.26}_{-0.13}$ & $15.50^{+0.23}_{-0.91}$ & $16.3^{+0.4}_{-0.4}$ & $14.90^{+0.30}_{-0.14}$\\
    ${\rm H_2\, J=4}$ & b\,km/s & -- & -- & $6.3^{+2.0}_{-0.5}$ & -- \\
                      & $\log N$ & $14.11^{+0.21}_{-0.19}$ &$14.3^{+0.4}_{-0.4}$ &  $14.63^{+0.16}_{-0.26}$ &$14.18^{+0.23}_{-0.18}$ \\
    \hline 
         & $\log N_{\rm tot}$ & $17.79^{+0.05}_{-0.08}$ & $18.00^{+0.36}_{-0.34}$ & $18.49^{+0.10}_{-0.11}$ & $17.03^{+0.28}_{0.20}$ \\
     \hline
     HD J=0 & b\,km/s &  $0.53^{+0.88}_{-0.03}$ & $4.1^{+2.7}_{-1.1}$ & $4.8^{+1.1}_{-0.8}$ & $0.9^{+0.6}_{-0.4}$ \\
            & $\log N$ & $\lesssim 15.2$ & $\lesssim 15.5$ & $\lesssim 14.2$ & $\lesssim 16.1$ \\
    \hline   
    \end{tabular}
    \begin{tablenotes}
    \item Doppler parameters H$_2$ $\rm J=4$ in the 1, 2 and 4 components  were tied to H$_2$ $\rm J=3$.
    \end{tablenotes}
\end{table*}

\begin{figure*}
    \centering
    \includegraphics[width=\linewidth]{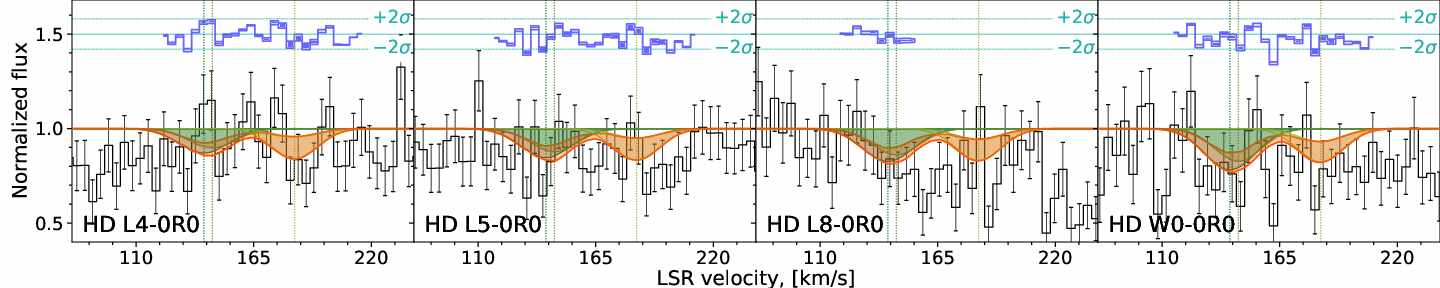}
    \caption{Fit to HD absorption lines towards AV 423 in SMC. Lines are the same as for \ref{fig:lines_HD_Sk67_2}.
    }
    \label{fig:lines_HD_AV423}
\end{figure*}

\begin{figure*}
    \centering
    \includegraphics[width=\linewidth]{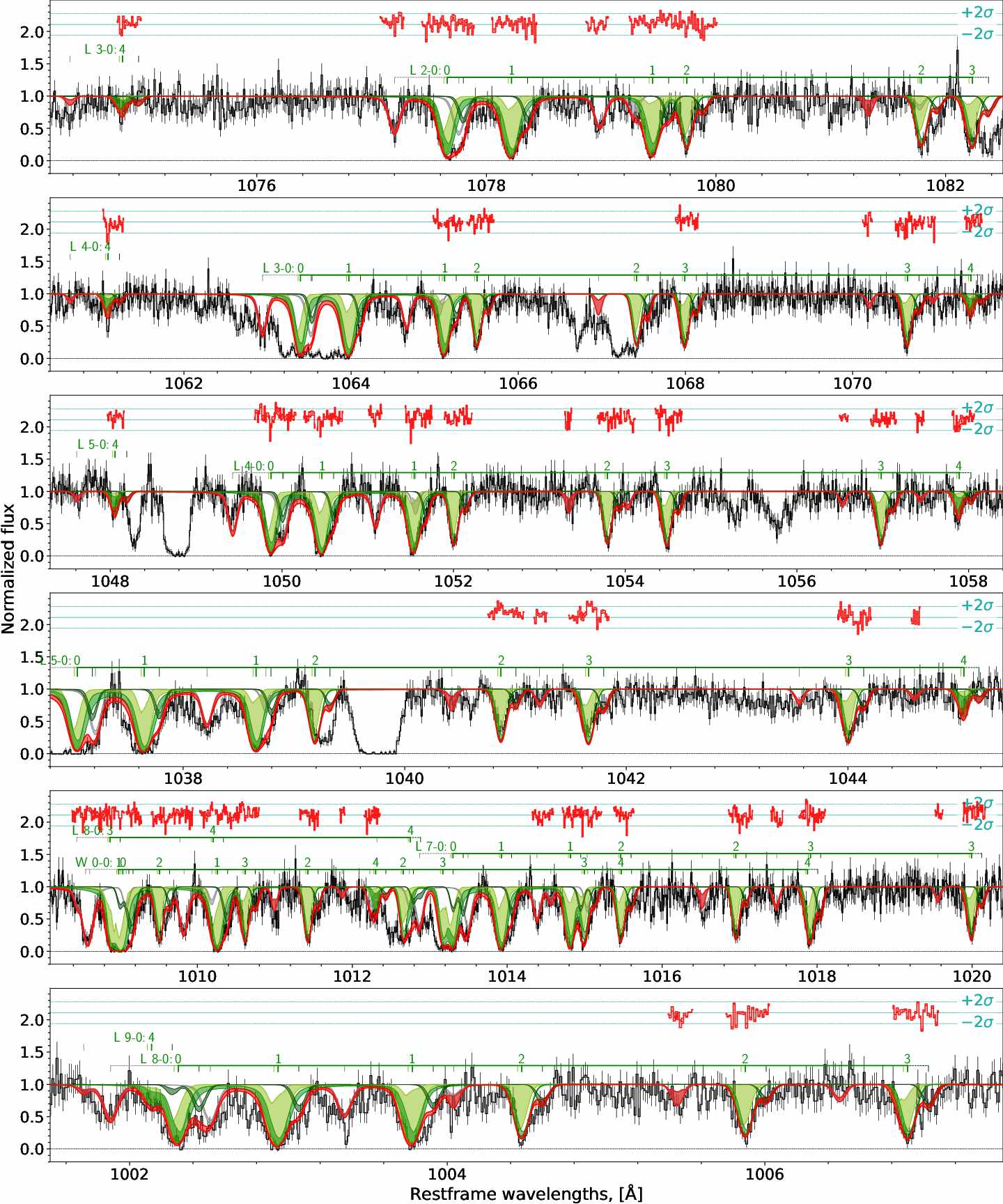}
    \caption{Fit to H2 absorption lines towards AV 423 in SMC. Lines are the same as for \ref{fig:lines_H2_Sk67_2}.
    }
    \label{fig:lines_H2_AV423}
\end{figure*}

\begin{table*}
    \caption{Fit results of H$_2$ lines towards AV 435}
    \label{tab:AV435}
    \begin{tabular}{ccccc}
    \hline
    \hline
    species & comp & 1 & 2 & 3  \\
            & z & $0.0000425(^{+19}_{-17})$ & $0.0000959(^{+21}_{-34})$ & $0.0005917(^{+6}_{-3})$ \\
    \hline 
     ${\rm H_2\, J=0}$ & b\,km/s & $1.38^{+0.27}_{-0.51}$ & $2.0^{+0.4}_{-0.5}$ & $0.8^{+0.5}_{-0.3}$\\
                       & $\log N$ & $18.224^{+0.033}_{-0.024}$ & $16.16^{+0.28}_{-0.47}$ & $19.795^{+0.004}_{-0.009}$\\
    ${\rm H_2\, J=1}$ & b\,km/s & $1.79^{+0.19}_{-0.43}$ & $2.28^{+0.27}_{-0.49}$ & $1.4^{+0.9}_{-0.4}$\\
                      & $\log N$ & $17.87^{+0.08}_{-0.04}$ & $15.00^{+0.41}_{-0.20}$ & $19.3137^{+0.0021}_{-0.0079}$\\
    ${\rm H_2\, J=2}$ & b\,km/s & $1.94^{+0.12}_{-0.18}$ & $2.32^{+0.26}_{-0.29}$ & $2.60^{+0.30}_{-0.19}$\\
                      & $\log N$ & $16.38^{+0.18}_{-0.20}$ & $14.18^{+0.16}_{-0.07}$ & $16.80^{+0.16}_{-0.19}$\\
    ${\rm H_2\, J=3}$ & b\,km/s & $1.96^{+0.09}_{-0.22}$ & $2.52^{+0.16}_{-0.20}$ & $2.68^{+0.18}_{-0.24}$\\
                      & $\log N$ & $14.60^{+0.19}_{-0.15}$ & $13.94^{+0.18}_{-0.19}$ & $15.91^{+0.19}_{-0.21}$\\
    ${\rm H_2\, J=4}$ & b\,km/s & -- & -- & $2.93^{+0.35}_{-0.31}$\\
    				  & $\log N$ & $13.53^{+0.17}_{-0.38}$ & $13.53^{+0.15}_{-0.29}$ & $14.33^{+0.07}_{-0.05}$\\
    ${\rm H_2\, J=5}$ & $\log N$ & -- & -- & $13.52^{+0.12}_{-0.30}$\\
    \hline 
         & $\log N_{\rm tot}$ & $18.39^{+0.03}_{-0.02}$ & $16.20^{+0.26}_{-0.41}$ & $19.919^{+0.003}_{-0.007}$ \\
     \hline
     HD J=0 & b\,km/s & $1.36^{+0.30}_{-0.49}$ & $2.0^{+0.5}_{-0.5}$ & $0.86^{+0.26}_{-0.29}$ \\
            & $\log N$ & $\lesssim 14.1$ & $\lesssim 13.8$ & $15.62^{+0.22}_{-0.86}$ \\
    \hline   
    \end{tabular}
    \begin{tablenotes}
    \item Doppler parameters H$_2$ $\rm J=4$ in 1 and 2 components and $\rm J=5$ in 3 component were tied to H$_2$ $\rm J=3$ and $\rm J=4$, respectively.
    \end{tablenotes}
\end{table*}

\clearpage 
\begin{figure*}
    \centering
    \includegraphics[width=\linewidth]{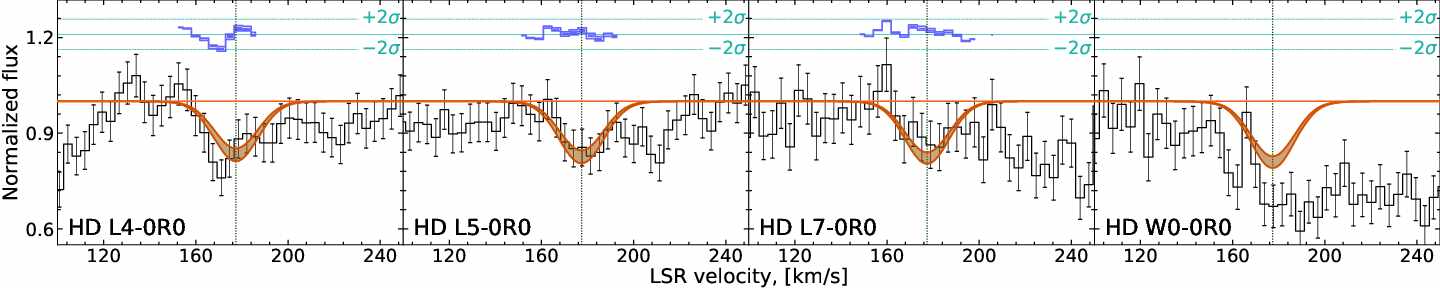}
    \caption{Fit to HD absorption lines towards AV 435 in SMC. Lines are the same as for \ref{fig:lines_HD_Sk67_2}.
    }
    \label{fig:lines_HD_AV435}
\end{figure*}

\begin{figure*}
    \centering
    \includegraphics[width=\linewidth]{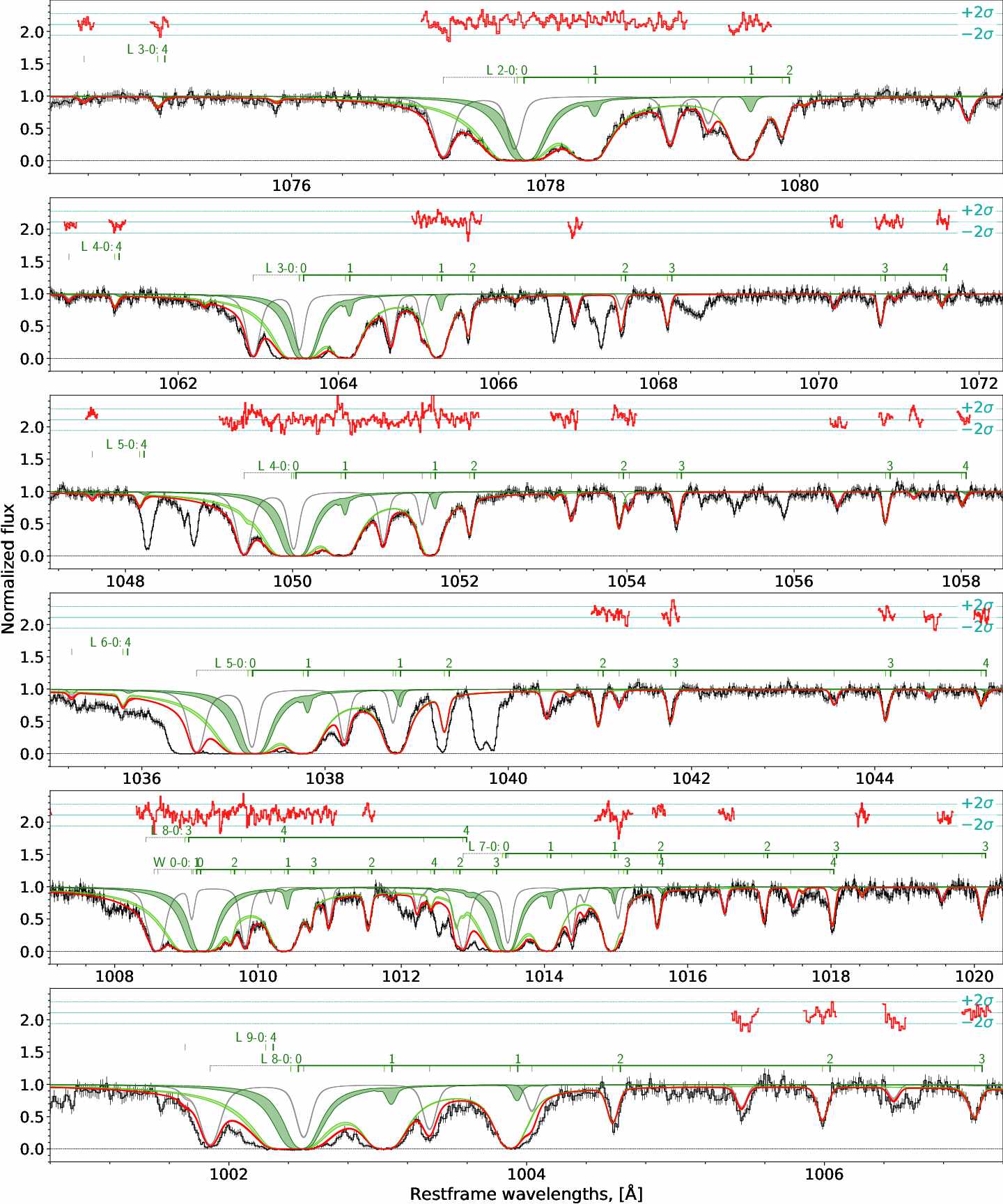}
    \caption{Fit to H2 absorption lines towards AV 435 in SMC. Lines are the same as for \ref{fig:lines_H2_Sk67_2}.
    }
    \label{fig:lines_H2_AV435}
\end{figure*}

\begin{table*}
    \caption{Fit results of H$_2$ lines towards AV 440}
    \label{tab:AV440}
    \begin{tabular}{ccccc}
    \hline
    \hline
    species & comp & 1 & 2 & 3   \\
            & z & $0.0000472(^{+6}_{-5})$ & $0.0004352(^{+14}_{-9})$ & $0.0005786(^{+4}_{-4})$  \\
    \hline 
     ${\rm H_2\, J=0}$ & b\,km/s & $2.38^{+0.22}_{-0.41}$ & $0.507^{+0.073}_{-0.007}$ & $0.62^{+0.19}_{-0.12}$  \\
                       & $\log N$ & $17.42^{+0.05}_{-0.04}$ & $16.75^{+0.09}_{-0.10}$ & $18.780^{+0.011}_{-0.009}$ \\
    ${\rm H_2\, J=1}$ & b\,km/s & $2.60^{+0.08}_{-0.14}$ & $0.56^{+0.09}_{-0.06}$ & $0.72^{+0.31}_{-0.14}$ \\
                      & $\log N$ & $17.531^{+0.044}_{-0.030}$ & $17.121^{+0.039}_{-0.023}$ & $18.707^{+0.007}_{-0.007}$ \\
    ${\rm H_2\, J=2}$ & b\,km/s & $2.56^{+0.10}_{-0.13}$ & $0.64^{+0.18}_{-0.08}$ &  $0.83^{+0.32}_{-0.19}$ \\
                      & $\log N$ & $15.87^{+0.14}_{-0.15}$ & $15.69^{+0.20}_{-0.30}$ & $17.349^{+0.020}_{-0.026}$ \\
    ${\rm H_2\, J=3}$ & b\,km/s & $2.56^{+0.13}_{-0.11}$ & $1.20^{+0.07}_{-0.33}$ & $0.96^{+0.22}_{-0.26}$\\
                      & $\log N$ & $14.72^{+0.14}_{-0.14}$ & $14.63^{+0.38}_{-0.20}$ & $17.233^{+0.030}_{-0.049}$\\
    ${\rm H_2\, J=4}$ & b\,km/s & -- & -- & $1.04^{+0.25}_{-0.20}$ \\
                      & $\log N$ & $13.89^{+0.15}_{-0.12}$ &$13.4^{+0.3}_{-0.5}$ &  $15.49^{+0.24}_{-0.34}$ \\
    ${\rm H_2\, J=4}$ & $\log N$ & -- & -- & $13.88^{+0.21}_{-0.17}$ \\
    \hline 
         & $\log N_{\rm tot}$ & $17.79^{+0.03}_{-0.02}$ & $17.29^{+0.04}_{-0.03}$ & $19.06^{+0.01}_{-0.01}$ \\
     \hline
     HD J=0 & b\,km/s & $2.44^{+0.20}_{-0.57}$ & $0.503^{+0.080}_{-0.003}$ & $0.63^{+0.15}_{-0.13}$ \\
            & $\log N$ & $\lesssim 14.8$ & $\lesssim 14.8$ & $\lesssim 15.8$ \\
    \hline   
    \end{tabular}
    \begin{tablenotes}
    \item Doppler parameters H$_2$ $\rm J=4$ in the 1 and 2 components and $\rm J=5$ in 3 component were tied to H$_2$ $\rm J=3$ and $\rm J=4$, respectively.
    \end{tablenotes}
\end{table*}

\begin{figure*}
    \centering
    \includegraphics[width=\linewidth]{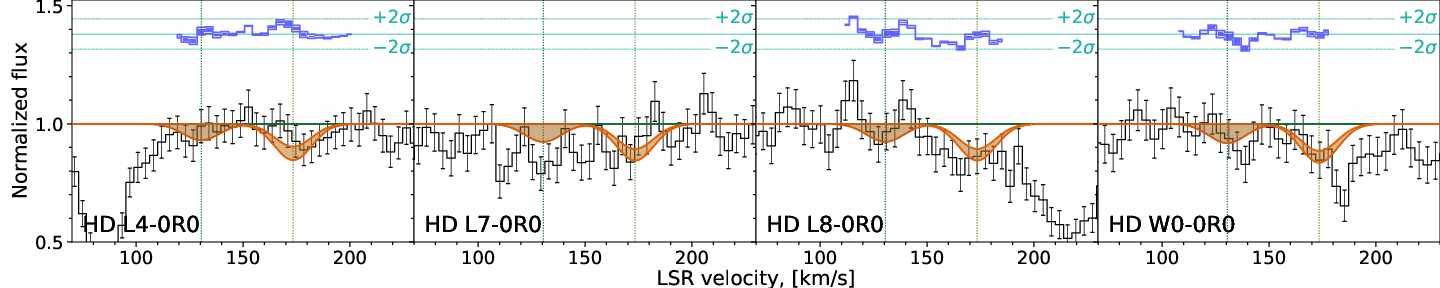}
    \caption{Fit to HD absorption lines towards AV 440 in SMC. Lines are the same as for \ref{fig:lines_HD_Sk67_2}.
    }
    \label{fig:lines_HD_AV440}
\end{figure*}

\begin{figure*}
    \centering
    \includegraphics[width=\linewidth]{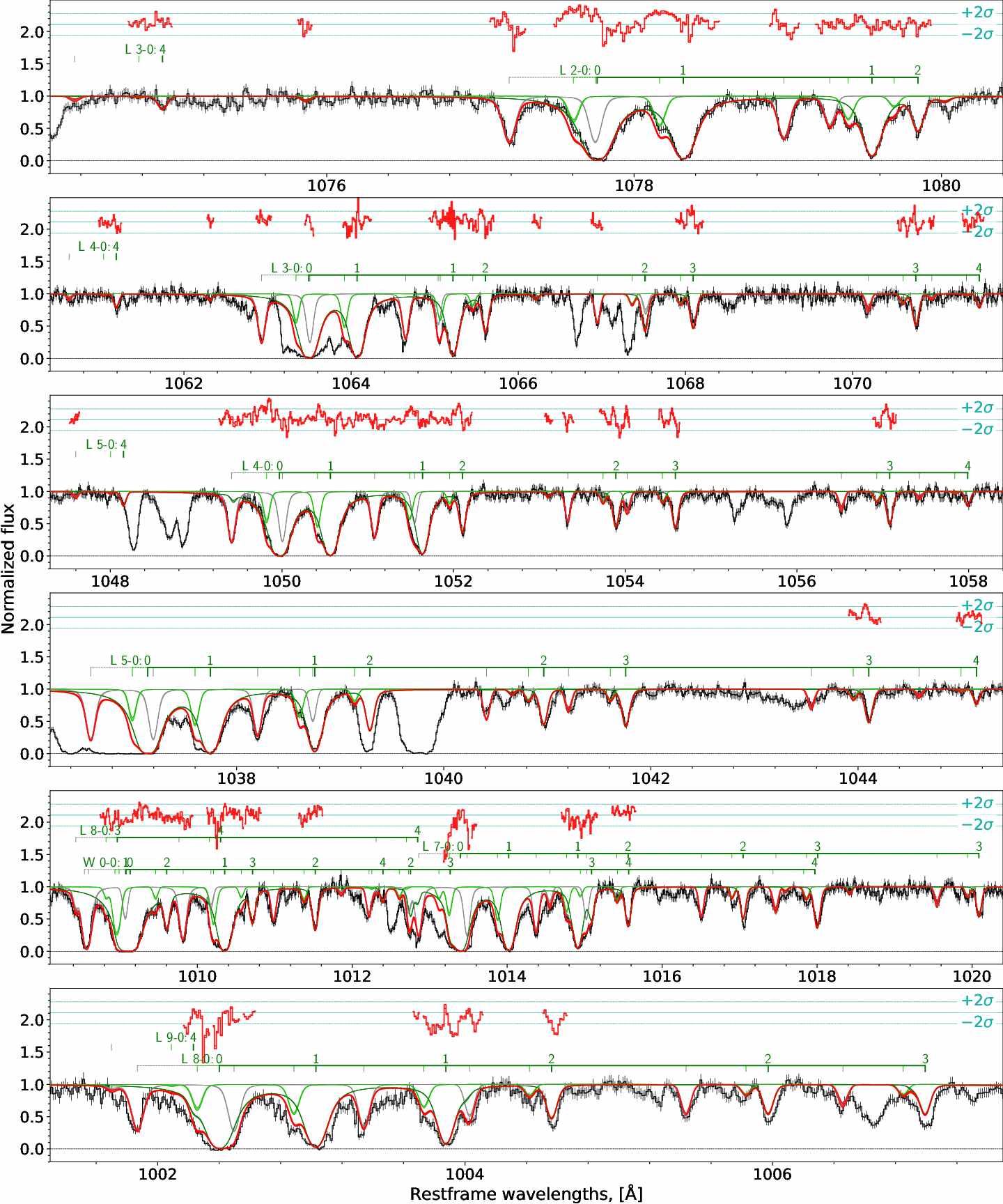}
    \caption{Fit to H2 absorption lines towards AV 440 in SMC. Lines are the same as for \ref{fig:lines_H2_Sk67_2}.
    }
    \label{fig:lines_H2_AV440}
\end{figure*}

\begin{table*}
    \caption{Fit results of H$_2$ lines towards AV 472}
    \label{tab:AV472}
    \begin{tabular}{cccc}
    \hline
    \hline
    species & comp & 1 & 2  \\
            & z & $0.0000403(^{+18}_{-15})$ & $0.0004598(^{+5}_{-8})$ \\
    \hline 
     ${\rm H_2\, J=0}$ & b\,km/s & $0.70^{+0.40}_{-0.20}$ & $0.516^{+0.564}_{-0.016}$\\
                       & $\log N$ & $17.69^{+0.08}_{-0.07}$ & $20.110^{+0.009}_{-0.011}$\\
    ${\rm H_2\, J=1}$ & b\,km/s & $1.05^{+0.51}_{-0.27}$ & $0.84^{+0.65}_{-0.16}$\\
                      & $\log N$ & $17.59^{+0.07}_{-0.09}$ & $20.035^{+0.006}_{-0.006}$ \\
    ${\rm H_2\, J=2}$ & b\,km/s & $2.03^{+0.10}_{-0.18}$ & $1.72^{+0.28}_{-0.52}$\\
                      & $\log N$ & $16.24^{+0.17}_{-0.18}$ & $17.798^{+0.026}_{-0.025}$\\
    ${\rm H_2\, J=3}$ & b\,km/s & $2.02^{+0.13}_{-0.16}$ & $2.02^{+0.11}_{-0.24}$\\
                      & $\log N$ & $14.22^{+0.16}_{-0.13}$ & $17.19^{+0.09}_{-0.04}$\\
    ${\rm H_2\, J=4}$ & b\,km/s & -- &  $2.01^{+0.11}_{-0.23}$ \\
    				  & $\log N$ &  $12.8^{+0.5}_{-1.5}$ &  $14.43^{+0.15}_{-0.11}$\\
    ${\rm H_2\, J=5}$ & $\log N$ &  -- &$11.6^{+1.2}_{-0.7}$ \\
    \hline 
         & $\log N_{\rm tot}$ &  $17.96^{+0.05}_{-0.05}$ & $20.377^{+0.005}_{-0.007}$\\
    \hline
    HD J=0 & b\,km/s & $0.520^{+0.513}_{-0.020}$ & $0.523^{+0.564}_{-0.023}$ \\
            & $\log N$ & $\lesssim 14.8$ & $15.90^{+0.31}_{-0.85}$ \\
    \hline   
    \end{tabular}
    \begin{tablenotes}
    \item Doppler parameters H$_2$ $\rm J=4$ in 1 component and $\rm J=5$ in 2 component were tied to H$_2$ $\rm J=3$ and $\rm J=4$, respectively.
    \end{tablenotes}
\end{table*}

\begin{figure*}
    \centering
    \includegraphics[width=\linewidth]{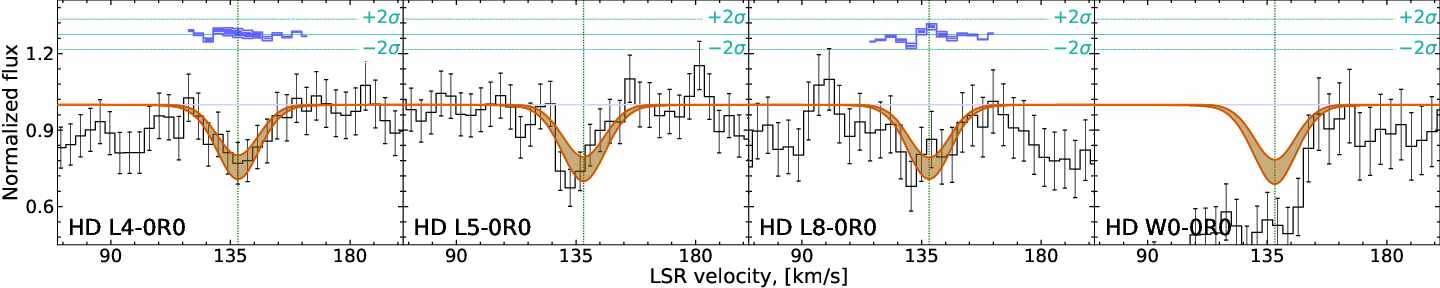}
    \caption{Fit to HD absorption lines towards AV 472 in SMC. Lines are the same as for \ref{fig:lines_HD_Sk67_2}.
    }
    \label{fig:lines_HD_AV472_appendix}
\end{figure*}

\begin{figure*}
    \centering
    \includegraphics[width=\linewidth]{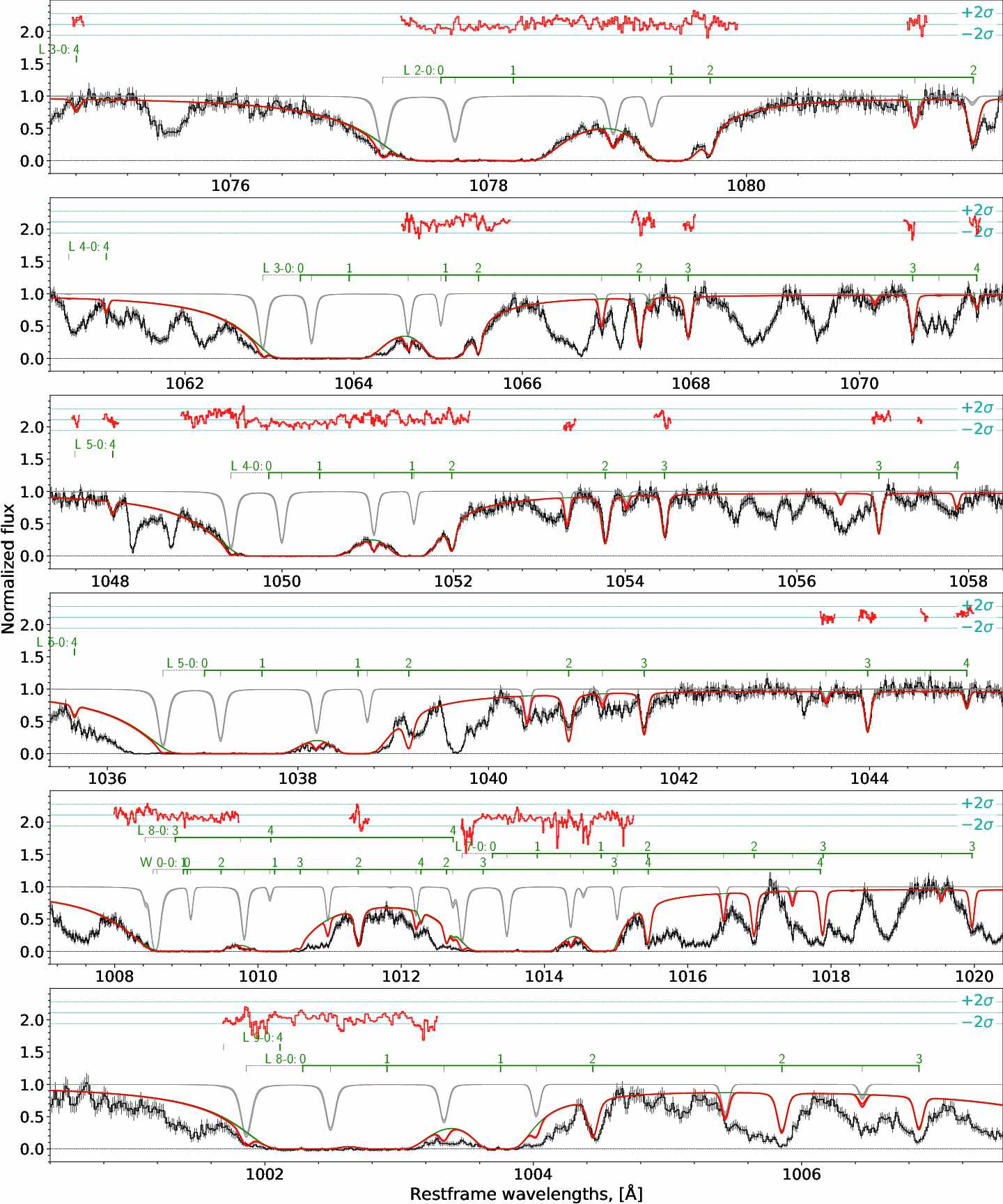}
    \caption{Fit to H2 absorption lines towards AV 472 in SMC. Lines are the same as for \ref{fig:lines_H2_Sk67_2}.
    }
    \label{fig:lines_H2_AV472}
\end{figure*}

\begin{table*}
    \caption{Fit results of H$_2$ lines towards AV 476}
    \label{tab:AV476}
    \begin{tabular}{ccccc}
    \hline
    \hline
    species & comp & 1 & 2 & 3  \\
            & z & $-0.000007(^{+36}_{-6})$ & $0.0005249(^{+16}_{-22})$ & $0.0005575(^{+24}_{-23})$ \\
    \hline 
     ${\rm H_2\, J=0}$ & b\,km/s & $0.77^{+0.19}_{-0.25}$ & $0.54^{+0.14}_{-0.04}$ &  $1.14^{+0.27}_{-0.41}$\\
                       & $\log N$ & $19.40^{+0.12}_{-0.25}$ & $20.48^{+0.06}_{-0.11}$ & $19.64^{+0.21}_{-0.61}$\\
    ${\rm H_2\, J=1}$ & b\,km/s & $0.89^{+0.32}_{-0.23}$ & $0.80^{+0.10}_{-0.20}$ &$1.44^{+0.84}_{-0.28}$ \\
                      & $\log N$ & $19.15^{+0.37}_{-0.12}$ & $20.591^{+0.024}_{-0.045}$ & $19.2^{+0.4}_{-0.3}$ \\
    ${\rm H_2\, J=2}$ & b\,km/s & $1.18^{+0.44}_{-0.24}$ & $0.87^{+0.33}_{-0.20}$ & $2.4^{+0.5}_{-0.5}$ \\
                      & $\log N$ & $16.72^{+0.26}_{-0.44}$ &$18.68^{+0.06}_{-0.04}$ & $17.89^{+0.20}_{-0.26}$\\
    ${\rm H_2\, J=3}$ & b\,km/s & $1.82^{+0.09}_{-0.38}$ & $2.04^{+0.14}_{-0.39}$ &$2.99^{+0.06}_{-0.37}$ \\
                      & $\log N$ & $14.5^{+0.6}_{-2.0}$ &$18.15^{+0.04}_{-0.04}$ &  $16.96^{+0.21}_{-0.38}$\\
    ${\rm H_2\, J=4}$ & b\,km/s & -- &$2.13^{+0.14}_{-0.13}$ & -- \\
    				  & $\log N$ & $13.8^{+0.4}_{-0.6}$ & $16.76^{+0.18}_{-0.14}$  & $15.45^{+0.28}_{-0.16}$ \\
    ${\rm H_2\, J=5}$ & b\,km/s & -- & $2.14^{+0.07}_{-0.20}$ & -- \\
                      & $\log N$ &  -- & $15.96^{+0.37}_{-0.24}$ & $14.98^{+0.17}_{-0.14}$\\
    ${\rm H_2\, J=6}$ & $\log N$ & -- & $15.08^{+0.32}_{-0.30}$ & $13.67^{+0.29}_{-0.17}$\\
    \hline 
         & $\log N_{\rm tot}$ & $19.56^{+0.18}_{-0.15}$ & $20.84^{+0.03}_{-0.05}$ & $19.79^{+0.21}_{-0.35}$ \\
    \hline
    HD J=0 & b\,km/s & $0.89^{+0.20}_{-0.16}$ & $0.508^{+0.170}_{-0.008}$ & $1.18^{+0.27}_{-0.37}$ \\
            & $\log N$ & $\lesssim 16.5$ & $\lesssim 17.1$ & $\lesssim 17.1$ \\
    \hline   
    \end{tabular}
    \begin{tablenotes}
    \item Doppler parameters H$_2$ $\rm J=4$ in 1 component and $\rm J = 4, 5, 6$ in 3 component and $\rm J=6$ in 2 component were tied to H$_2$ $\rm J=3$ and $\rm J=5$, respectively.
    \end{tablenotes}
\end{table*}

\begin{figure*}
    \centering
    \includegraphics[width=\linewidth]{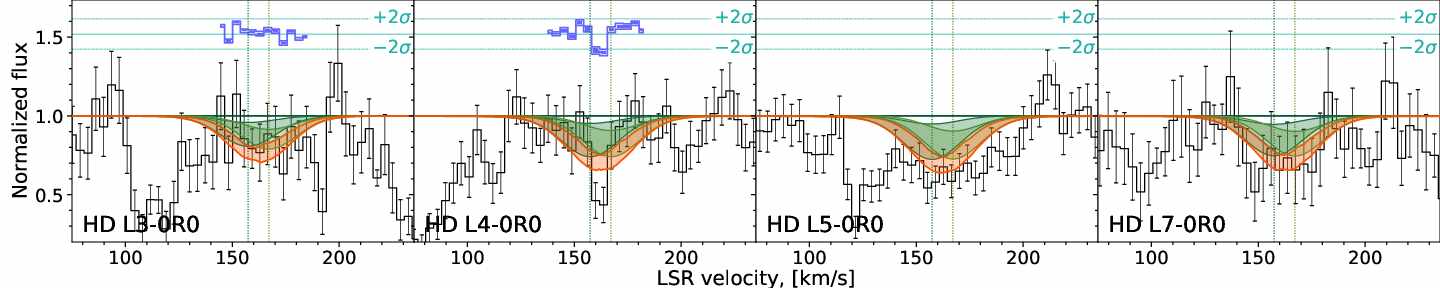}
    \caption{Fit to HD absorption lines towards AV 476 in SMC. Lines are the same as for \ref{fig:lines_HD_Sk67_2}.
    }
    \label{fig:lines_HD_AV476}
\end{figure*}

\begin{figure*}
    \centering
    \includegraphics[width=\linewidth]{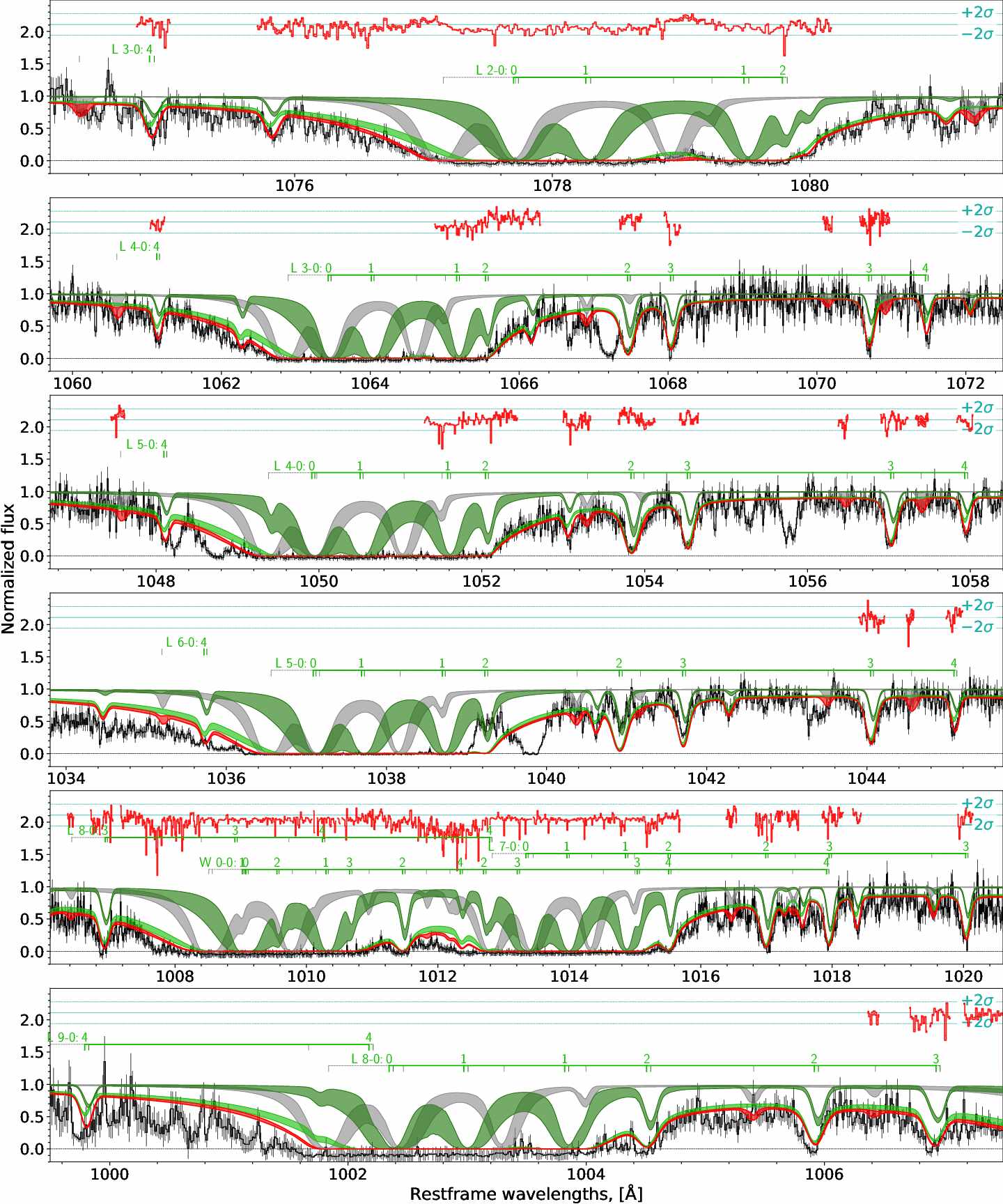}
    \caption{Fit to H2 absorption lines towards AV 476 in SMC. Lines are the same as for \ref{fig:lines_H2_Sk67_2}.
    }
    \label{fig:lines_H2_AV476}
\end{figure*}

\begin{table*}
    \caption{Fit results of H$_2$ lines towards AV 479}
    \label{tab:AV479}
    \begin{tabular}{ccccc}
    \hline
    \hline
    species & comp & 1 & 2 & 3 \\
            & z & $0.0000435(^{+15}_{-14})$ & $0.0005074(^{+4}_{-6})$ &  $0.00055958(^{+81}_{-30})$ \\
    \hline 
     ${\rm H_2\, J=0}$ & b\,km/s &$1.03^{+0.12}_{-0.47}$ & $2.65^{+0.26}_{-0.62}$ &$2.97^{+0.18}_{-0.77}$ \\
                       & $\log N$ & $16.85^{+0.05}_{-0.04}$ & $18.503^{+0.019}_{-0.039}$ &  $18.627^{+0.016}_{-0.016}$\\
    ${\rm H_2\, J=1}$ & b\,km/s & $1.41^{+0.23}_{-0.15}$ & $2.75^{+0.13}_{-0.18}$ & $2.86^{+0.16}_{-0.44}$\\
                      & $\log N$ & $17.21^{+0.06}_{-0.04}$ & $18.279^{+0.023}_{-0.014}$ & $18.715^{+0.012}_{-0.011}$\\
    ${\rm H_2\, J=2}$ & b\,km/s & $3.70^{+0.22}_{-0.15}$ & $2.77^{+0.11}_{-0.12}$ & $2.95^{+0.17}_{-0.06}$\\
                      & $\log N$ & $14.63^{+0.07}_{-0.07}$ & $17.31^{+0.04}_{-0.05}$ & $17.409^{+0.027}_{-0.102}$\\
    ${\rm H_2\, J=3}$ & b\,km/s & $3.18^{+0.22}_{-0.04}$ & $3.49^{+0.12}_{-0.09}$ &  $3.49^{+0.11}_{-0.11}$\\
                      & $\log N$ &$14.32^{+0.06}_{-0.04}$ & $16.59^{+0.07}_{-0.03}$ & $16.585^{+0.028}_{-0.061}$\\
    ${\rm H_2\, J=4}$ & b\,km/s & -- &  $3.02^{+0.10}_{-0.16}$ &$3.27^{+0.22}_{-0.14}$ \\
    				  & $\log N$ & $13.86^{+0.07}_{-0.11}$ & $13.95^{+0.07}_{-0.09}$ &$14.83^{+0.07}_{-0.06}$ \\
    ${\rm H_2\, J=5}$ & b\,km/s & -- &  $4.53^{+0.24}_{-0.24}$ & $5.62^{+0.14}_{-0.32}$\\
    				  & $\log N$ & -- &$14.18^{+0.09}_{-0.09}$ &$13.88^{+0.14}_{-0.11}$ \\
    ${\rm H_2\, J=6}$ & $\log N$ & -- &$13.62^{+0.10}_{-0.13}$ & $13.28^{+0.19}_{-1.51}$\\
    \hline 
         & $\log N_{\rm tot}$ & $17.37^{+0.04}_{-0.03}$ & $18.73^{+0.01}_{-0.02}$ & $18.99^{+0.01}_{-0.01}$\\
    \hline
    HD J=0 & b\,km/s & $1.04^{+0.14}_{-0.25}$ & $2.67^{+0.27}_{-0.76}$ & $2.80^{+0.30}_{-0.80}$ \\
           & $\log N$ & $\lesssim 15.0$ & $\lesssim 15.2$ & $\lesssim 15.4$ \\  
    \hline   
    \end{tabular}
    \begin{tablenotes}
    \item Doppler parameters H$_2$ $\rm J=4$ in 1 component and $\rm J=6$ in 2 and 3 components were tied to H$_2$ $\rm J=3$ and $\rm J=5$, respectively.
    \end{tablenotes}
\end{table*}

\begin{figure*}
    \centering
    \includegraphics[width=\linewidth]{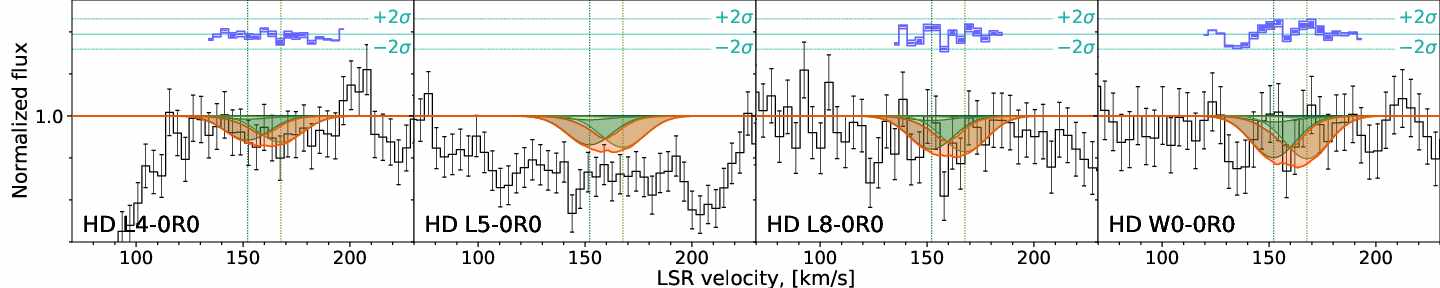}
    \caption{Fit to HD absorption lines towards AV 479 in SMC. Lines are the same as for \ref{fig:lines_HD_Sk67_2}.
    }
    \label{fig:lines_HD_AV479}
\end{figure*}

\begin{figure*}
    \centering
    \includegraphics[width=\linewidth]{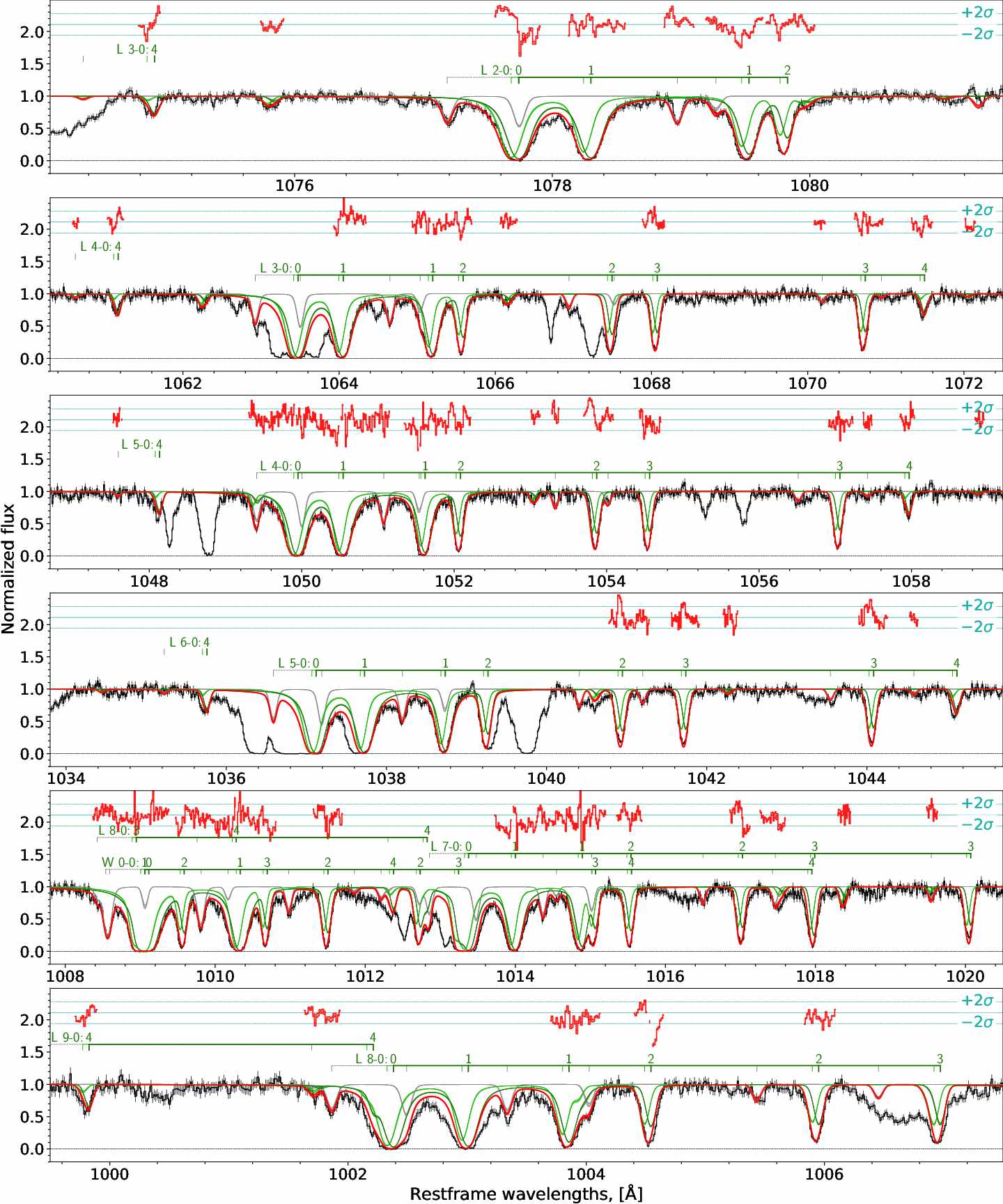}
    \caption{Fit to H2 absorption lines towards AV 479 in SMC. Lines are the same as for \ref{fig:lines_H2_Sk67_2}.
    }
    \label{fig:lines_H2_AV479}
\end{figure*}

\begin{table*}
    \caption{Fit results of H$_2$ lines towards AV 480}
    \label{tab:AV480}
    \begin{tabular}{cccccc}
    \hline
    \hline
    species & comp & 1 & 2 & 3 & 4 \\
            & z & $0.000074(^{+3}_{-3})$ & $0.0000982(^{+29}_{-24})$ & $0.000613(^{+4}_{-3})$ & $0.0006450(^{+27}_{-40})$ \\
    \hline 
     ${\rm H_2\, J=0}$ & b\,km/s & $1.50^{+0.20}_{-0.75}$ & $1.24^{+0.21}_{-0.53}$ & $1.0^{+0.3}_{-0.4}$ & $0.81^{+0.63}_{-0.21}$\\
                       & $\log N$ &$17.65^{+0.12}_{-0.18}$ & $18.06^{+0.07}_{-0.07}$ & $17.66^{+0.10}_{-0.29}$ & $18.17^{+0.06}_{-0.07}$\\
    ${\rm H_2\, J=1}$ & b\,km/s & $1.7^{+0.3}_{-0.4}$ & $1.34^{+0.48}_{-0.29}$ & $1.2^{+0.6}_{-0.4}$ & $1.65^{+0.21}_{-0.69}$\\
                      & $\log N$ &  $17.47^{+0.18}_{-0.18}$ & $17.72^{+0.09}_{-0.12}$ & $18.04^{+0.13}_{-0.23}$ &$18.27^{+0.11}_{-0.11}$ \\
    ${\rm H_2\, J=2}$ & b\,km/s & $2.04^{+0.10}_{-0.26}$ & $1.79^{+0.28}_{-0.29}$ & $2.04^{+0.13}_{-0.26}$ & $1.99^{+0.10}_{-0.27}$ \\
                      & $\log N$ & $15.9^{+0.4}_{-0.4}$ & $15.3^{+0.4}_{-0.4}$ & $16.90^{+0.21}_{-0.56}$& $16.8^{+0.3}_{-0.5}$ \\
    ${\rm H_2\, J=3}$ & b\,km/s & $2.14^{+0.14}_{-0.14}$ & $2.07^{+0.15}_{-0.24}$ & $2.05^{+0.10}_{-0.18}$ &$2.05^{+0.08}_{-0.24}$ \\
                      & $\log N$ & $14.61^{+0.25}_{-0.48}$ &$14.1^{+0.5}_{-0.5}$ &  $15.22^{+1.26}_{-0.12}$ & $16.5^{+0.4}_{-1.0}$\\
    ${\rm H_2\, J=4}$ & $\log N$ & $13.56^{+0.31}_{-0.93}$ & $13.2^{+0.4}_{-1.6}$ & $13.9^{+0.7}_{-0.9}$ & $14.8^{+0.4}_{-0.5}$\\
    \hline 
         & $\log N_{\rm tot}$ & $17.88^{+0.11}_{-0.12}$ & $18.22^{+0.06}_{-0.06}$ & $18.21^{+0.10}_{-0.16}$ & $18.53^{+0.07}_{-0.07}$ \\
     \hline
     HD J=0 &  b\,km/s & $1.54^{+0.17}_{-0.69}$ & $0.94^{+0.47}_{-0.22}$ & $1.13^{+0.15}_{-0.43}$ & $0.81^{+0.58}_{-0.31}$ \\
             & $\log N$ & $\lesssim 16.2$ & $\lesssim 16.2$ & $\lesssim 16.1$ & $\lesssim 16.2$ \\ 
    \hline   
    \end{tabular}
    \begin{tablenotes}
    \item Doppler parameters H$_2$ $\rm J=4$  were tied to H$_2$ $\rm J=3$.
    \end{tablenotes}
\end{table*}

\begin{figure*}
    \centering
    \includegraphics[width=\linewidth]{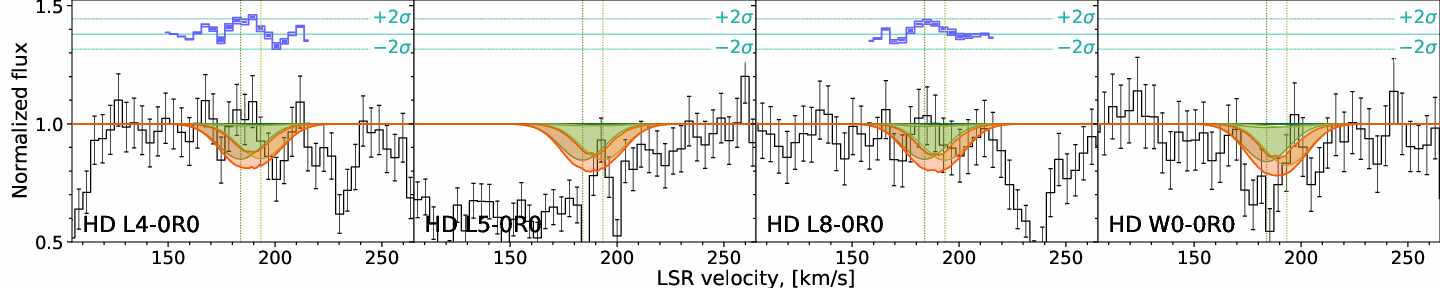}
    \caption{Fit to HD absorption lines towards AV 480 in SMC. Lines are the same as for \ref{fig:lines_HD_Sk67_2}.
    }
    \label{fig:lines_HD_AV480}
\end{figure*}

\begin{figure*}
    \centering
    \includegraphics[width=\linewidth]{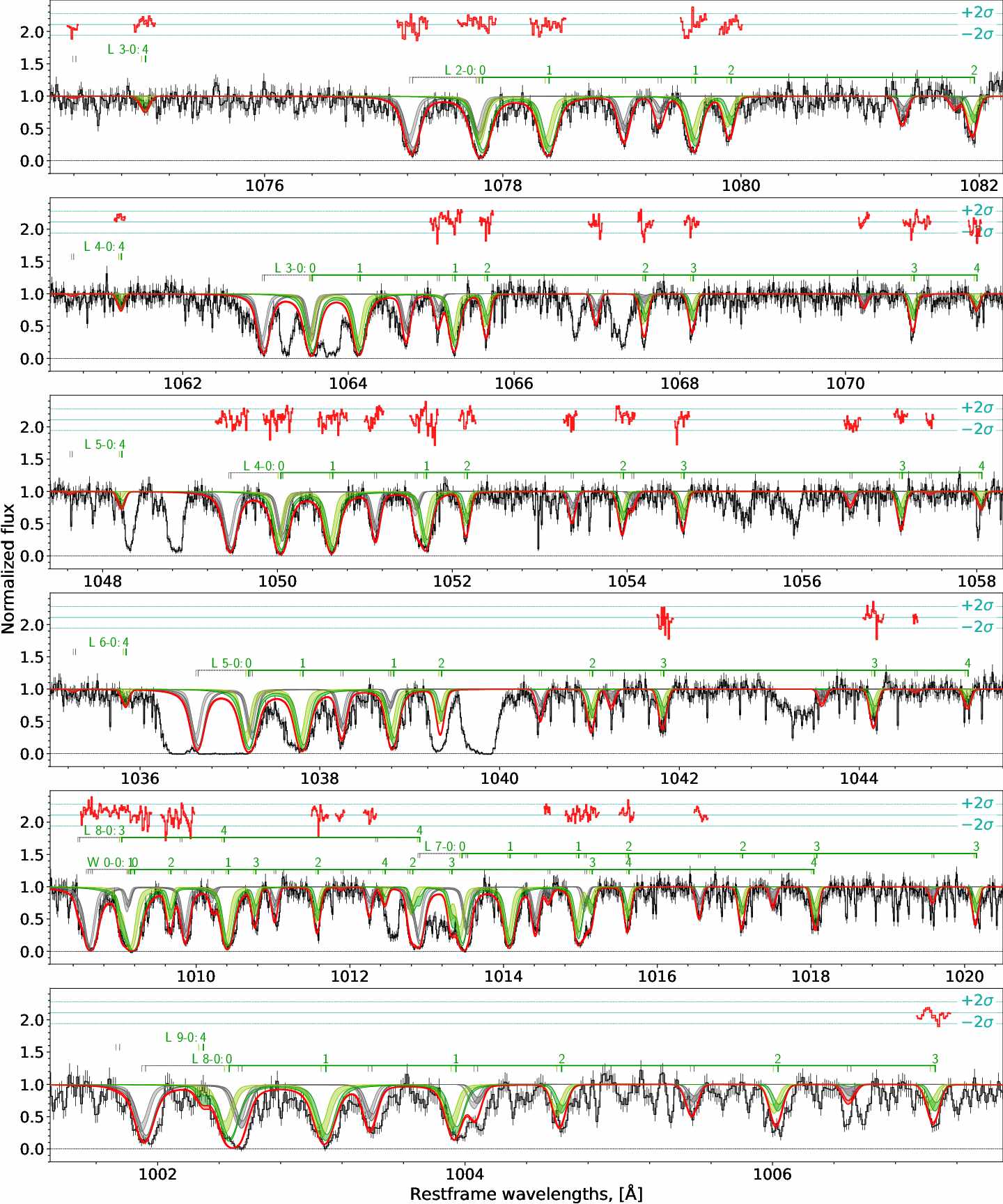}
    \caption{Fit to H2 absorption lines towards AV 480 in SMC. Lines are the same as for \ref{fig:lines_H2_Sk67_2}.
    }
    \label{fig:lines_H2_AV480}
\end{figure*}

\begin{table*}
    \caption{Fit results of H$_2$ lines towards AV 483}
    \label{tab:AV483}
    \begin{tabular}{cccccc}
    \hline
    \hline
    species & comp & 1 & 2 & 3 & 4 \\
            & z & $0.0000446(^{+12}_{-20})$ & $0.0001496(^{+29}_{-40})$ & $0.0004868(^{+11}_{-22})$ & $0.0005244(^{+18}_{-33})$ \\
    \hline 
     ${\rm H_2\, J=0}$ & b\,km/s & $0.534^{+0.161}_{-0.031}$ & $0.63^{+0.15}_{-0.12}$ & $0.67^{+0.78}_{-0.15}$ & $1.0^{+0.4}_{-0.4}$\\
                       & $\log N$ & $16.81^{+0.03}_{-0.05}$ & $15.84^{+0.17}_{-0.42}$ &$18.33^{+0.05}_{-0.05}$ & $18.19^{+0.04}_{-0.08}$\\
    ${\rm H_2\, J=1}$ & b\,km/s & $0.55^{+0.21}_{-0.04}$ & $1.74^{+0.31}_{-0.62}$ & $1.4^{+0.6}_{-0.5}$ & $1.3^{+0.6}_{-0.5}$\\
                      & $\log N$ & $17.214^{+0.039}_{-0.023}$ & $14.74^{+0.63}_{-0.19}$ & $18.44^{+0.04}_{-0.06}$ & $18.18^{+0.11}_{-0.04}$\\
    ${\rm H_2\, J=2}$ & b\,km/s & $1.03^{+0.16}_{-0.19}$ & $9.1^{+0.9}_{-2.3}$ & $1.7^{+0.6}_{-0.3}$ & $2.1^{+0.6}_{-0.4}$\\
                      & $\log N$ & $15.31^{+0.29}_{-0.21}$ & $13.94^{+0.06}_{-0.07}$ & $17.22^{+0.08}_{-0.15}$ & $15.76^{+0.75}_{-0.28}$ \\
    ${\rm H_2\, J=3}$ & b\,km/s & $9.0^{+0.9}_{-7.8}$ & $9.82^{+0.18}_{-1.86}$ & $2.80^{+0.22}_{-0.29}$ & $9.93^{+0.07}_{-0.52}$ \\
                      & $\log N$ & $13.4^{+0.4}_{-0.5}$ & $13.52^{+0.24}_{-0.45}$ &$16.63^{+0.17}_{-0.21}$ & $14.59^{+0.09}_{-0.03}$ \\
    ${\rm H_2\, J=4}$ & b\,km/s & -- & -- & $2.8^{+1.8}_{-0.6}$ & $14.6^{+0.4}_{-2.8}$\\
                      & $\log N$ & $11.2^{+0.6}_{-1.2}$ &$13.08^{+0.30}_{-1.54}$ & $14.18^{+0.07}_{-0.07}$ & $13.41^{+0.25}_{-0.88}$\\
    ${\rm H_2\, J=4}$ & $\log N$ & -- & -- &$13.66^{+0.10}_{-0.25}$ & $13.4^{+0.4}_{-1.0}$\\
    \hline 
         & $\log N_{\rm tot}$ & $17.36^{+0.03}_{-0.02}$ & $15.88^{+0.17}_{-0.36}$ & $18.71^{+0.03}_{-0.04}$ & $18.48^{+0.06}_{-0.04}$ \\
    \hline
    HD J=0 & b\,km/s & $0.508^{+0.218}_{-0.008}$ & $0.509^{+0.273}_{-0.009}$ & $0.54^{+1.02}_{-0.04}$ & $0.524^{+0.840}_{-0.024}$ \\
           & $\log N$ & $\lesssim 14.3$ & $\lesssim 14.1$ & $\lesssim 16.1$ & $\lesssim 16.4$ \\ 
    \hline   
    \end{tabular}
    \begin{tablenotes}
    \item Doppler parameters H$_2$ $\rm J=4$ in 1 and 2 components and $\rm J=5$ in 3 and 4 components  were tied to H$_2$ $\rm J=3$ and $\rm J=4$, respectively.
    \end{tablenotes}
\end{table*}

\begin{figure*}
    \centering
    \includegraphics[width=\linewidth]{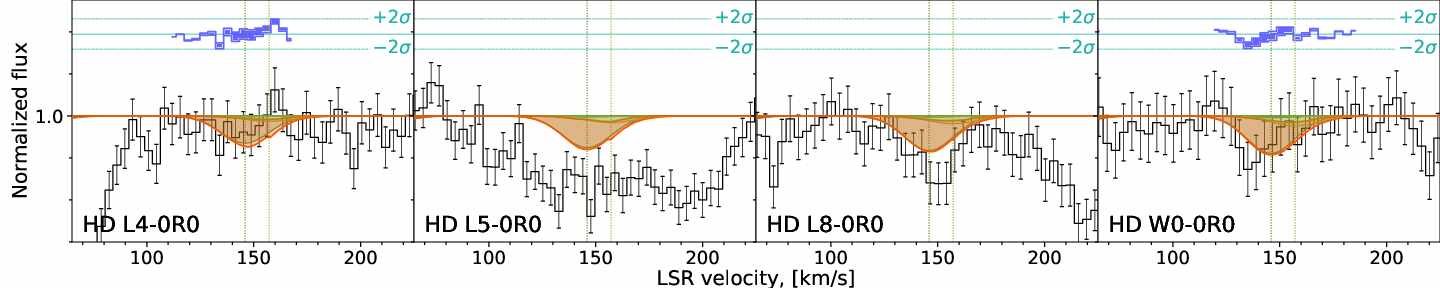}
    \caption{Fit to HD absorption lines towards AV 483 in SMC. Lines are the same as for \ref{fig:lines_HD_Sk67_2}.
    }
    \label{fig:lines_HD_AV483}
\end{figure*}

\begin{figure*}
    \centering
    \includegraphics[width=\linewidth]{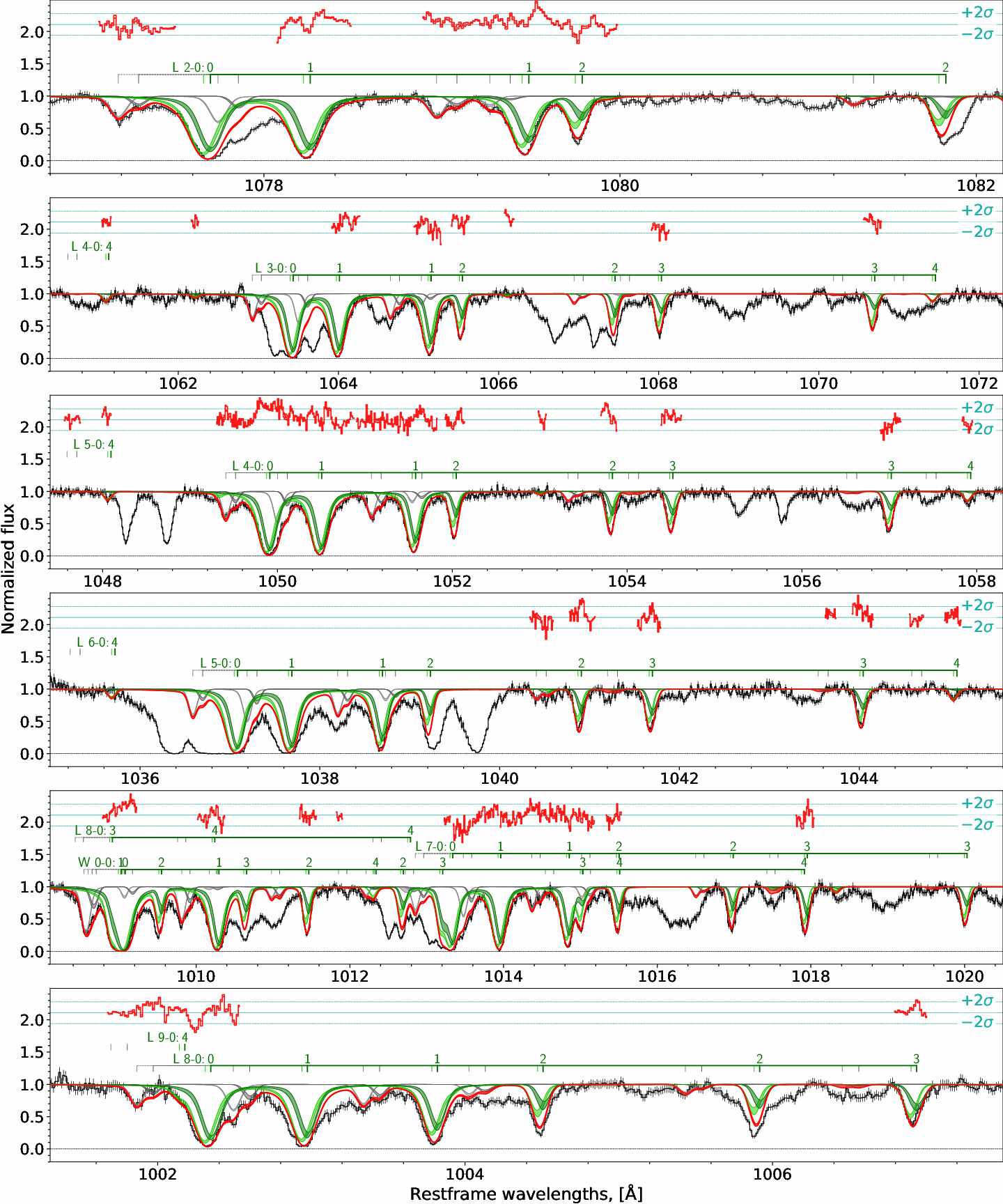}
    \caption{Fit to H2 absorption lines towards AV 483 in SMC. Lines are the same as for \ref{fig:lines_H2_Sk67_2}.
    }
    \label{fig:lines_H2_AV483}
\end{figure*}

\begin{table*}
    \caption{Fit results of H$_2$ lines towards AV 486}
    \label{tab:AV486}
    \begin{tabular}{ccccc}
    \hline
    \hline
    species & comp & 1 & 2 & 3 \\
            & z & $0.0000550(^{+5}_{-24})$ & $0.0004852(^{+28}_{-21})$ & $0.0005256(^{+21}_{-14})$\\
    \hline 
     ${\rm H_2\, J=0}$ & b\,km/s &$1.64^{+0.21}_{-0.28}$ & $0.80^{+0.39}_{-0.29}$ &$1.0^{+0.4}_{-0.4}$ \\
                       & $\log N$ &  $16.36^{+0.20}_{-0.09}$ & $18.25^{+0.10}_{-0.12}$ & $18.82^{+0.03}_{-0.04}$ \\
    ${\rm H_2\, J=1}$ & b\,km/s & $2.03^{+0.07}_{-0.25}$ &  $1.27^{+0.51}_{-0.28}$ &  $1.90^{+0.11}_{-0.89}$\\
                      & $\log N$ &  $16.72^{+0.19}_{-0.11}$ &  $18.24^{+0.09}_{-0.08}$ & $18.70^{+0.04}_{-0.04}$\\
    ${\rm H_2\, J=2}$ & b\,km/s & $1.98^{+0.13}_{-0.17}$ & $1.78^{+0.15}_{-0.22}$ & $2.08^{+0.11}_{-0.13}$\\
                      & $\log N$ & $14.87^{+0.20}_{-0.14}$ & $16.74^{+0.22}_{-0.15}$ &$16.89^{+0.18}_{-0.20}$ \\
    ${\rm H_2\, J=3}$ & b\,km/s & $2.04^{+0.11}_{-0.15}$ & $2.16^{+0.09}_{-0.09}$ & $2.06^{+0.09}_{-0.11}$ \\
                      & $\log N$ & $14.04^{+0.12}_{-0.08}$ & $16.25^{+0.22}_{-0.17}$ & $16.25^{+0.28}_{-0.35}$\\
    ${\rm H_2\, J=4}$ & b\,km/s & -- &  $2.05^{+0.10}_{-0.11}$ & -- \\
    				  & $\log N$ & $13.40^{+0.25}_{-1.26}$ & $13.97^{+0.12}_{-0.25}$ & $14.11^{+0.20}_{-0.20}$ \\
    ${\rm H_2\, J=5}$ & $\log N$ & -- &  $14.03^{+0.09}_{-0.08}$ & -- \\
    \hline 
         & $\log N_{\rm tot}$ & $16.88^{+0.15}_{-0.08}$ & $18.56^{+0.07}_{-0.07}$ & $19.07^{+0.02}_{-0.03}$  \\
    \hline
    HD J=0 & b\,km/s & $1.62^{+0.27}_{-0.29}$ & $0.520^{+0.563}_{-0.020}$ & $0.80^{+0.40}_{-0.30}$ \\
            & $\log N$ & $\lesssim 14.0$ & $\lesssim 15.6$ & $14.0^{+0.9}_{-0.4}$ \\
    \hline   
    \end{tabular}
    \begin{tablenotes}
    \item Doppler parameters H$_2$ $\rm J=4$ in 1 and 3 components and $\rm J=5$ in 2 component were tied to H$_2$ $\rm J=3$ and $\rm J=4$, respectively.
    \end{tablenotes}
\end{table*}

\begin{figure*}
    \centering
    \includegraphics[width=\linewidth]{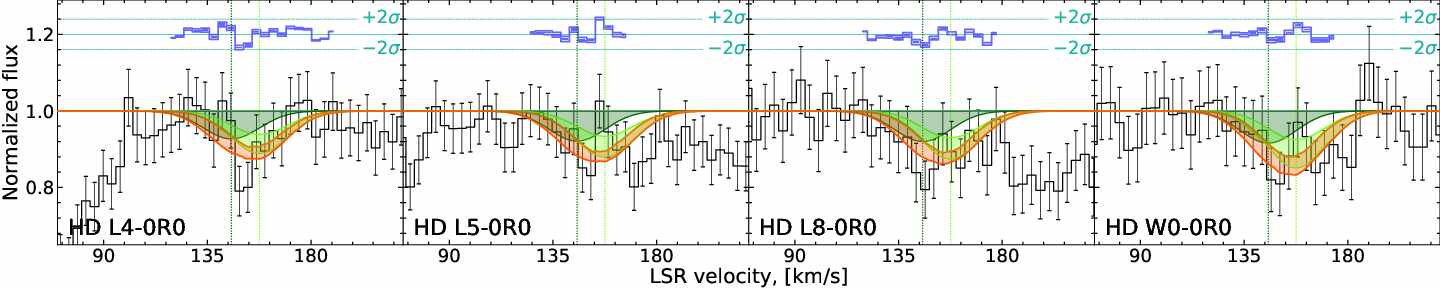}
    \caption{Fit to HD absorption lines towards AV 486 in SMC. Lines are the same as for \ref{fig:lines_HD_Sk67_2}.
    }
    \label{fig:lines_H2_AV486}
\end{figure*}

\begin{figure*}
    \centering
    \includegraphics[width=\linewidth]{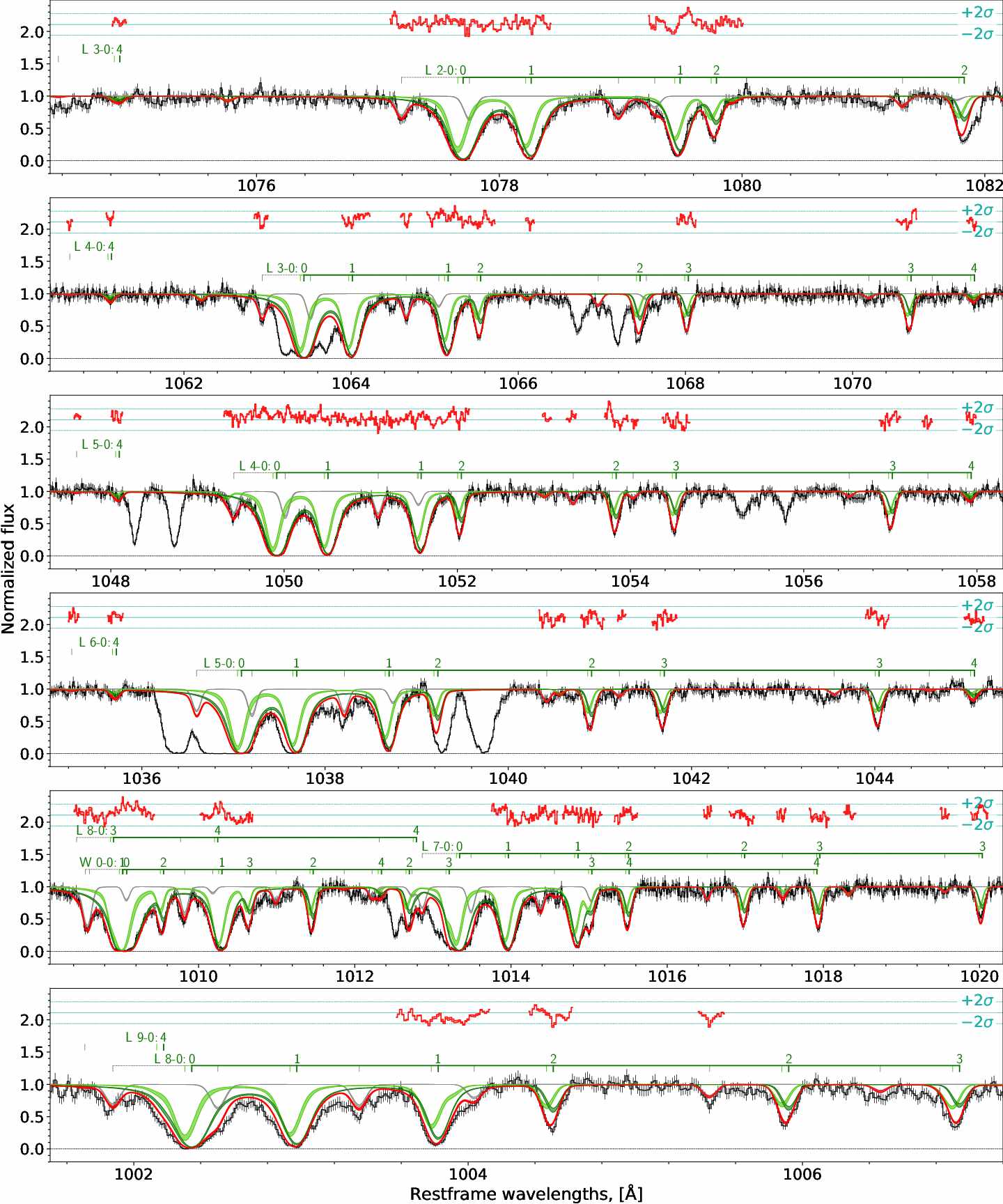}
    \caption{Fit to H2 absorption lines towards AV 486 in SMC. Lines are the same as for \ref{fig:lines_H2_Sk67_2}.
    }
    \label{fig:lines_HD_AV486}
\end{figure*}

\begin{table*}
    \caption{Fit results of H$_2$ lines towards AV 488}
    \label{tab:AV488}
    \begin{tabular}{ccccc}
    \hline
    \hline
    species & comp & 1 & 2 & 3 \\
            & z & $0.0000546(^{+7}_{-8})$ & $0.0004889(^{+11}_{-15})$ &  $0.0005286(^{+11}_{-14})$\\
    \hline 
     ${\rm H_2\, J=0}$ & b\,km/s & $1.85^{+0.27}_{-0.15}$ & $0.80^{+0.62}_{-0.30}$ & $0.511^{+0.150}_{-0.011}$\\
                       & $\log N$ &  $15.76^{+0.31}_{-0.21}$ & $18.818^{+0.014}_{-0.035}$ & $18.48^{+0.06}_{-0.03}$\\
    ${\rm H_2\, J=1}$ & b\,km/s & $5.12^{+0.13}_{-0.33}$ &$2.1^{+0.4}_{-0.9}$ & $3.0^{+0.4}_{-0.9}$\\
                      & $\log N$ & $15.18^{+0.05}_{-0.04}$ & $18.64^{+0.03}_{-0.04}$ & $18.787^{+0.026}_{-0.032}$\\
    ${\rm H_2\, J=2}$ & b\,km/s & $4.97^{+0.30}_{-0.25}$ & $3.12^{+0.17}_{-0.61}$ &$3.10^{+0.41}_{-0.27}$ \\
                      & $\log N$ & $14.30^{+0.04}_{-0.04}$ & $15.43^{+0.20}_{-0.24}$ &$16.38^{+0.24}_{-0.19}$ \\
    ${\rm H_2\, J=3}$ & b\,km/s & $5.06^{+0.39}_{-0.17}$ &$3.4^{+0.4}_{-0.4}$ &$3.7^{+0.5}_{-0.5}$ \\
                      & $\log N$ &$14.156^{+0.021}_{-0.038}$ &$14.86^{+0.12}_{-0.05}$ &$15.23^{+0.30}_{-0.13}$ \\
    ${\rm H_2\, J=4}$ & b\,km/s & -- & $6.4^{+11.1}_{-2.7}$ & $4.1^{+1.3}_{-0.6}$\\
    				  & $\log N$ & $13.57^{+0.06}_{-0.13}$ & $13.51^{+0.14}_{-0.18}$ & $13.82^{+0.09}_{-0.10}$\\
    ${\rm H_2\, J=5}$ & b\,km/s & -- & $10.78^{+11.84}_{-3.22}$ & $27.24^{+10.31}_{-15.12}$ \\
                      & $\log N$ & -- &  $13.83^{+0.10}_{-0.36}$ & $13.86^{+0.17}_{-0.30}$\\
   
    \hline 
         & $\log N_{\rm tot}$ & $15.88^{+0.25}_{-0.15}$ & $19.04^{+0.01}_{-0.03}$ & $18.96^{+0.03}_{-0.02}$ \\
     \hline
    HD J=0 & b\,km/s & $1.87^{+0.31}_{-0.22}$ & $0.531^{+0.699}_{-0.031}$ & $0.508^{+0.184}_{-0.008}$ \\    
           & $\log N$ & $\lesssim 13.4$ &  $13.6^{+0.6}_{-0.5}$ & $\lesssim 14.6$ \\
    \hline   
    \end{tabular}
    \begin{tablenotes}
    \item Doppler parameter H$_2$ $\rm J=4$ in 1 component was tied to H$_2$ $\rm J=3$.
    \end{tablenotes}
\end{table*}

\begin{figure*}
    \centering
    \includegraphics[width=\linewidth]{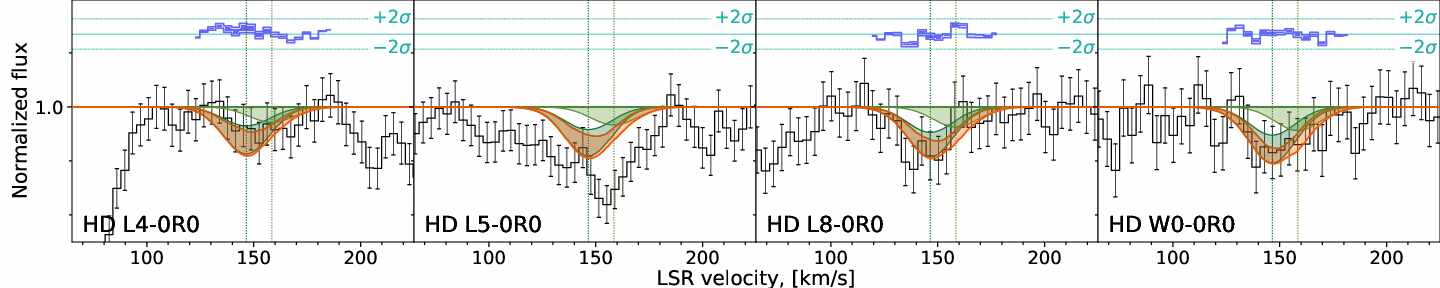}
    \caption{Fit to HD absorption lines towards AV 488 in SMC. Lines are the same as for \ref{fig:lines_HD_Sk67_2}.
    }
    \label{fig:lines_HD_AV488}
\end{figure*}

\begin{figure*}
    \centering
    \includegraphics[width=\linewidth]{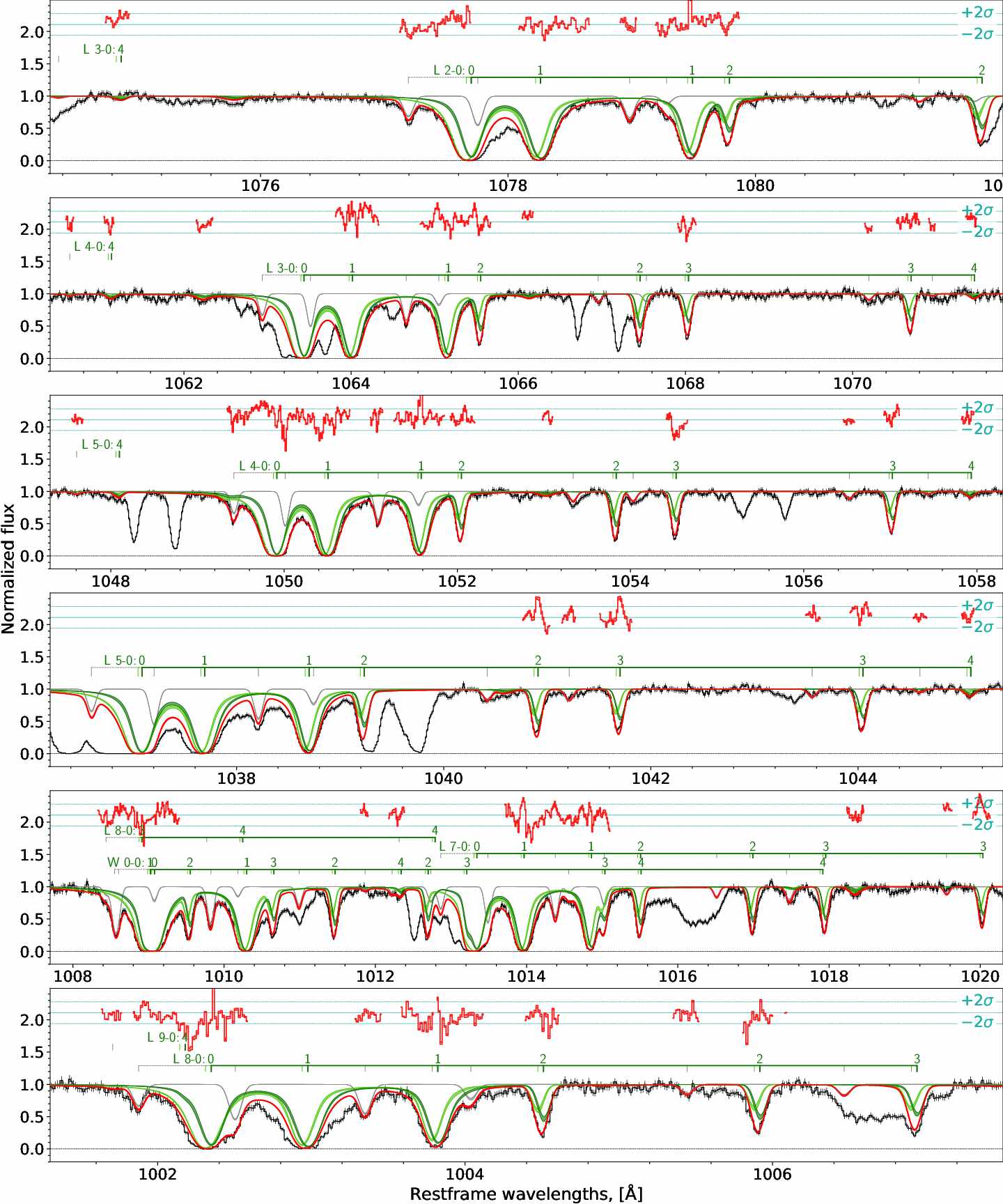}
    \caption{Fit to H2 absorption lines towards AV488 in SMC. Lines are the same as for \ref{fig:lines_H2_Sk67_2}.
    }
    \label{fig:lines_H2_AV488}
\end{figure*}

\clearpage
\begin{table*}
    \caption{Fit results of H$_2$ lines towards AV 490}
    \label{tab:AV490}
    \begin{tabular}{ccccc}
    \hline
    \hline
    species & comp & 1 & 2 & 3 \\
            & z & $0.0000189(^{+13}_{-14})$ &  $0.0004403(^{+3}_{-5})$ & $0.0005428(^{+19}_{-24})$\\
    \hline 
     ${\rm H_2\, J=0}$ & b\,km/s & $0.60^{+0.22}_{-0.10}$ & $0.73^{+0.64}_{-0.23}$ & $0.526^{+0.021}_{-0.025}$ \\
                       & $\log N$ & $16.974^{+0.028}_{-0.092}$ & $19.540^{+0.013}_{-0.011}$ & $17.93^{+0.29}_{-0.39}$ \\
    ${\rm H_2\, J=1}$ & b\,km/s &$0.91^{+0.13}_{-0.31}$ & $1.2^{+1.1}_{-0.4}$ & $0.54^{+0.04}_{-0.04}$ \\
                      & $\log N$ & $17.12^{+0.04}_{-0.06}$ &$19.609^{+0.004}_{-0.005}$ & $15.27^{+0.34}_{-0.22}$ \\
    ${\rm H_2\, J=2}$ & b\,km/s &$1.11^{+0.47}_{-0.18}$ & $4.78^{+0.34}_{-0.18}$ &  $2.0^{+0.4}_{-0.5}$\\
                      & $\log N$ & $15.55^{+0.30}_{-0.50}$ & $17.74^{+0.07}_{-0.13}$ & $14.42^{+0.15}_{-0.08}$\\
    ${\rm H_2\, J=3}$ & b\,km/s & $2.98^{+0.21}_{-1.01}$ & $5.1^{+0.5}_{-0.4}$ & $4.80^{+0.25}_{-0.53}$ \\
                      & $\log N$ & $14.11^{+0.12}_{-0.13}$ & $17.17^{+0.20}_{-0.32}$ &$14.59^{+0.04}_{-0.05}$  \\
    ${\rm H_2\, J=4}$ & b\,km/s &-- &  $8.0^{+0.8}_{-1.9}$ & -- \\
    				  & $\log N$ & $11.2^{+1.0}_{-0.8}$ &  $14.665^{+0.046}_{-0.031}$ & $13.66^{+0.08}_{-0.14}$ \\
    ${\rm H_2\, J=5}$ & b\,km/s & -- & $8.4^{+1.1}_{-0.9}$ & -- \\
                      & $\log N$ & -- & $14.31^{+0.04}_{-0.05}$ & $13.67^{+0.13}_{-0.17}$\\
    
    \hline 
         & $\log N_{\rm tot}$ & $17.36^{+0.03}_{-0.05}$ & $19.88^{+0.01}_{-0.01}$ & $17.94^{+0.29}_{-0.39}$ \\
    \hline
    HD J=0 & b\,km/s & $0.50^{+16.48}_{-0.00}$ &$0.78^{+0.82}_{-0.28}$ & $0.531^{+0.012}_{-0.031}$ \\
           & $\log N$ & $\lesssim 15.7$ &  $16.10^{+0.26}_{-0.66}$ & $\lesssim 15.6$  \\
    \hline   
    \end{tabular}
    \begin{tablenotes}
    \item Doppler parameters H$_2$ $\rm J=4$ in 1 component and $\rm J=4, 5$ in 3 component were tied to H$_2$ $\rm J=3$.
    \end{tablenotes}
\end{table*}

\begin{figure*}
    \centering
    \includegraphics[width=\linewidth]{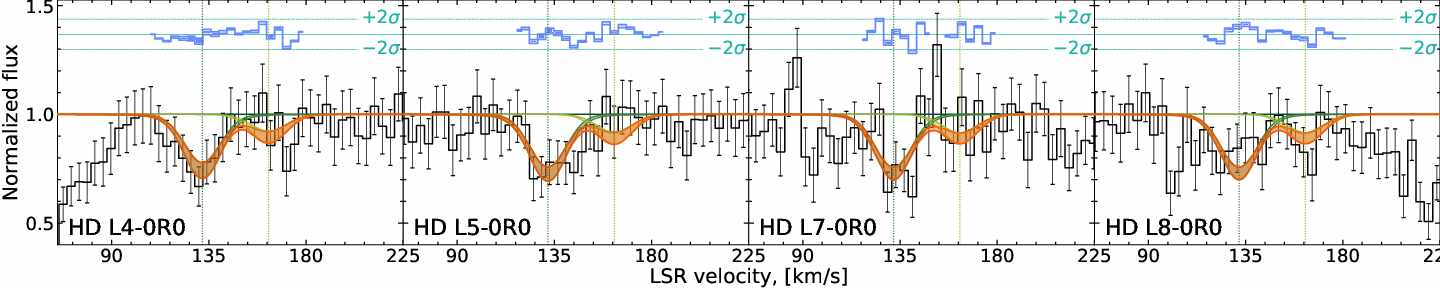}
    \caption{Fit to HD absorption lines towards AV 490 in SMC. Lines are the same as for \ref{fig:lines_HD_Sk67_2}.
    }
    \label{fig:lines_HD_AV490_appendix}
\end{figure*}

\begin{figure*}
    \centering
    \includegraphics[width=\linewidth]{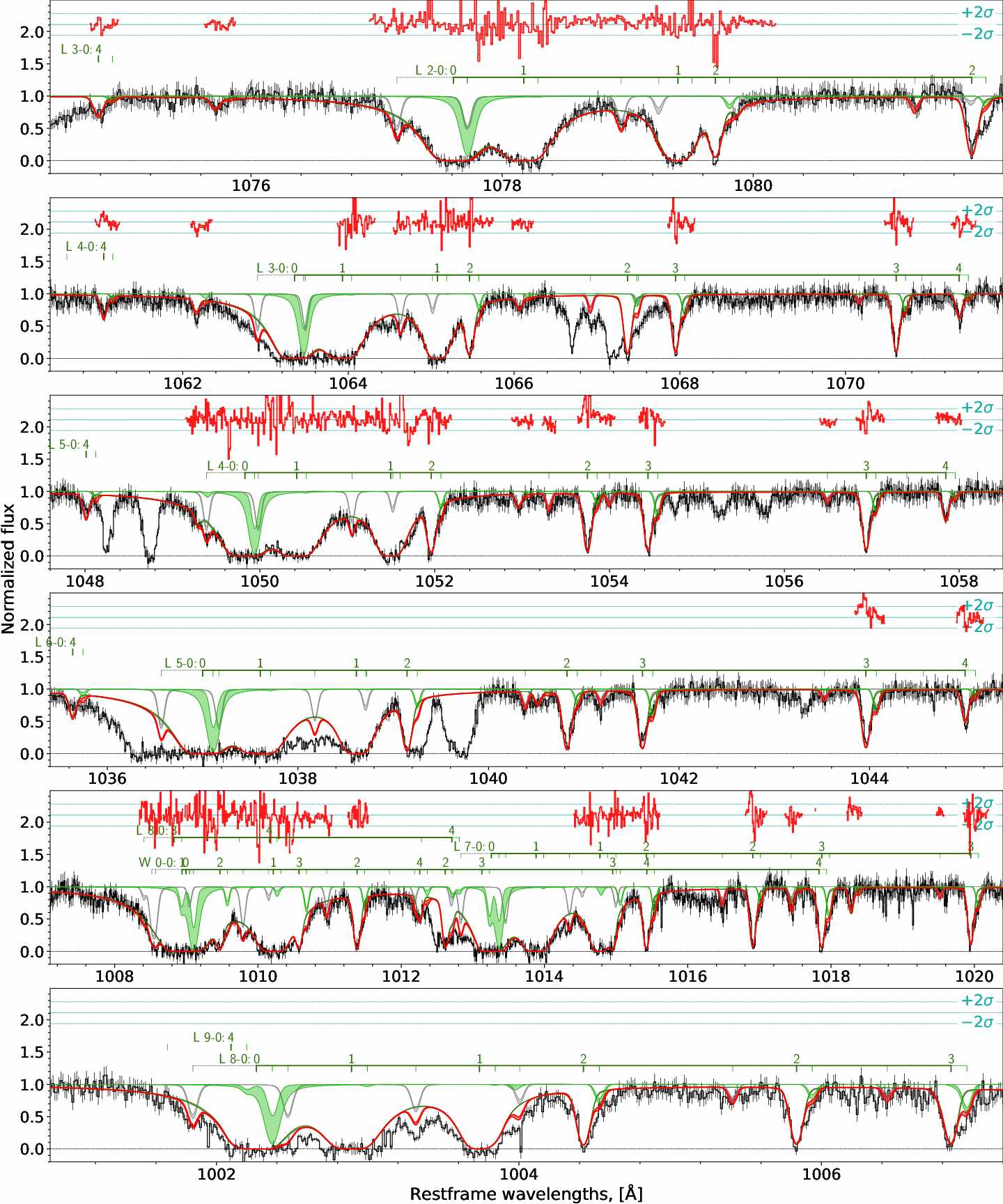}
    \caption{Fit to H2 absorption lines towards AV 490 in SMC. Lines are the same as for \ref{fig:lines_H2_Sk67_2}.
    }
    \label{fig:lines_H2_AV490}
\end{figure*}

\begin{table*}
    \caption{Fit results of H$_2$ lines towards AV 491}
    \label{tab:AV491}
    \begin{tabular}{ccccccc}
    \hline
    \hline
    species & comp & 1 & 2 & 3 & 4 & 5 \\
            & z &$0.0000463(^{+6}_{-54})$ & $0.000326(^{+6}_{-6})$ & $0.000479(^{+3}_{-5})$ & $0.000519(^{+26}_{-4})$ & $0.000615(^{+7}_{-6})$ \\
    \hline 
     ${\rm H_2\, J=0}$ & b\,km/s & $1.5^{+0.4}_{-0.7}$ & $0.57^{+0.55}_{-0.07}$ & $2.2^{+0.6}_{-1.5}$ & $0.61^{+0.52}_{-0.10}$ & $0.71^{+1.97}_{-0.18}$ \\
                       & $\log N$ & $16.2^{+0.3}_{-0.9}$ & $15.6^{+0.5}_{-1.0}$& $18.46^{+0.11}_{-0.19}$ & $18.730^{+0.020}_{-0.249}$ & $14.2^{+1.9}_{-0.3}$ \\
    ${\rm H_2\, J=1}$ & b\,km/s & $2.7^{+0.6}_{-0.8}$ & $0.70^{+0.79}_{-0.20}$ & $3.3^{+1.0}_{-1.5}$ & $1.1^{+0.5}_{-0.5}$ & $9.4^{+0.6}_{-2.8}$\\
                      & $\log N$ & $15.6^{+0.5}_{-0.5}$ & $14.21^{+0.53}_{-0.25}$ &$18.83^{+0.06}_{-0.12}$ & $18.61^{+0.10}_{-0.93}$ & $14.82^{+0.13}_{-0.07}$ \\
    ${\rm H_2\, J=2}$ & b\,km/s & $2.4^{+1.3}_{-0.5}$ & $1.3^{+2.4}_{-0.6}$ &  $4.7^{+0.7}_{-0.7}$ & $1.3^{+0.9}_{-0.4}$ &$8.7^{+1.6}_{-1.0}$ \\
                      & $\log N$ & $14.60^{+0.28}_{-0.19}$ & $13.77^{+0.13}_{-0.16}$ & $16.88^{+0.35}_{-0.19}$ &$15.6^{+1.1}_{-0.7}$ & $14.13^{+0.09}_{-0.21}$ \\
    ${\rm H_2\, J=3}$ & b\,km/s &  $3.1^{+3.4}_{-1.0}$ & $2.1^{+2.8}_{-0.7}$ &  $5.1^{+1.0}_{-1.0}$ & $4.3^{+2.7}_{-1.2}$ & $9.4^{+0.5}_{-1.0}$\\
                      & $\log N$ & $14.33^{+0.09}_{-0.07}$ & $14.26^{+0.14}_{-0.09}$ & $15.84^{+0.31}_{-0.37}$ &$14.51^{+0.32}_{-0.15}$ & $14.29^{+0.10}_{-0.12}$ \\
    ${\rm H_2\, J=4}$ & $\log N$ & $13.85^{+0.12}_{-0.16}$ & $13.89^{+0.11}_{-0.13}$ & $14.33^{+0.09}_{-0.22}$ & $14.04^{+0.26}_{-0.08}$ & $13.84^{+0.10}_{-0.28}$\\
    
    \hline 
         & $\log N_{\rm tot}$ & $16.31^{+0.27}_{-0.52}$ & $15.66^{+0.47}_{-0.71}$ & $18.99^{+0.05}_{-0.10}$ & $18.98^{+0.05}_{-0.26}$ & $15.09^{+0.81}_{-0.06}$ \\
    \hline
    HD J=0 & b\,km/s & $2.0^{+0.4}_{-0.4}$ & $0.56^{+0.82}_{-0.06}$ & $0.5^{+2.3}_{-0.0}$ & $0.528^{+0.698}_{-0.028}$ & $0.61^{+4.17}_{-0.11}$ \\
           & $\log N$ & $\lesssim 15.0$ & $\lesssim 14.7$ & $\lesssim 15.5$ & $\lesssim 15.0$ & $\lesssim 14.6$ \\ 
    \hline   
    \end{tabular}
    \begin{tablenotes}
    \item Doppler parameters H$_2$ $\rm J=4$ were tied to H$_2$ $\rm J=3$.
    \end{tablenotes}
\end{table*}

\begin{figure*}
    \centering
    \includegraphics[width=\linewidth]{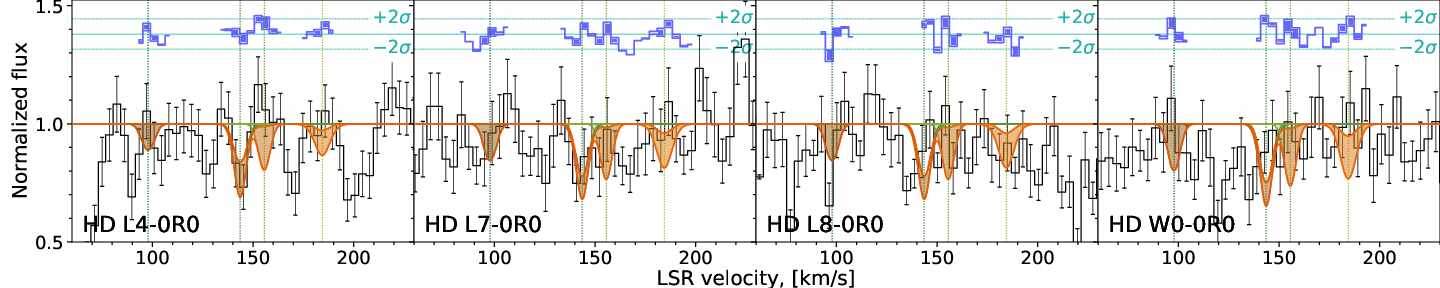}
    \caption{Fit to HD absorption lines towards AV 491 in SMC. Lines are the same as for \ref{fig:lines_HD_Sk67_2}.
    }
    \label{fig:lines_HD_AV491}
\end{figure*}

\begin{figure*}
    \centering
    \includegraphics[width=\linewidth]{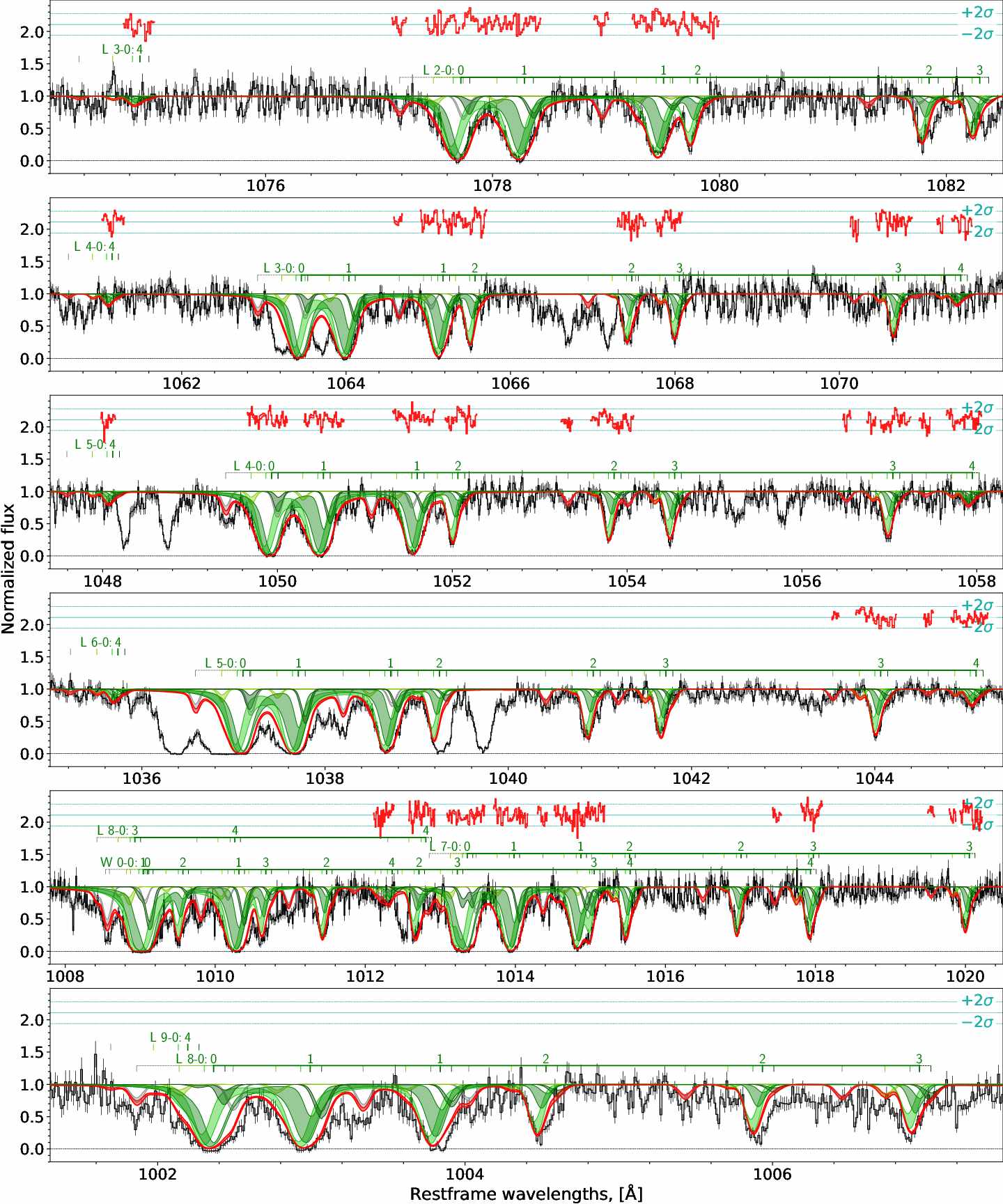}
    \caption{Fit to H2 absorption lines towards AV 491 in SMC. Lines are the same as for \ref{fig:lines_H2_Sk67_2}.
    }
    \label{fig:lines_H2_AV491}
\end{figure*}

\begin{table*}
    \caption{Fit results of H$_2$ lines towards AV 506}
    \label{tab:AV506}
    \begin{tabular}{ccccc}
    \hline
    \hline
    species & comp & 1 & 2 & 3 \\
            & z &$0.0000177(^{+7}_{-10})$ & $0.0004414(^{+8}_{-12})$ & $0.0005252(^{+5}_{-9})$ \\
    \hline 
     ${\rm H_2\, J=0}$ & b\,km/s &$0.99^{+0.67}_{-0.19}$ & $3.3^{+0.5}_{-0.6}$ & $0.9^{+0.5}_{-0.4}$\\
                       & $\log N$ &$17.693^{+0.021}_{-0.036}$ & $15.8^{+0.5}_{-0.3}$ & $18.605^{+0.025}_{-0.028}$\\
    ${\rm H_2\, J=1}$ & b\,km/s &$1.91^{+0.13}_{-0.58}$ & $4.1^{+0.4}_{-0.4}$ & $1.21^{+0.89}_{-0.31}$\\
                      & $\log N$ &  $17.72^{+0.04}_{-0.04}$ & $16.13^{+0.34}_{-0.16}$ & $18.819^{+0.010}_{-0.012}$\\
    ${\rm H_2\, J=2}$ & b\,km/s &$2.07^{+0.09}_{-0.21}$ & $4.74^{+0.75}_{-0.21}$ & $4.6^{+0.5}_{-0.8}$\\
                      & $\log N$ &$16.81^{+0.11}_{-0.10}$ & $15.22^{+0.17}_{-0.08}$ & $16.25^{+1.01}_{-0.14}$\\
    ${\rm H_2\, J=3}$ & b\,km/s & $2.02^{+0.13}_{-0.17}$ &$5.65^{+0.32}_{-0.28}$ & $6.6^{+0.7}_{-0.5}$\\
                      & $\log N$ & $16.13^{+0.24}_{-0.19}$ & $14.799^{+0.034}_{-0.028}$ & $15.24^{+0.10}_{-0.05}$\\
    ${\rm H_2\, J=4}$ & $\log N$ & $13.90^{+0.17}_{-0.14}$ &  $13.77^{+0.14}_{-0.11}$ &$14.15^{+0.07}_{-0.08}$ \\
    \hline 
         & $\log N_{\rm tot}$ & $18.04^{+0.02}_{-0.03}$ & $16.36^{+0.29}_{-0.12}$ & $19.03^{+0.01}_{-0.01}$  \\
    \hline
    HD J=0 & b\,km/s & $0.96^{+0.65}_{-0.30}$ & $3.3^{+0.6}_{-0.6}$ & $0.523^{+0.562}_{-0.023}$ \\
           & $\log N$ & $\lesssim 15.1$ & $\lesssim 14.0$ & $\lesssim 15.7$ \\ 
    \hline   
    \end{tabular}
    \begin{tablenotes}
    \item Doppler parameters H$_2$ $\rm J=4$ in all of the components were tied to H$_2$ $\rm J=3$.
    \end{tablenotes}
\end{table*}

\begin{figure*}
    \centering
    \includegraphics[width=\linewidth]{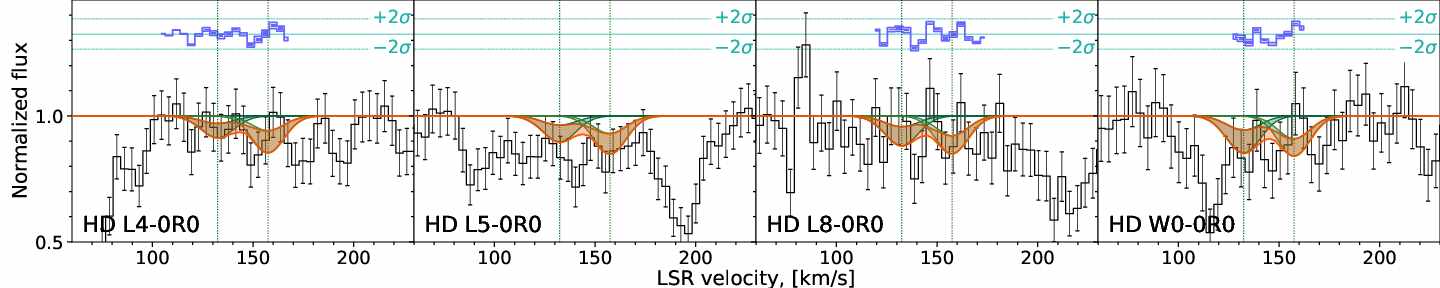}
    \caption{Fit to HD absorption lines towards AV 506 in SMC. Lines are the same as for \ref{fig:lines_HD_Sk67_2}.
    }
    \label{fig:lines_HD_AV506}
\end{figure*}

\begin{figure*}
    \centering
    \includegraphics[width=\linewidth]{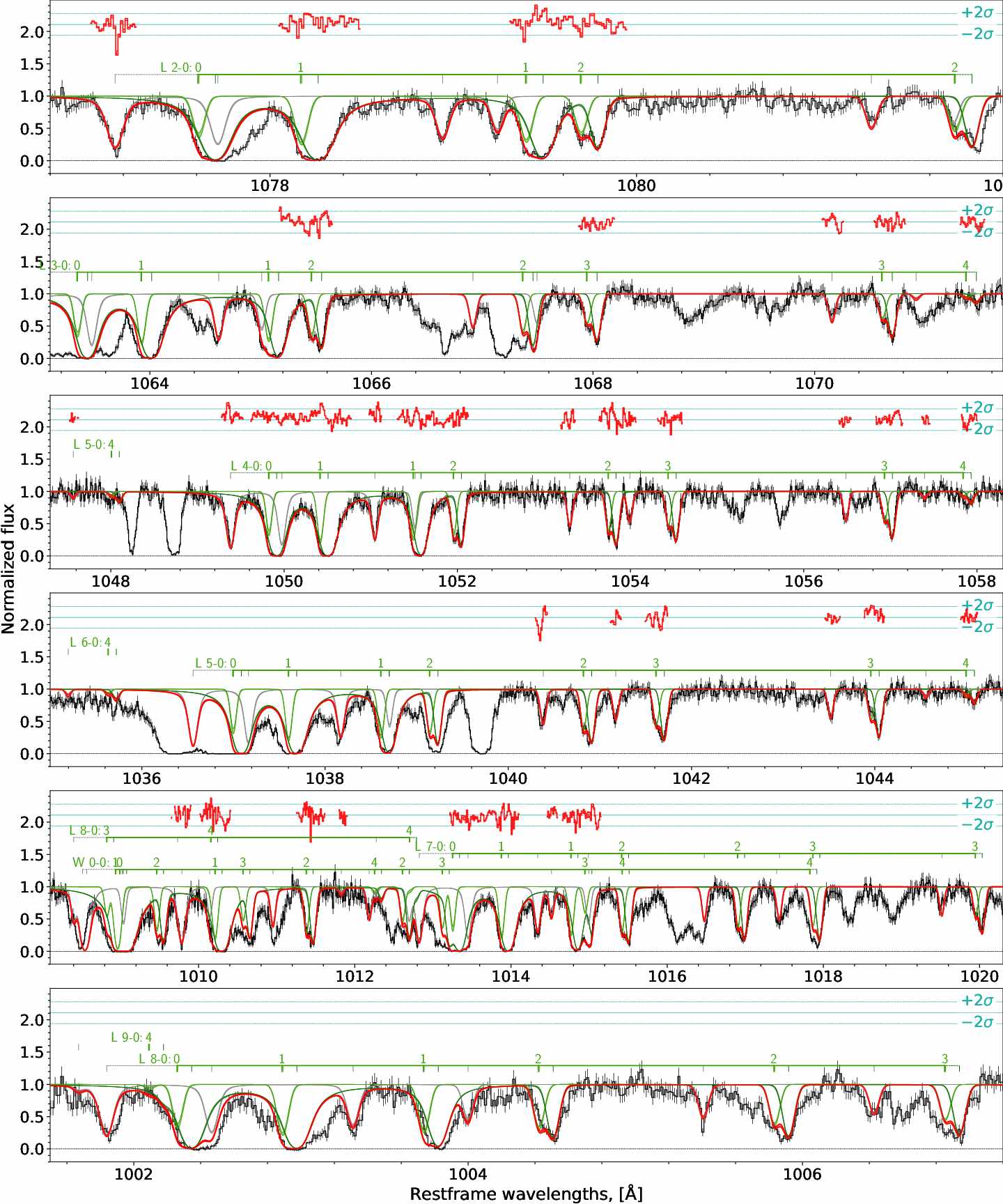}
    \caption{Fit to H2 absorption lines towards AV 506 in SMC. Lines are the same as for \ref{fig:lines_H2_Sk67_2}.
    }
    \label{fig:lines_H2_AV506}
\end{figure*}

\begin{table*}
    \caption{Fit results of H$_2$ lines towards Sk 191}
    \label{tab:Sk191}
    \begin{tabular}{cccc}
    \hline
    \hline
    species & comp & 1 & 2  \\
            & z & $0.0000494(^{+17}_{-14})$ &  $0.0005121(^{+14}_{-11})$\\
    \hline 
     ${\rm H_2\, J=0}$ & b\,km/s &$1.4^{+1.9}_{-0.9}$ & $0.81^{+0.74}_{-0.31}$\\
                       & $\log N$ &$19.11^{+0.18}_{-0.19}$ & $20.708^{+0.029}_{-0.030}$ \\
    ${\rm H_2\, J=1}$ & b\,km/s & $4.3^{+1.4}_{-1.4}$ & $1.2^{+1.1}_{-0.4}$\\
                      & $\log N$ & $18.09^{+0.20}_{-0.89}$ &$19.980^{+0.016}_{-0.021}$ \\
    ${\rm H_2\, J=2}$ & b\,km/s &$7.7^{+1.6}_{-1.7}$ & $2.4^{+0.6}_{-0.9}$\\
                      & $\log N$ & $15.50^{+0.37}_{-0.14}$ & $17.71^{+0.07}_{-0.15}$\\
    ${\rm H_2\, J=3}$ & b\,km/s & $12.8^{+0.7}_{-1.7}$ &$3.3^{+1.1}_{-0.8}$ \\
                      & $\log N$ &$15.009^{+0.039}_{-0.030}$ & $16.86^{+0.24}_{-1.07}$\\
    ${\rm H_2\, J=4}$ & b\,km/s & -- &   $4.5^{+2.0}_{-1.3}$ \\
    				  & $\log N$ & $13.84^{+0.23}_{-0.40}$ &$14.56^{+0.11}_{-0.09}$ \\
    ${\rm H_2\, J=5}$ & b\,km/s & -- &  $6.2^{+5.0}_{-2.0}$ \\
    				  & $\log N$ & $14.01^{+0.28}_{-0.66}$ & $14.52^{+0.11}_{-0.10}$\\
    \hline 
         & $\log N_{\rm tot}$ & $19.15^{+0.17}_{-0.18}$ & $20.78^{+0.02}_{-0.03}$  \\
    \hline
    HD J=0 & b\,km/s & $0.58^{+2.29}_{-0.08}$ &  $2.4^{+1.7}_{-0.9}$ \\
            & $\log N$ & $\lesssim 15.6$ & $14.71^{+1.08}_{-0.19}$ \\
    \hline   
    \end{tabular}
    \begin{tablenotes}
    \item Doppler parameters H$_2$ $\rm J=4$ in 1 and 4 components were tied to H$_2$ $\rm J=3$.
    \end{tablenotes}
\end{table*}

\begin{figure*}
    \centering
    \includegraphics[width=\linewidth]{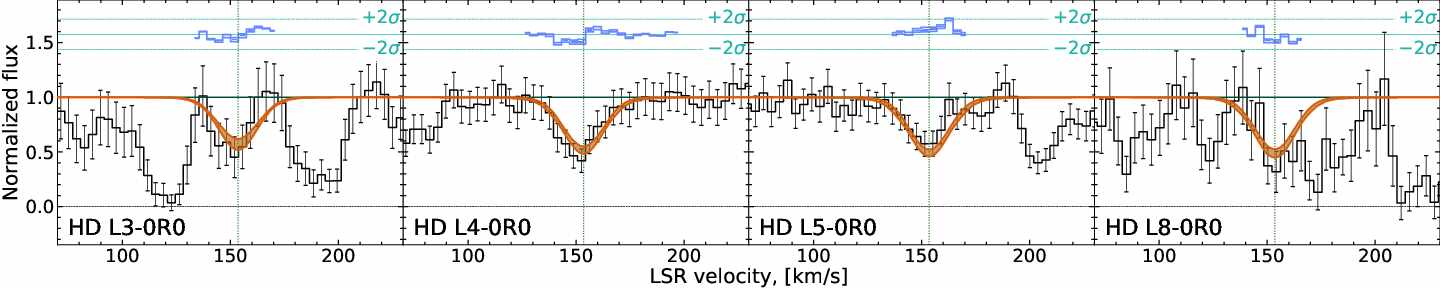}
    \caption{Fit to HD absorption lines towards Sk 191 in SMC. Lines are the same as for \ref{fig:lines_HD_Sk67_2}.
    }
    \label{fig:lines_HD_Sk191_appendix}
\end{figure*}

\begin{figure*}
    \centering
    \includegraphics[width=\linewidth]{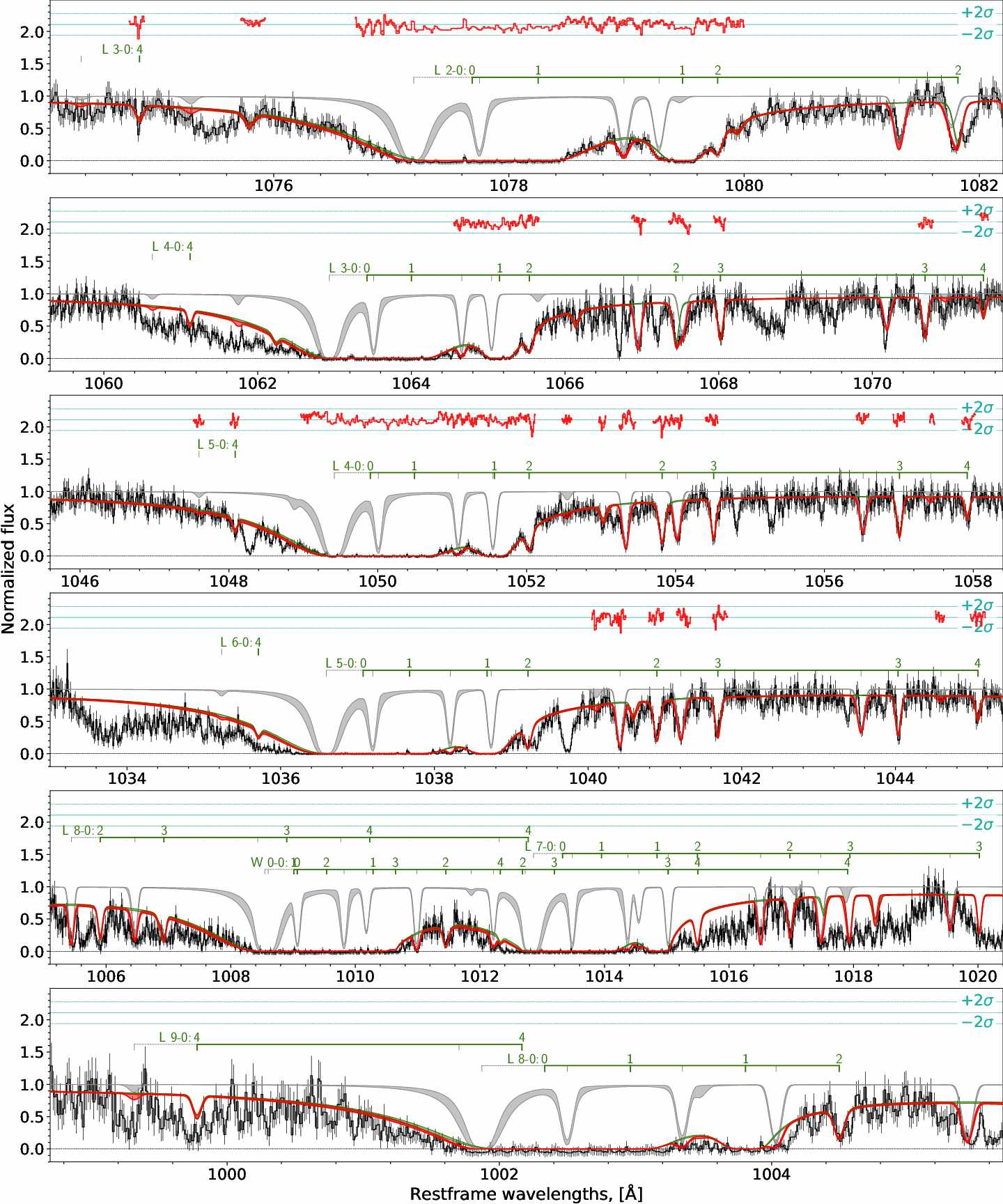}
    \caption{Fit to H2 absorption lines towards Sk 191 in SMC. Lines are the same as for \ref{fig:lines_H2_Sk67_2}.
    }
    \label{fig:lines_H2_Sk191}
\end{figure*}

\label{lastpage}
\end{document}